\documentclass[seceq]{ptptex}
\usepackage{wrapft}
\notypesetlogo                       
\usepackage{ifpdf}
\usepackage{psfrag} 
\usepackage{graphicx}
\usepackage{epstopdf}
\usepackage{color}
\usepackage{dcolumn}
\usepackage{mathrsfs}
\usepackage{amsfonts}
\usepackage{amssymb}
\usepackage{amsmath}
\usepackage{amsthm} 
\usepackage{dsfont}
\usepackage{bm}
\usepackage{bbm}
\usepackage{longtable}
\usepackage{fixmath}
\usepackage{upgreek}
\usepackage{bbding}
\usepackage{bold-extra}
\usepackage{lmodern} 
\usepackage[T1]{fontenc}
\usepackage{ae,aecompl}
\usepackage[bbgreekl]{mathbbol}
\usepackage{slashed}
\usepackage{pdfsync}
\usepackage[dvipsnames*,svgnames]{xcolor}
\usepackage[pdfauthor={Behnam Farid},%
           pdftitle={On the Luttinger-Ward functional and the convergence of skeleton diagrammatic series expansion of the self-energy for Hubbard-like models},%
           plainpages=false,pdfpagelabels,linktocpage,hypertexnames=false,%
           breaklinks=true,%
           hyperfootnotes=true,%
           bookmarks=true]{hyperref}
\hypersetup{backref=true,pageanchor=false,%
            colorlinks=true,%
            citecolor=DarkBlue,%
            filecolor=black,%
            linkcolor=DarkBlue,%
            urlcolor=blue,
            pdfstartview={FitH}}
\usepackage{breakcites}
\usepackage{cite}
\usepackage{perpage} 
\MakePerPage{footnote} 
\usepackage[perpage]{footmisc}
\renewcommand{\thefootnote}{\fnsymbol{footnote}}
\renewcommand{\thefootnote}{\ifcase\value{footnote}\or*\or \S \or \flat \or \natural \or \| \or \sharp\fi}
\usepackage{scrextend} 
\usepackage{accents}
\newlength{\dhatheight}

\def\t#1{\tilde{#1}}

\def\h#1{\hat{#1}}
\def\b#1{\bar{#1}}

\usepackage{xspace}
\makeatletter
\DeclareRobustCommand\onedot{\futurelet\@let@token\@onedot}
\def\@onedot{\ifx\@let@token.\else.\null\fi\xspace}
\def\ae{a.e\onedot}
\makeatother
\usepackage{scalerel}
\usepackage[usestackEOL]{stackengine}

\DeclareMathOperator{\re}{Re}
\DeclareMathOperator{\im}{Im}
\DeclareMathOperator{\e}{e}
\DeclareMathOperator{\rd}{d\!}
\DeclareMathOperator{\sgn}{sgn}

\DeclareMathOperator{\1BZ}{1BZ}
\DeclareMathOperator{\fbz}{\scriptscriptstyle{1BZ}}
\DeclareMathOperator{\ii}{\hspace{-0.2pt}\mathsf{i}\hspace{-1.0pt}}

\usepackage{stackengine,wasysym}

\newcommand{\X}[1]{
  \mathchoice
    {{\scriptstyle #1}}
    {{\scriptstyle #1}}
    {{\scriptscriptstyle #1}}
    {\scalebox{.7}{$\scriptscriptstyle #1$}} }
\newcommand\p[1]{\mathchoice{#1^\prime}{#1^\prime}{#1^\prime}%
  {#1^{\scalebox{.7}{$\scriptscriptstyle\prime$}}} } 
\makeatletter
\newcommand{\raisemath}[1]{\mathpalette{\raisem@th{#1}}}
\newcommand{\raisem@th}[3]{\raisebox{#1}{$#2#3$}}
\makeatother
\DeclareFontFamily{OT1}{pzc}{}
\DeclareFontShape{OT1}{pzc}{m}{it}{<-> s * [1.10] pzcmi7t}{}
\DeclareMathAlphabet{\mathpzc}{OT1}{pzc}{m}{it} 
\usepackage{euscript}
\DeclareMathAlphabet\EuScript{U}{eus}{m}{n}\SetMathAlphabet\EuScript{bold}{U}{eus}{b}{n}
\usepackage{eufrak}
\usepackage[utf8]{inputenc}
\DeclareUnicodeCharacter{2014}{\dash}
\DeclareRobustCommand\dash{%
  \unskip\nobreak\thinspace\textemdash\allowbreak\thinspace\ignorespaces}
\makeatletter
 \renewcommand*\l@subsection{\@dottedtocline{2}{2.0em}{2.6em}}
\makeatother
\newcounter{dummy}
\newcounter{dummyX}
\usepackage[usestackEOL]{stackengine}
\usepackage{titlesec}
\usepackage[british]{babel}
\usepackage{geometry}
\usepackage{setspace}
\usepackage{stmaryrd} 
\usepackage{alphabeta}
\newcommand{\pG}[1]{\text{#1}} 
\DeclareMathOperator{\ptau}{\pG{\tau}}
\title{
On the Luttinger-Ward functional and the convergence of skeleton diagrammatic series expansion of the self-energy for Hubbard-like models\hspace{0.6pt}\footnote{Dedicated to my brother Behzad.}
}

\author{
Behnam \textsc{Farid}\footnote{\href{mailto:behnam.farid@btinternet.com}{behnam.farid@btinternet.com}}%
}

\abst{
We consider a number of questions regarding the Luttinger-Ward functional and the many-body perturbation series expansion of the proper self-energy $\Sigma(\bm{k};z)$ specific to uniform ground states (ensemble of states) of interacting fermion systems in terms of skeleton self-energy diagrams and the interacting Green function $G(\bm{k};z)$. Utilising a link between the latter series expansion and the classical moment problem (of the Hamburger type), along with the associated continued-fraction expansion, we reaffirm our earlier observation (2007) that for lattice models of fermions interacting through short-range two-body potentials (\emph{i.e.} for Hubbard-like models) this series is uniformly convergent for almost all wave vectors $\bm{k}$ and complex energies $z$. The limit of this series is unique. We inquire into the reasons underlying the contrary observation by Kozik \emph{et al.} (2015) regarding skeleton-diagrammatic perturbation series expansion of $\Sigma$. In doing so, we make a number of observations of general interest. Chief amongst these, we observe that contrary to general belief thermal correlation functions calculated in the energy domain (\emph{i.e.} over the set of the relevant Matsubara frequencies) are generally unreliable, in contrast to those calculated in the imaginary-time domain. In an appendix we present a symbolic computational formalism and short programs for singling out topologically distinct proper self-energy diagrams that are algebraically equivalent up to determinate multiplicative constants. This complements the work presented in our previous publication (2019). {\scriptsize [Abridged abstract]}
}

\begin{document}

\maketitle

\renewcommand{\thefootnote}{\textrm{\alph{footnote}}}
{\scriptsize{\tableofcontents}}

\newgeometry{width=13cm, left=3.4cm}
\begin{spacing}{0.75}
\section*{Abstract}{\fontsize{10pt}{12pt}\selectfont
We consider a number of questions regarding the Luttinger-Ward functional and the many-body perturbation series expansions of the proper self-energy $\Sigma(\bm{k};z)$ specific to uniform ground states (ensemble of states) of interacting fermion systems, specifically the one in terms of \textsl{skeleton} self-energy diagrams and the \textsl{interacting} one-particle Green function $G(\bm{k};z)$. Utilising a link between the latter perturbation series expansion and the classical moment problem (of the Hamburger type), along with the associated continued-fraction expansion, we reaffirm our earlier observation (2007) that for the uniform ground states of lattice models (in spatial dimensions $d <\infty$) in which particles further interact through short-range two-body potentials (\emph{i.e.} for `Hubbard-like' models) this perturbation series is {uniformly convergent} to a unique limit for almost all $\bm{k}$ and $z$. In spite of this, the possibility cannot be \emph{a priori} ruled out that in practice self-consistent calculations of $\Sigma(\bm{k};z)$ in terms of skeleton self-energy diagrams may result in multiple self-energies, depending on the adopted self-consistency scheme and the choice of the trial function for $G$. The multiplicity of self-consistent solutions may possibly be entirely avoided through limiting the solution space to that of functions possessing some of the crucial analytic properties of the exact $\Sigma(\bm{k};z)$. The exact $-G(\bm{k};z)$ and $-\Sigma(\bm{k};z)$ being Nevanlinna functions of $z$, $z\in\mathds{C}$, we propose a general approach based on the Riesz-Herglotz representation of these functions, ensuring that $-G(\bm{k};z)$ and $-\Sigma(\bm{k};z)$ are Nevanlinna at all stages of self-consistent calculations. We inquire into the reasons underlying the contrary observation by Kozik \emph{et al.} (2015) regarding skeleton-diagrammatic perturbation series expansion of $\Sigma$. In doing so we make a number of observations, of which some are specific to the latter work and others general, with wider implications. Our observations include: \textbf{(i)} the perturbation series expansion of $\Sigma$ in terms of skeleton self-energy diagrams and $G$ is \textsl{artificial} for the cases where the perturbation series expansion of $\Sigma$ in terms of the non-interacting Green function $G_{\X{0}}$ and \textsl{proper} (consisting of skeleton and non-skeleton) self-energy diagrams is terminating; this is notably the case for the `Hubbard atom' dealt with by Kozik \emph{et al.}; \textbf{(ii)} finite-order perturbation series expansions of $G$ in terms of $G_{\X{0}}$ are pathological in the case of the `Hubbard atom', the $G$ thus obtained possessing prohibited zeros in the complex $z$-plane; \textbf{(iii)} the diagrammatic expansion of $\Sigma(\bm{k};z)$ in terms of skeleton self-energy diagrams and $G$ is problematical in the local / atomic limit (including the limit of infinite coordination number $\EuScript{Z}\equiv 2 d$ on a $d$-cubic lattice) in that for an arbitrary order $\nu$ of the perturbation expansion $\Sigma^{\X{(\nu)}}(z)/\lambda^{\nu}$ can be singular for $\lambda\downarrow 0$, where $\lambda$ stands for the coupling constant of the two-body interaction potential; \textbf{(iv)}, related to \textsl{(iii)}, the negative of the \textsl{local} self-energy $\Sigma(z)$ (in contrast to the negative of the corresponding \textsl{local} Green function $G(z)$) in the framework of the self-consistent dynamical mean-field theory on a Bethe lattice in $d=\infty$, specifically the one associated with a semi-circular density of states, fails in general to be a Nevanlinna function of $z$; \textbf{(v)} at a fundamental level, thermal many-body correlation functions calculated in the energy domain (\emph{i.e.} over the set of the relevant Matsubara frequencies) are generally \textsl{unreliable}, in contrast to those calculated in the imaginary-time domain; we explicitly demonstrate that contrary to general belief, correct calculation of the self-energy (and by extension, of any many-body correlation function) at fermionic Matsubara frequencies $\{\omega_m\| m\in\mathds{Z}\}$ requires more information regarding, say, the polarisation function $P$ than its values at bosonic Matsubara frequencies $\{\nu_m\|m\in\mathds{Z}\}$; information regarding $P$ in the infinitesimal neighbourhoods of the frequencies $\{\nu_m\|m\in\mathds{Z}\}$ proves vital for a correct calculation. In an appendix we present a symbolic computational formalism and short programs for singling out topologically distinct proper self-energy diagrams that are algebraically equivalent up to determinate multiplicative constants. This complements the work presented in our earlier publication (2019).
}
\end{spacing}
\restoregeometry

\refstepcounter{dummyX}
\section{Introduction}
\phantomsection
\label{sec1}
This publication\,\refstepcounter{dummy}\label{WorkOnThe}\footnote{Work on the present publication (cited as reference 28 in Ref.\,\protect\citen{BF19}) began in late July 2014, shortly after the publication of Ref.\,\protect\citen{KFG14} on \protect\href{http://arxiv.org/abs/1407.5687}{\textsf{arXiv}} (21 July 2014). It was nearly completed by early May 2015, however due to unforeseen circumstances its publication has been delayed until now. We have maintained the structure of the 2015 version of the present publication, and thus have made \textsl{no} effort (barring some minor textual alterations, including an updating of the references) to bring its form into full conformity with the contents of our recent publication, Ref.\,\protect\citen{BF19}. \label{notes}} is the outgrowth of what began as a Comment on the publication by Kozik \emph{et al.} \cite{KFG14} in which the authors have made several remarkable observations regarding the Luttinger-Ward functional $\Upphi[G]$ and the expansion of the proper self-energy \cite{FW03,BF19}\footnote{Also known as \textsl{one-particle irreducible} (1PI) self-energy diagrams.} $\Sigma$ in terms of skeleton self-energy diagrams \cite{LW60}\footnote{Also known as \textsl{two-particle irreducible} (2PI) self-energy diagrams.} and the exact one-particle Green function $G$. As the scope of our investigations expanded, we decided on the present format in preference to a Comment whose focus would have to be narrower than is the case with the present publication. Nonetheless, our presentation retains some characteristics of a Comment, its focus remaining the observations by Kozik \emph{et al.} \cite{KFG14}.

The work by Kozik \emph{et al.} \cite{KFG14} has prompted one Comment, by Eder \cite{RE14}, and two complementary publications, by Stan \emph{et al.} \cite{SRRRB15}, and Rossi and Werner \cite{RW15}. There may be further similar publications that have escaped our attention. Very briefly, we believe that the points as raised by Stan \emph{et al.} \cite{SRRRB15}, and Rossi and Werner \cite{RW15} are \textsl{not} directly relevant to the \textsl{main} observation by Kozik \emph{et al.} \cite{KFG14}, namely that the self-energy ``$\Sigma[G]$'' ``is \emph{not} a single-valued functional of $G$'', which we in this publication explicitly demonstrate to be a \textsl{false} assertion. In contrast, Eder's \cite{RE14} identification of a problem with the self-energy corresponding to $U=4$ as presented in Fig.\,3 of Ref.\,\citen{KFG14} points to an actual deep problem, fully clarified in the present publication. We should point out however that Stan \emph{et al.} \cite{SRRRB15} appropriately emphasise the importance of the `domain of definition' of functions to the problem at hand, and Rossi and Werner \cite{RW15} explicitly declare the phenomena observed by them, corresponding to their adopted zero space-time model (which in essence is identical to the `one-point model', OPM, considered by Stan \emph{et al.} \cite{SRRRB15}), as not being necessarily due to a similar mechanism as that underlying the observed phenomena by Kozik \emph{et al.} \cite{KFG14}.

\refstepcounter{dummyX}
\section{The Luttinger-Ward functional and the skeleton diagrammatic series expansion of the self-energy}
\label{sec2}
\refstepcounter{dummyX}
\subsection{Background and generalities}
\phantomsection
\label{sec2.a}
Kozik \emph{et al.} \cite{KFG14} have observed that the self-energy $\Sigma$, as expressed in terms of \textsl{skeleton} self-energy diagrams \cite{LW60} and the interacting one-particle Green function $G$, is \textsl{not} a single-valued functional in the case of Hubbard-like models; that for these models this expansion results in at least one non-physical self-energy in addition to the physical one. This observation is open to immediate criticism on the grounds that uniqueness, or multiplicity, of solutions of mathematical equations is not solely a property of these equations on their own, but also of the (function) spaces in which they are defined and are to be solved. For instance, the simple algebraic equation $x^2 = a$ has \textsl{two} solutions $x = \pm\sqrt{a}$ over $[-\infty,0) \cup (0,\infty]$ for $a \in (0,\infty]$, \textsl{one} unique solution in either of the two intervals $[-\infty,0)$ and $(0,\infty]$ for $a \in (0,\infty]$, and \textsl{no} solution over any of the three intervals $[-\infty,0) \cup (0,\infty]$, $[-\infty,0)$, and $(0,\infty]$ for $a \in [-\infty,0]$. This basic consideration applies in general, where defining a mapping, and its inverse (notably, such mappings as $G_{\X{0}} \mapsto G$ and $G \mapsto G_{\X{0}}$, where $G_{\X{0}}$ denotes the non-interacting one-particle Green function, that have been considered in Ref.\,\citen{KFG14}), is demanding of the specification of its \textsl{domain} and \textsl{range} \cite{AA14}.\footnote{As we have indicated above, Stan \emph{et al.} \protect\cite{SRRRB15} emphasise the importance of the `\textsl{domain} of definition' of functions to the problem at hand. They do no however refer to the significance of the \textsl{range} of functions.} Nowhere in Ref.\,\citen{KFG14} have the authors specified the function spaces in which for instance the one-particle Green functions $G_{\X{0}}$ and $G$ are to be varied and the self-energy $\Sigma$ is to be sought; only after solving the relevant equations have they qualified one of the calculated solutions as `physical' and the other as `unphysical'. Clearly, the non-uniqueness problem as observed in Ref.\,\citen{KFG14} (and by Stan \emph{et al.} \cite{SRRRB15}, and Rossi and Werner \cite{RW15}) would not have arisen, had the authors in advance excluded the space of the functions that they in retrospect have identified as `unphysical'.

For\refstepcounter{dummy}\label{ForIllustration} illustration, let us consider the `Hubbard atom',\footnote{Short for \textsl{the `Hubbard atom' of spin-$\tfrac{1}{2}$ particles at half-filling}.} which has been considered by Kozik \emph{et al.} \cite{KFG14} and which we shall discuss in some detail later in this publication, \S\,\ref{sec3}. For this model, the non-interacting Green function $G_{\X{0}}(z)$ can be expressed in terms of the interacting Green function $G(z)$ as
\begin{equation}\label{e1a}
G_{\X{0}}(z) = \mathcal{G}_{\mp}(G(z))\;\;\text{for}\;\; \vert z\vert \lessgtr U/2,
\end{equation}
where (\emph{cf.} Eqs\,(\ref{e25}) and (\ref{e15a}) below)
\begin{equation}\label{e1b}
\mathcal{G}_{\mp}(\zeta) \doteq \frac{2\hbar^2}{U^2 \zeta} \big(\hspace{-1.2pt}\mp\sqrt{1+ U^2 \zeta^2/\hbar^2} - 1\big).
\end{equation}
The result in Eq.\,(\ref{e1a}) is obtained through substituting the exact expression (Eqs\,(\ref{e25}) and (\ref{e15a}) below)
\begin{equation}\label{e1e}
\Sigma(z) = \frac{U^2}{4\hbar^2}\hspace{1.0pt} G_{\X{0}}(z)
\end{equation}
for the exact proper self-energy $\Sigma(z)$ in the Dyson equation \cite{FW03} and subsequently solving the resulting quadratic equation for $G_{\X{0}}(z)$, leading to the above two distinct solutions $\mathcal{G}_{\mp}(G(z))$ for $U \not= 0$. Although the suppression of the conditions $\vert z\vert \lessgtr U/2$ on the right-hand side (RHS) of Eq.\,(\ref{e1a}) would suggest that the mapping $G \mapsto G_{\X{0}}$ were two-valued, neither of the two functions \emph{a priori} qualifies as a one-particle Green function; whereas the latter function must be analytic everywhere on the complex $z$-plane away from the real axis of this plane \cite{JML61,BF07},\footnote{For the detrimental consequences of false analytic properties of the Green function and the self-energy for the Luttinger theorem, we refer the interested to \S\,6.2 [\emph{Case II}] of Ref.\,\protect\citen{BF07}.} both functions $\mathcal{G}_{\mp}(G(z))$ are non-analytic on the circle $\vert z\vert = U/2$.\footnote{In \S\,\protect\ref{s4xb} we shall encounter a precursor of the same circle. See in particular Eqs\,(\ref{e41a}) and (\protect\ref{e41b}) and the related discussions.} Although using the conditions $\vert z\vert \gtrless U/2$ on the RHS of Eq.\,(\ref{e1a}), instead of the present $\vert z\vert \lessgtr U/2$, naturally also results in an analytic function, this function does \textsl{not} satisfy the significant requirement of decaying to leading-order like $\hbar/z$ in the asymptotic region \cite{ETC65,RBD73} $z\to\infty$ of the complex $z$-plane,\footnote{Unless we indicate otherwise, throughout this publication the condition $\protect\im[z] \not=0$, or equivalently $0 < \vert\hspace{-1.0pt}\arg(z)\vert < \pi$, is implicit in the notation $z\to \infty$, signifying the approach of $z$ towards the point of infinity of the complex $z$-plane from outside the real axis of this plane.} as expected of a one-particle Green function, whether interacting or non-interacting \cite{BF02,BF07}. One verifies that, for the exact Green function $G(z)$ in Eq.\,(\ref{e25}) below,
\begin{equation}\label{e1c}
\Theta(U/2 - \vert z\vert)\hspace{0.8pt}\mathcal{G}_{-}(G(z)) + \Theta(\vert z\vert -U/2)\hspace{0.8pt} \mathcal{G}_{+}(G(z)) \equiv \frac{\hbar}{z},
\end{equation}
and
\begin{equation}\label{e1d}
\Theta(\vert z\vert -U/2)\hspace{0.8pt} \mathcal{G}_{-}(G(z)) + \Theta(U/2-\vert z\vert)\hspace{0.8pt} \mathcal{G}_{+}(G(z)) \equiv -\frac{4\hbar}{U^2}\hspace{0.6pt} z.
\end{equation}
The function on the RHS of Eq.\,(\ref{e1c}) indeed coincides with the exact non-interacting Green function $G_{\X{0}}(z)$ of the `Hubbard atom' under consideration, Eq.\,(\ref{e15a}) below. We thus conclude that at least in the case of the `Hubbard atom' the mapping $G \mapsto G_{\X{0}}$ to the space of functions that (i) are analytic outside the region $\im[z]=0$ of the complex $z$-plane, and (ii) decay to leading order like $\hbar/z$ for $z \to\infty$, \textsl{is} indeed single-valued. We note in passing that the latter leading-order asymptotic behaviour for $z\to\infty$ plays a vital role in rendering the analytic continuation of the thermal one-particle Green function, from the discrete set of the Matsubara frequencies to the entire complex $\omega$-plane, \textsl{unique} \cite{BM61}. This despite the latter discrete set not containing an accumulation point on the complex $\omega$-plane.\footnote{See \S\,4.12, p.\,139, in Ref.\,\protect\citen{ECT52} and Theorem 17.1, p.\,369, in Ref.\,\protect\citen{AIM65}.}

We should emphasize that the above approach of constructing the mapping $G \mapsto G_{\X{0}}$, based crucially on the equality in Eq.\,(\ref{e1e}), is \textsl{not} the one adopted by Kozik \emph{et al.} \cite{KFG14}. Nonetheless, our approach transparently illustrates the way in which one can possibly arrive at an apparently double-valued -- and more generally, multi-valued, inverse of the mapping $G_{\X{0}}\mapsto G$. The adopted approach here is however consistent with that underlying the iterative calculation of the interacting Green function $y$ and that of the non-interacting Green function $y_{\X{0}}$ within the Hartree-Fock approximation of the self-energy $\t{s}$ in the case of the one-point model, OPM, of Stan \emph{et al.} \cite{SRRRB15}. With $u$ denoting the interaction potential, in this model one has \cite{SRRRB15} $\t{s} = -\frac{1}{2}u y_{\X{0}}$ and $\t{s}^{\textsc{hf}} = -\frac{1}{2} u y$. Similarly as regards the treatment by Rossi and Werner \cite{RW15}, relying on the expression $\Sigma = U G_{\X{0}}$, deduced from the equalities in Eqs\,(6) and (7) of Ref.\,\citen{RW15}. The models of Refs\;\citen{SRRRB15} and \citen{RW15} being zero space-time ones, characterisation of specific $y_{\X{0}}$ and $G_{\X{0}}$, respectively, as being \emph{a priori} inadmissible clearly requires use of different criteria than those that we have invoked above in considering the function $G_{\X{0}}(z)$ corresponding to the `Hubbard atom'.\footnote{For instance, Stan \emph{et al.} \protect\cite{SRRRB15} use the requirement of the adiabatic connection \protect\cite{FW03} between the non-interacting and interacting Green functions, respectively $y_{\protect\X{0}}$ and $y$.}

Disregarding the above aspect concerning the definitions of the \textsl{domain} and the \textsl{range} of mappings, the non-uniqueness problem as observed by Kozik \emph{et al.} \cite{KFG14} is in its generality contrary to our earlier observation in Ref.\,\citen{BF07} (see in particular \S\hspace{0.6pt}5.3.2 herein), according to which for the given \textsl{exact} Green function, corresponding to the $N$-particle uniform ground state (GS) of a Hubbard-like model, the perturbation series for the self-energy in terms of skeleton self-energy diagrams and the exact Green function has a \textsl{unique} limit.\footnote{Although this statement may appear tautological at first glance, it is not, as will become evident in discussing the phenomenon of `oscillation' later in this publication (see in particular the paragraph following Eq.\,(\protect\ref{e4a}) below).} Here is an abridged presentation of a more general observation made in Ref.\,\citen{BF07} for whose description we need first to recapitulate the notational conventions of Ref.\,\citen{BF07}, which we shall do below (\S\,\ref{s2.2}):\,\footnote{With reference to appendix \protect\ref{sacx}, for spin-$\mathsf{s}$ particles one deals with an $(2\mathsf{s}+1)\times (2\mathsf{s}+1)$ Green matrix $\protect\t{\mathbb{G}}(\bm{k};z)$ with elements $\protect\t{G}_{\sigma,\sigma'}(\bm{k};z) \equiv (\protect\t{\mathbb{G}}(\bm{k};z))_{\sigma,\sigma'}$. The Green function $\protect\t{G}_{\sigma}(\bm{k};z)$ referred to here stands for the diagonal element $\protect\t{G}_{\sigma,\sigma}(\bm{k};z)$ of $\protect\t{\mathbb{G}}(\bm{k};z)$. Similarly, $\protect\t{\Sigma}_{\sigma}(\bm{k};z)$ stands for a diagonal element of the $(2\mathsf{s}+1)\times (2\mathsf{s}+1)$ matrix $\protect\t{\mathbb{\Sigma}}(\bm{k};z)$. The considerations in this publication, especially those in \S\,\protect\ref{sec2.b}, are mostly (but not entirely) centred on the assumption that $\protect\t{\mathbb{G}}$ and $\protect\t{\mathbb{\Sigma}}$ are diagonal (see \S\,2.2.2 in Ref.\,\protect\citen{BF19}). For these assumptions to be mutually compatible, the two-body interaction potentials considered must belong to categories (2) and (3), specified in appendix \protect\ref{sacx}, p.\,\protect\pageref{ThreeCases}. Some specific two-body potentials from category (1) are also admissible, as we explain in footnote \raisebox{-1.0ex}{\normalsize{\protect\footref{notes1}}} on p.\,\protect\pageref{ConsideringSpin}.} for $\t{G}_{\sigma}(\bm{k};z)$ and $\t{\Sigma}_{\sigma}(\bm{k};z)$ denoting respectively the interacting one-particle Green function and the self-energy corresponding to the $N$-particle uniform GS of a Hubbard-like Hamiltonian,\footnote{Here $\sigma$ denotes the spin index, $\bm{k}$ the wave vector, and $z$ the complex energy.} in Ref.\,\citen{BF07} we have shown that the perturbation series expansion of $\t{\Sigma}_{\sigma}(\bm{k};z)$ in terms of skeleton self-energy diagrams and $\{\t{G}_{\sigma}(\bm{k};z)\| \sigma\}$ is \textsl{uniformly convergent} to a \textsl{unique} limit for almost all $\bm{k}$ and $z$. The significance of the `almost' in this statement will be clarified below.

\refstepcounter{dummyX}
\subsection{Notational conventions and some essential properties}
\phantomsection
\label{s2.2}
In the notation that we employ throughout the remaining part of this publication,\footnote{Barring \S\,\protect\ref{sec3}, where, because of the use of additional indices, we opt for a lighter notation by suppressing the spin index $\sigma$.} the spin index $\sigma$ of particles is made explicit, whereby the zero-temperature self-energy \textsl{operator} (in the one-particle Hilbert space of the problem at hand) is denoted by $\h{\Sigma}_{\sigma}$. The more extensive notation $\h{\Sigma}_{\sigma}[\{\h{G}_{\sigma'}\}]$ emphasizes the functional dependence of $\h{\Sigma}_{\sigma}$ on the interacting one-particle Green operators $\{\h{G}_{\sigma} \| \sigma\}$ [\S\,III.A.2 in Ref.\,\citen{BF13}].\footnote{According to the notation adopted in Ref.\,\protect\citen{BF19}, the self-energy operator considered here is $\protect\h{\Sigma}_{\protect\X{01}}$.}\refstepcounter{dummy}\label{ItProvesAlso}\footnote{It proves also convenient to denote $\h{X}_{\sigma}$, where $\h{X}_{\sigma} = \protect\h{\Sigma}_{\sigma}, \protect\h{G}_{\sigma}$, more extensively as $\protect\h{X}_{\sigma}(z)$. Hereby, for the \textsl{energy}-momentum representation of $\protect\h{X}_{\sigma}$ it suffices to write $\langle\bm{k}\vert \protect\h{X}_{\sigma}(z)\vert\bm{k}'\rangle \equiv \t{X}_{\sigma}(\bm{k},\bm{k}';z)$, where $\vert\bm{k}\rangle$ and $\vert\bm{k}'\rangle$ stand for the eigenstates of the single-particle momentum operator $\protect\h{\bm{p}} \equiv \hbar \protect\h{\bm{k}}$, corresponding to eigenvalues $\hbar\bm{k}$ and $\hbar\bm{k}'$, respectively, and normalised according to $\langle\bm{k}\vert \bm{k}'\rangle = \delta_{\bm{k},\bm{k}'}$. For uniform GSs (ensemble of states, ESs) one has $\t{X}_{\sigma}(\bm{k},\bm{k}';z) = \t{X}_{\sigma}(\bm{k};z)\hspace{0.6pt} \delta_{\bm{k},\bm{k}'}$, where $\t{X}_{\sigma}(\bm{k};z) \doteq \t{X}_{\sigma}(\bm{k},\bm{k};z)$. We shall rely on similar conventions when dealing with other correlation functions encountered in this publication, such as the various screening functions in appendix \protect\ref{sa}. We should point out that while attaching $\sigma$ to $\protect\h{X}$ is merely a matter of convenience, attaching $(z)$ to $\protect\h{X}_{\sigma}$ in contrast amounts to an \emph{ad hoc} measure for bypassing the need to deal with a \textsl{time operator} (on the same footing as the above $\protect\h{\bm{p}}$ and $\protect\h{\bm{k}}$). For some relevant details, we refer the reader to \S\,2.5 of Ref.\,\protect\citen{BF19} where we explicitly deal with such functions as $G(1,2)$ and $\Sigma(1,2)$, where $i = \bm{r}_it_i\sigma_i$: writing, \emph{e.g.}, $G(1,2)$ as $\langle 1\vert\protect\h{G}\vert 2\rangle$, one observes that for expressing $\vert i\rangle$ as a product state, the variables $\bm{r}_i$ and $\sigma_i$ pose no problem, however treating $t_i$ on the same footing as $\bm{r}_i$ and $\sigma_i$ is not straightforward. One also observes why attaching $\sigma$ to $\h{G}$ is indeed merely a matter of convenience. \label{notea1}}

Assuming\refstepcounter{dummy}\label{AssumingUniformGSs} \textsl{uniform} GSs, in the (complex) energy-momentum representation one has $\t{\Sigma}_{\sigma}(\bm{k};z)$ and $\t{G}_{\sigma}(\bm{k};z)$ for respectively $\h{\Sigma}_{\sigma}$ and $\h{G}_{\sigma}$, where for $d$-dimensional systems $\bm{k} \in \mathds{R}^d$ or $\bm{k} \in \1BZ \subset \mathds{R}^d$ ($\1BZ$, the first Brillouin zone \cite{AM76}), depending on whether the system under investigation is defined over respectively the continuum $\mathds{R}^d$ \textsl{or} on a Bravais lattice \cite{AM76}  $\{\bm{R}_i \| i =1,2,\dots, N_{\textsc{s}}\}$ embedded in $\mathds{R}^d$.\footnote{The lattice $\{\bm{R}_i \| i\}$ need not be a Bravais one; our restriction to Bravais lattices is motivated by the desire to avoid the technical complexities that are not necessary to our discussions.} Here the tilde over the symbols $\Sigma_{\sigma}$ and $G_{\sigma}$ marks the distinction between $\t{\Sigma}_{\sigma}(\bm{k};z)$ and $\t{G}_{\sigma}(\bm{k};z)$, where $z\in \mathds{C}$, and their `physical' counterparts, namely $\Sigma_{\sigma}(\bm{k};\varepsilon)$ and $G_{\sigma}(\bm{k};\varepsilon)$, where $\varepsilon \in \mathds{R}$. These functions are related according to
\begin{equation}\label{e1}
X_{\sigma}(\bm{k};\varepsilon) = \lim_{\eta\downarrow 0} \t{X}_{\sigma}(\bm{k};\varepsilon \pm \ii\eta)\;\; \text{for}\;\; \varepsilon \gtrless \mu,
\end{equation}
where $X_{\sigma} = \Sigma_{\sigma}, G_{\sigma}$, $\t{X}_{\sigma} = \t{\Sigma}_{\sigma}, \t{G}_{\sigma}$, and $\mu$ is the chemical potential specific to the $N$-particle uniform GS of the system under investigation.\footnote{In dealing with finite-temperature correlation functions corresponding to equilibrium grand-canonical ensemble of states, to be encountered in \S\S\,\protect\ref{sec3.c}, \protect\ref{sec3.d}, and \protect\ref{s4xb}, $\mu$ corresponds to the equality of the ensemble average $\protect\b{N}$ of the number of particles with $N$. In other words, $\mu$ is the solution of the equation of state $\protect\b{N} = N$.} The relationship in Eq.\,(\ref{e1}) is relevant in the light of the fact that the functions $\t{\Sigma}_{\sigma}(\bm{k};z)$ and $\t{G}_{\sigma}(\bm{k};z)$ can be discontinuous for $z$ crossing an arbitrary point $\varepsilon$ of the real axis of the $z$-plane \cite{JML61,BF07}. Although use of such notation is not strictly necessary in the cases where the function $\t{X}_{\sigma}(\bm{k};z)$ is described in terms of isolated singularities on the real axis of the $z$-plane (as in the case of the `Hubbard atom'), nonetheless for the most general case such use leads to a more concise notation than $X_{\sigma}(\bm{k};\varepsilon \pm \ii 0^{+})$.

Since [Eqs\,(B.53), (B.55), (B.59) in Ref.\,\citen{BF07}] [Eq.\,(50) in Ref.\,\citen{JML61}]
\begin{equation}\label{e2}
\sgn(\im[\t{X}_{\sigma}(\bm{k};z)]) = - \sgn(\im[z]),\;\; \im[z]\not= 0,
\end{equation}
from the expression in Eq.\,(\ref{e1}), it follows that in particular [Eq.\,(2.16) in Ref.\,\citen{BF07}]
\begin{equation}\label{e3}
\im[\Sigma_{\sigma}(\bm{k};\varepsilon)] \lessgtr 0 \;\; \text{for}\;\; \varepsilon \gtrless \mu,\; \forall \bm{k}.
\end{equation}
These conditions reflect the assumed stability of the GS under consideration, a fact that can be explicitly demonstrated on the basis of the Lehmann representation of $\t{G}_{\sigma}(\bm{k};z)$ and the convexity of the GS energy as a function of the number of particles, leading to the Jensen inequality regarding the $M$-particle GS energy $E_{M;0}$ for\,\footnote{As we have discussed in \S\,B.1.1 of Ref.\,\protect\citen{BF07}, working in the canonical ensemble, the problem of $M$ taking solely integer values can be bypassed by considering the number density $n \doteq N/\Omega$, where $\Omega$ denotes the volume of the system; in the thermodynamic limit the number density $n$ transforms into a continuous variable. Working in the grand-canonical ensemble, one deals with the ensemble average particle number $\protect\b{N}$, which can take any prescribed non-negative value, real and integer, by an appropriate choice for the value of the chemical potential $\mu$.} $N-1 \le M \le N+1$ [App.\,B in Ref.\,\citen{BF07}] \cite{BF13}.

Both\refstepcounter{dummy}\label{BothSigmaAndG} $\t{\Sigma}_{\sigma}(\bm{k};z)$ and $\t{G}_{\sigma}(\bm{k};z)$ are \textsl{analytic} (or \textsl{holomorphic} \cite{ECT52}) over the entire $z$-plane outside the real axis \cite{JML61,BF07}. This and the property in Eq.\,(\ref{e2}) render $-\t{\Sigma}_{\sigma}(\bm{k};z)$ and $-\t{G}_{\sigma}(\bm{k};z)$ Nevanlinna functions [App. C in Ref.\,\citen{HMN72}] [Ch.\,3 in Ref.\,\citen{NIA65}], also \textsl{incorrectly} known as Herglotz functions.\footnote{In a recent publication, Fei \emph{et al.} \protect\cite{FYG20} introduce Nevanlinna analytic continuation of the retarded one-particle Green function.} The minus signs are in accordance with the prevailing convention in the literature regarding Nevanlinna functions, which are \textsl{endofunctions}.\footnote{Mapping each point of the upper/lower part of the $z$-plane into the same half-plane. Less accurately, one may also refer to these functions as \textsl{automorphisms} [Definition 2.3.3, p.\,79, in Ref.\,\protect\citen{AA14}]. Note that (non-zero) complex numbers form an Abelian \textsl{group} under addition (multiplication).} We note in passing that the equality in Eq.\,(\ref{e2}) is in conformity with the reflection property [\S\,4.5, p.\,155, in Ref.\,\citen{ECT52}]
\begin{equation}\label{e3b}
\t{X}_{\sigma}(\bm{k};z^*) = \t{X}_{\sigma}^*(\bm{k};z),\;\; \im[z] \not= 0,
\end{equation}
expected of the exact $\t{X}_{\sigma}(\bm{k};z)$ (see, e.g., Eq.\,(B.12) in Ref.\,\citen{BF07}).\refstepcounter{dummy}\label{ThisReflection}\footnote{The reflection principle as applied to the problem at hand requires the function $\protect\t{X}_{\sigma}(\bm{k};z)$ to be real over \textsl{some} open interval of the real axis of the $z$-plane [\S\,4.5, p.\,155, in Ref.\,\protect\citen{ECT52}]. This is the case on account of $\protect\im[\protect\t{X}_{\sigma}(\bm{k};\mu)] \equiv 0$, for all $\mu \in (\mu_{N;\sigma}^-,\mu_{N;\sigma}^+)$ \protect\cite{BF07,BF13}, where $\mu_{N;\sigma}^+ - \mu_{N;\sigma}^- > 0$ (strictly positive); even for metallic $N$-particle GSs, $\mu_{N;\sigma}^+ - \mu_{N;\sigma}^- = O(1/N)$. \label{noteh}} Similar reflection property applies to the inverse, or reciprocal, of $\t{X}_{\sigma}(\bm{k};z)$. It should however be noted that
\begin{equation}\label{e3a}
\sgn(\im[1/\t{X}_{\sigma}(\bm{k};z)]) = -\sgn(\im[\t{X}_{\sigma}(\bm{k};z)]),\;\; \im[z] \not=0.
\end{equation}
Thus, for $-\t{X}_{\sigma}(\bm{k};z)$ a Nevanlinna function of $z$, $1/\t{X}_{\sigma}(\bm{k};z)$ is a Nevanlinna function of $z$, and \emph{vice versa}.\refstepcounter{dummy}\label{ThisObservation}\footnote{This observation is of interest here since following the Dyson equation, Eq.\,(\protect\ref{e4a}) below, one has $\protect\t{\Sigma}_{\sigma}(\bm{k};z) = 1/\protect\t{G}_{\protect\X{0};\sigma}(\bm{k};z) - 1/\protect\t{G}_{\sigma}(\bm{k};z)$, where both $-\protect\t{G}_{\protect\X{0};\sigma}(\bm{k};z)$ and $-\protect\t{G}_{\sigma}(\bm{k};z)$ are Nevanlinna functions of $z$. It follows that without an appropriate cancellation, the contribution of $1/\protect\t{G}_{\protect\X{0};\sigma}(\bm{k};z)$ can potentially prevent $-\protect\t{\Sigma}_{\sigma}(\bm{k};z)$ from being also a Nevanlinna functions of $z$. See \S\,\protect\ref{sec3.f} and appendix \protect\ref{sd}. We note that a Nevanlinna function of $z$ is free from zeros in the finite part of the region $\protect\im[z]\not=0$ of the complex $z$-plane: for $\protect\t{\varphi}(z)$ such a function, $\mathrm{sgn}(\protect\im[\protect\t{\varphi}(z)]) = \mathrm{sgn}(\protect\im[z])$ implies that while the real part of $\protect\t{\varphi}(z)$ may be vanishing at some points in the finite part of the region $\protect\im[z] \not=0$, its imaginary part cannot. Thus, $1/\protect\t{\varphi}(z)$ cannot have poles in the region $\protect\im[z]\not=0$ of the $z$-plane; more generally, $1/\protect\t{\varphi}(z)$ cannot be unbounded in the finite part of the region $\protect\im[z]\not=0$ of the $z$-plane. In particular, $\protect\t{G}_{\protect\X{0};\sigma}(\bm{k};z)$ and $\protect\t{G}_{\sigma}(\bm{k};z)$ are free from zeros in the finite part of the region $\protect\im[z] \not=0$ [\S\,2 in Ref.\,\citen{JML61}] [\S\,4.7, p.\,137, in Ref.\,\citen{BF99a}]; following the asymptotic relations $\protect\t{G}_{\protect\X{0};\sigma}(\bm{k};z)\sim \hbar/z$ and $\protect\t{G}_{\sigma}(\bm{k};z)\sim \hbar/z$ for $z\to\infty$, Eq.\,(\protect\ref{e4n}), both functions are vanishing at the point of infinity of the $z$-plane. \label{notee1}}

Above we have explicitly considered the zero-temperature, $T=0$, functions $\t{G}_{\sigma}(\bm{k};z)$ and $\t{\Sigma}_{\sigma}(\bm{k};z)$. In view of the considerations later in this publication, we remark that the equality in Eq.\,(\ref{e2}) can be shown similarly to apply to the $T > 0$ counterparts of these functions, respectively $\t{\mathscr{G}}_{\sigma}(\bm{k};z)$ and $\t{\mathscr{S}}_{\sigma}(\bm{k};z)$.\footnote{It is common practice to refer to these functions as corresponding to `finite temperatures'. This is unfortunate, since $0$ is unquestionably \textsl{finite}.}  This is achieved along the same lines as in the cases of $\t{G}_{\sigma}(\bm{k};z)$ and $\t{\Sigma}_{\sigma}(\bm{k};z)$, on the basis of the spectral representation of $\t{\mathscr{G}}_{\sigma}(\bm{k};z)$ [App. C in Ref.\,\citen{BF07}].

Finally, with reference to the remarks at the outset of \S\,\ref{sec2} regarding the problem of uniqueness, we emphasise that already the transition from, for instance, the operator $\h{G}_{\sigma}$ to its energy-momentum representation $\t{G}_{\sigma}(\bm{k};z)$ involves two non-trivial restrictive assumptions: firstly, that $\h{G}_{\sigma}$ acts on functions of a one-particle Hilbert space,\footnote{The space of square-integrable functions defined over the one-particle configuration space of the problem at hand, where functions are both locally and globally constrained.} and, secondly, that the underlying $N$-particle GS is uniform. In the case of lattice models, for which in particular the one-particle configuration space consists of a Bravais lattice \cite{AM76} embedded in $\mathds{R}^d$ (see above), the latter Hilbert space is further limited by the requirement that the Green function $\t{G}_{\sigma}(\bm{k};z)$ be periodic over the entire $\bm{k}$ space, with the underlying $\1BZ$ forming the fundamental region of periodicity [App. B in Ref.\,\citen{FT09}]. Clearly, without the totality of these restrictions in mind, the question of \textsl{uniqueness} would be a misplaced one.

\refstepcounter{dummyX}
\section{On the uniform convergence of the series expansion of \texorpdfstring{$\t{\Sigma}_{\sigma}(\bm{k};z)$}{} in terms of skeleton self-energy diagrams for almost all \texorpdfstring{$\bm{k}$}{} and \texorpdfstring{$z$}{}}
\phantomsection
\label{sec2.b}
In this section we consider the perturbation series expansion of the self-energy $\t{\Sigma}_{\sigma}(\bm{k};z)$ in terms of skeleton (or two-particle irreducible, 2PI) self-energy diagrams and the exact one-particle Green functions $\{\t{G}_{\sigma}(\bm{k};z)\| \sigma\}$ and demonstrate its uniform convergence for almost all $\bm{k}$ and $z$. This we do by presenting a modified version of the proof of this property originally presented in Ref.\,\citen{BF07}. The modification concerns the substitution of an intuitive argument, based on `physical grounds', with a mathematically rigorous one. Notably, the new argument rests on a \textsl{weaker} condition to be satisfied by the self-energy than that on which the original intuitive argument rests.

An outline of the considerations to be encountered in this section is as follows. We begin with the original proof of the uniform convergence of the perturbation series under consideration \cite{BF07}, which consists of two main parts. In the first part, we rule out the possibility of the \textsl{divergence} of this perturbation series, leaving one with the two remaining possibilities of this series either \textsl{converging} or \textsl{oscillating}.\footnote{For the precise definition, see later, p.\,\protect\pageref{RegardingThePossibility}.} In the second part of the proof, the possibility of the perturbation series under consideration oscillating is also ruled out. The property that underlies the convergence of this series (as opposed to its oscillation) simultaneously ensures its \textsl{uniform} convergence for almost all $\bm{k}$ and $z$. The relevant property is the positivity of the non-vanishing terms of the infinite perturbation series for $\im[\t{\Sigma}_{\sigma}(\bm{k};\varepsilon-\ii\hspace{0.6pt}0^+)]$, $\forall\varepsilon \in \mathds{R}$, deduced from the original perturbation series for $\t{\Sigma}_{\sigma}(\bm{k};z)$, Eqs\,(\ref{e4}) and (\ref{e4b}) below. In Ref.\,\citen{BF07} we asserted the validity of $\im[\t{\Sigma}_{\sigma}^{\X{(\nu)}}(\bm{k};\varepsilon-\ii\hspace{0.6pt}0^+)] \ge 0$ for all $\nu \ge 2$ on `physical grounds',\footnote{See Eqs\,(5.42) and (5.43) in Ref.\,\protect\citen{BF07} and compare with Eq.\,(\protect\ref{e2}) above. As we shall emphasise later, the condition $\nu\ge 2$ is unnecessarily too restrictive; for the original proof, $\protect\im[\t{\Sigma}_{\sigma}^{\X{(\nu)}}(\bm{k};\varepsilon-\ii\hspace{0.6pt}0^+)] \ge 0$ is only to apply for $\nu\to\infty$, or, for all $\nu > \nu_{\protect\X{0}}$, where $\nu_{\protect\X{0}}$ is arbitrarily large but finite.} where $\nu$ denotes the order of the perturbation theory, that is the order of the skeleton self-energy diagrams contributing to the self-energy in their \textsl{explicit} dependence on the \textsl{bare} two-body interaction potential.

The new element that we inject into the original proof of the uniform convergence of the perturbation series under consideration for almost all $\bm{k}$ and $z$, to be presented below, bypasses the need to rely on the property $\im[\t{\Sigma}_{\sigma}^{\X{(\nu)}}(\bm{k};\varepsilon-\ii\hspace{0.6pt}0^+)] \ge 0$ for $\nu \ge 2$. It is based on the asymptotic series expansions \cite{ETC65,RBD73} of $\t{G}_{\sigma}(\bm{k};z)$ and $\t{\Sigma}_{\sigma}(\bm{k};z)$ in the asymptotic region $z \to \infty$.\cite{BF02,BF07} By demonstrating the \textsl{positivity}\,\footnote{See Eq.\,(\protect\ref{e4s}) \emph{et seq.}} of the sequences of the relevant coefficients in the latter asymptotic series, and bringing these coefficients into direct contact with the terms in the perturbation series expansion of the self-energy in terms of \textsl{skeleton} self-energy diagrams and the interacting Green function, we achieve what we had originally achieved in Ref.\,\citen{BF07}, this however \textsl{without} direct reliance on the property $\im[\t{\Sigma}_{\sigma}^{\X{(\nu)}}(\bm{k};\varepsilon-\ii\hspace{0.6pt}0^+)] \ge 0$ for $\nu \ge 2$.

The formalism underlying the above-mentioned new element of our proof is rooted in the classical \textsl{problem of moments} and the associated theories of orthogonal polynomials \cite{GS75,BS04,BG96} and continued-fraction expansion \cite{NIA65,ST70,BG96,CPVWJ08,LW08}. The property of the uniform convergence of the perturbation series under consideration proves to be intimately related to both the assumed stability of the GSs under consideration, as reflected by the inequalities in Eq.\,(\ref{e3}), and the way in which the above-mentioned sequences of coefficients of the asymptotic series for $\t{G}_{\sigma}(\bm{k};z)$ and $\t{\Sigma}_{\sigma}(\bm{k};z)$ behave at high orders, rendering the relevant moment problems both well-defined and \textsl{determinate}.\footnote{See Eqs\,(\protect\ref{e4oh}) and (\protect\ref{e7m}) and the attendant discussions.} The proofs of these properties (\emph{i.e.} that the mentioned asymptotic coefficients exist and are \textsl{determinate}), which we provide in appendix \ref{sab}, amount to the proof of the perturbation series expansion under consideration for the self-energy \textsl{converging}, ruling out \textsl{oscillation}. Some additional properties of two sequences of functions of $z$, corresponding to the measure functions \cite{PRH50,APM11} associated with two truncated moment problems, ultimately lead us, via a theorem due to Vitali, \cite{ECT52,RR98,LW08,AIM65} to the conclusion that the latter series is in addition \textsl{uniformly convergent} for almost all $\bm{k}$ and $z$.

\refstepcounter{dummyX}
\subsection{Remarks and discussions}
\phantomsection
\label{s.301}
Before proceeding with the main subject of this section, three remarks and some discussions are in order.

Firstly, as we shall discuss in some detail later in this publication, perturbation series expansion of the self-energy in terms of \textsl{skeleton} self-energy diagrams and the interacting one-particle Green functions $\{\t{G}_{\sigma} \| \sigma\}$ is meaningful \textsl{only if} the underlying formal perturbation series for the self-energy in terms of proper self-energy diagrams \cite{FW03} (including both skeleton and non-skeleton ones) and the non-interacting one-particle Green functions $\{\t{G}_{\X{0};\sigma}\| \sigma\}$ is \textsl{non-terminating}.\footnote{Note for instance that for the exact self-energy in Ref.\,\protect\citen{SRRRB15} one has $\protect\t{s} = -\frac{1}{2} u y_{\protect\X{0}}$, to be contrasted with the Hartree-Fock self-energy $\protect\t{s}^{\textsc{hf}} = -\frac{1}{2} u y$.} This, which may not be widely appreciated,\footnote{This is at least the case when limiting oneself to the physics literature. Viewing the problem from a wider perspective, for instance in Ref.\,\protect\citen{BS98} one encounters the remarks: ``We will also demand that $d\rho$ have infinite support, that is, that $\rho$ not be a pure point measure supported on a finite set. This eliminates certain degenerate cases.'' With reference to the connection that we establish in this publication between the perturbation series expansion of the self-energy $\protect\t{\Sigma}_{\sigma}$ in terms of \textsl{skeleton} self-energy diagrams (and $\{\protect\t{G}_{\sigma}\| \sigma\}$) and the positive sequence of moments of the one-particle spectral function $A_{\sigma}$, and in the light of the considerations later in this publication regarding the `Hubbard atom', we observe that ``certain degenerate cases'' includes those that in our considerations correspond to terminating perturbation series expansions of $\protect\t{\Sigma}_{\sigma}$ in term of $\{\protect\t{G}_{\protect\X{0};\sigma}\| \sigma\}$.} follows from the fact that the transition from the perturbation series expansion of the self-energy in terms of the non-interacting Green functions $\{\t{G}_{\X{0};\sigma} \| \sigma\}$ to that in terms of the interacting Green functions $\{\t{G}_{\sigma} \| \sigma\}$ relies on the assumption of the existence of an \textsl{infinite} hierarchy of terms in the former series that are systematically summed over [pp. 105 and 106 of Ref.\,\citen{FW03}]. Reversing this process, \emph{i.e.} starting from the perturbation series expansion of the self-energy in terms of skeleton self-energy diagrams and the interacting Green functions $\{\t{G}_{\sigma}\| \sigma\}$ and expanding the $\t{G}_{\sigma}$s internal to all skeleton self-energy diagrams in terms of $\{\t{G}_{\X{0};\sigma}\| \sigma\}$, results in an \textsl{infinite series}.\footnote{Following the Dyson equation, one has $\protect\h{G}_{\sigma}(z) = \big(\protect\h{I} - \protect\h{G}_{\protect\X{0};\sigma}(z)\protect\h{\Sigma}_{\sigma}(z)\big)^{-1} \protect\h{G}_{\protect\X{0};\sigma}(z)$, from which one observes that even a finite-order self-energy contribution is associated with an infinite-order perturbation series expansion for $\protect\h{G}_{\sigma}(z)$. Here $\protect\h{I}$ stands for the unit operator in the single-particle Hilbert space of the problem at hand.} For this series to reduce to a finite one, a systematic cancellation of various contributions is to take place, which in principle may \textsl{not} materialise at any finite order of the perturbation expansion for $\t{\Sigma}_{\sigma}$ in terms of skeleton self-energy diagrams and the interacting Green functions $\{\t{G}_{\sigma} \| \sigma\}$. More explicitly, the expected cancellation at any finite order in the coupling constant of interaction may depend on the contributions arising from arbitrary high orders of the perturbation expansion in terms of skeleton self-energy diagrams and $\{\t{G}_{\sigma} \| \sigma\}$. In this publication we explicitly show how in the case of the `Hubbard atom' the strict locality of the underlying interacting one-particle Green function gives rise to skeleton-diagrammatic contributions whose \textsl{implicit} dependence on the on-site interaction energy $U$ is singular at $U=0$, thus systematically reducing the orders of the perturbational contributions from in principle arbitrary high orders to arbitrary low orders. This process can be glimpsed from the $4$th-order self-energy contribution\,\footnote{Corresponding to the skeleton self-energy diagram in Fig.\,\protect\ref{f10}, p.\,\protect\pageref{A4thOr}, below.} in Eq.\,(\ref{e52b}) below (with the $\epsilon$ herein to be identified with $1$), which in addition to a contribution proportional to $U^4$ contains contributions proportional to $U^3$ and $U^2$. We should mention however that the singular behaviour just indicated does not show up at the $4$th-order, and that the \textsl{total} contribution of the $4$th-order skeleton self-energy diagrams divided by $U^4$ is bounded for $U \downarrow 0$, Eqs\,(\ref{ex0s}) and (\ref{ex0t}).\footnote{For the total contribution of the $4$th-order skeleton self-energy diagrams and its behaviour for $z\to\infty$ see Eqs\,(\protect\ref{ex0e}) and (\ref{ex0f}), respectively. Compare the expression in the latter equation with that in Eq.\,(\protect\ref{e7g}).} Our analysis reveals that this is generally not the case for the total contribution of the $\nu$th-order skeleton self-energy diagrams with $\nu \ge 6$, \S\,\ref{sec.d53}.

Secondly, with regard to the condition $\im[\t{\Sigma}_{\sigma}^{\X{(\nu)}}(\bm{k};\varepsilon-\ii\hspace{0.6pt}0^+)] \ge 0$, $\forall\nu \ge 2$, referred to above, we should emphasise that throughout Ref.\,\citen{BF07}, as well as the present publication, the function $\t{\Sigma}_{\sigma}^{\X{(\nu)}}(\bm{k};z)$ stands for the \textsl{total} contribution of \textsl{all} skeleton self-energy diagrams of order $\nu$, and not the contribution of a particular selection of these diagrams. The above inequality can be explicitly shown to be valid for $\nu=2$ [reference 50 in Ref.\,\citen{BF07}], however our efforts to prove or disprove it have been inconclusive even for the specific cases of $\nu = 3$ and $\nu = 4$ (more about this below).\footnote{With reference to Eq.\,(\protect\ref{e7h}) and to the positivity of the sequence $\{\Sigma_{\sigma;\infty_j}(\bm{k}) \| j \in \mathds{N}\}$, the case $\nu=2$ is seen to be very special; not only $\Sigma_{\sigma;\infty_1}^{\protect\X{(2)}}(\bm{k}) \equiv \Sigma_{\sigma;\infty_1}(\bm{k}) > 0$, Eq.\,(\protect\ref{e7l}), but also $\protect\t{\Sigma}_{\sigma}^{\protect\X{(2)}}(\bm{k};z)$ contributes in principle to \textsl{all} elements of the above positive sequence (note the lower bound of the sum on the RHS of Eq.\,(\protect\ref{e7h}) and compare with Eq.\,(\protect\ref{e50a}) below).}\footnote{We note that while the $2$nd-order self-energy $\t{\Sigma}_{\sigma}^{\protect\X{(2)}}(z)$ in Eq.\,(\ref{ex2c}) is a Nevanlinna function of $z$, the $4$th-order contribution $\t{\Sigma}_{\sigma}^{\protect\X{(4)}}(z)$ in Eq.\,(\protect\ref{ex0e}) is \textsl{not}. Both are analytic away from the real axis of the $z$-plane. The failure of $\t{\Sigma}_{\sigma}^{\protect\X{(4)}}(z)$ to be Nevanlinna would constitute a proof that in general the inequality in Eq.\,(\protect\ref{e4c}) below is false (here at least for $\nu=4$). However, the locality of the `Hubbard atom' and the consequent breakdown of the equality in Eq.\,(\ref{e7h}) (amongst others, later in this publication we show that this breakdown accounts for the coefficient of the leading-order asymptotic term on the RHS of Eq.\,(\protect\ref{ex0f}), namely $45/128$, deviating from the expected value of $9/16$, Eqs\,(\protect\ref{e7xg}) and (\protect\ref{ex0i})) calls for caution in considering the inequality in Eq.\,(\protect\ref{e4c}) below as having been shown to be generally invalid.} What is evident however is that the self-energy $\t{\Sigma}_{\sigma}(\bm{k};\varepsilon-\ii\hspace{0.6pt}0^+)$ as calculated on the basis of a \textsl{restricted} (even possibly unbounded) set of self-energy diagrams in general violates the inequality $\t{\Sigma}_{\sigma}(\bm{k};\varepsilon-\ii\hspace{0.6pt}0^+) \ge 0$, and thus the inequalities in Eq.\,(\ref{e3}).

We note that the Cutkosky method \cite{REC60,JSL61,PS18} of determining the analytic expression for the imaginary part of a diagrammatic contribution is of a very limited applicability here, since while the expression for the imaginary part of the contribution $\t{\Sigma}_{\sigma}^{\X{(\nu.j)}}(\bm{k};\varepsilon-\ii\hspace{0.6pt}0^+)$ corresponding to the $j$th $\nu$th-order self-energy diagram\,\footnote{\emph{Cf.} Eqs\,(\ref{ex1a}), (\ref{ex1b}), and (\ref{ex1c}). See also \S\,\protect\ref{sd21}.} can in principle be relatively easily written down for arbitrary $j$, without a quantitative calculation of the elements of $\{\im[\t{\Sigma}_{\sigma}^{\X{(\nu.j)}}(\bm{k};\varepsilon-\ii\hspace{0.6pt}0^+)] \| j\}$, each of which may be of either sign for arbitrary $\bm{k}$ and $\varepsilon$, the sign of $\im[\t{\Sigma}_{\sigma}^{\X{(\nu)}}(\bm{k};\varepsilon-\ii\hspace{0.6pt}0^+)]$ cannot be established.\footnote{We note that in the quantum field theory of scattering, one encounters the (generalised) \textsl{optical theorem} \protect\cite{PS18}, which can be considered within the framework of the perturbation series expansion of the $S$ matrix and in conjunction with the method of Cutkosky \protect\cite{PS18}. The optical theorem expressing the imaginary part of the forward scattering amplitude in terms of a sum over the squares of the absolute values of the appropriate matrix elements of the $S$ matrix (following the unitarity of the $S$ matrix, $S^{\dag} S = \protect\h{1}$, from which the optical theorem is deduced) \protect\cite{PS18}, the question of the latter imaginary part being negative within the framework of perturbation theory, does not arise (the diagram 7.5, p.\,231, in Ref.\,\protect\citen{PS18} makes this point apparent).} Furthermore, even if this sign could be established without any quantitative calculations for some finite values of $\nu$, determination of this sign for \textsl{arbitrarily} large values of $\nu$ is clearly out of practical reach.

Thirdly, throughout this publication, as in Ref.\,\citen{BF07}, the qualification `for \textsl{almost} all $\bm{k}$ and $z$' signifies exclusion of possible non-empty subsets of \textsl{measure zero} \cite{PRH50} of the relevant sets of $\bm{k}$ and $z$ over which $\t{\Sigma}_{\sigma}(\bm{k};z)$ is \emph{a priori} defined.\footnote{Our considerations leaving out sets of measure zero, they place the `Hubbard atom' in an exceptional category: the self-energy, \emph{i.a.}, pertaining to this system being non-dispersive, \textsl{independent} of $\bm{k}$, for this system \textsl{no} zero-measure subset of the $\bm{k}$-space can be singled out and discarded as such.} The exclusion concerns the subsets of the latter sets over which $\t{\Sigma}_{\sigma}(\bm{k};z)$ is possibly \textsl{discontinuous} or \textsl{unbounded} \cite{BF07}. The analyticity of $\t{\Sigma}_{\sigma}(\bm{k};z)$ over the entire region $\im[z] \not= 0$ of the $z$-plane \cite{JML61,BF07} implies that for a given $\bm{k}$ the latter possible zero-measure subset of the $z$-plane over which the self-energy can be discontinuous or unbounded must be necessarily confined to the real axis of this plane. In the tradition of in particular the literature on the measure theory \cite{PRH50,APM11}, \emph{in what follows we shall also employ the term `almost everywhere', and its abbreviation `\ae', instead of `for almost all $\bm{k}$ and $z$'.}

\refstepcounter{dummyX}
\subsection{The original proof}
\phantomsection
\label{s3.1}
Expanding the \textsl{proper}\,\footnote{The \textsl{proper} self-energy is denoted by $\protect\t{\Sigma}^{\star}$ in Ref.\,\protect\citen{FW03}, and the \textsl{improper} self-energy by $\protect\t{\Sigma}$ (the superscript $\star$ is not to be confused with the symbol $*$ for complex conjugation). We note that \textsl{proper} self-energy diagrams are one-particle irreducible (1PI). See, \emph{e.g.}, Figs\,9.13 and 9.14, p.\,106, in Ref.\,\protect\citen{FW03}.} self-energy $\t{\Sigma}_{\sigma}(\bm{k};z)$ in terms of the contributions of \textsl{skeleton} (or \textsl{two-particle irreducible}, 2PI) self-energy diagrams \cite{LW60} described in terms of the exact Green functions $\{\t{G}_{\sigma}(\bm{k};z) \| \sigma\}$ and the \textsl{bare} two-body interaction potential, one formally has
\begin{equation}\label{e4}
\t{\Sigma}_{\sigma}(\bm{k};z) = \sum_{\nu=1}^{\infty} \t{\Sigma}_{\sigma}^{\X{(\nu)}}(\bm{k};z),
\end{equation}
where $\t{\Sigma}_{\sigma}^{\X{(\nu)}}(\bm{k};z)$ denotes the \textsl{total} contribution to the self-energy $\t{\Sigma}_{\sigma}(\bm{k};z)$ of all skeleton self-energy diagrams of order $\nu$, with $\nu$ counting the number of the \textsl{bare} interaction lines in the diagrams. For Hubbard-like models, \textsl{all} terms of the sequence $\{ \t{\Sigma}_{\sigma}^{\X{(\nu)}}(\bm{k};z) \| \nu\in \mathds{N}\}$ are bounded for almost all $\bm{k}$ and $z$ [\S\S\,B.3-B.7 in Ref.\,\citen{BF07}]. Explicitly, for lattice models \textsl{ultraviolet} divergences are ruled out, and for short-range interaction potentials \textsl{infrared} divergences.

The infinite sequence $\{\t{\Sigma}_{\sigma}^{\X{(\nu)}}(\bm{k};z) \| \nu\in \mathds{N}\}$ consists of unbounded terms for models defined on unbounded \textsl{continuum} subsets of $\mathds{R}^d$ and/or in terms of long-range two-body interaction potentials (such as the Coulomb potential) \cite{BF02}. The former (latter) gives rise to \textsl{ultraviolet}- (\textsl{infrared}-) divergent contributions to the functions of the sequence $\{ \t{\Sigma}_{\sigma}^{\X{(\nu)}}(\bm{k};z) \| \nu\}$, rendering the equality in Eq.\,(\ref{e4}) ill-defined. This problem is circumvented by constructing a new, regularized, sequence of functions $\{ \t{\Sigma}_{\sigma}^{\X{\prime\hspace{0.4pt} (\nu)}}(\bm{k};z)\| \nu \}$ determined by means of effecting partial infinite summations over infinite classes of terms of the sequence $\{\t{\Sigma}_{\sigma}^{\X{(\nu)}}(\bm{k};z) \| \nu\}$. We describe the details underlying this well-known procedure in appendix \ref{sa}.\footnote{See also Ref.\,\protect\citen{BF19}, in particular \S\S\,2.7, 2.8, and 3 herein.} \emph{In the following considerations, the functions $\t{\Sigma}_{\sigma}^{\X{(\nu)}}(\bm{k};z)$ and $\t{\Sigma}_{\sigma}^{\X{\prime\hspace{0.4pt} (\nu)}}(\bm{k};z)$ are therefore formally interchangeable,} although in this publication, as in Ref.\,\citen{BF07}, we do \textsl{not} make any statement regarding the series expansion of $\t{\Sigma}_{\sigma}(\bm{k};z)$ in terms of the sequence $\{\t{\Sigma}_{\sigma}^{\X{\prime\hspace{0.4pt} (\nu)}}(\bm{k};z)\| \nu\}$ for non-Hubbard-like models, the treatment of such models strictly falling outside the scope of the considerations of this publication.\footnote{We have \textsl{not} investigated the problem in any depth in order to be very specific about the possible unsurmountable difficulties to be encountered in its treatment. We can however mention that in dealing with this problem, the crucial equality in Eq.\,(\protect\ref{e7h}), which is rooted in the expression in Eq.\,(\protect\ref{e4p}), cannot be relied upon, not least by the fact that for general systems $\Sigma_{\sigma;\infty_j}(\bm{k})$ is unbounded for arbitrary values of $j$ \protect\cite{BF02}. In the absence of the mentioned equality, those in Eq.\,(\protect\ref{e7t}) become unfounded. We should point out however that since for Hubbard-like models the perturbation series expansion in terms of  $\{\protect\t{\Sigma}_{\sigma}^{\protect\X{\prime\hspace{0.4pt} (\nu)}}(\bm{k};z)\| \nu\}$ is necessarily uniformly convergent for almost all $\bm{k}$ and $z$ (the sequence of the partial sums of this sequence being a \textsl{subsequence} of the convergent sequence of partial sums of $\{\protect\t{\Sigma}_{\sigma}^{\protect\X{(\nu)}}(\bm{k};z)\| \nu\}$), by \textsl{formally} expressing this series in terms of $\{\protect\t{\Sigma}_{\sigma}^{\protect\X{(\nu)}}(\bm{k};z) \| \nu\}$ one may be able to infer a constructive approach whereby possibly to demonstrate the uniform convergence of the perturbation series expansion of $\protect\t{\Sigma}_{\sigma}(\bm{k};z)$ in terms of $\{\protect\t{\Sigma}_{\sigma}^{\protect\X{\prime\hspace{0.4pt} (\nu)}}(\bm{k};z)\| \nu\}$ for almost all $\bm{k}$ and $z$ for general models.}

The\refstepcounter{dummy}\label{InPrinciple} series on the RHS of Eq.\,(\ref{e4}) can in principle (i) converge, (ii) diverge, or (iii) oscillate (to be specified below, p.\,\pageref{RegardingThePossibility}).\footnote{The Note on \S\,1.1 in Ref.\,\protect\citen{GHH73} (p.\,21 herein) is illuminating.} Disregarding the possible non-empty subsets of measure zero of the relevant sets of $\bm{k}$ and $z$, specified above, \S\,\ref{s.301}, \emph{the possibility (ii) is \emph{a priori} ruled out on account of self-consistency} [\S\,5.3.2, p.\,34,  in Ref.\,\citen{BF07}].\footnote{As the proof of the uniform convergence of this series \ae implies its convergence, reference to this possibility here is only of a formal character.} In this connection, note that each term of the summand on the RHS of Eq.\,(\ref{e4}) is a functional of the function on the left-hand side (LHS). This is to be compared with the more transparent case of the fixed-point equation $x = f(x)$, where the variable on the LHS is the argument of the function on the RHS, whereby the equality cannot be satisfied for any finite $x$ at which the function $f(x)$ is unbounded, \S\,\ref{sec.3.2.3}. Importantly, supposing $\t{\Sigma}_{\sigma}(\bm{k};z)$ to be unbounded over non-zero-measure subsets of the relevant spaces of $\bm{k}$ and $z$, the union of which we denote by $\EuScript{S}$, for $\t{G}_{\sigma}(\bm{k};z)$ \textsl{bounded} \ae, in the light of the Dyson equation \cite{FW03},
\begin{equation}\label{e4a}
\t{G}_{\sigma}(\bm{k};z) = \t{G}_{\X{0};\sigma}(\bm{k};z) + \t{G}_{\X{0};\sigma}(\bm{k};z) \t{\Sigma}_{\sigma}(\bm{k};z) \t{G}_{\sigma}(\bm{k};z),
\end{equation}
one observes that $\t{G}_{\sigma}(\bm{k};z)$ must be identically vanishing over $\EuScript{S}$. With the function $\t{G}_{\X{0};\sigma}(\bm{k};z)$ clearly bounded and non-vanishing \ae, over the set $\EuScript{S}$ one more explicitly has $\t{G}_{\sigma}(\bm{k};z) = -1/\t{\Sigma}_{\sigma}(\bm{k};z) \equiv 0$. This observation is relevant in that \textsl{absence} of contribution from the region $\EuScript{S}$ to the self-energy, described in terms of skeleton self-energy diagrams and the interacting Green functions $\{\t{G}_{\sigma}(\bm{k};z)\|\sigma\}$, can clearly \textsl{not} self-consistently lead to the unboundedness of the calculated $\t{\Sigma}_{\sigma}(\bm{k};z)$ over $\EuScript{S}$.\footnote{For the algebraic expression corresponding to $\Sigma_{\sigma}^{\protect\X{(\nu)}}(\bm{k};\varepsilon)$, see \S\,\protect\ref{sd21}.}

Regarding\refstepcounter{dummy}\label{RegardingThePossibility} the possibility of \textsl{oscillation} of the series in Eq.\,(\ref{e4}), the possibility (iii) above, according to the theory of infinite series [Ch.\,VI in Ref.\,\citen{EWH07}] [\S\S\hspace{0.6pt}52-54 in Ref.\,\citen{EWH27}] the set $\mathpzc{S} = \{s_1,s_2,\dots\}$ of the partial sums $s_1 = u_1$, $s_2 = u_1+u_2$, \dots of the \textsl{oscillating} series $u_1+u_2+u_3+\dots$ of the real sequence $\{u_1, u_2, u_3, \dots\}$ has a \textsl{closed} subset $\mathpzc{S}'$ of limit points, or cluster points \cite{RR91,RR98}, called the \textsl{derived} (or \textsl{derivative}) set.\footnote{Note that the elements of the \textsl{infinite} set $\mathpzc{S}$ being bounded (since by assumption the series $u_1+u_2+u_3+\dots$ is not divergent), by the Bolzano-Weierstrass theorem [\S\,2.11, p.\,12, in Ref.\,\protect\citen{WW62}] it must contain at least one limit point.} The cardinal number $\vert\mathpzc{S}'\vert$ of $\mathpzc{S}'$ may be finite ($\mathpzc{S}'$ consisting of at least two elements) or infinite. It being \textsl{closed}, $\mathpzc{S}'$ has a lower and an upper bound, $\mathpzc{l}$ and $\mathpzc{u}$ respectively, referred to as \textsl{limits of indeterminacy} of the series under consideration. The set $\mathpzc{S}'$ may consist of the entire \textsl{closed} real interval $[\mathpzc{l},\mathpzc{u}]$, \textsl{or} of a non-dense subset of this interval. For every element $s$ of $\mathpzc{S}'$, including $\mathpzc{l}$ and $\mathpzc{u}$, an ordered sequence of integers $\{n_i\| i\in \mathds{N}\}$, satisfying $n_1 < n_2 < \dots$, can be produced for which the \textsl{subsequence} $\{s_{n_1}, s_{n_2},\dots\}$ of $\mathpzc{S}$ converges to $s$;\,\footnote{Convergent infinite \textsl{subsequences} of infinite sequences feature prominently in the context of the theorems of Montel and Vitali \protect\cite{ECT52,RR98} and the first and second theorems of Helly \protect\cite{ST70}, which in turn are relevant to the classical moment problem \protect\cite{NIA65,ST70}, to be encountered later in this publication, in particular in appendix \protect\ref{sab}. This is rooted in Stieltjes' original insight with regard to the ``principle of propagation of convergence'' [Ch. 7 in Ref.\,\protect\citen{RR98}]. } any subsequence of this form corresponds to a specific `system of bracketing' of the sequence  $\{u_1, u_2, u_3, \dots\}$ [\S\hspace{0.6pt}331 in Ref.\,\citen{EWH07}]. When convergent, the series $u_1 + u_2 + u_3 + \dots$ may be viewed as the limiting case of an oscillating series for which the $\mathpzc{l}$ and $\mathpzc{u}$ of the set $\mathpzc{S}'$ coincide [\S\hspace{0.6pt}5.3.3 in Ref.\,\citen{BF07}].\footnote{For illustration, consult for instance Ex.\,1 on p.\,20 of Ref.\,\protect\citen{TB65}.}

In the light of the introductory remarks in the previous paragraph, the observation by Kozik \emph{et al.} \cite{KFG14}, regarding the series expansion of the self-energy in terms of skeleton self-energy diagrams, convergence of this series to different self-energies is equivalent to the series in Eq.\,(\ref{e4}) being an \textsl{oscillating} one.\footnote{Fig.\,2a in Ref.\,\protect\citen{KFG14} shows existence of \textsl{two} self-energies at all but one value of $U$ for $U \in [1,5)$.} Below we show that this \textsl{cannot} be the case for stable $N$-particle GSs of Hubbard-like models [\S\S\hspace{0.6pt}5.3.2-11 in Ref.\,\citen{BF07}]. We remark that Kozik \emph{et al.} \cite{KFG14} do \textsl{not} report divergence of the series under consideration, which in principle may be attributable to the practical limitation on the maximum order of the skeleton self-energy diagrams that program ``DiagMC'' had been capable of taking into account (in Ref.\,\citen{KFG14} this order is reported as being $8$).

The approach in Ref.\,\citen{BF07} in ruling out the possibility of the series in Eq.\,(\ref{e4}) oscillating, relies on the consideration that an infinite convergent complex series of the form $\sum_{\nu} f_{\nu}$ can be expressed as
\begin{equation}\label{e4ba}
\sum_{\nu} f_{\nu} = \sum_{\nu} \re[f_{\nu}] + \ii\hspace{1.4pt} \sum_{\nu} \im[f_{\nu}],
\end{equation}
owing to the fact that convergence of $\sum_{\nu} f_{\nu}$ is contingent on the convergence of both $\sum_{\nu} \re[f_{\nu}]$ and $\sum_{\nu} \im[f_{\nu}]$ [\S\,75 in Ref.\,\citen{TB65}] [\S\,5.3.5 in Ref.\,\citen{BF07}]. This is contrary to the general case where a convergent \textsl{infinite} series of the form $\sum_{\nu} (a_{\nu} + b_{\nu})$ \textsl{cannot} in general be expressed as $\sum_{\nu} a_{\nu} + \sum_{\nu} b_{\nu}$, since in general the latter two series may not be convergent.\footnote{In view of the later considerations in this publication, it is important to emphasize that the sequence of functions $\{\protect\t{\Sigma}_{\sigma}^{\protect\X{(\nu)}}(\bm{k};z)\| \nu\}$ considered here are \textsl{complex} and \textsl{analytic} over the entire complex $z$-plane outside the real axis. This is relevant in that the Montel theorem [p.\,149 in Ref.\,\protect\citen{RR98}] [\S\,5.23, p.\,170, in Ref.\,\protect\citen{ECT52}], which is closely related to the Stieltjes-Vitali theorem [pp.\,150 and 151 in Ref.\,\protect\citen{RR98}] [\S\,3.1.5, p.\,114, in Ref.\,\protect\citen{LW08}] to be relied upon later in this section, does \textsl{not} in general apply to `\textsl{real-analytic}' sequences of functions [p.\,149 in Ref.\,\protect\citen{RR98}].}

With reference to the above remarks, as in Ref.\,\citen{BF07} we first consider the following series [Eq.\,(5.40) in Ref.\,\citen{BF07}]:
\begin{equation}\label{e4b}
\im[\t{\Sigma}_{\sigma}(\bm{k};\varepsilon-\ii\hspace{0.6pt}0^+)] = \sum_{\nu=1}^{\infty} \im[\t{\Sigma}_{\sigma}^{\X{(\nu)}}(\bm{k};\varepsilon-\ii\hspace{0.6pt}0^+)]\;\; \text{for}\;\; \varepsilon\in\mathds{R},
\end{equation}
where, on account of the relationship in Eq.\,(\ref{e2}), $\im[\t{\Sigma}_{\sigma}(\bm{k};\varepsilon-\ii\hspace{0.6pt}0^+)] \ge 0$ for all $\varepsilon \in \mathds{R}$.\footnote{Since $\protect\im[\protect\t{\Sigma}_{\sigma}^{\protect\X{(1)}}(\bm{k};z)] \equiv 0$, the lower bound of the sum in Eq.\,(\protect\ref{e4b}) can be identified with $2$.} The equality in Eq.\,(\ref{e4b}) is a limiting case of the more general equality
\begin{equation}\label{e5}
\im[\t{\Sigma}_{\sigma}(\bm{k};z)] = \sum_{\nu=1}^{\infty} \im[\t{\Sigma}_{\sigma}^{\X{(\nu)}}(\bm{k};z)]\;\; \text{for}\;\; z \in \mathds{C}\backslash\mathds{R},
\end{equation}
in which, following the equality in Eq.\,(\ref{e2}), the sign of the LHS (when non-vanishing) is fixed, depending \textsl{only} on whether $z$ is located in the lower or the upper half of the $z$-plane. In the light of the remarks in the previous paragraph, the series in Eq.\,(\ref{e4}) not diverging, the series in Eq.\,(\ref{e4b}), and that in Eq.\,(\ref{e5}), can either (i$'$) converge or (ii$'$) oscillate.

In Ref.\,\citen{BF07} we have argued that \textsl{on physical grounds} [Eq.\,(5.43) in Ref.\,\citen{BF07}]
\begin{equation}\label{e4c}
\im[\t{\Sigma}_{\sigma}^{(\nu)}(\bm{k};\varepsilon-\ii\hspace{0.6pt}0^+)] \ge 0,\;\; \forall \nu \ge 2,
\end{equation}
whereby the series in Eq.\,(\ref{e4b}) \textsl{cannot} oscillate [\S\,7 in Ref.\,\citen{TB65}]. This series can therefore only (i$'$) converge to a \textsl{unique} limit function. By the same reasoning, the series in Eq.\,(\ref{e5}) \textsl{cannot} oscillate and therefore similarly (i$'$) converges to a \textsl{unique} limit function. Moreover, since $\im[\t{\Sigma}_{\sigma}(\bm{k};z)]$ is a \textsl{continuous} function [Ch.\,III in Ref.\,\citen{WW62}] of $\bm{k}$ and $z$ for almost all $\bm{k}$ and $z$ [\S\hspace{0.6pt}5.3.6 in Ref.\,\citen{BF07}], by the Dini theorem [\S\hspace{0.6pt}347 in Ref.\,\citen{EWH07}] [\S\hspace{0.6pt}49.2 in Ref.\,\citen{TB65}] [\S\hspace{0.6pt}79 in Ref.\,\citen{EWH26}] the series in Eq.\,(\ref{e4b}) (Eq.\,(\ref{e5})) is \textsl{uniformly} convergent for almost all $\bm{k}$ and $\varepsilon$ ($z$) [\S\hspace{0.6pt}5.3.5 in Ref.\,\citen{BF07}].

We shall not go into further details here, for which we refer the reader to \S\,5.3.7 of Ref.\,\citen{BF07}, and suffice to mention that through a variety of ways one can rigorously establish that the uniform convergence of the series in Eq.\,(\ref{e5}) \ae to a unique function implies a similar property for $\re[\t{\Sigma}_{\sigma}(\bm{k};z)]$ expanded in terms of $\re[\t{\Sigma}_{\sigma}^{\X{(\nu)}}(\bm{k};z)]$. One thus establishes that \emph{the series in Eq.\,(\ref{e4}) converges uniformly for all $\bm{k}$ and $z$ where the function $\t{\Sigma}_{\sigma}(\bm{k};z)$ is continuous} \cite{BF07}. Because of the continuity of $\t{\Sigma}_{\sigma}(\bm{k};z)$ for almost all $\bm{k}$ and $z$, it follows that, for the sequence $\{\t{\Sigma}_{\sigma}^{\X{(\nu)}}(\bm{k};z)\| \nu\}$ as calculated in terms of the \textsl{exact} interacting Green functions $\{\t{G}_{\sigma}(\bm{k};z)\|\sigma\}$, \emph{the series in Eq.\,(\ref{e4}) converges uniformly to the exact self-energy $\t{\Sigma}_{\sigma}(\bm{k};z)$ for almost all $\bm{k}$ and $z$ \cite{BF07}.}

\refstepcounter{dummyX}
\subsection{An alternative proof based on two basic properties of the single-particle spectral function \texorpdfstring{$A_{\sigma}(\bm{k};\varepsilon)$}{}}
\phantomsection
\label{sec.3.2}
The work by Kozik \emph{et al.} in Ref.\,\citen{KFG14} led us to re-examine the relationship in Eq.\,(\ref{e4c}). In the process we have established that although the relationship in Eq.\,(\ref{e4c}) is a \textsl{sufficient} condition for the uniform convergence of the series in Eq.\,(\ref{e4}) \ae,\footnote{Appendix \protect\ref{sae}.} it is \textsl{not} a necessary one. Rather, the non-negativity of the one-particle spectral function $A_{\sigma}(\bm{k};\varepsilon)$ corresponding to the \textsl{exact} Green function $\t{G}_{\sigma}(\bm{k};z)$, Eq.\,(\ref{e4d}), in conjunction with the demonstrably faster decay of $A_{\sigma}(\bm{k};\varepsilon)$ towards zero than any finite power of $1/\varepsilon$ for $\vert\varepsilon\vert \to \infty$ in the case of Hubbard-like models -- implying the existence of the elements of the \textsl{positive sequence}\,\footnote{For definition, see Eq.\,(\protect\ref{e4s}) \emph{et seq.}} of the moments $\{G_{\sigma;\infty_j}(\bm{k}) \| j\}$ of this function for arbitrary finite values of $j$, Eq.\,(\ref{e4o}), suffice for the series in Eq.\,(\ref{e4}) to be uniformly convergent for almost all $\bm{k}$ and $z$. This is in part rooted in the fact that each element of the sequence $\{G_{\sigma;\infty_j}(\bm{k}) \| j\}$ is directly related to some perturbational contributions to the self-energy $\t{\Sigma}_{\sigma}(\bm{k};z)$ as arising from the expansion of this function in terms of \textsl{skeleton} self-energy diagrams and the \textsl{interacting} Green functions $\{\t{G}_{\sigma}(\bm{k};z)\| \sigma\}$.\footnote{See the discussions centred on the relationship in Eq.\,(\protect\ref{e4p}).}

The alternative proof to be presented below, \S\,\ref{sec.3.2.1}, relies on a detailed discussion regarding the functions $\t{G}_{\sigma}(\bm{k};z)$ and $\t{\Sigma}_{\sigma}(\bm{k};z)$ in the context of the classical moment problem \cite{NIA65,ST70,BG96}, which we present in appendix \ref{sab}.

\refstepcounter{dummyX}
\subsubsection{Details}
\phantomsection
\label{sec.3.2.1}
We begin with the partial-sum function $\t{\mathfrak{S}}_{\sigma}^{\X{(n)}}(\bm{k};z)$ associated with the infinite series in Eq.\,(\ref{e4}), defined according to
\begin{equation}\label{e7r}
\t{\mathfrak{S}}_{\sigma}^{\X{(n)}}(\bm{k};z) \doteq \sum_{\nu=1}^{\mathcal{I}(n)+1} \t{\Sigma}_{\sigma}^{\X{(\nu)}}(\bm{k};z) \equiv \Sigma_{\sigma}^{\textsc{hf}}(\bm{k}) + \sum_{\nu=2}^{\mathcal{I}(n)+1} \t{\Sigma}_{\sigma}^{\X{(\nu)}}(\bm{k};z),\;\; n\in \mathds{N},
\end{equation}
where $\mathcal{I}(n)$ is an integer-valued function satisfying
\begin{equation}\label{e7ra}
\mathcal{I}(n) \ge 2 n + \varsigma,
\end{equation}
in which $\varsigma \in \{0,1\}$ (\emph{cf.} Eqs\,(\ref{e6oa}) and (\ref{e6q})).\footnote{As is evident from the $\ge$, we could have safely identified $\varsigma$ with $1$, or any larger integer for that matter. We have nonetheless employed $\varsigma$ here so as to preserve a visible connection with the considerations in appendix \protect\ref{sab}. In fact, this is the sole purpose of introducing $\mathcal{I}(n)$, instead of merely using $n$, in which case $\protect\t{\mathfrak{S}}_{\sigma}^{\protect\X{(n)}}(\bm{k};z)$ would have to be denoted as, say, $\protect\t{\mathfrak{S}}_{\sigma}^{\protect\X{(2n+\varsigma)}}(\bm{k};z)$.} Further, $\Sigma_{\sigma}^{\textsc{hf}}(\bm{k})$ denotes the \textsl{exact} Hartree-Fock self-energy,\footnote{See \S\,\protect\ref{sec.3a.1} below. Also consult Ref.\,\protect\citen{BF13}, in particular Eqs\,(2.34) -- (2.36) herein.} appendix \ref{sab}. Before clarifying the reason underlying the condition in Eq.\,(\ref{e7ra}), we point out that the function $\t{\mathfrak{S}}_{\sigma}^{\X{(n)}}(\bm{k};z)$ is the counterpart of the approximants
$\t{f}^{\X{(n)}}(z)$ and $-\t{w}_{n}(z,\ptau)$ discussed in appendix \ref{sab}.\footnote{See in particular Eqs\,(\protect\ref{e6qb}) and (\protect\ref{e6qe}).}

The function $\t{\mathfrak{S}}_{\sigma}^{\X{(n)}}(\bm{k};z)$ has the asymptotic series expansion \cite{ETC65}
\begin{equation}\label{e7s}
\t{\mathfrak{S}}_{\sigma}^{\X{(n)}}(\bm{k};z) \sim \mathfrak{S}_{\sigma;\infty_0}^{\X{(n)}}(\bm{k}) + \frac{\mathfrak{S}_{\sigma;\infty_1}^{\X{(n)}}(\bm{k})}{z} + \frac{\mathfrak{S}_{\sigma;\infty_2}^{\X{(n)}}(\bm{k})}{z^2} + \ldots\;\; \text{for}\;\; z\to\infty,
\end{equation}
where, following Eq.\,(\ref{e7r}),
\begin{equation}\label{e7sa}
\mathfrak{S}_{\sigma;\infty_0}^{\X{(n)}}(\bm{k}) \equiv \Sigma_{\sigma}^{\textsc{hf}}(\bm{k}),\;\;\;\;
\mathfrak{S}_{\sigma;\infty_j}^{\X{(n)}}(\bm{k}) \equiv \sum_{\nu=2}^{\mathcal{I}(n)+1} \Sigma_{\sigma;\infty_j}^{\X{(\nu)}}(\bm{k}),\;\; j\in\mathds{N}.
\end{equation}
In the light of the equalities in Eqs\,(\ref{e7ba}) and (\ref{e7h}),
\begin{equation}\label{e7t}
\mathfrak{S}_{\sigma;\infty_j}^{\X{(n)}}(\bm{k}) \equiv \Sigma_{\sigma;\infty_j}(\bm{k}),\;\; j \in \{0,1,\dots, \mathcal{I}(n)\},
\end{equation}
in which $\Sigma_{\sigma;\infty_j}(\bm{k})$ is the coefficient of the term decaying like $1/z^j$ in the asymptotic series expansion of $\t{\Sigma}_{\sigma}(\bm{k};z)$ for $z\to \infty$, Eqs\,(\ref{e7b}), (\ref{e7ba}), and (\ref{e7i}). In the light of the leading-order asymptotic term on the RHS of Eq.\,(\ref{e7g}), one in general has
\begin{equation}\label{e7u}
\mathfrak{S}_{\sigma;\infty_j}^{\X{(n)}}(\bm{k}) \not\equiv \Sigma_{\sigma;\infty_j}(\bm{k})\;\; \text{for}\;\; j > \mathcal{I}(n).
\end{equation}
In appendix \ref{sab} we demonstrate that for Hubbard-like models the function $\Sigma_{\sigma;\infty_{j}}(\bm{k})$ is bounded for arbitrary finite values of $j$, from which and the equality in Eq.\,(\ref{e7t}) it follows that $\mathfrak{S}_{\sigma;\infty_j}^{\X{(n)}}(\bm{k})$ is bounded for arbitrary finite values of $n$ and $0 \le j \le \mathcal{I}(n)$.

As\refstepcounter{dummy}\label{WeDiscussIn} we discuss in \S\,\ref{sec.b2.1}, for the `Hubbard atom' the equivalence in Eq.\,(\ref{e7t}) breaks down in general. We identify the origin of this breakdown with the \textsl{locality} of the underlying one-particle Green function, whereby, for $\t{\Sigma}^{\X{(\nu)}}(z)$ evaluated in terms of the $\nu$th-order skeleton self-energy diagrams, the bare interaction potential $U$, and the interacting Green function $\t{G}(z)$, the function $\t{\Sigma}^{\X{(\nu)}}(z)/U^{\nu}$ can diverge for $U\downarrow 0$. To appreciate the role of the above-mentioned \textsl{locality}, one should bear in mind that our proof of the uniform convergence of the series in Eq.\,(\ref{e4}) leaves out subsets of measure zero in the $\bm{k}$-space and the $z$-space.\footnote{See the last paragraph in \S\,\protect\ref{s.301}.} For the case of a local Green function, all points of the $\bm{k}$-space become equivalent\,\footnote{For the purpose of the present discussions, one could equally state that the $\bm{k}$-space disappears.} and thereby the notion of a zero-measure subset of this space becomes devoid of meaning. In order to make the considerations of this section applicable to Hubbard-like \textsl{local} models, in principle one would have to equate the upper bound $\mathcal{I}(n)$ in Eq.\,(\ref{e7r}) with $\infty$, which transforms the series in Eq.\,(\ref{e7r}) into that in Eq.\,(\ref{e4}), rendering the arguments of this section circular. \emph{We shall therefore exclude Hubbard-like \textsl{local} models from the considerations of this section.}

The result in Eq.\,(\ref{e7t}) is to be contrasted with that in Eq.\,(\ref{e6oa}). Both asymptotic coefficients $s_j$ and $\mathfrak{S}_{\sigma;\infty_j}^{\X{(n)}}(\bm{k})$ correspond to truncated moment problems \cite{NIA65,ST70}, which however are non-trivially different in character. Consequently, it is \emph{a priori} not evident that the general results pertinent to the classical moment problem, presented in appendix \ref{sab}, may be directly applicable to the function $\t{\mathfrak{S}}_{\sigma}^{\X{(n)}}(\bm{k};z)$ considered here. Notably, whereas the function $\t{f}^{\X{(n)}}(z)$, or $-\t{w}_{n}(z)$, Eqs\,(\ref{e5tc}) and (\ref{e6k}), is describable as the ratio of two finite-order polynomials of $z$ for any finite value of $n \ge 1$, the function $\t{\mathfrak{S}}_{\sigma}^{\X{(n)}}(\bm{k};z)$ is \textsl{not} in general describable as such. Below we establish that \emph{the differences between the two moment problems are of \textsl{no} consequence insofar as the specific considerations of this section are concerned.}

On considering the equivalence expression for the function $\t{f}^{\X{(n)}}(z)$ in Eq.\,(\ref{e5p}), one observes that the difference between the two functions $\t{f}^{\X{(n)}}(z)$ and $\t{\mathfrak{S}}_{\sigma}^{\X{(n)}}(\bm{k};z)$ is \textsl{fully} accounted for through substituting the \textsl{analytic} function $\t{g}_n(z) - g_{n;\infty_0}$ for the $0$ in the argument of the function $\mathsf{S}_n(z,0)$ in Eq.\,(\ref{e5p}), where\,\footnote{Evidently, for $\t{\mathfrak{S}}_{\sigma}^{\protect\X{(n)}}(\bm{k};z)$ a non-trivial function of $\bm{k}$, the corresponding functions $\t{g}_n(z)$ and $g_{n;\infty_0}$ are in general similarly non-trivial functions of $\bm{k}$. We note that the notation $g_{n;\infty_0}$ is in keeping with that employed in Eqs\,(\protect\ref{e4n}) and (\protect\ref{e7b}).}\footnote{For the small-$o$ symbol see for instance \S\,1 in Ch. 2, p.\,5, of  Ref.\,\protect\citen{ETC65}. Briefly, $\protect\t{\upphi}(z) = o(\protect\t{\uppsi}(z))$ for, \emph{e.g.}, $z\to\infty$ signifies that to leading order $\protect\t{\upphi}(z)/\protect\t{\uppsi}(z) \sim 0$ as $z\to\infty$, or $\lim_{z\to\infty}\protect\t{\upphi}(z)/\protect\t{\uppsi}(z) =0$.}
\begin{equation}\label{e7v}
\t{g}_n(z) - g_{n;\infty_0} = o(1)\;\; \text{for}\;\; z\to\infty,
\end{equation}
however in general\,\footnote{Note that one also has $\t{f}_n(z) - \alpha_n = o(1)$ for $z\to\infty$, conform the asymptotic series expansions in Eqs\,(\ref{e5c}) and (\protect\ref{e5h}). Similarly, the expression in Eq.\,(\ref{e7v}) is in conformity with the asymptotic series expansion in Eq.\,(\ref{e7s}).} (\emph{cf.} Eq.\,(\ref{e5o}))
\begin{equation}\label{e7w}
\t{g}_n(z) - g_{n;\infty_0} \not\equiv \t{f}_n(z) - \alpha_n.
\end{equation}
To appreciate the reason underlying the equality
\begin{equation}\label{e7ya}
\t{\mathfrak{S}}_{\sigma}^{\X{(n)}}(\bm{k};z) = \mathsf{S}_n(z,\t{g}_n(z) - g_{n;\infty_0}),
\end{equation}
one should consider the exact equality in Eq.\,(\ref{e5o}) with $j$ herein identified with $n$, according to which $\mathsf{S}_n(z,\t{f}_n(z) - \alpha_n)$ identically coincides with the exact self-energy $\t{\Sigma}_{\sigma}(\bm{k};z)$.\footnote{As we have indicated earlier, on account of the equality in Eq.\,(\protect\ref{e7m}), the moment problem associated with $\t{\Sigma}_{\sigma}(\bm{k};z)$ is \textsl{determinate} in the case of Hubbard-like models.} The validity of the functional form in Eq.\,(\ref{e7ya}) rests on the defining expression in Eq.\,(\ref{e7r}) and the infinite asymptotic series in Eq.\,(\ref{e7g}),\footnote{The equality in Eq.\,(\protect\ref{e7sa}) is in conformity with the expressions in Eqs\,(\protect\ref{e7r}) and (\protect\ref{e7g}).} resulting in an infinite asymptotic series expansion for $\t{\mathfrak{S}}_{\sigma}^{\X{(n)}}(\bm{k};z)$ corresponding to $z\to\infty$ (for \textsl{any} finite value of $n$) that \textsl{functionally} is equivalent to that for $\t{\Sigma}_{\sigma}(\bm{k};z)$ in Eq.\,(\ref{e7b}). In this connection, the expression in Eq.\,(\ref{e7w}) reflects the fact that for any finite value of $n$ the sequence $\{\mathfrak{S}_{\sigma;\infty_j}^{\X{(n)}}(\bm{k}) \| j \in \mathds{N}_0\}$ does \textsl{not} correspond to the exact self-energy $\t{\Sigma}_{\sigma}(\bm{k};z)$ (as a given function),\footnote{\emph{Cf.} Eq.\,(\protect\ref{e7u}).} but to a partial sum of its perturbational constituents, determined according to the expression in Eq.\,(\ref{e7r}).

On account of the fact that $\t{\mathfrak{S}}_{\sigma}^{\X{(n)}}(\bm{k};z)$ differs from $\t{\Sigma}_{\sigma}(\bm{k};z)$, for any $n < \infty$, the sequence $\{\mathfrak{S}_{\sigma;\infty_j}^{\X{(n)}}(\bm{k})\| j \in \mathds{N}_0\}$ cannot be assumed to be \textsl{positive} in the sense specified in appendix \ref{sab}.\footnote{See in particular Eq.\,(\protect\ref{e4s}).} Thus the function $-\t{g}_n(z) \equiv -\t{g}_{\sigma;n}(\bm{k};z)$ is not necessarily Nevanlinna for arbitrary finite $n$. However, since $\t{\mathfrak{S}}_{\sigma}^{\X{(n)}}(\bm{k};z)$ is analytic everywhere over the complex $z$-plane outside the region $\im[z] =0$ (see later), the function $\t{g}_{n}(z)$ is similarly \textsl{analytic} in the region $\im[z] \not=0$. By the reflection property  (see the relevant discussions on p.\,\pageref{BothSigmaAndG})
\begin{equation}\label{e7wj}
\t{\mathfrak{S}}_{\sigma}^{\X{(n)}}(\bm{k};z^*) \equiv \t{\mathfrak{S}}_{\sigma}^{\X{(n)}\hspace{0.6pt}*}(\bm{k};z)\;\; \text{for}\;\; \im[z] \not=0,
\end{equation}
the constant $g_{n;\infty_0}$ is also similar to the constant $\alpha_n$ real.\footnote{Regarding $\alpha_n$, see the remark following Eq.\,(\protect\ref{e5i}).}

The\refstepcounter{dummy}\label{TheFunctionMinusG} most fundamental difference between the functions $\t{f}_n(z)$ and $\t{g}_n(z)$ is that whereas $-\t{f}_n(z)$ is Nevanlinna,\footnote{See the paragraph following Eq.\,(\protect\ref{e5ea}).} the function $-\t{g}_n(z)$ may or may not be Nevanlinna. Both functions \textsl{are} however analytic over the entire $z$-plane outside the region $\im[z] =0$.\footnote{This is related to both functions $\protect\t{\Sigma}_{\sigma}(\bm{k};z)$ and $\protect\t{\mathfrak{S}}_{\sigma}^{\protect\X{(n)}}(\bm{k};z)$ being analytic in the region $\protect\im[z]\not=0$. As regards the analyticity of the function $\protect\t{\mathfrak{S}}_{\sigma}^{\protect\X{(n)}}(\bm{k};z)$, see later.} The mentioned difference between the two functions amounts to the possibility of the failure of the equality $\sgn(\im[-\t{g}_n(z)]) = \sgn(\im[z])$, Eq.\,(\protect\ref{e2}), for some region of the $z$-plane outside the region $\im[z] =0$, to be contrasted with $\sgn(\im[-\t{f}_n(z)]) = \sgn(\im[z])$, which is valid for arbitrary finite values of $n$ and \textsl{all} $z$ over the region $\im[z] \not=0$. The region of the $z$-plane outside the real axis where the function $-\t{g}_{n}(z)$ fails to satisfy the equality $\sgn(\im[-\t{g}_n(z)]) = \sgn(\im[z])$, if such region exists, coincides with the region where the function $-\t{\mathfrak{S}}_{\sigma}^{\X{(n)}}(\bm{k};z)$ fails to satisfy a similar equality, and thus is not Nevanlinna. Conversely, for any value of $n$ for which $-\t{g}_n(z)$ is Nevanlinna, the function $-\t{\mathfrak{S}}_{\sigma}^{\X{(n)}}(\bm{k};z)$ is similarly Nevanlinna. Satisfaction of the conditions in Eq.\,(\ref{e4c}) is \textsl{sufficient} for $-\t{\mathfrak{S}}_{\sigma}^{\X{(n)}}(\bm{k};z)$ to be a Nevanlinna function for arbitrary finite value of $n \ge 1$. In this connection, as we have mentioned above, the function $\t{\mathfrak{S}}_{\sigma}^{\X{(n)}}(\bm{k};z)$ and the associated function $\t{g}_n(z)$ are analytic over the entire $z$-plane outside $\im[z] = 0$ for arbitrary \textsl{finite} values of $n$. Consequently, provided the sequence $\{\t{\mathfrak{S}}_{\sigma}^{\X{(n)}}(\bm{k};z)\| n\in \mathds{N}\}$ is convergent, by the Vitali theorem \cite{ECT52,RR98,LW08,AIM65}, this analyticity applies also for $n =\infty$.

The function $-\t{f}_n(z)$ being Nevanlinna simplifies the proof of the uniform convergence of the function $\t{f}^{\X{(n)}}(z)$ towards $\t{f}(z)$ for $n\to \infty$ in the region $\im[z] \not=0$ in the cases of \textsl{determinate}\,\footnote{See Eqs\,(\protect\ref{e4oh}) and (\protect\ref{e7m}) and the attendant discussions.} moment problems. The same simplification would apply if one could establish existence of a \textsl{finite} integer $n_{\X{0}}$, \textsl{independent} of $z$, for which $-\t{g}_n(z)$ would be a Nevanlinna function for all $n \ge n_{\X{0}}$. Identifying $\mathcal{I}(n)$ with $2n +\varsigma$ (\emph{cf.} Eq.\,(\ref{e7ra})), the strict satisfaction of the condition in Eq.\,(\ref{e4c}) would imply that $n_{\X{0}} = 1-\varsigma$. Not wishing to rely on the condition in Eq.\,(\ref{e4c}), which would take us back to the approach in Ref.\,\citen{BF07} described above, \S\,\ref{s3.1}, we now consider whether non-existence of a finite $z$-independent integer $n_{\X{0}}$ can adversely influence the convergence behaviour of the partial sum $\t{\mathfrak{S}}_{\sigma}^{\X{(n)}}(\bm{k};z)$ for $n\to\infty$. To this end, we proceed by considering the function $\t{w}_{n+1}(z,\ptau)$ introduced in Eq.\,(\ref{e6k}) and discussed in some detail in appendix \ref{sab}. For what follows, it proves convenient to introduce the following functions of $z$:
\begin{equation}\label{e7wa}
a_n(z) \doteq Q_n(z),\;\; b_n(z) \doteq - Q_{n+1}(z),\;\; c_n(z) \doteq -P_n(z),\;\; d_n(z) \doteq P_{n+1}(z).
\end{equation}

Following Eqs\,(\ref{e6j}) and (\ref{e6k}), one can write
\begin{equation}\label{e7wb}
\t{w}_{n+1}(z,\zeta) \equiv \frac{a_n(z) \zeta + b_n(z)}{c_n(z) \zeta + d_n(z)},
\end{equation}
where $\zeta$ is a complex variable, introduced here instead of $\ptau$ so as to preserve the convention in appendix \ref{sab}, where $\ptau \in \mathds{R}$ parameterises the circle $\mathsf{K}_{n+1}(z)$.\footnote{The main symbol $\mathsf{K}$ as employed here, traditionally signifies [der] \emph{Kreis}, the \emph{circle} in German.} Expressing the function $\t{w}_{n+1}(z,\zeta)$ as in Eq.\,(\ref{e7wb}) signifies this function as a M\"{o}bius transformation [Ch. 2, \S\,2, p.\,80, in Ref.\,\citen{RR91}] [Ch. 8, \S\,33, in Ref.\,\citen{AIM65}] [Ch. 3 in Ref.\,\citen{TN00}], conformally \cite{TN00} transforming the points of the complex $\zeta$-plane into those of the complex $w$-plane. Since by the Liouville-Ostrogradskii formula [p.\,9 in Ref.\,\citen{NIA65}] one has\,\footnote{See footnote \raisebox{-1.0ex}{\normalsize{\protect\footref{noteq1}}} on p.\,\protect\pageref{TheLS}. Note that the $s_0\hspace{0.6pt}\sqrt{\beta_{n+1}}$ on the RHS of Eq.\,(\protect\ref{e7wc}), instead of $\beta_n$ according to Eq.\,[1.15] in Ref.\,\protect\citen{NIA65}, is tied to the expressions in Eqs\,(\protect\ref{e5y}), (\protect\ref{e5u}), and (\protect\ref{e5x}). The \textsl{positive} constant $s_0$, Eq.\,(\protect\ref{e6c}), is the $0$th-order moment of the relevant measure function; since here we are dealing with $\protect\t{\Sigma}_{\sigma}(\bm{k};z)$, the latter measure function is $\upsigma_{\sigma}(\bm{k};\varepsilon)$, Eq.\,(\protect\ref{e20c}), whereby $s_0 = \Sigma_{\sigma;\infty_1}(\bm{k})$, Eqs\,(\protect\ref{e7j}), (\protect\ref{e7k}), and (\protect\ref{e7c}). One has $\llbracket \Sigma_{\sigma;\infty_1}(\bm{k})\rrbracket = \mathrm{J}\mathrm{s}^{-1}$. For the `Hubbard atom', $\Sigma_{\infty_1} = U^2/(4\hbar)$, Eq.\,(\protect\ref{e7xb}).}
\begin{equation}\label{e7wc}
b_n(z) c_n(z) - a_n(z) d_n(z) = \frac{1}{s_0\hspace{0.6pt}\sqrt{\beta_{n+1}}},\;\; n\in \mathds{N}_0,
\end{equation}
the positivity of $\beta_{n+1}$, Eq.\,(\ref{e5ia}),\footnote{For $\Delta_j$, see Eq.\,(\protect\ref{e6e}).} establishes that the M\"{o}bius transformation in Eq.\,(\ref{e7wb}) is non-singular for all $z$ and $n$ [p.\,124 in Ref.\,\protect\citen{TN00}].\footnote{See Eqs\,(\protect\ref{e5kd}), (\protect\ref{e5kb}), and (\protect\ref{e5j}), as well as Eq.\,(\protect\ref{e5i}). Note that since here by assumption we are considering the cases where the function $\protect\t{\Sigma}_{\sigma}(\bm{k};z)$ and the associated functions cannot be described in terms of a ratio of two finite-order polynomials, we have $\beta_j > 0$ for all $j \in \protect\mathds{N}_{0}$.} With reference to the exact equality in Eq.\,(\ref{e5te}), it should be evident that the $\zeta$-plane, introduced here as the \textsl{domain} of the mapping in Eq.\,(\ref{e7wb}), is to accommodate the set of values taken by the function $\t{f}_{n+1}(z) -\alpha_{n+1}$, $\forall n\in\mathds{N}_{0}$, for all $z\in\mathds{C}$, $\im[z] \not=0$. With $-\t{f}_{n+1}(z)$ a Nevanlinna function, for $z$ varying in the region $\im[z] \gtrless 0$, the $\zeta$ in Eq.\,(\ref{e7wb}), being identified with $\t{f}_{n+1}(z) - \alpha_{n+1}$, will necessarily vary in the region $\im[\zeta] \lessgtr 0$. In contrast, since $-\t{g}_{n+1}(z)$ may not be a Nevanlinna function, account has to be taken of the possibility that for $z$ varying in the region $\im[z] \gtrless 0$, the function $\t{g}_{n+1}(z) - g_{n+1;\infty_0}$ may be everywhere in the $\zeta$-plane, and not necessarily in the region $\im[\zeta] \lessgtr 0$. This is a complicating factor specific to the considerations of this section regarding the self-energy $\t{\Sigma}_{\sigma}(\bm{k};z)$.

With
\begin{equation}\label{e7wd}
v_n(z) \doteq -d_n(z)/c_n(z) \equiv P_{n+1}(z)/P_n(z),
\end{equation}
from Eq.\,(\ref{e7wb}) one observes that $\t{w}_{n+1}(z,\zeta)$ maps $\zeta = v_n(z)$ to the point of infinity of the $w$-plane. Before discussing the relevance of the point $v_n(z)$ in the $\zeta$-plane, it is important to point out that $v_n(z)$ is a Nevanlinna function of $z$, which is established as follows: firstly, since the $n$ zeros of the polynomial $P_n(z)$ are real [Theorem 1.2.2, p.\,10, in Ref.\,\citen{NIA65}], the function $v_n(z)$ is analytic over the entire $z$-plane away from the region $\im[z] = 0$, and, secondly, since the $n$ real zeroes of $P_n(z)$ are bracketed by the $n+1$ real zeroes of $P_{n+1}(z)$ [Theorem 1.2.2, p.\,10, in Ref.\,\citen{NIA65}], a geometrical construction immediately reveals that\,\footnote{Drawing vectors originating from the zeros of the polynomials $P_n(z)$ and $P_{n+1}(z)$ on the real axis to the point $z$ in the $z$-plane, one immediately notices that the above-mentioned bracketing of the $n$ real zeros of $P_n(z)$ by the $n+1$ real zeros of $P_{n+1}(z)$ leads to $\arg(P_{n+1}(z)) \gtrless \arg(P_n(z))$ for $\im[z] \gtrless 0$, establishing the validity of the statement. In this connection, note that, following Eq.\,(\protect\ref{e7wg}) below, the coefficient of $z^m$ in $P_m(z)$, $\forall m$, is positive, so that $P_{n+1}(z)/P_n(z)$ is equal to a \textsl{positive} constant, equal to $A_n \equiv 1/\sqrt{\beta_{n+1}}$, times a ratio of two \textsl{monic} polynomials of $z$. Thus, $v_n(z)$ can be expressed as $A_n \prod_{i=1}^{n+1} (z - \lambda_i^{\protect\X{(n+1)}})/\prod_{i=1}^n (z - \lambda_i^{\protect\X{(n)}})$, where $\lambda_i^{\protect\X{(m)}}$, $\lambda_i^{\protect\X{(m)}}\in \mathds{R}$, denotes the $i$th zero of $P_m(z)$.}
\begin{equation}\label{e7we}
\sgn(\im[v_n(z)]) = \sgn(\im[z])\;\; \text{for}\;\; \im[z] \not=0.
\end{equation}
Thus, \emph{the point $v_n(z)$ is located in the same half-plane of the $\zeta$-plane as $z$ is in the $z$-plane.} Note that from the defining expression for $v_n(z)$ it follows that to leading order
\begin{equation}\label{e7wex}
v_{n}(z) \sim A_n z\;\; \text{for}\;\; z\to\infty,
\end{equation}
where $A_n$ is a \textsl{positive} constant, which follows from the equality [p.\,3 in Ref.\,\citen{NIA65}]
\begin{equation}\label{e7wg}
P_n(z) = \sqrt{\frac{\Delta_{n-1}}{\Delta_n}}\hspace{0.8pt} z^n + R_{n-1}(z),
\end{equation}
where $\Delta_j >0$ is the $j$th \textsl{leading} minor of the underlying Hankel moment matrix, Eqs\,(\ref{e6c}) and (\ref{e6e}), and $R_{n-1}(z)$ some polynomial of degree $n-1$. Thus one has (\emph{cf.} Eq.\,(\ref{e7wc}) and see Eq.\,(\ref{e5ia}))
\begin{equation}\label{e7wgx}
A_n = \frac{1}{\sqrt{\beta_{n+1}}}.
\end{equation}

For\refstepcounter{dummy}\label{ForTheConsiderations} the considerations of this section, the significance of the equality in Eq.\,(\ref{e7we}) lies in the fact that the interior of any disc in the $\zeta$-plane that does (not) contain $v_n(z)$ is conformally mapped by $\t{w}_{n+1}(z,\zeta)$ into the exterior (interior) of a disc in the $w$-plane whose circular boundary coincides with the image of the circular boundary of the disc in the $\zeta$-plane.\refstepcounter{dummy}\label{ConsideringA}\footnote{Considering a disc $\mathscr{D}_{\zeta}$ in the $\zeta$-plane whose circular boundary $\mathscr{C}_{\zeta}$ is oriented clockwise (counter-clockwise), the \textsl{interior} of $\mathscr{D}_{\zeta}$ is the region in the $\zeta$-plane to one's right (left) on walking along $\mathscr{C}_{\zeta}$ in the direction of its orientation. Similarly as regards the association between the disc $\mathscr{D}_w$ and its oriented circular boundary $\mathscr{C}_{w}$ in the $w$-plane. If $v_n(z)$ is not located inside $\mathscr{D}_{\zeta}$, then the M\"{o}bius transformation $\t{w}_{n+1}(z,\zeta)$ maps the interior (exterior) of $\mathscr{D}_{\zeta}$ into the interior (exterior) of $\mathscr{D}_{w}$; otherwise, it maps the interior (exterior) of $\mathscr{D}_{\zeta}$ into the exterior (interior) of $\mathscr{D}_{w}$ [pp.\,148 and 155 in Ref.\,\protect\citen{TN00}]. It follows that in the former (latter) case $\mathscr{C}_{w}$ has the same (opposite) orientation, clockwise or counterclockwise, as $\mathscr{C}_{\zeta}$. Note that here straight infinite lines in the $\zeta$- and $w$-planes are identified as closed circles of infinite radius. \label{notex}} A straight line in the $\zeta$-plane being a circle of infinite radius \cite{TN00}, for $\zeta$ traversing the real axis of the $\zeta$-plane (that is, the $\ptau$-axis of appendix \ref{sab}) from $-\infty$ to $+\infty$, in the case at hand for $\im[z] \gtrless 0$ the entire lower/upper half-plane of the $\zeta$-plane is by $\t{w}_{n+1}(z,\zeta)$ conformally mapped, on account of the equality in Eq.\,(\ref{e7we}), into the interior of the disk bounded by the circle $\mathsf{K}_{n+1}(z)$, appendix \ref{sab}. For \textsl{determinate} moment problems, the radius of $\mathsf{K}_{n+1}(z)$ uniformly converges to $0$ for $n\to\infty$. With reference to the expressions in Eqs\,(\ref{e5tc}), (\ref{e5te}), and (\ref{e6k}) (see also Eq.\,(\ref{e6qb})), the complex parameter $\zeta$ representing the function $\t{f}_{n+1}(z) -\alpha_{n+1}$, with $-\t{f}_{n+1}(z)$ a Nevanlinna function, here the region $\im[z] \gtrless 0$ is pertinent to the region $\im[\zeta] \lessgtr 0$ of the $\zeta$-plane. In the light of the above statement with regard to $\mathsf{K}_{n+1}(z)$, the Nevanlinna nature of the function $-\t{f}_{n+1}(z)$ signals the progressive \textsl{insignificance} of $\t{f}_{n+1}(z) -\alpha_{n+1}$ to the deviation of the approximant $\t{f}^{\X{(n+1)}}(z)$ from $\t{f}(z)$ in the region $\im[z] \not=0$ for increasing values of $n$, leading to the conclusion that for the case of the moment problem at hand being \textsl{determinate}, for $n\to \infty$ the approximant $\t{f}^{\X{(n+1)}}(z)$ converges uniformly to $\t{f}(z)$ over the entire finite part of the $z$-plane outside the region $\im[z] = 0$. In appendix \ref{sab} we demonstrate that the moment problem associated with the \textsl{exact} self-energy $\t{\Sigma}_{\sigma}(\bm{k};z)$ pertaining to Hubbard-like models is indeed \textsl{determinate}, Eq.\,(\ref{e7m}).

With the function $-\t{g}_n(z)$ associated with the partial sum $\t{\mathfrak{S}}_{\sigma}^{\X{(n)}}(\bm{k};z)$ not being necessarily a Nevanlinna function for arbitrary values of $n \in\mathds{N}$, the question arises as to the behaviour of this function for $n\to\infty$. To discuss this problem, it proves convenient to express the function $\t{w}_{n+1}(z,\zeta)$ in Eq.\,(\ref{e7wb}) in the following equivalent but more appealing form [p.\,124 in Ref.\,\citen{TN00}] [p.\,85 in Ref.\,\citen{RR91}]:\,\footnote{\emph{E.g.}, $\protect\t{w}_1(z,\zeta) = -1/(z - \alpha_1 - \sqrt{\beta}_1\hspace{0.6pt} \zeta) \equiv (1/\sqrt{\beta_1})/\big(\zeta - (z-\alpha_1)/\sqrt{\beta_1}\big)$, which is independent of $s_0$.}
\begin{equation}\label{e7wf}
\t{w}_{n+1}(z,\zeta) = \frac{1}{s_0\hspace{0.6pt}\sqrt{\beta_{n+1}}\hspace{0.6pt} c_n^2(z)} \frac{1}{\zeta - v_n(z)} + \frac{a_n(z)}{c_n(z)},
\end{equation}
where we have used the equality in Eq.\,(\ref{e7wc}). The expression in Eq.\,(\ref{e7wf}) makes in particular explicit the way in which the asymptotic region $\zeta \to v_n(z)$ in the $\zeta$-plane is conformally mapped by $\t{w}_{n+1}(z,\zeta)$ into the asymptotic region $w\to\infty$ in the $w$-plane.

The asymptotic expressions $v_n(z) \sim A_n z$ and $\t{g}_n(z) \sim g_{n;\infty_0}$ for $z\to\infty$ imply that for $\zeta \equiv \zeta_n(z)$, where
\begin{equation}\label{e7wh}
\zeta_n(z) \doteq \t{g}_n(z) - g_{n;\infty},
\end{equation}
the equation
\begin{equation}\label{e7wi}
\zeta_n(z) = v_n(z)
\end{equation}
\textsl{cannot} be satisfied in the asymptotic region $z\to\infty$ of the $z$-plane, which is not surprising in the light of the expressions in Eqs\,(\ref{e7s}), (\ref{e7t}), and (\ref{e7b}). Importantly, Eq.\,(\ref{e7wi}) \textsl{cannot} be satisfied anywhere in the region $\im[z] \not=0$ of the $z$-plane, since a solution in this region would imply violation of the analyticity of the function $\t{\mathfrak{S}}_{\sigma}^{\X{(n)}}(\bm{k};z)$ in the region $\im[z] \not=0$ (see above, p.\,\pageref{TheFunctionMinusG}). The possible solutions of the equation in Eq.\,(\ref{e7wi}), if any, are therefore real and countable,\footnote{By the `interior uniqueness theorem' of analytic functions [\S\,4.12, p.\,139, in Ref.\,\protect\citen{ECT52}] [Theorem 17.1, p.\,369, in Ref.\,\protect\citen{AIM65}], satisfaction of the equality in Eq.\,(\protect\ref{e7wi}) over a line segment or a discrete set of points with an accumulation point over the $z$-plane implies the identity of $\zeta_n(z)$ with $v_n(z)$ over this plane. Even admitting the possibility of $\zeta_n(z) \equiv v_n(z)$ for some $n$, the two functions \textsl{cannot} be identical for all $n \in \protect\mathds{N}$, for the two evidently depend differently on the coupling-constant of interaction.} and thus comprise a subset of measure zero of the $z$-plane $\mathds{C}$. We have thus established that although for finite values of $n$ the function $-\t{\mathfrak{S}}_{\sigma}^{\X{(n)}}(\bm{k};z)$ may or may not be a Nevanlinna function of $z$, implying the associated function $-\t{g}_{n}(z)$ to be accordingly respectively Nevanlinna or non-Nevanlinna, the analyticity of $\t{\mathfrak{S}}_{\sigma}^{\X{(n)}}(\bm{k};z)$, and therefore of  $\t{g}_{n}(z)$, in the region $\im[z] \not=0$ of the $z$-plane guarantees that, similarly to the approximant $\t{f}^{\X{(n)}}(z)$, for $\im[z] \not=0$ the function $\t{\mathfrak{S}}_{\sigma}^{\X{(n)}}(\bm{k};z)$ will be confined to the disc\,\footnote{See p.\,\protect\pageref{ForTheConsiderations}.} circumscribed by the circle $\mathsf{K}_{n}(z)$ whose radius uniformly converges towards $0$ for $n \to\infty$ in the case the moment problem associated with the self-energy $\t{\Sigma}_{\sigma}(\bm{k};z)$, described in terms of the \textsl{positive sequence}\,\footnote{For definition, see Eq.\,(\protect\ref{e4s}) \emph{et seq.}} of moments $\{\Sigma_{\sigma;\infty_j}(\bm{k})\| j\}$, is \textsl{determinate}.\footnote{See Eq.\,(\protect\ref{e7m}) and the attendant discussions. See also \S\,\protect\ref{sec.3.2.2} below.} In such case, $\t{\mathfrak{S}}_{\sigma}^{\X{(n)}}(\bm{k};z)$ converges uniformly to the exact self-energy $\t{\Sigma}_{\sigma}(\bm{k};z)$ over the entire $z$-plane outside the region $\im[z] =0$ for $n\to\infty$. Thus, while the truncated moment problems associated with the sequences of approximants $\{\t{f}^{\X{(n)}}(z)\| n\}$ and $\{\t{\mathfrak{S}}_{\sigma}^{\X{(n)}}(\bm{k};z)\| n\}$ are fundamentally distinct,\footnote{In the context of the present considerations, this distinction is fundamental \textsl{only} if $-\t{\mathfrak{S}}_{\sigma}^{\X{(n)}}(\bm{k};z)$ is not a Nevanlinna function of $z$ for $n$ over an unbounded subset of $\mathds{N}$.} nonetheless the implication of this distinction is strictly confined to a countable set of points with \textsl{no} accumulation points on the real axis of the $z$-plane. For a given $\bm{k}$, this set may or may not be empty. For some additional details relevant to the discussion at hand, see \S\,\ref{sec.3.2.2} below.

The above observations shed light on the fact that the requirement of the function $-\t{f}_{n+1}(z)$ being Nevanlinna is unnecessarily too strong for the function $\t{f}^{\X{(n)}}(z)$ uniformly converging towards $\t{f}(z)$ for $n\to\infty$ in the region $\im[z] \not=0$. With reference to the remarks on p.\,\pageref{ForTheConsiderations} (in particular those concerning the discs $\mathscr{D}_{\zeta}$ and $\mathscr{D}_{w}$),\footnote{See footnote \raisebox{-1.0ex}{\normalsize{\protect\footref{notex}}} on p.\,\protect\pageref{ConsideringA}.} for the latter convergence the requirement of the equation (\emph{cf.} Eqs\,(\ref{e7w}), (\ref{e7wh}), and (\ref{e7wi}))
\begin{equation}\label{e7wh1}
\t{f}_n(z) -\alpha_n = v_n(z)
\end{equation}
not possessing any solution in the region $\im[z] \not=0$ would suffice. With $-\t{f}_n(z)$ and $v_n(z)$ being Nevanlinna, for $\im[z] \not=0$ the sign of the imaginary part of the LHS of Eq.\,(\ref{e7wh1}) is opposite to that of the imaginary part of the RHS, whereby Eq.\,(\ref{e7wh1}) has indeed no solution in the region $\im[z] \not=0$.

In appendix \ref{sab} we demonstrate that for Hubbard-like models the moment problem associated with the \textsl{positive sequence} $\{\Sigma_{\sigma;\infty_j}(\bm{k})\| j\}$ is a \textsl{determinate} one (see in particular Eq.\,(\ref{e7m})).\footnote{See also \S\,\protect\ref{sec.3.2.2} below.} In the light of the above observations, in particular of the identity in Eq.\,(\ref{e7t}) and the analytic properties of the sequence of functions $\{\t{\mathfrak{S}}_{\sigma}^{\X{(n)}}(\bm{k};z)\| n\}$, p.\,\pageref{TheFunctionMinusG}, the latter observation implies that the sequence $\{\t{\mathfrak{S}}_{\sigma}^{\X{(n)}}(\bm{k};z)\| n\}$ converges uniformly \ae to the exact self-energy $\t{\Sigma}_{\sigma}(\bm{k};z)$. \emph{This completes the proof of the uniform convergence of the perturbation series in Eq.\,(\ref{e4}) for almost all $\bm{k}$ and $z$ in the case of the uniform GSs of Hubbard-like models.}\footnote{``A series $\sum f_{\nu}$ of functions converges uniformly in $A$ if the sequence $s_n = \sum^n f_{\nu}$ of partial sums converges uniformly in $A$'' [Ch. 3, \S\,1, p. 93, in Ref.\,\protect\citen{RR91}].}

The considerations of this section have exposed deep connections between
\vspace{0.1cm}
\begin{itemize}
\item[(i)] the convergence of the perturbation series expansion for the self-energy $\t{\Sigma}_{\sigma}(\bm{k};z)$ (more explicitly, the \textsl{uniform} convergence of this series \ae) in Eq.\,(\ref{e4}),
\item[(ii)] the Nevanlinna nature of the function $-\t{\Sigma}_{\sigma}(\bm{k};z)$, implying its analyticity over the complex $z$-plane outside the region $\im[z] =0$ \textsl{and} a specific way it decays in the asymptotic region $z\to\infty$ (\emph{cf.} Eqs\,(\ref{e4da}) and (\ref{e7b})), and
\item[(iii)] the analyticity of the elements of the sequence $\{\t{\Sigma}_{\sigma}^{\X{(\nu)}}(\bm{k};z)\| \nu\}$ of the perturbational terms (where $-\t{\Sigma}_{\sigma}^{\X{(\nu)}}(\bm{k};z)$, $\forall \nu$, may or may not be a Nevanlinna function of $z$) over the complex $z$-plane away from the region $\im[z] = 0$.
\end{itemize}
\vspace{0.1cm}
\noindent
The significance of the connections between (i), (ii), and (iii) is most clearly illustrated by the simple example that we have presented in the paragraph containing Eqs\,(\ref{e1a}) -- (\ref{e1d}), beginning on p.\,\pageref{ForIllustration}. Earlier, we have explicitly shown how failure of $\t{\Sigma}_{\sigma}(\bm{k};z)$ to be analytic in the region $\im[z] \not=0$ of the $z$-plane directly leads to the failure of the Luttinger-Ward identity and the Luttinger theorem [\S\,6.2 in Ref.\,\citen{BF07}]. It stands therefore to reason that by employing the Riesz-Herglotz representation\,\footnote{See Eq.\,(\protect\ref{e20h}). See also our remark on p.\,\protect\pageref{WeShouldEmphasise}, in the concluding part of the paragraph preceding that beginning with `The perturbation series ...'.} of the relevant functions (the Green functions $\{\t{G}_{\sigma}(\bm{k};z)\| \sigma\}$ and the self-energy $\t{\Sigma}_{\sigma}(\bm{k};z)$, Eqs\,(\ref{e4j}), (\ref{e4l}), (\ref{e20d}), and (\ref{e20g})), and solving for the relevant non-decreasing bounded measure functions  \cite{PRH50,APM11}, in practice one can entirely avoid `non-physical' solutions in self-consistent many-body calculations.

\refstepcounter{dummyX}
\subsubsection{Additional details}
\phantomsection
\label{sec.3.2.2}
For completeness, defining
\begin{equation}\label{e7wk}
\rho_n(\bm{k}) \doteq \sum_{j=1}^n \frac{1}{\sqrt[2j]{\mathfrak{S}_{\sigma;\infty_{2j+1}}^{\protect\X{(n)}}(\bm{k})}},
\end{equation}
for $\mathcal{I}(n) \ge 2n+1$, Eq.\,(\protect\ref{e7ra}), in the light of Eq.\,(\protect\ref{e7t}), one has (\emph{cf.} Eq.\,(\ref{e7m}))
\begin{equation}\label{e7wl}
\rho_n(\bm{k}) \equiv \sum_{j=1}^n \frac{1}{\sqrt[2j]{\Sigma_{\sigma;\infty_{2j+1}}(\bm{k})}}.
\end{equation}
On account of the positive sequence $\{\Sigma_{\sigma;\infty_j}(\bm{k}) \| j\}$ being \textsl{determinate}, Eq.\,(\ref{e7m}), it follows that
\begin{equation}\label{e7wm}
\lim_{n\to\infty} \rho_n(\bm{k}) = \infty,\;\; \forall\bm{k}.
\end{equation}
We note that the sum on the LHS of Eq.\,(\ref{e7m}) is a \textsl{lower} bound of the sum
\begin{equation}\label{e7wn}
\mathcal{S} \doteq\sum_{j=1}^{\infty} \frac{1}{\sqrt{\beta_j}},
\end{equation}
where $\{\sqrt{\beta_j}\| j \in \mathds{N}\}$ are the diagonal elements above or below the main diagonal of the symmetric tridiagonal Jacobi matrix $\mathscr{J}^{\X{\t{\Sigma}_{\sigma}}}$ corresponding to the self-energy $\t{\Sigma}_{\sigma}(\bm{k};z)$, Eqs\,(\ref{e5j}) and (\ref{e5z}). The moment problem at hand is \textsl{determinate} if and only if the latter Jacobi matrix is of the $\mathfrak{D}$ type, appendix \ref{sab}. This is the case for $\mathcal{S} = \infty$ [p.\,24 in Ref.\,\citen{NIA65}]. The above-mentioned lower bound is obtained through the application of the Carleman inequality [p.\,85 in Ref.\,\citen{NIA65}] [\S\,9.12, p.\,249, in Ref.\,\citen{HLP34}]. Carleman's inequality having been deduced for an \textsl{infinite} sum, the question arises as to the validity of the process of considering the sum
\begin{equation}\label{e7wo}
\mathcal{S}_n \doteq\sum_{j=1}^{n} \frac{1}{\sqrt{\beta_j}},
\end{equation}
and on the basis of  (\emph{cf.} Eq.\,(\ref{e7wm}))
\begin{equation}\label{e7wp}
\lim_{n\to\infty} \mathcal{S}_n = \infty
\end{equation}
arriving at the conclusion that the moment problem at hand is determinate. \emph{The answer to this question is that this process is indeed valid.} With reference to Eq.\,(\ref{e4og}), this follows from the fact that [\emph{cf.} p.\,86 in Ref.\,\citen{NIA65}]\,\footnote{In Eq.\,(\protect\ref{e7wq}), $\mathrm{e} \equiv \ln^{-1}(1)= 2.718\dots\;$.}\footnote{We note that while $\sum_{k=j}^{\infty} 1/[k (k+1)] = 1/j$, one has $\sum_{k=j}^{n} 1/[k (k+1)] = 1/j - 1/(n+1)$. As regards the former sum and its relation to the Carleman inequality, see p.\,250 of Ref.\,\protect\citen{HLP34}.}
\begin{equation}\label{e7wq}
\sum_{j=1}^{n} \frac{1}{\sqrt[2j]{s_{2j}}} \le \sum_{j=1}^{n} \frac{1}{\sqrt[2j]{\beta_1 \beta_2\dots \beta_{j}}} < \mathrm{e} \sum_{j=1}^{n} \big(1 - \frac{j}{n+1}\big) \frac{1}{\sqrt{\beta_{j}}} < \mathrm{e}\hspace{0.8pt}\mathcal{S}_n,
\end{equation}
where we have used the fact that $\beta_j > 0$ for $j\in \{1,2,\dots, n\}$, Eq.\,(\ref{e5kb}),\footnote{See also Eqs\,(\protect\ref{e6e}) and (\protect\ref{e5ia}).} and that $0 < j/(n+1) < 1$ for $j \in \{1,2,\dots,n\}$. The contribution $j/(n+1)$ is identically vanishing in dealing with $n=\infty$ from the outset. Clearly,
\begin{equation}\label{e7wr}
\lim_{n\to\infty}  \sum_{j=1}^{n} \frac{1}{\sqrt[2j]{s_{2j}}}= \infty\; \Longrightarrow\; \lim_{n\to\infty} \mathcal{S}_n = \infty,
\end{equation}
signifying that indeed the relevant Jacobi matrix is of the $\mathfrak{D}$ type.\footnote{Alternatively, with reference to Eq.\,(\ref{e6ub}), we note that for $\protect\im[z]\not=0$ the radius $\rho_{n+1}^{\mathrm{c}}(z) \to 0$ for $n\to \infty$ when $\sum_{j=0}^n \vert P_j(z)\vert^2 \to \infty$ for $n\to\infty$ [Eq.\,[2.23], p.\,49, in Ref.\,\protect\citen{NIA65}]. With reference to the explicit expression for $P_j(z)$, $j\in\mathds{N}$, in terms of the determinants of the Hankel moment matrices $\mathbb{H}_{j-1}$ and $\mathbb{H}_{j}$, Eq.\,(\protect\ref{e6c}), and the determinant of the \textsl{bordered} Hankel moment matrix $\mathbb{H}_{j}$ [Eq.\,[1.4], p.\,3, in Ref.\,\protect\citen{NIA65}], one observes that $\{P_j(z) \| j=1,2,\dots,n\}$ is fully determined in terms of the moments $\{s_0, s_1, \dots, s_{2n}\}$. This aspect is indeed reflected in the sum on the LHS of Eq.\,(\protect\ref{e7wq}). Therefore, starting with the function $\rho_{n}(\bm{k})$, Eq.\,(\protect\ref{e7wk}), for an arbitrary \textsl{finite} value of $n$, the result in Eq.\,(\protect\ref{e7wm}) establishes that the moment problem at hand, corresponding to the function $\protect\t{\mathfrak{S}}_{\sigma}^{\protect\X{(n)}}(\bm{k};z)$, is indeed \textsl{determinate}.}

\refstepcounter{dummyX}
\subsubsection{Remarks}
\phantomsection
\label{sec.3.2.3}
A textbook example often advanced in asserting that perturbation series were generally asymptotically divergent (in line with the argument by Dyson in Ref.\,\protect\citen{FJD52}, criticised by Simon \cite{SD70} as constituting a `folk theorem') corresponds to the following function [Eqs\,(2.24) and (7.117) in Ref.\,\protect\citen{NO98}], [Eq.\,(5.1) in Ref.\,\protect\citen{AS10}]:
\begin{equation}\label{e7ws}
\EuScript{I}(g) \doteq \int_{-\infty}^{\infty} \frac{\rd x}{\sqrt{2\pi}}\, \e^{-\frac{1}{2} x^2 - g x^4}.
\end{equation}
The `perturbation expansion' corresponding to small values of $g$ is obtained through Taylor expanding $\e^{-g x^4}$ in powers of $g\hspace{0.4pt} x^4$ and subsequently exchanging the order of summation and integration. As we shall see below, this `perturbation expansion' is \emph{a priori} problematical, however what is most relevant from the perspective of the considerations of this publication, is that the RHS of the above expression is \textsl{not} a functional of $\EuScript{I}(g)$, to be contrasted with the perturbation series expansion of $\t{\Sigma}_{\sigma}(\bm{k};z)$ in terms of skeleton self-energy diagrams, Eq.\,(\ref{e4}), where the contribution of each diagram is a functional of $\{\t{G}_{\sigma}(\bm{k};z)\| \sigma\}$ and thus of $\{\t{\Sigma}_{\sigma}(\bm{k};z)\| \sigma\}$.

As for the problematic nature of the above-mentioned `perturbation expansion' of $\EuScript{I}(g)$ in powers of $g$, importantly $g$ is multiplied by the function $x^4$, which is \textsl{unbounded} over the unbounded interval $(-\infty,\infty)$, rendering the notion of a \textsl{non-vanishing} $g$ as being `small' meaningless. In this connection, note that no matter how small $\vert g\vert$ is, for $g < 0$ the unperturbed part of the integrand is less than the perturbation part, as one has $\e^{-\frac{1}{2} x^2} \le \e^{- g x^4}$, $\forall x$.\,\footnote{Leaving aside that the integral in Eq.\,(\protect\ref{e7ws}) does not exist for $g<0$, although $\EuScript{I}(g)$ can be analytically continued to the region $\re[g] <0$, as attested by the exact expression in Eq.\,(\protect\ref{e7wt}) below.} Even for $g>0$, the latter inequality applies for \textsl{all} $\vert x\vert \le 1/\sqrt{2 g}$. Note that $(-1/\sqrt{2 g}, +1/\sqrt{2g}) \to (-\infty,+\infty)$ for $g\downarrow 0$.

For $\re[g] \ge 0$ one has
\begin{equation}\label{e7wt}
\EuScript{I}(g) = \sqrt{2\gamma/\pi} \e^{\gamma}\hspace{-1.4pt} K_{1/4}(\gamma),
\end{equation}
where $\gamma \doteq 1/(32 g)$, and $K_{\nu}(z)$ is the modified Bessel function of the second kind  [\S\,9.6, p.\,374, in Ref.\,\protect\citen{AS72}]. In particular [\S\,9.6.9, p.\,375, in Ref.\,\protect\citen{AS72}],
\begin{equation}\label{e7wu}
\EuScript{I}(g) \sim \frac{\Gamma(1/4)}{2\sqrt{2\pi}}\hspace{0.6pt}\frac{1}{g^{1/4}} + \dots\;\; \text{for} \;\; g\to \infty,
\end{equation}
which clearly \textsl{cannot} be described in terms of the asymptotic sequence [Ch. 2 in Ref.\,\protect\citen{ETC65}] $\{1, g, g^2, \dots\}$ specific to the region $g \to 0$ (for some relevant details, consult appendix C in Ref.\,\protect\citen{BF13}). We note that the series expansion of $\EuScript{I}(g)$ in terms of the latter asymptotic sequence is Borel summable \cite{GHH73} [p.\,373 in Ref.\,\protect\citen{NO98}]. The function obtained in this way identically coincides with the exact expression on the RHS of Eq.\,(\ref{e7wt}).

\refstepcounter{dummyX}
\subsection{On the multiplicity of self-energies in self-consistent calculations and how to enforce appropriate function spaces}
\phantomsection
\label{s3.3}
In the light of the considerations of the previous sections, given the \textsl{exact} one-particle Green functions $\{\t{G}_{\sigma}(\bm{k};z)\|\sigma\}$ corresponding to the $N$-particle uniform GS (ES) of a Hubbard-like Hamiltonian,  it is strictly ruled out that the series in Eq.\,(\ref{e4}) would yield any other function but the exact self-energy $\t{\Sigma}_{\sigma}(\bm{k};z)$ for almost all $\bm{k}$ and $z$.\footnote{As we have indicated in \S\,\protect\ref{sec.3.2.1}, p.\,\pageref{WeDiscussIn}, this conclusion does \textsl{not} apply to the cases where the underlying Green functions are \textsl{local} (that is, non-dispersive), in particular as in these cases a zero-measure subset of the underlying $\bm{k}$-space is devoid of meaning.} Consequently, the general observation by Kozik \emph{et al.} \cite{KFG14}, namely that ``$\Sigma[G]$'' ``is not a single-valued functional of $G$'', is unquestionably false. The same applies insofar as the observations by Stan \emph{et al.} \cite{SRRRB15}, and Rossi and Werner \cite{RW15} to the same effect are concerned.

Multiplicity of self-energies are however \textsl{not} \emph{a priori} ruled out in attempts to solve the self-energy in a self-consistent fashion, for instance within the following framework. Denoting the function $\t{\Sigma}_{\sigma}^{\X{(\nu)}}(\bm{k};z)$ as encountered in Eq.\,(\ref{e4}) more explicitly as $\t{\Sigma}_{\sigma}^{\X{(\nu)}}(\bm{k};z;[\{\X{\t{G}_{\sigma'}}\}])$ \cite{BF19}, with reference to the expressions in Eqs\,(\ref{e4}) and (\ref{e7r}), one can consider the following \textsl{equation}\,\footnote{With reference to the remark at the outset of \S\,\protect\ref{s.301} (p.\,\protect\pageref{s.301}), the expression in Eq.\,(\protect\ref{e19a}) is to be used \textsl{only} if the formal perturbation series expansion of $\protect\t{\Sigma}_{\sigma}(\bm{k};z)$ in terms of the proper self-energy diagrams (including both skeleton and non-skeleton diagrams) and the non-interacting Green functions $\{\protect\t{G}_{\protect\X{0};\sigma}(\bm{k};z)\|\sigma\}$ is \textsl{non-terminating}.} [Eq.\,(5.37), p.\,35, in Ref.\,\citen{BF07}]:\,\footnote{\emph{Cf.} Eq.\,(\protect\ref{e119b}) below.}
\begin{equation}\label{e19a}
\t{\Sigma}_{\sigma\phantom{\prime}}^{\X{[m]}}(\bm{k};z) = \sum_{\nu=1}^{\mathpzc{I}(m)} \t{\Sigma}_{\sigma}^{(\nu)}(\bm{k};z;[\{\t{G}_{\sigma'}^{\X{[m]}}\}]),
\end{equation}
where, as in Eq.\,(\ref{e7r}), $\mathpzc{I}(m)$ is an increasing integer-valued function of $m$, without however being necessarily subject to the restriction imposed by the inequality in Eq.\,(\ref{e7ra}); thus, while one may identify $\mathpzc{I}(m)$ with $\mathcal{I}(m)$, one may for instance also opt for $\mathpzc{I}(m) \doteq m$. As for the function $\t{\Sigma}_{\sigma}^{\X{(\nu)}}(\bm{k};z;[\{\X{\t{G}_{\sigma'}^{[m]}}\}])$ on the RHS, it denotes the total contributions of all skeleton self-energy diagrams of order $\nu$ evaluated in terms of the Green functions $\{\t{G}_{\sigma}^{\X{[m]}}(\bm{k};z)\| \sigma\}$ that through the Dyson equation, Eq.\,(\ref{e4a}), correspond to the self-energies $\{\t{\Sigma}_{\sigma}^{\X{[m]}}(\bm{k};z)\| \sigma\}$:\,\footnote{For a discussion of the importance of self-consistency to the Luttinger theorem, see Ref.\,\protect\citen{BF13}.}
\begin{equation}\label{e19b}
\t{G}_{\sigma}^{\X{[m]}}(\bm{k};z) \doteq \frac{\hbar}{\hbar/\t{G}_{\X{0};\sigma}(\bm{k};z) - \hbar\t{\Sigma}_{\sigma}^{\X{[m]}}(\bm{k};z)},\;\; \forall \sigma.
\end{equation}
The combination of the expressions in Eqs\,(\ref{e19a}) and (\ref{e19b}) constitute a highly non-linear self-consistency equation the uniqueness of whose solution is not guaranteed. In fact, in dealing with the `Hubbard atom' in \S\,\ref{sec2.a} we have already encountered a nonlinearity of a quadratic form in calculating the non-interacting Green function corresponding to the exact interacting Green function of this system, leading to two solutions, neither of which proves to qualify as a Green function; only after identifying the circle $\vert z\vert =U/2$ in the complex $z$-plane over which the two solutions are discontinuous, and a matching of a specific \textsl{unique} pair of solutions over this circle, Eq.\,(\ref{e1c}), have we been able to construct the appropriate solution. In doing so, the two requirements regarding \textbf{(i)} the analyticity of the non-interacting Green function over the region $\im[z] \not=0$ of the $z$-plane, and \textbf{(ii)} the decay of this function to leading order like $\hbar/z$ for $z\to\infty$, have proved sufficient for determining the correct, and unique, combination of the two solutions of the non-linear equation for the non-interacting Green function. Interestingly, the above two conditions equally apply to the exact interacting Green function \cite{JML61,BF07}, appendix \ref{sab}.

At this stage, it is not known to us whether imposition of similar conditions as above would render the solution of the coupled non-linear equations in Eqs\,(\ref{e19a}) and (\ref{e19b}) unique. Nonetheless, it is important to realise that the above-mentioned conditions \textbf{(i)} and \textbf{(ii)} regarding the one-particle Green function (whether interacting or non-interacting) are fully accounted for by expressing this function in the form presented in Eq.\,(\ref{e4j})\footnote{Regarding the measure function specific to $\protect\t{G}_{\protect\X{0};\sigma}(\bm{k};z)$, see the expression in Eq.\,(\protect\ref{e6s}).} or Eq.\,(\ref{e4l}). In adopting for instance the expression in Eq.\,(\ref{e4j}), any non-decreasing bounded real function of $\varepsilon$, approximating the actual measure function $\upgamma_{\sigma}(\bm{k};\varepsilon)$, that varies between $0$ and $\hbar$, Eq.\,(\ref{e4i}), guarantees not only that the corresponding function $-\t{G}_{\sigma}(\bm{k};z)$ is Nevanlinna, but also that this $\t{G}_{\sigma}(\bm{k};z)$ to leading order decays like $\hbar/z$ in the asymptotic region $z\to\infty$. Similarly as regards the self-energy $\t{\Sigma}_{\sigma}(\bm{k};z)$, for which the appropriate representations are given in Eqs\,(\ref{e20d}) and (\ref{e20g}).

In solving the coupled non-linear equations in Eqs\,(\ref{e19a}) and (\ref{e19b}), one could employ the representation in Eq.\,(\ref{e20d}) and solve for the measure function $\upsigma_{\sigma}^{\X{[m]}}(\bm{k};\varepsilon)$ is the space of non-decreasing bounded real functions of $\varepsilon$. In the light of the considerations of in particular the preceding section, it is expected that by fixing an increasing number of the moments of this function (see Eqs\,(\ref{e7b}) -- (\ref{e7k})), making use of the exact results in Eqs\,(\ref{e7ba}), (\ref{e7c}) -- (\ref{e7e}), \emph{etc.}, \cite{Note11} in which the functions $\{G_{\sigma;\infty_j}(\bm{k})\| j\}$ are determined on the basis of the exact expression in Eq.\,(\ref{e4p}), only one solution will obtain. Note that the latter sequence of functions is similarly to be used to constrain the sought-after measure function $\upgamma_{\sigma}(\bm{k};\varepsilon)$ corresponding to the Green function, Eqs\,(\ref{e4oa}) and (\ref{e4ob}). In the light of the expression in Eq.\,(\ref{e7g}),\footnote{See also Eqs\,(\protect\ref{e7ra}), (\protect\ref{e7t}), and (\protect\ref{e7h}).} it is reasonable to expect that constraining $\upsigma_{\sigma}^{\X{[m]}}(\bm{k};\varepsilon)$ to reproduce the moments $\{\Sigma_{\sigma;\infty_j}(\bm{k}) \| j=0,1,\dots, \mathpzc{I}(m)-1\}$ should result in a unique solution to Eqs\,(\ref{e19a}) and (\ref{e19b}), and that additional constraints (that is those beyond these and that $\upsigma_{\sigma}^{\X{[m]}}(\bm{k};\varepsilon)$ be real, bounded and non-decreasing) should amount to over-specification, resulting in the possibility of Eqs\,(\ref{e19a}) and (\ref{e19b}) having no solution. It is naturally possible to solve such an over-specified problem in some optimal sense (requiring introduction of a norm for quantifying the closeness of the LHS of Eq.\,(\ref{e19a}) to its RHS), thus bypassing the problem of the non-existence of solution.

In \S\,\ref{sec.5} we present a formalism for the construction of the measure function $\upgamma_{\sigma}(\bm{k};\varepsilon)$ pertaining to the one-particle Green function $\t{G}_{\sigma}(\bm{k};z)$, Eq.\,(\ref{e4j}), on the basis of the sequence of the perturbational contributions $\{\t{G}_{\sigma}^{\X{(\nu)}}(\bm{k};z)\| \nu\}$ to this function. This formalism can similarly be applied for the calculation of the measure function $\upsigma_{\sigma}(\bm{k};\varepsilon)$ pertaining to the self-energy $\t{\Sigma}_{\sigma}(\bm{k};z)$,\footnote{See footnote \raisebox{-1.0ex}{\normalsize{\protect\footref{notey}}} on p.\,\protect\pageref{InDealingWith}.} Eq.\,(\ref{e20d}).

\refstepcounter{dummyX}
\section{On the multiplicity of self-energy functions as observed by Kozik \emph{et al.}}
\phantomsection
\label{sec3}
In this section we subject the observations by Kozik \emph{et al.} \cite{KFG14} to a critical analysis.\footnote{For simplicity of notation, in this section we generally suppress the spin index $\sigma$ (thus, \emph{e.g.}, $\protect\t{G}(z)$ denotes $\protect\t{G}_{\sigma}(z)$). This is possible because for the `Hubbard atom' all the relevant correlation functions are invariant under the transformation of $\sigma$ into $\protect\b\sigma$.} The order of the following subsections is mainly determined by the order in which the mathematical expressions underlying the discussions are most naturally introduced, and not necessarily by the significance that we attach to the subject matters dealt with in these subsections.

We begin in \S\,\ref{sec3.a} by calculating the site double-occupancy $D$ corresponding to the atomic limit of the Hubbard Hamiltonian for spin-$\tfrac{1}{2}$ particles at half-filling (abbreviated \textsl{the `Hubbard atom'}, \S\,\ref{sec3.b}) on the basis of the expressions for $D$ and the exact one-particle Green $\t{G}(z)$ function and self-energy $\t{\Sigma}(z)$ as presented in Ref.\,\citen{KFG14}.\footnote{Eqs\,(\ref{e23}), (\ref{e24}), and (\ref{e25}) below.} We show that the indicated expressions for $D$, $\t{G}(z)$, and $\t{\Sigma}(z)$ are incompatible, resulting in \textsl{negative} values for $D$ for \textsl{all} $\beta U > 0$, where $\beta \equiv 1/(k_{\textsc{b}} T)$, with $T$ denoting the absolute temperature, and $U$ the on-site interaction energy. Although the appropriate expression for $D$ can be straightforwardly deduced, \S\,\ref{sec3.c}, the question arises as to why this shortcoming has not been detected by Kozik \emph{et al.} \cite{KFG14} -- the computer code underlying the calculations reported in Ref.\,\citen{KFG14} must unquestionably have shifted the calculated $D$ upwards by the correct constant amount, without this being part of the underlying formalism.\footnote{See the discussion in \protect\ref{sec.3a.1} regarding reference [28] in Ref.\,\protect\citen{KFG14}, as well as Eqs\,(\protect\ref{e89y2}) and (\protect\ref{e89y3}) below and the ensuing remarks.}

We devote \S\,\ref{sec3.b} to the Hubbard Hamiltonian for spin-$\tfrac{1}{2}$ particles and the details relevant to this Hamiltonian. Amongst others, we deduce the exact closed expression for $D$ corresponding to this system.

In \S\,\ref{sec3.f} we focus on the above-mentioned asserted non-invertibility of the mapping $G_{\X{0}} \mapsto G$ and identify a fundamental problem in the relevant calculations reported by Kozik \emph{et al.} \cite{KFG14}. This problem is associated with the iterative schemes A and B employed by Kozik \emph{et al.} \cite{KFG14} for effecting the inversion $G \mapsto G_{\X{0}}$. For clarity, on the basis of these schemes a ``non-interacting'' Green function $\t{G}_{\X{0};n}(z)$ is iteratively determined for use in the evaluation of the interacting Green function $\t{G}_{\X{(n)}}(z)$ on the basis of the perturbation series expansion for this function, where $n \in \mathds{N}$ denotes the iteration level.\footnote{More explicitly, $\t{G}_{\protect\X{0};n}(z)$ denotes the ``non-interacting'' one-particle Green function at the conclusion of the $n$th step of iteration.} To describe the mentioned problem, we point out that unless for the $\t{G}_{\X{0};n}(z)$ determined with the aid of the iterative schemes A and B the associated function $\hbar\hspace{0.4pt}\t{G}_{\X{0};n}^{-1}(z) - z$ is \textsl{independent} of $z$, \textsl{or} unless $\hbar\hspace{0.4pt}\t{G}_{\X{0};n}^{-1}(z)$ is a first-order \textsl{monic} polynomial\,\footnote{The coefficient of the leading term is equal to unity [Definition 4.5.4, p.\,182, in Ref.\,\protect\citen{AA14}].} of $z$, evaluation of the contributions to the last-mentioned perturbation series expansion in terms of $\t{G}_{\X{0};n}(z)$ is fundamentally incorrect. In \S\,\ref{s4x} we rigorously establish that the function $\hbar\hspace{0.4pt}\t{G}_{\X{0};n}^{-1}(z)$ as calculated on the basis of the above-mentioned schemes A and B violates the latter requirement.\footnote{Note in addition that since $-\protect\t{G}(z)$ is a Nevanlinna function of $z$, the negative of the self-energy deduced from the Dyson equation $\protect\t{\Sigma}(z) = 1/\protect\t{G}_{\protect\X{0}}(z) - 1/\protect\t{G}(z)$ can fail to be a Nevanlinna function of $z$ for an inappropriate $\protect\t{G}_{\protect\X{0}}(z)$ (see the footnote associated with the remark following Eq.\,(\protect\ref{e3a})).}

In \S\,\ref{s4xb} we illustrate the validity of the observations in \S\,\ref{s4x} by explicitly dealing with an $\mathpzc{n}$th-order perturbation series expansion of $\t{G}(z)$ in terms of the bare on-site interaction energy $U$ and the non-interacting Green function $\t{G}_{\X{0}}(z)$, showing that for any finite value of $\mathpzc{n}$ this perturbation expansion is pathological,\footnote{As we point out in \S\,\protect\ref{scon}, this pathology may not be a peculiarity of the `Hubbard atom', but a general property of \textsl{any} truncated perturbation series for the one-particle Green function. A remedy for avoiding the possible spurious zeros of the one-particle Green functions corresponding to truncated perturbation series is provided in \S\,\ref{sec.5}. Note that an approximation of the one-particle Green function, no matter how crude this approximation may be, whose negative is a Nevanlinna function of $z$ \textsl{cannot} possess zeros in the region $\protect\im[z]\not=0$ of the complex $z$-plane.} in that the perturbational Green function possesses zeros outside the real axis of the $z$-plane, in violation of an exact property expected of non-interacting and interacting one-particle Green functions alike. We also establish that for $\hbar\omega_0 \gtrsim U/2 \Leftrightarrow U \lesssim 2\pi/\beta$ it is practically impossible to detect the mentioned pathology when the calculations in the complex $z$-plane are limited to the Matsubara energies $\{\zeta_m\| m\in \mathds{Z}\}$, where $\zeta_m = \ii\hbar\omega_m + \mu$, in which $\mu$ denotes the chemical potential associated with the ensemble average of the number of particles, that is $\b{N}$, satisfying $\b{N} = N$. For the half-filled `Hubbard atom' under consideration,\footnote{At half-filling number of particles $N$ coincides with the number of lattice sites $N_{\textsc{s}}$.} one has $\mu=0$, \S\,\ref{sec3.d}.

In \S\,\ref{s4xa} we demonstrate that the conventional \textsl{zero-temperature} perturbation series expansion of the \textsl{proper} self-energy in terms of the non-interacting Green function and proper self-energy diagrams is ill-defined in the case of the `Hubbard atom'. We show that for this perturbation expansion to become applicable, it is necessary to apply a spectral `hair-splitting' operation (see the next paragraph), which is not necessary for the perturbation series expansion of the self-energy in terms of the interacting Green function and skeleton self-energy diagrams. Following this, we discuss the fundamental \textsl{inferiority} of the latter perturbation expansion in the case of the `Hubbard atom'. This can be directly attributed to the formal expansion of the self-energy for this model in terms of proper self-energy diagrams and the non-interacting Green function $\t{G}_{\epsilon\X{\downarrow 0}}(z)$, Eq.\,(\ref{e44}), being terminating, as evidenced by the exact expression in Eq.\,(\ref{e1e}). As we discuss in detail, on the basis of explicit calculations,\footnote{The bulk of these calculations is presented in \S\,\protect\ref{sd4}.} the \textsl{locality} of the problem at hand leads to the breakdown of the result in Eq.\,(\ref{e7h}) for $j=3$, and most likely for all $j\ge 3$.\footnote{We referred to this problem in the first part of \S\,\protect\ref{s.301}. We recall that Eq.\,(\protect\ref{e7h}) underlies the equalities in Eq.\,(\protect\ref{e7sa}).} This aspect compounds the problem of the perturbation series expansion of the self-energy of the `Hubbard atom' in terms of skeleton self-energy diagrams and the exact one-particle Green function.

With $\h{\mathcal{H}} = \h{\mathcal{H}}_{\X{0}} + \h{\mathcal{H}}_{\X{1}}$ expressing the full Hubbard Hamiltonian in terms of the non-interacting Hamiltonian $\h{\mathcal{H}}_{\X{0}}$ and the Hamiltonian $\h{\mathcal{H}}_{\X{1}}$ due to interaction, for the case of the `Hubbard atom' the above-mentioned `hair-splitting' process amounts to expressing $\h{\mathcal{H}}$ as $\h{\mathcal{H}} = \h{\mathcal{H}}_{\X{0}}' + \h{\mathcal{H}}_{\X{1}}'$, where $\h{\mathcal{H}}_{\X{0}}' \doteq \h{\mathcal{H}}_{\X{0}} + \epsilon \hspace{0.4pt}\h{\mathcal{H}}_{\X{1}}$ and $\h{\mathcal{H}}_{\X{1}}' \doteq (1-\epsilon) \h{\mathcal{H}}_{\X{1}}$, with $\epsilon$ infinitesimally small (see later); $\epsilon$ can be identified with zero only \textsl{after} evaluating the integrals underlying the expressions for the sought-after perturbational contributions to the self-energy. Thus, and remarkably, one observes that the `non-interacting' Hamiltonian required for bringing about the above-mentioned spectral hair-splitting proves to be the interacting Hamiltonian, albeit with the actual interaction potential energy $U$ being multiplied by the non-vanishing constant $\epsilon$. Consequently, in order for the self-energy to be describable in terms of the one-particle Green function associated with $\h{\mathcal{H}}_{\X{0}}'$ and the 1PI self-energy diagrams, the constant $\epsilon$ must be infinitesimally small.\footnote{The \textsl{contractions} arising from the application of the Wick theorem \protect\cite{BF19} underlying the diagrammatic series expansion of the self-energy in terms of 1PI self-energy diagrams are necessarily non-interacting Green functions, \emph{i.e.} Green functions corresponding to a \textsl{quadratic} Hamiltonian. In this connection, $\protect\h{\mathcal{H}}_{\protect\X{0}}'$ consists of a \textsl{quartic} part, in addition to a quadratic one, for any non-vanishing value of $\epsilon$.} Alternatively, the spectral `hair-splitting' associated with the non-interacting problem can be bypassed by means of effecting the atomic limit of the Hubbard Hamiltonian subsequent to evaluating the self-energy. This may be achieved by introducing a non-vanishing nearest-neighbour hopping term\,\refstepcounter{dummy}\label{TheSingleParticle}\footnote{The single-particle energy dispersion $\varepsilon_{\bm{k}}$ in Eq.\,(\protect\ref{ex01bx}) is related to the hopping matrix elements $\{T_{i,j}\| i,j\}$ according to $T_{i,j} = N_{\textsc{s}}^{-1} \sum_{\bm{k} \in \protect\1BZ} \varepsilon_{\bm{k}} \exp(\protect\ii \bm{k} \cdot [\bm{R}_i -\bm{R}_j])$, \S\,\protect\ref{s2.2}. For $\varepsilon_{\bm{k}} \equiv T_0$, Eq.\,(\protect\ref{e12}), with $T_{\protect\X{0}}$ \textsl{independent} of $\bm{k}$, one has $T_{i,j} = T_{\protect\X{0}}\hspace{0.8pt} \delta_{i,j}$. \label{noteb1}} in the Hamiltonian of the `Hubbard atom', to be equated with zero \textsl{after} the evaluation of the self-energy. In numerical calculations, this is achieved by means of \textsl{extrapolation}.

In investigating the consequences of the parameter $\epsilon$ referred to above, in \S\,\ref{s4xc} we consider the second-order self-energy within the finite-temperature Matsubara \cite{FW03} framework. We observe that while within this formalism $\epsilon$ can be identified with $0$ in advance of the calculations\,\footnote{See Eqs\,(\protect\ref{e59}), (\protect\ref{e62}), and (\protect\ref{e62a}) below.} (in contrast to the zero-temperature framework), \emph{calculations in the energy/frequency domain are in general unreliable}. For clarity, at non-zero temperatures, corresponding to $\beta <\infty$, where $\beta \equiv 1/(k_{\textsc{b}}T)$, the imaginary-time Green and associated functions (such as the self-energy and the polarization function) being periodic/anti-periodic functions of their imaginary-time argument $\tau$, with $\hbar\beta$ the length of the fundamental period along the $\tau$-axis [Eq.\,(24.14), p.\,236, in Ref.\,\citen{FW03}],\footnote{See also Eq.\,(2.38) in Ref.\,\protect\citen{BF19}. Note that these functions are \textsl{periodic} with fundamental period $2\hbar\beta$.} in the energy domain they are conveniently calculated at $\{\ii \hbar \omega_m + \mu \| m\in\mathds{Z}\}$ when anti-periodic, and at $\{\ii\hbar\nu_m \| m\in\mathds{Z}\}$ when periodic, where $\{\omega_m \| m\in\mathds{Z}\}$ are the fermionic and $\{\nu_m\| m\in\mathds{Z}\}$ the bosonic Matsubara frequencies,\footnote{Here $\mathds{Z} = \{\dots, -2,-1,0,1,2,\dots\}$, appendix \protect\ref{sae}.} for which one has [Eq.\,(25.15), p.\,245, in Ref.\,\citen{FW03}]
\begin{equation}\label{e70}
\omega_m \doteq \frac{(2 m + 1) \pi}{\hbar\beta},\;\;\; \nu_{m} \doteq \frac{2 m \pi}{\hbar\beta}.
\end{equation}
Under certain conditions \cite{BM61}, the relevant correlation functions are \textsl{uniquely} determined over the entire $z$-plane by their values over the \textsl{discrete} sets $\{\ii\hbar\omega_m + \mu\| m\}$ and $\{\ii\hbar\nu_m\| m\}$, which have no limit points on the $z$-plane.\footnote{See \S\,4.12, p.\,139, in Ref.\,\citen{ECT52} and Theorem 17.1, p.\,369, in Ref.\,\protect\citen{AIM65}.} In spite of this fact, we explicitly demonstrate that the procedure of calculating one many-body correlation function (such as the self-energy\,\footnote{Here we are using the term `correlation function' rather loosely, since technically self-energy, in contrast to the hierarchy of Green functions, does not qualify as a correlation function.}) from the knowledge of another many-body correlation function (such as the polarisation function) at the Matsubara frequencies is an \textsl{unsafe} one, and that in general for the correct determination of the former function knowledge of the latter function in the infinitesimal neighbourhoods of the Matsubara frequencies may be required. \emph{We arrive at the conclusion that it is safest to calculate finite-temperature many-body correlation functions in the imaginary-time domain}, evaluating the Fourier transforms of these functions at the relevant Matsubara frequencies as the last step of the calculations.

\refstepcounter{dummyX}
\subsection{The `Hubbard atom' as considered by Kozik \emph{et al.}}
\phantomsection
\label{sec3.a}
In dealing with the `Hubbard atom', \S\,\ref{sec3.b}, Kozik \emph{et al.} \cite{KFG14} have employed the following expression for the site double-occupancy\,\footnote{See \S\,\protect\ref{sec3.b} below.} [Fig.\,2 in Ref.\,\citen{KFG14}]:
\begin{equation}\label{e23}
D \equiv D(\beta,\mu,U) \doteq \langle \h{n}_{\uparrow} \h{n}_{\downarrow}\rangle = \frac{1}{U\beta}\hspace{0.8pt} \mathrm{tr} \Sigma G \equiv  \frac{1}{U\beta} \sum_m \t{\Sigma}(\zeta_m) \t{G}(\zeta_m),
\end{equation}
where\,\footnote{Appendix \protect\ref{sae}.} (\emph{cf.} Eq.\,(\ref{e70}))
\begin{equation}\label{e24}
\zeta_m \doteq \ii\hbar\omega_m + \mu,,\;\; m \in \mathds{Z}.
\end{equation}
Further \cite{KFG14}
\begin{equation}\label{e25}
\t{G}(z) = \frac{\hbar}{2} \Big(\frac{1}{z + U/2} + \frac{1}{z - U/2}\Big),\;\; \t{\Sigma}(z) = \frac{U^2}{4\hbar z}.
\end{equation}
From the Dyson equation, Eq.\,(\ref{e4a}), for the non-interacting Green function $\t{G}_{\X{0}}(z)$ corresponding to the functions in Eq.\,(\ref{e25}), one obtains
\begin{equation}\label{e15a}
\t{G}_{\X{0}}(z) = \frac{\hbar}{z},
\end{equation}
which indeed coincides with the function $\t{G}(z)$ in Eq.\,(\ref{e25}) for $U=0$.\footnote{With reference to the considerations centred on Eq.\,(\protect\ref{e12}) below, note however that for $T_0 = -U/2$ the Green function $\protect\t{G}_{\protect\X{0}}(z)$ in Eq.\,(\protect\ref{e15a}) coincides with $\protect\t{G}^{\textsc{hf}}(z)$, the \textsl{exact} Hartree-Fock one-particle Green function corresponding to the `Hubbard atom' for spin-$\tfrac{1}{2}$ particles at half-filling, \S\,\protect\ref{sec.3a.1} below.} One observes that the functions $\t{G}(z)$ and $\t{G}_{\X{0}}(z)$ in Eqs\,(\ref{e25}) and (\ref{e15a}) correctly describe the expected leading-order behaviour $\hbar/z$ in the asymptotic region $z \to\infty$, Eqs\,(\ref{e4n}) and (\ref{e4qa}). At first glance, the function in Eq.\,(\ref{e15a}) corresponds to the bare atomic energy $\varepsilon_{\textrm{at}}$ (\emph{i.e.} the energy $T_{\X{0}}$ in Eq.\,(\ref{e12}) below) identified with zero (\emph{cf.} Eq.\,(\ref{e14b}) below). This is however at odds with the fact that the lowest-lying pole of $\t{G}(z)$ is located at $z = -U/2$, instead of $z=0$ in the case of $T_{\X{0}} = 0$ (\emph{cf.} Eq.\,(\ref{e14a}) below).\footnote{Throughout this publication we assume $U\ge 0$.} Considering the fact that the Hartree-Fock self-energy $\Sigma^{\textsc{hf}}$ corresponding to the `Hubbard atom' for spin-$\tfrac{1}{2}$ particles at half-filling is equal to $U/(2\hbar)$ \cite{BF13},\footnote{See also \S\,\protect\ref{sec.3a.1} below.} one is led to suspect that $\varepsilon_{\textrm{at}}$ has been chosen to coincide with $-U/2$ \textsl{and} that in addition the non-interacting Hamiltonian to which the function $\t{G}_{\X{0}}(z)$ in Eq.\,(\ref{e15a}) corresponds takes account of the Hartree-Fock self-energy $\Sigma^{\textsc{hf}}$. In this way, the self-energy in Eq.\,(\ref{e25}) corresponds to the exact self-energy \textsl{minus} $\Sigma^{\textsc{hf}}$, brought about by expressing $\h{\mathcal{H}} = \h{\mathcal{H}}_{\X{0}} + \h{\mathcal{H}}_{\X{1}}$ as $\h{\mathcal{H}} = (\h{\mathcal{H}}_{\X{0}} + \h{\EuScript{H}}^{\textsc{hf}}) + (\h{\mathcal{H}}_{\X{1}} -\h{\EuScript{H}}^{\textsc{hf}})$, \S\,\ref{sec.3a.1},\refstepcounter{dummy}\label{AsWeIndicate}\footnote{As we indicate in \S\,\protect\ref{sec.3a.1}, in Ref.\,\protect\citen{KFG14} the underlying Hubbard Hamiltonian is either $\hspace{0.28cm}\h{\hspace{-0.28cm}\mathpzc{H}}$, Eq.\,(\ref{ex01f}), or $\h{\mathsf{H}}$, Eq.\,(\ref{ex01k}), and that the perturbation series expansions in Ref.\,\protect\citen{KFG14} are centred on the `non-interacting' Hamiltonian $\protect\h{\mathcal{H}}_{\protect\X{0}} + \protect\h{\EuScript{H}}^{\textsc{h}}$. Here we have chosen to identify the latter Hamiltonian with $\h{\mathcal{H}}_{\X{0}} + \h{\EuScript{H}}^{\textsc{hf}}$ for two reasons. First, the Hartree self-energy corresponding to $\hspace{0.28cm}\h{\hspace{-0.28cm}\mathpzc{H}}$ and $\h{\mathsf{H}}$, Eq.\,(\protect\ref{e27d}), identically coincides with the Hartree-Fock self-energy corresponding to $\protect\h{\mathcal{H}}$, Eq.\,(\protect\ref{e27b}). Second, with $\t{\Sigma}(z) = o(1)$ as $z\to\infty$, Eq.\,(\protect\ref{e25}), we wish to remain consistent with the expansions in Eqs\,(\protect\ref{e7b}) and (\protect\ref{e7ba}). We note in passing that in the considerations of
Ref.\,\protect\citen{KFG14} $n_{\uparrow} = n_{\downarrow} = n/2$, with $n=1$ (half-filling). \label{notel}} thus identifying $\h{\mathcal{H}}_{\X{0}} + \h{\EuScript{H}}^{\textsc{hf}}$ as the unperturbed and $\h{\mathcal{H}}_{\X{1}} - \h{\EuScript{H}}^{\textsc{hf}}$ as the perturbation Hamiltonian.\footnote{This possibility is in fact corroborated by the remark in reference $[28]$ of Ref.\,\protect\citen{KFG14}.} This view is consistent with the fact that for $z\to \infty$ to leading order the \textsl{exact} self-energy is to approach $\Sigma^{\textsc{hf}} = U/(2\hbar)$, Eqs\,(\ref{e7b}) and (\ref{e7ba}), while the self-energy in Eq.\,(\ref{e25}) approaches zero for $z \to \infty$. One explicitly verifies that indeed this is how the functions in Eqs\,(\ref{e25}) and (\ref{e15a}) are to be understood, \S\S\,\ref{sec.3a.1}, \ref{sec3.b}.

Making use of the standard approach [p.\,248 in Ref.\,\citen{FW03}], from the expressions in Eqs\,(\ref{e23}) and (\ref{e25}) one obtains the following exact equality (\emph{cf.} Eq.\,(6.7) in Ref.\,\citen{BF07}):
\begin{equation}\label{e26}
D(\beta,\mu,U) = \frac{1}{8} \Big(\hspace{-1.0pt}\tanh\big(\frac{\beta}{2} (\mu - U/2)\big) - \tanh\big(\frac{\beta}{2} (\mu + U/2)\big)\hspace{-0.8pt}\Big).
\end{equation}
By particle-hole (p-h) symmetry, at half-filling one has $\mu=0$,\footnote{We note that for \textsl{no} real value of $\mu$ can the right-hand side of Eq.\,(\protect\ref{e26}) be made to coincide with the exact result in Eq.\,(\protect\ref{e10e}) below. Therefore, the specific choice $\mu=0$, even if incorrect, cannot be considered as being the cause of the problem discussed here.} \S\,\ref{sec3.d}, resulting in the following expression for the double-occupancy at this filling:
\begin{equation}\label{e27}
D(\beta,0,U) = -\frac{1}{4} \tanh(\beta U/4),
\end{equation}
which is \textsl{negative} for all $\beta U>0$. Since $D(\beta,0,U)\to -1/4$ for $\beta U \to\infty$, the ``exact'' $D$ in Fig.\,2 of Ref.\,\citen{KFG14} seems to have inexplicably been shifted upwards by the amount $1/4$. \emph{We have therefore established that for the repulsive `Hubbard atom' for spin-$\tfrac{1}{2}$ particles at half-filling the expression for the double-occupancy $D$ as adopted by Kozik \emph{et al.} \cite{KFG14}, Eq.\,(\ref{e23}), is incompatible with the employed expressions for the exact Green function and self-energy, Eq.\,(\ref{e25}).} In \S\,\ref{sec3.b} we consider the `Hubbard atom' in some detail and deduce the appropriate expression for $D$.

\refstepcounter{dummyX}
\subsubsection{Remarks}
\phantomsection
\label{sec.3a.1}
In the light of the above considerations and in view of later references in this publication, some remarks are in order. \emph{Unless we indicate otherwise, the particles in this section are assumed to be spin-$\tfrac{1}{2}$ ones.}

In dealing with the perturbation series expansion of the self-energy $\Sigma_{\sigma}$ in terms of the bare two-body interaction potential and the non-interacting Green functions $\{G_{\X{0};\sigma}\| \sigma\}$,\footnote{In the notation of Ref.\,\protect\citen{BF19}, the self-energy calculated thus is $\Sigma_{\protect\X{00};\sigma}$.} all proper (\emph{i.e.} 1PI) self-energy diagrams with tadpole\,\footnote{The tadpole diagram \protect\cite{FW03} representing the exact Hartree self-energy $\Sigma_{\sigma}^{\textsc{h}}$ ($\Sigma^{\textsc{h}}$ when dealing with $\protect\h{\mathcal{H}}$) consists of a fermion loop with one vertex. Expanding the exact Green function in the expression for the Hartree self-energy in terms of its non-interacting counterpart, in the case of \textsl{two-body} interaction potentials [photon lines] one obtains an infinite set of fermion loops each of which consisting of an \textsl{odd} number of vertices. In quantum electrodynamics, the contributions of these loops vanish by Furry's theorem \protect\cite{WHF37,SSS66}. Furry's theorem, concerning electron and positron loops, is more general and is dealt with in some illustrative detail in Ref.\,\protect\citen{AB65}. See also in particular \S\,C.1 in appendix C of Ref.\,\protect\citen{BF19}.} insertions are to be discarded when $G_{\X{0};\sigma} \equiv G_{\sigma}^{\textsc{h}}$,\footnote{For uniform GSs (ensemble of states, ESs), one has $\protect\t{G}_{\sigma}^{\textsc{h}}(\bm{k};z) = \hbar/(z - \varepsilon_{\bm{k}} - \hbar\Sigma_{\sigma}^{\textsc{h}}(\bm{k}))$.} the `non-interacting' Green function that takes account of the \textsl{exact} Hartree self-energy $\Sigma_{\sigma}^{\textsc{h}}$.\footnote{The Hartree self-energy in Eq.\,(\protect\ref{e27b}) below is clearly independent of $\sigma$. That in Eq.\,(\protect\ref{e27d}) below is however dependent of $\sigma$.}\footnote{See Eq.\,(\protect\ref{e27i}) below.} In doing so, one can disregard also the Hartree diagram for being a tadpole itself, in which case the calculated self-energy amounts to
\begin{equation}\label{e27a}
\Sigma_{\sigma}' \doteq \Sigma_{\sigma} - \Sigma_{\sigma}^{\textsc{h}}.
\end{equation}
In this case, to leading order one has $\Sigma_{\sigma}'(z) \sim \Sigma_{\sigma}^{\textsc{f}}$ as $z\to\infty$ (\emph{cf.} Eqs\,(\ref{e7b}) and (\ref{e7ba})). In dealing with the uniform GSs (ensemble of states, ESs) of the Hubbard Hamiltonian $\h{\mathcal{H}}$ in Eq.\,(\ref{ex01bx}), one has
\begin{equation}\label{e27b}
\Sigma_{\sigma}^{\textsc{h}}(\bm{k}) = U n/\hbar,\;\; \Sigma_{\sigma}^{\textsc{f}}(\bm{k}) = -U n_{\sigma}/\hbar, \;\; \Sigma_{\sigma}^{\textsc{hf}}(\bm{k}) \equiv \Sigma_{\sigma}^{\textsc{h}}(\bm{k}) + \Sigma_{\sigma}^{\textsc{f}}(\bm{k}) = U n_{\b{\sigma}}/\hbar,
\end{equation}
where $n = n_{\sigma} + n_{\b{\sigma}}$, the total site-number density. In this case, to leading order one has
\begin{equation}\label{e27c}
\Sigma_{\sigma}'(\bm{k};z) \sim -U n_{\sigma}/\hbar \;\; \text{for}\;\; z\to\infty.
\end{equation}
In contrast, in dealing with the uniform GSs (ESs) of the Hubbard Hamiltonian $\hspace{0.28cm}\h{\hspace{-0.28cm}\mathpzc{H}}$ in Eq.\,(\ref{ex01f}), or $\h{\mathsf{H}}$ in Eq.\,(\ref{ex01k}), one has\,\footnote{In the light of the considerations in \S\,\protect\ref{sd2}, we emphasise that in dealing with the Hamiltonian $\hspace{0.28cm}\protect\h{\hspace{-0.28cm}\mathpzc{H}}$ the Fock diagram and its insertions do \textsl{not} occur in the diagrammatic expansion of the self-energy. In dealing with $\protect\h{\mathsf{H}}$ the latter do occur, however their contributions are identically vanishing. This is relevant here in that, in dealing with  $\hspace{0.28cm}\protect\h{\hspace{-0.28cm}\mathpzc{H}}$ the LHS of $\Sigma_{\sigma}^{\textsc{f}}(\bm{k}) \equiv 0$ is non-existent. Nonetheless, use of $\Sigma_{\sigma}^{\textsc{f}}(\bm{k}) \equiv 0$ is \textsl{quantitatively} consistent with $\Sigma_{\sigma}^{\textsc{f}}(\bm{k})$ being non-existent. For this reason, for brevity in this section we disregard the conceptual difference between a quantity not existing and a quantity existing however being numerically identically vanishing.}
\begin{equation}\label{e27d}
\Sigma_{\sigma}^{\textsc{h}}(\bm{k}) = U n_{\b{\sigma}}/\hbar,\;\; \Sigma_{\sigma}^{\textsc{f}}(\bm{k}) \equiv 0,\;\; \Sigma_{\sigma}^{\textsc{hf}}(\bm{k}) \equiv \Sigma_{\sigma}^{\textsc{h}}(\bm{k}) + \Sigma_{\sigma}^{\textsc{f}}(\bm{k}) = U n_{\b{\sigma}}/\hbar.
\end{equation}
However, in this case to leading order one has\,\footnote{Appendix \protect\ref{sae}.}
\begin{equation}\label{e27e}
\Sigma_{\sigma}'(\bm{k};z) \sim 0 \,\Longleftrightarrow\, \Sigma_{\sigma}'(\bm{k};z) = o(1)\;\; \text{for}\;\; z\to\infty.
\end{equation}
One observes that whereas the Hartree and Fock self-energy contributions, $\Sigma_{\sigma}^{\textsc{h}}$ and $\Sigma_{\sigma}^{\textsc{f}}$, depend on whether one has chosen to work with $\h{\mathcal{H}}$, or $\hspace{0.28cm}\protect\h{\hspace{-0.28cm}\mathpzc{H}}$ / $\h{\mathsf{H}}$, the Hartree-Fock self-energy contribution, $\Sigma_{\sigma}^{\textsc{hf}}$, is independent of the choice. This is in agreement with the fact that the \textsl{total} perturbational contribution to the self-energy at each order of the perturbation expansion is independent of whether one adopts $\h{\mathcal{H}}$, $\hspace{0.28cm}\protect\h{\hspace{-0.28cm}\mathpzc{H}}$, or $\h{\mathsf{H}}$.

In order for the leading-order asymptotic expression  $\Sigma_{\sigma}'(\bm{k};z) \sim 0$ for $z\to\infty$ to apply to the self-energy as calculated on the basis of the Hamiltonian $\h{\mathcal{H}}$ in Eq.\,(\ref{ex01bx}), the self-energy $\Sigma_{\sigma}'$ must be defined according to (\emph{cf.} Eqs\,(\ref{e7b}) and (\ref{e7ba}))
\begin{equation}\label{e27f}
\Sigma_{\sigma}' \doteq \Sigma_{\sigma} - \Sigma_{\sigma}^{\textsc{hf}}.
\end{equation}
This definition is suggestive that the perturbation series expansion of $\Sigma_{\sigma}'$ should be based on $G_{\X{0};\sigma} \equiv G_{\sigma}^{\textsc{hf}}$,\footnote{For uniform GSs (ESs), one has $\protect\t{G}_{\sigma}^{\textsc{hf}}(\bm{k};z) = \hbar/(z - \varepsilon_{\bm{k}} - \hbar\Sigma_{\sigma}^{\textsc{hf}}(\bm{k}))$, Eq.\,(\protect\ref{e7da}).} the `non-interacting' one-particle Green function that takes account of the \textsl{exact} Hartree-Fock self-energy $\Sigma_{\sigma}^{\textsc{hf}} \equiv \Sigma_{\sigma}^{\textsc{h}} + \Sigma_{\sigma}^{\textsc{f}}$.\footnote{See Eq.\,(\protect\ref{e27m}) below.} While possible (see later), this is not necessary: on writing
\begin{equation}\label{e27g}
\Sigma_{\sigma}' \equiv (\Sigma_{\sigma} - \Sigma_{\sigma}^{\textsc{h}}) - \Sigma_{\sigma}^{\textsc{f}},
\end{equation}
one can as before calculate the perturbational contributions to $\Sigma_{\sigma} - \Sigma_{\sigma}^{\textsc{h}}$ in terms of $G_{\X{0};\sigma} \equiv G_{\sigma}^{\textsc{h}}$ and determine $\Sigma_{\sigma}'$ by subtracting $\Sigma_{\sigma}^{\textsc{f}}$ from the thus-calculated $\Sigma_{\sigma} - \Sigma_{\sigma}^{\textsc{h}}$. If, on the other hand, one performs the calculations in terms of $G_{\X{0};\sigma} \equiv G_{\sigma}^{\textsc{hf}}$, one should discard from the set of the proper self-energy diagrams all those that in addition to the Hartree self-energy insertions (tadpoles) contain the Fock self-energy insertions. This is relevant, as in the case of $\h{\mathcal{H}}$, Eq.\,(\ref{ex01bx}), the Fock self-energy $\Sigma_{\sigma}^{\textsc{f}}$ is non-vanishing, Eq.\,(\ref{e27d}). Note that since any proper self-energy diagram of order $\nu \ge 2$ that contains the Hartree and/or the Fock self-energy insertion is non-skeleton, none of the self-energy diagrams of order $\nu \ge 2$ discarded as a result of using $G_{\X{0};\sigma} \equiv G_{\sigma}^{\textsc{h}}$ or $G_{\X{0};\sigma} \equiv G_{\sigma}^{\textsc{hf}}$ is from the set of skeleton self-energy diagrams. In the case of $\hspace{0.28cm}\h{\hspace{-0.28cm}\mathpzc{H}}$, Eq.\,(\ref{ex01f}), there are no diagrams with Fock insertions to reckon with, and in the case of $\h{\mathsf{H}}$, Eq.\,(\ref{ex01k}), the contributions of the diagrams with Fock insertions are identically vanishing, Eq.\,(\ref{e27d}).\footnote{See the (i) and (ii) on p.\,\protect\pageref{SelfEnergyDiagrams}. See also \S\,\protect\ref{sd21}.}

For completeness, let\,\footnote{With reference to the zeroth-order Hamiltonian on the RHSs of Eqs\,(\protect\ref{ex01bx}), (\protect\ref{ex01f}), and (\protect\ref{ex01k}), one observes that the unperturbed Hamiltonians $\protect\h{\mathcal{H}}_{\protect\X{0}} + \protect\h{\EuScript{H}}^{\textsc{x}}$, $\hspace{0.28cm}\protect\h{\hspace{-0.28cm}\mathpzc{H}}_{\protect\X{0}} + \protect\h{\EuScript{H}}^{\textsc{x}}$, and $\protect\h{\mathsf{H}}_{\protect\X{0}} + \protect\h{\EuScript{H}}^{\textsc{x}}$ are described in terms of the one-particle energy dispersion $\varepsilon_{\bm{k};\sigma}^{\textsc{x}} \doteq\varepsilon_{\bm{k}} + \hbar\Sigma_{\sigma}^{\textsc{x}}(\bm{k})$. For instance, in the case of $\protect\h{\mathcal{H}}_{\protect\X{0}} + \protect\h{\EuScript{H}}^{\textsc{x}}$ the latter energy dispersion is spin-split when $\textsc{x} = \textsc{hf}$ and $n_{\sigma} \not= n_{\protect\b{\sigma}}$.}
\begin{equation}\label{e27h}
\h{\EuScript{H}}^{\textsc{x}} \doteq \sum_{\bm{k},\sigma} \hbar\Sigma_{\sigma}^{\textsc{x}}(\bm{k})\, \h{a}_{\bm{k};\sigma}^{\dag} \h{a}_{\bm{k};\sigma}^{\phantom{\dag}},\;\; \textsc{x} = \textsc{h}, \textsc{hf},
\end{equation}
where $\h{a}_{\bm{k};\sigma}^{\dag}$ and $\h{a}_{\bm{k};\sigma}^{\phantom{\dag}}$ are the canonical creation and annihilation operators \cite{FW03}, respectively, \S\,\ref{sd2}, and \emph{for consistency the self-energy contribution $\Sigma_{\sigma}^{\textsc{x}}(\bm{k})$ is to be that corresponding to the specific representation of the Hubbard Hamiltonian under consideration, which may be $\h{\mathcal{H}}$, Eq.\,(\ref{ex01bx}), $\hspace{0.28cm}\protect\h{\hspace{-0.28cm}\mathpzc{H}}$, (\ref{ex01f}), or $\h{\mathsf{H}}$, Eq.\,(\ref{ex01k})}.  Let further\,\footnote{With reference to the preceding remark, since in Eq.\,(\protect\ref{e27i}) $\protect\h{\EuScript{H}}^{\textsc{h}}$ is subtracted from $\protect\h{\mathcal{H}}$, it follows that the Hartree self-energy in terms of which $\protect\h{\EuScript{H}}^{\textsc{h}}$ is defined is that corresponding to $\protect\h{\mathcal{H}}$.}
\begin{equation}\label{e27i}
\h{\mathcal{H}}' \doteq \h{\mathcal{H}} - \h{\EuScript{H}}^{\textsc{h}}.
\end{equation}
Importantly, since the Hartree self-energy in Eq.\,(\ref{e27b}) is independent of $\sigma$, one has
\begin{equation}\label{e27j}
\h{\EuScript{H}}^{\textsc{h}} = U n\h{N},
\end{equation}
where
\begin{equation}\label{e27ja}
\h{N} \doteq \sum_{\bm{k},\sigma} \h{a}_{\bm{k};\sigma}^{\dag} \h{a}_{\bm{k};\sigma}^{\phantom{\dag}} \equiv \sum_{\sigma} \h{N}_{\sigma}
\end{equation}
is the total-number operator. Thus, in dealing with the thermodynamic Hamiltonian [Eq.\,(23.1), p.\,228, in Ref.\,\citen{FW03}]
\begin{equation}\label{e27k}
\h{\mathcal{K}} \doteq \h{\mathcal{H}} - \mu \h{N},
\end{equation}
the contribution of $\h{\EuScript{H}}^{\textsc{h}}$ can be absorbed in the chemical potential $\mu$:
\begin{equation}\label{e27l}
\h{\mathcal{K}} =  \h{\mathcal{H}}' - \mu' \h{N},\;\; \text{where}\;\; \mu' \doteq \mu - Un.
\end{equation}
Let now $\h{\mathcal{H}}'$ be defined according to
\begin{equation}\label{e27m}
\h{\mathcal{H}}' \doteq \h{\mathcal{H}} - \h{\EuScript{H}}^{\textsc{hf}}.
\end{equation}
The equality on the left in Eq.\,(\ref{e27l}) applies for the $\h{\mathcal{H}}'$ herein defined according to Eq.\,(\ref{e27m}) and the $\mu'$ according to $\mu' \doteq \mu - Un/2$ \textsl{only} if $n_{\sigma} = n_{\b{\sigma}} = n/2$.\footnote{With $\protect\h{N}_{\sigma} \doteq \sum_{\bm{k}}  \h{a}_{\bm{k};\sigma}^{\dag} \h{a}_{\bm{k};\sigma}^{\phantom{\dag}}$, in general $\protect\h{\EuScript{H}}^{\textsc{hf}} = U \sum_{\sigma} n_{\protect\b{\sigma}}\protect\h{N}_{\sigma} \equiv U (n_{\downarrow} \protect\h{N}_{\uparrow} + n_{\uparrow} \protect\h{N}_{\downarrow})$. For the specific case of $n_{\sigma} = n_{\protect\b{\sigma}} = n/2$, $\protect\h{\EuScript{H}}^{\textsc{hf}}$ reduces to $\protect\h{\EuScript{H}}^{\textsc{hf}} = U n \protect\h{N}/2 \equiv \tfrac{1}{2} \protect\h{\EuScript{H}}^{\textsc{h}}$.} Defining, in analogy with Eq.\,(\ref{e27i}),\footnote{See the remark following Eq.\,(\protect\ref{e27h}).}
\begin{equation}\label{e27n}
\hspace{0.28cm}\h{\hspace{-0.28cm}\mathpzc{H}}\hspace{0.4pt}' \doteq \hspace{0.28cm}\h{\hspace{-0.28cm}\mathpzc{H}} - \h{\EuScript{H}}^{\textsc{x}},\;\; \h{\mathsf{H}}' \doteq \h{\mathsf{H}} - \h{\EuScript{H}}^{\textsc{x}},\;\; \textsc{x} = \textsc{h}, \textsc{hf},
\end{equation}
in the light of Eq.\,(\ref{e27d}), again a process similar to that in Eq.\,(\ref{e27l}) is \textsl{only} possible if $n_{\sigma} = n_{\b{\sigma}} = n/2$, irrespective of whether $\textsc{x} = \textsc{h}$ or $\textsc{x} = \textsc{hf}$. The adjusted chemical potential in this case is $\mu' \doteq \mu - U n/2$.

The above observations make evident that without explicitly specifying the adopted representation of a named Hamiltonian (in the case at hand, the Hubbard Hamiltonian), there is some room for confusion regarding the calculated self-energy. As regards the work by Kozik \emph{et al.} \cite{KFG14}, the diagrammatic expansion for $\Sigma[G]$ in the lower part of Fig.~1 in Ref.\,\citen{KFG14} lacking the Fock diagram, it follows that the Hubbard Hamiltonian as adopted in this work is either $\hspace{0.28cm}\h{\hspace{-0.28cm}\mathpzc{H}}$ or $\h{\mathsf{H}}$. This is consistent with the exact self-energy $\Sigma^{\textrm{exact}}(z)$ for the `Hubbard atom' as presented in Ref.\,\citen{KFG14}, namely $\Sigma^{\textrm{exact}}(z) = U^2/(4z)$ (in the units where $\hbar=1$), appropriately satisfying $\Sigma^{\textrm{exact}}(z) = o(1)$ as $z\to\infty$, Eqs\,(\ref{e27e}).\footnote{This is corroborated by the data in at least the two left-most panels in Fig.\,3 of Ref.\,\protect\citen{KFG14}.} Furthermore, under reference [28] of Ref.\,\citen{KFG14}, citing Ref.\,\citen{vHKPS10}, it is explicitly mentioned that
\vspace{0.2cm}
\begin{itemize}{\footnotesize
\item[{}] ``we include the mean-field Hartree contribution $\Sigma_{\textsc{h}\uparrow,\downarrow} = U n_{\uparrow,\downarrow}$, where $n_{\uparrow,\downarrow}$ is the particle density per spin component, into $G_{\X{0}}$ by appropriately shifting the chemical potential ...\,.''}
\end{itemize}
\vspace{0.2cm}
\noindent Leaving aside the fact that (in the units where $\hbar = 1$) $\Sigma_{\textsc{h}\uparrow,\downarrow} = U n_{\downarrow,\uparrow}$,\footnote{Note the orders of the spin indices.} Eq.\,(\ref{e27d}), as we have indicated above, in the case at hand the process of subsuming the Hartree self-energy in the chemical potential is conditional on the equality $n_{\uparrow} = n_{\downarrow}$.

\refstepcounter{dummyX}
\subsection{The `Hubbard atom' of spin-\texorpdfstring{$\tfrac{1}{2}$}{} particles}
\phantomsection
\label{sec3.b}
In the atomic limit, for the Hubbard Hamiltonian corresponding to spin-$\tfrac{1}{2}$ particles defined on the lattice $\{\bm{R}_i \| i=1,2,\dots,N_{\textsc{s}}\}$, p.\,\pageref{AssumingUniformGSs}, one has [Eq.\,(33) in Ref.\,\citen{JH63}] [Eq.\,(4.32) in Ref.\,\citen{PF03}]\,\footnote{With $N$ denoting the number of spin-$\tfrac{1}{2}$ particles, for $N \le N_{\textsc{s}}$ the GS energy of the Hubbard Hamiltonian in the atomic limit is $\upnu_{\textrm{g}}$-fold degenerate, where [\S\,4.4.2 in Ref.\,\protect\citen{PF03}] $\upnu_{\textrm{g}} = 2^{N} \binom{N_{\textsc{s}}}{N\phantom{.}}$, in which the factor $2^{N}$ accounts for the spin degeneracy specific to spin-$\tfrac{1}{2}$ particles. At half-filling, where $N/N_{\textsc{s}}=1$, only the spin-degeneracy of the GS energy obtains. In Ref.\,\protect\citen{BF12} we have discussed \textsl{and} explicitly demonstrated (for a related model) that the Luttinger theorem \protect\cite{LW60,BF07} is in general \textsl{not} applicable to systems with degenerate GS energy and that its apparent validity for systems with p-h symmetry is accidental. This failure can be amended by considering the one-particle Green \textsl{matrix} $\mathds{G}$ (not to be confused with one-particle Green matrix $\mathbb{G}$ in the spin space) whose element $(\mathds{G})_{i,j}$ is the matrix element of the  time-ordered product of the creation and annihilation field operators in the Heisenberg picture with respect to the $i$th and the $j$th $N$-particle GS of the Hamiltonian in the degenerate manifold, respectively $\vert\Psi_{N;0_i}\rangle$ and $\vert\Psi_{N;0_j}\rangle$ \protect\cite{BF21a}. In Ref.\,\protect\citen{BF12} we have restored the validity of the Luttinger theorem by first lifting the degeneracy of the GS energy by means of an external perturbation, turning the strength of this perturbation to zero subsequent to the completion of the relevant calculations.}
\begin{equation}\label{e12}
\h{\mathcal{H}} =  T_{\X{0}} \sum_{\sigma} \sum_{i=1}^{N_{\textsc{s}}} \h{n}_{i;\sigma} + \frac{U}{2} \sum_{\sigma} \sum_{i=1}^{N_{\textsc{s}}} \h{n}_{i;\sigma} \h{n}_{i;\b{\sigma}},
\end{equation}
where
\begin{equation}\label{e12a}
\h{n}_{i;\sigma} \doteq \h{c}_{i;\sigma}^{\dag} \h{c}_{i;\sigma}^{\phantom{\dag}}
\end{equation}
is the site-occupation operator for particles with spin index $\sigma$ in terms of the canonical site creation and annihilation operators in the Schr\"odinger picture [p.\,53 in Ref.\,\citen{FW03}], respectively $\h{c}_{i;\sigma}^{\dag}$ and $\h{c}_{i;\sigma}^{\phantom{\dag}}$. The subscript $\b{\sigma}$ denotes the spin index complementary to $\sigma$,\footnote{For $\sigma = \uparrow$ ($\downarrow$), $\b{\sigma} = \downarrow$ ($\uparrow$).} and $T_{\X{0}}$ the atom energy $\varepsilon_{\textrm{at}}$ in the absence of interaction. Since $\h{n}_{i;\sigma}$ and $\h{n}_{i;\b{\sigma}}$ commute, for spin-$\tfrac{1}{2}$ particles one has
\begin{equation}\label{e12b}
\sum_{\sigma}\h{n}_{i;\sigma} \h{n}_{i;\b{\sigma}} = 2\hspace{0.4pt} \h{n}_{i;\uparrow} \h{n}_{i;\downarrow},
\end{equation}
whereby
\begin{equation}\label{e13}
\h{\mathcal{H}} = T_{\X{0}}\hspace{0.4pt} \h{N} + U \h{D},
\end{equation}
where
\begin{align}\label{e13a}
\h{N} &\doteq \sum_{\sigma} \h{N}_{\sigma} \equiv \sum_{\sigma}\sum_{i=1}^{N_{\textsc{s}}}  \h{n}_{i;\sigma},\nonumber\\
\h{D} &\doteq  \sum_{i=1}^{N_{\textsc{s}}} \h{D}_i \equiv \sum_{i=1}^{N_{\textsc{s}}}\h{n}_{i;\uparrow} \h{n}_{i;\downarrow}.
\end{align}
The operator $\h{D}$ describes the \textsl{total} site double-occupancy.

In order to avoid confusion with the notation elsewhere in this publication, in the remaining part of \textsl{this} subsection we shall denote the interacting (non-interacting) one-particle Green function corresponding to the `Hubbard atom' by $\t{\mathsf{G}}(z)$ ($\t{\mathsf{G}}_{\X{0}}(z)$) and the corresponding self-energy by $\t{\Upsigma}(z)$.

For the one-particle Green function $\t{\mathsf{G}}(z)$ corresponding to the Hamiltonian $\h{\mathcal{H}}$ in Eq.\,(\ref{e12}) and half-filling, $n \doteq N/N_{\textsc{s}} = 1$, where $N \doteq \langle \h{N}\rangle$ (see later),\footnote{For the ensemble average $\langle \h{N}\rangle$, compare with Eq.\,(\protect\ref{e34}) below.} one has [Eq.\,(39) in Ref.\,\citen{JH63}]
\begin{equation}\label{e14a}
\t{\mathsf{G}}(z) = \frac{\hbar}{2} \Big(\frac{1}{z - T_{\X{0}}} + \frac{1}{z - T_{\X{0}} - U}\Big),
\end{equation}
whereby, identifying $U$ with $0$,
\begin{equation}\label{e14b}
\t{\mathsf{G}}_{\X{0}}(z) = \frac{\hbar}{z - T_{\X{0}}}.
\end{equation}
From the Dyson equation, Eq.\,(\ref{e4a}), one thus obtains the following expression for the self-energy:
\begin{equation}\label{e14c}
\t{\Upsigma}(z) = \frac{U}{2\hbar} \frac{z - T_{\X{0}}}{z - T_{\X{0}} - U/2}.
\end{equation}
One observes that\,\footnote{With reference to Eqs\,(\protect\ref{e7b}) and (\ref{e7ba}), and in the light of the fact that for the Hubbard Hamiltonian under consideration $\Sigma^{\textsc{hf}} = U/(2\hbar)$, the asymptotic series expansion in Eq.\,(\protect\ref{e14c1}) reveals that $\Sigma_{\infty_j}$, \emph{i.e.} the coefficient of $1/z^j$, is a polynomial of order $j+1$ in $U$, in conformity with the observation in \S\,\protect\ref{sec.b2} (see p.\,\protect\pageref{FurtherTheFact}). It turns out that for the `Hubbard atom' $\Sigma_{\infty_j}$ is equal to $U^2$ times a polynomial of order $j-1$ in $U$, which is however identically vanishing for $T_{\protect\X{0}} = -U/2$, $\forall j\ge 2$.}
\begin{equation}\label{e14c1}
\t{\Upsigma}(z) \sim \frac{U}{2\hbar} + \frac{U^2}{4\hbar z} + \frac{U^2 (U + 2 T_{\X{0}})}{8\hbar z^2} + \frac{U^2 (U^2 + 4 T_{\X{0}} U + 4 T_{\X{0}}^2)}{16\hbar z^3}+\dots\;\; \text{for}\;\; z \to \infty,
\end{equation}
where $U/(2\hbar)$ appropriately coincides with the Hartree-Fock self-energy $\Sigma_{\sigma}^{\textsc{hf}}(\bm{k})$ corresponding to the uniform GS of the Hubbard Hamiltonian at half-filling \cite{BF13}.\footnote{See \S\,\protect\ref{sec.3a.1} and \S\,\protect\ref{sec3.c} below, as well as Eqs\,(\protect\ref{e7b}) and (\protect\ref{e7ba}).} The Green function $\t{\mathsf{G}}(z)$ in Eq.\,(\protect\ref{e14a}) identically coincides with that in Eq.\,(\protect\ref{e25}) for $T_{\protect\X{0}} = -U/2$, however for this value of $T_{\protect\X{0}}$ the self-energy in Eq.\,(\protect\ref{e14c}) deviates from that in Eq.\,(\protect\ref{e25}); one has $\t{\Upsigma}(z) = \protect\t{\Sigma}(z) + U/(2\hbar) \equiv \protect\t{\Sigma}(z) + \Sigma^{\textsc{hf}}$, as expected. Not surprisingly, for $T_{\protect\X{0}} = -U/2$, with $U\not= 0$, the non-interacting Green function $\t{\mathsf{G}}_{\protect\X{0}}(z)$ in Eq.\,(\protect\ref{e14b}) deviates from that in Eq.\,(\protect\ref{e15a}); as we have indicated earlier, $\t{G}_{\X{0}}(z)$ takes account of the Hartree-Fock self-energy $\Sigma^{\textsc{hf}}$: $\t{G}_{\X{0}}^{-1}(z) = \t{\mathsf{G}}_{\X{0}}^{-1}(z) - \Sigma^{\textsc{hf}}$.\footnote{See footnote \raisebox{-1.0ex}{\normalsize{\protect\footref{notel}}} on p.\,\protect\pageref{AsWeIndicate} and \S\,\protect\ref{sec.3a.1} for some relevant details.}

We note that since to leading order $\t{\Upsigma}(z)\hspace{0.4pt} \t{\mathsf{G}}(z) \sim U/(2 z)$ as $z\to \infty$, the sum in Eq.\,(\ref{e23}) is \textsl{conditionally} convergent [\S\,2.32 in Ref.\,\citen{WW62}] when the $\t{G}(z)$ and $\t{\Sigma}(z)$ herein are replaced by respectively $\t{\mathsf{G}}(z)$ and $\t{\Upsigma}(z)$. The above asymptotic expression is to be contrasted with $\t{\Sigma}(z)\hspace{0.4pt} \t{G}(z) \sim U^2/(4 z^2)$ as $z \to \infty$ for the functions $\t{G}(z)$ and $\t{\Sigma}(z)$ in Eq.\,(\ref{e25}), so that the sum in Eq.\,(\ref{e23}) is absolutely convergent. It appears therefore that we have uncovered the origin of the pathological result for the site double occupancy in Eq.\,(\ref{e27}), however for arriving at a definite conclusion, we first deduce the relevant expression for the site double-occupancy pertaining to the Hamiltonian in Eq.\,(\ref{e12}).

\refstepcounter{dummyX}
\subsubsection{The site double-occupancy for the `Hubbard atom' of spin-\texorpdfstring{$\tfrac{1}{2}$}{} particles}
\phantomsection
\label{sec3.c}
Here we deduce the expression for the single-site double-occupancy (\emph{cf.} Eq.\,(\ref{e13a}))
\begin{equation}\label{e34}
D_i = \langle \h{D_i}\rangle \doteq \mathrm{Tr}\{\h{\varrho}_{\textsc{g}}\hspace{0.4pt} \h{n}_{i;\uparrow} \h{n}_{i;\downarrow}\},
\end{equation}
where
\begin{equation}\label{eg3}
\h{\varrho}_{\textsc{g}} \doteq \e^{-\beta \h{\mathcal{K}}}/\mathrm{Tr}[\e^{-\beta \h{\mathcal{K}}}]
\end{equation}
is the statistical operator, with (\emph{cf.} Eq.\,(\ref{e27k}))
\begin{equation}\label{eg1}
\h{\mathcal{K}} = U \h{D} - (\mu - T_{\X{0}}) \h{N}
\end{equation}
the grand-canonical Hamiltonian [Eq.\,(23.1), p.\,228, in Ref.\,\citen{FW03}] for the `Hubbard atom' under consideration, Eqs\,(\ref{e12}) and (\ref{e13}). By the assumed uniformity of the ESs under consideration, the single-site double occupancy $D_i$ is independent of $i$, whereby $D_i$ is equal to $\sum_{i=1}^{N_{\textsc{s}}} D_i/N_{\textsc{s}}$, the average site double occupancy. \emph{To avoid confusion with the \textsl{extensive} quantity $D = \langle \h{D} \rangle$, in the following we denote the \textsl{intensive} quantity $D_i$ by $\mathpzc{D}$.}

With $\h{c}_{i;\sigma}(\tau)$ denoting the imaginary-time Heisenberg-picture counterpart of the Schr\"odinger-picture site annihilation operator $\h{c}_{i;\sigma}$, Eq.\,(\ref{e12a}), for the site-representation of the one-particle Green function corresponding to $\h{\mathcal{K}}$, \emph{i.e.} $\mathsf{G}_{i;\sigma}(\tau,\tau')$, one has [Eq.\,(23.6), p.\,228, in Ref.\,\citen{FW03}]\,\footnote{See also \S\,2.2.2 in Ref.\,\protect\citen{BF19}.}
\begin{equation}\label{eg2}
\mathsf{G}_{i;\sigma}(\tau,\tau') \doteq - \mathrm{Tr}\{\h{\varrho}_{\textsc{g}}\hspace{0.4pt}\mathcal{T}[\h{c}_{i;\sigma}^{\phantom{\dag}}(\tau) \h{c}_{i;\sigma}^{\dag}(\tau')]\},
\end{equation}
where $\mathcal{T}$ is the fermion imaginary-time-ordering operator. The function $\mathsf{G}_{i;\sigma}(\tau,\tau')$ can be more generally denoted by $\mathcal{G}_{i,i;\sigma}(\tau,\tau')$, representing the site-diagonal element of the following more general one-particle Green function [Eq.\,(7.1.21), p.\,134, in Ref.\,\citen{PF02}]:
\begin{equation}\label{eg3x}
\mathcal{G}_{i,j;\sigma}(\tau,\tau') \doteq - \mathrm{Tr}\{\h{\varrho}_{\textsc{g}}\hspace{0.4pt}\mathcal{T}[\h{c}_{i;\sigma}^{\phantom{\dag}}(\tau) \h{c}_{j;\sigma}^{\dag}(\tau')]\}.
\end{equation}
On employing the Fourier representation\,\footnote{For the lattice vector $\bm{R}_i$, see p.\,\protect\pageref{AssumingUniformGSs}.}
\begin{equation}\label{eg3a}
\h{c}_{i;\sigma} = \frac{1}{\sqrt{N_{\textsc{s}}\phantom{|}\!}} \sum_{\bm{k}\in \fbz} \e^{-\ii\bm{k}\cdot \bm{R}_i} \h{a}_{\bm{k};\sigma},
\end{equation}
from the expression in Eq.\,(\ref{eg3x}) one obtains
\begin{equation}\label{eg3b}
\mathcal{G}_{i,j;\sigma}(\tau,\tau') = \frac{1}{N_{\textsc{s}}} \sum_{\bm{k},\bm{k}' \in\fbz} \e^{-\ii\bm{k}\cdot \bm{R}_i + \ii\bm{k}'\cdot \bm{R}_j} \mathpzc{G}_{\sigma}(\bm{k},\bm{k}';\tau,\tau'),
\end{equation}
where
\begin{equation}\label{eg3c}
\mathpzc{G}_{\sigma}(\bm{k},\bm{k}';\tau,\tau') \doteq  - \mathrm{Tr}\{\h{\varrho}_{\textsc{g}}\hspace{0.4pt} \mathcal{T}[\h{a}_{\bm{k};\sigma}^{\phantom{\dag}}(\tau) \h{a}_{\bm{k}';\sigma}^{\dag}(\tau')]\}.
\end{equation}
For uniform ESs\,\footnote{For some relevant technical details, see footnote \raisebox{-1.0ex}{\normalsize{\protect\footref{notez}}} on p.\,\pageref{WithP}.}
\begin{equation}\label{eg3d}
\mathpzc{G}_{\sigma}(\bm{k},\bm{k}';\tau,\tau') \equiv \hspace{4.0pt} \EuScript{G}_{\sigma}(\bm{k};\tau,\tau') \hspace{0.6pt} \delta_{\bm{k},\bm{k}'},
\end{equation}
whereby
\begin{equation}\label{eg3e}
\mathcal{G}_{i,j;\sigma}(\tau,\tau') = \frac{1}{N_{\textsc{s}}} \sum_{\bm{k} \in\fbz} \e^{-\ii\bm{k}\cdot (\bm{R}_i -\bm{R}_j)}\hspace{0.6pt} \EuScript{G}_{\sigma}(\bm{k};\tau,\tau'),
\end{equation}
making explicit that for uniform ESs $\mathcal{G}_{i,j;\sigma}(\tau,\tau')$ indeed depends on $\bm{R}_i - \bm{R}_j$. With reference to Eqs\,(\ref{eg2}) and (\ref{eg3x}), from the expression in Eq.\,(\ref{eg3e}) one thus obtains
\begin{equation}\label{eg3f}
\mathsf{G}_{i;\sigma}(\tau,\tau') = \frac{1}{N_{\textsc{s}}} \sum_{\bm{k} \in\fbz} \EuScript{G}_{\sigma}(\bm{k};\tau,\tau').
\end{equation}
The independence of the RHS of this equality from $i$ reflects the fact that the expression in Eq.\,(\ref{eg3e}) is specific to \textsl{unform} ESs. In the \textsl{atomic limit}, where the function $\EuScript{G}_{\sigma}(\bm{k};\tau,\tau')$ is independent of $\bm{k}$, to be denoted by $\EuScript{G}_{\sigma}(\tau,\tau')$, the expression in Eq.\,(\ref{eg3f}) reduces to\,\footnote{See Eq.\,(\protect\ref{ex08x}).}
\begin{equation}\label{eg3g}
\mathsf{G}_{i;\sigma}(\tau,\tau') = \EuScript{G}_{\sigma}(\tau,\tau'),\;\; \forall i \in \{1,2,\dots,N_{\textsc{s}}\}.\;\;\; \text{(Atomic limit)}
\end{equation}
Inverting the relationship in Eq.\,(\ref{eg3e}), one has
\begin{equation}\label{eg3h}
\EuScript{G}_{\sigma}(\bm{k};\tau,\tau') = \sum_{i=1}^{N_{\textsc{s}}} \e^{\ii\bm{k}\cdot (\bm{R}_i - \bm{R}_j)} \mathcal{G}_{i,j;\sigma}(\tau,\tau') \equiv \sum_{i=1}^{N_{\textsc{s}}} \e^{\ii\bm{k}\cdot \bm{R}_i} \mathcal{G}_{i,1;\sigma}(\tau,\tau'),
\end{equation}
where in the last expression we have used the convention $\bm{R}_1 = \bm{0}$. The independence of $\EuScript{G}_{\sigma}(\bm{k};\tau,\tau')$ from $\bm{k}$ in the atomic limit implies that (\emph{cf.} Eqs\,(\ref{eg2}) and (\ref{eg3x}))
\begin{equation}\label{eg3i}
\mathcal{G}_{i,j;\sigma}(\tau,\tau') = \mathsf{G}_{i;\sigma}(\tau,\tau')\hspace{0.6pt} \delta_{i,j} \equiv
 \EuScript{G}_{\sigma}(\tau,\tau')\hspace{0.6pt} \delta_{i,j}.\;\;\; \text{(Atomic limit)}
\end{equation}

With reference to Eq.\,(\ref{eg1}), the operator $\h{c}_{i;\sigma}(\tau)$ satisfies the following Heisenberg equation of motion [Eq.\,(23.12), p.\,230, in Ref.\,\citen{FW03}]:
\begin{equation}\label{eg4}
\hbar \frac{\partial}{\partial\tau} \h{c}_{i;\sigma}(\tau) = [\h{\mathcal{K}},\h{c}_{i;\sigma}(\tau)]_{-},
\end{equation}
where $[\,,\,]_{-}$ denotes commutation. With
\begin{equation}\label{eg7}
 [\h{D},\h{c}_{i;\sigma}(\tau)]_{-} = - \delta_{\sigma,\downarrow} \h{n}_{i;\uparrow}(\tau) \h{c}_{i;\downarrow}(\tau) -\delta_{\sigma,\uparrow} \h{c}_{i;\uparrow}(\tau) \h{n}_{i;\downarrow}(\tau),
\end{equation}
in which $\h{n}_{i;\sigma}(\tau) \equiv \h{c}_{i;\sigma}^{\dag}(\tau) \h{c}_{i;\sigma}^{\phantom{\dag}}(\tau)$ (\emph{cf.} Eq.\,(\ref{e12a})), and
\begin{equation}\label{eg8}
[\h{N},\h{c}_{i;\sigma}(\tau)]_{-} = -\h{c}_{i;\sigma}(\tau),
\end{equation}
both deduced on the basis of the equal-time canonical anti-commutation relations\,\footnote{See appendix \protect\ref{sae}.}
\begin{equation}\label{eg6}
[\h{c}_{i;\sigma}^{\phantom{\dag}}(\tau),\h{c}_{j;\sigma'}^{\phantom{\dag}}(\tau)]_{+} = \h{0},\;\;\;
[\h{c}_{i;\sigma}^{\phantom{\dag}}(\tau),\h{c}_{j;\sigma'}^{\dag}(\tau)]_{+} = \delta_{\sigma,\sigma'} \delta_{i,j}\hspace{0.6pt} \h{1},
\end{equation}
one obtains the expression for $\partial\mathsf{G}_{i;\sigma}(\tau,\tau')/\partial\tau \equiv \partial\EuScript{G}_{\sigma}(\tau,\tau')/\partial\tau$, Eq.\,(\ref{eg3g}), and thereby, on account of (\emph{cf.} Eq.\,(\ref{eg2}))
\begin{equation}\label{eg9}
\mathsf{G}_{i;\sigma}(\tau,\tau') = \mathrm{Tr}\{\h{\varrho}_{\textsc{g}}\hspace{0.4pt} \h{c}_{i;\sigma}^{\dag}(\tau') \h{c}_{i;\sigma}^{\phantom{\dag}}(\tau)\}\;\; \text{for}\;\; \tau' > \tau,
\end{equation}
arrives at the result (\emph{cf.} Eq.\,(\ref{e34}))
\begin{equation}\label{eg10}
\mathpzc{D} = \hspace{0.4pt} \mathrm{Tr}\{\h{\varrho}_{\textsc{g}}\hspace{0.4pt} \h{n}_{i;\uparrow}(\tau) \h{n}_{i;\downarrow}(\tau)\} = \frac{1}{U} \lim_{\tau' \downarrow \tau}\Big(\mu - T_{\X{0}} -\hbar \frac{\partial}{\partial\tau}\Big) \EuScript{G}_{\sigma}(\tau,\tau').
\end{equation}
In equilibrium, $\mathsf{G}_{i;\sigma}(\tau,\tau')$, and therefore $\EuScript{G}_{\sigma}(\tau,\tau')$, is a function $\tau-\tau'$. In the following we shall therefore employ the function $\mathpzc{G}_{\sigma}(\tau)$, defined according to
\begin{equation}\label{eg9a}
\mathpzc{G}_{\sigma}(\tau-\tau') \doteq \EuScript{G}_{\sigma}(\tau,\tau'). \;\;\;\; \text{(Equilibrium ESs in the atomic limit)}
\end{equation}

Making use of the Fourier series representation [Eq.\,(25.10), p.\,244, in Ref.\,\citen{FW03}]
\begin{equation}\label{eg11}
\mathpzc{G}_{\sigma}(\tau) = \frac{1}{\beta\hbar} \sum_{m=-\infty}^{\infty} \e^{-\ii\omega_m \tau} \mathsf{g}_{\sigma}(\omega_m),
\end{equation}
where $\{\omega_m\| m\in \mathds{Z}\}$ are the fermionic Matsubara frequencies, Eq.\,(\ref{e70}), the expression in Eq.\,(\ref{eg10}) can be written as
\begin{equation}\label{eg13}
\mathpzc{D} = \frac{1}{U\beta} \sum_{m=-\infty}^{\infty} \e^{\ii\omega_m 0^+} \frac{1}{\hbar}\big(\hspace{-0.6pt}\ii\hbar\omega_m + \mu -T_{\X{0}}\big)\hspace{0.6pt} \mathsf{g}_{\sigma}(\omega_m),
\end{equation}
where the converging factor $\e^{+\ii\omega_m 0^+}$ originates from the limit $\tau' \downarrow \tau$ on the right-hand side of Eq.\,(\ref{eg10}). With (\emph{cf.} Eq.\,(\ref{e24}))
\begin{equation}\label{eg14}
\t{\mathscr{G}}(\zeta_m) \doteq \mathsf{g}_{\sigma}(\omega_m),\;\;\forall\sigma,
\end{equation}
the expression in Eq.\,(\ref{eg13}) can be equivalently written as
\begin{equation}\label{eg15}
\mathpzc{D} = \frac{1}{U\beta} \sum_{m=-\infty}^{\infty} \e^{\zeta_m 0^+/\hbar} \t{\mathscr{G}}_{\X{0}}^{-1}(\zeta_m)\hspace{0.4pt} \t{\mathscr{G}}(\zeta_m),
\end{equation}
where $\t{\mathscr{G}}(z) \equiv \t{\mathsf{G}}(z)$ and $\t{\mathscr{G}}_{\X{0}}(z) \equiv \t{\mathsf{G}}_{\X{0}}(z)$, Eqs\,(\ref{e14a}) and (\ref{e14b}).\footnote{We have introduced the new symbols $\protect\t{\mathscr{G}}_{\protect\X{0}}(z)$ and $\protect\t{\mathscr{G}}(z)$ in order to conform with our convention, both in this publication and in \emph{e.g.} Ref.\,\protect\citen{BF19}, of distinguishing between the zero-temperature Green functions and the non-zero temperature Green functions in the Matsubara formalism. For the case under consideration this distinction is entirely formal. For the same reason, in Eq.\,(\protect\ref{eg15ax}) below we introduce the function $\protect\t{\mathscr{S}}(z)$ for the Matsubara-formalism counter-part of the self-energy $\protect\t{\Upsigma}(z)$ in Eq.\,(\protect\ref{e14c}).} On account of the Dyson equation (\emph{cf.} Eq.\,(\ref{e4a}))
\begin{equation}\label{eg15ax}
\t{\mathscr{S}}(z) = \t{\mathscr{G}}_{\X{0}}^{-1}(z) - \t{\mathscr{G}}^{-1}(z),
\end{equation}
where $\t{\mathscr{S}}(z) \equiv \t{\Upsigma}(z)$, Eq.\,(\ref{e14c}), and the equality [Eq.\,(26.9), p.\,252, in Ref.\,\citen{FW03}]
\begin{equation}\label{e9}
\sum_{m=-\infty}^{\infty} \e^{\zeta_m 0^+/\hbar} = 0,
\end{equation}
the expression in Eq.\,(\ref{eg15}) can be equivalently written as
\begin{equation}\label{eg15a}
\mathpzc{D} = \frac{1}{U\beta} \sum_{m=-\infty}^{\infty} \e^{\zeta_m 0^+/\hbar} \t{\mathscr{S}}(\zeta_m)\hspace{0.4pt} \t{\mathscr{G}}(\zeta_m),
\end{equation}
which, disregarding for the moment the \textsl{important} converging factor $\e^{\zeta_m 0^+/\hbar}$, formally coincides with the expression in Eq.\,(\ref{e23}). The latter converging factor is relevant, since, for the functions $\t{\mathscr{G}}(z)$ and $\t{\mathscr{S}}(z)$ in Eqs\,(\ref{e14a}) and (\ref{e14c}), to leading order one has
\begin{equation}\label{eg15b}
\t{\mathscr{S}}(z)\hspace{0.4pt} \t{\mathscr{G}}(z) \sim \frac{U}{2 z}\;\;\text{as}\;\; z \to \infty.
\end{equation}
The product $\t{\mathscr{S}}(z)\hspace{0.4pt} \t{\mathscr{G}}(z)$ decaying not faster than $1/z$ for $z \to \infty$, without the converging factor $\e^{\zeta_m 0^+/\hbar}$ the expression in Eq.\,(\ref{eg15a}) would be ambiguous, reflecting the distinction between the limit $\tau'\downarrow \tau$ in Eq.\,(\ref{eg10}) and the limits $\tau'\uparrow \tau$ and $\tau' \to \tau$.

To evaluate $\mathpzc{D}$ on the basis of the expression in Eq.\,(\ref{eg15a}), corresponding to the functions in Eqs\,(\ref{e14a}) and (\ref{e14c}), we employ the identity
\begin{equation}\label{e14d}
\t{\mathscr{S}}(z)\hspace{0.4pt} \t{\mathscr{G}}(z) \equiv \frac{U}{2 z} + \frac{U}{2} \Big(\frac{1}{z - T_{\X{0}} - U} -\frac{1}{z}\Big),
\end{equation}
where, in the light of the asymptotic expression in Eq.\,(\ref{eg15b}), the second term on the RHS decays \textsl{faster} than $1/z$ for $z\to\infty$. This implies that in calculating the contribution of this term to $\mathpzc{D}$, the converging factor $\e^{\zeta_m 0^+/\hbar}$ can be safely discarded.

Making use of [Eq.\,(25.37), p.\,250, in Ref.\,\citen{FW03}]
\begin{equation}\label{e10b}
\frac{1}{\beta}\sum_{m=-\infty}^{\infty} \frac{\e^{\zeta_m 0^+/\hbar}}{\zeta_m} = \frac{1}{\e^{-\beta\mu} + 1},
\end{equation}
and
\begin{align}\label{e10c}
\Phi(\beta,\mu,T_{\X{0}},U) &\doteq \frac{1}{\beta}\sum_{m=-\infty}^{\infty}  \Big(\frac{1}{\zeta_m - T_{\X{0}} - U} -\frac{1}{\zeta_m}\Big)\nonumber\\
&= \frac{1}{2} \Big(\hspace{-1.2pt}\tanh\big(\frac{\beta}{2} (\mu - T_{\X{0}} - U)\big) - \tanh\big(\frac{\beta}{2} \mu\big)\Big),
\end{align}
obtained through applying the standard technique [p.\,248 in Ref.\,\citen{FW03}] [\S\S\,14.5 and 16.4 in Ref.\,\citen{WW62}], from the expression in Eq.\,(\ref{eg15a}) one arrives at the exact result
\begin{equation}\label{e10d}
\mathpzc{D}(\beta,\mu,T_{\X{0}},U) = \frac{1}{2}\hspace{0.6pt} \frac{1}{\e^{-\beta\mu} + 1} +\frac{1}{2} \hspace{0.6pt}\Phi(\beta,\mu,T_{\X{0}},U).
\end{equation}
At half-filling, corresponding to $n \doteq N/N_{\textsc{s}} = 1$, by p-h symmetry one has $\mu = T_{\X{0}} + U/2$, \S\,\ref{sec3.d}, whereby
\begin{equation}\label{e10e}
\mathpzc{D} \doteq \mathpzc{D}(\beta,T_{\X{0}}+U/2,T_{\X{0}},U) =  \frac{1}{4} - \frac{1}{4} \tanh\big(\beta U/4).\;\;\;\; \text{(At half-filling)}
\end{equation}
This is to be contrasted with the expression in Eq.\,(\ref{e27}). As expected, $\mathpzc{D}$ is \textsl{independent} of $T_{\X{0}}$. Interestingly, the function $\mathpzc{D}(\beta,T_{\X{0}}+U/2,T_{\X{0}},U)$ coincides with the function $D(\beta,0,U)$ in Eq.\,(\ref{e27}) \textsl{increased} by the constant $1/4$, the latter rendering the resulting function non-negative for \textsl{all} $\beta U \ge 0$. The function on the RHS of Eq.\,(\ref{e10e}) monotonically decreases from $1/4$ at $\beta U =0$ towards $0$ (to leading order like $\tfrac{1}{2} \e^{-\beta U/2}$) for $\beta U \to\infty$.

\refstepcounter{dummyX}
\subsubsection{The chemical potential}
\phantomsection
\label{sec3.d}
To calculate the chemical potential $\mu = \mu(\beta,n)$ as a function of $\beta$ and the site-occupation number $n \equiv N/N_{\textsc{s}}$, we note that for the expectation value of $\h{n} \equiv \sum_{\sigma} \h{n}_{\sigma}$ in the grand canonical ensemble of uniform states, one has (\emph{cf.} Eqs\,(\ref{eg3g}), (\ref{eg9}), and (\ref{eg9a}))
\begin{equation}\label{e28}
n \doteq \sum_{\sigma} \mathrm{Tr}\{\h{\varrho}_{\textsc{g}}\hspace{0.4pt} \h{n}_{i;\sigma}\} = \sum_{\sigma} \lim_{\tau'\downarrow \tau} \mathsf{G}_{i;\sigma}(\tau,\tau')\equiv \sum_{\sigma} \lim_{\tau\hspace{0.4pt}\uparrow\hspace{0.4pt}0} \mathpzc{G}_{\sigma}(\tau),
\end{equation}
leading to (\emph{cf.} Eqs\,(\ref{eg11}) and (\ref{eg14}))
\begin{equation}\label{e29}
n = \frac{2}{\hbar\beta} \sum_{m=-\infty}^{\infty} \e^{\zeta_m 0^+/\hbar} \t{\mathscr{G}}(\zeta_m).
\end{equation}
Note that $n \in [0,2]$. From the expression for $\t{\mathscr{G}}(z) \equiv \t{\mathsf{G}}(z)$ in Eq.\,(\ref{e14a}), making use of the equality in Eq.\,(\ref{e10b}), one obtains
\begin{equation}\label{e30}
n = \frac{1}{\e^{\beta (T_{\X{0}} - \mu)} +1} + \frac{1}{\e^{(T_{\X{0}} + U -\mu)}+1}.
\end{equation}
One verifies that the RHS of this expression is identically equal to $1$ (corresponding to half-filling) for \textsl{all} $\beta$ and $T_{\X{0}}$ when $\mu = T_{\X{0}} + U/2$. More generally, with
\begin{equation}\label{e30a}
\varsigma \doteq \e^{-\beta U},
\end{equation}
from the expression in Eq.\,(\ref{e30}) one obtains
\begin{equation}\label{e31}
\mu = T_{\X{0}} -\frac{1}{\beta} \ln(\phi(n,\varsigma)),
\end{equation}
where
\begin{equation}\label{e32}
\phi(n,\varsigma) \doteq \frac{1+\varsigma}{2n} \Big(\hspace{-1.2pt}1-n + \sqrt{(1-n)^2 +4 n (2-n)\varsigma/(1+\varsigma)^2}\Big).
\end{equation}
One has $\phi(1,\varsigma) = \sqrt{\varsigma}$, so that for $n=1$ the expression in Eq.\,(\ref{e31}) leads to the expected result $\mu = T_{\X{0}} + U/2$. Since the expression in Eq.\,(\ref{e31}), in conjunction with the function $\phi(n,\varsigma)$ in  Eq.\,(\ref{e32}), corresponds to the Green function $\t{\mathscr{G}}(z)$ specific to half-filling (compare this Green function with that in Eq.\,(39) of Ref.\,\citen{JH63}), the dependence of the $\mu$ in Eq.\,(\ref{e31}) on $n$ outside $n=1$ is of \textsl{no} physical significance. The significance the expression in Eq.\,(\ref{e31}) lies in the fact that it establishes $\mu = T_{\X{0}} + U/2$ as the \textsl{only} chemical potential appropriate in the context of the problem at hand, where $n=1$.

In other to deduce a physically-significant expression for $\mu(\beta,n)$, applicable to the physical range of site densities, that is applicable to $n \in [0,2]$, with $n= 2 n_{\sigma} \equiv 2\langle \h{n}_{\sigma}\rangle$, $\sigma\in\{\uparrow,\downarrow\}$,\footnote{Concerning uniform spin-\textsl{unpolarised} cases.} one should employ the appropriate Green function, as presented in Eq.\,(39) of Ref.\,\citen{JH63}. By doing so, along the same lines as above one obtains a similar expression for $\mu$ as in Eq.\,(\ref{e31}), with however the function $\phi(n,\varsigma)$ herein replaced by the function $\psi(n,\varsigma)$, where
\begin{equation}\label{e34x}
\psi(n,\varsigma) \doteq \frac{1}{n} \Big(1-n + \sqrt{(1-n)^2 + (2 -n) n \varsigma}\Big),
\end{equation}
in which $\varsigma$ is the same quantity as in Eq.\,(\ref{e30a}). Evidently, $\psi(1,\varsigma) \equiv \phi(1,\varsigma) \equiv\sqrt{\varsigma}$.

\refstepcounter{dummyX}
\subsection{On the iterative schemes A and B for inverting the mapping
\texorpdfstring{$G_{\protect\X{0}} \mapsto G$}{}}
\phantomsection
\label{sec3.f}
We begin by recalling that the considerations of \S\,\ref{sec2.a} have exposed the way in which in the case of the `Hubbard atom' the exact inversion of the mapping $G_{\X{0}} \mapsto G$ seemingly results in two non-interacting Green functions corresponding to the interacting Green function $\t{G}(z)$. As we have shown, only a unique combination of these two Green functions, Eq.\,(\ref{e1c}), matching on the circle $\vert z\vert = U/2$, satisfies the fundamental analytic conditions required of the one-particle Green functions (whether interacting or non-interacting), establishing that in the case of the `Hubbard atom' the mapping $G_{\X{0}} \mapsto G$ is one-to-one in the relevant space of functions.

In Ref.\,\citen{KFG14} two iterative schemes have been introduced and employed, referred to as schemes A and B, for inverting the mapping $G_{\X{0}} \mapsto G$. In a notation that without further specification, to be presented below, is incomplete, these schemes are based on the following recursive relations [Eq.\,(3) in Ref.\,\citen{KFG14}]:
\begin{equation}\label{e5a}
\t{G}_{\X{0};n+1}^{-1}(z) = \t{G}_{\X{0};n}^{-1}(z) \pm \big(\t{G}^{-1}(z) -\t{G}_{\X{(n)}}^{-1}(z)\big),\;\; z \in \{ \zeta_m\| m\},
\end{equation}
where $\zeta_m$ is defined in Eq.\,(\ref{e24}), and the index $n$ denotes the recursion level. Restricting $z$ to $\zeta_0$, the upper sign corresponds to scheme A, and the lower sign to scheme B. The upper sign applies to both schemes for $z=\zeta_m$, $\forall m \in \mathds{Z}\backslash\{0\}$.\footnote{Appendix \protect\ref{sae}.} The function $\t{G}_{\X{(n)}}(z)$ on the RHS of Eq.\,(\ref{e5a}) is by Kozik \emph{et al.} \cite{KFG14} calculated on the basis of the perturbation series expansion of $\t{G}(z)$ in terms of $\t{G}_{\X{0};n}(z)$ with the aid of the continuous-time Monte-Carlo method \cite{GMLRTW11}, ``using an interaction-expansion continuous-time quantum Monte Carlo solver [31] implemented with the TRIQS [32] toolbox.'' \cite{KFG14}. One can therefore write\,\footnote{See \S\,1.2 in Ref.\,\protect\citen{BF19}.}
\begin{equation}\label{e5a1}
\t{G}_{\X{(n)}}(z) \equiv \t{G}(z;[G_{\X{0};n}]).
\end{equation}

In the case of the half-filled `Hubbard atom' of spin-$\tfrac{1}{2}$ particles for which the exact Green function $\t{G}(z)$ is known, \S\,\ref{sec3.a}, Kozik \emph{et al.} \cite{KFG14} have found that schemes A and B in general fail to reproduce the exact non-interacting Green function $\t{G}_{\X{0}}(z)$ in the limit of $n\to \infty$ for the range of the values of the on-site interaction energy $U$ considered,\footnote{See in particular Fig.\,2 of Ref.\,\protect\citen{KFG14} and \S\,\protect\ref{sec3.a} for a discussion of the data displayed herein.} thus concluding that the mapping $G_{\X{0}} \mapsto G$ were non-invertible. This conclusion is arrived at on the basis of the consideration that with the exact Green function $\t{G}(z)$ given, the approach of $\t{G}_{\X{0};n}(z)$, for $n\to\infty$, to different functions representing $\t{G}_{\X{0}}(z)$, signals, through the Dyson equation $\t{\Sigma}(z) = \t{G}_{\X{0}}^{-1}(z) - \t{G}^{-1}(z)$, Eq.\,(\ref{e4a}), the existence of different branches for the mapping $G \mapsto \Sigma$. Hence the general observation by Kozik \emph{et al.} \cite{KFG14}, that ``the self-energy $\Sigma[G] \propto \delta \Upphi[G]/\delta G$ is not a single-valued functional of $G$'', and that ``the functional $\Sigma[G]$ has at least two branches''.

\refstepcounter{dummyX}
\subsubsection{Defects of schemes A and B}
\phantomsection
\label{s4x}
It is not difficult to recognize that the adopted schemes by Kozik \emph{et al.} \cite{KFG14}, described above, are defective. Most importantly, for the function $\t{G}_{\X{0};n}(z)$, $\forall n$, to qualify as a \textsl{non-interacting} Green function to be used in a perturbation series expansion scheme that relies on the application of the Wick theorem\,\footnote{See appendix A in Ref.\,\protect\citen{BF19}.}\footnote{We emphasise that the formalism in Ref.\,\protect\citen{GMLRTW11} \textsl{does} rely on the use of the Wick theorem.} and the discarding of the expectation values of the terms that are not fully contracted [pp.\,83-92, 234-241 in Ref.\,\citen{FW03}], it is required that $\t{G}_{\X{0};n}(z)$ correspond to an effective one-particle Hamiltonian (explicitly, one that does \textsl{not} depend on $z$), which may be spatially local or non-local.\footnote{One aspect that is of particular relevance here, is the way in which two particular contractions prove to be $c$-numbers and directly related to the non-interacting Green function, an aspect that is very explicitly described in Ch. 4 of Ref.\,\protect\citen{MYS95}.} It can be verified that this can be the case \textsl{only} if $\hbar\hspace{0.4pt}\t{G}_{\X{0};n}^{-1}(z) - z$, $\forall n$, is \textsl{independent} of $z$, that is $\hbar\hspace{0.4pt} \t{G}_{\X{0};n}^{-1}(z)$ is a first-order \textsl{monic} polynomial of $z$.\footnote{Note for instance that while for the $\t{G}_{\protect\X{0}}(z)$ in Eq.\,(\protect\ref{e15a}) one has $\hbar\hspace{0.4pt} \protect\t{G}_{\protect\X{0}}^{-1}= z$, for the $\protect\t{G}(z)$ in Eq.\,(\protect\ref{e25}) one has $\hbar\hspace{0.4pt}\protect\t{G}^{-1}(z) = z - U^2/(4 z)$, in which the second term is a non-trivial function of $z$.} Deviation from this property implies that $\t{G}_{\X{0};n}(z)$ takes account of some dynamical interaction effects, which are beyond the reach of a mean-field description;\footnote{These effects are generally referred to as \textsl{quantum fluctuations}.} we recall that in dealing with the `Hubbard atom', Kozik \emph{et al.} \cite{KFG14} take account of the Hartree-Fock self-energy $\Sigma^{\textsc{hf}} = U/(2\hbar)$ at the zeroth-order of the perturbation expansion, \S\,\ref{sec3.a}. The mentioned dynamical interaction effects not being describable in terms of an energy-independent effective one-particle potential, the function $\t{G}_{\X{0};n}(z)$, with $\hbar\hspace{0.4pt}\t{G}_{\X{0};n}^{-1}(z) - z$ depending on $z$, has \textsl{no} rightful place within the framework of the perturbation series expansion as employed by Kozik \emph{et al.} \cite{KFG14,GMLRTW11}.

The iterative schemes A and B, described above, in general do not guarantee the satisfaction of the above-mentioned requirement by the sequence of functions $\{\t{G}_{\X{0};n}(z) \| n\}$. In principle, \S\,\ref{s4xa}, one exception corresponds to the case where one initiates the iterative procedures in schemes A and B with the \textsl{exact} non-interacting Green function $\t{G}_{\X{0}}(z)$, that is with $\t{G}_{\X{0};\X{0}}(z) \equiv \t{G}_{\X{0}}(z)$, in which case the $\t{G}_{\X{(1)}}(z)$ calculated with the aid of the continuous-time Monte-Carlo method \cite{GMLRTW11}, with the calculations carried out to infinite accuracy, \textsl{formally} (see \S\,\ref{s4xa}) identically coincides with the exact Green function $\t{G}(z)$, resulting in $\t{G}_{\X{0;1}}(z) = \t{G}_{\X{0}}(z)$. Consequently, $\t{G}_{\X{0};n}(z) \equiv \t{G}_{\X{0}}(z)$, $\forall n$, implying that $\t{G}_{\X{0}}(z)$ is a `fixed-point' of the iterative schemes A and B, provided that $\t{G}_{\X{(n)}}(z)$, $\forall n$, is calculated to infinite accuracy. In practice, however, where $\t{G}_{\X{(n)}}(z)$ is calculated to finite accuracy, even initiating the iterative schemes A and B with the exact non-interacting Green function $\t{G}_{\X{0}}(z)$, the errors inherent to the calculation (arising from the unavoidable truncation of the underlying perturbation series as well as the round-off errors associated with the use of the floating-point arithmetic) give rise to a $z$-dependent contribution, arising from $\t{G}^{-1}(z) - \t{G}_{\X{(0)}}^{-1}(z)$, to $\t{G}_{\X{0;1}}(z)$, leading to this function not being appropriate for use as the non-interacting Green function in the diagrammatic expansion of the Green function $\t{G}(z)$. This problem proliferates throughout the entire iterative process.

\refstepcounter{dummyX}
\subsubsection{Illustration}
\phantomsection
\label{s4xb}
The perturbation series expansion of the function $\t{G}(z)$ in Eq.\,(\ref{e25}) in terms of the non-interacting Green function $\t{G}_{\X{0}}(z)$ in Eq.\,(\ref{e15a}) is trivially obtained by expanding the former function in powers of $U$. One formally obtains\,\footnote{Recall that a $\nu$th-order connected Feynman diagram contributing to the one-particle Green functions is comprised of $2\nu+1$ lines representing the underlying non-interacting one-particle Green functions \protect\cite{FW03,BF19}.}
\begin{equation}\label{e38}
\t{G}(z) = \frac{\hbar}{z} \sum_{\nu=0}^{\infty} \Big(\frac{U}{2 z}\Big)^{2\nu} \equiv \sum_{\nu=0}^{\infty} \Big(\frac{U}{2\hbar}\Big)^{2\nu} \t{G}_{\X{0}}^{2\nu+1}(z).
\end{equation}
The dependence of $\t{G}(z)$ on only the even powers of $U$ follows from the p-h symmetry of the `Hubbard atom'. We note that technically the series in Eq.\,(\ref{e38}) is the weak-coupling perturbation series expansion of $\t{G}(z)$. As we shall discuss later, since $z\hspace{0.6pt} \t{G}(z)$ is a function of $U/z$, the notion of `weak-coupling' is not applicable here, since for any non-vanishing value of $U$, there is a finite neighbourhood of $z=0$ in the $z$-plane where the series in Eq.\,(\ref{e38}) is not convergent. Given the fact that for the case at hand, corresponding to half-filling, for the chemical potential one has $\mu = 0$, \S\,\ref{sec3.d}, here we are facing a specific case of a more general problem discussed in detail in Ref.\,\citen{BF13} corresponding to the perturbation series expansion of $\t{G}_{\sigma}(\bm{k};z)$ in terms of $\{\t{G}_{\X{0};\sigma}(\bm{k};z)\|\sigma\}$ for $\bm{k}$ approaching the associated non-interacting Fermi surface\,\footnote{For single-band models and in the thermodynamic limit, $\protect\t{G}_{\protect\X{0};\sigma}(\bm{k};z)$ invariably corresponds to a \textsl{metallic} GS, even though $\protect\t{G}_{\sigma}(\bm{k};z)$ may not. The considerations in Ref.\,\protect\citen{BF13} are mainly focussed on the $N$-particle uniform \textsl{metallic} GS of in particular the single-band Hubbard Hamiltonian. In the light of the results in Eqs\,(2.29) and (2.41) of Ref.\,\protect\citen{BF13}, the specifications ``$\bm{k}$ approaching the associated non-interacting Fermi surface'' and ``$z$ approaching $\mu$'' are of especial significance.} and $z$ approaching $\mu$.

For completeness, with the same caveat as indicated above, for the strong-coupling perturbation series expansion of $\t{G}(z)$ one has
\begin{equation}\label{e38a}
\t{G}(z) = -\frac{2\hbar}{U} \sum_{\nu=0}^{\infty} \Big(\frac{2 z}{U}\Big)^{2\nu+1}.
\end{equation}
With $U\hspace{0.2pt} \t{G}(z)$ being a function of $z/U$, clearly the notion of strong coupling cannot apply for sufficiently large values of $\vert z\vert$. One verifies that the boundary in the $z$-plane separating the regions of convergence of the weak- and strong-coupling perturbation series expansions in respectively Eq.\,(\ref{e38}) and Eq.\,(\ref{e38a}) is the circle $\vert z\vert = U/2$ [\S\,2.6, p.\,29, in Ref.\,\citen{WW62}]. This is the same boundary on one side of which the non-interacting Green function $G_{\X{0}}(z)$ in Eq.\,(\ref{e1a}) coincides with the function $\mathcal{G}_{-}(G(z))$ and on the other side with $\mathcal{G}_{+}(G(z))$.\footnote{The Green functions are here stripped of the usual tilde for consistency with the notation of \S\,\protect\ref{sec2.a}.}

In the light of the above observations, considering the iterative schemes A and B discussed in \S\,\ref{sec3.f}, initiated by the non-interacting Green function considered here (that is, by identifying $\t{G}_{\X{0;0}}(z)$ with $\t{G}_{\X{0}}(z)$), it follows that irrespective of the accuracy with which one calculates $\t{G}_{\X{(0)}}(z)$ (say, with the aid of the continuous-time Monte Carlo method \cite{GMLRTW11} as in Ref.\,\citen{KFG14}), this function cannot describe $\t{G}(z)$ in the region $\vert z \vert < U/2$. It should be noted that because of the even powers of $\zeta \doteq U/(2 z)$ in the series in Eq.\,(\ref{e38}), the terms of this series are \textsl{positive} (thus not \textsl{alternating}) along the real axis of the $z$-plane. This implies that along the real axis of the $z$-plane the series in Eq.\,(\ref{e38}) either converges or diverges, but does not oscillate, \S\,\ref{sec2.b}.

With the assumption $\t{G}_{\X{0;0}}(z) = \t{G}_{\X{0}}(z)$, making use of the $\mathpzc{n}$th-order expansion of $\t{G}(z)$ in powers of $U$, Eq.\,(\ref{e38}), where we assume $\mathpzc{n}$ to be \textsl{even} (solely for avoiding unnecessary notational complication), according to schemes A and B, Eq.\,(\ref{e5a}) one has
\begin{equation}\label{e40}
\t{G}_{\X{0;1}}^{-1}(z) = \t{G}_{\X{0}}^{-1}(z) \pm \t{G}^{-1}(z) \mp \t{G}_{\X{(0)}}^{-1}(z),
\end{equation}
where
\begin{equation}\label{e41}
\t{G}_{\X{(0)}}(z) \equiv \frac{\hbar}{z} \sum_{\nu=0}^{\mathpzc{n}/2} \Big(\frac{U}{2 z}\Big)^{2\nu} = \frac{\hbar}{z} \frac{1- (U/2 z)^{\mathpzc{n}+2}}{1 - (U/2 z)^2}.
\end{equation}
For any finite \textsl{even} value of $\mathpzc{n}$, this function can be identically expressed as
\begin{equation}\label{e41a}
\t{G}_{\X{(0)}}(z) = \frac{\hbar}{z^{\mathpzc{n}+1}} \prod_{\nu=1}^{\mathpzc{n}/2} (z-z_{\nu}) (z-z_{\nu}^*),
\end{equation}
where\,\footnote{Clearly, $\theta_{\nu}$ and $z_{\nu}$ depend on $\mathpzc{n}$, which we do not display explicitly for economy of notation.}
\begin{equation}\label{e41b}
z_{\nu} \doteq \frac{U}{2} \e^{\ii \theta_{\nu}},\;\; \text{with}\;\;\; \theta_{\nu} \equiv \frac{\pi\nu}{\mathpzc{n}/2+1}.
\end{equation}
One notes that $\im[z_{\nu}] \not=0$ for any finite value of $\mathpzc{n}$ (here assumed to be \textsl{even}), implying that the function $\t{G}_{\X{(0)}}(z)$ in Eq.\,(\ref{e41}) violates an important analytic property expected of the exact Green function, namely that $\t{G}(z)$ \textsl{must} be free from zeros in the finite part of the region $\im[z]\not=0$ of the complex the $z$-plane [\S\,2 in Ref.\,\citen{JML61}] [\S\,4.7, p.\,137, in Ref.\,\citen{BF99a}].

\begin{figure}[t!]
\centerline{\includegraphics[angle=0, width=0.6\textwidth]{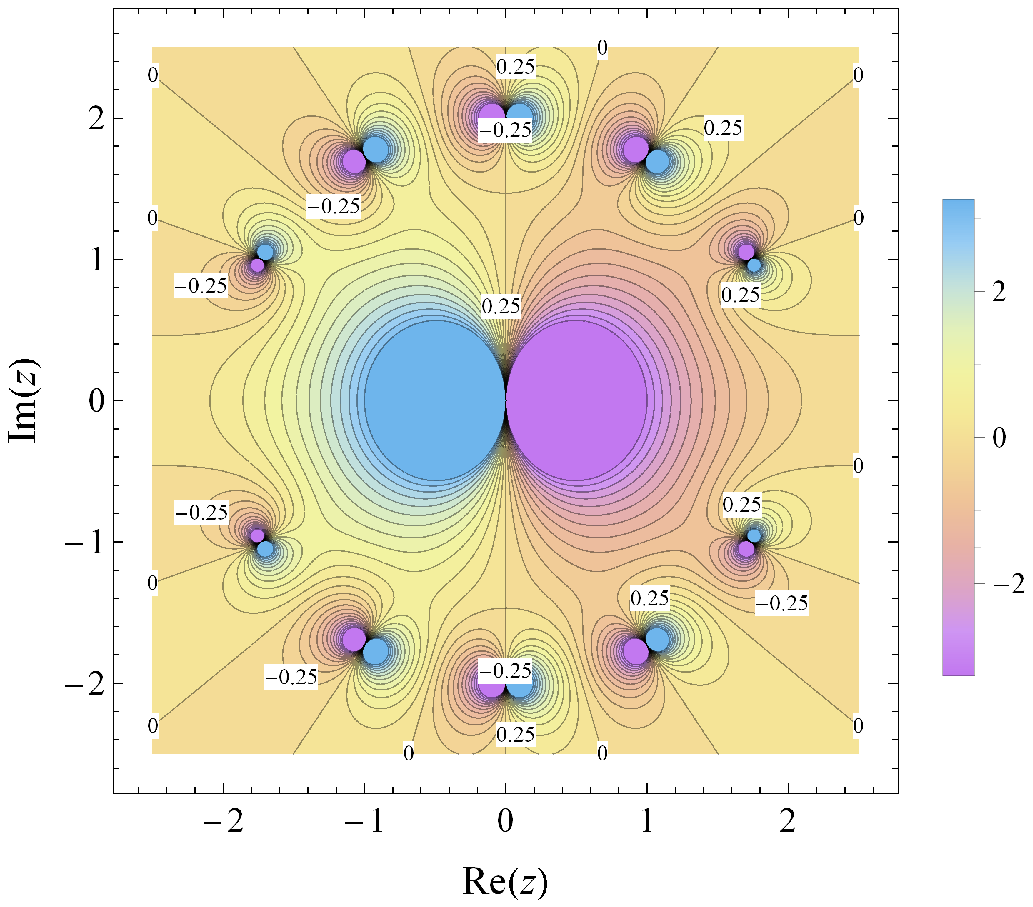}}
\caption{(Colour) \protect\refstepcounter{dummy}\label{TheContour1}The contour plot of $\re[\t{G}^{-1}(z) -\t{G}_{\protect\X{(n)}}^{-1}(z)]$, Eq.\,(\protect\ref{e5a}), for $n=0$, with  $\t{G}_{\protect\X{(n)}}(z)$ calculated to $10$th-order in $U$ in terms of the non-interacting Green function $\t{G}_{\protect\X{0}}(z)$ in Eq.\,(\protect\ref{e15a}). Here $\hbar =1$ and $U=4$. The locations of the ten simple poles $\{z_{\nu}, z_{\nu}^*\| \nu=1,\dots,5\}$ of $\t{G}^{-1}(z) -\t{G}_{\protect\X{(n)}}^{-1}(z)$, arising from the simple zeros of $\t{G}_{\protect\X{(0)}}(z)$, Eqs\,(\protect\ref{e41a}) and (\protect\ref{e41b}), are clearly visible. The two poles on the imaginary axis reflect the fact that here $\mathpzc{n}/2 = 5$ is an odd integer (see the caption of Fig.\,\protect\ref{f9}, p.\,\protect\pageref{SolidLine2}, below).}
\label{f5}
\end{figure}

\begin{figure}[t!]
\centerline{\includegraphics[angle=0, width=0.6\textwidth]{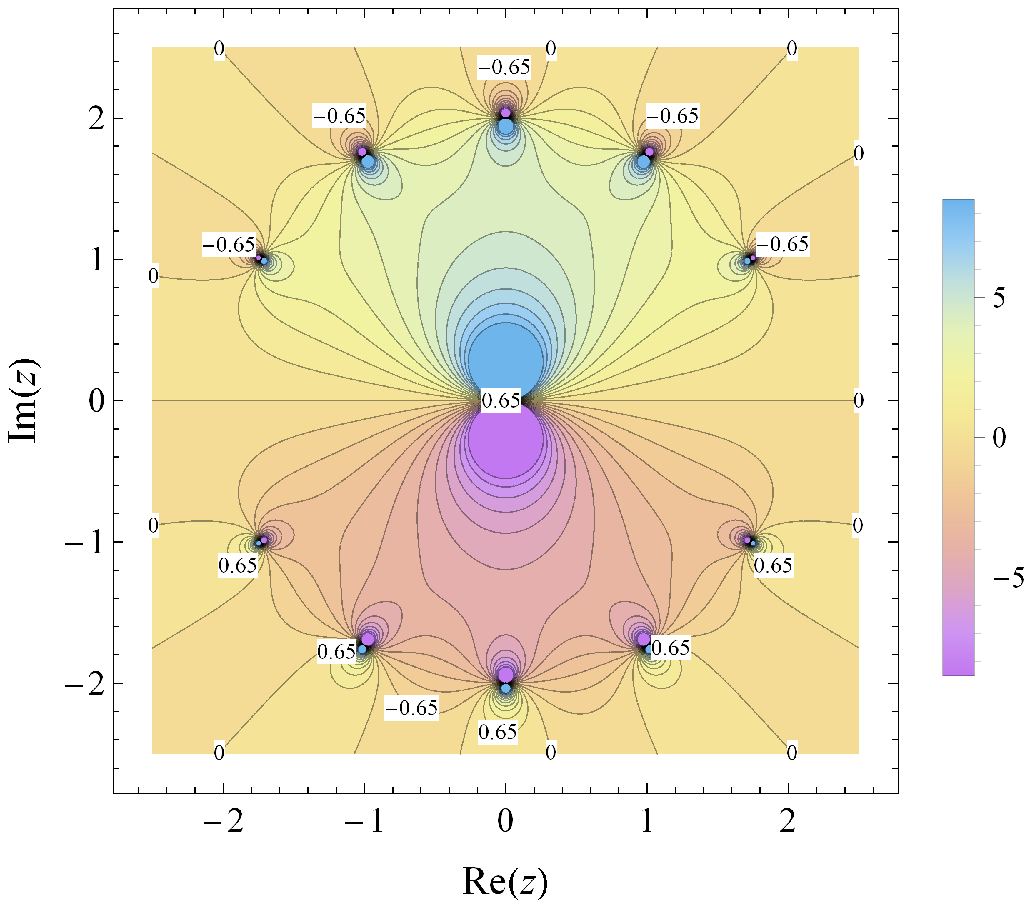}}
\caption{(Colour) \protect\refstepcounter{dummy}\label{TheContour2}The contour plot of $\im[\t{G}^{-1}(z) -\t{G}_{\protect\X{(n)}}^{-1}(z)]$ for $n=0$. Further details are identical to those indicated in the caption of Fig.\,\protect\ref{f5}.}
\label{f6}
\end{figure}

Figures \ref{f5} and \ref{f6} depict the deviation of the function $\t{G}_{\X{(0)}}^{-1}(z)$ specific to $\mathpzc{n} = 10$ from $\t{G}^{-1}(z)$, for $U=4$. One clearly observes that the function $\t{G}_{\X{0;1}}^{-1}(z)$, calculated according to the expression in Eq.\,(\ref{e40}), is \textsl{not} a first-order monic polynomial of $z$ (for more details, see the following paragraph). \emph{We have therefore established that at least in the case of the half-filled GS of the Hubbard Hamiltonian for spin-$\tfrac{1}{2}$ particles in the atomic limit to any finite order the zero-temperature perturbation series expansion of $\t{G}(z)$ in terms of the non-interacting Green function $\t{G}_{\X{0}}(z)$ results in a function that violates an essential analytic property expected of the exact $\t{G}(z)$.} This observation is in conformity with a more general earlier observation in Ref.\,\citen{BF13} (see in particular \S\,III.5 herein). In this connection, recall that for the chemical potential of the half-filled `Hubbard atom' one has $\mu =0$, \S\,\ref{sec3.d}, so that indeed a finite-order perturbation series expansion of $\t{G}(z)$ in terms of $\t{G}_{\X{0}}(z)$ is most problematic for $z$ in the neighbourhood of $\mu$. It is noteworthy that in the limit $\mathpzc{n} = \infty$ the zeroes $\{z_{\nu}, z_{\nu}^* \| \nu = 1,2,\dots,\mathpzc{n}/2 \}$ in Eq.\,(\ref{e41b}) condense into a continuous circle of radius $U/2$ in the $z$-plane, over which the functions $\mathcal{G}_{\mp}(G(z))$ in Eq.\,(\ref{e1a}) are discontinuous.

With reference to Figs.\,\ref{f5} and \ref{f6}, since the function $\hbar\hspace{0.4pt}\t{G}_{\X{0;0}}^{-1}(x) \equiv \hbar \t{G}_{\X{0}}^{-1}(z)$ is a first-order \textsl{monic} polynomial of $z$, in order for the function $G_{\X{0;1}}(z)$ obtained according to the expression in Eq.\,(\ref{e5a}) (irrespective of whether on the basis of scheme A or scheme B) to qualify as a `non-interacting' Green function to be used in the perturbational calculation of $\t{G}_{\X{(1)}}(z)$, the function $\t{G}^{-1}(z) -\t{G}_{\X{(0)}}^{-1}(z)$ would have to be a constant, independent of $z$. This is evidently \textsl{not} the case. Qualitatively similar results as displayed in Figs.\,\ref{f5} and \ref{f6} are obtained for arbitrary non-vanishing values of $U$, and $\t{G}_{\X{(0)}}^{-1}(z)$ calculated to arbitrary \textsl{finite} order in $U$. Interestingly, the contour plot in Fig.\,\ref{f6} clearly demonstrates that the function $-\t{G}_{\X{0;1}}(z)$ is \textsl{not} a Nevanlinna function of $z$, in violation of what is expected of any one-particle Green function, whether interacting or `non-interacting', \S\,\ref{s2.2}. For clarity, in order for $-\t{G}_{\X{0;1}}(z)$ to be a Nevanlinna function of $z$, it is \textsl{necessary} that the function $\t{G}_{\X{0;1}}^{-1}(z)$ be a Nevanlinna function of $z$, Eq.\,(\ref{e3a}).\footnote{See also footnote \raisebox{-1.0ex}{\normalsize{\protect\footref{notee1}}} on p.\,\protect\pageref{ThisObservation}.} This is however clearly \textsl{not} the case, since $\im[\t{G}^{-1}(z) -\t{G}_{\X{(n)}}^{-1}(z)]$ takes negative values less than $-2$ in the region $0 <\im[z] < 2$ (note the colour coding in the neighbourhoods of the points $\{z_{\nu}\| \nu\}$). Such value cannot be compensated by the $z$ arising from the function $\t{G}_{\X{0}}^{-1}(z)$ on the RHS of Eq.\,(\ref{e40}).

Following Eq.\,(\ref{e41}), for $\vert z\vert/U \to 0$ to leading order one has
\begin{equation}\label{e42}
\t{G}_{\X{(0)}}^{-1}(z) \sim \frac{z}{\hbar} \Big(\frac{2 z}{U}\Big)^{\mathpzc{n}},
\end{equation}
from which it follows that for a fixed finite $z$ and $\mathpzc{n}\ge 2$, one can choose $U$ so large as to make the last term on the RHS of Eq.\,(\ref{e40}) negligibly small in comparison with the other two terms. Since $\t{G}^{-1}(z) = \t{G}_{\X{0}}^{-1}(z) - \t{\Sigma}(z)$, Eq.\,(\ref{eg15ax}), where $\t{G}_{\X{0}}(z)$ is independent of $U$ and in the case at hand $\t{\Sigma}(z)$ is proportional to $U^2/z$, Eq.\,(\ref{e25}), it indisputably follows that for sufficiently large $U$ the function $\t{G}_{\X{0;1}}^{-1}(z)$, according to both schemes A and B, bears \textsl{no} resemblance to what one expects of a `non-interacting' Green function.

In view of the above observations, we have explicitly demonstrated that by initiating the iterative schemes A and B with the function $\t{G}_{\X{0}}(z)$ and calculating $\t{G}_{\X{(0)}}(z)$ on the basis of the perturbation series expansion of $\t{G}(z)$ in terms of $\t{G}_{\X{0}}(z)$, in practice the function $\t{G}^{-1}(z) -\t{G}_{\X{(0)}}^{-1}(z)$ on the RHS of Eq.\,(\ref{e5a}) proves to be a highly non-trivial function of $z$ (this in particular in a finite neighbourhood of the origin of the $z$-plane), whereby the resulting function $\hbar\hspace{0.4pt}\t{G}_{\X{0;1}}^{-1}(z)$ \textsl{cannot} be a first-order \textsl{monic} polynomial of $z$ in order for $\t{G}_{\X{0;1}}(z)$ to be employable in the perturbation series expansion of $\t{G}(z)$ in terms of a `non-interacting' Green function, \S\,\ref{s4x}. Nonetheless, this is exactly the way the function $\t{G}_{\X{(1)}}(z)$, and by extension the function $\t{G}_{\X{(n)}}(z)$ for arbitrary integer values of $n$, is calculated, according to the schemes A and B, in Ref.\,\citen{KFG14}.

From the Dyson equation, Eq.\,(\ref{e4a}), and the expression for $\t{G}_{\X{(0)}}(z)$ in Eq.\,(\ref{e41}), for the associated self-energy one obtains
\begin{equation}\label{e54}
\t{\Sigma}_{\X{(0)}}(z) = \frac{z}{\hbar}\frac{\sum_{\nu=1}^{\mathpzc{n}/2} \big(U/2 z\big)^{2\nu}}{1 + \sum_{\nu=1}^{\mathpzc{n}/2} \big(U/2 z\big)^{2\nu}} \equiv \frac{U^2}{4\hbar z} \frac{1- (U/2z)^{\mathpzc{n}}}{1-(U/2z)^{\mathpzc{n}+2}},
\end{equation}
so that\,\footnote{\emph{Cf.} Eqs\,(2.150) and (2.151) in Ref.\,\protect\citen{BF19}.}
\begin{equation}\label{e55}
\t{\Sigma}_{\X{(0)}}(z) = \t{\Sigma}(z) + \frac{z}{\hbar} \hspace{1.2pt} O\big((U/2 z)^{\mathpzc{n}+2}\big)\;\; \text{for}\;\, 2 \le \mathpzc{n} < \infty,
\end{equation}
where $\t{\Sigma}(z)$ is the exact self-energy, Eq.\,(\ref{e25}). On the other hand, from the perspective of the strong-coupling perturbation theory, one has
\begin{equation}\label{e55a}
\t{\Sigma}_{\X{(0)}}(z) = \frac{z}{\hbar} + \frac{z}{\hbar}\hspace{1.2pt} O\big((2 z/U)^{\mathpzc{n}}\big)\;\; \text{for}\;\, 2 \le \mathpzc{n} < \infty.
\end{equation}
We should emphasise that the result in Eq.\,(\ref{e55a}) is \textsl{not} to be identified with the strong-coupling perturbation series expansion of $\t{\Sigma}(z)$. In the strong-coupling, the counterpart of the expansion in Eq.\,(\ref{e54}) is to be obtained from a truncated expression (similarly to that in Eq.\,(\ref{e41})), deduced from the infinite series in Eq.\,(\ref{e38a}).

We note that for $\mathpzc{n}$ \textsl{even} and $\mathpzc{n} \ge 4$ the self-energy in Eq.\,(\ref{e54}) can be expressed as
\begin{equation}\label{e55b}
\t{\Sigma}_{\X{(0)}}(z) = \frac{U^2 z}{4\hbar}\hspace{0.6pt}  \prod_{\nu=1}^{\mathpzc{n}/2-1} (z-z_{\nu}') (z-{z_{\nu}'}^*)\Big/
\prod_{\nu=1}^{\mathpzc{n}/2} (z-z_{\nu}) (z-z_{\nu}^*),
\end{equation}
where the $z_{\nu}$ is introduced in Eq.\,(\ref{e41b}), and
\begin{equation}\label{e55c}
z_{\nu}' \doteq \frac{U}{2} \e^{\ii \theta_{\nu}'},\;\; \text{with}\;\;\; \theta_{\nu}' \equiv \frac{2\pi\nu}{\mathpzc{n}}.
\end{equation}
The poles $\{z_{\nu},z_{\nu}^*\| \nu = 1,\dots, \mathpzc{n}/2\}$, all of which are located strictly away from the real axis of the $z$-plane for any finite \textsl{even} value of $\mathpzc{n}$, signify the major shortcoming of the function  $\t{\Sigma}_{\X{(0)}}(z) $ as representing the self-energy.\footnote{Note that as $\mathpzc{n}$ increases towards $\infty$, the distinction between $\theta_{\nu}$ and $\theta_{\nu}'$, and therefore that between $z_{\nu}$ and $z_{\nu}'$, increasingly diminishes for any $\nu \ll \mathpzc{n}$, ultimately resulting in the ratio of the two products on the RHS of Eq.\,(\protect\ref{e55b}) reducing to the function $1/z^2$ for $\mathpzc{n} =\infty$. Clearly, the latter ratio is equal to $1/z^2$ for $U = 0$ when $\mathpzc{n}$ is \textsl{even} and $\mathpzc{n} \ge 4$.}

\begin{figure}[t!]
\psfrag{x}[c]{\huge $\varepsilon$}
\psfrag{y}[c]{\huge $\t{\Sigma}_{\protect\X{(0)}}(\varepsilon + \protect\ii\eta)$}
\centerline{\includegraphics[angle=0, width=0.6\textwidth]{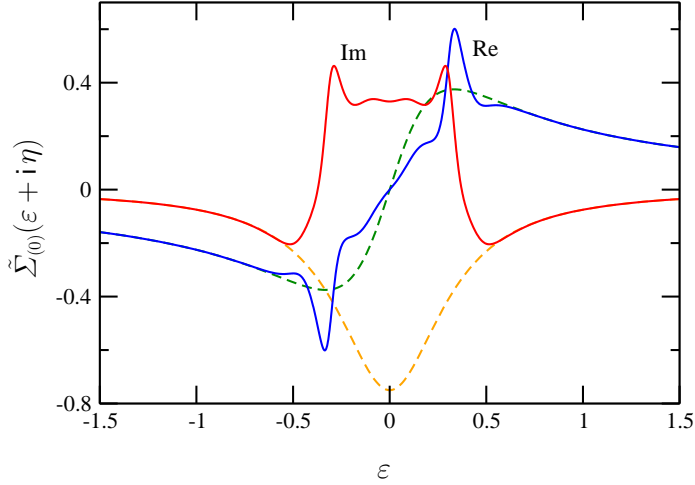}}
\caption{(Colour) \protect\refstepcounter{dummy}\label{SolidLine1}\emph{Solid lines}: The real and imaginary parts of $\t{\Sigma}_{\protect\X{(0)}}(\varepsilon+\protect\ii\eta)$, Eq.\,(\protect\ref{e54}), corresponding to $U=1$, $\mathpzc{n}=12$ and $\eta = 1/3$. \emph{Broken lines}: The real and imaginary parts of the exact $\t{\Sigma}(\varepsilon+\protect\ii\eta)$, Eq.\,(\protect\ref{e25}). With reference to the remarks concerning the expression in Eq.\,(\protect\ref{e54}), $\t{\Sigma}_{\protect\X{(0)}}(\varepsilon+\protect\ii\eta)$ does not converge to the exact $\t{\Sigma}(\varepsilon+\protect\ii\eta)$ for $\vert\varepsilon + i\eta\vert < 1/2$, instead for increasing values of $\mathpzc{n}$, with $\mathpzc{n} <\infty$, it converges towards $\varepsilon + \protect\ii\eta$ for $\vert \varepsilon + \protect\ii\eta\vert \to 0$, Eq.\,(\protect\ref{e54a}) (identifying $\hbar$ with unity). Note that the exact $\t{\Sigma}(\varepsilon + \protect\ii \eta)$, in contrast to $\t{\Sigma}_{\protect\X{(0)}}(\varepsilon + \protect\ii\eta)$, satisfies the equality in Eq.\,(\protect\ref{e2}) for the range of $\varepsilon$ considered. }
\label{f8}
\end{figure}

\begin{figure}[t!]
\psfrag{x}[c]{\huge $y$}
\psfrag{y}[c]{\huge $\im[\t{\Sigma}_{\protect\X{(0)}}(\protect\ii y)]$}
\centerline{\includegraphics[angle=0, width=0.6\textwidth]{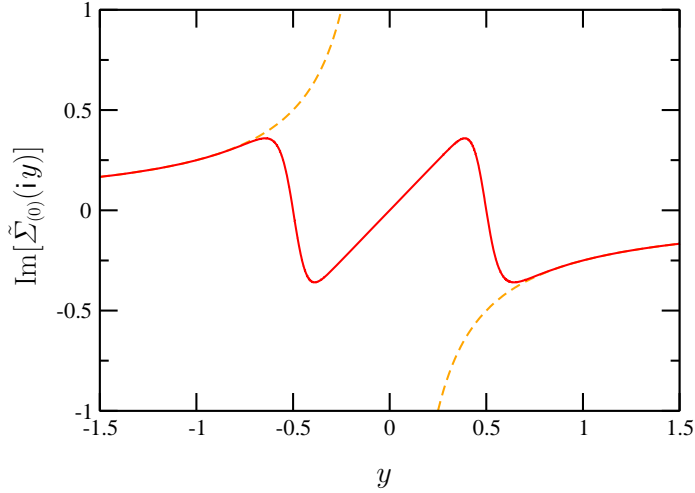}}
\caption{(Colour) \protect\refstepcounter{dummy}\label{SolidLine2}\emph{Solid line}: The imaginary part of $\t{\Sigma}_{\protect\X{(0)}}(\protect\ii y)$, Eq.\,(\protect\ref{e54}), corresponding to $U=1$ and $\mathpzc{n}=12$. \emph{Broken line}: The imaginary part of the exact $\t{\Sigma}(\protect\ii y)$, Eq.\,(\protect\ref{e25}). For $z = \protect\ii y$, $y \in \mathds{R}$, the real parts of these functions are identically vanishing. Note that the exact $\t{\Sigma}(\protect\ii y)$, in contrast to $\t{\Sigma}_{\protect\X{(0)}}(\protect\ii y)$, satisfies the equality in Eq.\,(\protect\ref{e2}) for the range of $y$ considered. With reference to the remarks concerning the expression in Eq.\,(\protect\ref{e54}), $\t{\Sigma}_{\protect\X{(0)}}(\protect\ii y)$ does not converge to the exact $\t{\Sigma}(\protect\ii y)$, Eq.\,(\protect\ref{e25}), for $\vert y\vert < 1/2$ and increasing values of $\mathpzc{n}$, with $\mathpzc{n} <\infty$, as evidenced by the fact that it converges towards $\protect\ii y$ for $y\to 0$, Eq.\,(\protect\ref{e54a}) (identifying $\hbar$ with unity). In order for the deviations of the two functions displayed here to become noticeable at fermionic Matsubara energies $\{\zeta_{m}\| m\}$, Eq.\,(\protect\ref{e24}), with $\mu=0$ (corresponding to half-filling, \S\,\protect\ref{sec3.d}), one must have (for $\hbar=1$ and $k_{\textsc{b}} =1$) $\omega_0 \equiv \pi/\beta\lesssim 0.7$ (\emph{cf.} Eq.\,(\protect\ref{e71})). For $\beta = 2$ one has $\omega_0 = \pi/2 \approx 1.57$, which lies outside the frame of this figure. For the cases where $\mathpzc{n}/2$ is \textsl{odd}, the function $\t{\Sigma}_{\protect\X{(0)}}(\protect\ii y)$ has two poles along the $y$-axis, located at $y = \pm U/2$; with $\mathpzc{n}/2 = 2 j +1$, $j \in \mathds{N}_{0}$, from the expressions in Eq.\,(\protect\ref{e41b}) one has $\theta_{\nu} = \pi/2$ for $\nu =  j+1$.}
\label{f9}
\end{figure}

The result in Eq.\,(\ref{e55}) is not only in conformity with that in Eq.\,(\ref{e49a}) below, but also shows that the self-energy as calculated in terms of the non-interacting Green function $\t{G}_{\X{0}}(z)$ (under the condition to be specified in \S\,\ref{s4xa} below) is identically vanishing order-by-order beyond the second order in $U$. Note that while for any non-vanishing even value of $\mathpzc{n}$ and finite value of $U$, $0 < U < \infty$, the function $\t{\Sigma}_{\X{(0)}}(z)$ in Eq.\,(\ref{e54}) behaves correctly in the asymptotic region $z \to \infty$, it behaves incorrectly in the asymptotic region $z \to 0$; for this function to leading order one has
\begin{equation}\label{e54a}
\t{\Sigma}_{\X{(0)}}(z) \sim \frac{z}{\hbar}\;\,\text{for}\;\, z \to 0\;\;\, (\mathpzc{n} < \infty),
\end{equation}
to be contrasted with the exact result $\t{\Sigma}(z) = U^2/(4\hbar z)$, $\forall z$, Eq.\,(\ref{e25}) (see Figs\,\ref{f8} and \ref{f9}).

Since for the `Hubbard atom' the thermal self-energy $\t{\mathscr{S}}(z)$ identically coincides with its zero-temperature counterpart $\t{\Sigma}(z)$,\footnote{See Eqs\,(\protect\ref{e63}) and (\protect\ref{e67x3}) below.} on the basis of the above observations one directly deduces the behaviour of the thermal counterpart of $\t{\Sigma}_{\X{(0)}}(z)$ in Eq.\,(\ref{e54}), that is $\t{\mathscr{S}}_{\X{(0)}}(z)$ -- for which one has $\t{\mathscr{S}}_{\X{(0)}}(z) \equiv \t{\Sigma}_{\X{(0)}}(z)$, in particular at the Matsubara energies $\{\zeta_{m} \| m\}$, Eq.\,(\ref{e24}). Thus, from the right-most expression in Eq.\,(\ref{e54}) one observes that for $\vert z\vert > U/2$ the function $\t{\mathscr{S}}_{\X{(0)}}(z)$ converges rapidly towards the exact self-energy $\t{\mathscr{S}}(z)$ for increasing values of $\mathpzc{n}$. Sampling the function $\t{\mathscr{S}}_{\X{(0)}}(z)$ at the Matsubara energies $\{\zeta_{m} \| m\}$, Eq.\,(\ref{e24}), one would therefore observe a fast convergence of $\t{\mathscr{S}}_{\X{(0)}}(\zeta_m)$ towards the exact value $\t{\mathscr{S}}(\zeta_m)$ for \textsl{all} $m$ for which $\vert \zeta_m \vert > U/2$ as $\mathpzc{n} \to \infty$. This is in particular the case when (since by p-h symmetry $\mu=0$ for all $\beta$, \S\,\ref{sec3.d})
\begin{equation}\label{e71}
\hbar\omega_0 > U/2 \iff U< 2\pi/\beta.
\end{equation}
Introducing
\begin{equation}\label{e71a}
U_{\beta} \doteq \frac{2\pi}{\beta} \equiv 2\pi k_{\textsc{b}} T,
\end{equation}
for $U < U_{\beta}$ the function $\t{\mathscr{S}}_{\X{(0)}}(\zeta_m)$ must therefore be almost indistinguishable from $\t{\mathscr{S}}(\zeta_m)$ for \textsl{all} $m \in \mathds{Z}$ and sufficiently large $\mathpzc{n}$.\footnote{Appendix \protect\ref{sae}. \label{notea}} By the same reasoning, $\t{\mathscr{S}}_{\X{(0)}}(\zeta_m)$ must be almost indistinguishable from $\t{\mathscr{S}}(\zeta_m)$, $\forall m\in \mathds{Z}\backslash \{-1,0\}$, for $U < U_{2\beta/3}$.\footnote{Note that $\hbar\omega_{-1} = -\pi/\beta$ and $\hbar\omega_{1} = -\hbar\omega_{-2} = 3\pi/\beta$, Eq.\,(\protect\ref{e70}).} One has $U_{2} = \pi \approx 3.14$ and $U_{4/3} = 3\pi/2 \approx 4.71$, where $\beta = 2$ corresponds the temperature $T = 0.5$ (in the units where $k_{\textsc{b}} = 1$) in the calculations by Kozik \emph{et al.} \cite{KFG14}. Noting that the data in Fig.\,3 of Ref.\,\citen{KFG14} correspond to $U \in \{1,2,3,4\}$, it follows that restricting oneself to $\beta = 2$ and $U$ to $\{1,2,3,4\}$, for sufficiently large $\mathpzc{n}$ the function $\t{\mathscr{S}}_{\X{(0)}}(\zeta_m)$ must be almost indistinguishable from $\t{\mathscr{S}}(\zeta_m)$ for all $m\in \mathds{Z}\backslash \{-1,0\}$.\footref{notea} Such and similar observations would \textsl{incorrectly} suggest fast convergence of $\t{\mathscr{S}}_{\X{(0)}}(z)$ towards the exact function $\t{\mathscr{S}}(z)$, $\forall z \in \mathds{C}$, for increasing values of  $\mathpzc{n}$, in contradiction with the fact that $\t{\mathscr{S}}_{\X{(0)}}(z)$, Eq.\,(\ref{e54}), \textsl{cannot} converge towards the exact $\t{\mathscr{S}}(z)$ inside the disc $\vert z\vert < U/2$ for any $\mathpzc{n} < \infty$, as evidenced by the asymptotic expression in Eq.\,(\ref{e54a}). See Eq.\,(\ref{e54a}) as well as Figs\,\ref{f8} and \ref{f9}, where $\t{\Sigma}_{\X{(0)}}(z) \equiv \t{\mathscr{S}}_{\X{(0)}}(z)$ and $\t{\Sigma}(z) \equiv \t{\mathscr{S}}(z)$.

\refstepcounter{dummyX}
\subsubsection{The self-energy at zero temperature}
\phantomsection
\label{s4xa}
The fact that the single pole of the non-interacting Green function $\t{G}_{\X{0}}(z)$  in Eq.\,(\ref{e15a}) at $z=0$ is to represent both a half-filled and a half-empty non-dispersive band of energies, clearly signifies the problematic nature of this function.\footnote{Recall that at half-filling $\mu = 0$, \S\,\protect\ref{sec3.d}.} Consequently, without a redefinition of $\t{G}_{\X{0}}(z)$ as the limit of an appropriate sequence of Green functions, it cannot be used in the framework of the \textsl{zero-temperature} perturbation series expansion of the Green function $\t{G}(z)$ and the self-energy $\t{\Sigma}(z)$. Most evidently, the Green function in Eq.\,(\ref{e15a}) \textsl{cannot} be the time-Fourier transform of a function of time-ordered operators. This is in contrast to the Green function $\t{G}(z)$ in Eq.\,(\ref{e25}) for all $U\not=0$. For the `physical' function $G(\varepsilon)$, $\varepsilon \in \mathds{R}$, associated with $\t{G}(z)$ (\emph{cf.} Eq.\,(\ref{e1})), one has\,\footnote{The corresponding spectral function $A(\varepsilon)$ is presented in Eq.\,(\protect\ref{exc1}) below (\emph{cf.} Eq.\,(\protect\ref{e4oj})).}
\begin{equation}\label{e53}
G(\varepsilon) = \frac{\hbar}{2}\Big(\frac{1}{\varepsilon + U/2 - \ii 0^+} + \frac{1}{\varepsilon - U/2 + \ii 0^+}\Big).
\end{equation}
For $\epsilon\hspace{0.6pt} U > 0$, we thus define
\begin{equation}\label{e44}
\t{G}_{\epsilon}(z) \doteq \frac{\hbar}{2}\hspace{0.8pt}\frac{1}{z + \epsilon\hspace{0.6pt} U/2} + \frac{\hbar}{2}\hspace{0.8pt}\frac{1}{z -\epsilon\hspace{0.6pt} U/2} \equiv \t{G}_{\epsilon}^-(z) + \t{G}_{\epsilon}^+(z),
\end{equation}
which identically coincides with $\t{G}_{\X{0}}(z)$ for $\epsilon=0$ and with $\t{G}(z)$ for $\epsilon=1$. Clearly, the above-mentioned problem with regard to $\t{G}_{\X{0}}(z)$ is \textsl{not} shared by $\t{G}_{\X{0^+}}(z)$. For $\epsilon = 0^+$, the single-particle spectrum of the non-interacting atom is `hair-split', with non-trivial consequences. It is this function, to be distinguished from $\t{G}_{\X{0}}(z)$, that is to be employed for the perturbational calculation of $\t{\Sigma}(z)$ in terms of the non-interacting Green function at zero temperature.\footnote{See Ref.\,\protect\citen{BF12} for the detrimental effect of the GS degeneracy on the Luttinger theorem.}

On reflection, the problem indicated above with regard to the function $\t{G}_{\X{0}}(z)$ is directly attributable to the \textsl{locality} of the functions considered here, arising from the atomic limit of the Hubbard Hamiltonian having been effected prior to calculating the terms of the relevant perturbation series; in the thermodynamic limit, outside the atomic limit a single singular point on the energy axis ($z=0$, or $z=T_{\X{0}}$, Eq.\,(\ref{e14b})) would be associated with a zero-measure subset of the $\bm{k}$ points of the underlying first Brillouin zone.

For the reasons discussed in \S\,\ref{s4x}, perturbational calculation of the self-energy in terms of $\t{G}_{\epsilon}(z)$ is in principle only meaningful in the limits $\epsilon = 0^+$ and $\epsilon =1$. The relevant terms in the perturbation series expansion of $\t{\Sigma}(z)$ in terms of $\t{G}_{\X{0^+}}(z)$ correspond to \textsl{proper}, 1PI, self-energy diagrams, and to \textsl{skeleton}, 2PI, self-energy diagrams when $\t{\Sigma}(z)$ is evaluated in terms of $\t{G}_{\X{1}}(z) \equiv \t{G}(z)$. Since the Hartree-Fock self-energy $\Sigma^{\textsc{hf}}$ corresponding to half-filling as evaluated in terms of $\t{G}_{\epsilon}(z)$ does not depend on the value of $\epsilon$, it follows that similarly to the Green function $\t{G}_{\X{0}}(z)$ in Eq.\,(\ref{e15a}) the function $\t{G}_{\epsilon}(z)$ in Eq.\,(\ref{e44}) takes explicit account of $\Sigma^{\textsc{hf}}$ for all values of $\epsilon$. Consequently, at the second order in the bare interaction potential the sets of the self-energy diagrams are the same irrespective of whether the perturbation expansion is in terms of $\t{G}_{\X{0^+}}(z)$ or $\t{G}_{\X{1}}(z)$, Fig.\,\ref{f7}, p.\,\pageref{SecondOrderS}, below.\footnote{With reference to \S\,\protect\ref{sec.3a.1}, since at the zeroth-order of perturbation theory $\Sigma_{\sigma}^{\textsc{hf}}$, and not merely $\Sigma^{\textsc{f}}$, has been taken into account, in dealing with the 1PI (\emph{i.e.} \textsl{proper}) self-energy diagrams, describing the self-energy $\protect\t{\Sigma}_{\sigma}$ in terms of $\protect\t{G}_{\protect\X{0}}$, also diagrams containing Fock self-energy insertions are to be discarded. This is why in calculating $\protect\t{\Sigma}_{\epsilon}^{(\protect\X{2})}(z)$ in terms of $\protect\t{G}_{\protect\X{0}}$ we do \textsl{not} take into account the 1PI second-order self-energy diagram (2.0) besides the two 2PI second-order self-energy diagrams (2.1) and (2.2) in Fig.\,\protect\ref{f7}, p.\,\protect\pageref{SecondOrderS}, below. We emphasize that, insofar as the Hubbard Hamiltonian is concerned, unless we indicate otherwise the basis of the perturbation series expansions in this publication is the representation $\protect\h{\mathcal{H}}$ in Eq.\,(\ref{ex01bx}), to be distinguished from the alternative representations in Eqs\,(\protect\ref{ex01f}) and (\protect\ref{ex01k}).} \emph{In evaluating the second-order self-energy diagrams in the following, we shall therefore employ the Green function $\t{G}_{\epsilon}(z)$ for an unspecified value of $\epsilon > 0$, with the understanding that the second-order self-energy thus calculated is only physically meaningful for $\epsilon = 0^+$ and $\epsilon=1$.}

\emph{Unless}\refstepcounter{dummy}\label{UnlessWe} \emph{we explicitly indicate otherwise, the diagrammatic contributions (to the self-energy and the polarization function) to be presented in this publication correspond to the Hamiltonian $\h{\mathcal{H}}$, Eq.\,(\ref{ex01bx}) (as opposed to the equivalent Hamiltonians $\hspace{0.28cm}\h{\hspace{-0.28cm}\mathpzc{H}}$, Eq.\,(\ref{ex01f}), and $\h{\mathsf{H}}$, Eq.\,(\ref{ex01k})), in the atomic limit.} As regards the self-energy, we emphasize that at any given order of the perturbation theory the distinction between the indicated three Hamiltonians is immaterial only if one considers the \textsl{total} contribution of the self-energy diagrams of that order. In this connection, note the equalities in Eqs\,(\ref{ex01fx}) and (\ref{ex01l}).

We denote the contributions of the second-order skeleton self-energy diagrams $(2.2)$ and $(2.1)$ depicted in Fig.\,\ref{f7} below, p.\,\pageref{SecondOrderS}, evaluated in terms of $\t{G}_{\epsilon}(z)$, by $\Sigma_{\epsilon}^{\X{(2.2)}}(\varepsilon)$ and $\Sigma_{\epsilon}^{\X{(2.1)}}(\varepsilon)$, representing respectively $\Sigma^{\X{(2.2)}}(\varepsilon;[G_{\epsilon}])$ and $\Sigma^{\X{(2.1)}}(\varepsilon;[G_{\epsilon}])$.\footnote{For the notation, consult \S\,1.2 in Ref.\,\protect\citen{BF19}.} One obtains, \S\,\ref{sd21} [pp.\,102-103 in Ref.\,\citen{FW03}],
\begin{align}\label{e46}
\Sigma_{\epsilon}^{\X{(2.2)}}(\varepsilon) &= \frac{U^2}{4\hbar} \Big(\frac{1}{\varepsilon +3\epsilon\hspace{0.6pt} U/2 - \ii 0^+} + \frac{1}{\varepsilon -3\epsilon\hspace{0.6pt} U/2 + \ii 0^+}\Big), \\
\label{e47}
\Sigma_{\epsilon}^{\X{(2.1)}}(\varepsilon) &= -\frac{1}{2} \Sigma_{\epsilon}^{\X{(2.2)}}(\varepsilon),
\end{align}
resulting in
\begin{equation}\label{e48}
\Sigma_{\epsilon}^{\X{(2)}}(\varepsilon) \equiv \Sigma_{\epsilon}^{\X{(2.2)}}(\varepsilon) + \Sigma_{\epsilon}^{\X{(2.1)}}(\varepsilon) = \frac{1}{2} \Sigma_{\epsilon}^{\X{(2.2)}}(\varepsilon).
\end{equation}
Hence (\emph{cf.} Eq.\,(\ref{e1}) and recall that here $\mu=0$)
\begin{equation}\label{e49}
\t{\Sigma}_{\epsilon}^{\X{(2)}}(z) = \frac{U^2}{8\hbar} \Big(\frac{1}{z + 3\epsilon\hspace{0.6pt} U/2} + \frac{1}{z - 3\epsilon\hspace{0.6pt} U/2}\Big).
\end{equation}
We note that the self-energy $\Sigma_{\epsilon}^{\X{(2.2)}}(\varepsilon)$ contains a factor of $2$ arising from the sum over the spin index of the polarization loop\,\footnote{Appendix \protect\ref{sa}. See \S\,\protect\ref{sd21}, in particular Eq.\,(\protect\ref{ex04a}). See also the remarks following Eq.\,(\protect\ref{e73}) below.} in the diagram associated with it, Fig.\,\ref{f7}\,(2.2), p.\,\pageref{SecondOrderS} (\emph{cf.} Eq.\,(\ref{e57}) below).\refstepcounter{dummy}\label{WithReference}\footnote{With reference to the considerations of \S\,\protect\ref{sd2}, the self-energy contributions $\Sigma_{\epsilon}^{\protect\X{(2.2)}}(\varepsilon)$ and $\Sigma_{\epsilon}^{\protect\X{(2.1)}}(\varepsilon)$ correspond to the atomic limit of the Hubbard Hamiltonian in Eq.\,(\protect\ref{ex01bx}), \S\,\protect\ref{sec3.b}. On employing the Hamiltonian in Eq.\,(\protect\ref{ex01f}) (Eq.\,(\protect\ref{ex01k})), the self-energy contribution $\Sigma_{\epsilon}^{\protect\X{(2.1)}}(\varepsilon)$ does not occur (is identically vanishing), and the contribution of $\Sigma_{\epsilon}^{\protect\X{(2.2)}}(\varepsilon)$ is qual to one-half of that in Eq.\,(\protect\ref{e46}). Naturally, for the above-mentioned three representations of the Hubbard Hamiltonians the total second-order self-energy $\Sigma_{\epsilon}^{\protect\X{(2)}}(\varepsilon)$ is the same, equal to that in Eq.\,(\protect\ref{e49}). Similar equalities apply for the \textsl{total} contribution of the self-energy diagrams at any arbitrary order $\nu$ of the perturbation theory. \label{notet}} With reference to Eq.\,(\ref{e25}), for $\vert U\vert < \infty$ one has (see \S\,\ref{s4xb})
\begin{equation}\label{e49a}
\lim_{\epsilon\downarrow 0} \t{\Sigma}_{\epsilon}^{\X{(2)}}(z) = \frac{U^2}{4\hbar z} \equiv \t{\Sigma}(z),
\end{equation}
to be contrasted with\,\footnote{According to the notation of Ref.\,\protect\citen{BF19}, $\t{\Sigma}^{\protect\X{(2)}}(z) \equiv \t{\Sigma}_{\protect\X{01};\sigma}^{\protect\X{(2)}}(z) \equiv \t{\Sigma}_{\protect\X{01};\sigma}^{\protect\X{(2)}}(z;[\{G_{\sigma'}\}])$.}
\begin{equation}\label{e50}
\t{\Sigma}^{\X{(2)}}(z) \equiv \Sigma_{\X{1}}^{\X{(2)}}(z) = \frac{U^2}{8\hbar} \Big(\frac{1}{z + 3 U/2} + \frac{1}{z - 3 U/2}\Big) \not\equiv \t{\Sigma}(z).
\end{equation}
The remarkable result in Eq.\,(\ref{e49a}), identified in Ref.\,\citen{KFG14}, should not distract from the fact that $\t{\Sigma}_{\X{0}}^{\X{(2)}}(z)$ (evaluated in terms of $\t{G}_{\X{0}}(z)$ at zero temperature), to be distinguished from $\t{\Sigma}_{\X{0^+}}^{\X{(2)}}(z)$ (evaluated in terms of $\t{G}_{\X{0^+}}(z)$), does \textsl{not} exist. The existence of $\t{\Sigma}_{\X{0}}^{\X{(2)}}(z)$ according to the expression in Eq.\,(\ref{e49})\,\footnote{That is, the equality of the \textsl{value} of $\t{\Sigma}_{\epsilon}^{\protect\X{(2)}}(z)$ at $\epsilon = 0$ with its \textsl{limit} for $\epsilon\downarrow 0$.} corresponds to the process of the underlying integrations with respect to the internal energy variables having been carried out \textsl{prior} to identifying $\epsilon$ with $0$. The non-existence of $\t{\Sigma}_{\X{0}}^{\X{(2)}}(z)$, to which we have referred above, corresponds to the process of evaluating these integrals \textsl{following} the identification of $\epsilon$ with $0$.

For the self-energy in Eq.\,(\ref{e50}), one has (\emph{cf.} Eq.\,(\ref{e55}))
\begin{equation}\label{e50a}
\t{\Sigma}^{\X{(2)}}(z) = \t{\Sigma}(z) + \frac{3^2 z}{\hbar} \Big(\frac{U}{2z}\Big)^4 + \frac{3^4 z}{\hbar} \Big(\frac{U}{2z}\Big)^6 + \frac{3^6 z}{\hbar} \Big(\frac{U}{2z}\Big)^8  + \frac{z}{\hbar}\hspace{0.6pt} O\Big(\Big(\frac{U}{2z}\Big)^{10}\Big).
\end{equation}
We shall discuss the significance of the infinite series on the RHS of Eq.\,(\ref{e50a}),
viewed as the asymptotic series expansion of $\t{\Sigma}^{\X{(2)}}(z) -\t{\Sigma}(z)$ in the asymptotic region $z\to\infty$, or the weak-coupling perturbation series expansion of this function, in \S\S\,\ref{sdis1} and \ref{sd4}. For now we only point out that one ideally expects the infinite series on the RHS of Eq.\,(\ref{e50a}) to be cancelled by the contributions of higher-order skeleton self-energy diagrams evaluated in terms of the interacting Green function, however in the case of the `Hubbard atom' under consideration this proves \textsl{not} to materialise in the way it is expected on the basis of the equality in Eq.\,(\ref{e7h}). See in particular the discussions in \S\,\ref{sec.b2.1} for the violation of the equality in Eq.\,(\ref{e7h}) in this case, arising from the strict locality of the Green function $\t{G}(z)$.

In the light of the above observations, the apparent superiority in the present case of the perturbation series expansion of $\t{\Sigma}(z)$ in terms of $\t{G}_{\X{0^+}}(z)$ to that of $\t{\Sigma}(z)$ in terms of $\t{G}_{\X{1}}(z)$ may be understood by realising that the general perturbation series expansion of $\t{\Sigma}_{\sigma}(z)$ in terms of skeleton (\emph{i.e.} 2PI) self-energy diagrams and the interacting Green functions $\{\t{G}_{\sigma}(z)\| \sigma\}$ is formally deduced from the perturbation series expansion of $\t{\Sigma}_{\sigma}(z)$ in terms of proper (\emph{i.e.} 1PI) self-energy diagrams and the non-interacting Green functions $\{\t{G}_{\X{0};\sigma}(z)\| \sigma\}$ through a partial summation over \textsl{infinite} perturbational contributions. Clearly, when the latter series is terminating, the former series is artificially comprised of infinite number of contributions that ideally must cancel amongst each other on expanding $G_{\sigma}(z)$ in terms of $\{\t{G}_{\X{0};\sigma}(z)\| \sigma\}$, thus leading to the original terminating perturbation series of $\t{\Sigma}(z)$ in terms of $\{\t{G}_{\X{0};\sigma}(z)\| \sigma\}$. As we demonstrate in \S\,\ref{sd4} (appendix \ref{sacx}), this cancellation does \textsl{not} obtain in the case of the `Hubbard atom' without the violation of the equality in Eq.\,(\ref{e7h}), which is exact \ae\footnote{Appendix \protect\ref{sae}.} Explicit calculations in \S\,\ref{sd4} reveal that this failure is due to the strict \textsl{locality} (\emph{i.e.} the strict independence from $\bm{k}$) of the interacting one-particle Green function corresponding to this model, whereby possible failures over subsets of measure zero of the $\bm{k}$-space in the general case manifest themselves as total failures in the case of the `Hubbard atom'. Note that, all $\bm{k}$ points being equivalent in the local limit, in this limit no subset of the $\bm{k}$-space corresponding to a macroscopic system can qualify as being of measure zero.

\begin{figure}[t!]
\centerline{\includegraphics*[angle=0, width=0.4\textwidth]{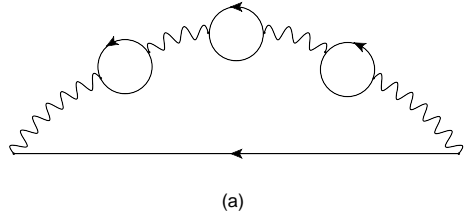}}
\caption{A\protect\refstepcounter{dummy}\label{A4thOr} $4$th-order skeleton self-energy diagram consisting of `bubbles'. The solid wavy lines represent the two-body interaction potential.}
\label{f10}
\end{figure}

To illustrate the significance of $\epsilon\not=0$ to the zero-temperature calculations, we note that for the contribution of the $4$th-order self-energy diagram in Fig.\,\ref{f10},\footnote{This diagram coincides with that specified in Eq.\,(\protect\ref{e93a}). The contribution of this diagram in the case of the `Hubbard atom' and in terms of $\protect\t{G}(z)$ (\emph{i.e.} in terms of $\protect\t{G}_{\epsilon}(z)$ for $\epsilon=1$) is presented in Eq.\,(\protect\ref{ex82}).} which we denote by $\t{\Sigma}_{\epsilon}^{\X{(4\textrm{a})}}(z)$, one has\,\footnote{See \S\,\protect\ref{sd21}. Note that unless indicated otherwise the expression in Eq.\,(\protect\ref{e52b}) and similar expressions correspond to the Hamiltonian $\protect\h{\mathcal{H}}$, Eq.\,(\protect\ref{ex01bx}) (as opposed to $\hspace{0.28cm}\protect\h{\hspace{-0.28cm}\mathpzc{H}}$, Eq.\,(\protect\ref{ex01f}), or $\protect\h{\mathsf{H}}$, Eq.\,(\protect\ref{ex01k})), in the atomic limit.}
\begin{align}\label{e52b}
\t{\Sigma}_{\epsilon}^{\X{(4\textrm{a})}}(z) &= \frac{U^4}{16\hbar} \Big\{\frac{1}{(z + 3\epsilon\hspace{0.6pt} U/2)^3} + \frac{1}{(z - 3\epsilon\hspace{0.6pt} U/2)^3}\Big\} \nonumber\\
&+ \frac{3 U^3}{32\hbar\epsilon} \Big\{\frac{1}{(z + 3\epsilon\hspace{0.6pt} U/2)^2} - \frac{1}{(z - 3\epsilon\hspace{0.6pt} U/2)^2}\Big\}\nonumber\\
&+ \frac{3 U^2}{32\hbar \epsilon^2}\Big\{\frac{1}{z + 3\epsilon\hspace{0.6pt} U/2} + \frac{1}{z - 3\epsilon\hspace{0.6pt} U/2}\Big\}.
\end{align}
Clearly, on account of the equalities in Eqs\,(\ref{e55}) and (\ref{e49a}), the contributions of all $4$th-order \textsl{proper} self-energy diagrams in terms of $\t{G}_{\epsilon}(z)$ must add to zero in the limit $\epsilon=0^+$, a fact that we explicitly establish in \S\,\ref{sec.d53}. Neglecting this for now, it is interesting to note that $\t{\Sigma}_{\epsilon}^{\X{(4\textrm{a})}}(z)$ is comprised of two terms that are singular at $\epsilon = 0$: the second (third) term on the RHS of Eq.\,(\ref{e52b}) has a simple (double) pole at $\epsilon=0$. Further, the second (third) term scales like $U^3$ ($U^2$). This is to be contrasted with the first term on the RHS of Eq.\,(\ref{e52b}), which is both regular at $\epsilon=0$ and scales like $U^4$, conform the expectation that a $\nu$th-order self-energy contribution should directly scale like $U^{\nu}$. Note that for each term on the RHS of Eq.\,(\ref{e52b}) the powers of $U$ and $\epsilon$ add up to $4$. This is immediately appreciated by realising that in the expression for $\t{G}_{\epsilon}(z)$, Eq.\,(\ref{e44}), one encounters $U$ and $\epsilon$ in the form $\epsilon\hspace{0.4pt}U$.

\refstepcounter{dummyX}
\subsubsection{The self-energy at non-zero temperatures}
\phantomsection
\label{s4xc}
Here we calculate $\t{\mathscr{S}}_{\epsilon}^{\X{(2)}}(z)$, the thermal counterpart of $\t{\Sigma}_{\epsilon}^{\X{(2)}}(z)$, Eq.\,(\ref{e49}), denoting the contributions of diagrams $(2.2)$ and $(2.1)$ in Fig.\,\ref{f7} below, p.\,\pageref{SecondOrderS}, by respectively $\t{\mathscr{S}}_{\epsilon}^{\X{(2\textrm{a})}}(z)$ and $\t{\mathscr{S}}_{\epsilon}^{\X{(2\textrm{b})}}(z)$. These are the thermal counterparts of $\t{\Sigma}_{\epsilon}^{\X{(2.2)}}(z)$ and $\t{\Sigma}_{\epsilon}^{\X{(2.1)}}(z)$ considered above, \S\,\ref{s4xa}.\footnote{See Eqs\,(\protect\ref{e46}), (\protect\ref{e47}), and (\protect\ref{e48}).} The calculations of this section uncover a number of interesting aspects that to our knowledge have never earlier been discussed in the literature. Importantly, \emph{calculation of the finite-temperature correlation functions in the energy/frequency domain is fraught with a mathematical problem.} In the following we denote the finite-temperature counterpart of $\t{G}_{\epsilon}(z)$ by $\t{\mathscr{G}}_{\epsilon}(z)$, even though we rely on $\t{\mathscr{G}}_{\epsilon}(z) \equiv \t{G}_{\epsilon}(z)$, where $\t{G}_{\epsilon}(z)$ is presented in Eq.\,(\ref{e44}).

With
\begin{equation}\label{e57}
 \alpha \doteq \left\{\begin{array}{lc} -2, & \textrm{s} = \textrm{a},\\
+1, & \textrm{s} = \textrm{b}, \end{array}\right.
\end{equation}
one has [p.\,247 in Ref.\,\citen{FW03}]
\begin{equation}\label{e59}
\t{\mathscr{S}}_{\epsilon}^{\X{(2\textrm{s})}}(\zeta_m) = \frac{\alpha U^2}{\hbar^4 \beta^2} \sum_{m',m''=-\infty}^{\infty} \t{\mathscr{G}}_{\epsilon}(\ii\hbar \omega_{m'} +\mu)\hspace{0.6pt}\t{\mathscr{G}}_{\epsilon}(\ii\hbar (\omega_{m}-\omega_{m'} +\omega_{m''}) +\mu)\hspace{0.6pt} \t{\mathscr{G}}_{\epsilon}(\ii\hbar\omega_{m''} +\mu),
\end{equation}
where, as required, the number of the Matsubara frequencies in the arguments of all fermion Green functions is odd. \emph{In the following we consider half-filling and thus explicitly identify $\mu$ with $0$}, \S\,\ref{sec3.d}. Before presenting the general expression for $\t{\mathscr{S}}_{\epsilon}^{\X{(2\textrm{s})}}(\zeta_m)$ for arbitrary $\epsilon$, we note that
\begin{align}\label{e62}
\frac{1}{\hbar^2\beta^2}\hspace{-3.0pt}\sum_{m''=-\infty}^{\infty}\hspace{-3.0pt} \t{\mathscr{G}}_{\X{0}}(\ii\hbar (\omega_m-\omega_{m'} +\omega_{m''}))\hspace{0.6pt} \t{\mathscr{G}}_{\X{0}}(\ii\hbar \omega_{m''}) &= -\frac{\tan((m-m') \pi)}{4\pi(m-m')}\nonumber\\
&\equiv -\frac{1}{4}\hspace{0.6pt}\delta_{m,m'}\;\,\text{for}\;\, m,m' \in \mathds{Z},\hspace{1.2cm}
\end{align}
whereby (\emph{cf.} Eq.\,(\ref{e24}))
\begin{equation}\label{e62a}
\t{\mathscr{S}}_{\X{0}}^{\X{(2\textrm{s})}}(\zeta_m) = -\frac{\alpha U^2}{4\hbar \zeta_m}.
\end{equation}
With reference to the definition of $\alpha$ in Eq.\,(\ref{e57}), following the analytic continuation\,\footnote{Following Eq.\,(\protect\ref{e24}), at half-filling, where $\mu = 0$, \S\,\protect\ref{sec3.d}, $\zeta_m = \protect\ii\hbar\omega_m$.}
\begin{equation}\label{e72}
\zeta_m \rightharpoonup z,
\end{equation}
for
\begin{equation}\label{e72a}
\t{\mathscr{S}}_{\X{0}}^{\X{(2)}}(z) \equiv \t{\mathscr{S}}_{\X{0}}^{\X{(2\textrm{a})}}(z) + \t{\mathscr{S}}_{\X{0}}^{\X{(2\textrm{b})}}(z)
\end{equation}
one thus obtains the expected result (\emph{cf.} Eq.\,(\ref{e49a}))
\begin{equation}\label{e63}
\t{\mathscr{S}}_{\X{0}}^{\X{(2)}}(z) = \frac{U^2}{4\hbar z} \equiv \t{\Sigma}(z).
\end{equation}
Thus for any $\beta <\infty$, the self-energy $\t{\mathscr{S}}_{\epsilon}^{\X{(2)}}(z)$ is well-defined for $\epsilon=0$.

With
\begin{equation}\label{e64}
\upsilon \doteq \epsilon\hspace{0.6pt} U/2,
\end{equation}
employing the standard strategy of expressing the sums in Eq.\,(\ref{e59}) as contour integrals [p.\,248 in Ref.\,\citen{FW03}] [\S\S\,14.5 and 16.4 in Ref.\,\citen{WW62}], for arbitrary $\epsilon$ we obtain\,\footnote{The expression in Eq.\,(\protect\ref{e67}) is obtained from that in Eq.\,(\protect\ref{e59}) by first performing the summation with respect to $m''$ and subsequently that with respect to $m'$.}
\begin{align}\label{e67}
\t{\mathscr{S}}_{\epsilon}^{\X{(2\textrm{s})}}(\zeta_m) &= -\frac{\alpha U^2}{4\hbar \ii} \Big\{\hbar\omega_m \big[\cosh(\beta\upsilon) \big((\hbar\omega_m)^2 + 5\upsilon^2\big)\nonumber\\
&\hspace{2.3cm} -\cos(\beta\hbar\omega_m) \big((\hbar\omega_m)^2 + 7\upsilon^2\big) + 2\upsilon^2\big] \nonumber\\
&\hspace{-0.25cm} -2\upsilon \tanh(\beta\upsilon/2) \big[\upsilon\hbar\omega_m \tanh(\beta\upsilon/2) \big(2\cosh(\beta\upsilon) -\cos(\beta\hbar\omega_m) + 1\big)\nonumber\\
&\hspace{2.3cm} -\sin(\beta\hbar\omega_m) \big((\hbar\omega_m)^2 + 3 \upsilon^2 \big)\big]\Big\}\nonumber\\
&\hspace{-0.4cm}{\Big/}\Big\{\big((\hbar\omega_m)^4 + 10 \upsilon^2 (\hbar\omega_m)^2 + 9 \upsilon^4\big) \big(\cosh(\beta\upsilon)-\cos(\beta\hbar\omega_m)\big) \Big\}.\hspace{0.2cm}
\end{align}
With $\zeta_m = \ii\hbar\omega_m$ for $\mu=0$,  Eq.\,(\ref{e24}), it is tempting to obtain $\t{\mathscr{S}}_{\epsilon}^{\X{(2\textrm{s})}}(z)$, the analytic continuation of the self-energy under consideration into the complex $z$-plane, by effecting the transformation $\zeta_m\vert_{\mu=0} \equiv \ii\hbar\omega_m \rightharpoonup z$, Eq.\,(\ref{e72}), in the expression on the RHS of Eq.\,(\ref{e67}),  thus obtaining
\begin{align}\label{e67x}
\t{\mathscr{S}}_{\epsilon}^{\X{(2\textrm{s})}}(z) &= -\frac{\alpha U^2}{4\hbar} \Big\{z \cosh (\beta  v) \Big(z^2+4 v^2 \tanh ^2(\beta  v/2)-5 v^2\Big)\nonumber\\
&\hspace{1.35cm}-z \cosh(\beta  z) \Big(z^2+ 2 v^2 \tanh ^2(\beta  v/2)-7 v^2\Big) \nonumber\\
&\hspace{1.35cm}+2 v \big( (z^2-3 v^2) \tanh(\beta  v/2) \sinh (\beta  z)-v z\hspace{2.0pt} \text{sech}^2(\beta v/2)\big)\Big\}\nonumber\\
&{\Big/} \Big\{(z^2 - v^2) (z^2 - 9 v^2) \big(\cosh (\beta  v)-\cosh (\beta  z)\big)\Big\}.\;\;\;\; \text{(Incorrect)}
\end{align}
This expression is \emph{a priori} incorrect on account of containing the functions $\sinh(\beta z)$ and $\cosh(\beta z)$, which are non-analytically (or \textsl{essentially}) singular \cite{RR98,RR91,WW62} functions of $\zeta \doteq \beta z$ at the point of infinity of the complex $\zeta$-plane. In this connection, while the singularities of the self-energy are located on the real axis of the $z$-plane \cite{JML61}, in principle this function may be non-analytically singular at the point of infinity of the $z$-plain.\footnote{This possible singularity is a branch point, associated with a branch point in the finite part of the real energy axis, the two points being the end-points of a branch cut along the real axis across which, while remaining on the physical Riemann sheet of the $z$-plane, the self-energy is discontinuous.} However, such singularity is an inherent property of the single-particle excitation spectrum of the underlying many-body Hamiltonian, which is \textsl{independent} of the external parameter $\beta$.\,\footnote{See appendix C in Ref.\,\protect\citen{BF07}. In particular, from the spectral decompositions in Eqs\,(C.7) and (C.8) of the latter reference one observes that the parameter $\beta$, which occurs in the Boltzmann factor, is entirely decoupled from the external energy $\varepsilon$.} These observations imply that prior to the analytic continuation of the finite-temperature self-energy (as well as other many-body correlation functions) into the complex $z$-plane, the functions $\sin(\beta\hbar\omega_m)$ and $\cos(\beta\hbar\omega_m)$ are to be replaced by their values, for which one has\,\refstepcounter{dummy}\label{AsWeShall}\footnote{As we shall establish below, this procedure is generally \textsl{unsafe}. To be on the safe side, such straightforward identification, as in Eq.\,(\protect\ref{e67x1}) (as well as Eq.\,(\ref{e78x1}) below), must be effected only when the many-body correlation function at hand is the function of interest, and not an intermediate function in the calculation of another many-body correlation function. For clarity, for the calculation of the self-energy, it matters greatly whether one employs the polarisation function in Eq.\,(\ref{e76}) below or that in Eq.\,(\ref{e89b}) below (\emph{cf.} Eqs\,(\protect\ref{e67x2}) and (\protect\ref{e89d3}) below). \label{notef}} (\emph{cf.} Eq.\,(\ref{e70}))
\begin{equation}\label{e67x1}
\sin(\beta\hbar\omega_m) = 0,\;\; \cos(\beta\hbar\omega_m) = -1,\;\; \forall m \in \mathds{Z}.
\end{equation}
On doing so, effecting the transformation $\zeta_m\vert_{\mu =0} \equiv \ii\hbar\omega_m \rightharpoonup z$, for
\begin{equation}\label{e67b}
\t{\mathscr{S}}_{\epsilon}^{\X{(2)}}(z) \equiv \t{\mathscr{S}}_{\epsilon}^{\X{(2\textrm{a})}}(z) + \t{\mathscr{S}}_{\epsilon}^{\X{(2\textrm{b})}}(z)
\end{equation}
one obtains\,\footnote{With reference to Eqs\,(\protect\ref{e57}) and (\protect\ref{e67b}), note that $\t{\mathscr{S}}_{\epsilon}^{\protect\X{(2)}}(z)$ is obtained from $\t{\mathscr{S}}_{\epsilon}^{\X{(2\textrm{s})}}(z)$ on identifying the $\alpha$ in the expression for the latter function with $-1$. Conversely, $\t{\mathscr{S}}_{\epsilon}^{\X{(2\textrm{s})}}(z)$ is obtained from $\t{\mathscr{S}}_{\epsilon}^{\protect\X{(2)}}(z)$ on multiplying the expression for the latter function with $-\alpha$.}
\begin{equation}\label{e67x2}
\t{\mathscr{S}}_{\epsilon}^{\X{(2)}}(z) = \frac{U^2}{4\hbar} \frac{z \big((z^2 - \upsilon^2) \cosh(\beta\upsilon) + z^2 - 13 \upsilon^2\big)}{\big(\hspace{-1.4pt}\cosh(\beta\upsilon) + 1\big) (z^2 - \upsilon^2) (z^2 - 9 \upsilon^2)},\;\;\forall z\in\mathds{C}.
\end{equation}
Evidently, for $\upsilon = 0$, Eq.\,(\ref{e64}), the above expression reproduces the exact result in Eq.\,(\ref{e63}). For $\beta < \infty$ and $0 < U < \infty$, Eq.\,(\ref{e64}), from the expression in Eq.\,(\ref{e67x2}) one obtains
\begin{equation}\label{e67x3}
\t{\mathscr{S}}_{\epsilon}^{\X{(2)}}(z) \sim \Big\{ 1 + 3 \Big(\frac{\upsilon}{z}\Big)^2 + \frac{3}{2} \big(14 + (\beta z)^2\big) \Big(\frac{\upsilon}{z}\Big)^4 + \dots\Big\}\,\hspace{0.6pt}\t{\mathscr{S}}_{\X{0}}^{\X{(2)}}(z)\;\;\text{for}\;\; \epsilon\to 0,
\end{equation}
implying that the limit processes $\epsilon\to 0$ and $z\to 0$ do \textsl{not} commute, and that to obtain the correct expression for the self-energy the limit $z\to 0$ is to be taken \textsl{after} having effected the limit $\epsilon \to 0$.\footnote{For $\upsilon > C >0$, to leading order one has $\protect\t{\mathscr{S}}_{\epsilon}^{\protect\X{(2)}}(z) \sim -U^2 (\cosh(\beta\upsilon) + 13) z/\big(36 \hbar \upsilon^2 (\cosh(\beta\upsilon) +1)\big)$ as $z\to 0$, where the coefficient of $z$ to leading order diverges like $1/\upsilon^2$ for $\upsilon\to 0$. Because the coefficient of the latter $1/\upsilon^2$ is a function of $\beta\upsilon$, its behaviour for $\beta\to\infty$ depends on the order of the limits $\upsilon\to 0$ and $\beta\to\infty$.} This is in accord with the result in Eq.\,(\ref{e63}). It is interesting to note that the function $-\t{\mathscr{S}}_{\epsilon}^{\X{(2)}}(z)$ can be explicitly shown\,\footnote{The proof of this statement is straightforward, however somewhat cumbersome. One readily verifies that with $z\equiv x + \protect\ii y$, $x, y\in \mathds{R}$, the function $\protect\im[\t{\mathscr{S}}_{\epsilon}^{\X{(2)}}(z)]$ is an odd function of $y$ for all $x$ and that it is negative (positive) for $y > 0$ ($y < 0$) in the neighbourhood of $y=0$. The proof is completed upon showing that $\protect\im[\t{\mathscr{S}}_{\epsilon}^{\X{(2)}}(z)]$ has no \textsl{real} zero away from $y=0$ and $y=\infty$. This part of the proof involves investigation of the behaviour of a $3$rd-order polynomial of $y' \equiv y^2$. For $\cosh(\beta\upsilon) \ge 1$, the coefficients of ${y'}^3$, ${y'}^2$, and ${y'}$ in the latter polynomial prove to be negative, whereby for ruling out the existence of a zero of $\protect\im[\t{\mathscr{S}}_{\epsilon}^{\X{(2)}}(z)]$ in the region $0 < y' < \infty$ it suffices to establish the negativity of the constant term (that is, the coefficient of ${y'}^0$) in the latter polynomial, which in turn is a $3$rd-order polynomial of $x' \equiv x^2$. The coefficient of ${x'}^3$ in this polynomial being $-(\cosh(\beta\upsilon)+1) <0 $, this polynomial is negative for $x'\to\infty$. All its coefficients being also negative for $29/7 < \cosh(\beta\upsilon) < 103/17$, this polynomial is evidently also negative for this range of values of $\cosh(\beta\upsilon)$. It turns of that for $(83-18 \sqrt{11})/25 < \cosh(\beta\upsilon) < (83+18 \sqrt{11})/25$ this polynomial has a maximum equal to $-9 \big(\hspace{-1.2pt}\cosh(\beta\upsilon)+ 13\big)\hspace{0.4pt} \upsilon^6 < 0$ at $x'=0$, and, since $\cosh(\beta\upsilon) \ge 1$, for $\cosh(\beta\upsilon) > (83+18 \sqrt{11})/25 = 5.707\dots$ at its maximum away from $x' = 0$ it is invariably negative (and proportional to $\upsilon^6$) for $\upsilon > 0$.} to be a Nevanlinna function of $z$. Further
\begin{align}\label{e67x4}
\t{\mathscr{S}}_{\epsilon}^{\X{(2)}}(z) &\sim \frac{3 U^2}{16\hbar\hspace{0.6pt} (\cosh(\beta\upsilon) +1)} \frac{1}{z\mp \upsilon}\;\;\text{for}\;\; z\to \pm\upsilon,\nonumber\\
\t{\mathscr{S}}_{\epsilon}^{\X{(2)}}(z) &\sim \frac{U^2 (2\cosh(\beta\upsilon) -1)}{16\hbar\hspace{0.6pt} (\cosh(\beta\upsilon)+1)} \frac{1}{z\mp 3\upsilon}\;\;\text{for}\;\; z\to \pm 3\upsilon.
\end{align}
Thus, in contrast to $\t{\mathscr{S}}_{\X{0}}(z)$ which has a simple pole at $z=0$, Eq.\,(\ref{e63}), for $\epsilon\not=0$ the function $\t{\mathscr{S}}_{\epsilon}^{\X{(2)}}(z)$ in Eq.\,(\ref{e67x2}) has four simple poles, at $z = \pm\upsilon,\pm 3\upsilon$. Note that for $\beta <\infty$ the sum of the residues of the latter four poles of $\t{\mathscr{S}}_{\epsilon}^{\X{(2)}}(z)$ appropriately approaches $U^2/(4\hbar)$ for $\upsilon\to 0$, in which limit the four simple poles merge into a single simple pole at $z=0$, Eq.\,(\ref{e63}).

\emph{The following considerations will shed light on some aspects regarding the evaluation of the finite-temperature correlation functions that to our best knowledge have not been touched upon hitherto elsewhere.}

We begin by considering the contribution of the lowest-order polarisation insertion\,\footnote{See appendix \protect\ref{sa}.} that contributes to the self-energy $\t{\mathscr{S}}_{\epsilon}^{\X{(2\mathrm{a})}}(z)$, diagram $(2.2)$ in Fig.\,\ref{f7} below, p.\,\pageref{SecondOrderS}. Denoting this \textsl{zeroth-order} polarization insertion by $\t{\mathscr{P}}^{\X{(0)}}_{\epsilon}(\ii\hbar\nu_{m})$, Eq.\,(\ref{e70}), one has\,\footnote{With reference to the general remarks on p.\,\protect\pageref{UnlessWe}, recall that the expression in Eq.\,(\protect\ref{e73}) is specific to spin-$\tfrac{1}{2}$ particles at half-filling and corresponds to the atomic limit of the Hubbard Hamiltonian $\protect\h{\mathcal{H}}$ in Eq.\,(\protect\ref{ex01bx}). Hence the pre-factor $2$ to which we refer below.}
\begin{equation}\label{e73}
\t{\mathscr{P}}^{\X{(0)}}_{\epsilon}(\ii\hbar\nu_m) = \frac{2}{\hbar^2 \beta} \sum_{m'=-\infty}^{\infty} \t{\mathscr{G}}_{\epsilon}(\ii\hbar\omega_{m'} + \ii\hbar\nu_{m})\hspace{0.6pt} \t{\mathscr{G}}_{\epsilon}(\ii\hbar\omega_{m'}),
\end{equation}
where the $2$ on the RHS is associated with the sums on the RHS of Eq.\,(\ref{ea3a}), the property $\t{P}_{\sigma,\sigma'} =  \t{P}_{\sigma}\delta_{\sigma,\sigma'}$ in the \textsl{lowest} order of the perturbation theory \cite{BF19},\footnote{See also Fig.\,\protect\ref{f12}, p.\,\protect\pageref{Two2ndO}, below.} and that for spin-unpolarised GSs (ESs) of spin-$\tfrac{1}{2}$ particles $\t{P}_{\sigma}\equiv \t{P}_{\b{\sigma}}$. Thus for $\mu=0$ (specific to half-filling, \S\,\ref{sec3.d}) the expression in Eq.\,(\ref{e59}) can be expressed as
\begin{equation}\label{e74}
\t{\mathscr{S}}_{\epsilon}^{\X{(2\mathrm{s})}}(\zeta_m) = \frac{\alpha U^2}{2\hbar^2 \beta}\hspace{-1.0pt} \sum_{m'=-\infty}^{\infty} \t{\mathscr{G}}_{\epsilon}(\ii\hbar\omega_{m'}) \t{\mathscr{P}}^{\X{(0)}}_{\epsilon}(\ii\hbar\omega_m -\ii\hbar\omega_{m'}),
\end{equation}
in which $\omega_m -\omega_{m'} = \nu_{m-m'}$, Eq.\,(\ref{e70}).

Along the same lines as leading to the expression in Eq.\,(\ref{e67}), we obtain
\begin{align}\label{e76}
\t{\mathscr{P}}^{\X{(0)}}_{\epsilon}(\ii\hbar\nu_m) &= -\Big\{\frac{\hbar\nu_m \upsilon \sinh(\beta\upsilon) + 2\sin(\beta\hbar\nu_m) \big((\hbar\nu_m)^2/2 +\upsilon^2\big)}{\cosh(\beta\upsilon) + \cos(\beta\hbar\nu_m)} + \hbar\nu_m\upsilon \tanh(\beta\upsilon/2) \Big\}\nonumber\\
&\hspace{0.1cm} {\Big/}\Big\{(\hbar\nu_m)^3 + 4\upsilon^2 (\hbar\nu_m)\Big\}.
\end{align}
Effecting the analytic continuation
\begin{equation}\label{e77}
\ii\hbar\nu_m \rightharpoonup z,
\end{equation}
one obtains
\begin{align}\label{e78}
\t{\mathscr{P}}^{\X{(0)}}_{\epsilon}(z) &= \Big\{\frac{\upsilon \sinh(\beta\upsilon) - \sinh(\beta z) \big(z^2 - 2\upsilon^2\big)/z}{\cosh(\beta\upsilon) + \cosh(\beta z)} + \upsilon \tanh(\beta\upsilon/2) \Big\}{\Big/}\Big\{z^2 - 4\upsilon^2\Big\}.\nonumber\\
&\hspace{8.6cm} \text{(Incorrect)}
\end{align}
Similarly to the expression in Eq.\,(\ref{e67x}), the expression in Eq.\,(\ref{e78}) is \emph{a priori} incorrect on account of containing the functions $\sinh(\beta z)$ and $\cosh(\beta z)$, which are non-analytically singular at the point of infinity of the $\zeta$-plane, where $\zeta \doteq\beta z$.\footnote{See the remarks following Eq.\,(\protect\ref{e67x}).} Equating the functions $\sin(\beta\hbar\nu_m)$ and $\cos(\beta\hbar\nu_m)$ in Eq.\,(\ref{e76}) with their numerical values, that is using\,\footnote{See footnote \raisebox{-1.0ex}{\normalsize{\protect\footref{notef}}} on page \protect\pageref{AsWeShall}.} (\emph{cf.} Eq.\,(\ref{e70}))
\begin{equation}\label{e78x1}
\sin(\beta\hbar\nu_m) = 0,\;\; \cos(\beta\hbar\nu_m) = 1,\;\; \forall m \in \mathds{Z},
\end{equation}
and subsequently analytically continuing the resulting expression according to the prescription in Eq.\,(\ref{e77}), one obtains
\begin{equation}\label{e89b}
\t{\mathscr{P}}^{\X{(0)}}_{\epsilon}(z) = \tanh(\beta\upsilon/2)\,\t{P}_{\epsilon}^{\X{(0)}}(z),
\end{equation}
where $\t{P}_{\epsilon}^{\X{(0)}}(z)$ is the zero-temperature counterpart of $\t{\mathscr{P}}_{\epsilon}^{\X{(0)}}(z)$ introduced in Eq.\,(\ref{e79}) below and whose explicit form is given in Eq.\,(\ref{e81}) below. One observes that whereas along the $\upsilon$-axis $\t{P}_{\epsilon}^{\X{(0)}}(z)$ has a single zero at $\upsilon=0$, for $\beta < \infty$ the function $\t{\mathscr{P}}_{\epsilon}^{\X{(0)}}(z)$ has a double-zero at $\upsilon$. It follows that in evaluating the zero-temperature limit of $\t{\mathscr{P}}_{\epsilon}^{\X{(0)}}(z)$, the limit process $\beta\to\infty$ is to be effected for $\epsilon > 0$, as
\begin{equation}\label{e89bx}
\lim_{\beta\to\infty}\tanh(\beta\upsilon/2) = 1
\end{equation}
only for $\upsilon > 0$, or $\epsilon U > 0$, Eq.\,(\ref{e64}).

For the `physical' polarisation function at zero temperature, $P_{\epsilon}^{\X{(0)}}(\varepsilon)$, $\varepsilon\in\mathds{R}$, one has (appendix \ref{sa})
\begin{equation}\label{e79}
P_{\epsilon}^{\X{(0)}}(\varepsilon) \equiv 2\int_{-\infty}^{\infty}\frac{\rd\varepsilon}{2\pi i\hbar^2}\; G_{\epsilon}(\varepsilon'+\varepsilon) G_{\epsilon}(\varepsilon').
\end{equation}
This function is related to the polarisation function $\t{P}_{\epsilon}^{\X{(0)}}(z)$, $z \in\mathds{C}$, according to (\emph{cf.} Eq.\,(\ref{e1}))
\begin{equation}\label{e80}
P_{\epsilon}^{\X{(0)}}(\varepsilon) = \lim_{\eta\downarrow 0} \t{P}_{\epsilon}(\varepsilon \pm \ii\eta)\;\, \text{for}\;\, \varepsilon \gtrless 0.
\end{equation}
Making use of the expression for $G_{\epsilon}(z)$ in Eq.\,(\ref{e44}), by the residue theorem \cite{WW62} one obtains
\begin{equation}\label{e81}
\t{P}_{\epsilon}^{\X{(0)}}(z) = \frac{1}{2} \Big\{ \frac{1}{z - 2\upsilon} - \frac{1}{z + 2\upsilon}\Big\} \equiv \frac{2\upsilon}{z^2 -4\upsilon^2},
\end{equation}
where we have used the equality in Eq.\,(\ref{e64}). With reference to Eq.\,(\ref{e64}), note that
\begin{equation}\label{e81a}
\lim_{\epsilon\downarrow 0} \t{P}_{\epsilon}^{\X{(0)}}(z) = 0,\;\; \forall z\in \mathds{C}\backslash\{0\},
\end{equation}
implying that in calculating diagrammatic contributions (for instance to the self-energy) involving polarisation insertions, the limit $\epsilon\downarrow 0$ is to be taken \textsl{after} the evaluation of the underlying integrals; the first-order zero of $\t{P}_{\epsilon}^{\X{(0)}}(z)$ at $\epsilon=0$ generally cancels against a divergence of the form $1/\epsilon$.

With (\emph{cf.} Eq.\,(\ref{e70}))
\begin{equation}\label{e82}
\rd\hspace{2.0pt}(\ii\hbar\omega_m) \doteq \ii\hbar\omega_{m+1} - \ii\hbar\omega_m \equiv \frac{2\pi\!\ii}{\beta},
\end{equation}
from the expression in Eq\,(\ref{e73}) one obtains
\begin{equation}\label{e83}
\lim_{\beta\to\infty} \t{\mathscr{P}}^{\X{(0)}}_{\epsilon}(z) = 2\int_{-\ii\infty}^{\ii\infty} \frac{\rd z'}{2\pi\!\ii\hbar^2}\; \t{G}_{\epsilon}(z' + z) \t{G}_{\epsilon}(z'),
\end{equation}
where the integral on the RHS represents the Riemann sum on the RHS of Eq.\,(\ref{e73}). In arriving at the expression on the RHS of Eq.\,(\ref{e83}), we have further used the property
\begin{equation}\label{e84}
\t{G}_{\epsilon}(z) = \lim_{\beta\to\infty} \t{\mathscr{G}}_{\epsilon}(z),
\end{equation}
which in the case at hand applies by the convention $\t{\mathscr{G}}_{\epsilon}(z) \equiv \t{G}_{\epsilon}(z)$.\footnote{See the opening paragraph of \S\,\protect\ref{s4xc}.} One explicitly demonstrates that for a general $z$ the function on the RHS of Eq.\,(\ref{e83}) does not coincide with the function $\t{P}_{\epsilon}^{\X{(0)}}(z)$ associated (through the expression in Eq.\,(\ref{e80})) with the function $P_{\epsilon}^{\X{(0)}}(\varepsilon)$ in Eq.\,(\ref{e79}). In other words, the function on the RHS of Eq.\,(\ref{e83}) is \textsl{not} identically equal to the function in Eq.\,(\ref{e81}). This is established by the residue theorem \cite{WW62}, whereby (\emph{cf.} Eq.\,(\ref{e81}))
\begin{align}\label{e85}
\lim_{\beta\to\infty} \t{\mathscr{P}}^{\X{(0)}}_{\epsilon}(z) &= \frac{1}{2} \Big\{\frac{\Theta(\upsilon-\re[z])}{z -2\upsilon} - \frac{\Theta(\upsilon+\re[z])}{z + 2\upsilon}\Big\}\nonumber\\
&+ \frac{\Theta(\upsilon-\re[z]) -\Theta(\upsilon+\re[z])}{2 z} \not\equiv \t{P}_{\epsilon}^{\X{(0)}}(z).
\end{align}
From the equality in Eq.\,(\ref{e85}) one observes however that
\begin{equation}\label{e86}
\lim_{\beta\to\infty} \t{\mathscr{P}}^{\X{(0)}}_{\epsilon}(z) \equiv \t{P}_{\epsilon}^{\X{(0)}}(z)\;\, \text{for}\;\, \vert\hspace{-1.2pt}\re[z]\vert < \upsilon.
\end{equation}
This result is interesting in particular because for $\upsilon\to 0$ the condition $\vert\hspace{-1.2pt}\re[z]\vert < \upsilon$ results in the equality, to exponential accuracy, of the incorrect expression in Eq.\,(\ref{e78}) with the correct one in Eq.\,(\ref{e89b}) in the asymptotic region $\beta\to\infty$.\footnote{Note that for $\protect\re[z] = 0$ one has $\vert\sinh(\beta z)\vert \le 1$ and $\vert\cosh(\beta z)\vert \le 1$ for all $\beta\in\mathds{R}$.}

We now proceed with evaluating the function $\t{\mathscr{P}}^{\X{(0)}}_{\epsilon}(z) $ in an alternative way, one along which other finite-temperature correlation functions may in principle be evaluated. This alternative way suggested itself to us by the appearance of the non-analytically singular functions of $\zeta \doteq\beta z$ in the incorrect expressions in Eqs\,(\ref{e67x}) and (\ref{e78}). Clearly, the functions $\sinh(\beta z)$ and $\cosh(\beta z)$ in the latter equations, which originate from the functions $\sin(\beta\hbar\omega_m)$, $\sin(\beta\hbar\nu_m)$, and $\cos(\beta\hbar\omega_m)$, $\cos(\beta\hbar\nu_m)$ in Eqs\,(\ref{e67}) and (\ref{e76}), would have been avoided if through a judicious combination of the Matsubara frequencies prior to evaluating the relevant Matsubara-frequency sums the transfers of $\beta\hbar\omega_m$ and $\beta\hbar\nu_m$ into the arguments of trigonometric functions had been avoided. To our knowledge, the extant literature is \textsl{not} sufficiently explicit with regard to the evaluation of the non-zero-temperature correlation functions at the Matsubara energies and the subsequent analytic continuation of these functions into the entire $z$-plane.\footnote{Exception can be made, however only in a limited sense, to the text by Negele and Orland \protect\cite{NO98}; on page 285 of this text (Eq.\,(5.155)) one encounters an explicit hint as to the way in which the polarisation function in the random-phase approximation [RPA] (denoted by $\mathcal{D}(k,\nu_m)$ in the latter text) is to be calculated at the bosonic Matsubara frequencies $\{ \protect\ii\nu_m\| m\}$. This is of course the conventional approach adopted universally (see for instance Eq.\,(5.152) in Ref.\,\protect\citen{PN66}, Eq.\,(30.9) in Ref.\,\protect\citen{FW03} and Eqs\,(3.216) -- (3.218) in Ref.\,\protect\citen{GDM00}), however to our knowledge the significance of the adopted approach has nowhere been spelled out. Later in this section we shall consider the formalism of Taheridehkordi \emph{et al.} \protect\cite{TCLB19,TCLB20,VF20}, which by several years postdates the date the present text was originally written. As we have indicated above, the considerations of this part of this section have their roots in our attempt to bypass such transcendental functions as $\sin(\beta\hbar\omega_m)$, $\sin(\beta\hbar\nu_m)$, $\cos(\beta\hbar\omega_m)$, and $\cos(\beta\hbar\nu_m)$. The above-mentioned details in the text by Negele and Orland \protect\cite{NO98} led us to the construction of the partial-fraction-decomposition (PFD) approach to be discussed here and in the following parts of this section. The approach by Taheridehkordi \emph{et al.} \protect\cite{TCLB19,TCLB20,VF20} proves to be a systematic generalisation of our approach, implying that the approach by the latter authors suffers from the same shortcoming as our approach. We shall explicitly establish this shortcoming below.} For instance, in Eq.\,(5.152) in Ref.\,\citen{PN66} one encounters only a substitution prescription; in Ref.\,\citen{GDM00} [\S\,3.5 herein] the derivation of the equality in Eq.\,(3.217) (relevant to the calculation of the function $\t{\mathscr{P}}^{\X{(0)}}_{\epsilon}(\ii\hbar\nu_m)$ in this section) is ``left as an exercise for the student''. From Eq.\,(30.9) in Ref.\,\citen{FW03} and Eq.\,(5.152) in Ref.\,\citen{PN66} one can gain a glimpse of how ``the student'' is to proceed in order to obtain the correct expression for the polarisation function in the random-phase approximation, RPA (functionally the equivalent of the function $\t{\mathscr{P}}^{\X{(0)}}_{\epsilon}(\ii\hbar\nu_m)$ considered here).\footnote{We note that the techniques described in \S\,3.5 of Ref.\,\protect\citen{GDM00} are standard and are the same as those described in, Ref.\,\protect\citen{FW03}, pp. 248-250, relying on the Mellin-Barnes integral representation of specific sums [\S\S 14.5 and 16.4 in Ref.\,\protect\citen{WW62}].}

On the basis of the above considerations, for avoiding the appearance of the above-mentioned non-analytically singular functions of $\zeta \doteq \beta z$ in the expressions for thermal many-body correlation functions, we propose the following prescription for the evaluation of the underlying sums over the Matsubara energies:
\vspace{0.1cm}
\begin{itemize}
\item[{}]\emph{With the understanding that a linear combination, with coefficients $\pm 1$, of an \text{even} (odd) number of fermionic Matsubara energies corresponds to a bosonic (fermionic) Matsubara energy, the sums involving bosonic Matsubara energies must be evaluated separately from those involving fermionic Matsubara energies.}
\end{itemize}
\vspace{0.1cm}
This can be achieved through the application of appropriate Mittag-Leffler partial-fraction decompositions \cite{RR98,KK47,MF53}\footnote{See in particular Ch.\,2, \S\,4, in Ref.\,\protect\citen{KK47}.} of the relevant functions.\footnote{The approach as presented here proves to be implicit in the framework of Taheridehkordi \emph{et al.} \protect\cite{TCLB19,TCLB20,VF20} (as we have indicated earlier [footnote \raisebox{-1.0ex}{\normalsize{\protect\footref{notes}}} on p.\,\protect\pageref{WorkOnThe}], the text of the present publication almost entirely dates back to early May 2015). For clarity, the fact that the $\upsigma$ in Eq.\,(13) of Ref.\,\protect\citen{TCLB19} is \textsl{independent} of the Matsubara frequencies ensures that summation over these frequencies according to the method of Taheridehkordi \emph{et al.} \protect\cite{TCLB19} does \textsl{not} result in such transcendental functions of $\zeta \doteq\beta z$ as $\sinh(\beta z)$ and $\cosh(\beta z)$, encountered in Eqs\,(\protect\ref{e67x}) and (\ref{e78}) above. Further, the `residue theorem' in Eq.\,(10) of Ref.\,\protect\citen{TCLB19} achieves what the partial-fraction decomposition considered in this section does. In this connection, compare the latter equation in Ref.\,\protect\citen{TCLB19} with those in Eqs\,(\protect\ref{e89}) and (\ref{e89h}) below. Note that the relevant `residues' are the multiplicative coefficients in the partial-fraction decompositions considered in this section (for instance, the coefficients $\pm 1/(\protect\ii\hbar\nu_m)$ on the RHS of Eq.\,(\protect\ref{e88}) below are such `residues'). \emph{As we shall see later in this section, neither the partial-fraction-decomposition method as described in this section, nor the method of Taheridehkordi \emph{et al.} \protect\cite{TCLB19,TCLB20,VF20} is safe.}}

In applying the above prescription, we begin with the polarisation function $\t{\mathscr{P}}^{\X{(0)}}_{\epsilon}(\ii\hbar\nu_m)$ as introduced in Eq.\,(\ref{e73}). With $\t{\mathscr{G}}_{\epsilon}(z) \equiv \t{G}_{\epsilon}(z)$, Eq.\,(\ref{e44}), one obtains the following extensive expression:
\begin{align}\label{e87}
\t{\mathscr{P}}^{\X{(0)}}_{\epsilon}(\ii\hbar\nu_m) &= \frac{1}{2\beta} \sum_{m'} \frac{1}{\ii\hbar\omega_{m'} + \upsilon} \frac{1}{\ii\hbar\omega_{m'} + \ii\hbar\nu_{m} +\upsilon}\nonumber\\
&+  \frac{1}{2\beta} \sum_{m'} \frac{1}{\ii\hbar\omega_{m'} + \upsilon} \frac{1}{\ii\hbar\omega_{m'} + \ii\hbar\nu_{m} -\upsilon}\nonumber\\
&+ \frac{1}{2\beta} \sum_{m'} \frac{1}{\ii\hbar\omega_{m'} - \upsilon} \frac{1}{\ii\hbar\omega_{m'} + \ii\hbar\nu_{m} +\upsilon} \nonumber\\
&+  \frac{1}{2\beta} \sum_{m'} \frac{1}{\ii\hbar\omega_{m'} - \upsilon} \frac{1}{\ii\hbar\omega_{m'} + \ii\hbar\nu_{m} -\upsilon}.
\end{align}
One has the following partial-fraction decompositions \cite{RR98,KK47,MF53}:
\begin{equation}\label{e88}
\frac{1}{\ii\hbar\omega_{m'} \pm \upsilon}\, \frac{1}{\ii\hbar\omega_{m'} + \ii\hbar\nu_{m} \pm \upsilon} = \frac{1}{\ii\hbar\nu_m}\Big\{ \frac{1}{\ii\hbar\omega_{m'} \pm \upsilon} -  \frac{1}{\ii\hbar\omega_{m'} +\ii\hbar\nu_{m} \pm \upsilon}\Big\},
\end{equation}
so that
\begin{align}\label{e88a}
&\frac{1}{2\beta} \sum_{m'} \frac{1}{\ii\hbar\omega_{m'} \pm \upsilon}\, \frac{1}{\ii\hbar\omega_{m'} + \ii\hbar\nu_{m} \pm \upsilon} \nonumber\\
&\hspace{1.2cm} = \frac{1}{\ii\hbar\nu_m}\frac{1}{2\beta} \sum_{m'} \frac{\e^{\ii \omega_{m'} 0^+}}{\ii\hbar\omega_{m'} \pm \upsilon} -\frac{1}{\ii\hbar\nu_m}\frac{1}{2\beta} \sum_{m'}  \frac{\e^{\ii \omega_{m'} 0^+}}{\ii\hbar\omega_{m'} +\ii\hbar\nu_{m} \pm \upsilon},
\end{align}
where the converging factor $\e^{\ii \omega_{m'} 0^+}$ multiplying the summands on the RHS have been introduced in accordance with the general prescription.\footnote{See, for instance, the remark following Eq.\,(\protect\ref{eg13}) above. The reason underlying the $0^+$, rather than a $0^-$, is the general prescription that in evaluating contributions of Feynman diagrams in the time domain, the internal one-particle Green function $\mathcal{G}_{i,j;\sigma}(\tau,\tau)$, Eq.\,(\protect\ref{eg3x}), is to be identified with $\mathcal{G}_{i,j;\sigma}(\tau,\tau^+)$, where $\tau^+ \doteq \tau+ 0^+$ (see step 7 on p.\,243 of Ref.\,\protect\citen{FW03}; for $T=0$, see step $(i)$ on p.\,99 of Ref.\,\protect\citen{FW03}). Physically, this accounts for an infinitesimally small retardation involved in the process of two particles interacting through a mathematically instantaneous two-body interaction potential. Note that the $0^+$ in $\tau+ 0^+$ has the dimensionality of \textsl{time} (in the SI base units, $\llbracket 0^+\rrbracket = \mathrm{s}$, second).} Following Eq.\,(\ref{e70}), one has
\begin{equation}\label{e88b}
\omega_{m'} + \nu_{m} = \omega_{m+m'} \equiv \omega_{m''},
\end{equation}
implying that the two sums on the RHS of Eq.\,(\ref{e88a}) identically cancel.\footnote{The vanishing of the sum on the LHS of Eq.\,(\protect\ref{e88a}) stands on a par with the vanishing of the contribution of $G_{\epsilon}^{\varsigma}(\varepsilon'+\varepsilon) G_{\sigma}^{\varsigma}(\varepsilon')$, where $\varsigma \in \protect\X{\{-,+\}}$, Eq.\,(\protect\ref{e44}), to the function $P_{\epsilon}^{\X{(0)}}(\varepsilon)$ in Eq.\,(\protect\ref{e79}).} Using the partial-fraction decompositions
\begin{equation}\label{e88c}
\frac{1}{\ii\hbar\omega_{m'} \pm \upsilon}\, \frac{1}{\ii\hbar\omega_{m'} + \ii\hbar\nu_{m} \mp \upsilon} = \frac{1}{\ii\hbar\nu_m \mp 2 \upsilon}\Big\{ \frac{1}{\ii\hbar\omega_{m'} \pm \upsilon} -  \frac{1}{\ii\hbar\omega_{m'} +\ii\hbar\nu_{m} \mp \upsilon}\Big\},
\end{equation}
on account of the equality in Eq.\,(\ref{e88b}) and (\emph{cf.} Eq.\,(\ref{e10b}))
\begin{equation}\label{e89}
\frac{1}{\beta}\sum_{m'=-\infty}^{\infty} \frac{\e^{\ii \omega_{m'} 0^+}}{\ii\hbar\omega_{m'} \pm \upsilon} = \frac{1}{\e^{\mp\beta\upsilon} +1},
\end{equation}
from the expression in Eq.\,(\ref{e87}) one obtains (\emph{cf.} Eq.\,(\ref{e76}))
\begin{equation}\label{e89a}
\t{\mathscr{P}}^{\X{(0)}}_{\epsilon}(\ii\hbar\nu_m) = \frac{1}{2} \tanh(\beta\upsilon/2)\hspace{0.6pt} \Big\{ \frac{1}{\ii \hbar\nu_m - 2\upsilon} - \frac{1}{\ii\hbar\nu_m +2\upsilon}\Big\}.
\end{equation}
Effecting the analytic continuation $\ii\hbar\nu_m \rightharpoonup z$, Eq.\,(\ref{e77}), one thus recovers the expression in Eq.\,(\ref{e89b}). Clearly, the prescription described above has proved effective in avoiding the functions that upon \textsl{straightforward} analytic continuation (that is, an analytic continuation without employing in advance such equalities as in Eqs\,(\ref{e67x1}) and (\ref{e78x1})) would result in the spurious contributions encountered in Eqs\,(\ref{e67x}) and (\ref{e78}). In this connection, note that on neglecting the relations in Eq.\,(\ref{e88b}), following Eq.\,(\ref{e89}) one would have
\begin{equation}\label{e89d}
\frac{1}{\beta}\sum_{m'=-\infty}^{\infty} \frac{\e^{\ii \omega_{m'} 0^+}}{\ii\hbar\omega_{m'} + \ii \hbar\nu_m \pm \upsilon} = \frac{1}{\e^{\beta(\mp\upsilon - \ii \hbar\nu_m)} +1},
\end{equation}
which result accounts for the functions $\sin(\beta\hbar\nu_m)$ and $\cos(\beta\hbar\nu_m)$ in the expression in Eq.\,(\ref{e76}).

Having determined the correct thermal polarisation function $\t{\mathscr{P}}^{\X{(0)}}_{\epsilon}(z)$, we now proceed with the determination of the self-energy contribution $\t{\mathscr{S}}_{\epsilon}^{\X{(2\mathrm{s})}}(\zeta_m) $ on the basis of the expression in Eq.\,(\ref{e74}). A direct evaluation of the sum in Eq.\,(\ref{e74}) yields
\begin{align}\label{e89d1}
\t{\mathscr{S}}_{\epsilon}^{\X{(2\mathrm{s})}}(\zeta_m) &= -\frac{\alpha U^2}{4\hbar\ii} \tanh(\beta v/2) \nonumber\\ &\times \Big\{\frac{\hbar\omega_m ((\hbar\omega_m)^2 + 5\upsilon^2) \sinh(2\beta\upsilon) - 2\upsilon ((\hbar\omega_m)^2 + 3\upsilon^2) \sin(\beta\hbar\omega_m)}{\cosh(2\beta\upsilon) + \cos(\beta\hbar\omega_m)}\nonumber\\
&\hspace{0.6cm} - 4\hbar\omega_m \upsilon^2 \tanh(\beta\upsilon/2)\Big\} {\Big/} \Big\{(\hbar\omega_m)^4 + 10 \upsilon^2 (\hbar\omega_m)^2 + 9\upsilon^4\Big\}.
\end{align}
Effecting the analytic continuation $\zeta_m\vert_{\mu=0} \equiv \ii\hbar\omega_m \rightharpoonup z$, one obtains (\emph{cf.} Eq.\,(\ref{e67x}))
\begin{align}\label{e89d2}
\t{\mathscr{S}}_{\epsilon}^{\X{(2\mathrm{s})}}(z) &= \frac{\alpha U^2}{4\hbar} \tanh(\beta v/2) \Big\{\frac{2\upsilon (z^2 - 3\upsilon^2) \sinh(\beta z) - z (z^2 - 5\upsilon^2) \sinh(2\beta\upsilon)}{\cosh(2\beta\upsilon) + \cosh(\beta z)}\nonumber\\
&\hspace{3.5cm} - 4 \upsilon^2 z \tanh(\beta\upsilon/2)\Big\}{\Big/} \Big\{(z^2 -\upsilon^2) (z^2 - 9\upsilon^2)\Big\}.\nonumber\\
&\hspace{8.4cm}\text{(Incorrect)}
\end{align}
On the other hand, effecting the analytic continuation $\zeta_m\vert_{\mu=0} \equiv \ii\hbar\omega_m \rightharpoonup z$ \textsl{subsequent to} employing the equalities in Eq.\,(\ref{e67x1}), from Eq.\,(\ref{e89d1}) one arrives at (\emph{cf.} Eq.\,(\ref{e67x2}))
\begin{equation}\label{e89e1}
\t{\mathscr{S}}_{\epsilon}^{\X{(2\mathrm{s})}}(z) = -\frac{\alpha U^2}{4\hbar} \frac{z\big((z^2 - \upsilon^2) \cosh(\beta\upsilon) - 4\upsilon^2\big)}{\big(\cosh(\beta\upsilon) + 1\big) (z^2 - \upsilon^2) (z^2 - 9 \upsilon^2)},\;\;\forall z\in\mathds{C},
\end{equation}
and thus, following Eq.\,(\ref{e57}) and (\ref{e67b}), at
\begin{equation}\label{e89d3}
\t{\mathscr{S}}_{\epsilon}^{\X{(2)}}(z) =  \frac{U^2}{4\hbar} \frac{z\big((z^2 - \upsilon^2) \cosh(\beta\upsilon) - 4\upsilon^2\big)}{\big(\cosh(\beta\upsilon) + 1\big) (z^2 - \upsilon^2) (z^2 - 9 \upsilon^2)},\;\;\forall z\in\mathds{C}.
\end{equation}
This result clearly deviates fundamentally from the exact\,\footnote{As regards the exactness, see the considerations based on the expression in Eq.\,(\protect\ref{e89x1}) below.} result in Eq.\,(\ref{e67x2}). Importantly, for $\t{\mathscr{S}}_{\X{0}}^{\X{(2)}}(z)$ according to the expression in Eq.\,(\ref{e89d3}) one has (\emph{cf.} Eq.\,(\ref{e63}))
\begin{equation}\label{e89k}
\t{\mathscr{S}}_{\X{0}}^{\X{(2)}}(z) =  \frac{U^2}{8\hbar z} \equiv \frac{1}{2}\hspace{0.6pt} \t{\Sigma}(z).
\end{equation}
More generally,
\begin{equation}\label{e89kx1}
\t{\mathscr{S}}_{\epsilon}^{\X{(2)}}(z) \sim \frac{U^2}{8\hbar z} \Big\{1 + \Big(\frac{5}{z^2}+\frac{\beta^2}{4}\Big) \upsilon^2 + \Big(\frac{41}{z^4} + \frac{13 \beta^2}{4 z^2}- \frac{\beta^4}{24}\Big)\upsilon^4 + \dots\Big\}\;\, \text{for}\;\, \upsilon\to 0\,\wedge\, \beta <\infty.
\end{equation}
Remarkably, however
\begin{align}\label{e89kx2}
\t{\mathscr{S}}_{\epsilon}^{\X{(2)}}(z) &\sim \frac{U^2}{4\hbar z}\Big(1 + \frac{9\upsilon^2}{z^2 -\upsilon^2}\Big) \hspace{0.6pt}\Big\{1 - \Big(2 + \frac{8\upsilon^2}{z^2 - \upsilon^2}\Big)\hspace{0.6pt} \e^{-\beta\upsilon} + \Big(4 +\frac{16 \upsilon^2}{z^2 - \upsilon^2}\Big) \hspace{0.6pt}\e^{-2\beta\upsilon} -\dots\Big\}\hspace{0.4cm}\nonumber\\
&\hspace{8.4cm}\text{for}\;\; \beta\upsilon \to\infty.
\end{align}
This asymptotic expression reveals that the expression in Eq.\,(\ref{e89d3}) yields the exact self-energy at zero temperature on effecting the limit $\upsilon\downarrow 0$ \textsl{subsequent to} effecting the limit $\beta\to\infty$. We note that,\footnote{With $x \equiv \protect\e^{a}$, $\mathrm{cosh}(a)/(\cosh(a) +1) \equiv (x^2 +1)/(x^2 + 2 x + 2) \sim 1 - 2/x + 4/x^2 - 6/x^4 +\dots$ for $x\to \infty$ (corresponding to $a\to\infty$).} for $a > 0$,
\begin{align}\label{e89k1}
\t{\mathscr{S}}_{\epsilon}^{\X{(2)}}(z) &\sim \frac{U^2}{4\hbar z}\hspace{0.6pt}\frac{\mathrm{cosh}(a)}{\cosh(a) + 1} \sim \frac{U^2}{4\hbar z}\hspace{0.6pt} \big( 1 -2 \e^{-a} + 4 \e^{-2 a} - \dots\big),\nonumber\\
&\hspace{2.5cm}\text{for}\;\; \beta \equiv \frac{a}{\upsilon}\;\, \text{as}\;\; \upsilon\downarrow 0 \iff
\upsilon \equiv \frac{a}{\beta}\;\, \text{as}\;\; \beta \uparrow \infty.\;\;
\end{align}
It is seen that the right-most asymptotic expression in Eq.\,(\ref{e89k1}) coincides with that in Eq.\,(\ref{e89kx2}) wherein the $\upsilon$ in the multiplicative coefficients of $\e^{-\beta\upsilon}$, $\e^{-2\beta\upsilon}$, $\dots$, are identified with zero. Since $\upsilon \doteq \epsilon U/2$, Eq.\,(\ref{e64}), one thus has
\begin{equation}\label{e89l}
\t{\mathscr{S}}_{\epsilon}^{\X{(2)}}(z) \sim  \frac{U^2}{4\hbar z} \equiv \t{\Sigma}(z)\;\; \text{only if}\;\; \epsilon\hspace{0.4pt}U\beta \to \infty.
\end{equation}
The condition on the RHS of Eq.\,(\ref{e89l}) amounts to a major restriction in practical finite-temperature calculations where $\beta > 0 $ is finite, whereby effecting the limit $\epsilon \downarrow 0$ results in $\epsilon\hspace{0.4pt}U\beta \downarrow 0$ for $0< U < \infty$. In this connection, it is interesting to note that calculations in Ref.\,\citen{KFG14} have been carried out for $T = 0.5$, or $\beta = 2$ (in the units where $k_{\textsc{b}} = 1$ and $\hbar =1$). For comparison, for the self-energy in Eq.\,(\ref{e89d3}) one further has (\emph{cf.} Eq.\,(\ref{e67x4}))
\begin{align}\label{e89n}
\t{\mathscr{S}}_{\epsilon}^{\X{(2)}}(z) &\sim \frac{U^2}{16\hbar\hspace{0.6pt} (\cosh(\beta\upsilon) + 1} \frac{1}{z \mp\upsilon}\;\;\text{for}\;\; z\to \pm \upsilon,  \nonumber\\
\t{\mathscr{S}}_{\epsilon}^{\X{(2)}}(z) &\sim\frac{U^2 (2\cosh(\beta\upsilon) -1)}{16\hbar\hspace{0.6pt} (\cosh(\beta\upsilon)+1)} \frac{1}{z\mp 3\upsilon}\;\;\text{for}\;\; z\to \pm 3\upsilon.
\end{align}
One observes that the residues of the poles at $z = \pm\upsilon$ of the $\t{\mathscr{S}}_{\epsilon}^{\X{(2)}}(z)$ according the expression in Eq.\,(\ref{e89d3}) are three times \textsl{smaller} than those of the $\t{\mathscr{S}}_{\epsilon}^{\X{(2)}}(z)$ according the expression in Eq.\,(\ref{e67x2}). Note that for $\beta <\infty$ the sum of the residues of the four simple poles of the function $\t{\mathscr{S}}_{\epsilon}^{\X{(2)}}(z)$ under consideration appropriately approaches $U^2/(8\hbar)$ for $\upsilon\to 0$, in which limit these poles merge into a single simple pole at $z=0$, Eq.\,(\ref{e89k}).

In the light of the above observations, we now proceed with the calculation of the function $\t{\mathscr{S}}_{\epsilon}^{\X{(2\mathrm{s})}}(\zeta_m)$ by evaluating the sum with respect to $m'$ in Eq.\,(\ref{e74}) with the aid of the Mittag-Leffler partial-fraction decomposition that we have employed above for evaluating the polarization function $\t{\mathscr{P}}^{\X{(0)}}_{\epsilon}(z)$.

With reference to Eq.\,(\ref{e74}) and in the light of Eqs\,(\ref{e25}) and (\ref{e89b}), for $\zeta_{m}\vert_{\mu=0} = \ii\hbar\omega_m$ one has (\emph{cf.} Eq.\,(\ref{e87}))
\begin{align}\label{e89f}
\t{\mathscr{S}}_{\epsilon}^{\X{(2\mathrm{s})}}(\zeta_m) &= \frac{\alpha U^2}{4\hbar} \tanh(\beta\upsilon/2) \Big\{ \frac{1}{\beta}\sum_{m'} \frac{1}{\ii\hbar\omega_{m'} - v}\, \frac{1}{\ii\hbar\omega_{m} - \ii\hbar\omega_{m'} - 2 v}\nonumber\\
&\hspace{3.1cm} -\frac{1}{\beta}\sum_{m'} \frac{1}{\ii\hbar\omega_{m'} - v}\, \frac{1}{\ii\hbar\omega_{m} - \ii\hbar\omega_{m'} + 2 v}\nonumber\\
&\hspace{3.1cm} +\frac{1}{\beta}\sum_{m'}\frac{1}{\ii\hbar\omega_{m'} + v}\, \frac{1}{\ii\hbar\omega_{m} - \ii\hbar\omega_{m'} - 2 v}\nonumber\\
&\hspace{3.1cm} -\frac{1}{\beta}\sum_{m'}\frac{1}{\ii\hbar\omega_{m'} + v}\, \frac{1}{\ii\hbar\omega_{m} - \ii\hbar\omega_{m'} + 2 v}\Big\}.
\end{align}
On account of the partial-fraction decompositions
\begin{align}\label{e89g}
\frac{1}{\ii\hbar\omega_{m'} \mp v}\, \frac{1}{\ii\hbar\omega_{m} - \ii\hbar\omega_{m'} \mp 2 v} &= \frac{1}{\ii\hbar\omega_m \mp 3 v} \Big\{\frac{1}{\ii\hbar\omega_{m'} \mp v} + \frac{1}{\ii\hbar\omega_{m} - \ii\hbar\omega_{m'} \mp 2 v} \Big\},\nonumber\\
\frac{1}{\ii\hbar\omega_{m'} \mp v}\, \frac{1}{\ii\hbar\omega_{m} - \ii\hbar\omega_{m'} \pm 2 v} &= \frac{1}{\ii\hbar\omega_m \pm v}\Big\{\frac{1}{\ii\hbar\omega_{m'} \mp v}+ \frac{1}{\ii\hbar\omega_{m} - \ii\hbar\omega_{m'} \pm 2 v} \Big\},
\end{align}
making use of Eq.\,(\ref{e89}) and\,\footnote{As regards the converging factor $\protect\e^{\protect\ii \nu_{m'} 0^+}$ to be added to the sums in consequence of the partial-fraction decompositions in Eq.\,(\protect\ref{e89g}), the reasoning is the same as that given in relation to Eqs\,(\protect\ref{e88}) and (\protect\ref{e88a}).} [Eq.\,(25.35), p. 249, in Ref.\,\citen{FW03})]
\begin{equation}\label{e89h}
\frac{1}{\beta}\sum_{m'=-\infty}^{\infty} \frac{\e^{\ii \nu_{m'} 0^+}}{\ii\hbar\nu_{m'} \pm \upsilon} = \frac{-1}{\e^{\mp\beta\upsilon} -1},
\end{equation}
from Eqs\,(\ref{e89f}), (\ref{e89g}), and (\ref{e89h}) one obtains the same expression for $\t{\mathscr{S}}_{\epsilon}^{\X{(2\mathrm{s})}}(\zeta_m)$ as in Eq.\,(\ref{e89e1}) wherein $z \equiv \zeta_m\vert_{\mu=0} = \ii\hbar\omega_m$. Recall that the partial-fraction-decomposition method as adopted here avoids generation of such non-analytically-singular functions as $\sin(\beta\hbar\omega_m)$, $\sin(\beta\hbar\nu_m)$, $\cos(\beta\hbar\omega_m)$, and $\cos(\beta\hbar\nu_m)$, ordinarily to be replaced by their values as presented in Eqs\,(\ref{e67x1}) and (\ref{e78x1}) in advance of effecting the analytic continuations $\ii\hbar\omega_m \rightharpoonup z$ and $\ii\hbar\nu_m\rightharpoonup z$. As regards the use of the equality in Eq.\,(\ref{e89h}), note that, in view of Eq.\,(\ref{e70}) (\emph{cf.} Eq.\,(\ref{e88b})),
\begin{equation}\label{e89j}
\omega_m - \omega_{m'} = \nu_{m-m'} \equiv \nu_{m''}.
\end{equation}

The above observations raises the question as to whether the expression for $\t{\mathscr{S}}_{\epsilon}^{\X{(2)}}(z)$ in Eq.\,(\ref{e67x2}) is reliable. To answer this question, we consider the counterpart of $\t{\mathscr{S}}_{\epsilon}^{\X{(2)}}(z)$ in the imaginary-time domain, to be denoted by $\mathpzc{S}_{\epsilon}^{\X{(2)}}(\tau)$, and calculate $\t{\mathscr{S}}_{\epsilon}^{\X{(2)}}(\zeta_m)$, with $\zeta_m \equiv \ii\hbar\omega_m$ (assuming $\mu=0$) on the basis of the expression [Eq.\,(25.14), p.\,245, in Ref.\,\citen{FW03}]
\begin{equation}\label{e89x1}
\t{\mathscr{S}}_{\epsilon}^{\X{(2)}}(\zeta_m) = \int_{0}^{\hbar\beta} \mathrm{d}\tau\, \e^{\ii\omega_m\tau} \mathpzc{S}_{\epsilon}^{\X{(2)}}(\tau).
\end{equation}
With $\mathpzc{G}_{\epsilon}(\tau)$ denting the counterpart of $\t{\mathscr{G}}_{\epsilon}(z)$ in the $\tau$-domain, \S\,\ref{sec3.c},\footnote{See in particular Eqs\,(\protect\ref{eg9a}) and (\protect\ref{eg11}).} using the standard rules for expressing the contribution of Feynman diagrams [pp.\,242 and 243 in Ref.\,\citen{FW03}], taking account of the instantaneous nature of the two-body interaction potential considered here, one obtains
\begin{equation}\label{e89x2}
\mathpzc{S}_{\epsilon}^{\X{(2)}}(\tau) = -\frac{U^2}{\hbar^2}\hspace{0.6pt} \mathpzc{G}_{\epsilon}^2(\tau) \mathpzc{G}_{\epsilon}(-\tau).
\end{equation}
Making use of the Heisenberg equation of motion, \S\,\ref{sec3.c}, for the `Hubbard atom' under consideration one obtains\,\footnote{Note that $\t{\mathscr{G}}_{\epsilon}(z)$ coincides with the \textsl{interacting} one-particle Green function for $\epsilon=1$.}\footnote{The expression on the RHS of Eq.\,(\protect\ref{e89x3}) satisfies the Kubo-Martin-Schwinger (KMS) relation for fermion Green functions [Eq.\,(24.15), p.\,236, in Ref.\,\protect\citen{FW03}] \protect\cite{BF19}: for $-\beta\hbar < \tau < 0$ one has $\mathpzc{G}_{\epsilon}(\tau) = -\mathpzc{G}_{\epsilon}(\tau + \hbar\beta)$. Let $\Lambda_{a}(x) \doteq x -2 a (\lfloor (x-a)/(2a)\rfloor +1)$, where $\lfloor x\rfloor$ is the floor function. For the extension of $\mathpzc{G}_{\epsilon}(\tau)$ to the entire $\tau$-axis, to be denoted by $\mathpzc{G}_{\epsilon}'(\tau)$, one has $\mathpzc{G}_{\epsilon}'(\tau) = \mathpzc{G}_{\epsilon}(\Lambda_{\hbar\beta}(\tau))$.}
\begin{align}\label{e89x3}
\mathpzc{G}_{\epsilon}(\tau) &= \frac{-\X{1/2}}{\e^{\beta\upsilon}+1} \Big\{\e^{(\beta-\tau/\hbar) \upsilon} + \e^{\upsilon\tau/\hbar}\Big\}\hspace{0.6pt} \Theta(\tau) + \frac{\X{1/2}}{\e^{\beta\upsilon}+1} \Big\{\e^{(\beta+\tau/\hbar) \upsilon} + \e^{-\upsilon\tau/\hbar}\Big\}\hspace{0.6pt} \Theta(-\tau),\nonumber\\
&\hspace{8.0cm} -\hbar\beta \le \tau \le \hbar\beta.
\end{align}
One verifies that
\begin{equation}\label{e89x4}
\mathpzc{G}_{\epsilon}(-\tau) \equiv -\mathpzc{G}_{\epsilon}(\tau)\;\;\text{for}\;\; -\hbar\beta \le \tau \le \hbar\beta,
\end{equation}
whereby
\begin{equation}\label{e89x5}
\mathpzc{S}_{\epsilon}^{\X{(2)}}(\tau) = \frac{U^2}{\hbar^2}\hspace{0.6pt} \mathpzc{G}_{\epsilon}^3(\tau)\;\;\text{for}\;\; -\hbar\beta \le \tau \le \hbar\beta.
\end{equation}
From Eq.\,(\ref{e89x1}), one thus obtains
\begin{align}\label{e89x6}
\t{\mathscr{S}}_{\epsilon}^{\X{(2)}}(\zeta_m) &= \frac{U^2}{8\hbar}\hspace{0.6pt}\frac{1}{(\e^{\beta\upsilon} +1)^3}\hspace{0.6pt} \Big\{\frac{\e^{3\beta\upsilon}-\e^{\ii\beta\hbar\omega_m}}{\ii\hbar\omega_m - 3 \upsilon}+ \frac{1-\e^{\ii\beta\hbar\omega_m + 3\beta\upsilon}}{\ii\hbar\omega_m + 3 \upsilon}\nonumber\\
&\hspace{2.55cm} + \frac{3 \e^{\beta\upsilon}(\e^{\beta\upsilon} - \e^{\ii\hbar\beta\omega_m})}{\ii\hbar\omega_m -\upsilon} + \frac{3 \e^{\beta\upsilon}(1-\e^{\ii\beta\hbar\omega_m + \beta\upsilon})}{\ii\hbar\omega_m +\upsilon} \Big\}.\;\;\;\;
\end{align}
Using (see Eq.\,(\ref{e67x1}))
\begin{equation}\label{e89x7}
\e^{\ii\hbar\omega_m} = -1,\;\;\forall m \in\mathds{Z},
\end{equation}
followed by the analytic continuation $\zeta_m\vert_{\mu=0} \equiv \ii\hbar\omega_m \rightharpoonup z$, Eq.\,(\ref{e72}), from the expression in Eq.\,(\ref{e89x6}) one arrives at\,\footnote{From the expression in Eq.\,(\protect\ref{e89x8}) one immediately observes that for $\upsilon\beta\to\infty$ and $\vert z \mp\upsilon\vert > C > 0$ (or $\vert\hspace{-1.0pt}\protect\im[z]\vert > C > 0$) the contributions of the last two terms on the RHS are exponentially smaller than those of the first two terms (\emph{cf.} Eqs\,(\protect\ref{e49}) and (\protect\ref{e64})).}
\begin{align}\label{e89x8}
\t{\mathscr{S}}_{\epsilon}^{\X{(2)}}(z) &= \frac{U^2}{8\hbar}\hspace{0.6pt}\frac{1}{(\e^{\beta\upsilon} +1)^3}\hspace{0.6pt} \Big\{\frac{\e^{3\beta\upsilon}+1}{z - 3 \upsilon}+ \frac{1+\e^{3\beta\upsilon}}{z + 3 \upsilon} + \frac{3 \e^{\beta\upsilon}(\e^{\beta\upsilon} +1)}{z -\upsilon} + \frac{3 \e^{\beta\upsilon}(1+\e^{\beta\upsilon})}{z +\upsilon} \Big\},\nonumber\\
&\hspace{9.5cm} \forall z\in\mathds{C},
\end{align}
which can be shown to be identical to the expression in Eq.\,(\ref{e67x2}).\footnote{For $\epsilon = 0 \Rightarrow \upsilon =0$ one immediately observes that the expression in Eq.\,(\protect\ref{e89x8}) yields the exact result $\protect\t{\mathscr{S}}_{\protect\X{0}}(z) = U^2/(4\hbar z)$, Eq.\,(\protect\ref{e63}).} This may be achieved by verifying that the residues of the simple poles of the expression in Eq.\,(\ref{e89x8}) at $z = \pm \upsilon, \pm 3\upsilon$ identically coincide with those in Eq.\,(\ref{e67x4}). \emph{We have thus established that the expression for $\mathscr{S}_{\epsilon}^{\X{(2)}}(z)$ in Eq.\,(\ref{e67x2}) is indeed correct.}

To uncover the reason for the deviation of the expression in Eq.\,(\ref{e89d3}) from that in Eq.\,(\ref{e67x2}), we first note that the correct calculation of the polarisation function $\t{\mathscr{P}}^{\X{(0)}}_{\epsilon}(z)$, irrespective of the method of calculation, leads one to the conclusion that the deviation under considerations is rooted in the summation with respect to the $m'$ in Eq.\,(\ref{e59}) and the use of the expression in Eq.\,(\ref{e89b}), instead of the expression in Eq.\,(\ref{e76}), in evaluating the latter sum. At first glance there is no apparent reason for the choice between the latter two expressions for the polarisation function to result in any difference, however inspection of the relevant expressions proves otherwise. Explicitly, evaluating\,\footnote{Here we use $\protect\ii\hbar\omega_m$ as the argument of $\t{\mathscr{S}}_{\epsilon}^{\protect\X{(2\textrm{s})}}$ instead of $\zeta_m$, for which one has $\zeta_m\vert_{\mu=0} \equiv \protect\ii\hbar\omega_m$.} $\t{\mathscr{S}}_{\epsilon}^{\protect\X{(2\textrm{s})}}(\ii\hbar\omega_m)$ in terms of the polarisation function in Eq.\,(\ref{e76}), which leads to the correct expression for $\t{\mathscr{S}}_{\epsilon}^{\protect\X{(2\textrm{s})}}(\ii\hbar\omega_m)$ in Eq.\,(\ref{e67}), one encounters the following function
\begin{equation}\label{e89x9}
\varphi(\ii\hbar\omega_m) \doteq \Big(\!\coth\big(\beta(\ii\hbar\omega_m -\upsilon)/2\big) + \coth\big(\beta(\ii\hbar\omega_m +\upsilon)/2\big)\!\Big) \tanh(\beta\ii\hbar\omega_m/2).
\end{equation}
The RHS of this function amounts to $0/0$ for $\ii\hbar\omega_m = (2m+1)\pi\ii/\beta$, $\forall m\in \mathds{Z}$, Eq.\,(\ref{e70}). To obtain the value of $\varphi(z)$ at $z = \ii\hbar\omega_m$, one therefore will have to evaluate $\varphi(z)$ in a neighbourhood of $z = \ii\hbar\omega_m$, for which one has\,\footnote{Let $h(z) \equiv f(z)/g(z)$. For $z_0$ such that $f(z_0) = g(z_0) = 0$ however $g'(z_0)\not=0$, one has $h(z_0+\delta z) = f'(z_0)/g'(z_0) + \tfrac{1}{2} \big\{\big(g'(z_0) f''(z_0) - f'(z_0) g''(z_0)\big)/\big(g'(z_0)\big)^2\big\}\hspace{0.6pt}\delta z + O\big((\delta z)^2\big)$ as $\delta z\to 0$. In the case at hand $h(z) = \varphi(z)$, $f(z) = \coth\!\big(\beta (z-\upsilon)/2\big) + \coth\!\big(\beta (z + \upsilon)/2\big)$, $g(z) = \coth\!\big(\beta z/2\big)$, and $z_0 = \protect\ii\hbar\omega_m$, $m\in\mathds{Z}$.}
\begin{equation}\label{e89y1}
\varphi(\ii\hbar\omega_m + \delta z) = \frac{4}{\cosh(\beta\upsilon) +1} + O\big((\delta z)^2\big)\;\; \text{for}\;\; \ii\hbar\omega_m = \frac{(2 m+1)\pi\ii}{\beta},\;\; m \in \mathds{Z}.
\end{equation}
The contribution $4/(\cosh(\beta\upsilon+1)$ is clearly lost on prematurely setting the contribution of the terms enclosed by the round brackets on the RHS of Eq.\,(\ref{e89x9}) equal to zero.\footnote{Importantly, by identifying $\varphi(\protect\ii\hbar\omega_n)$ with zero in the expression leading to that in Eq.\,(\protect\ref{e67}), we obtain the expression in Eq.\,(\protect\ref{e89d1}).} This does not take place when evaluating $\t{\mathscr{S}}_{\epsilon}^{\protect\X{(2\textrm{s})}}(\ii\hbar\omega_m)$ without enforcing the equalities in Eq.\,(\ref{e67x1}) (and similarly as regards those in Eq.\,(\ref{e78x1})). In other words, for the correct evaluation of $\t{\mathscr{S}}_{\epsilon}^{\protect\X{(2\textrm{s})}}(\ii\hbar\omega_m)$, as well as other finite-temperature correlation functions, the argument $\ii\hbar\omega_m$ is to be viewed as an arbitrary complex variable that \textsl{in general} is to be identified with $(2 m+1)\pi\ii/\beta$, $m\in\mathds{Z}$, only \textsl{after} the conclusion of the calculations.\footnote{With reference to Eq.\,(\protect\ref{e74}), note that it is only for $\omega_m = (2m+1) \pi/(\hbar\beta)$, with $m\in\mathds{Z}$, that $\omega_m - \omega_{m'} = \nu_{m-m'}$, where $m'$ as a summation variable naturally varies over $\mathds{Z}$.} The prospect of otherwise missing nontrivial contributions does not bode well for the finite-temperature calculations in the energy/frequency domain. \emph{The above calculations in the imaginary-time domain (Eqs\,(\ref{e89x1}) -- (\ref{e89x8})) establish that calculations in this domain are in contrast safe, that is free from the problem of inappropriately discounting non-trivial contributions.}

In view of the above observations, it may be helpful to realise that the equalities in Eqs\,(\ref{e89}) and (\ref{e89h}) apply by virtue of the functions $\e^{\ii\omega_{m'} 0^+}$ and $\e^{\ii\nu_{m'} 0^+}$, respectively, which, despite being equal to unity for any finite value of $\vert m'\vert$, encode crucial information necessary for assigning definite values to conditionally convergent sums. In a way, the $\delta z$ in Eq.\,(\ref{e89y1}) plays a role similar to the $0^+$ in the arguments of the latter two exponential functions. It is instructive to note the following formal operations, signified by $\rightharpoonup$ (\emph{cf.} Eqs\,(\ref{e89}) and (\ref{e89h})):\,\footnote{We use $\sum_{k=-\infty}^{\infty} f(k) = f(0) + \sum_{k=-\infty}^{-1} f(k) + \sum_{k=1}^{\infty} f(k) = f(0) + \sum_{k=1}^{\infty} \big\{f(k) + f(-k)\big\}$, where in the case at hand neither $\sum_{k=-\infty}^{-1} f(k)$ nor $\sum_{k=1}^{\infty} f(k)$ exists, in contrast to $ \sum_{k=1}^{\infty} \big\{f(k) + f(-k)\big\}$.}
\begin{align}\label{e89y2}
\frac{1}{\beta} \sum_{m'=-\infty}^{\infty} \frac{\e^{\ii\omega_{m'} 0^+}}{\ii\hbar\omega_{m'} \pm \upsilon} &\rightharpoonup \frac{1}{\pi\hspace{-1.2pt}\ii \pm\beta\upsilon} + \sum_{m'=1}^{\infty} \Big\{\frac{1}{(2m'+1) \pi\hspace{-1.2pt}\ii \pm\beta\upsilon} - \frac{1}{(2m'-1)\pi\hspace{-1.2pt}\ii \mp\beta\upsilon}\Big\}\nonumber\\
&= \pm\frac{1}{2}\tanh(\beta\upsilon/2) \equiv \frac{1}{\e^{\mp\beta\upsilon}+1} -\frac{1}{2}, \\
\label{e89y3}
\frac{1}{\beta} \sum_{m'=-\infty}^{\infty} \frac{\e^{\ii\nu_{m'} 0^+}}{\ii\hbar\nu_{m'} \pm \upsilon} &\rightharpoonup \frac{1}{\pm\beta\upsilon} + \sum_{m'=1}^{\infty} \Big\{\frac{1}{2m' \pi\hspace{-1.2pt}\ii \pm\beta\upsilon} - \frac{1}{2m'\pi\hspace{-1.2pt}\ii \mp\beta\upsilon}\Big\}\nonumber\\
&= \pm\frac{1}{2}\coth(\beta\upsilon/2) \equiv \frac{-1}{\e^{\mp\beta\upsilon}-1} -\frac{1}{2}.
\end{align}
Clearly, these operations had been justified if the sums on the LHSs of Eqs\,(\ref{e89y2}) and (\ref{e89y3}) were over the set $\{-M, -M+1,\dots,-1,0, 1,\dots, M-1,M\}$, with $M$ an arbitrary large but \textsl{finite} integer. This observation sheds light on the problem of the numerical evaluation of the left-most \textsl{infinite} sums in Eqs\,(\ref{e89y2}) and (\ref{e89y3}), which amounts to an implicit truncation of these sums to finite sums, resulting in demonstrably incorrect limits.\footnote{With $\epsilon \doteq 1/M$, the values of the present sums for $\epsilon = 0$ do not coincide with their limits for $\epsilon\downarrow 0$.}

In the light of the above observations, we should emphasise that calculations in the imaginary-time domain and obtaining the relevant correlation functions of interest, such as the self-energy, at Matsubara energies (possibly with the aim of a numerical analytic continuation of these functions into the $z$-plain) through a numerical Fourier transformation (\emph{cf.} Eq.\,(\ref{e89x1})) is generally non-trivial.\footnote{The practical problem corresponds to the cases where convolution-type integrals are to be evaluated, whereby the above-mentioned cusps shift through the interval $[0,\hbar\beta]$, necessitating a careful bookkeeping in order to identify their locations as boundaries of the integrals to be evaluated numerically. For clarity, let $f(\tau)$ be a function with cusps at $\tau=0$, $\tau=\tau_1$, and $\tau=\hbar\beta$, where $0 < \tau_1 < \hbar\beta$. Efficient and reliable numerical evaluation of the integral of $f(\tau)$ over $[0,\hbar\beta]$ requires expressing this integral as the sum of the integrals of $f(\tau)$ over $[0,\tau_1]$ and $[\tau_1,\hbar\beta]$. Clearly, the complexity of this approach increases when evaluating the integral of $f(g(\tau))$ over $[0,\hbar\beta]$. Cusps in the \textsl{interior} of intervals of integration are detrimental to the convergence rate of all (Gaussian) quadrature rules \protect\cite{PTVF01}, which formally rely on the expansion of the underlying integrands in terms of appropriate orthogonal \textsl{polynomials}.} As can be inferred from the expression in Eq.\,(\ref{e89x3}), the Green function $\mathpzc{G}_{\epsilon}(\tau)$, with $\upsilon> 0$, has cups for $\tau \downarrow 0$ and $\tau\uparrow \hbar\beta$, and similarly for $\tau\uparrow 0$ and $\tau\downarrow -\hbar\beta$. This and the oscillating nature of $\exp(\ii\omega_m\tau)$ as a function of $\tau$ render reliable numerical evaluation of the integral on the RHS of Eq.\,(\ref{e89x1}) difficult and time-consuming. Note that the power $3$ in the expression for $\mathpzc{S}_{\epsilon}^{\X{(2)}}(\tau)$ in Eq.\,(\ref{e89x5}) coincides with $2\nu-1$, the number of Green-function lines of which a $\nu$th-order proper self-energy diagram is comprised. The consequences of the above-mentioned cusps in $\mathpzc{G}_{\epsilon}(\tau)$ for an accurate evaluation of the Fourier integral in Eq.\,(\ref{e89x1}) are therefore exacerbated for increasing values of $\nu$. Note that while the expression in Eq.\,(\ref{e89x3}) is specific to the `Hubbard atom', it is not dissimilar to the expression for the non-interacting Green function $\mathpzc{G}_{\X{0};\sigma}(\bm{k};\tau)$ away from the atomic limit.\footnote{With reference to Eqs\,(\ref{e44}) and (\protect\ref{e64}), the expression in Eq.\,(\protect\ref{e89x3}) amounts to $\tfrac{1}{2} \big(\mathfrak{g}_{\epsilon}^+(\tau)+ \mathfrak{g}_{\epsilon}^-(\tau)\big)$, where $\mathfrak{g}_{\epsilon}^{\mp}(\tau)$ corresponds to $\protect\t{G}_{\epsilon}^{\mp}(z)$. With $f_{\beta}(\upsilon) \equiv 1/(\protect\e^{\beta\upsilon} +1)$ denoting the Fermi function, one has $\mathfrak{g}_{\epsilon}^{\mp}(\tau) = \big\{\big(f_{\beta}(\mp\upsilon)-1\big) \Theta(\tau) + f_{\beta}(\mp\upsilon) \hspace{0.4pt}\Theta(-\tau)\big\} \protect\e^{\pm\upsilon\tau/\hbar}$. Thus, for instance, the expressions for the imaginary-time counterparts of the functions in Eq.\,(\protect\ref{e98y4}) below are obtained from the expression for $\mathfrak{g}_{\epsilon}^{\mp}(\tau)$ on identifying $\upsilon$ with respectively $\mp 4\varepsilon_{\protect\X{0}}$, $\mp \varepsilon_{\protect\X{0}}$, and $\pm\varepsilon_{\protect\X{0}}$. Compare with Eq.\,(6) in Ref.\,\protect\citen{vHKPS10}.}

We note in passing that Fourier transforming the function $\mathpzc{G}_{\epsilon}(\tau)$ in Eq.\,(\ref{e89x3}) according to the expression in Eq.\,(\ref{e89x1}), with $\zeta_{m}\vert_{\mu=0} \equiv \ii\hbar\omega_m$, one obtains
\begin{equation}\label{e98y0}
\t{\mathscr{G}}_{\epsilon}(\zeta_m) = \frac{\hbar/2}{(\hbar\omega_m)^2 + \upsilon^2}\hspace{0.6pt} \Big\{\ii\hbar\omega_m \big(\hspace{-1.2pt}\e^{\ii\beta\hbar\omega_m}-1\big) - \upsilon \tanh(\beta\upsilon/2) \big(\hspace{-1.2pt}\e^{\ii\beta\hbar\omega_m}+1\big)\Big\}.
\end{equation}
Using the equality in Eq.\,(\ref{e89x7}) and subsequently effecting the analytic continuation $\ii\hbar\omega_m \rightharpoonup z$, the RHS of Eq.\,(\ref{e98y0}) reduces to the function $\t{G}_{\epsilon}(z)$ in Eq.\,(\ref{e44}) wherein $\epsilon U/2 \equiv \upsilon$, Eq.\,(\ref{e64}). The question arises as to whether the expression on the RHS of Eq.\,(\ref{e98y0}) is to be used in the calculations of thermal correlation functions pertaining to the `Hubbard atom', instead of the one in Eq.\,(\ref{e44}) employed in the above calculations. This question is to be investigated, an undertaking that we shall not pursue in this publication; \emph{note that this question is relevant \textsl{only} for the Green functions whose arguments involve the \textsl{external} energy (or energies\hspace{0.4pt}\footnote{In the case of higher-order correlation functions than the self-energy.}) in linear combinations with some internal energies.} We suffice to mentioned that the expression for $\t{G}_{\epsilon}(z)$ in Eq.\,(\ref{e44}) has proved appropriate for the calculation of the correct expressions in Eqs\,(\ref{e62a}) and (\ref{e67}) on the basis of the expression in Eq.\,(\ref{e59}).

We close this section by considering the following function (\emph{cf.} Eq.\,(\ref{e59})):
\begin{align}\label{e98y2}
S^{\X{(2\mathrm{s})}}(\bm{k},\bm{k}',\bm{k}'';\ii\hbar\omega_m) &\doteq \frac{\alpha U^2}{\hbar^4\beta^2} \sum_{m',m''=-\infty}^{\infty} \t{\mathscr{G}}_{\X{0}}(\bm{k}';\ii\hbar\omega_{m'})\nonumber\\
&\hspace{-1.4cm}\times \t{\mathscr{G}}_{\X{0}}(\bm{k}-\bm{k}' +\bm{k}'';\ii\hbar(\omega_{m}-\omega_{m'} + \omega_{m''})) \hspace{1.2pt}\t{\mathscr{G}}_{\X{0}}(\bm{k}'';\ii\hbar\omega_{m''}),\;\;\;
\end{align}
from which the second-order self-energy contribution $\t{\mathscr{S}}^{\X{(2\mathrm{s})}}(\bm{k};\ii\hbar\omega_m)$ corresponding to a uniform thermal ESs of the single-band Hubbard Hamiltonian, Eq.\,(\ref{ex01bx}), is obtained according to (see \S\,\ref{sd21})
\begin{equation}\label{e98y3}
\t{\mathscr{S}}^{\X{(2\mathrm{s})}}(\bm{k};\ii\hbar\omega_m) = \frac{1}{N_{\textsc{s}}^2}\sum_{\bm{k}',\bm{k}''} S^{\X{(2\mathrm{s})}}(\bm{k},\bm{k}',\bm{k}'';\ii\hbar\omega_m),
\end{equation}
where $N_{\textsc{s}}$ denotes the number of lattice sites, or, the number of $\bm{k}$-points from which the underlying $\1BZ$ is comprised, Eq.\,(\ref{ex08x}). For the given non-interacting energy dispersion $\varepsilon_{\bm{k}}$, Eq.\,(\ref{ex01bx}), let the wave vectors $\bm{k}$, $\bm{k}'$, and $\bm{k}''$ be such that
\begin{align}\label{e98y4}
&\t{\mathscr{G}}_{\X{0}}(\bm{k}';\ii\hbar\omega_m) = \frac{\hbar}{\ii\hbar\omega_m -4\hspace{0.6pt} \varepsilon_{\X{0}}},\nonumber\\
&\t{\mathscr{G}}_{\X{0}}(\bm{k}-\bm{k}'+\bm{k}'';\ii\hbar\omega_m) = \frac{\hbar}{\ii\hbar\omega_m - \varepsilon_{\X{0}}},\nonumber\\
&\t{\mathscr{G}}_{\X{0}}(\bm{k}'';\ii\hbar\omega_m) = \frac{\hbar}{\ii\hbar\omega_m + \varepsilon_{\X{0}}},
\end{align}
where $\varepsilon_{\X{0}}$ is some constant energy, for instance the nearest-neighbour hopping integral\,\footnote{See footnote \raisebox{-1.0ex}{\normalsize{\protect\footref{notec1}}} on p.\,\protect\pageref{SeeFootnote}.} $t$. On the basis of \textsl{all} the methods described and applied above, for the wave vectors under consideration we obtain\,\footnote{Making use of the partial-fraction-decomposition method, described above, one obtains the equivalent expression $\alpha U^2 \tanh(\beta \varepsilon_{\protect\X{0}}/2) \big\{1/(\protect\e^{4\beta\varepsilon_{\protect\X{0}}} + 1) + 1/(\protect\e^{-2\beta\varepsilon_{\protect\X{0}}} -1)\big\}/\hbar\hspace{0.4pt}(z- 6\hspace{0.6pt}\varepsilon_{\protect\X{0}})$.}
\begin{equation}\label{e98y5}
S^{\X{(2\mathrm{s})}}(\bm{k},\bm{k}',\bm{k}'';z) = -\frac{\alpha U^2}{4\hbar}\hspace{0.6pt}\frac{\cosh(3\beta \varepsilon_{\X{0}})\hspace{1.2pt} \mathrm{sech}(2\beta \varepsilon_{\X{0}})\hspace{1.2pt}\mathrm{sech}^2(\beta \varepsilon_{\X{0}}/2)}{z - 6\hspace{0.6pt} \varepsilon_{\X{0}}}.
\end{equation}
Interestingly, in the expression for $S^{\X{(2\mathrm{s})}}(\bm{k},\bm{k}',\bm{k}'';\ii\hbar\omega_m)$ \textsl{no} $0/0$ condition, as indicated above, Eq.\,(\ref{e89x9}), occurs in identifying $\ii\hbar\omega_m$ with $(2m+1)\pi\ii/\beta$, $m\in\mathds{Z}$. It should be noted that for $\beta <\infty$ the RHS of Eq.\,(\ref{e98y5}) approaches towards $-\alpha U^2/(4\hbar z)$ for $\varepsilon_{\X{0}} \to 0$, so that $S^{\X{(2\mathrm{a})}} + S^{\X{(2\mathrm{b})}} \to U^2/(4\hbar z)$ for $\varepsilon_{\X{0}} \to 0$ (\emph{cf.} Eqs\,(\ref{e72a}) and (\ref{e63})). The absence of a $0/0$ condition in the case at hand may be attributable to the Green functions in Eq.\,(\ref{e98y4}) each possessing a single pole on either side of the chemical potential $\mu =0$, in contrast to the Green function underlying the calculation of the self-energy contribution in Eq.\,(\ref{e67}).\footnote{\emph{Cf.} Eq.\,(\protect\ref{e44}).} If so, then a $0/0$ condition is to be invariably expected in the event of evaluating the contributions of skeleton self-energy diagrams in terms of the \textsl{interacting} Green functions $\{\t{\mathscr{G}}_{\sigma}(\bm{k};z)\| \sigma\}$, since for arbitrary $\bm{k}$ the chemical potential $\mu$ (corresponding to $\b{N}=N$, where $\b{N}$ stands for the ensemble average of the number of particles) is inevitably sandwiched between two branch points\,\footnote{For instance, for macroscopic systems, the points $z=\mu_{N;\sigma}^{\pm}$ referred to in footnote \raisebox{-1.0ex}{\normalsize{\protect\footref{noteh}}} on p.\,\protect\pageref{ThisReflection} are branch points of the interacting one-particle Green function $\protect\t{G}_{\sigma}(\bm{k};z)$.} (each associated with a branch cut on the real axis of the $z$-plane) of $\t{\mathscr{G}}_{\sigma}(\bm{k};z)$,\footnote{See Eq.\,(\protect\ref{e4oc}).} which in the case of metallic GSs (ESs) are infinitesimally close to the chemical potential \cite{BF07,BF13}.\footnote{See in particular the discussions in \S\S\,III.4, III.5, and appendix B of Ref.\,\protect\citen{BF13}.}

We note that for $\alpha=1$ and in the units where $\hbar=1$ and $U = \varepsilon_{\X{0}} = 1$, the function $S^{\X{(2\mathrm{s})}}(\bm{k},\bm{k}',\bm{k}'';\ii\hbar\omega_m)$ identically coincides with the function $\Upsigma^{\X{(2)}}$ in Eq.\,(24) of Ref.\,\citen{TCLB19}. One can verify that for the mentioned units and $\beta\varepsilon_{\X{0}} =10$ the RHS of Eq.\,(\ref{e98y5}) reproduces the upper two graphs in Fig.\,1 of Ref.\,\citen{TCLB19}.\footnote{Similarly as regards the graphs in Fig.\,S1 of the Supplemental Material of Ref.\,\protect\citen{TCLB19}.}

\refstepcounter{dummyX}
\subsection{Discussion}
\phantomsection
\label{sdis1}
The considerations of \S\S\,\ref{s4xb}, \ref{s4xa}, and \ref{s4xc} have shown that the self-energy corresponding to the `Hubbard atom' as expanded in terms of proper self-energy diagrams and the non-interacting Green function is \textsl{terminating}, Eqs\,(\ref{e55}) and (\ref{e49a}). More explicitly, on evaluating these diagrams in terms of the Hartree-Fock Green function (subject to a hair-splitting of the degenerate excitation energy at the chemical potential $\mu=0$), the only contribution to the self-energy is second order in the on-site interaction energy $U$. Realising that on account of the Dyson equation, Eq.\,(\ref{e4a}), even a terminating perturbation series for the self-energy gives rise to an infinite perturbation series for the one-particle Green function, it becomes immediately evident that a terminating perturbation series expansion for the self-energy in terms of proper self-energy diagrams and the non-interacting Green function \textsl{cannot} be naturally transformed into one in terms of skeleton self-energy diagrams and the interacting Green function. The artificiality of such a series is clearly reflected in the asymptotic series in Eq.\,(\ref{e50a}), where all terms beyond the first one are artifacts of replacing the non-interacting Green function by the interacting one in the second-order proper self-energy diagrams, which in the case at hand are skeleton (see the diagrams $(2.1)$ and $(2.2)$ in Fig.\,\ref{f7} below, p.\,\pageref{SecondOrderS}). The present observation is general, \emph{i.e.} it is \textsl{not} specific to the self-energy corresponding to the half-filled GS of the Hubbard Hamiltonian of spin-$\tfrac{1}{2}$ particles in the atomic limit: in all cases where the perturbation series for the self-energy in terms of proper self-energy diagrams and the non-interacting Green function is terminating, the perturbation series expansion of the self-energy in terms of skeleton self-energy diagrams and the interacting Green function is artificial; on using this perturbation series, one will be spending effort on the meaningless task of calculating an \textsl{infinite} sequence of non-trivial contributions the sum of which is to reproduce the sum of the terms of a \textsl{finite} sequence. Importantly, the rigorous calculations in \S\,\ref{sd4} make explicit that in the case of the `Hubbard atom' the equality in Eq.\,(\ref{e7h}) fails to be satisfied for $j = 3$.\footnote{For a brief summary of the relevant observations, see \S\,\protect\ref{sec.b2.1}.} With reference to Eq.\,(\ref{e50a}), this implies that in order to reproduce the exact asymptotic behaviour of the self-energy in the region $z\to \infty$ to order $1/z^3$ (inclusive),\footnote{That is, to reproduce the \textsl{exact} $\Sigma_{\infty_3}$.} one has to calculate the self-energy in terms of skeleton self-energy diagrams and the interacting Green function to \textsl{at least} order $\nu = 6$.\footnote{See in particular the closing part of \S\,\protect\ref{sec.d53}.} We should emphasize that at present we have no reason to believe that in the case at hand $\Sigma_{\infty_3}$ is fully reproduced on taking account of the contributions of the $6$th-order skeleton self-energy diagrams in terms of the interacting Green function; arbitrary higher-order terms may contribute to $\Sigma_{\infty_3}$, albeit with progressively diminishing amounts for increasing values of $\nu$.

\refstepcounter{dummyX}
\section{(Re-)Summation of the perturbation series}
\phantomsection
\label{sec.5}
There are a variety of ways in which to sum the perturbation series for many-body correlation functions.\footnote{See \S\,\protect\ref{sec.5.2} below.} One method that fits naturally into the theoretical considerations of this publication, falls under the category of \emph{moment constant methods} [\S\,4.13, p.\,81, in Ref.\,\citen{GHH73}]. In this section we focus on the summation of the perturbation series for the one-particle Green function \cite{BF19}. This summation approach applies similarly to the perturbation series expansion of the self-energy\,\refstepcounter{dummy}\label{InDealingWith}\footnote{In dealing with the self-energy, while focussing on $\protect\t{\Sigma}_{\sigma}'(\bm{k};z) \doteq\protect\t{\Sigma}_{\sigma}(\bm{k};z) -\Sigma_{\sigma}^{\protect\textsc{hf}}(\bm{k})$, Eqs\,(\ref{e4}) and (\protect\ref{e20a}), the counterpart of the function $\protect\t{\mathfrak{G}}_{\sigma}(\bm{k};z,\varepsilon)$ in Eq.\,(\protect\ref{e101}) below is the function $\protect\t{\mathfrak{S}}_{\sigma}'(\bm{k};z,\varepsilon) \doteq \sum_{\nu=0}^{\infty} [\protect\t{\Sigma}_{\sigma}^{\protect\X{(\nu+2)}}(\bm{k};z)/\Sigma_{\sigma;\infty_{\nu+1}}(\bm{k})] \hspace{0.6pt}\varepsilon^{\nu}$ (\emph{cf.} Eqs\,(\protect\ref{e4}) and (\protect\ref{e7i})). With $\protect\t{\Sigma}_{\sigma}^{\protect\X{(\nu+2)}}(\bm{k};z)$ denoting the total contribution of the $(\nu+2)$th-order skeleton self-energy diagrams evaluated in terms of the interacting Green functions $\{\protect\t{G}_{\sigma}(\bm{k};z) \| \sigma\}$, it is generally (\emph{i.e.} barring such exceptional case as that of the `Hubbard atom') explicitly proportional to $\lambda^{\nu+2}$, where $\lambda$ denotes the dimensionless coupling constant of the two-body interaction potential (to be identified with $1$ in actual calculations). As we show in appendix \protect\ref{sab} (\S\,\protect\ref{sec.b2}, p.\,\protect\pageref{FurtherTheFact} -- see in particular Eq.\,(\protect\ref{e7h})), $\Sigma_{\sigma;\infty_{\nu+1}}(\bm{k})$ is a polynomial of order $\nu+2$ in $\lambda$. Thus, for $\lambda\to\infty$ (in the case of the Hubbard Hamiltonian, for $U\to\infty$) the ratio $\protect\t{\Sigma}_{\sigma}^{\protect\X{(\nu+2)}}(\bm{k};z)/\Sigma_{\sigma;\infty_{\nu+1}}(\bm{k})$ becomes explicitly independent of $\lambda$. We note that a similar remark as in the case of the perturbation series expansion of $\protect\t{G}_{\sigma}(\bm{k};z)$ in terms of $\{\protect\t{G}_{\protect\X{0};\sigma}(\bm{k};z) \| \sigma\}$ applies in regard to the possible zeros $\{\protect\p{\lambda}_i^{\protect\X{(j)}}\| i\}$ of $\Sigma_{\sigma;\infty_{2j}}(\bm{k})$, $\forall j \in \mathds{N}$, over the interval $[0,1]$ (as regards $\Sigma_{\sigma;\infty_{2j+1}}(\bm{k})$, see Eq.\,(\protect\ref{e7l})). \label{notey}} as well as those of other many-body correlation functions, such as, for instance, the polarisation function \cite{BF19},\footnote{As regards the density-density response function $\protect\t{\protect\b{\chi}}(\bm{k};z)$, see footnote \raisebox{-1.0ex}{\normalsize{\protect\footref{notei1}}} on p.\,\protect\pageref{SimilarToChi}.} appendix \ref{sa}.

\refstepcounter{dummyX}
\subsection{Formalism}
\phantomsection
\label{sec.5.1}
We begin by expressing the perturbation series expansion for the one-particle Green function pertaining to the uniform $N$-particle GS of a Hubbard-like Hamiltonian as \cite{BF19}\footnote{\emph{Cf.} Eq.\,(2.106) in Ref.\,\protect\citen{BF19}. For brevity, here we identify the dimensionless coupling constant $\lambda$ of the two-body interaction potential with unity.}
\begin{equation}\label{e100}
\t{G}_{\sigma}(\bm{k};z) = \sum_{\nu=0}^{\infty} \t{G}_{\sigma}^{\X{(\nu)}}(\bm{k};z),
\end{equation}
where $\t{G}_{\sigma}^{\X{(0)}}(\bm{k};z) \equiv \t{G}_{\X{0};\sigma}(\bm{k};z)$, the non-interacting one-particle Green function in terms of which the perturbational terms $\{\t{G}_{\sigma}^{\X{(\nu)}}(\bm{k};z)\| \nu\in \mathds{N}\}$ have been determined. With reference to Eq.\,(\ref{e4o}), and assuming that $G_{\sigma;\infty_{\nu+1}}(\bm{k}) \not\equiv 0$, $\forall\nu\in \mathds{N}_0$, we introduce the auxiliary function\,\footnote{\emph{Cf.} Eq.\,(4.12.2), p.\,81, in Ref.\,\protect\citen{GHH73}.}
\begin{equation}\label{e101}
\t{\mathfrak{G}}_{\sigma}(\bm{k};z,\varepsilon) \doteq \sum_{\nu=0}^{\infty} \frac{\t{G}_{\sigma}^{\X{(\nu)}}(\bm{k};z)}{G_{\sigma;\infty_{\nu+1}}(\bm{k})}\hspace{1.2pt} \varepsilon^{\nu}.
\end{equation}
With $\lambda$ denoting the dimensionless coupling constant of the two-body interaction potential,\footnote{The dimensionless coupling constant $\lambda$ is to be identified with $1$ in the actual calculations. Nonetheless, the instance of $\lambda\to\infty$ is of particular relevance to the case of the single-band Hubbard Hamiltonian where one has the dimensionless coupling constant $\protect\b{\lambda} \doteq \lambda U/\vert t\vert$, with $U$ the on-site interaction energy, and $t$ a hopping integral (for instance, and notably, the nearest-neighbour hopping integral). Thus, whereas the fully-interacting system corresponds to $\lambda=1$, for the physically-relevant coupling constant $\protect\b{\lambda}$ one has $\protect\b{\lambda} \to \infty$ as $U\to \infty$.} the function $\t{G}_{\sigma}^{\X{(\nu)}}(\bm{k};z)$, as determined in terms of $\{\t{G}_{\X{0};\sigma}(\bm{k};z)\| \sigma\}$ and the bare interaction potential, is explicitly proportional to $\lambda^{\nu}$.\footnote{For the $\t{G}_{\X{0};\sigma}(\bm{k};z)$ corresponding to the fully non-interacting Hamiltonian, $\t{G}_{\sigma}^{\protect\X{(\nu)}}(\bm{k};z)/\lambda^{\nu}$ is independent of $\lambda$. This is however not the case when $\t{G}_{\X{0};\sigma}(\bm{k};z)$ corresponds to a mean-field Hamiltonian, whereby $\t{G}_{\sigma}^{\protect\X{(\nu)}}(\bm{k};z)/\lambda^{\nu}$ depends implicitly on $\lambda$. In this publication $\t{G}_{\X{0};\sigma}(\bm{k};z)$ generally (but not invariably) takes account of the Hartree or the Hartree-Fock self-energy, $\Sigma^{\textsc{h}}(\bm{k})$ (or $\Sigma_{\sigma}^{\protect\textsc{h}}(\bm{k})$) and $\Sigma_{\sigma}^{\protect\textsc{hf}}(\bm{k})$, respectively, \S\,\protect\ref{sec.3a.1}.} With reference to the remark following Eq.\,(\ref{e4r}), insofar as its \textsl{explicit} dependence on $\lambda$ is concerned, for Hubbard-like Hamiltonians the \textsl{exact} $G_{\sigma;\infty_{\nu+1}}(\bm{k})$ is a $\nu$th-order \textsl{polynomial} of $\lambda$.\footnote{With reference to Eq.\,(\protect\ref{e7xa}), in the case of the `Hubbard atom' $G_{\infty_{\nu+1}} \equiv 0$ for $\nu$ \textsl{odd}, and $G_{\infty_{\nu+1}} = \hbar\hspace{0.6pt} (U/2)^{\nu}$ for $\nu$ \textsl{even}. Therefore, in this case $G_{\infty_{\nu+1}}$ is peculiarly either vanishing or a \textsl{monomial} of order $\nu$ in $\lambda$.} It follows therefore that for Hubbard-like Hamiltonians the ratio $\t{G}_{\sigma}^{\X{(\nu)}}(\bm{k};z)/G_{\sigma;\infty_{\nu+1}}(\bm{k})$ is generally a smooth function of $\lambda$ that in the region $\lambda\to\infty$ becomes explicitly independent of $\lambda$.\footnote{As we shall see, Eq.\,(\protect\ref{e107}) below, in the case of the `Hubbard atom' this ratio is \textsl{independent} of $U$ for all $\nu$.} The structure of the function $\t{\mathfrak{G}}_{\sigma}(\bm{k};z,\varepsilon)$ and its relation in particular to the one-particle Green function and the associated spectral function, Eqs\,(\ref{e102}) and (\ref{e104}) below, suggest that possible zeroes $\{\lambda_i^{\X{(\nu)}} \| i \}$ of $G_{\sigma;\infty_{2j}}(\bm{k})$, $\forall j \in \mathds{N}$, over the interval $[0,1]$ signify breakdown of the adiabatic assumption \cite{FW03,BF19} that underlies the zero-temperature perturbation series expansion of $\t{G}_{\sigma}(\bm{k};z)$ in terms of $\{\t{G}_{\X{0};\sigma}(\bm{k};z) \| \sigma\}$.\footnote{Note that $G_{\sigma;\infty_1}(\bm{k}) \equiv \hbar$, Eq.\,(\protect\ref{e4qa}), and that, barring the specific case discussed in relation to Eq.\,(\protect\ref{e4oi}), $G_{\sigma;\infty_{2j+1}}(\bm{k}) > 0$, $\forall j \in\mathds{N}$.} With reference to Eq.\,(\ref{e4p}), we remark that in practice one can calculate $G_{\sigma;\infty_{\nu+1}}(\bm{k})$ without recourse to the calculation of the perturbational contributions to $\t{G}_{\sigma}(\bm{k};z)$.\footnote{See Ref.\,\protect\citen{BF02}. See also \S\,A.5 (in particular the second footnote of this Section) of Ref.\,\protect\citen{BF19}.}

With reference to Eqs\,(\ref{e4g}), (\ref{e4oa}), and (\ref{e4ob}), one has
\begin{equation}\label{e102}
\t{G}_{\sigma}(\bm{k};z) = \int_{-\infty}^{\infty} \mathrm{d}\varepsilon'\; A_{\sigma}(\bm{k};\varepsilon')\hspace{0.6pt} \t{\mathfrak{G}}_{\sigma}(\bm{k};z,\varepsilon') \equiv
\int_{-\infty}^{\infty}\mathrm{d}\upgamma_{\sigma}(\bm{k};\varepsilon')\hspace{0.6pt} \t{\mathfrak{G}}_{\sigma}(\bm{k};z,\varepsilon').
\end{equation}
Defining
\begin{equation}\label{e103}
K_{\sigma}(\bm{k};\varepsilon;\varepsilon') \doteq \pm\frac{1}{\pi}\im[\t{\mathfrak{G}}_{\sigma}(\bm{k};\varepsilon\mp \ii 0^+,\varepsilon')],
\end{equation}
following Eq.\,(\ref{e4d}) and on account of $A_{\sigma}(\bm{k};\varepsilon) \in \mathds{R}$, from Eq.\,(\ref{e102}) one obtains
\begin{align}\label{e104}
A_{\sigma}(\bm{k};\varepsilon) &= \int_{-\infty}^{\infty} \mathrm{d}\varepsilon'\; K_{\sigma}(\bm{k};\varepsilon,\varepsilon')\hspace{0.6pt} A_{\sigma}(\bm{k};\varepsilon') \nonumber\\
&\equiv \int_{-\infty}^{\infty}\mathrm{d}\upgamma_{\sigma}(\bm{k};\varepsilon')
\hspace{0.6pt}K_{\sigma}(\bm{k};\varepsilon,\varepsilon')\hspace{0.6pt}.
\end{align}
For the function $K_{\sigma}(\bm{k};\varepsilon,\varepsilon')$ assumed as given, the first equality in Eq.\,(\ref{e104}) amounts to an integral equation for $A_{\sigma}(\bm{k};\varepsilon)$.\footnote{Technically, a homogeneous Fredholm integral equation of the first kind [Ch. XI, p.\,211, in Ref.\,\protect\citen{WW62}] [Ch. 8, p.\,896, in Ref.\,\citen{MF53}]. For \textsl{inhomogeneous} integral equations of this type, the solution is unique provided that the Fredholm determinant of the relevant kernel (here $K(\bm{k};\varepsilon,\varepsilon')$) is non-vanishing \cite{WW62}. In contrast, for \textsl{homogeneous} integral equations of this type, the latter determinant is to be vanishing in order for the equation to have a non-trivial solution, which in general is not unique [\S\,11.21, p.\,215, in Ref.\,\protect\citen{WW62}].} As will become evident below, this equation has in general no unique solution. Further, while it is tempting to suspect that $K_{\sigma}(\bm{k};\varepsilon,\varepsilon') = \delta(\varepsilon-\varepsilon')$, below we explicitly show that this is \textsl{not} the case for even the `Hubbard atom' for which the relevant kernel $K_{\sigma}(\bm{k};\varepsilon,\varepsilon')$ is independent of $\bm{k}$, Eq.\,(\ref{e108}) below. This may be immediately appreciated by realising that $K_{\sigma}(\bm{k};\varepsilon,\varepsilon')$ is a functional of the function $A_{\sigma}(\bm{k};\varepsilon)$.

Multiplying both sides of the first equality in Eq.\,(\ref{e104}) with $\varepsilon^{j-1}$, $j\in \mathds{N}$, and integrating the resulting equality with respect to $\varepsilon$, in view of Eq.\,(\ref{e4o}) one obtains
\begin{equation}\label{e104a}
\int_{-\infty}^{\infty} \mathrm{d}\varepsilon\, A_{\sigma}(\bm{k};\varepsilon)\hspace{0.6pt} L_j(\bm{k};\varepsilon) = G_{\sigma;\infty_j}(\bm{k})\Longleftrightarrow \int_{-\infty}^{\infty} \mathrm{d}\upgamma_{\sigma}(\bm{k};\varepsilon)\hspace{0.6pt}L_j(\bm{k};\varepsilon) = G_{\sigma;\infty_j}(\bm{k}),
\end{equation}
where\,\footnote{By the same reasoning that for Hubbard-like models $G_{\sigma;\infty_j}(\bm{k})$ is bounded for arbitrary finite values of $j$, \S\,\ref{sec.b.1}, for these models the function $L_{\sigma;j}(\bm{k};\varepsilon)$ is bounded for arbitrary finite values of $j$ and $\varepsilon$ (\emph{cf.} Eq.\,(\protect\ref{e111}) below).}
\begin{equation}\label{e104b}
L_{\sigma;j}(\bm{k};\varepsilon) \doteq \int_{-\infty}^{\infty} \mathrm{d}\varepsilon'\; \p{\varepsilon}^{j-1} K_{\sigma}(\bm{k};\varepsilon',\varepsilon),\;\; j\in \mathds{N}.
\end{equation}
Comparison of the left-most equality in Eq.\,(\ref{e104a}) with that in Eq.\,(\ref{e4o}), it is tempting to suspect that $L_{\sigma;j}(\bm{k};\varepsilon) = \varepsilon^{j-1}$, which below we explicitly show not to be the case even for the `Hubbard atom' for which $L_{\sigma;j}(\bm{k};\varepsilon)$ is independent of $\bm{k}$, Eq.\,(\ref{e111}) below.

Defining
\begin{equation}\label{e105}
\mathcal{K}_{\sigma}(\bm{k};\varepsilon,\varepsilon') \doteq\int_{-\infty}^{\varepsilon} \mathrm{d}\varepsilon''\; K_{\sigma}(\bm{k};\varepsilon'',\varepsilon'),
\end{equation}
with reference to Eq.\,(\ref{e4g}), from the second equality in Eq.\,(\ref{e104}) one obtains
\begin{equation}\label{e106}
\upgamma_{\sigma}(\bm{k};\varepsilon) = \int_{-\infty}^{\infty} \mathrm{d}\upgamma_{\sigma}(\bm{k};\varepsilon')\hspace{0.6pt} \mathcal{K}_{\sigma}(\bm{k};\varepsilon,\varepsilon').
\end{equation}
With
\begin{equation}\label{e106a}
\kappa_{\sigma}(\bm{k};\varepsilon) \doteq \int_{-\infty}^{\infty} \mathrm{d}\varepsilon'\; K_{\sigma}(\bm{k};\varepsilon',\varepsilon),
\end{equation}
following Eq.\,(\ref{e4h}), from Eq.\,(\ref{e101}) one arrives at the exact sum-rule
\begin{equation}\label{e106b}
\kappa_{\sigma}(\bm{k};0) = 1,\;\; \forall \bm{k}.
\end{equation}
One further deduces that\,\footnote{With $G_{\sigma;\infty_1}(\bm{k}) \equiv \hbar$, Eq.\,(\protect\ref{e4qa}), in the light of Eqs\,(\protect\ref{e4d}) and (\protect\ref{e4h}), note that $\kappa_{\sigma}^{\X{(0)}}(\bm{k};0) \equiv \kappa_{\sigma}(\bm{k};0)$.}
\begin{equation}\label{e106c}
\kappa_{\sigma}^{\X{(\nu)}}(\bm{k};0) \doteq \left.\frac{\partial^{\nu}}{\partial\varepsilon^{\nu}}\hspace{0.6pt}
\kappa_{\sigma}(\bm{k};\varepsilon)\right|_{\varepsilon=0} = \pm\frac{\nu !}{\pi} \frac{\int_{-\infty}^{\infty} \mathrm{d}\varepsilon\; \im[\t{G}_{\sigma}^{\X{(\nu)}}(\bm{k};\varepsilon\mp \ii 0^+)]}{G_{\sigma;\infty_{\nu+1}}(\bm{k})},
\end{equation}
which satisfies the exact sum-rule
\begin{equation}\label{e106d}
\sum_{\nu=1}^{\infty} \frac{G_{\sigma;\infty_{\nu+1}}(\bm{k})}{\nu !}\hspace{1.2pt}\kappa_{\sigma}^{\X{(\nu)}}(\bm{k};0) = 0,\;\; \forall\bm{k}.
\end{equation}
With $A_{\sigma}^{\X{(0)}}(\bm{k};\varepsilon)$ denoting the single-particle spectral function corresponding to $\t{G}_{\sigma}^{\X{(0)}}(\bm{k};z) \equiv \t{G}_{\X{0};\sigma}(\bm{k};z)$, this sum-rule reflects the fact that $A_{\sigma}^{\X{(0)}}(\bm{k};\varepsilon)$ satisfies the same normalisation condition as $A_{\sigma}(\bm{k};\varepsilon)$ in Eq.\,(\protect\ref{e4h}). This observation is immediately appreciated by considering the expression on the RHS of Eq.\,(\ref{e106c}). The sum-rule in Eq.\,(\ref{e106d}) is alternatively deduced from the equality
\begin{equation}\label{e106e}
\frac{1}{\hbar} \int_{-\infty}^{\infty} \mathrm{d}\varepsilon\, A_{\sigma}(\bm{k};\varepsilon) \hspace{0.6pt}\kappa_{\sigma}(\bm{k};\varepsilon) = 1,
\end{equation}
which is obtained by integrating both sides of the first equality in Eq.\,(\ref{e104}) with respect to $\varepsilon$ and making use of Eq.\,(\ref{e4h}). Taylor expanding the $\kappa_{\sigma}(\bm{k};\varepsilon)$ on the LHS of Eq.\,(\ref{e106e}) around $\varepsilon=0$, the sum-rule in Eq.\,(\ref{e106d}) follows on account of the equalities in Eqs\,(\ref{e106b}), (\ref{e4h}), and (\ref{e4o}). The accuracy with which the sum-rules in Eqs\,(\ref{e106b}) and (\ref{e106d}) are reproduced in practical calculations may be relied upon for quantifying the accuracy of the calculations. This is relevant in particular in the light of the fact that in practice the infinite sum in Eq.\,(\ref{e101}) is generally bound to be evaluated approximately.

Applying the above approach to the `Hubbard atom' is instructive. Following the expressions in Eqs\,(\ref{e38}) and (\ref{e7xa}), for the `Hubbard atom' one obtains\,\footnote{With reference to Eq.\,(\ref{e38}) and Eq.\,(\protect\ref{e7xa}), the summands corresponding to terms of the form $0/0$ are discarded.}\footnote{It is significant that in the first equality on the second line of Eq.\,(\protect\ref{e107}) we explicitly assume $\vert\varepsilon/z\vert < 1$, however in the subsequent calculations the ratio $\varepsilon/z$ can take any arbitrary value. This observation gives strong credence to the process of re-summing the series in Eq.\,(\protect\ref{e101}) in the neighbourhood of $\varepsilon = 0$ and employing the resulting closed expression for arbitrary $\varepsilon$.}
\begin{align}\label{e107}
\t{\mathfrak{G}}(z,\varepsilon) &= \sum_{\nu=0}^{\infty} \frac{(\hbar/z) (U/2z)^{2\nu}}{\hbar (U/2)^{2\nu}}\hspace{0.6pt} \varepsilon^{2\nu} \equiv \frac{1}{z} \sum_{\nu=0}^{\infty} \Big(\frac{\varepsilon}{z}\Big)^{2\nu}\nonumber\\
&= \frac{1}{z} \frac{1}{1 - (\varepsilon/z)^2} \equiv \frac{1}{2} \Big\{\frac{1}{z-\varepsilon} + \frac{1}{z+\varepsilon}\Big\},
\end{align}
whereby, following Eq.\,(\ref{e103}),\footnote{Following Eq.\,(\protect\ref{e106a}), one has $\kappa(\varepsilon) \equiv 1$, whereby indeed $\kappa(0) =1$ and $\kappa^{\protect\X{(\nu)}}(0) = 0$, $\forall\nu \in \mathds{N}$ (\emph{cf.} Eqs\,(\protect\ref{e106b}) and (\protect\ref{e106d})).}
\begin{equation}\label{e108}
K(\varepsilon,\varepsilon') = \frac{1}{2} \big\{\delta(\varepsilon'-\varepsilon) + \delta(\varepsilon'+\varepsilon)\big\}.
\end{equation}
As we have indicated above, even in the present case $K(\varepsilon,\varepsilon') \not\equiv \delta(\varepsilon-\varepsilon')$. Note that $\t{\mathfrak{G}}(z,\varepsilon)$ is independent of $U$, so that in equating the series expansion of $\t{\mathfrak{G}}(z,\varepsilon)$ with the closed expression on the second line of Eq.\,(\ref{e107}) the magnitude of $U$ has played no role. Further, the first closed expression on the second line of Eq.\,(\ref{e107}) is identically reproduced by the $[p/q]$ Pad\'{e} approximant \cite{BG96},\footnote{For a concise however very informative introduction to the Pad\'{e} approximation, consult \S\,19.7, p.\,349, of Ref.\,\protect\citen{BD02}.} with $q=p+1$, of the truncated power series, centred on $\varepsilon=0$,
\begin{equation}\label{e108a}
\t{\mathfrak{G}}^{\X{[m]}}(z,\varepsilon) \doteq \frac{1}{z} \sum_{\nu=0}^{m} \frac{1 + (-1)^{\nu}}{2 z^{\nu}}\hspace{1.2pt} \varepsilon^{\nu},\;\;\forall m \ge 2,
\end{equation}
where, on account of the requirement $p+q \le m+1$, for a given $m \ge 2$ the integer $p$ is to satisfy $0 < p \le \lfloor m/2\rfloor$ \cite{Note18}.\footnote{For $p+q$, with $q=p+1$, one thus has $1 < p+q \le 2 \lfloor m/2\rfloor +1$, where $2 \lfloor m/2\rfloor + 1$ is equal to $m$ for $m$ \textsl{odd}, and equal to $m+1$ for $m$ \textsl{even}. We note that to the $[p/q]$ Pad\'{e} approximant of a function, with $p, q\in \mathds{N}_0$, is also assigned the $(p,q)$ location of a matrix (positioned at the intersection of the $p$th row and the $q$th column of this matrix) referred to as the \textsl{Pad\'{e} table} \protect\cite{BG96}. The sequence of the Pad\'{e} approximants relevant to $\protect\t{\mathfrak{G}}(z,\varepsilon)$ consists of the upper-diagonal elements $\{(p,p+1)\| p\in \mathds{N}_0\}$ of this table.}

Following Eq.\,(\ref{e105}), from Eq.\,(\ref{e108}) one obtains
\begin{equation}\label{e109}
\mathcal{K}(\varepsilon,\varepsilon') = -\frac{1}{2} \Big\{ \Theta(\varepsilon' -\varepsilon) - \Theta(\varepsilon'+\varepsilon)\Big\}.
\end{equation}
From Eq.\,(\ref{e108}) and the first equality in Eq.\,(\ref{e104}), one arrives at
\begin{equation}\label{e110}
A(\varepsilon) = \frac{1}{2} \big\{ A(\varepsilon) + A(-\varepsilon)\big\}.
\end{equation}
This equality implies, correctly (\emph{cf.} Eq.\,(\ref{exc1})), that the single-particle spectral function $A(\varepsilon)$ of the `Hubbard atom' must be an \textsl{even} function of $\varepsilon$, without specifying any further details regarding this function. In the light of the \textsl{independence} from $U$ of the function $\t{\mathfrak{G}}(z,\varepsilon)$ in Eq.\,(\ref{e107}), and consequently of the function $K(\varepsilon,\varepsilon')$ in Eq.\,(\ref{e108}), the result in Eq.\,(\ref{e110}) is not unexpected: considering the first equality in Eq.\,(\ref{e104a}) as an integral equation from which to determine $A_{\sigma}(\bm{k};\varepsilon)$, in the case of the `Hubbard atom' the kernel $K(\varepsilon,\varepsilon')$ contains \textsl{no} information regarding the strength of the on-site interaction energy $U$. While for a general Hubbard-like model the first equality in Eq.\,(\ref{e104a}), viewed as a homogeneous Fredholm integral equation \cite{WW62,MF53} for $A_{\sigma}(\bm{k};\varepsilon)$, may also not have a unique solution, any solution of this equation will depend on the strength of the interaction potential. The peculiarity of the `Hubbard atom' in the case at hand lies in the fact that its corresponding $G_{\sigma;\infty_{\nu+1}}(\bm{k})$, with $\nu$ \textsl{even}, is a \textsl{monomial} of $U$ (of degree $\nu$), Eq.\,(\ref{e7xa}), thus rendering the ratio $\t{G}_{\sigma}^{\X{(\nu)}}(\bm{k};z)/G_{\sigma;\infty_{\nu+1}}(\bm{k})$ in Eq.\,(\ref{e101}) independent of $U$. In general, $G_{\sigma;\infty_{\nu+1}}(\bm{k})$ is a \textsl{polynomial} of degree $\nu$ in the coupling constant of the two-body interaction potential.\footnote{See the discussions following Eq.\,(\protect\ref{e4r}).}

From Eqs\,(\ref{e104b}) and (\ref{e108}), for the `Hubbard atom' one obtains
\begin{equation}\label{e111}
L_j(\varepsilon) = \frac{1}{2} \big(1 + (-1)^{j-1}\big)\hspace{0.6pt} \varepsilon^{j-1},\;\; j \in \mathds{N},
\end{equation}
whereby, following Eqs\,(\ref{e104a}) and (\ref{exc1}),
\begin{equation}\label{e112}
\int_{-\infty}^{\infty} \mathrm{d}\varepsilon\, A(\varepsilon)\hspace{0.6pt} L_j(\varepsilon) = \frac{\hbar}{4} \big(1+(-1)^{j-1}\big)^2\hspace{0.6pt} \Big(\frac{U}{2}\Big)^{j-1} \equiv G_{\infty_j},\;\; j\in\mathds{N}.
\end{equation}
The second equality in Eq.\,(\ref{e112}) is in conformity with the results in Eq.\,(\ref{e7xa}).

\refstepcounter{dummyX}
\subsection{A practical approach}
\phantomsection
\label{sec.5.2}
In this section we present a practical method for calculating $A_{\sigma}(\bm{k};\varepsilon)$, and thereby $\t{G}_{\sigma}(\bm{k};z)$, for the function $\t{\mathfrak{G}}_{\sigma}(\bm{k};z;\varepsilon)$ assumed as given.\footnote{Contrast the present method with that in Ref.\,\protect\citen{FLvH88}, and those in Refs\,\protect\citen{EFNM91} and \protect\citen{DvHF92}.} In practice, this function may have been determined on the basis of a finite-order perturbation series expansion \cite{BF19} of $\t{G}_{\sigma}(\bm{k};z)$ in conjunction with an appropriate summation technique \cite{GHH73}, such as Pad\'{e}'s, and more generally rational, approximation \cite{PTVF01,BG96,BD02,SB91}, or Ces\`{a}ro's.\footnote{Relatively recently, the Pad\'{e} resummation method for the perturbation series has been employed in Refs\,\protect\citen{YP17} and \protect\citen{TDiS19}, and the Ce\`{a}ro method in Refs\,\protect\citen{vHKPS10,PS08a}, and \protect\citen{PS08b}. Summation of the perturbation series has a long history in the context of the calculation of the critical exponents/indices of models of interest to statistical mechanics \cite{GAB61,DG724,BNGM76,LGZJ77,JZJ02}.} In this connection, we draw the attention of the reader to the close link between such methods for accelerating convergence of sums as Aitken's $\Delta^2$ and Shanks' \cite{DS55,PW56} and the Pad\'{e} approximation [\S\,3.2, p. 71, in Ref.\,\citen{BG96}].

Let
\begin{equation}\label{e113}
\Pi_{\eta}(x, \updelta) \doteq \Theta_{\eta}(x + \updelta/2) -\Theta_{\eta}(x-\updelta/2),
\end{equation}
where
\begin{equation}\label{e114}
\Theta_{\eta}(x) \doteq \frac{1}{2} \big(1 + \tanh(x/\eta)\big),\;\; \eta >0,
\end{equation}
for which one has\,\footnote{Following the equality in Eq.\,(\protect\ref{e115}), $\Theta(0) = \tfrac{1}{2}$ provided that $x$ is identified with $0$ prior to effecting the limit $\eta\downarrow 0$.}
\begin{equation}\label{e115}
\Theta(x) = \lim_{\eta\downarrow 0} \Theta_{\eta}(x).
\end{equation}
We write
\begin{equation}\label{e116}
A_{\sigma}(\bm{k};\varepsilon) = \lim_{\substack{\X{\Delta,M} \to\infty \\ \X{\Delta/M}\to 0 \\ \eta\downarrow 0}} \mathcal{A}_{\sigma;\eta}^{\X{\Delta,M}}(\bm{k};\varepsilon),
\end{equation}
where
\begin{equation}\label{e117}
\mathcal{A}_{\sigma;\eta}^{\X{\Delta,M}}(\bm{k};\varepsilon) \doteq \sum_{i=1}^{M} \alpha_i(\bm{k})\hspace{0.6pt} \Pi_{\eta}(\varepsilon-\varepsilon_i,\Delta/M),
\end{equation}
in which, with $\mu$ denoting the chemical potential,
\begin{equation}\label{e118}
\varepsilon_i \doteq \mu+ \frac{\Delta}{2} \Big\{ \frac{2}{M}\hspace{0.6pt}\big(i -\frac{1}{2}\big) -1\Big\},\;\; i \in \{1,2,\dots,M\}.
\end{equation}
Since $A_{\sigma}(\bm{k};\varepsilon) \ge 0$, $\forall\varepsilon$, and $\Pi_{\eta}(x,\updelta) > 0$, $\forall x$ and $\forall\eta, \updelta >0$, it is required that
\begin{equation}\label{e119}
\alpha_i(\bm{k}) \ge 0,\;\; \forall i \in \{1,2,\dots,M\}.
\end{equation}
With reference to the condition $\Delta/M \to 0$ in Eq.\,(\ref{e116}), it is reasonable that $\eta$ should satisfy the condition
\begin{equation}\label{e119a}
\eta \approx \frac{\Delta}{2 M}.
\end{equation}
Since for Hubbard-like models $A_{\sigma}(\bm{k};\varepsilon)$ decays faster than any positive power of $1/\varepsilon$ for $\vert\varepsilon\vert\to \infty$, appendix \ref{sab}, with $\eta$ satisfying the relationship in Eq.\,(\ref{e119a}) and $\Delta$ sufficiently large, in practice for these models the function $\mathcal{A}_{\sigma;\eta}^{\X{\Delta,M}}(\bm{k};\varepsilon)$ is to exponential accuracy equal to $A_{\sigma}(\bm{k};\varepsilon)$ for $M\to\infty$. In other words, with $\eta$ satisfying the approximate equality in Eq.\,(\ref{e119a}) and $\Delta$ sufficiently large, however finite, for Hubbard-like models the three limits in Eq.\,(\ref{e116}) can be safely limited to the one corresponding to $M\to\infty$.

The function $\mathcal{A}_{\sigma;\eta}^{\X{\Delta,M}}(\bm{k};\varepsilon)$ as defined according to Eq.\,(\ref{e117}), with the coefficients $\{\alpha_i(\bm{k})\| i\}$ satisfying Eq.\,(\ref{e119}), is by construction appropriately positive semi-definite. The same applies to the associated single-particle spectral function $A_{\sigma}(\bm{k};\varepsilon)$ determined on the basis of the expression in Eq.\,(\ref{e116}). This is the case even while the negative of the function (\emph{cf.} Eq.\,(\ref{e100}) and, as regards the `Hubbard atom', \emph{cf.} Eq.\,(\ref{e41}))
\begin{equation}\label{e119b}
\t{G}_{\sigma}^{\X{[m]}}(\bm{k};z) \doteq \sum_{\nu=0}^{m} \t{G}_{\sigma}^{\X{(\nu)}}(\bm{k};z),\;\; 0 \le m < \infty,
\end{equation}
may not be a Nevanlinna function of $z$.\footnote{This possibility is explicitly illustrated in \S\,\protect\ref{s4xb} for the case of the `Hubbard atom'.} Here $m$ denotes the largest value of $\nu$ for which the perturbational terms $\{\t{G}_{\sigma}^{\X{(\nu)}}(\bm{k};z)\| \nu\}$ in Eq.\,(\ref{e101}) have been explicitly calculated.\footnote{Here we are referring to the fact that in general the \textsl{infinite} series in Eq.\,(\protect\ref{e101}) is determined on the basis of some appropriate \textsl{extrapolation} of the partial sums of the first $m$ leading terms of this series to infinity. In this connection, we refer the reader to our discussion of the $[p/q]$ Pad\'{e} approximant of the truncated series in Eq.\,(\protect\ref{e108a}) \protect\cite{Note18}.}

\begin{figure}[t!]
\psfrag{x}[c]{\huge $\varepsilon$}
\psfrag{y}[c]{\huge $\mathcal{A}_{\eta}^{\protect\X{\Delta,M}}(\varepsilon)$}
\centerline{\includegraphics[angle=0, width=0.62\textwidth]{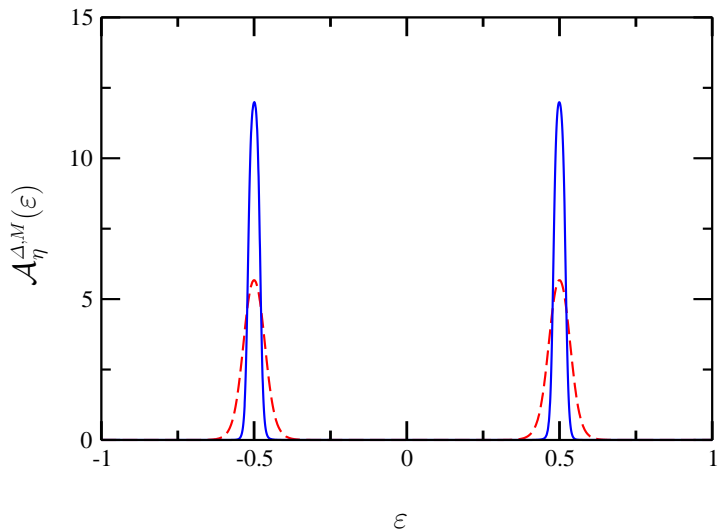}}
\caption{(Colour) \protect\refstepcounter{dummy}\label{TheSPSF}The single-particle spectral function $\mathcal{A}_{\eta}^{\protect\X{\Delta,M}}(\varepsilon)$, Eq.\,(\protect\ref{e117}), corresponding to the `Hubbard atom' as calculated by the formalism described in this section (in the units where $\hbar = 1$) \protect\cite{Note13}. The relevant parameters are $\mu=0$, $U=1$, $\Delta = 2$, and $\eta = \Delta/(2 M)$, Eq.\,(\protect\ref{e119a}). Solid/Blue (Broken/Red) line: $M =100$ ($30$). The constraint in Eq.\,(\protect\ref{e112}) has been taken into account for $j=1,2,3,4,5$. Taking account of this constraint for $j > 5$ does \textsl{not} perceptibly change the outcome. Calculations have \textsl{not} taken explicit account of the equality in Eq.\,(\protect\ref{e110}). We note that, with $\eta = O(1/M)$, the peaks in this figure appropriately approach the functions $\delta(\varepsilon \pm U/2)$ for $M\to \infty$ (\emph{cf.} Eq.\,(\protect\ref{exc1})).}
\label{f11}
\end{figure}

In Fig.\,\ref{f11} we present the single-particle spectral function $\mathcal{A}_{\eta}^{\X{\Delta,M}}(\varepsilon)$ corresponding to the `Hubbard atom', Eq.\,(\ref{exc1}), as calculated by means of the formalism described above \cite{Note13}. The calculated $\mathcal{A}_{\eta}^{\X{\Delta,M}}(\varepsilon)$ is based on minimising \cite{PTVF01} the sum of the squares of the difference between the integral on the LHS of Eq.\,(\ref{e112}) and $G_{\infty_j}$ for $j=1,2,3,4,5$.\footnote{As regards the $\mathcal{A}_{\eta}^{\protect\X{\Delta,M}}(\varepsilon)$ corresponding to $M=100$, the square root of the sum of these $5$ squares at the solution is less than $1.6\times 10^{-4}$ (this and the following data correspond to calculations with \texttt{iex = 0} as the input parameter of the program \texttt{Minim} reproduced in Ref.\,\protect\citen{Note13}). The integral of the spectral function displayed in Fig.\,\protect\ref{f11} is approximately equal to $1.00001$, to be contrasted to the exact value of $1$, Eq.\,(\protect\ref{e4h}). Its first moment is approximately equal to $-1.7\times 10^{-15}$ (to be contrasted with exact value of $0$), and its second moment approximately equal to $0.25001$ (to be contrasted with the exact value of $0.25$ for $U=1$), \emph{etc.}} Taking account of more residual errors, corresponding to $j> 5$, does not affect the outcome in any perceptible way. The key to this success of the approach is the summation to infinite order of the series for $\t{\mathfrak{G}}(z,\varepsilon)$ in Eq.\,(\ref{e107}), which, as we have indicated following Eq.\,(\ref{e108}) above, is notably accomplished by employing the $[p/q] = [1/2]$ approximant of this series (that is, by taking explicit account of merely the $3$ leading terms of this series, corresponding to $m=2$) \cite{Note18}. Importantly, this approach has in the case of the `Hubbard atom' clearly bypassed the possibility of summing the perturbation series expansion of the Green function to an unphysical function (\emph{cf.} Eqs\,(\ref{e41}) and (\ref{e41a})).

We remark that one can ensure the satisfaction of the equalities in Eq.\,(\ref{e119}) by describing $\alpha_i(\bm{k})$ in terms of an $M$th-order \textsl{positive (semidefinite) polynomial} \cite{MM08}. A necessary and sufficient condition for a polynomial of a single real variable to be positive semidefinite is that it be expressible as a sum of the squares of two polynomials of the same variable [\S\,1.2.1, p.\,4, in Ref.\,\citen{MM08}]. With $\EuScript{P}_{n}(x)$ and $\EuScript{Q}_{n}(x)$ denoting two arbitrary polynomials of order $n$, specified in terms of the real coefficients $\{p_0, p_1,\dots, p_n\}$ and $\{q_0,q_1,\dots,q_n\}$, respectively, we thus introduce the following $\mathcal{M}$th-order positive semi-definite polynomial:
\begin{equation}\label{e120}
\EuScript{R}_{\X{\mathcal{M}}}(x) = \EuScript{P}^2_{m}(x) + \EuScript{Q}^2_{n}(x),\;\, m+n \ge M -2,
\end{equation}
where
\begin{equation}\label{e120a}
\mathcal{M} \equiv \mathcal{M}(m,n) \doteq \max(m^2,n^2).
\end{equation}
The positive semi-definite polynomial $\EuScript{R}_{\X{\mathcal{M}}}(x)$ is specified in terms of $m+n+2$ real variables\,\footnote{The sets in Eq.\,(\protect\ref{e121}) are assumed to be \textsl{ordered}, so that, $r_i = p_i$ for $i \in \{0,1,\dots,m\}$, and $r_i = q_{i-m-1}$ for $i \in \{m+1,m+2,\dots, m+n+1\}$.}
\begin{equation}\label{e121}
\{r_0,r_1,\dots, r_{m+n+1}\} \equiv \{p_0,p_1,\dots,p_{m},q_0,q_1,\dots,q_{n}\}.
\end{equation}
The strict equality $m+n = M-2$ corresponds to the case of exchanging $M$ non-negative (variational) variables $\{\alpha_i(\bm{k}) \| i=1,2,\dots,M\}$ with strictly $M$ new (variational) variables $\{r_i \| i = 0,1,\dots,M-1\}$.

With $\EuScript{R}_{\mathcal{M}}(x)$ specified as above, for a given $\bm{k}$ we write
\begin{equation}\label{e122}
\alpha_i(\bm{k}) = \EuScript{R}_{\X{\mathcal{M}}}(u\hspace{0.6pt}(\varepsilon_i-\mu)/\Delta),
\end{equation}
where $u$ is a positive constant whose value is to be chosen with an eye on the numerical stability of the calculations.\footnote{The specific form of the argument of the polynomial $\EuScript{R}_{\mathcal{M}}(x)$ in Eq.\,(\protect\ref{e122}) is mainly motivated by the representations in Eq.\,(\protect\ref{e122a}) below. It is also suitable for $\EuScript{P}_m(x)$ and $\EuScript{Q}_n(x)$ expressed in terms of powers of $x$, as numerical evaluation of high powers of $x$, based on floating-point arithmetic, suffers from loss of accuracy for $\vert x\vert > 1$.} In this light, following the expression in Eq.\,(\ref{e118}), the choice $u=2$ implies that the argument $u \hspace{0.6pt}(\varepsilon_i-\mu)/\Delta$ of the polynomial on the RHS of Eq.\,(\ref{e122}) varies over the interval $(-1,1)$. Numerical tests on the `Hubbard atom' have shown the description according to Eq.\,(\ref{e122}) to be robust in the case of describing the polynomials $\EuScript{P}_m(x)$ and $\EuScript{Q}_n(x)$ in Eq.\,(\ref{e120}) in terms of the Chebyshev polynomials of the first kind $\{T_j(x)\| j \in \mathds{N}_0\}$ [Ch.\,22, p.\,773, in Ref.\,\citen{AS72}], for which $u$ must be identified with $2$. In this case, the real constants $\{p_0,p_1, \dots,p_m\}$ and $\{q_0,q_1,\dots,q_n\}$ referred to above are those in the following expressions:
\begin{equation}\label{e122a}
\EuScript{P}_m(x) = \sum_{j=0}^{m} p_j\hspace{0.6pt} T_j(x),\;\;\; \EuScript{Q}_n(x) = \sum_{j=0}^{n} q_j\hspace{0.6pt} T_j(x).
\end{equation}
The only drawback of the representation in Eq.\,(\ref{e122}) is the requirement of repeated evaluation of the sums in Eq.\,(\ref{e122a}), which becomes time-consuming for large values of $m$ and $n$. Since, however, the arguments $\{u (\varepsilon_i-\mu)/\Delta\| i\}$ are fixed throughout the calculation, the computational overload can be minimised by tabulating the required values of the Chebyshev polynomials.

With reference to Eqs\,(\ref{e120}) and (\ref{e122}), given $\eta$, $\Delta$, and $M$, the expression for $\mathcal{A}_{\sigma;\eta}^{\X{\Delta,M}}(\bm{k};\varepsilon)$ is a quadratic function of the real variables $\{r_0,r_1,\dots,r_{m+n+1}\}$, Eq.\,(\ref{e121}). These variables are to be obtained through a minimisation \cite{PTVF01} of a positive semi-definite penalty function defined in terms of the residual errors of the equalities in Eqs\,(\ref{e104}) and (\ref{e104a}). Because of this, rather than varying $m$ and $n$, subject to the condition $m+n \ge M-2$, one can choose $m = n \ge M-2$ at the outset and perform the calculations with these fixed values of $m$ and $n$.\footnote{Since these parameters are \textsl{variational} ones in the framework of the above-mentioned minimisation process, their number $m+n+2$ can be both less and more than $M$. For the cases where $A_{\sigma}(\bm{k};\varepsilon)$ is free from very narrow peaks, use of Eqs\,(\protect\ref{e122}) and (\protect\ref{e122a}) can yield accurate results even for $m+n < M -2$.} We note that with the coefficients $\{\alpha_i(\bm{k})\| i\}$ expressed in terms of (positive semi-definite) polynomials, Eqs\,(\ref{e122}) and (\ref{e122a}), the partial derivatives of these coefficients with respect to the variational parameters $\{r_0,r_1,\dots,r_{m+n+1}\}$ (as may be required in the above-mentioned minimization process \cite{PTVF01}) are straightforwardly determined.\footnote{Note also that $\partial T_n(x)/\partial x = n U_{n-1}(x)$, where $U_j(x)$ is the $j$th-order Chebyshev polynomial of the second kind [\S\S\,22.3.15 and 22.3.16, p.\,776, in Ref.\,\citen{AS72}]. See \S\,\protect\ref{sabx}.}

\refstepcounter{dummyX}
\section{Summary and concluding remarks}
\phantomsection
\label{scon}
In this publication we have returned to an earlier work \cite{BF07} regarding the convergence properties of the perturbation series expansion of the self-energy $\t{\Sigma}_{\sigma}(\bm{k};z)$ corresponding to the GS (ensemble of states, ESs) of Hubbard-like Hamiltonians as evaluated in terms of skeleton self-energy diagrams and the \textsl{exact} one-particle Green functions $\{G_{\sigma}(\bm{k};\varepsilon)\| \sigma\}$, Eq.\,(\ref{e4}). This undertaking was prompted by the observation by Kozik \emph{et al.} \cite{KFG14} that ``at strong interactions'' this series converged to an ``unphysical'' self-energy. This being at variance with the main observation in Ref.\,\citen{BF07}, namely that the mentioned series converges uniformly almost everywhere (\ae) in the $\bm{k}$- and $z$-space\,\footnote{Explicitly, converging everywhere barring possible subsets of measure zero of the underlying $\bm{k}$-space and the complex energy plane $\mathds{C}$.} to the exact self-energy,\footnote{That is, the self-energy that through the Dyson equation corresponds to the exact Green function $\protect\t{G}_{\sigma}(\bm{k};z)$ underlying the calculation of the perturbational contributions to the self-energy.} we reconsidered a key component of the proof of the above-mentioned uniform convergence in Ref.\,\citen{BF07}: that $\im[\t{\Sigma}_{\sigma}^{\X{(\nu)}}(\bm{k};\varepsilon -\ii 0^+)] \ge 0$, $\forall \varepsilon\in\mathds{R}$ and $\nu\ge 2$,\footnote{For the proof under discussion the condition $\im[\protect\t{\Sigma}_{\sigma}^{\protect\X{(\nu)}}(\bm{k};\varepsilon -\ii 0^+)] \ge 0$ is to apply for $\nu\to\infty$, or for $\nu\ge \nu_0$, where $\nu_0$ is an arbitrary finite integer (independent of $\bm{k}$ and $\varepsilon$). \label{notee}} for the \textsl{total} contribution of all $\nu$th-order skeleton self-energy diagrams as evaluated in terms of the exact $\{G_{\sigma}(\bm{k};\varepsilon)\| \sigma\}$. In Ref.\,\citen{BF07} this non-negativity of $\im[\t{\Sigma}_{\sigma}^{\X{(\nu)}}(\bm{k};\varepsilon -\ii 0^+)]$ is taken for granted on account of the assumed stability of the underlying GS (ESs).\footnote{See Eq.\,(5.43) and the subsequent discussions in Ref.\,\protect\citen{BF07}. \label{noted}} While this non-negativity condition is demonstrably satisfied for $\nu=2$,\footref{noted}\footnote{We note that the violation of the non-negativity of the single-particle spectral function as observed in Ref.\,\protect\citen{PM74}, is ``due to the incorrect analytic properties of the approximate Green functions'' \protect\cite{PM74}; as shown in Ref.\,\protect\citen{PM74}, the latter functions are inappropriately non-analytic away from the real axis of the $z$-plane. We point out that the second-order self-energy contributions as considered in Ref.\,\protect\citen{PM74} are in terms of the screened interaction potential $W$ (dealt with within the framework of the plasmon-pole approximation \protect\cite{EF93} of this potential), appendix \protect\ref{sa}, unlike the $2$nd-order self-energy contributions considered here, which are in terms of the bare interaction potential $v$.} our attempts to prove or disprove it for $\nu> 2$ remained inconclusive. For this reason, we focussed on determining the necessity of $\im[\t{\Sigma}_{\sigma}^{\X{(\nu)}}(\bm{k};\varepsilon -\ii 0^+)] \ge 0$, $\forall \varepsilon\in\mathds{R}$ and $\nu\ge 2$,\footnote{We emphasize once more, that in the present context the condition $\nu\ge 2$ is unnecessarily too stringent; the non-negativity of $\protect\im[\t{\Sigma}_{\sigma}^{\protect\X{(\nu)}}(\bm{k};\varepsilon -\protect\ii 0^+)]$ for $\nu \ge \nu_0$, with $\nu_0$ an arbitrary but finite integer, would suffice.} for the proof of the above-mentioned uniform convergence. In \S\,\ref{sec.3.2} we show that this uniform convergence can be established on the basis of a far weaker condition, which we explicitly demonstrate to be satisfied for the uniform GSs (ESs) of Hubbard-like Hamiltonians, appendix \ref{sab}. This weaker condition concerns the \textsl{positive sequence}\,\footnote{See Eq.\,(\protect\ref{e4s}) and the subsequent considerations.} of the asymptotic coefficients $\{\Sigma_{\sigma;\infty_j}(\bm{k}) \| j\}$, Eq.\,(\ref{e7b}), which in principle may or not be \textsl{determinate} according to the specification in appendix \ref{sab}. We show that for the GSs (ESs) of Hubbard-like Hamiltonians the latter sequence is indeed \textsl{determinate}, Eq.\,(\ref{e7m}). The connection between the perturbation series for $\t{\Sigma}_{\sigma}(\bm{k};z)$ in Eq.\,(\ref{e4}) and the positive sequence $\{\Sigma_{\sigma;\infty_j}(\bm{k}) \| j\}$ is established through Eq.\,(\ref{e7h}), which underlies the identities in Eq.\,(\ref{e7sa}). In \S\,\ref{sec.3.2.1} we show how the convergence of the continued-fraction expansion of the function $\t{\mathfrak{S}}_{\sigma}^{\X{(n)}}(\bm{k};z)$ for $n\to\infty$, Eq.\,(\ref{e7r}), which is linked to the property of the positive sequence $\{\Sigma_{\sigma;\infty_j}(\bm{k}) \| j\}$ being \textsl{determinate} in the case of the GSs (ESs) of Hubbard-like models, amounts to the uniform convergence of the series in Eq.\,(\ref{e4}) for almost all $\bm{k}$ and $z$. Importantly, in \S\,\ref{sec.3.2.1} we establish that even though the function $-\t{\mathfrak{S}}_{\sigma}^{\X{(n)}}(\bm{k};z)$ may fail to be a Nevanlinna function of $z$ for $n\to\infty$, its \textsl{analyticity} in the complex $z$-plane away from the real axis of this plane (\emph{i.e.} away from the $\varepsilon$-axis) suffices to compensate for this possible failure; we show that the possible drawback of such failure is reflected only in the existence of a \textsl{countable} set of points on the real axis of the $z$-plane\,\footnote{Such set, if non-empty and even infinite, is thus a measure-zero subset of the real energy axis, and by extension, of the complex $z$-plane.} where the perturbation series in Eq.\,(\ref{e4}) fails to be uniformly convergent for almost all $\bm{k}$. Remarkably, the property $\im[\t{\Sigma}_{\sigma}^{\X{(\nu)}}(\bm{k};\varepsilon -\ii 0^+)] \ge 0$ for $\nu\ge 2$ corresponds to that of $-\t{\mathfrak{S}}_{\sigma}^{\X{(n)}}(\bm{k};z)$ being a Nevanlinna function of $z$ for $n\in \mathds{N}$. Further, the uniform convergence of $\t{\mathfrak{S}}_{\sigma}^{\X{(n)}}(\bm{k};z)$ to $\t{\Sigma}_{\sigma}(\bm{k};z)$ a.e. for $n\to\infty$ in a self-consistent way implies that the function $-\t{\mathfrak{S}}_{\sigma}^{\X{(n)}}(\bm{k};z)$, if non-Nevanlinna for some finite values of $n$, must uniformly transform into a Nevanlinna function of $z$ for increasing values of $n$.

With reference to the relationship in Eq.\,(\ref{e7h}), to which we have referred above, in this publication we establish that this relationship fails to be generally valid in the atomic (or local) limit. The reason underlying this failure is the singularity of $\t{\Sigma}_{\sigma}^{\X{(\nu)}}(\bm{k};z)/\lambda^{\nu}$ at $\lambda = 0$, where $\lambda$ denotes the dimensionless coupling constant of the two-body interaction potential. While this is demonstrably not the case for $\nu < 6$ insofar as the `Hubbard atom' of spin-$\tfrac{1}{2}$ particles at half-filling is concerned, \S\S\,\ref{sec3.b} and \ref{sd4}, the possibility of this singularity at $\lambda = 0$ can be gleaned from the $4$th-order self-energy contribution in Eq.\,(\ref{e52b}) whose diagrammatic representation in presented in Fig.\,\ref{f10}, p.\,\pageref{A4thOr}; dividing both sides of Eq.\,(\ref{e52b}) by $U^4$, the resulting second and third terms on the RHS diverge like respectively $1/U$ and $1/U^2$ for $U\to 0$. Considering macroscopic systems, clearly the \textsl{zero-measure} subset of the $\bm{k}$-points over which $\t{\Sigma}_{\sigma}^{\X{(\nu)}}(\bm{k};z)/\lambda^{\nu}$ would be singular at $\lambda = 0$ away from the atomic / local limit, no longer qualifies as such in the atomic / local limit; in the absence of dispersion, a single singularity is one associated with \textsl{all} points of the underlying $\bm{k}$-space. Systems in the atomic / local limit display some peculiarities that we have discussed earlier in Refs\,\citen{BF07,BF07a}, and \citen{KDP13}.

Having established the validity of our earlier main observation in Ref.\,\citen{BF07}, we have inquired into the reasons underlying the above-mentioned observation in Ref.\,\citen{KFG14}. The results of these investigations are presented in \S\S\,\ref{sec2.a} and \ref{sec3}. In \S\,\ref{sec2.a} we show that insofar as the `Hubbard atom'\,\footnote{More explicitly, the system of spin-$\tfrac{1}{2}$ particles described by the Hubbard Hamiltonian at half-filling in the atomic limit.} is concerned, the mapping $G\mapsto G_{\X{0}}$ is single-valued provided that its range \cite{AA14} is limited to analytic functions in the region $\im[z] \not=0$ that to leading order decay like $\hbar/z$ in the asymptotic region $z\to\infty$.\footnote{Note that the indicated leading-order term $\hbar/z$ is directly tied to the sum-rule in Eq.\,(\protect\ref{e4h}), which applies to both interacting and non-interacting single-particle spectral functions. See Eqs\,(\protect\ref{e4n}), (\protect\ref{e4o}), and (\protect\ref{e4qa}).} The former condition is guaranteed by limiting the range of the latter mapping to the space of functions whose negative (in the light of Eq.\,(\ref{e2}), which is applicable to both $G$ and $G_{\X{0}}$) are Nevanlinna functions of $z$, most generally described according to the Riesz-Herglotz representation in Eq.\,(\ref{e20h}).\footnote{See Eq.\,(\protect\ref{e4l}).} The latter condition implies that as regards $G_{\X{0}}$, as well as $G$, the constants $\upmu$ and $\upnu$ in the latter representation are to satisfy the equalities in Eq.\,(\ref{e4m1}).\footnote{Thus for both $G_{\protect\X{0}}$ and $G$ one must have $\upmu= 0$. As regards the $\upnu$, for $G_{\protect\X{0}}$ the measures $\upeta_{\sigma}(\bm{k};u)$ and $\upgamma_{\sigma}(\bm{k};\varepsilon)$ in Eq.\,(\protect\ref{e4m1}) are to be replaced by their non-interacting counterparts, respectively $\upeta_{\protect\X{0};\sigma}(\bm{k};u)$ and $\upgamma_{\protect\X{0};\sigma}(\bm{k};\varepsilon)$. These are step-wise constant functions of $\varepsilon$ (\emph{cf.} Eq.\, (\protect\ref{e6s}) and see footnote \raisebox{-1.0ex}{\normalsize{\protect\footref{notew}}} on p.\,\protect\pageref{Conversely}).} It is worth mentioning that the observations as presented in \S\,\ref{sec2.a}, regarding the significance of the analyticity of the one-particle Green function $\t{G}_{\sigma}(\bm{k};z)$ for $\im[z] \not= 0$ and the behaviour of this function in the asymptotic region $z\to\infty$, gave the initial impulse that led to the development of the approach employed in \S\,\ref{sec.3.2}.\footnote{The details underlying the consideration in \S\,\protect\ref{sec.3.2} constitute the contents of appendix \protect\ref{sab}.}

In \S\,\ref{sec3} we show that the non-interacting Hartree (or Hartree-Fock)\,\footnote{See \S\,\protect\ref{sec.3a.1} for some relevant details.} Green function $\t{G}_{\X{0}}(z)$ in Eq.\,(\ref{e15a}) is not suitable for the calculation of the contributions of the proper self-energy diagrams to the self-energy at zero temperature, this on account of the pole of $\t{G}_{\X{0}}(z)$ at $z=0$ coinciding with the chemical potential $\mu=0$ corresponding to the half-filling of the underlying GS, \S\,\ref{sec3.d}, and that the calculations are to be carried out with the Green function $\t{G}_{\epsilon}(z)$ in Eq.\,(\ref{e44}), where $\epsilon$ is to be identified with $0$ subsequent to evaluating the contributions of the self-energy diagrams. We further show that the weak-coupling perturbation series expansion of $\t{G}(z)$ to any \textsl{finite} order in $U$ is pathological, it giving rise to a Green function with zeros in the region $\im[z] \not=0$ of the complex $z$-plane, in violation of the exact property that zeros of one-particle Green functions are restricted to be located on the real axis (that is, the $\varepsilon$-axis) of this plane, \S\,\ref{s4xb}.\footnote{Remarkably, while the infinite geometric series in Eq.\,(\protect\ref{e38}) reproduces the function  in Eq.\,(\protect\ref{e25}), for $\mathpzc{n} \to\infty$ the zeros of the function $\protect\t{G}_{\protect\X{(0)}}(z)$ in Eq.\,(\protect\ref{e41}) condense into the circle of radius $U/2$ centred at $z=0$ that we have encountered in \S\,\protect\ref{sec2.a}. This circle delineates the regions of convergence, in the $z$-plane, of the weak- and strong-coupling perturbation expansions in respectively Eq.\,(\protect\ref{e38}) and Eq.\,(\protect\ref{e38a}).} The pathological nature of the perturbation series expansion of $\protect\t{G}(z)$ in terms of $\t{G}_{\X{0}}(z)$ (or $\t{G}_{\epsilon\downarrow \X{0}}(z)$) raises the immediate question as to why this pathology has not been observed in the finite-order numerical calculations reported in Ref.\,\citen{KFG14}. The answer to this question is summarised in Figs\,\ref{f8} and \ref{f9} (pp.\,\pageref{SolidLine1} and \pageref{SolidLine2}). Defining the energy parameter $U_{\beta}$ according to Eq.\,(\ref{e71a}), for $U \lesssim U_{\beta}$ the pathology at issue is practically difficult, if not impossible, to detect for $\t{G}(z)$ calculated only at the Matsubara energies/frequencies.

We point out that the above-mentioned pathology is not necessarily a peculiarity of the `Hubbard atom', that even away from the local / atomic limit any truncated perturbation series for the one-particle Green function is likely to have zeros in the region $\im[z]\not=0$ of the complex $z$-plane. Such spurious zeros are automatically discarded by relying on this series as a basis for deducing an optimised (from the perspective of some appropriate norm function to be defined) one-particle spectral function $A_{\sigma}(\bm{k};\varepsilon)$, or, what is the same, bounded non-decreasing measure function $\upgamma_{\sigma}(\bm{k};\varepsilon)$, or $\upeta_{\sigma}(\bm{k};u)$, to be discussed below; this instead of using the sum of the perturbational terms, Eq.\,(\ref{e119b}), as an approximation of the exact one-particle Green function, possibly to be used in further calculations. In \S\,\ref{sec.5.1} we introduce a formalism for summing the perturbation series for the one-particle Green function, and in \S\,\ref{sec.5.2} we put forward a method for obtaining an optimised positive-semi-definite one-particle spectral function $A_{\sigma}(\bm{k};\varepsilon)$ whose corresponding one-particle Green function optimally matches the latter sum. The negative of this one-particle Green function is a Nevanlinna function of $z$. This Green function is therefore free from zeros in the finite part of the region $\im[z]\not=0$ of the complex $z$-plane.\footnote{See footnote \raisebox{-1.0ex}{\normalsize{\protect\footref{notee1}}} on p.\,\protect\pageref{ThisObservation}.} The above-mentioned measure functions associated with this spectral function are invariably bounded and non-decreasing functions of $\varepsilon$ and $u$.

A further observation discussed in detail in \S\,\ref{sec3} is that the iterative schemes A and B as employed in Ref.\,\citen{KFG14} are defective. This we show to be related to the sequence $\{\t{G}_{\X{0};n}(z)\| n\in \mathds{N}\}$ of the `non-interacting' Green functions as calculated within these schemes failing to correspond to mean-field Hamiltonians. In order for $\t{G}_{\X{0};n}(z)$ to correspond to a mean-field Hamiltonian, one must have $\hbar \t{G}_{\X{0};n}^{-1}(z) - z$ equal to a constant, which is demonstrably not the case for the Green functions $\t{G}_{\X{0};n}(z)$ according to the above-mentioned schemes A and B, Figs\,\ref{f5} and \ref{f6} (pp.\,\pageref{TheContour1} and \pageref{TheContour2}). Use of such Green functions as $\t{G}_{\X{0};n}(z)$ in the evaluation of the contributions of the Green-function diagrams is in violation of the Wick theorem \cite{BF19} underlying the perturbation series expansion of the interacting Green function, which violation in turn undermines the perturbation series expansion of the self-energy in terms of the non-interacting Green function.

The findings in Ref.\,\citen{KFG14} regarding the two-dimensional Hubbard Hamiltonian at half-filling, insofar as they indicate an anomalous behaviour in the self-energy as calculated in terms of skeleton self-energy diagrams and the interacting Green function\,\footnote{The technical details of these calculations (concerning DiagMC, \emph{etc.}) are described in Ref.\,\protect\citen{vHKPS10}. See in particular the section \textsf{Bold propagators} herein.} (observation of a non-dispersive behaviour for $U/t = 8$), are in our view to be understood foremost along the lines described in \S\,\ref{s3.3}.\footnote{Note that in the caption of Fig.\,4 in Ref.\,\protect\citen{KFG14} `$\protect\re\Upsigma$ ($\protect\im\Upsigma$)' should be `$\protect\im\Upsigma$ ($\protect\re\Upsigma$)'.} Very briefly, there is no \emph{a priori} reason why a solution of a highly \textsl{non-linear} equation for the self-energy is to be physical without having imposed strict analytic constraints on the solution function-space. In \S\,\ref{sec2.a} we explicitly discuss a related problem (concerning the mapping $G\mapsto G_{\X{0}}$) for the case of the `Hubbard atom' where one obtains an un-physical solution on solving merely a \textsl{quadratic} equation without restricting the solution function-space.

The most fundamental observation discussed in \S\,\ref{sec3} is that calculation of the thermal self-energy, and by extension of any other thermal many-body correlation function, in the energy/frequency domain is fraught with a mathematical problem that to our knowledge has hitherto not been recognised in the literature. The problem corresponds to the fact that while for the purpose of the analytic continuation of the self-energy (as well as other thermal many-body correlation functions) evaluated at the Matsubara energies/frequencies it is \textsl{necessary} to identify such transcendental functions as $\sin(\beta\hbar\omega_m)$, $\cos(\beta\hbar\omega_m)$, $\sin(\beta\hbar\nu_m)$, and $\cos(\beta\hbar\nu_m)$ with their values, Eqs\,(\ref{e67x1}) and (\ref{e78x1}), the functions thus obtained may not be suitable as intermediate functions for the calculation of other many-body correlation functions.\refstepcounter{dummy}\label{ThisPoblemIs}\footnote{This problem is specific to the cases where the \textsl{external} Matsubara energy (energies) associated with the correlation function of interest occurs (occur) in the argument of the intermediary correlation function referred to here. We have established that not until the underlying sums with respect to the \textsl{internal} Matsubara energies have been carried out, should the \textsl{external} Matsubara energy (energies) be viewed as coinciding with $\hbar\omega_m$ or $\hbar\nu_m$, as the case may be, with $m \in \mathds{Z}$; the external energy (energies) is (are) to be viewed as real variable(s) in some \textsl{finite neighbourhood(s)} of $\hbar\omega_m$ and/or $\hbar\nu_m$ with $m\in\mathds{Z}$. \label{noteg}} In the case of the `Hubbard atom', we explicitly show that the polarization function thus obtained, that is that in Eq.\,(\ref{e89b}) -- to be contrasted with its original form in Eq.\,(\ref{e76}), gives rise to a demonstrably erroneous self-energy (that is, to the self-energy in Eq.\,(\ref{e89d3}), to be contrasted with the correct self-energy in Eq.\,(\ref{e67x2})).\footnote{With reference to footnote \raisebox{-1.0ex}{\normalsize{\protect\footref{noteg}}} on p.\,\protect\pageref{ThisPoblemIs}, note that the \textsl{external} energy $\hbar\omega_m$ occurs in the argument of the polarisation function $\t{\mathscr{P}}^{\X{(0)}}_{\epsilon}$ on the RHS of Eq.\,(\protect\ref{e74}).} Examining the underlying mathematical expressions, we have established that the behaviour of thermal correlation functions in the infinitesimal \textsl{neighbourhoods} of the Matsubara energies contains consequential information that is lost following the above-mentioned identifications (that is those in Eqs\,(\ref{e67x1}) and (\ref{e78x1})). This information can be crucially important, as in the case of the calculation of the second-order thermal self-energy corresponding to the `Hubbard atom'. Explicitly, this information is decisive for the determination of the correct value of the function $\varphi(\ii\hbar\omega_m)$ in Eq.\,(\ref{e89x9}) for $m\in \mathds{Z}$; while use of the polarisation function in Eq.\,(\ref{e89b}) amounts to the identification of $\varphi(\ii\hbar\omega_m)$ with $0$, $\forall m\in \mathds{Z}$, use of the polarisation function in Eq.\,(\ref{e76}) leads to $\varphi(\ii\hbar\omega_m) = 4/(\cosh(\beta\upsilon)+1)$, $\forall m\in \mathds{Z}$, Eq.\,(\ref{e89y1}), following the limiting process $\delta z\downarrow 0$. This observation makes explicit the crucial role played by the information contained in the non-vanishing neighbourhood of the external Matsubara energy $\hbar\omega_m$, defined by $\delta z \not=0$, for correctly calculating the self-energy contribution under discussion.

In the process of searching for the source(s) of the deviation between the self-energies in Eqs\,(\ref{e67x2}) and (\ref{e89d3}), we have independently devised a method for the evaluation of sums over the Matsubara energies/frequencies based on the Mittag-Leffler partial-fraction decomposition (PFD) \cite{RR98,KK47} of rational functions, \S\,\ref{s4xc},\footnote{As we have indicated earlier, the present publication was nearly completed by early May 2015.} an approach that has in a more systematic and elegant form been devised and published by Taheridehkordi \emph{et al.} \cite{TCLB19}, under the heading \textsl{algorithmic Matsubara integration} (AMI).\footnote{In \S\,\protect\ref{s4xc} we briefly discuss the mathematical equivalence of the two approaches.} \emph{For the reason indicated above, the PFD and AMI approaches are both equally unsafe.} We propose that the safest approach for the calculation of thermal many-body correlation functions is the performance of the calculations in the imaginary-time domain, relying on an appropriate Fourier transformation when requiring the values of these functions in the Fourier domain at the Matsubara energies/frequencies.\footnote{\emph{Cf.} Eq.\,(\ref{e89x1}). Note that the $\omega_m$ in Eq.\,(\protect\ref{e89x1}) would have been $\nu_m$, Eq.\,(\ref{e70}), had the one-particle Green function under discussion corresponded to bosons.} In this connection, note that in the \textsl{numerical} calculations performed in the Fourier domain the above-mentioned transcendental functions are automatically identified with their constant values at the relevant Matsubara energies/frequencies. The information regarding the behaviours of the functions of interest in the infinitesimal neighbourhoods of the relevant Matsubara energies/frequecies thus lost cannot be retrieved,\footnote{Note that even analytically there is no unique way of reconstructing the function in Eq.\,(\protect\ref{e67}) [(\protect\ref{e76})] from that in Eq.\,(\protect\ref{e67x2}) [(\protect\ref{e89b})]. This observation is not at variance with that by Baym and Mermin \protect\cite{BM61}, which concerns the analytic continuation of the thermal Green function from the set of the Matsubara energies into the complex $z$-plane, not the other way around, the subject of the discussion here.} unless one repeats the calculations at temperatures in a close neighbourhood of the temperature of interest\,\footnote{This task places a greater demand on the absolute accuracy of the calculations than may be practically feasible. It is conceivable that the problem discussed here can be bypassed through the explicit calculation of the values of the derivative of the relevant thermal correlation function, in addition to its values, at the relevant Matsubara energies/frequencies. While this approach appears theoretically reasonable at first glance, it is at present not clear to us how it can be implemented in practice. In this connection, note that for a correction to be non-trivial a divergent contribution must compete against a vanishing one, whereby the question arises as to how numerically to deal with these contributions.} and thus deduces the relevant information from the values of the said functions at displaced Matsubara energies/frequencies.

In the light of the significance of the \textsl{positive sequences}\,\footnote{For definition, see Eq.\,(\protect\ref{e4s}).} of asymptotic coefficients $\{G_{\sigma;\infty_j}(\bm{k})\| j\}$ and $\{\Sigma_{\sigma;\infty_j}(\bm{k})\| j\}$, Eqs\,(\ref{e4n}) and (\ref{e7b}), in the context of the many-body perturbation expansion of the self-energy, \S\,\ref{sec.3.2}, in \S\,\ref{sec.5} we introduce a general method of summing the perturbational terms of the perturbation series corresponding to $\t{G}_{\sigma}(\bm{k};z)$ that takes explicit account of the coefficients $\{G_{\sigma;\infty_j}(\bm{k})\| j\}$ and is especially suited for the construction of the single-particle spectral function $A_{\sigma}(\bm{k};\varepsilon)$ corresponding to $\t{G}_{\sigma}(\bm{k};z)$. This spectral function is strictly positive semi-definite and can thus be \textsl{essentially uniquely}\,\footnote{This notion is specified in the paragraph following Eq.\,(\protect\ref{e6e9}), p.\,\protect\pageref{Followingtheabove}.} associated with the measure $\upgamma_{\sigma}(\bm{k};\varepsilon)$, or $\upeta_{\sigma}(\bm{k};u)$, encountered in the Riesz-Herglotz representation of $\t{G}_{\sigma}(\bm{k};z)$, Eqs\,(\ref{e4g}), (\ref{e4j}), (\ref{e4k}), and (\ref{e4l}). Importantly, the proposed summation approach is structurally weakly dependent on the coupling constant of the two-body interaction potential, and thus in the specific case of the conventional Hubbard Hamiltonian described in terms of an on-site interaction energy parameter $U$, Eq.\,(\ref{ex01bx}), is weakly dependent on the magnitude of $U$. In the extreme case of the `Hubbard atom', \S\,\ref{sec3.b}, the terms of the series in the proposed summation approach are \textsl{independent} of $U$, Eq.\,(\ref{e107}). The proposed summation approach can be directly applied to the perturbation series expansions of the self-energy, \S\,\ref{sec.5},\footnote{See footnote \raisebox{-1.0ex}{\normalsize{\protect\footref{notey}}} on p.\,\protect\pageref{InDealingWith}.} and the polarisation function, appendix \ref{sa}.\footnote{See footnote \raisebox{-1.0ex}{\normalsize{\protect\footref{notei1}}} on p.\,\protect\pageref{SimilarToChi}.} \refstepcounter{dummy}\label{WeShouldEmphasise} We should emphasise that from the perspective of avoiding unphysical functions, the representations in Eqs\,(\ref{e4j}) and (\ref{e20d}), in terms of the measures $\upgamma_{\sigma}(\bm{k};\varepsilon)$ and $\upsigma_{\sigma}(\bm{k};\varepsilon)$, respectively, are as good as the Riesz-Herglotz representations in Eqs\,(\ref{e4l}) and (\ref{e20g}), in terms of the measures $\upeta_{\sigma}(\bm{k};u)$ and $\uptau_{\sigma}(\bm{k};u)$, respectively. \emph{Therefore, in this publication the term `Riesz-Herglotz representation' equally applies to the representations in Eqs\,(\ref{e4j}) and (\ref{e20d}).}

The perturbation series expansion of the self-energy considered in the main body of this publication, Eq.\,(\ref{e4}), is in terms of the \textsl{bare} two-body interaction potential. For systems defined on unbounded continuum subsets of $\mathds{R}^d$ and/or in terms of long-range two-body interaction potentials, this expansion is not viable to arbitrary order. For these systems the perturbation series must be in terms of the dynamically \textsl{screened} interaction potential \cite{BF19}. With this in mind, in appendix \ref{sa} we consider such a series, Eq.\,(\ref{ea1}), and focus on some essential elements underlying the calculation of the said screened interaction potential. This leads us to considerations regarding the perturbation series expansion of the polarisation function $\t{P}(\bm{k};z)$, which along the same lines as described in \S\,\ref{sec.3.2} can be demonstrated to be uniformly convergent for almost all $\bm{k}$ and $z$ in the case of the uniform GSs (ESs) of Hubbard-like models. We establish the details relevant to this demonstration (such as the boundedness of the elements of the sequence $\{P_{\infty_j}(\bm{k})\| j\}$ for any finite value of $j$, Eqs\,(\ref{ea53}) and (\ref{ea61})) in appendix \ref{sa}. Appendix \ref{sa} also contains some details and exact results that should prove useful in contexts other than the context of the present publication. Amongst others, in appendix \ref{sa} we establish that $-\t{P}(\bm{k};z)$, while analytic in the region $\im[z]\not=0$ of the complex $z$-plane, is unlike $-\t{G}_{\sigma}(\bm{k};z)$ and $-\t{\Sigma}_{\sigma}(\bm{k};z)$ \textsl{not} a Nevanlinna function of $z$, Eq.\,(\ref{ea64}).

In appendix \ref{sab} we deal with the classical moment problem in relation to the functions $\t{G}_{\sigma}(\bm{k};z)$ and $\t{\Sigma}_{\sigma}(\bm{k};z)$. The details in this appendix are arranged and presented in such a way as to make the considerations of in particular \S\,\ref{sec.3.2} relatively short and light. The equality in Eq.\,(\ref{e4p}) plays a key role in our considerations throughout this publication, for two main reasons. Firstly, the expectation value on the RHS of this equation being with respect to the \textsl{interacting} $N$-particle GS of the system, it provides a link, via the equalities of the kind presented in Eqs\,(\ref{e7c}), (\ref{e7d}), and (\ref{e7e}) \cite{Note11}, between the elements of the sequence $\{G_{\sigma;\infty_j}(\bm{k})\| j\}$ and the contributions of the perturbation series expansion of the self-energy in terms of \textsl{skeleton} self-energy diagrams and the \textsl{interacting} one-particle Green functions $\{\t{G}_{\sigma}(\bm{k};z)\| \sigma\}$ (see \S\,\ref{sec.3.2.1}). Secondly, it forms a basis, via the equality in Eq.\,(\ref{e4q}), for demonstrating that for Hubbard-like models the asymptotic coefficient $G_{\sigma;\infty_j}(\bm{k})$ is bounded for arbitrary finite values of $j$, $j\in \mathds{N}$, implying in turn that for these models the asymptotic coefficient $\Sigma_{\sigma;\infty_j}(\bm{k})$ is similarly bounded for arbitrary finite values of $j$. In view of Eqs\,(\ref{e4o}) and (\ref{e7i}), the finite values of $G_{\sigma;\infty_j}(\bm{k})$ and $\Sigma_{\sigma;\infty_j}(\bm{k})$ for arbitrary finite values of $j$ in the case of Hubbard-like models imply that the relevant spectral functions $A_{\sigma}(\bm{k};\varepsilon)$ and $B_{\sigma}(\bm{k};\varepsilon)$, respectively, Eqs\,(\ref{e4d}) and (\ref{e20b}), decay faster than any finite power of $1/\varepsilon$ for $\vert\varepsilon\vert\to \infty$. This in turn establishes that the classical moment problems associated with the sequences $\{G_{\sigma;\infty_j}(\bm{k}) \| j\}$ and $\{\Sigma_{\sigma;\infty_j}(\bm{k})\| j\}$ are \textsl{determinate}, Eqs\,(\ref{e4oh}) and (\ref{e7m}).\footnote{As we discuss in appendix \ref{sab} (see in particular p.\,\protect\pageref{FollowingAkhiezer}), the fact that the Jacobi matrix $\mathscr{J}^{\protect\X{\protect\t{\Sigma}_{\sigma}}}$ associated with the self-energy $\protect\t{\Sigma}_{\sigma}(\bm{k};z)$ is the \textsl{$1$st abbreviated matrix} in relation to the Jacobi matrix $\mathscr{J}^{\protect\X{\protect\t{G}_{\sigma}}}$ associated with the one-particle Green function $\protect\t{G}_{\sigma}(\bm{k};z)$, implies that the moment problem associated with $\{G_{\sigma;\infty_j}(\bm{k}) \| j\}$ being determinate leads to the immediate conclusion that the moment problem associated with $\{\Sigma_{\sigma;\infty_j}(\bm{k})\| j\}$ is similarly determinate. This result can be directly surmised from a comparison of the expressions in Eqs\,(\protect\ref{e5kc}) and (\protect\ref{e5kd}).} The fact that for Hubbard-like models the sequence $\{\Sigma_{\sigma;\infty_j}(\bm{k})\| j\}$ is determinate plays a significant role in the considerations of \S\,\ref{sec.3.2}.

The locality of the `Hubbard atom', \S\,\ref{sec3.b}, giving rise to some anomalous properties, referred to above (such as the violation of the equality in Eq.\,(\ref{e7h})), in appendix \ref{sd} we consider the self-energy $\t{\Sigma}(z)$ on a Bethe lattice in infinite dimensions as calculated within the framework of the dynamical mean-field theory, DMFT. Here we show that $-\t{\Sigma}(z)$ is generally \textsl{not} a Nevanlinna function of $z$, even though by construction the negative of the underlying one-particle Green function $\t{G}(z)$ is invariably a Nevanlinna function of $z$.\footnote{See footnote \raisebox{-1.0ex}{\normalsize{\protect\footref{notee1}}} on p.\,\protect\pageref{ThisObservation}.}

Appendix \ref{sacx} is dedicated to a number of subjects related to the many-body perturbation series expansion of the self-energy pertaining to normal states of interacting fermion systems, in particular in terms of skeleton self-energy diagrams and the interacting one-particle Green functions $\{\t{G}_{\sigma}\|\sigma\}$. Some sections of this appendix are general, applicable to arbitrary systems of fermions interacting through an unrestricted two-body potential, and some specific, dealing with uniform systems, which in principle support GSs of lower symmetry. In this appendix we consider three separate but equivalent representations of the single-band Hubbard Hamiltonian for spin-$\tfrac{1}{2}$ particles, denoted by $\h{\mathcal{H}}$, $\hspace{0.28cm}\h{\hspace{-0.28cm}\mathpzc{H}}$, and $\h{\mathsf{H}} $ (Eqs\,(\ref{ex01bx}), (\ref{ex01f}), and (\ref{ex01k}), respectively), which however are \textsl{not} equivalent from the perspective of diagrammatic perturbation expansion, although this inequivalence is rather formal than actual insofar as the latter two representations are concerned.\footnote{The general treatment of the Hubbard Hamiltonian for spin-$\tfrac{1}{2}$ particles in Ref.\,\protect\citen{BF19} is based on the representation $\protect\h{\mathsf{H}}$ (\emph{cf.} Eqs\,(2.64) and (2.70) herein). The choice has been necessary in order to have the Hamiltonian in a form satisfying the technical needs of the general diagram-free perturbation-expansion formalism advanced in Ref.\,\protect\citen{BF19}. The simplifications associated with the representations $\protect\h{\mathcal{H}}$ and $\hspace{0.28cm}\protect\h{\hspace{-0.28cm}\mathpzc{H}}$ are effected in \S\,2.2.7 and appendix D of Ref.\,\protect\citen{BF19}.} In this appendix we also introduce a formal representation of Feynman diagrams in terms of pairs of integers, \S\,\ref{sx1}, that lends itself for use in symbolic computations\,\footnote{That is, computations that do not explicitly rely on floating-point arithmetic operations.} and present all skeleton self-energy diagrams of orders $2$, $3$, and $4$ in this representation, \S\S\,\ref{sd11}, \ref{sd12}, and \ref{sd13}. We also provide the relevant information linking these diagrams with their representations in terms of the cycle decompositions of the elements of the symmetric groups $S_{2\nu}$, $\nu = 2,3,4$, to which they correspond according to the formalism of diagram generation/representation introduced in Ref.\,\citen{BF19}.

In \S\,\ref{sd3} we introduce a method of symbolic computation for investigating the algebraic equivalence (up to \textsl{determinate} multiplicative constants as regards their algebraic contributions) of the proper self-energy diagrams that are topologically inequivalent. We also provide the programs, written in the Mathematica$^{\X{\circledR}}$ programming language,\footnote{The Mathematica programming language is also referred to as the \textsl{Wolfram Language} \protect\cite{SW16}. We shall hereafter generally suppress the mark $\protect\X{\circledR}$.} relevant to the task. The main program, named \texttt{Equiv}, p.\,\pageref{Equiv}, also outputs the data necessary for the calculation of the above-mentioned determinate constants for equivalent diagrams. Two input parameters of this program (\texttt{phs} and \texttt{all}) enable one to specify whether (\texttt{phs $\not=$ 0}) or not (\texttt{phs = 0}) the underlying GS (ESs) is particle-hole (p-h) symmetric, and whether (\texttt{all $\not=$ 0}) or not (\texttt{all = 0}) all proper self-energy diagrams equivalent to a given proper self-energy diagram of specified order (\texttt{nu}) are to be determined.  For transparency of presentation and programming, in \S\,\ref{sd3} we focus on systems in which the two-body interaction potential is local (as in the Hubbard model) and spin-independent, and further on the ground and ensemble of states (ESs) that are spin-unpolarised. We in addition indicate how these limitations can be relaxed thereby to deal with more general systems and GSs (ESs). In \S\,\ref{sd4} we present the algebraic expressions for all skeleton self-energy diagrams specified in \S\S\,\ref{sd11}, \ref{sd12}, and \ref{sd13} for the half-filled GS of the `Hubbard atom'. The contributions of some non-skeleton proper self-energy diagrams are also presented in this section -- specifically, in the last part of \S\,\ref{sec.d53}. The results in \S\,\ref{sd4} have been necessary for establishing the breakdown of the equality in Eq.\,(\ref{e7h}) in the local / atomic limit.

We present the Mathematica programs underlying those algebraic expressions presented in this publication that can be of general use as numbered references (p.\,\pageref{Refs} onwards).\footnote{See Refs\,\protect\citen{Note11}, \protect\citen{Note18}, \protect\citen{Note13}, \protect\citen{Note15}, \protect\citen{Note19}, \protect\citen{Note20}, \protect\citen{Note10}, \protect\citen{Note16}, \protect\citen{Note14}, \protect\citen{Note17}, and \protect\citen{Note12}. In particular the programs in Refs\,\protect\citen{Note19} and \protect\citen{Note12} (along with an extension of the programs in the latter reference [cited in \S\,\ref{sabx}] that we have included in the Mathematica notebook to which we refer below) can serve in the context of a practical introduction to the classical moment problem. These programs may also be used as teaching aids in academic courses on the latter problem.} A collection of all Mathematica programs relevant to the present publication, including those reproduced in \S\,\ref{sd3}, will be published alongside the present text on \textsf{arXiv} (in the section `Ancillary files') as a notebook.\footnote{This notebook will be in due course published also at \href{https://www.notebookarchive.org/search?q=Behnam Farid}{\textsf{Wolfram Notebook Archive}}.}

\refstepcounter{dummyX}
\section{Acknowledgement}
\phantomsection
\label{s7}
We have drawn the Feynman diagrams for this publication with the aid of the program \textsf{JaxoDraw}.\footnote{\href{http://jaxodraw.sourceforge.net/}{\textsf{JaxoDraw:\,\textsl{Feynman Diagrams with Java}}}.}

\begin{appendix}
\refstepcounter{dummyX}
\section{The screened interaction function \texorpdfstring{$\protect\t{W}(\bm{k};z)$}{} and the regularized
series expansion for \texorpdfstring{$\protect\t{\Sigma}_{\sigma}(\bm{k};z)$}{}}
\phantomsection
\label{sa}
In the cases where the uniform system of interest is defined on an unbounded continuum subset of $\mathds{R}^d$ and/or in terms of a long-range two-body interaction potential, the infinite sequence $\{ \t{\Sigma}_{\sigma}^{\X{(\nu)}}(\bm{k};z)\| \nu \}$, introduced in the main text, \S\,\ref{sec2.b}, consists of infinite number of unbounded terms. A particular systematic method of regularizing the terms of this sequence, resulting in the infinite sequence $\{ \t{\Sigma}_{\sigma}^{\X{\prime\hspace{0.4pt} (\nu)}}(\bm{k};z)\| \nu \}$ of functions that are bounded \ae,\footnote{Appendix \protect\ref{sae}.} consists of restricting the calculations to skeleton self-energy diagrams without polarization insertions.\footnote{An exposition of the underlying considerations, relying on elementary power-counting \protect\cite{JZJ02}, is presented in Ch.\,10 of Ref.\,\protect\citen{RDM92}.}\footnote{A \textsl{polarization insertion} is a connected subdiagram that can be detached from self-energy diagrams on severing two interaction lines [p.\,110 in Ref.\,\protect\citen{FW03}] \cite{JH57}. One may refer to a (self-energy) diagram without polarization insertions as \textsl{two-interaction irreducible} (2II) \protect\cite{BF19}, in analogy with the skeleton self-energy diagrams that are alternatively referred to as \textsl{two-particle irreducible} (2PI).} In exchange for this, the interaction lines, initially representing the static spin-independent bare two-body interaction potential $\b{v}(\bm{k})$,\footnote{In the case of the conventional Hubbard Hamiltonian, the function $\protect\b{v}(\bm{k})$ is identified with the constant $\Omega U/N_{\textsc{s}}$, Eq.\,(\protect\ref{ex03}), where $\Omega$ denotes the volume of the region in $\mathds{R}^d$ into which the underlying lattice sites $\{\bm{R}_i \| i = 1,2,\dots,N_{\textsc{s}}\}$ are embedded.} represent the dynamically-screened two-body interaction potential $\t{W}(\bm{k};z)$ \cite{JH57,BF19} in the reduced set of skeleton self-energy diagrams.\footnote{The dynamically-screened interaction potential $W$ is dealt with in some detail in \S\,3 of Ref.\,\protect\citen{BF19}. In particular, the $W$ corresponding to the Hubbard Hamiltonian is the focus of special attention in the mentioned section. See in particular Eqs\,(3.26) and (3.28) of Ref.\,\protect\citen{BF19}.} Thus the $\nu$ in $\t{\Sigma}_{\sigma}^{\X{\prime\hspace{0.4pt} (\nu)}}(\bm{k};z)$ counts the number of the dynamically-screened interaction lines in the underlying polarization-insertion-free skeleton self-energy diagrams. These diagrams are as in the case of $\t{\Sigma}_{\sigma}^{\X{(\nu)}}(\bm{k};z)$ determined in terms of the exact one-particle Green functions $\{G_{\sigma}(\bm{k};z) \|\sigma\}$. One has (\emph{cf.} Eq.\,(\ref{e4}))
\begin{equation}\label{ea1}
\t{\Sigma}_{\sigma}(\bm{k};z) = \sum_{\nu=1}^{\infty} \t{\Sigma}_{\sigma}^{\X{\prime\hspace{0.4pt} (\nu)}}(\bm{k};z).
\end{equation}
Following the notational convention in Ref.\,\citen{BF19} [\S\,1.2 herein], the self-energy as described by the perturbation series in Eq.\,(\ref{ea1}) is more explicitly denoted by $\t{\Sigma}_{\X{11};\sigma}(\bm{k};z)$, or $\t{\Sigma}_{\X{11};\sigma}(\bm{k};z;[W,G])$, to be contrasted with the self-energy in Eq.\,(\ref{e4}) that is more explicitly denoted by $\t{\Sigma}_{\X{01};\sigma}(\bm{k};z)$, or $\t{\Sigma}_{\X{01};\sigma}(\bm{k};z;[v,G])$.\,\footnote{Accordingly, $\protect\t{\Sigma}_{\sigma}^{\protect\X{(\nu)}}(\bm{k};z)\equiv \protect\t{\Sigma}_{\protect\X{01};\sigma}^{\protect\X{(\nu)}}(\bm{k};z)$, Eq.\,(\protect\ref{e4}), and $\protect\t{\Sigma}_{\sigma}^{\protect\X{\prime\hspace{0.4pt} (\nu)}}(\bm{k};z) \equiv \protect\t{\Sigma}_{\protect\X{11};\sigma}^{\protect\X{(\nu)}}(\bm{k};z)$, Eq.\,(\protect\ref{ea1}).} Because of the dependence of the function $\t{W}(\bm{k};z)$ on $z$, the analysis of the analytic properties of $\t{\Sigma}_{\sigma}^{\X{\prime\hspace{0.4pt} (\nu)}}(\bm{k};z)$, $\forall\nu$, is more involved than that of $\t{\Sigma}_{\sigma}^{\X{(\nu)}}(\bm{k};z)$ as presented in Ref.\,\citen{JML61}.\footnote{Luttinger's \protect\cite{JML61} relevant observations are summarised in \S\,5.3.1 of Ref.\,\protect\citen{BF07}.}

The determination of the function $\t{W}(\bm{k};z)$ is achieved by employing the following expression [p.\,110 in Ref.\,\citen{FW03}]:\,\footnote{See also in particular \S\,3 of Ref.\,\protect\citen{BF19}.}
\begin{equation}\label{ea2}
\h{W}(z) = \h{v} + \h{v}\hspace{0.6pt} \h{\chi}(z)\hspace{0.6pt} \h{v} \equiv \h{v} + \hspace{6.0pt}\h{\hspace{-6.0pt}\mathpzc{W}}(z),
\end{equation}
where $\h{W}(z)$ and $\h{v}$ denote the one-particle operators associated with respectively $\t{W}(\bm{k};z)$ and $\b{v}(\bm{k})$,\footnote{Compare with the one-particle operators $\protect\h{G}_{\sigma}(z)$ and $\protect\h{\Sigma}_{\sigma}(z)$ associated with respectively $\protect\t{G}_{\sigma}(\bm{k};z)$ and $\protect\t{\Sigma}_{\sigma}(\bm{k};z)$. See footnote \raisebox{-1.0ex}{\normalsize{\protect\footref{notea1}}} on p.\,\pageref{ItProvesAlso}.} and $\h{\chi}(z) \equiv \sum_{\sigma,\sigma'} \h{\chi}_{\sigma,\sigma'}(z)$ denotes the one-particle operator associated with the dynamical density-density correlation function $\t{\chi}(\bm{k};z)$, \S\,\ref{sa.1}.\refstepcounter{dummy}\label{TheAssumption}\footnote{The assumption here that the spin-$\mathsf{s}$ particles under consideration interact through the same bare two-body interaction potential $\b{v}(\bm{k})$, irrespective of their spin indices, and the fact that the one-particle Green matrices $\mathbb{G}_{\protect\X{0}}$ and $\mathbb{G}$ are diagonal in the spin space (see footnotes \raisebox{-1.0ex}{\normalsize{\protect\footref{notef1}}} and \raisebox{-1.0ex}{\normalsize{\protect\footref{noteg1}}} on pp.\,\protect\pageref{ForSpin} and \protect\pageref{WithRefTo}), are the reasons for explicitly dealing with $\protect\h{\chi}(z)$ and $\protect\h{P}(z)$, instead of respectively $\protect\h{\chi}_{\sigma,\sigma'}(z)$ and $\protect\h{P}_{\sigma,\sigma'}(z)$. In Ref.\,\protect\citen{BF19} we consider a more general two-body interaction potential (see in particular \S\,3.2 herein). See the remarks in footnote \raisebox{-1.0ex}{\normalsize{\protect\footref{noted1}}} on p.\,\protect\pageref{WeNoteThat}. \label{notep}} This operator is determined according to\,\footnote{The single-particle operator $\protect\h{\chi}$ coincides with the operator $\protect\h{P}^{\star}$ in Eq.\,(3.17) of Ref.\,\protect\citen{BF19}.}
\begin{equation}\label{ea3}
\h{\chi}(z) = \big(\h{I} - \h{P}(z)\hspace{0.6pt}\h{v}\big)^{-1} \h{P}(z) = \h{P}(z) \big(\h{I} - \h{v}\hspace{0.6pt}\h{P}(z)\big)^{-1} \equiv \h{P}(z)\hspace{0.6pt} \h{\upepsilon}^{-1}(z),
\end{equation}
where $\h{I}$ is the identity operator in the single-particle Hilbert space of the problem at hand, and $\h{P}(z)$ the single-particle operator associated with the \textsl{proper} polarization function [p.\,110 in Ref.\,\citen{FW03}] \cite{BF19}
\begin{equation}\label{ea3a}
\t{P}(\bm{k};z) \equiv \sum_{\sigma,\sigma'} \t{P}_{\sigma,\sigma'}(\bm{k};z)
\end{equation}
described in terms of proper polarisation diagrams, Eq.\,(\ref{ea6a}) and Fig.\,\ref{f12}, p.\,\pageref{Two2ndO}, below. Further, the operator $\h{\upepsilon}^{-1}(z)$ stands for the inverse of the single-particle dielectric operator
\begin{equation}\label{ea3b}
\h{\upepsilon}(z) \doteq \h{I} - \h{v}\hspace{0.6pt}\h{P}(z)
\end{equation}
associated with the dielectric function.\footnote{See \S\,3.1 in Ref.\,\protect\citen{BF19}.} For completeness, a polarization diagram is a connected diagram with two external vertices to (from) each one of which one internal Green-function line enters (leaves). A \textsl{proper} polarization diagram\,\footnote{Or a \textsl{one-interaction irreducible} (1II) diagram, in analogy with a \textsl{one-particle irreducible} (1P1) diagram \protect\cite{BF19}.} is one that does not fall into two separate diagrams by removing an interaction line, Fig.\,\ref{f12}, p.\,\pageref{Two2ndO}, below.

From the perspective of the considerations of this appendix, the proper polarization function $\t{P}(\bm{k};z)$ has two significant properties. Firstly, $\t{P}(\bm{k};z)$ can be expanded in terms of \textsl{skeleton} proper polarization diagrams, which are proper polarization diagrams without self-energy insertions, Eq.\,(\ref{ea6a}) below.\footnote{For various perturbation series expansions of the polarisation function without recourse to diagrams, consult \S\,3 of Ref.\,\protect\citen{BF19}.} These diagrams are thus expressed in terms of $\{\t{G}_{\sigma}(\bm{k};z)\| \sigma\}$ and $\b{v}(\bm{k})$. Secondly, the function $\t{P}(\bm{k};z)$ is analytic over the entire domain $\im[z] \not= 0$ of the $z$-plane, \S\,\ref{sa.1}. As we show in \S\,\ref{sa.1} below, unlike $-\t{\Sigma}_{\sigma}(\bm{k};z)$ (as well as $-\t{G}_{\sigma}(\bm{k};z)$), $-\t{P}(\bm{k};z)$ is however \textsl{not} a Nevanlinna function [App.\,C in Ref.\,\citen{HMN72}] [Ch.\,3 in Ref.\,\citen{NIA65}] of $z$. Instead, one has the property in Eq.\,(\ref{ea64}) below. The same applies to the functions $\t{\b{\chi}}(\bm{k};z)$,\footnote{We use the \textsl{bar} over $\chi$ in order to distinguish this function from its real-space counterpart (\emph{cf.} Eq.\,(\protect\ref{ea36a}) below).} Eq.\,(\ref{ea40e}), $\hspace{6.0pt}\t{\hspace{-6.0pt}\mathpzc{W}}(\bm{k};z)$,\footnote{With reference to Eq.\,(\protect\ref{ea2}), and assuming the two-body interaction potential $v_{\sigma,\sigma'}(\bm{r},\bm{r}')$, Eq.\,(\protect\ref{ea34k}), to be spin-independent and a function of $\bm{r} - \bm{r}'$ (\emph{i.e.}, assuming $v_{\sigma,\sigma'}(\bm{r},\bm{r}') \equiv v(\bm{r}-\bm{r}')$), one has $\hspace{6.0pt}\protect\t{\hspace{-6.0pt}\mathpzc{W}}(\bm{k};z) = (\protect\b{v}(\bm{k}))^2\hspace{0.6pt} \protect\t{\protect\b{\chi}}(\bm{k};z)$, where $\protect\t{\protect\b{\chi}}(\bm{k};z)$ is presented in Eq.\,(\protect\ref{ea40a}) below.} and $\t{W}(\bm{k};z)$. With reference to Eq.\,(\ref{e1}), we note that the `physical' function $P(\bm{k};\varepsilon)$, $\varepsilon \in\mathds{R}$, is related to $\t{P}(\bm{k};z)$, $z\in \mathds{C}$, according to the following relationship:
\begin{equation}\label{ea5}
P(\bm{k};\varepsilon) = \lim_{\eta\downarrow 0} \t{P}(\bm{k};\varepsilon \pm \ii\eta)\;\, \text{for}\;\, \varepsilon \gtrless 0.
\end{equation}
The $0$ on the RHS of Eq.\,(\ref{ea5}) is to be contrasted with the $\mu$ on the RHS of Eq.\,(\ref{e1}). Similar equalities as that in Eq.\,(\ref{ea5}) relate $\t{\b{\chi}}(\bm{k};z)$ to $\b{\chi}(\bm{k};\varepsilon)$, $\t{W}(\bm{k};z)$ to $W(\bm{k};\varepsilon)$, and $\hspace{6.0pt}\t{\hspace{-6.0pt}\mathpzc{W}}(\bm{k};z)$ to $\hspace{6.0pt}\hspace{-6.0pt}\mathpzc{W}(\bm{k};\varepsilon)$.

Expressing the perturbations series for $\t{P}(\bm{k};z)$ in terms of the bare two-body interaction potential $\b{v}(\bm{k})$ and the interacting one-particle Green functions $\{\t{G}_{\sigma}(\bm{k};z)\| \sigma\}$ as\,\footnote{In Eqs\,(\protect\ref{ea6a}) and (\protect\ref{ea7}), $\protect\t{P}^{\protect\X{(\nu)}}(\bm{k};z) \equiv \sum_{\sigma,\sigma'} \protect\t{P}_{\sigma,\sigma'}^{\protect\X{(\nu)}}(\bm{k};z)$, and $\protect\t{P}^{\protect\X{\prime\hspace{0.4pt}(\nu)}}(\bm{k};z) \equiv \sum_{\sigma,\sigma'} \protect\t{P}_{\sigma,\sigma'}^{\protect\X{\prime\hspace{0.4pt}(\nu)}}(\bm{k};z)$.} (\emph{cf.} Eq.\,(\ref{e4}))
\begin{equation}\label{ea6a}
\t{P}(\bm{k};z) = \sum_{\nu=0}^{\infty} \t{P}^{\X{(\nu)}}(\bm{k};z),
\end{equation}
in analogy with the expression in Eq.\,(\ref{ea1}), one has
\begin{equation}\label{ea7}
\t{P}(\bm{k};z) = \sum_{\nu=0}^{\infty} \t{P}^{\X{\prime\hspace{0.4pt}(\nu)}}(\bm{k};z),
\end{equation}
where $\t{P}^{\X{\prime\hspace{0.4pt}(\nu)}}(\bm{k};z)$ denotes the contributions of all skeleton \textsl{proper} polarization diagrams of order $\nu$ calculated in terms of $\{\t{G}_{\sigma}(\bm{k};z)\| \sigma\}$ and $\t{W}(\bm{k};z)$. Following the notational convention in Ref.\,\citen{BF19} [\S\,1.2 herein], the polarization function as described by the perturbation series in Eq.\,(\ref{ea6a}) is more explicitly denoted by $\t{P}_{\X{01}}(\bm{k};z)$, to be contrasted with the polarization function in Eq.\,(\ref{ea7}) that is more explicitly denoted by $\t{P}_{\X{11}}(\bm{k};z)$.\,\footnote{Accordingly, $\protect\t{P}^{\protect\X{(\nu)}}(\bm{k};z)\equiv \protect\t{P}_{\protect\X{01}}^{\protect\X{(\nu)}}(\bm{k};z)$, Eq.\,(\protect\ref{ea6a}), and $\protect\t{P}^{\protect\X{\prime\hspace{0.4pt} (\nu)}}(\bm{k};z) \equiv \protect\t{P}_{\protect\X{11}}^{\protect\X{\prime\hspace{0.4pt}(\nu)}}(\bm{k};z)$, Eq.\,(\protect\ref{ea7}).} All elements of the infinite sequence $\{\t{P}^{\X{\prime\hspace{0.4pt}(\nu)}}(\bm{k};z) \| \nu\}$ are bounded for almost all relevant $\bm{k}$ and $z$.

Although $-\t{P}(\bm{k};z)$ is \textsl{not} a Nevanlinna function of $z$, Eq.\,(\ref{ea64}) below, the fact that $\t{P}(\bm{k};z)$ and  $\t{P}^{\X{(\nu)}}(\bm{k};z)$ are analytic functions of $z$ in the region $\im[z] \not= 0$ of the $z$-plane and that the coefficients $\{P_{\infty_j}(\bm{k})\| j\}$ of the asymptotic series expansion of $\t{P}(\bm{k};z)$ in the region $z\to\infty$, Eqs\,(\ref{ea53}) and (\ref{ea61}) below (forming a \textsl{determinate} \textsl{positive sequence}),\footnote{For these notions, consult appendix \protect\ref{sab}.} are bounded for Hubbard-like models, \S\,\ref{sa.1}, on the basis of similar considerations as in \S\,\ref{sec.3.2.1} one can demonstrate that for the uniform GSs (ESs) of these models the series in Eq.\,(\ref{ea6a}) corresponding to the exact Green functions $\{\t{G}_{\sigma}(\bm{k};z)\| \sigma\}$ is uniformly convergent for almost all relevant $\bm{k}$ and $z$ (see in particular the discussions centred on Eq.\,(\ref{e7wi})).

In considering the perturbation series expansion in Eq.\,(\ref{ea7}), one has to calculate the screened interaction function $\t{W}(\bm{k};z)$, underlying the calculation of the elements of the sequence $\{\t{P}^{\X{\prime\hspace{0.4pt}(\nu)}}(\bm{k};z)\| \nu\}$, according to\,\footnote{With $\protect\h{W}(z) = \protect\h{\upepsilon}^{-1}(z)\hspace{0.4pt}\protect\h{v}$, from Eqs\,(\protect\ref{ea2}) and (\protect\ref{ea3}) one obtains $\protect\h{W}(z) = \protect\h{v} + \protect\h{v} \protect\h{P}(z) \protect\h{W}(z)$, whereby $\protect\h{W}(z) = \big(\h{I} -\protect\h{v} \protect\h{P}(z)\big)^{-1}\protect\h{v}$.}
\begin{equation}\label{ea8}
\t{W}(\bm{k};z) = \big(1 - \b{v}(\bm{k}) \t{P}(\bm{k};z)\big)^{-1} \b{v}(\bm{k}) \equiv \frac{\b{v}(\bm{k})}{1 - \b{v}(\bm{k}) \t{P}(\bm{k};z)},
\end{equation}
which is the counterpart of the Dyson equation for the one-particle Green function.\footnote{With $\protect\t{W}(\bm{k};z)$, $\b{v}(\bm{k})$, $\protect\t{P}(\bm{k};z)$ taking the place of respectively $\protect\t{G}_{\sigma}(\bm{k};z)$, $\protect\t{G}_{\protect\X{0};\sigma}(\bm{k};z)$, $\protect\t{\Sigma}_{\sigma}(\bm{k};z)$.}

\refstepcounter{dummyX}
\subsection{Details regarding the screening functions}
\phantomsection
\label{sa.1}
Detailed considerations regarding the \textsl{proper} polarisation function $P(1,2)$, where $j = \bm{r}_jt_j\sigma_j$, with $j \in \{1,2\}$, are to be based on considerations regarding the \textsl{improper} polarisation function\,\footnote{This follows from the fact that $\chi(1,2)$ is the density-density correlation function, Eq.\,(\protect\ref{ea12}) below, defined according to $\chi(1,2) \doteq -\protect\ii\hspace{0.6pt} \delta G(1,1^+)/\delta \varphi(2)\vert_{\varphi\equiv 0}$, where $\varphi(2)$ is the scalar classical field coupled \textsl{linearly} to the number-density operator $\protect\h{\psi}^{\dag}(2)\protect\h{\psi}(2)$ in the many-body Hamiltonian $\protect\h{H}$ of the system under consideration [Ch.\,5, \S\,13, p.\,172, in Ref.\,\protect\citen{FW03}]. Here $1^+ = \bm{r}_1^{\phantom{+}}t_1^+\sigma_1^{\phantom{+}}$, in which $t_1^+ \doteq t + 0^+$.} $P^{\star}(1,2)\equiv \chi(1,2)$, for which one has\,\footnote{See Eqs\,(3.14) and (3.17) of Ref.\,\protect\citen{BF19}.}
\begin{equation}\label{ea10}
\chi(1,2) = \frac{\ii}{\hbar} \big\{G_2(1,2;1^+,2^+) - G(1,1^+) G(2,2^+)\big\},
\end{equation}
where $G_2$ denotes the two-particle Green function. From the definitions of the functions on the RHS of Eq.\,(\ref{ea10}), making use of
\begin{equation}\label{ea11}
\langle \h{A} \h{B} \rangle -\langle\h{A}\rangle \langle\h{B}\rangle \equiv \langle (\h{A}-\langle\h{A}\rangle)(\h{B}-\langle\h{B}\rangle)\rangle,
\end{equation}
one obtains
\begin{equation}\label{ea12}
\chi(1,2) = -\frac{\ii}{\hbar}\, \langle\Psi_{N;0}\vert \mathcal{T}\big\{\h{n}_{\textsc{h}}'(1) \h{n}_{\textsc{h}}'(2)\big\}\vert\Psi_{N;0}\rangle,
\end{equation}
where $\vert\Psi_{N;0}\rangle$ denotes the normalised $N$-particle GS of the system under consideration,\footnote{That is, $\vert\Psi_{N;0}\rangle$ it is an eigenstate of the total number operator $\h{N}$ corresponding to eigenvalue $N$.} which we in addition assume to be an eigenstate of the $z$-component of the total spin operator,\footnote{For relevant details, consult \S\,2.2.2 of Ref.\,\protect\citen{BF19}. See also footnote \raisebox{-1.0ex}{\normalsize{\protect\footref{notem}}} on p.\,\protect\pageref{notem}.} $\mathcal{T}$ the fermionic time-ordering operator,\footnote{One however has $\mathcal{T}\big\{\protect\h{n}_{\textsc{h}}'(1) \protect\h{n}_{\textsc{h}}'(2)\big\} = \mathcal{T}\big\{\protect\h{n}_{\textsc{h}}'(2) \protect\h{n}_{\textsc{h}}'(1)\big\}$, on account of $\h{n}_{\textsc{h}}(j)$ consisting of a product of an \textsl{even} number of field operators. One thus has $\chi(1,2) \equiv \chi(2,1)$. Notably, the minus sign in Eq.\,(\protect\ref{ea18}) below arises from the $\Theta(t_2-t_1)$ in $\mathcal{T}\big\{\h{n}_{\textsc{h}}'(1) \h{n}_{\textsc{h}}'(2)\big\} \equiv \Theta(t_1-t_2)\hspace{0.6pt}\h{n}_{\textsc{h}}'(1) \h{n}_{\textsc{h}}'(2) + \Theta(t_2-t_1)\hspace{0.6pt} \h{n}_{\textsc{h}}'(2) \h{n}_{\textsc{h}}'(1)$.} and
\begin{equation}\label{eq13}
\h{n}_{\textsc{h}}'(j) \doteq \h{n}_{\textsc{h}}(j) - n_{\textsc{h}}(j)\hspace{0.6pt}\h{1},
\end{equation}
where $\h{1}$ stands for the identity operator in the Fock space of the system under consideration,
\begin{equation}\label{ea14}
\h{n}_{\textsc{h}}(j) \doteq \h{\psi}_{\textsc{h}}^{\dag}(j) \h{\psi}_{\textsc{h}}^{\phantom{\dag}}(j),
\end{equation}
and
\begin{equation}\label{ea15}
n_{\textsc{h}}(j) \doteq \langle\Psi_{N;0}\vert\h{n}_{\textsc{h}}(j)\vert\Psi_{N;0}\rangle.
\end{equation}
In Eq.\,(\ref{ea14}), $\h{\psi}_{\textsc{h}}^{\dag}$ and $\h{\psi}_{\textsc{h}}^{\phantom{\dag}}$ are the canonical creation and annihilation field operators, respectively, in the Heisenberg picture. Since $n_{\textsc{h}}(j)$ is an \textsl{expectation value} with respect to an \textsl{eigenstate} of the Hamiltonian $\h{H}$ of the system under consideration, it is independent of $t_j$, and is equal to $\langle\Psi_{N;0}\vert\h{n}(\bm{r}_j\sigma_j)\vert\Psi_{N;0}\rangle \equiv n_{\sigma_j}(\bm{r}_j)$, the GS number density of the particles with spin index $\sigma_j$ at the position $\bm{r}_j$.

\begin{figure}[t!]
\centerline{\includegraphics[angle=0, width=0.6\textwidth]{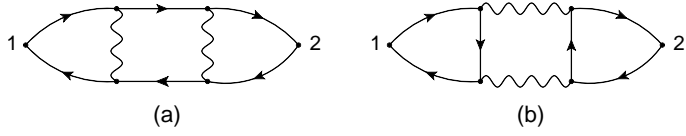}}
\caption{Two\protect\refstepcounter{dummy}\label{Two2ndO} second-order proper polarisation diagrams contributing to $P(1,2)$. Considering the many-body Hamiltonian $\protect\h{H}$ in Eq.\,(\protect\ref{ea34k}) below, the wavy lines represent the two-body potential $v_{\sigma,\sigma'}$, $\forall\sigma,\sigma'$. Assuming the one-particle Green matrix $\mathbb{G}$ corresponding to the $N$-particle GS of the above-mentioned Hamiltonian to be diagonal in the spin space, with $G_{\sigma}$ denoting the diagonal element $(\mathbb{G})_{\sigma,\sigma}$, the directed solid lines represent $G_{\sigma}$, $\forall\sigma$. With $1 \equiv \bm{r}t\sigma$ and $2\equiv \bm{r}'t'\sigma'$, for the $N$-particle GS of $\protect\h{H}$ whose corresponding $\mathbb{G}$ is diagonal in the spin space, the contribution of diagram (a) is proportional to $\delta_{\sigma,\sigma'}$ and that of diagram (b) is not. In general, the contribution of any proper polarisation diagram in which the external vertices $1$ and $2$ are linked through a contiguous set of directed solid lines is similar to that of diagram (a) proportional to $\delta_{\sigma,\sigma'}$.}
\label{f12}
\end{figure}

Expressing the Heisenberg field operators in terms of their Schr\"{o}dinger-picture counterparts,\footnote{Eq.\,(6.28), p.\,59, in Ref.\,\protect\citen{FW03}.} making use of the closure relation
\begin{equation}\label{ea16}
\sum_{s,M} \vert\Psi_{M;s}\rangle\langle\Psi_{M;s}\vert = \h{1},
\end{equation}
where $\vert\Psi_{M;s}\rangle$ denotes the $s$th normalised $M$-particle eigenstate of the many-body Hamiltonian $\h{H}$ of the system under consideration,\refstepcounter{dummy}\label{SimilarTo}\footnote{We assume that all eigenstates of $\protect\h{H}$ considered here are like $\vert\Psi_{N;0}\rangle$ the simultaneous eigenstates of the total-number operator $\protect\h{N}$ as well as of the $z$-component $\protect\h{S}^z$ of the total-spin operator $\protect\h{S}$ (\emph{e.g.} the one-particle Green matrix $\mathbb{G}$ as considered here is diagonal in the spin space as a results of $\vert\Psi_{N;0}\rangle$ being an eigenstate of $\protect\h{S}^z$). As a consequence, with $\protect\h{S}^z \vert\Psi_{N;s}\rangle = S_{N;s}^z \vert\Psi_{N;s}\rangle$, the amplitude $\varrho_{\sigma;s}(\bm{r})$ in Eq.\,(\ref{ea19}) would be identically vanishing if $S_{N;s}^z \not= S_{N;0}^z$ for $s\not= 0$, this on account of the fact that $\protect\h{n}_{\sigma}(\bm{r}) \equiv \protect\h{\psi}_{\sigma}^{\dag}(\bm{r}) \protect\h{\psi}_{\sigma}^{\phantom{\dag}}(\bm{r})$ commutes with $\protect\h{S}^z$, whereby $\protect\h{S}^z \protect\h{n}_{\sigma}(\bm{r}) \vert\Psi_{N;s}\rangle = S_{N;s}^z  \protect\h{n}_{\sigma}(\bm{r}) \vert\Psi_{N;s}\rangle$. We note that for spin-$\tfrac{1}{2}$ particles one has $\protect\h{S}^{z} = \hbar (\protect\h{N}_{\uparrow} - \protect\h{N}_{\downarrow})/2$ \protect\cite{BF19}, where $\protect\h{N}_{\sigma} = \int \mathrm{d}^dr\, \protect\h{n}_{\sigma}(\bm{r})$. Further, for the Hamiltonian $\protect\h{H}$ in Eq.\,(\protect\ref{ea34k}) below, $[\protect\h{H}, \protect\h{N}_{\sigma}]_{-} = \protect\h{0}$, $\forall\sigma$, so that $[\protect\h{H}, \protect\h{S}^z]_{-} = \protect\h{0}$. For some relevant details, consult \S\,2.2.2 of Ref.\,\protect\citen{BF19}. In dealing with the uniform GSs of uniform systems (\emph{i.e.} those described by the Hamiltonian $\protect\h{H}$ in Eq.\,(\protect\ref{ea34k}) below in which the external potential $u(\bm{r})$ is a constant, independent of $\bm{r}$) later in this appendix, we consider the states $\{\vert\Psi_{N;s}\rangle \| s \ge 0\}$ to be additionally the simultaneous eigenstates of the total momentum operator $\protect\h{\mathsf{P}}$. See footnote \raisebox{-1.0ex}{\normalsize{\protect\footref{notez}}} on p.\,\protect\pageref{WithP}. \label{notem}}\footnote{The index $s$ is a compound one \protect\cite{KP58}, and not merely an integer. By definition, we associate $s=0$ with the compound index specific to the GS under consideration, and $s>0$ with the compound indices associated with all other states. For degenerate $M$-particle GS energy, some $s > 0$ naturally correspond to the GSs in the degenerate manifold. Elsewhere in this publication we denote the states in this manifold by $\vert\Psi_{M;0_j}\rangle$, $j\in \{1,2,\dots,\upnu_{\mathrm{g}}\}$.}
\begin{equation}\label{ea16a}
\h{H} \vert\Psi_{M;s}\rangle = E_{M;s} \vert\Psi_{M;s}\rangle,
\end{equation}
from the expression in Eq.\,(\ref{ea12}), making use of\,\footnote{One has $\langle\Psi_{N;0}\vert \protect\h{n}_{\textsc{h}}(j)\vert\Psi_{M;s}\rangle = \protect\e^{-\protect\ii e_s t_j/\hbar} \langle\Psi_{N;0}\vert \protect\h{n}_{\sigma_j}(\bm{r}_j)\vert\Psi_{M;s}\rangle$, where $e_s$ is defined in Eq.\,(\protect\ref{ea20}) below. Since $\protect\h{n}_{\sigma_j}(\bm{r}_j)$ commutes with $\protect\h{N}$ (it also commutes with $\protect\h{N}_{\sigma}$), it follows that $\protect\h{n}_{\sigma_j}(\bm{r}_j)\vert\Psi_{M;s}\rangle$ is an eigenstate of $\protect\h{N}$ corresponding to eigenvalue $M$. Hereby follows the first relationship in Eq.\,(\protect\ref{ea16a1}).}
\begin{equation}\label{ea16a1}
\langle\Psi_{N;0}\vert \h{n}_{\textsc{h}}(j)\vert\Psi_{M;s}\rangle \propto \delta_{M,N}\;\;\;\text{and}\;\;\; \langle\Psi_{N;0}\vert\Psi_{M;s}\rangle = \delta_{s,0}\hspace{0.4pt} \delta_{M,N},
\end{equation}
with
\begin{equation}\label{ea16a2}
1 \equiv \bm{r}t\sigma,\;\;\; 2 \equiv \bm{r}'t'\sigma',
\end{equation}
for the time-Fourier transform of $\chi(1,2)$ one obtains the Lehmann-type representation\,\footnote{Eq.\,(\protect\ref{ea23}) implies $\protect\t{\chi}_{\sigma,\sigma'}(\bm{r},\bm{r}';-z) \equiv \protect\t{\chi}_{\sigma',\sigma}(\bm{r}',\bm{r};z)$. Thus, in order for $\protect\t{\chi}_{\sigma,\sigma'}(\bm{r},\bm{r}';z)$ to be an \textsl{even} functions of $z$ one must have  $\protect\t{\chi}_{\sigma,\sigma'}(\bm{r},\bm{r}';z) \equiv \protect\t{\chi}_{\sigma',\sigma}(\bm{r}',\bm{r};z)$. A sufficient condition for this is that presented in Eq.\,(\protect\ref{ea24b}) below.}
\begin{equation}\label{ea18}
\chi_{\sigma,\sigma'}(\bm{r},\bm{r}';\varepsilon) = \sum_{s} \Big\{\frac{\varrho_{\sigma;s}(\bm{r}) \varrho_{\sigma';s}^*(\bm{r}')}{\varepsilon - e_s + \ii 0^+} - \frac{\varrho_{\sigma;s}^*(\bm{r}) \varrho_{\sigma';s}(\bm{r}')}{\varepsilon + e_s -\ii 0^+}\Big\},\;\; \varepsilon\in\mathds{R},
\end{equation}
in which\,\footnote{With $1 \equiv \bm{r}t\sigma$, one has $\protect\h{n}_{\textsc{h}}'(1) = \exp(\hspace{0.4pt}\protect\ii\protect\h{H} t/\hbar)\hspace{0.6pt} \protect\h{n}'(\bm{r}\sigma)\exp(\protect-\ii\protect\h{H} t/\hbar)$, where $\protect\h{n}'(\bm{r}\sigma) \equiv \protect\h{n}_{\sigma}'(\bm{r})$.}\refstepcounter{dummy}\label{SinceCommut}\footnote{Since $[\protect\h{H}, \protect\h{N}_{\sigma}]_{-} = \protect\h{0}$ (footnote \raisebox{-1.0ex}{\normalsize{\protect\footref{notem}}} on p.\,\protect\pageref{SimilarTo}), with $\vert\Psi_{N;0}\rangle$ a simultaneous eigenstate of $\protect\h{N}_{\sigma}$, $\forall\sigma$, one has $\int \mathrm{d}^dr\, \varrho_{\sigma;s}(\bm{r}) \equiv 0$, $\forall s$, implying that $\int \mathrm{d}^dr\, \chi_{\sigma,\sigma'}(\bm{r},\bm{r}';z) \equiv 0$ and $\int \mathrm{d}^dr'\, \chi_{\sigma,\sigma'}(\bm{r},\bm{r}';z) \equiv 0$. \label{noten}}
\begin{equation}\label{ea19}
\varrho_{\sigma;s}(\bm{r}) \doteq \langle\Psi_{N;0}\vert \h{n}_{\sigma}'(\bm{r}) \vert\Psi_{N;s}\rangle,
\end{equation}
and
\begin{equation}\label{ea20}
e_s \doteq E_{N;s} - E_{N;0} \ge 0,\;\; \forall s.
\end{equation}
The set $\{e_s \| s\}$ thus comprises the spectrum of the neutral excitations of the system under consideration. Clearly, the equality $e_s = 0$ for $s\not=0$ signifies the GS energy of the $N$-particle system under consideration as being degenerate. We note that since the difference between $\h{n}_{\textsc{h}}'(j)$ and $\h{n}_{\textsc{h}}(j)$ is a $c$-number, on account $\langle\Psi_{N;0}\vert\Psi_{N;s}\rangle = \delta_{0,s}$, one has
\begin{equation}\label{ea21}
\varrho_{\sigma;0}(\bm{r}) \equiv 0,\;\,\text{and}\;\;\;
\varrho_{\sigma;s}(\bm{r}) = \langle\Psi_{N;0}\vert \h{n}_{\sigma}(\bm{r}) \vert\Psi_{N;s}\rangle\;\;\text{for}\;\; s\not= 0.
\end{equation}
For the analytic continuation of the `physical' function $\chi_{\sigma,\sigma'}(\bm{r},\bm{r}';\varepsilon)$ in Eq.\,(\ref{ea18}), one has\,\footnote{\emph{Cf.} Eq.\,(243), p.\,1501, in Ref.\,\protect\citen{BF02}.}
\begin{equation}\label{ea23}
\t{\chi}_{\sigma,\sigma'}(\bm{r},\bm{r}';z) = \sum_{s} \Big\{\frac{\varrho_{\sigma;s}(\bm{r}) \varrho_{\sigma';s}^*(\bm{r}')}{z - e_s} - \frac{\varrho_{\sigma;s}^*(\bm{r}) \varrho_{\sigma';s}(\bm{r}')}{z + e_s}\Big\},\;\; z\in \mathds{C},
\end{equation}
which satisfies the equality (\emph{cf.} Eq.\,(\ref{ea5}))
\begin{equation}\label{ea22}
\chi_{\sigma,\sigma'}(\bm{r},\bm{r}';\varepsilon) = \lim_{\eta\downarrow 0} \t{\chi}_{\sigma,\sigma'}(\bm{r},\bm{r}';\varepsilon \pm \ii \eta)\;\;\text{for}\;\; \varepsilon \gtrless 0.
\end{equation}

From the expression in Eq.\,(\ref{ea23}) one obtains the following \textsl{formal} asymptotic series expansion:
\begin{equation}\label{ea24}
\t{\chi}_{\sigma,\sigma'}(\bm{r},\bm{r}';z) \sim \sum_{j=1}^{\infty}\frac{\chi_{\sigma,\sigma';\infty_j}(\bm{r},\bm{r}')}{z^j}\;\;\text{for}\;\; z\to\infty,
\end{equation}
where
\begin{equation}\label{ea24a}
\chi_{\sigma,\sigma';\infty_{j}}(\bm{r},\bm{r}') \equiv \sum_{s} e_{s}^{j-1} \hspace{0.6pt} \big\{\varrho_{\sigma;s}(\bm{r}) \varrho_{\sigma';s}^*(\bm{r}') + (-1)^j \varrho_{\sigma;s}^*(\bm{r}) \varrho_{\sigma';s}(\bm{r}')\big\}.
\end{equation}
By time-reversal symmetry, one has\,\footnote{For details, consult Sec.\,4.4, p.\,132, in Ref.\,\protect\citen{BF99a}.}
\begin{equation}\label{ea24b}
\varrho_{\sigma;s}(\bm{r}) \varrho_{\sigma';s}^*(\bm{r}') \equiv \varrho_{\sigma;s}^*(\bm{r}) \varrho_{\sigma';s}(\bm{r}'),
\end{equation}
whereby
\begin{equation}\label{ea25}
\chi_{\sigma,\sigma';\infty_{2j-1}}(\bm{r},\bm{r}') \equiv 0,\;\;
\chi_{\sigma,\sigma';\infty_{2j}}(\bm{r},\bm{r}') \equiv 2\sum_{s} e_{s}^{2j-1} \hspace{0.6pt} \varrho_{\sigma;s}(\bm{r}) \varrho_{\sigma';s}^*(\bm{r}'),\;\; \forall j\in\mathds{N}.
\end{equation}
Neither the infinite series in Eq.\,(\ref{ea24}) is \emph{a priori} convergent, nor the asymptotic coefficient $\chi_{\sigma,\sigma';\infty_{2j}}(\bm{r},\bm{r}')$ exists for arbitrary large $j$ \cite{BF02}.\footnote{With $j_{\protect\X{0}}$ denoting the largest value of $j$ for which $\chi_{\sigma,\sigma';\infty_{2j}}(\bm{r},\bm{r}')$ is bounded, the term following $\chi_{\sigma,\sigma';\infty_{2j_{\protect\X{0}}}}(\bm{r},\bm{r}')/z^{2j_{\protect\X{0}}}$ in the asymptotic series expansion of $\chi_{\sigma,\sigma'}(\bm{r},\bm{r}';z)$ for $z\to\infty$ decays \textsl{faster} than $1/z^{2j_{\protect\X{0}}+2}$. For some relevant details, consult appendix C in Ref.\,\protect\citen{BF13}.} Therefore, \emph{unless we indicate otherwise, the following simplifying algebraic manipulations are of a formal character.}

To simplify the expression in Eq.\,(\ref{ea24a}), we employ the binomial expansion [\S\,3.1.1, p.\,10, in Ref.\,\citen{AS72}]
\begin{equation}\label{ea27}
e_s^{j-1} = \sum_{l=0}^{j-1} \binom{j-1}{l} (-1)^l E_{N;0}^l E_{N;s}^{j-1-l},\;\; j\in\mathds{N}.
\end{equation}
Making use of Eqs\,(\ref{ea16}) and (\ref{ea16a}), the expression in Eq.\,(\ref{ea24a}) can thus be equivalently written as
\begin{equation}\label{ea28}
\chi_{\sigma,\sigma';\infty_{j}}(\bm{r},\bm{r}') \equiv \sum_{l=0}^{j-1} \binom{j-1}{l} (-1)^l E_{N;0}^l\hspace{0.6pt} \langle\Psi_{N;0}\vert \h{n}_{\sigma}'(\bm{r}) \h{H}^{j-1-l} \h{n}_{\sigma'}'(\bm{r}')\vert\Psi_{N;0}\rangle + (-1)^j\hspace{0.6pt}\text{cc},
\end{equation}
where $\text{cc}$ stands for \textsl{complex conjugate}. To proceed, we use the identity
\begin{equation}\label{ea28a}
\h{n}_{\sigma}'(\bm{r}) \h{H}^{j-1-l} \equiv  \h{H}^{j-1-l} \h{n}_{\sigma}'(\bm{r}) + [\h{n}_{\sigma}'(\bm{r}), \h{H}^{j-1-l}]_{-},
\end{equation}
whereby
\begin{align}\label{ea28b}
&\hspace{-0.8cm}E_{N;0}^l\hspace{0.6pt} \langle\Psi_{N;0}\vert \h{n}_{\sigma}'(\bm{r}) \h{H}^{j-1-l} \h{n}_{\sigma'}'(\bm{r}')\vert\Psi_{N;0}\rangle \equiv E_{N;0}^{j-1}\hspace{0.6pt} \langle\Psi_{N;0}\vert \h{n}_{\sigma}'(\bm{r})\h{n}_{\sigma'}'(\bm{r}')\vert\Psi_{N;0}\rangle\nonumber\\
&\hspace{3.8cm} + E_{N;0}^l \hspace{0.6pt} \langle\Psi_{N;0}\vert  [\h{n}_{\sigma}'(\bm{r}), \h{H}^{j-1-l}]_{-} \h{n}_{\sigma'}'(\bm{r}')\vert\Psi_{N;0}\rangle.
\end{align}
Since
\begin{equation}\label{ea28c}
\sum_{l=0}^{j-1} \binom{j-1}{l} (-1)^l = (1-1)^{j-1} = 0,\;\;\forall j\in\mathds{N}\backslash\{1\},
\end{equation}
the first term on the RHS of Eq.\,(\ref{ea28b}), which is independent of $l$, does not contribute to $\chi_{\sigma,\sigma';\infty_{j}}(\bm{r},\bm{r}')$, $\forall j> 1$. Since $\h{H}^0 = \h{1}$, taking into account that $l\in \{0,1,\dots,j-1\}$, Eq.\,(\ref{ea28}), for the case of $j=1$ the second term on the RHS of Eq.\,(\ref{ea28b}) is identically vanishing. Since further $\h{n}_{\sigma}'(\bm{r})$ is Hermitian and $[\h{n}_{\sigma}'(\bm{r}),\h{n}_{\sigma'}'(\bm{r}')]_{-} =\h{0}$, it follows that
\begin{equation}\label{ea28d}
\chi_{\sigma,\sigma';\infty_{1}}(\bm{r},\bm{r}') \equiv 0,
\end{equation}
in conformity with the first identity in Eq.\,(\ref{ea25}). Therefore, \emph{unless we indicate otherwise, in what follows we shall assume $j \in \mathds{N}\backslash \{1\}$.}

Introducing the  Liouville superoperator (\emph{cf.} Eq.\,(\ref{e4q}))
\begin{equation}\label{ea29}
\h{L} \h{A} \doteq [\h{A},\h{H}]_{-},
\end{equation}
one has
\begin{equation}\label{ea29a}
\h{L}^k \h{A} =  \underbrace{[\dots [}_{k \times [}\h{A}\underbrace{,\h{H}]_{-},\protect\h{H}]_{-}, \dots,\h{H}]_{-}}_{k \times ,\h{H}]_{-}} \equiv \{(\h{A}),(\protect\h{H})^k\},\;\; \forall k \in\mathds{N},
\end{equation}
where the last term is the short-hand notation for the middle term according to the notation adopted in Ref.\,\citen{HCV68}. By convention
\begin{equation}\label{ea30}
\h{L}^0 \h{A} = \h{A}.
\end{equation}
Making use of the expansion \cite{HCV68}
\begin{equation}\label{ea31}
[\h{A},\h{H}^n]_{-} = \sum_{k=1}^{n} \binom{n}{k} \h{H}^{n-k} \h{L}^{k} \h{A},
\end{equation}
one obtains
\begin{align}\label{ea31x}
&\hspace{-0.8cm}E_{N;0}^l \hspace{0.6pt} \langle\Psi_{N;0}\vert  [\h{n}_{\sigma}'(\bm{r}), \h{H}^{j-1-l}]_{-} \h{n}_{\sigma'}'(\bm{r}')\vert\Psi_{N;0}\rangle\nonumber\\
&\hspace{1.0cm} = \sum_{k=1}^{j-1-l} \binom{j-1-l}{k}\hspace{0.6pt} E_{N;0}^{j-1-k}\hspace{0.6pt}\langle\Psi_{N;0}\vert \big(\h{L}^{k} \h{n}_{\sigma}'(\bm{r})\big) \hspace{0.6pt}\h{n}_{\sigma'}'(\bm{r}') \vert\Psi_{N;0}\rangle.
\end{align}
Since $\h{n}_{\sigma}'(\bm{r})$ differs from $\h{n}_{\sigma}(\bm{r})$ by a $c$-number, Eq.\,(\ref{eq13}), one has
\begin{equation}\label{ea32}
\langle\Psi_{N;0}\vert \big(\h{L}^{k} \h{n}_{\sigma}'(\bm{r})\big) \hspace{0.6pt}\h{n}_{\sigma'}'(\bm{r}') \vert\Psi_{N;0}\rangle \equiv \langle\Psi_{N;0}\vert \big(\h{L}^{k} \h{n}_{\sigma}(\bm{r})\big) \hspace{0.6pt}\h{n}_{\sigma'}(\bm{r}') \vert\Psi_{N;0}\rangle.
\end{equation}
One can thus write
\begin{align}\label{ea33}
&\hspace{-0.8cm}\chi_{\sigma,\sigma';\infty_{j}}(\bm{r},\bm{r}') = \sum_{l=0}^{j-2} \binom{j-1}{l} (-1)^l\nonumber\\
&\hspace{-0.2cm}\times \sum_{k=1}^{j-1-l} \binom{j-1-l}{k} E_{N;0}^{j-1-k}\hspace{0.6pt} \langle\Psi_{N;0}\vert \big(\h{L}^{k} \h{n}_{\sigma}(\bm{r})\big) \hspace{0.6pt}\h{n}_{\sigma'}(\bm{r}') \vert\Psi_{N;0}\rangle + (-1)^j\hspace{0.6pt}\text{cc},
\end{align}
where the change of the upper bound of the sum with respect to $l$ from $j-1$ to $j-2$ is a consequence of the fact that for $l = j-1$ the commutator on the RHS of Eq.\,(\ref{ea28b}) is identically vanishing (recall that here by assumption $j\ge 2$). Since $\chi_{\sigma,\sigma';\infty_{j}}(\bm{r},\bm{r}')$ is an intensive quantity,\footnote{See \emph{e.g.} Eq.\,(\protect\ref{ea50}) below.} and $E_{N;0}$ an extensive one, one expects that the only term contributing to the sum with respect to $k$ on the RHS of Eq.\,(\ref{ea33}) corresponds to $k=j -1$, whereby the only term to contribute to the sum with respect to $l$ on that RHS of Eq.\,(\ref{ea33}) is that corresponding to $l=0$. Assuming this to be the case, the equality in Eq.\,(\ref{ea33}) reduces to
\begin{equation}\label{ea34}
\chi_{\sigma,\sigma';\infty_{j}}(\bm{r},\bm{r}') = \langle\Psi_{N;0}\vert \big(\h{L}^{j-1} \h{n}_{\sigma}(\bm{r})\big) \hspace{0.6pt}\h{n}_{\sigma'}(\bm{r}') \vert\Psi_{N;0}\rangle + (-1)^j\hspace{0.6pt}\text{cc},\;\; \forall j \in \mathds{N}\backslash\{1\}.
\end{equation}
This result can be explicitly shown to be indeed correct.\footnote{One generally has $\sum_{l=0}^{j-2} \binom{j-1}{l} (-1)^l \sum_{k=1}^{j-1-l} \binom{j-1-l}{k} A^{j-1-k} B^k = B^{j-1}$ for $j\ge 2$.} Since both $\h{n}_{\sigma}(\bm{r})$ and $\h{H}$ are Hermitian, one has
\begin{equation}\label{ea34a}
\big(\h{L}^{k} \h{n}_{\sigma}(\bm{r})\big)^{\dag} = (-1)^k\hspace{0.6pt} \h{L}^{k} \h{n}_{\sigma}(\bm{r}),
\end{equation}
whereby
\begin{align}\label{ea34b}
&\hspace{-0.5cm} (\langle\Psi_{N;0}\vert \big(\h{L}^{j-1} \h{n}_{\sigma}(\bm{r})\big) \hspace{0.6pt}\h{n}_{\sigma'}(\bm{r}') \vert\Psi_{N;0}\rangle)^* = (-1)^{j-1} \langle\Psi_{N;0}\vert \h{n}_{\sigma'}(\bm{r}') \big(\h{L}^{j-1} \h{n}_{\sigma}(\bm{r})\big) \hspace{0.6pt}\vert\Psi_{N;0}\rangle,\nonumber\\
&\hspace{10.2cm} \forall j \in\mathds{N},
\end{align}
where the case of $j=1$ relies on the definition in Eq.\,(\ref{ea30}). Combining Eqs\,(\ref{ea34}) and (\ref{ea34b}), one arrives at
\begin{equation}\label{ea34c}
\chi_{\sigma,\sigma';\infty_{j}}(\bm{r},\bm{r}') = \langle\Psi_{N;0}\vert [\big(\h{L}^{j-1} \h{n}_{\sigma}(\bm{r})\big), \hspace{0.6pt}\h{n}_{\sigma'}(\bm{r}')]_{-} \vert\Psi_{N;0}\rangle,\;\; \forall j \in \mathds{N}.
\end{equation}
Following Eq.\,(\ref{ea30}), and since $\h{n}_{\sigma}(\bm{r})$ and $\h{n}_{\sigma'}(\bm{r}')$ commute, one observes that the equality in Eq.\,(\ref{ea34c}) is indeed in conformity with the first equality in Eq.\,(\ref{ea25}). On the basis of the general result in Eq.\,(\ref{ea34c}), from Eq.\,(\ref{ea24}) one obtains\,\footnote{Compare with the Mori-Zwanzig expression \protect\cite{HM65,RWZ61,PF02} for $\protect\t{G}_{\sigma}(\bm{r},\bm{r}';z)$ in Eq.\,(36), p.\,1440, of Ref.\,\protect\citen{BF02}. Note that the Green function is described in terms of an \textsl{anti-commutation}, unlike the density-density response function, which is described in terms of a \textsl{commutation}. The difference arises from the bosonic nature of the density operators, being composed of a product of two fermionic operators.} (\emph{cf.} Eq.\,(\ref{e4q1}))
\begin{equation}\label{ea36}
\t{\chi}_{\sigma,\sigma'}(\bm{r},\bm{r}';z) = \langle\Psi_{N;0}\vert [(z\h{1} - \h{L})^{-1}  \h{n}_{\sigma}(\bm{r}), \h{n}_{\sigma'}(\bm{r}')]_{-}\vert\Psi_{N;0}\rangle.
\end{equation}

It is instructive to consider in some detail the case of $j=2$, for which one has (\emph{cf.} Eq.\,(\ref{ea29}))
\begin{equation}\label{ea34c1}
\chi_{\sigma,\sigma';\infty_{2}}(\bm{r},\bm{r}') = \langle\Psi_{N;0}\vert [[\h{n}_{\sigma}(\bm{r}),\h{H}]_{-}, \h{n}_{\sigma'}(\bm{r}')]_{-} \vert\Psi_{N;0}\rangle.
\end{equation}
Since (see Eq.\,(\ref{ea34p}) below)
\begin{equation}\label{ea34h}
[[\h{n}_{\sigma}(\bm{r}),\h{H}]_{-},\h{n}_{\sigma'}(\bm{r}')]_{-} \propto \delta_{\sigma,\sigma'},
\end{equation}
it follows that
\begin{equation}\label{ea34i}
\chi_{\sigma,\sigma';\infty_2}(\bm{r},\bm{r}') = \langle\Psi_{N;0}\vert [[\h{n}_{\sigma}(\bm{r}),\h{H}]_{-},\h{n}_{\sigma}(\bm{r}')]_{-} \vert\Psi_{N;0}\rangle\hspace{0.6pt} \delta_{\sigma,\sigma'},
\end{equation}
whereby \cite{BF99a,BF99}\,\footnote{Apparently, the RHS of Eq.\,(\protect\ref{ea34j}) deviates from the relevant expressions in Refs\,\protect\citen{BF99a} and \protect\citen{BF99}. The discrepancy is due to the field operators in the latter references being stripped of spin indices (on account of the GSs under consideration being spin-compensated), whereby at places factors of $2$ and $1/2$ have been inserted on an \emph{ad hoc} basis. Considering, \emph{e.g.} Eq.\,(97), p.\,180, in Ref.\,\protect\citen{BF99a}, it becomes apparent that the $\protect\h{n}(\bm{r})$ in Eq.\,(118), p.\,188, of the latter reference coincides with the $\protect\h{n}_{\sigma}(\bm{r})$ of the present text, assumed in Ref.\,\protect\citen{BF99a} to operate in a Fock space where $\protect\h{n}_{\uparrow}(\bm{r}) \equiv \protect\h{n}_{\downarrow}(\bm{r})$. The factor of $2$ on the RHS of the just-mentioned Eq.\,(118) accounts for the sum $\sum_{\sigma}$ on the RHS of Eq.\,(\protect\ref{ea34j}).} (\emph{cf.} Eq.\,(\ref{ea34d1}) below)
\begin{equation}\label{ea34j}
\chi_{\infty_2}(\bm{r},\bm{r}')  \doteq \sum_{\sigma,\sigma'} \chi_{\sigma,\sigma';\infty_2}(\bm{r},\bm{r}') = \sum_{\sigma}\, \langle\Psi_{N;0}\vert [[\h{n}_{\sigma}(\bm{r}),\h{H}]_{-},\h{n}_{\sigma}(\bm{r}')]_{-} \vert\Psi_{N;0}\rangle.
\end{equation}

To establish the relationship in Eq.\,(\ref{ea34h}), we consider the many-body Hamiltonian\,\footnote{For some relevant details, consult \S\,2.2.1 of Ref.\,\protect\citen{BF19}. See also the categorisation of the two-body potentials in appendix \protect\ref{sacx}, p.\,\protect\pageref{ThreeCases}. For the reason spelled out in Ref.\,\protect\citen{BF19}, the two-body potential in this reference represents those in category (2), from which the two-body potentials in category (3) can be reached, but not those in category (1).}
\begin{align}\label{ea34k}
\h{H} &= \sum_{\sigma} \int \textrm{d}^d r\, \h{\psi}_{\sigma}^{\dag}(\bm{r}) \big(\uptau(\bm{r})+ u(\bm{r})\big) \h{\psi}_{\sigma}^{\phantom{\dag}}(\bm{r})\nonumber\\
&+ \hspace{-0.7pt} \frac{1}{2} \sum_{\sigma,\sigma'} \int \textrm{d}^d r \textrm{d}^d r'\, v_{\sigma,\sigma'}(\bm{r},\bm{r}')\hspace{0.6pt} \h{\psi}_{\sigma}^{\dag}(\bm{r}) \h{\psi}_{\sigma'}^{\dag}(\bm{r}') \h{\psi}_{\sigma'}^{\phantom{\dag}}(\bm{r}') \h{\psi}_{\sigma}^{\phantom{\dag}}(\bm{r}),\hspace{0.2cm}
\end{align}
where $\uptau(\bm{r})$ stand for the single-particle kinetic-energy operator (\emph{cf.} Eq.\,(\ref{ea34q}) below), $u(\bm{r})$ for the external potential (which we assume to be an ordinary scalar function), and $v_{\sigma,\sigma'}(\bm{r},\bm{r}')$ for the two-body interaction potential, satisfying
\begin{equation}\label{ea34l}
v_{\sigma,\sigma'}(\bm{r},\bm{r}') \equiv v_{\sigma',\sigma}(\bm{r}',\bm{r}).
\end{equation}
On the basis of the canonical anti-commutation relations satisfied by the Schr\"{o}dinger-picture field operators $\h{\psi}_{\sigma}^{\phantom{\dag}}$ and $\h{\psi}_{\sigma}^{\dag}$ \cite{FW03}, for the Hamiltonian in Eq.\,(\ref{ea34k}) one obtains\,\footnote{\emph{Cf.} Eq.\,(158), p.\,1472, in Ref.\,\protect\citen{BF02}.}
\begin{equation}\label{ea34m}
\h{A}_{\sigma}(\bm{r}) \doteq [\h{\psi}_{\sigma}(\bm{r}),\h{H}]_{-}
= \{\h{\upalpha}_{\sigma}(\bm{r}) \h{\psi}_{\sigma}(\bm{r})\},
\end{equation}
where
\begin{equation}\label{ea34n}
\h{\upalpha}_{\sigma}(\bm{r}) \doteq \uptau(\bm{r}) + u(\bm{r}) + \sum_{\sigma'} \int \mathrm{d}^dr'\; v_{\sigma,\sigma'}(\bm{r},\bm{r}')\hspace{0.6pt} \h{n}_{\sigma'}(\bm{r}').
\end{equation}
The curly braces on the RHS of Eq.\,(\ref{ea34m}) signify that the $\uptau(\bm{r})$ in $\h{\upalpha}_{\sigma}(\bm{r})$ operates \textsl{solely} on the $\h{\psi}_{\sigma}(\bm{r})$ and not on any function of $\bm{r}$ to its right with which the operator $\h{A}_{\sigma}(\bm{r})$ may be multiplied. This limitation is a consequence of the $\uptau(\bm{r})$ on the RHS of Eq.\,(\ref{ea34n}) having originated from the \textsl{integrand} of the first integral with respect to $\bm{r}$ on the RHS of Eq.\,(\ref{ea34k}), where $\uptau(\bm{r})$ acts solely on the $\h{\psi}_{\sigma}(\bm{r})$ to its right.\footnote{We shall continue to use this convention in the following several expressions.} One obtains\,\footnote{Using $[\protect\h{C}\protect\h{A},\protect\h{B}]_{-} = \protect\h{C} [\protect\h{A},\protect\h{B}]_{-} + [\protect\h{C},\protect\h{B}]_{-} \protect\h{A}$. With $\protect\h{D} \doteq [\protect\h{A},\protect\h{B}]_{-}$, in the case at hand where $[\protect\h{C},\protect\h{B}]_{-} = -\protect\h{D}^{\dag}$, one has $[\protect\h{C}\protect\h{A},\protect\h{B}]_{-} = \protect\h{C}\protect\h{D} - \protect\h{D}^{\dag} \protect\h{A}$.}
\begin{equation}\label{ea34o}
[\h{n}_{\sigma}(\bm{r}),\h{H}]_- = \h{\psi}_{\sigma}^{\dag}(\bm{r}) \{\uptau(\bm{r}) \h{\psi}_{\sigma}^{\phantom{\dag}}(\bm{r})\} - \{\uptau(\bm{r}) \h{\psi}_{\sigma}^{\dag}(\bm{r})\} \h{\psi}_{\sigma}^{\phantom{\dag}}(\bm{r}),
\end{equation}
which is independent of both the (scalar) external potential $u(\bm{r})$ and the two-body interaction potential $v_{\sigma,\sigma'}(\bm{r},\bm{r}')$. The relationship in Eq.\,(\ref{ea34h}) is a direct consequence of the expression on the RHS of Eq.\,(\ref{ea34o}) not involving any summation with respect to intermediate spin indices.\footnote{This property is \textsl{not} shared by all terms contributing to $\protect\h{L}^k \protect\h{n}_{\sigma}(\bm{r})$, $\forall k \ge 2$.} One explicitly obtains
\begin{align}\label{ea34p}
&\hspace{-0.02cm}[[\h{n}_{\sigma}(\bm{r}),\h{H}]_{-},\h{n}_{\sigma'}(\bm{r}')]_{-} = -\Big\{\h{\psi}_{\sigma}^{\dag}(\bm{r}) \{\uptau(\bm{r}) \h{\psi}_{\sigma}^{\phantom{\dag}}(\bm{r})\} + \{\uptau(\bm{r}) \h{\psi}_{\sigma}^{\dag}(\bm{r})\}\h{\psi}_{\sigma}^{\phantom{\dag}}(\bm{r})\Big\} \hspace{0.6pt}\delta(\bm{r}-\bm{r}')\hspace{0.6pt} \delta_{\sigma,\sigma'}\nonumber\\
&\hspace{3.5cm} + \Big\{\h{\psi}_{\sigma}^{\dag}(\bm{r}) \h{\psi}_{\sigma}^{\phantom{\dag}}(\bm{r}') + \psi_{\sigma}^{\dag}(\bm{r}') \h{\psi}_{\sigma}^{\phantom{\dag}}(\bm{r})\Big\}\hspace{0.6pt} \{\uptau(\bm{r}) \delta(\bm{r}-\bm{r}')\}\hspace{0.6pt} \delta_{\sigma,\sigma'},
\end{align}
establishing the validity of the relationship in Eq.\,(\ref{ea34h}). The RHS of Eq.\,(\ref{ea34p}) considerably simplifies in the specific case of
\begin{equation}\label{ea34q}
\uptau(\bm{r}) \equiv -\frac{\hbar^2}{2\mathsf{m}} \nabla_{\bm{r}}^2,
\end{equation}
where $\mathsf{m}$ stands for the bare particle mass. One obtains\,\footnote{Making use of such identities as $\protect\h{f}(\bm{r}) \big(\bm{\nabla}_{\bm{r}} \protect\h{g}(\bm{r})\big) \equiv \bm{\nabla}_{\bm{r}} \big(\protect\h{f}(\bm{r}) \protect\h{g}(\bm{r})\big) -\big(\bm{\nabla}_{\bm{r}} \protect\h{f}(\bm{r})\big) \protect\h{g}(\bm{r})$, \emph{etc}.}
\begin{equation}\label{ea34r}
[[\h{n}_{\sigma}(\bm{r}),\h{H}]_{-},\h{n}_{\sigma'}(\bm{r}')]_{-} = -\frac{\hbar^2}{\mathsf{m}}\hspace{0.6pt} \big\{\bm{\nabla}_{\bm{r}}\cdot \big(\h{n}_{\sigma}(\bm{r}) \{\bm{\nabla}_{\bm{r}}\delta(\bm{r}-\bm{r'})\}\big)\big\}\hspace{0.6pt}\delta_{\sigma,\sigma'},
\end{equation}
from which it follows that
\begin{equation}\label{ea34u}
\chi_{\sigma,\sigma';\infty_2}(\bm{r},\bm{r}') = -\frac{\hbar^2}{\mathsf{m}}\hspace{0.6pt} \big\{\bm{\nabla}_{\bm{r}}\cdot \big(n_{\sigma}(\bm{r}) \{\bm{\nabla}_{\bm{r}}\delta(\bm{r}-\bm{r'})\}\big)\big\}\hspace{0.6pt} \delta_{\sigma,\sigma'},
\end{equation}
and \cite{BF99a,BF99} (\emph{cf.} Eq.\,(\ref{ea34j}))
\begin{equation}\label{ea34s}
\chi_{\infty_2}(\bm{r},\bm{r}') = -\frac{\hbar^2}{\mathsf{m}}\hspace{0.6pt} \big\{\bm{\nabla}_{\bm{r}}\cdot \big(n(\bm{r}) \{\bm{\nabla}_{\bm{r}}\delta(\bm{r}-\bm{r'})\}\big)\big\},
\end{equation}
where $n(\bm{r})$ is the \textsl{total} GS number density, defined according to
\begin{equation}\label{ea34t}
n(\bm{r}) = \langle\Psi_{N;0}\vert\h{n}(\bm{r})\vert\Psi_{N;0}\rangle,
\end{equation}
in which
\begin{equation}\label{ea34f}
\h{n}(\bm{r}) \doteq \sum_{\sigma} \h{n}_{\sigma}(\bm{r}).
\end{equation}

With\,\footnote{With $f(\bm{r},\bm{r}') \equiv \langle\bm{r}\vert\protect\h{f}\vert\bm{r}'\rangle$, one has $\protect\b{f}(\bm{k},\bm{k}') \equiv \langle\bm{k}\vert\protect\h{f}\vert\bm{k}'\rangle$, where $\vert\bm{r}\rangle$ is the eigenstate of the $\protect\h{\bm{r}}$ operator corresponding to eigenvalue $\bm{r}$ and normalised according to $\langle\bm{r}\vert\bm{r}'\rangle = \delta(\bm{r}-\bm{r}')$, and $\vert\bm{k}\rangle$ the eigenvector of the $\protect\h{\bm{k}}$ operator corresponding to eigenvalue $\bm{k}$ and normalised according to $\langle\bm{k}\vert\bm{k}'\rangle = \delta_{\bm{k},\bm{k}'}$ (box normalisation condition). One thus has $\langle\bm{k}\vert\bm{r}\rangle = \protect\e^{-\protect\ii \bm{k}\cdot\bm{r}}/\Omega^{1/2}$, which accounts for the $1/\Omega$ on the RHS of Eq.\,(\protect\ref{ea36a}).}
\begin{equation}\label{ea36a}
\b{f}(\bm{k},\bm{k}') \doteq \frac{1}{\Omega} \int_{\b{\Omega}} \mathrm{d}^dr \mathrm{d}^dr'\, \e^{-\ii \bm{k}\cdot \bm{r}} f(\bm{r},\bm{r}') \e^{\ii \bm{k}'\cdot \bm{r}'},
\end{equation}
where $\b{\Omega} \subseteq \mathds{R}^d$ denotes the region of the space into which the system is confined, and $\Omega$ its volume, from the Lehmann-type representation in Eq.\,(\ref{ea23}), one obtains
\begin{equation}\label{ea39b}
\t{\b{\chi}}_{\sigma,\sigma'}(\bm{k},\bm{k}';z) = \frac{1}{\Omega} \sum_{s} \Big\{\frac{\b{\varrho}_{\sigma;s}(\bm{k}) \b{\varrho}_{\sigma';s}^{\hspace{0.4pt}*}(\bm{k}')}{z - e_s} - \frac{\b{\varrho}_{\sigma;s}^{\hspace{0.4pt}*}(-\bm{k}) \b{\varrho}_{\sigma';s}(-\bm{k}')}{z + e_s}\Big\},
\end{equation}
where\,\refstepcounter{dummy}\label{OnAccountOf}\footnote{With reference to footnote \raisebox{-1.0ex}{\normalsize{\protect\footref{noten}}} on p.\,\protect\pageref{SinceCommut}, $\b{\varrho}_{\sigma;s}(\bm{0}) \equiv 0$. Consequently, $\protect\t{\protect\b{\chi}}_{\sigma,\sigma'}(\bm{0},\bm{k}';z)\equiv \protect\t{\protect\b{\chi}}_{\sigma,\sigma'}(\bm{k},\bm{0};z) \equiv 0$. For macroscopic metallic systems this result does not necessarily obtain on identifying $z$ with zero prior to taking the limit $\bm{k}\to \bm{0}$ (\emph{cf.} Eq.\,(\protect\ref{ea52d}) below). \label{noteo}}
\begin{equation}\label{ea39c}
\b{\varrho}_{\sigma;s}(\bm{k}) = \int_{\b{\Omega}} \mathrm{d}^dr\, \e^{-\ii \bm{k}\cdot\bm{r}} \varrho_{\sigma;s}(\bm{r}).
\end{equation}
Accordingly, from the expression in Eq.\,(\ref{ea34c}) one obtains
\begin{equation}\label{ea37}
\b{\chi}_{\sigma,\sigma';\infty_{j}}(\bm{k},\bm{k}') = \frac{1}{\Omega}\hspace{0.6pt}\langle\Psi_{N;0}\vert [(\h{L}^{j-1} \h{n}_{\bm{k};\sigma}^{\phantom{\dag}}), \hspace{0.6pt}\h{n}_{\bm{k}';\sigma'}^{\dag}]_{-} \vert\Psi_{N;0}\rangle,\;\; \forall j \in \mathds{N},
\end{equation}
where
\begin{equation}\label{ea38}
\h{n}_{\bm{k};\sigma}^{\phantom{\dag}} \doteq \int_{\b{\Omega}} \mathrm{d}^dr\, \e^{-\ii \bm{k}\cdot\bm{r}} \h{n}_{\sigma}(\bm{r}),
\end{equation}
and
\begin{equation}\label{ea39}
\h{n}_{\bm{k};\sigma}^{\dag} \equiv \h{n}_{-\bm{k};\sigma}^{\phantom{\dag}}.
\end{equation}

With $\h{\mathsf{P}}$ denoting the total momentum operator, for $\vert\Psi_{N;0}\rangle$ an eigenstate of $\h{\mathsf{P}}$, one has\,\refstepcounter{dummy}\label{WithP}\footnote{With $\protect\h{\mathsf{P}} \equiv \sum_{\bm{k},\sigma} \hbar\bm{k}\hspace{0.6pt}\protect\h{a}_{\bm{k};\sigma}^{\dag} \protect\h{a}_{\bm{k};\sigma}^{\phantom{\dag}}$ [Eq.\,(7.50) in Ref.\,\protect\citen{FW03}], for the canonical operators $\{\protect\h{a}_{\bm{k};\sigma}^{\phantom{\dag}}, \protect\h{a}_{\bm{k};\sigma}^{\dag}\}$ one has $[\protect\h{\mathsf{P}},\protect\h{a}_{\bm{k};\sigma}^{\phantom{\dag}}]_{-} = -\hbar\bm{k}\hspace{0.8pt} \protect\h{a}_{\bm{k};\sigma}^{\phantom{\dag}}$ and $[\protect\h{\mathsf{P}},\protect\h{a}_{\bm{k};\sigma}^{\dag}]_{-} = +\hbar\bm{k}\hspace{0.8pt} \protect\h{a}_{\bm{k};\sigma}^{\dag}$ (note that $\protect\h{\mathsf{P}}^{\dag} = \protect\h{\mathsf{P}}$). Thus, on the basis of the expression for $\protect\h{n}_{\bm{k};\sigma}$ in Eq.\,(\protect\ref{ea44}) below, one obtains $[\protect\h{\mathsf{P}},\protect\h{n}_{\bm{k};\sigma}]_{-} = -\hbar\bm{k}\hspace{0.8pt} \protect\h{n}_{\bm{k};\sigma}$. It follows that for $\vert\Psi_{N;s}\rangle$ an eigenstate of $\protect\h{\mathsf{P}}$ corresponding to eigenvalue $\bm{P}_s$,  $\protect\h{n}_{\bm{k};\sigma}\vert\Psi_{N;s}\rangle$ is an eigenstate of $\protect\h{\mathsf{P}}$ corresponding to eigenvalue $\bm{P}_s -\hbar\bm{k}$. Consequently, following Eqs\,(\protect\ref{ea21}), (\protect\ref{ea39c}), and (\protect\ref{ea38}), whereby $\protect\b{\varrho}_{\sigma;s}(\bm{k}) = \langle\Psi_{N;0}\vert \protect\h{n}_{\bm{k};\sigma}\vert\Psi_{N;s}\rangle$ for $s\not=0$, unless $\bm{P}_s - \bm{P}_0 = \hbar \bm{k}$ one has $\protect\b{\varrho}_{\sigma;s}(\bm{k}) \equiv 0$. As a result, unless $\bm{k}=\bm{k}'$, the product $\protect\b{\varrho}_{\sigma;s}(\bm{k}) \protect\b{\varrho}_{\sigma;s}^*(\bm{k}')$ is identically vanishing. \label{notez}}\footnote{We note in passing that for periodic systems specified by the vectors of the Bravais lattice $\{\bm{R}_i\| i\}$, one has $\protect\t{\b{\chi}}_{\sigma,\sigma'}(\bm{k},\bm{k}';z) \equiv 0$ unless $\bm{k}-\bm{k}' = \bm{K}_j$ for some $j$, where $\{\bm{K}_i \| i\}$ is the Bravais lattice reciprocal to $\{\bm{R}_i\| i\}$ \protect\cite{AM76}. Since $\bm{K}_i -\bm{K}_j = \bm{K}_k$ for some $k$, $\forall i, j$, for these systems one thus deals with the matrix $\protect\t{\b{\mathbb{\chi}}}_{\sigma,\sigma'}(\bm{k};z)$ for which one has $(\protect\t{\b{\mathbb{\chi}}}_{\sigma,\sigma'}(\bm{k};z))_{i,j} \equiv \protect\t{\b{\chi}}_{\sigma,\sigma'}(\bm{k}+\bm{K}_i,\bm{k}+\bm{K}_j;z)$, where $\bm{k} \in \protect\1BZ$.}
\begin{equation}\label{ea39d}
\t{\b{\chi}}_{\sigma,\sigma'}(\bm{k},\bm{k}';z) = \t{\b{\chi}}_{\sigma,\sigma'}(\bm{k};z)\hspace{0.6pt}\delta_{\bm{k},\bm{k}'},
\end{equation}
and
\begin{equation}\label{ea40}
\b{\chi}_{\sigma,\sigma';\infty_{j}}(\bm{k},\bm{k}') = \b{\chi}_{\sigma,\sigma';\infty_{j}}(\bm{k})\hspace{0.6pt} \delta_{\bm{k},\bm{k}'}.
\end{equation}
We note in passing that a comparison of the expression in Eq.\,(\ref{ea37}) for $\bm{k}' = \bm{k}$ with that in Eq.\,(\ref{e4p}) makes evident that for Hubbard-like models $\b{\chi}_{\sigma,\sigma';\infty_j}(\bm{k})$ is bounded for arbitrary finite values of $j$. This is not the case for general systems \cite{BF02}. This implies that for Hubbard-like models the spectral function corresponding to the function $\b{\chi}_{\sigma}(\bm{k};\varepsilon)$, Eq.\,(\ref{ea40g1}) below, decays faster than any finite power of $1/\varepsilon$ as $\vert\varepsilon\vert \to \infty$.

Thus far we have considered the function $\t{\chi}_{\sigma,\sigma'}(\bm{r},\bm{r}';z)$ and the associated Fourier transform. In the cases where the two-body interaction potential $v_{\sigma,\sigma'}(\bm{r},\bm{r}')$ in Eq.\,(\ref{ea34k}) is independent of $\sigma$ and $\sigma'$,\footnote{See the discussions in appendix \ref{sacx} regarding different category of two-body potentials. The two-body potential to be considered here falls into category (3) (see p.\,\protect\pageref{ThreeCases}).} the function of interest is\,\footnote{See \S\,3.2 of Ref.\,\protect\citen{BF19}. See also pp.\,110 and 111 (Ch.\,4) in Ref.\,\protect\citen{FW03}, in particular Eqs\,(9.42), (9.43a), and (9.43b) herein.}
\begin{equation}\label{ea34d}
\t{\chi}(\bm{r},\bm{r}';z) \doteq \sum_{\sigma,\sigma'} \t{\chi}_{\sigma,\sigma'}(\bm{r},\bm{r}';z).
\end{equation}
Accordingly, for systems in which $v_{\sigma,\sigma'}(\bm{r},\bm{r}')$ is independent of $\sigma$ and $\sigma'$, the sequence of interest is $\{\chi_{\infty_j}(\bm{r},\bm{r'})\|j\}$, where
\begin{equation}\label{ea34d1}
\chi_{\infty_j}(\bm{r},\bm{r'}) \doteq \sum_{\sigma,\sigma'} \chi_{\sigma;\sigma';\infty_j}(\bm{r},\bm{r'}),
\end{equation}
for which, following Eqs\,(\ref{ea34c}) and (\ref{ea34f}), one has
\begin{equation}\label{ea34e}
\chi_{\infty_j}(\bm{r},\bm{r}') = \langle\Psi_{N;0}\vert [\big(\h{L}^{j-1} \h{n}(\bm{r})\big), \hspace{0.6pt}\h{n}(\bm{r}')]_{-} \vert\Psi_{N;0}\rangle,\;\; \forall j \in \mathds{N}.
\end{equation}
Thus (\emph{cf.} Eqs\,(\ref{ea36}) and (\ref{e4q1}))
\begin{equation}\label{ea34e0}
\t{\chi}(\bm{r},\bm{r}';z) = \langle\Psi_{N;0}\vert [(z\h{1} - \h{L})^{-1}  \h{n}(\bm{r}), \h{n}(\bm{r}')]_{-}\vert\Psi_{N;0}\rangle.
\end{equation}
Following Eq.\,(\ref{ea39b}), one has
\begin{equation}\label{ea34e1}
\t{\b{\chi}}(\bm{k},\bm{k}';z) \doteq \sum_{\sigma,\sigma'} \t{\b{\chi}}_{\sigma,\sigma'}(\bm{k},\bm{k}';z) = \frac{1}{\Omega} \sum_s \Big\{ \frac{\b{\varrho}_s(\bm{k}) \b{\varrho}_s^*(\bm{k}')}{z -e_s} - \frac{\b{\varrho}_s^*(-\bm{k}) \b{\varrho}_s(-\bm{k}')}{z + e_s}\Big\},
\end{equation}
where
\begin{equation}\label{ea34e2}
\b{\varrho}_s(\bm{k}) \doteq \sum_{\sigma} \b{\varrho}_{\sigma;s}(\bm{k}).
\end{equation}
Further,\footnote{We note that the frequency moments of the dielectric function, Eq.\,(\protect\ref{ea3b}), and its inverse have been considered by Taut \protect\cite{MT85}.} following Eq.\,(\ref{ea37}),
\begin{equation}\label{ea34e3}
\b{\chi}_{\infty_{j}}(\bm{k},\bm{k}') \doteq \sum_{\sigma,\sigma'} \b{\chi}_{\sigma,\sigma';\infty_{j}}(\bm{k},\bm{k}') = \frac{1}{\Omega}\hspace{0.6pt}\langle\Psi_{N;0}\vert [(\h{L}^{j-1} \h{n}_{\bm{k}}^{\phantom{\dag}}), \hspace{0.6pt}\h{n}_{\bm{k}'}^{\dag}]_{-} \vert\Psi_{N;0}\rangle,\;\; \forall j \in \mathds{N},
\end{equation}
where
\begin{equation}\label{ea34e4}
\h{n}_{\bm{k}}^{\phantom{\dag}} \doteq \sum_{\sigma} \h{n}_{\bm{k};\sigma}^{\phantom{\dag}},
\end{equation}
and\,\footnote{For instance, following Eqs\,(\protect\ref{ea36}), (\protect\ref{ea36a}), (\protect\ref{ea38}), (\protect\ref{ea39}), (\protect\ref{ea34d}), and (\protect\ref{ea34e4}).}
\begin{equation}\label{ea34e5}
\t{\b{\chi}}(\bm{k},\bm{k}';z) = \frac{1}{\Omega} \hspace{0.6pt}\langle\Psi_{N;0}\vert [(z\h{1} - \h{L})^{-1}   \h{n}_{\bm{k}}^{\phantom{\dag}}, \hspace{0.6pt}\h{n}_{\bm{k}'}^{\dag}]_{-} \vert\Psi_{N;0}\rangle,
\end{equation}
whereby for uniform GSs, Eq.\,(\ref{ea39d}),
\begin{equation}\label{ea34e6}
\t{\b{\chi}}(\bm{k};z) = \frac{1}{\Omega} \hspace{0.6pt}\langle\Psi_{N;0}\vert [(z\h{1} - \h{L})^{-1}   \h{n}_{\bm{k}}^{\phantom{\dag}}, \hspace{0.6pt}\h{n}_{\bm{k}}^{\dag}]_{-} \vert\Psi_{N;0}\rangle.
\end{equation}
Compare with Eq.\,(\ref{e4q1}).

With reference to Eq.\,(\ref{ea39d}), for uniform GSs from Eq.\,(\ref{ea34e1}) one has\,\footnote{From the first equality in Eq.\,(\protect\ref{ea40a}) one obtains $\t{\b{\chi}}(\bm{k};z) \equiv \t{\b{\chi}}(-\bm{k};-z)$, and from the second equality $\t{\b{\chi}}(\bm{k};z) \equiv \t{\b{\chi}}(\bm{k};-z)$.}
\begin{align}\label{ea40a}
\t{\b{\chi}}(\bm{k};z) &= \frac{1}{\Omega} \sum_{s} \Big\{\frac{\vert\b{\varrho}_{s}(\bm{k})\vert^2}{z - e_s} - \frac{\vert\b{\varrho}_{s}(-\bm{k})\vert^2}{z + e_s}\Big\}\nonumber\\
&=  \frac{1}{\Omega} \sum_{s} \Big\{\frac{1}{z - e_s} - \frac{1}{z + e_s}\Big\}\hspace{0.6pt} \vert\b{\varrho}_{s}(\bm{k})\vert^2 \equiv \t{\b{\chi}}^+(\bm{k};z) -\t{\b{\chi}}^-(\bm{k};z),\hspace{0.5cm}
\end{align}
where the second equality applies on account of the identity in Eq.\,(\ref{ea24b}). Clearly, one has the reflection property (\emph{cf.} Eq.\,(\ref{e3b}))
\begin{equation}\label{ea40b}
\t{\b{\chi}}(\bm{k};z^*) \equiv \t{\b{\chi}}^*(\bm{k};z),\;\;\im[z]\not= 0,
\end{equation}
and further
\begin{equation}\label{ea40b1}
\t{\b{\chi}}^{+}(\bm{k};z) \equiv -\t{\b{\chi}}^{-}(\bm{k};-z).
\end{equation}
Writing $\t{\b{\chi}}(\bm{k};z)$ as\,\footnote{\emph{Cf.} Eq.\,(B.52) in Ref.\,\protect\citen{BF07}.}
\begin{equation}\label{ea40c}
\t{\b{\chi}}(\bm{k};z) = \frac{1}{\Omega} \sum_{s} \Big\{\frac{z^* - e_s}{\vert z - e_s\vert^2} - \frac{z^* + e_s}{\vert z + e_s\vert^2}\Big\}\hspace{0.6pt} \vert\b{\varrho}_{s}(\bm{k})\vert^2,
\end{equation}
whereby
\begin{equation}\label{ea40c1}
\im[\t{\b{\chi}}(\bm{k};z)] = \frac{-4 \re[z]\hspace{0.0pt}\im[z]}{\Omega} \sum_{s} \frac{e_s \hspace{0.6pt}\vert\b{\varrho}_{s}(\bm{k})\vert^2}{\vert z - e_s\vert^2 \vert z + e_s\vert^2},
\end{equation}
on account of $e_s \ge 0$, Eq.\,(\ref{ea20}), one deduces that
\begin{equation}\label{ea40d}
\sgn(\im[-\t{\b{\chi}}(\bm{k};z)]) = \sgn(\re[z]\hspace{0.0pt} \im[z]),
\end{equation}
implying that
\begin{equation}\label{ea40e}
\mp\t{\b{\chi}}(\bm{k};z)\;\; \text{is Nevanlinna for}\; \re[z] \gtrless 0.
\end{equation}
Along the same line as above, one obtains
\begin{equation}\label{ea40f}
\sgn(\im[-\t{\b{\chi}}^{\pm}(\bm{k};z)]) = \sgn(\im[z]),
\end{equation}
implying that
\begin{equation}\label{ea40f1}
-\t{\b{\chi}}^{\pm}(\bm{k};z)\;\; \text{is Nevanlinna}.
\end{equation}

With reference to Eq.\,(\ref{ea18}), from Eq.\,(\ref{ea40a}) for the imaginary part of the `physical' counterpart of $\t{\b{\chi}}(\bm{k};z)$, one has  (\emph{cf.} Eq.\,(\ref{ea22}))
\begin{equation}\label{ea40g}
\im[\t{\b{\chi}}(\bm{k};\varepsilon \mp \ii 0^+)] = \pm\frac{\pi}{\Omega} \sum_s \big\{\delta(\varepsilon-e_s) - \delta(\varepsilon+e_s)\big\}\hspace{0.6pt} \vert\b{\varrho}_{s}(\bm{k})\vert^2,\;\; \varepsilon\in\mathds{R}.
\end{equation}
Thus, introducing the spectral function (\emph{cf.} Eq.\,(\ref{e4d}))
\begin{equation}\label{ea40g1}
\mathcal{A}_{\X{\chi}}(\bm{k};\varepsilon) \doteq \pm\frac{1}{\pi} \im[\t{\b{\chi}}(\bm{k};\varepsilon \mp \ii 0^+)],\;\; \varepsilon\in\mathds{R},
\end{equation}
one has (\emph{cf.} Eq.\,(\ref{ea40a}))
\begin{align}\label{ea40g2}
\t{\b{\chi}}^{+}(\bm{k};z) &= \int_{0}^{\infty} \mathrm{d}\varepsilon'\; \frac{\mathcal{A}_{\X{\chi}}(\bm{k};\varepsilon')}{z-\varepsilon'}, \nonumber\\
\t{\b{\chi}}^{-}(\bm{k};z) &= \int_{0}^{\infty} \mathrm{d}\varepsilon'\; \frac{\mathcal{A}_{\X{\chi}}(\bm{k};\varepsilon')}{z+\varepsilon'},
\end{align}
from which one obtains
\begin{equation}\label{ea40g3}
\t{\b{\chi}}(\bm{k};z) = \int_{0}^{\infty} \mathrm{d}\varepsilon'\; \Big\{\frac{1}{z-\varepsilon'} - \frac{1}{z+\varepsilon'}\Big\}\hspace{0.6pt} \mathcal{A}_{\X{\chi}}(\bm{k};\varepsilon').
\end{equation}
Note that, for all $\bm{k}$,
\begin{equation}\label{ea40g4}
\mathcal{A}_{\chi}(\bm{k};\varepsilon) \ge 0,\;\; \forall\varepsilon \ge 0,\;\;\;\;
\mathcal{A}_{\X{\chi}}(\bm{k};-\varepsilon) \equiv -\mathcal{A}_{\X{\chi}}(\bm{k};\varepsilon),\;\;\forall\varepsilon\in\mathds{R}.
\end{equation}
With $\mathcal{A}_{\X{\chi}}(\bm{k};\varepsilon) \ge 0$, $\forall\varepsilon \ge 0$ and $\forall \bm{k}$, the moment problems associated with the functions $\t{\b{\chi}}^{+}(\bm{k};z)$ and $\t{\b{\chi}}^{-}(\bm{k};z)$ are of the Stieltjes type \cite{NIA65,ST70}.

Writing (\emph{cf.} Eq.\,(\ref{ea24}))
\begin{equation}\label{ea40h}
\t{\b{\chi}}^{\pm}(\bm{k};z) \sim \sum_{j=1}^{\infty} \frac{\b{\chi}_{\infty_j}^{\pm}(\bm{k})}{z^j}\;\;\text{for}\;\; z\to\infty,
\end{equation}
from Eq.\,(\ref{ea34e3}) one has
\begin{align}\label{ea40i}
\b{\chi}_{\infty_j}^{+}(\bm{k}) &= \frac{1}{\Omega}\hspace{0.6pt}\langle\Psi_{N;0}\vert \big(\h{L}^{j-1} \h{n}_{\bm{k}}^{\phantom{\dag}}\big) \hspace{0.6pt}\h{n}_{\bm{k}}^{\dag} \vert\Psi_{N;0}\rangle, \nonumber\\
\b{\chi}_{\infty_j}^{-}(\bm{k}) &= \frac{1}{\Omega}\hspace{0.6pt}\langle\Psi_{N;0}\vert \h{n}_{\bm{k}}^{\dag}\hspace{0.6pt}\big(\h{L}^{j-1} \h{n}_{\bm{k}}^{\phantom{\dag}}\big) \vert\Psi_{N;0}\rangle.
\end{align}
From Eq.\,(\ref{ea40a}) one infers that\,\footnote{With $\protect\b{\chi}_{\infty_j}^{+}(\bm{k}) = \Omega^{-1} \sum_s e_s^{j-1} \vert\protect\b{\varrho}_s(\bm{k})\vert^2$, one clearly has $\protect\b{\chi}_{\infty_j}^{+}(\bm{k})\ge 0$, $\forall j, \bm{k}$ (recall that $e_s \ge 0$, Eq.\,(\protect\ref{ea20})). With $\protect\b{\chi}_{\infty_j}^{+}(\bm{k})\in \mathds{R}$, the equality in Eq.\,(\protect\ref{ea40j}) can also be obtained from $\protect\b{\chi}_{\infty_j}^{+}(\bm{k}) \equiv (\protect\b{\chi}_{\infty_j}^{+}(\bm{k}))^*$ in conjunction with $\b{\chi}_{\infty_j}^{\pm}(-\bm{k}) \equiv \b{\chi}_{\infty_j}^{\pm}(\bm{k})$ and Eqs\,(\protect\ref{ea34a}) and (\protect\ref{ea39}).}
\begin{equation}\label{ea40j}
\b{\chi}_{\infty_j}^{-}(\bm{k}) \equiv (-1)^{j-1}\hspace{0.6pt}\b{\chi}_{\infty_j}^{+}(\bm{k}),
\end{equation}
in conformity with the identity in Eq.\,(\ref{ea40b1}). Note that, following Eqs\,(\ref{ea40g2}) and (\ref{ea40g3}),
\begin{equation}\label{ea40l}
\b{\chi}_{\infty_j}^{\pm}(\bm{k}) = (\pm 1)^{j-1}\int_{0}^{\infty} \mathrm{d}\varepsilon\, \mathcal{A}_{\X{\chi}}(\bm{k};\varepsilon)\hspace{0.6pt} \varepsilon^{j-1},\;\; \forall j\in\mathds{N},
\end{equation}
and (\emph{cf.} Eq.\,(\ref{ea25}))
\begin{equation}\label{ea40k}
\b{\chi}_{\infty_j}(\bm{k}) = \big(1+(-1)^j\big) \int_{0}^{\infty} \mathrm{d}\varepsilon\, \mathcal{A}_{\X{\chi}}(\bm{k};\varepsilon)\hspace{0.6pt} \varepsilon^{j-1},\;\; \forall j\in\mathds{N}.
\end{equation}
Since for Hubbard-like models $\b{\chi}_{\infty_j}(\bm{k})$ and $\b{\chi}_{\infty_j}^{\pm}(\bm{k})$, Eqs\,(\ref{ea34e3}) and (\ref{ea40i}), are bounded for arbitrary finite values of $j$, the expressions in Eqs\,(\ref{ea40l}) and (\ref{ea40k}) establish that the spectral function $\mathcal{A}_{\X{\chi}}(\bm{k};\varepsilon)$ corresponding to these models decays faster than any finite power of $1/\varepsilon$ for $\varepsilon\to\infty$. In view of the second equality in Eq.\,(\ref{ea40g4}), similarly as regards $\varepsilon\to -\infty$.

For illustration, following Eq.\,(\ref{ea37}), for uniform GSs one has
\begin{equation}\label{ea41}
\b{\chi}_{\sigma,\sigma';\infty_{1}}(\bm{k}) \equiv 0,\;\;\;
\b{\chi}_{\sigma,\sigma';\infty_{2}}(\bm{k}) = \frac{1}{\Omega}\hspace{0.6pt}\langle\Psi_{N;0}\vert [[\h{n}_{\bm{k};\sigma}^{\phantom{\dag}},\h{H}]_{-},\h{n}_{\bm{k};\sigma'}^{\dag}]_{-} \vert\Psi_{N;0}\rangle.
\end{equation}
For these states of systems confined to $\b{\Omega} \subseteq\mathds{R}^d$ and subject to the box boundary condition,
\begin{equation}\label{ea42}
\h{\psi}_{\sigma}(\bm{r}) = \frac{1}{\Omega^{1/2}} \sum_{\bm{k}} \e^{\ii \bm{k}\cdot\bm{r}} \h{a}_{\bm{k};\sigma}^{\phantom{\dag}},
\end{equation}
whereby (\emph{cf.} Eq.\,(\ref{ea14}))
\begin{equation}\label{ea43}
\h{n}_{\sigma}(\bm{r}) = \frac{1}{\Omega} \sum_{\bm{k},\bm{k}'} \e^{-\ii\hspace{1.2pt} (\bm{k}-\bm{k}')\cdot \bm{r}} \h{a}_{\bm{k};\sigma}^{\dag} \h{a}_{\bm{k}';\sigma}^{\phantom{\dag}},
\end{equation}
so that, following Eq.\,(\ref{ea38}),
\begin{equation}\label{ea44}
\h{n}_{\bm{k};\sigma} = \sum_{\bm{k}'} \h{a}_{\bm{k}';\sigma}^{\dag} \h{a}_{\bm{k}+\bm{k}';\sigma}^{\phantom{\dag}}.
\end{equation}
For the Hamiltonian in Eq.\,(\ref{ex01}), on account of the canonical anti-commutation relations $[\h{a}_{\bm{k};\sigma}^{\phantom{\dag}},\h{a}_{\bm{k}';\sigma'}^{\dag}]_{+} = \delta_{\bm{k},\bm{k}'} \delta_{\sigma,\sigma'}$ and $[\h{a}_{\bm{k};\sigma}^{\phantom{\dag}},\h{a}_{\bm{k}';\sigma'}^{\phantom{\dag}}]_{+} = \h{0}$ \cite{FW03}, one thus has
\begin{equation}\label{ea45}
[\h{n}_{\bm{k};\sigma}^{\phantom{\dag}},\h{H}]_{-} = \sum_{\bm{k}'} \big\{\varepsilon_{\bm{k}+\bm{k}'} - \varepsilon_{\bm{k}'}\big\}\hspace{0.6pt} \h{a}_{\bm{k}';\sigma}^{\dag} \h{a}_{\bm{k}+\bm{k}';\sigma}^{\phantom{\dag}}.
\end{equation}
As expected from Eq.\,(\ref{ea34o}), the commutator in Eq.\,(\ref{ea45}) does not depend on the interaction potential $\b{v}$. For the single-particle energy dispersion $\varepsilon_{\bm{k}}$ associated with the single-particle operator in Eq.\,(\ref{ea34q}), one has
\begin{equation}\label{ea48}
\varepsilon_{\bm{k}} = \frac{\hbar^2 \|\bm{k}\|^2}{2\mathsf{m}}.
\end{equation}
From the perspective of the considerations in this section, the energy dispersion in Eq.\,(\ref{ea48}) has the advantage that for it (see Eq.\,(\ref{ea45}))
\begin{equation}\label{ea49}
\varepsilon_{\bm{k}+\bm{k}'} -\varepsilon_{\bm{k}'} = \varepsilon_{\bm{k}} + \frac{\hbar^2}{\mathsf{m}}\hspace{0.6pt} \bm{k}\cdot \bm{k}'.
\end{equation}

Following the equality in Eq.\,(\ref{ea45}), on the basis of the above-mentioned canonical anti-commutation relations, one arrives at (\emph{cf.} Eq.\,(\ref{ea34p}))
\begin{equation}\label{ea46a}
[[\h{n}_{\bm{k};\sigma}^{\phantom{\dag}},\h{H}]_{-},\h{n}_{\bm{k}';\sigma'}^{\dag}]_{-} = \sum_{\bm{k}''} \big\{\varepsilon_{\bm{k}+\bm{k}''} -\varepsilon_{\bm{k}''}\big\} \big(\h{a}_{\bm{k}'';\sigma}^{\dag} \h{a}_{\bm{k}-\bm{k}'+\bm{k}'';\sigma}^{\phantom{\dag}} - \h{a}_{\bm{k}'+\bm{k}'';\sigma}^{\dag} \h{a}_{\bm{k}+\bm{k}'';\sigma}^{\phantom{\dag}}\big)\hspace{0.6pt} \delta_{\sigma,\sigma'} .
\end{equation}
With $\vert\Psi_{N;0}\rangle$ an eigenstate of the total-momentum operator $\h{\mathsf{P}}$, one observes that unless $\bm{k} = \bm{k}'$ the expectation value with respect to $\vert\Psi_{N;0}\rangle$ of the expression on the RHS of Eq.\,(\ref{ea46a}) is indeed identically vanishing, Eq.\,(\ref{ea40}).\footnote{See footnote \raisebox{-1.0ex}{\normalsize{\protect\footref{notez}}} on p.\,\pageref{WithP}.} Defining the GS momentum-distribution function for particles with spin index $\sigma$,
\begin{equation}\label{ea47}
\mathsf{n}_{\sigma}(\bm{k}) \doteq \langle\Psi_{N;0}\vert \h{a}_{\bm{k};\sigma}^{\dag}\h{a}_{\bm{k};\sigma}^{\phantom{\dag}}\vert\Psi_{N;0}\rangle,
\end{equation}
one thus obtains
\begin{equation}\label{ea48a}
\b{\chi}_{\sigma,\sigma';\infty_2}(\bm{k}) = \frac{1}{\Omega}\sum_{\bm{k}'} \big\{\varepsilon_{\bm{k}+\bm{k}'} -\varepsilon_{\bm{k}'}\big\} \big(\mathsf{n}_{\sigma}(\bm{k}') - \mathsf{n}_{\sigma}(\bm{k}+\bm{k}')\big)\hspace{0.6pt} \delta_{\sigma,\sigma'}.
\end{equation}
This expression simplifies in the case of the energy dispersion in Eq.\,(\ref{ea48}). On the basis of the expression in Eq.\,(\ref{ea49}), for this energy dispersion one obtains\,\footnote{The last equality is obtained by writing the middle expression in Eq.\,(\protect\ref{ea50}) as the difference of two sums, and subsequently evaluating the second sum by employing $\bm{k}'' \doteq \bm{k} + \bm{k}'$ as the summation variable.}
\begin{equation}\label{ea50}
\b{\chi}_{\sigma,\sigma';\infty_2}(\bm{k}) = \frac{\hbar^2}{\mathsf{m}} \frac{1}{\Omega} \sum_{\bm{k}'} \bm{k}\cdot \bm{k}'\hspace{0.6pt} \big(\mathsf{n}_{\sigma}(\bm{k}') - \mathsf{n}_{\sigma}(\bm{k}+\bm{k}')\big) \hspace{0.6pt} \delta_{\sigma,\sigma'} \equiv
2 n_{\sigma}\hspace{0.6pt}\varepsilon_{\bm{k}}\hspace{0.6pt} \delta_{\sigma,\sigma'},
\end{equation}
where
\begin{equation}\label{ea50c}
n_{\sigma} \doteq \frac{N_{\sigma}}{\Omega}
\end{equation}
is the concentration of the particles of spin index $\sigma$ in the $N$-particle GS under consideration.\,\footnote{With $N = \sum_{\sigma} N_{\sigma}$, one has $N_{\sigma} = \sum_{\bm{k}} \mathsf{n}_{\sigma}(\bm{k})$.}\footnote{\emph{Cf.} Eqs\,(2.28), (2.29), and (2.34), p.\,92, in Ref.\,\protect\citen{PN66}.} One observes that indeed $\b{\chi}_{\sigma,\sigma';\infty_2}(\bm{k})$ is an intensive quantity. From Eq.\,(\ref{ea50}) one obtains (\emph{cf.} Eq.\,(\ref{ea34e3}))
\begin{equation}\label{ea50a}
\b{\chi}_{\infty_2}(\bm{k}) \doteq \sum_{\sigma,\sigma'} \b{\chi}_{\sigma,\sigma';\infty_2}(\bm{k}) = 2n\hspace{0.6pt} \varepsilon_{\bm{k}},
\end{equation}
where (\emph{cf.} Eq.\,(\ref{ea34f}))
\begin{equation}\label{ea50b}
n = \sum_{\sigma} n_{\sigma} \equiv \frac{N}{\Omega}.
\end{equation}

For the Lindhard function \cite{AM76,PN66} $\t{\b{\chi}}_{\sigma,\sigma'}^{\X{(0)}}(\bm{k};z)$, which coincides with the $\t{\b{\chi}}_{\sigma,\sigma'}(\bm{k};z)$ corresponding to the non-interacting uniform $N$-particle GS $\vert\Phi_{N;0}\rangle$,\footnote{That is, $\vert\Phi_{N;0}\rangle$ is the $N$-particle GS of $\protect\h{H}_{\protect\X{0}} \equiv \sum_{\bm{k},\sigma} \varepsilon_{\bm{k}}\hspace{1.0pt} \protect\h{a}_{\bm{k};\sigma}^{\dag} \protect\h{a}_{\bm{k};\sigma}^{\phantom{\dag}}$.} one has
\begin{equation}\label{ea51}
\t{\b{\chi}}_{\sigma,\sigma'}^{\X{(0)}}(\bm{k};z) = \frac{1}{\Omega} \sum_{\bm{k}'} \frac{\Theta(\varepsilon_{\textsc{f}} -\varepsilon_{\bm{k}'}) - \Theta(\varepsilon_{\textsc{f}} - \varepsilon_{\bm{k}+\bm{k}'})}{z - \varepsilon_{\bm{k}+\bm{k}'} +\varepsilon_{\bm{k}'}}\hspace{1.2pt} \delta_{\sigma,\sigma'},
\end{equation}
where $\varepsilon_{\textsc{f}}$ is the Fermi energy corresponding to $\vert\Phi_{N;0}\rangle$. Here,
\begin{equation}\label{ea52}
\Theta(\varepsilon_{\textsc{f}} - \varepsilon_{\bm{k}}) \equiv \mathsf{n}_{\sigma}^{\X{(0)}}(\bm{k}),
\end{equation}
where $\mathsf{n}_{\sigma}^{\X{(0)}}(\bm{k})$ denotes the momentum distribution function, Eq.\,(\ref{ea47}), corresponding to $\vert\Phi_{N;0}\rangle$. Although $\mathsf{n}_{\sigma}^{\X{(0)}}(\bm{k}) \not\equiv \mathsf{n}_{\sigma}(\bm{k})$, in the specific case of the single-particle energy dispersion $\varepsilon_{\bm{k}}$ in Eq.\,(\ref{ea48}), one has \cite{BF99,BF99a}\footnote{For the \textsl{uniform} $N$-particle GSs $\vert\Psi_{N;0}\rangle$ and $\vert\Phi_{N;0}\rangle$ considered here it is relevant that both correspond to the same particle numbers $\{N_{\sigma}\| \sigma\}$. In the considerations of Refs\,\protect\citen{BF99a} and \protect\citen{BF99} $\vert\Psi_{N;0}\rangle$ and $\vert\Phi_{N;0}\rangle$ are non-uniform $N$-particle GSs, and the counterpart of the equality in Eq.\,(\protect\ref{ea51x}) applies when for the corresponding number densities, respectively $n_{\sigma}(\bm{r})$ and $n_{\sigma}^{\protect\X{(0)}}(\bm{r})$, one has $n_{\sigma}(\bm{r}) \equiv n_{\sigma}^{\protect\X{(0)}}(\bm{r})$, $\forall\sigma$. This is the case when the `non-interacting' Hamiltonian $\protect\h{H}_{\protect\X{0}}$ coincides with the many-body Kohn-Sham \protect\cite{KS65} Hamiltonian. For such Kohn-Sham Hamiltonian to exist, it is required that $n_{\sigma}(\bm{r})$, $\forall\sigma$, be so-called pure-state non-interacting $v$-representable \protect\cite{BF97,BF97b,BF99}. In this connection, the expressions in Eqs\,(\protect\ref{ea34u}) and (\protect\ref{ea34s}) make evident the significance of $n_{\sigma}(\bm{r})$ and $n(\bm{r})$ to respectively $\chi_{\sigma,\sigma';\infty_2}(\bm{r},\bm{r}')$ and $\chi_{\infty_2}(\bm{r},\bm{r}')$.}
\begin{equation}\label{ea51x}
\b{\chi}_{\sigma,\sigma';\infty_2}^{\X{(0)}}(\bm{k}) \equiv \b{\chi}_{\sigma,\sigma';\infty_2}(\bm{k}).
\end{equation}

With reference to Eq.\,(\ref{ea40a}), on writing the function $\t{\b{\chi}}_{\sigma,\sigma'}^{\X{(0)}}(\bm{k};z)$ in Eq.\,(\ref{ea51}) as difference of two functions, the summand of the first one involving $\Theta(\varepsilon_{\textsc{f}}-\varepsilon_{\bm{k}'})$ and that of the second one  $\Theta(\varepsilon_{\textsc{f}}-\varepsilon_{\bm{k}+\bm{k}'})$, it is tempting to assume that the first and the second functions were respectively $\t{\b{\chi}}^{\X{(0)}\X{+}}(\bm{k};z)$ and $\t{\b{\chi}}^{\X{(0)}\X{+}}(\bm{k};z)$. This is however \textsl{not} the case, as we shall now demonstrate. Starting from the expression\,\footnote{See \S\,3.3 of Ref.\,\protect\citen{BF19}. The minus sign on the RHS of Eq.\,(\protect\ref{ea51a}) is specific to fermions.}
\begin{equation}\label{ea51a}
\chi^{\X{(0)}}(1,2) = -\frac{\ii}{\hbar} G_{\X{0}}(1,2) G_{\X{0}}(2,1),
\end{equation}
one obtains\,\footnote{With $1 \equiv \bm{r}t\sigma$ and $2\equiv \bm{r}'t'\sigma'$, one has $G_{\protect\X{0}}(1,2) \equiv G_{\protect\X{0};\sigma,\sigma'}(\bm{r}t,\bm{r}'t')$. The $\delta_{\sigma,\sigma'}$ on the RHS of Eq.\,(\protect\ref{ea51b}) follows on account of the equality $G_{\protect\X{0};\sigma,\sigma'}(\bm{r}t,\bm{r}'t') = G_{\protect\X{0};\sigma}(\bm{r}t,\bm{r}'t')\hspace{0.6pt} \delta_{\sigma,\sigma'}$ [\S\,2.2.2 in Ref.\,\protect\citen{BF19}], where $G_{\protect\X{0};\sigma}(\bm{r}t,\bm{r}'t') \equiv G_{\protect\X{0};\sigma,\sigma}(\bm{r}t,\bm{r}'t')$, or $G_{\protect\X{0};\sigma} = (\mathbb{G}_{\protect\X{0}})_{\sigma,\sigma}$. See footnotes \raisebox{-1.0ex}{\normalsize{\protect\footref{notef1}}} and \raisebox{-1.0ex}{\normalsize{\protect\footref{noteg1}}} on pp.\,\protect\pageref{ForSpin} and \protect\pageref{WithRefTo} as well as the caption of Fig.\,\protect\ref{f12}, p.\,\protect\pageref{Two2ndO}.}
\begin{equation}\label{ea51b}
\t{\chi}^{\X{(0)}}_{\sigma,\sigma'}(\bm{r},\bm{r}';z) =  \big\{\t{p}_{\sigma}(\bm{r},\bm{r}';z) +\t{p}_{\sigma}(\bm{r}',\bm{r};-z) \big\}\hspace{0.6pt} \delta_{\sigma,\sigma'},
\end{equation}
from which it follows that (\emph{cf.} Eq.\,(\ref{ea36a}))
\begin{equation}\label{ea51c}
\t{\b{\chi}}_{\sigma,\sigma'}^{\X{(0)}}(\bm{k},\bm{k}';z) = \big\{\t{\b{p}}_{\sigma}(\bm{k},\bm{k}';z) +\t{\b{p}}_{\sigma}(-\bm{k}',-\bm{k};-z) \big\}\hspace{0.6pt} \delta_{\sigma,\sigma'},
\end{equation}
where, for the uniform GS $\vert\Phi_{N;0}\rangle$ considered here,
\begin{equation}\label{ea51d}
\t{\b{p}}_{\sigma}(\bm{k},\bm{k}';z) = \t{\b{p}}_{\sigma}(\bm{k};z)\hspace{0.6pt} \delta_{\bm{k},\bm{k}'},
\end{equation}
in which
\begin{equation}\label{ea51e}
\t{\b{p}}_{\sigma}(\bm{k};z) = \frac{1}{\Omega} \sum_{\bm{k}'} \frac{\Theta(\varepsilon_{\textsc{f}}-\varepsilon_{\bm{k}'}) \Theta(\varepsilon_{\bm{k}+\bm{k}'}-\varepsilon_{\textsc{f}})}{z - \varepsilon_{\bm{k}+\bm{k}'} +\varepsilon_{\bm{k}}},\;\;\forall \sigma.
\end{equation}
As a consequence of the product the $\Theta$-functions in the numerator of the above expression, one observes that $\t{\b{p}}_{\sigma}(\bm{k};z)$ is analytic in the region $\re[z] < 0$. This is relevant, since on account of $e_s \ge 0$, Eq.\,(\ref{ea20}), the function $\t{\b{\chi}}^+(\bm{k};z)$ as defined in Eq.\,(\ref{ea40a}) is analytic in the region $\re[z] < 0$. One thus obtains
\begin{align}\label{ea52a}
\t{\b{\chi}}^{\X{(0)}\X{+}}(\bm{k};z) &=  \frac{2}{\Omega} \sum_{\bm{k}'} \frac{\Theta(\varepsilon_{\textsc{f}}-\varepsilon_{\bm{k}'}) \Theta(\varepsilon_{\bm{k}+\bm{k}'}-\varepsilon_{\textsc{f}})}{z - \varepsilon_{\bm{k}+\bm{k}'} +\varepsilon_{\bm{k}'}},\nonumber\\
\t{\b{\chi}}^{\X{(0)}\X{-}}(\bm{k};z) &= \frac{2}{\Omega} \sum_{\bm{k}'} \frac{\Theta(\varepsilon_{\bm{k}'}-\varepsilon_{\textsc{f}}) \Theta(\varepsilon_{\textsc{f}} -\varepsilon_{\bm{k}+\bm{k}'})}{z - \varepsilon_{\bm{k}+\bm{k}'} +\varepsilon_{\bm{k}'}}.
\end{align}
By a change of the variable of summation in one of the two expressions in Eq.\,(\ref{ea52a}), one establishes the following equality for the functions in Eq.\,(\ref{ea52a}):
\begin{equation}\label{ea52a1}
\t{\b{\chi}}^{\X{(0)}\X{+}}(\bm{k};z) \equiv -\t{\b{\chi}}^{\X{(0)}\X{-}}(-\bm{k};-z),
\end{equation}
and on account time-reversal symmetry, whereby $\varepsilon_{-\bm{k}} \equiv \varepsilon_{\bm{k}}$, $\forall \bm{k}$ (\emph{cf.} Eq.\,(\ref{eac6b})),
\begin{equation}\label{ea52b}
\t{\b{\chi}}^{\X{(0)}\X{+}}(\bm{k};z) \equiv -\t{\b{\chi}}^{\X{(0)}\X{-}}(\bm{k};-z),
\end{equation}
in conformity with Eq.\,(\ref{ea40b1}).

We note in passing that since
\begin{equation}\label{ea52c}
\frac{\Theta(\varepsilon_{\textsc{f}} -\varepsilon_{\bm{k}'}) - \Theta(\varepsilon_{\textsc{f}} -\varepsilon_{\bm{k}+\bm{k}'})}{-\varepsilon_{\bm{k}+\bm{k}'} +\varepsilon_{\bm{k}'}} \sim -\delta(\varepsilon_{\textsc{f}}-\varepsilon_{\bm{k}'})\;\; \text{for}\;\; \bm{k}\to \bm{0},
\end{equation}
whereby
\begin{equation}\label{ea52d}
\t{\b{\chi}}_{\sigma,\sigma}^{\X{(0)}}(\bm{k};0) \sim -\frac{1}{\Omega} \sum_{\bm{k}'} \delta(\varepsilon_{\textsc{f}}-\varepsilon_{\bm{k}'}) \underbrace{\sim}_{\Omega\to\infty} -\int \frac{\mathrm{d}^dk'}{(2\pi)^d}\,  \delta(\varepsilon_{\textsc{f}}-\varepsilon_{\bm{k}'})\equiv -\EuScript{D}(\varepsilon_{\textsc{f}})\;\;\text{for}\;\; \bm{k}\to \bm{0},
\end{equation}
where $\EuScript{D}(\varepsilon_{\textsc{f}})$ is the \textsl{spin-resolved} density of one-particle states associated with the energy dispersion $\varepsilon_{\bm{k}}$ at the Fermi energy $\varepsilon_{\textsc{f}}$. Sine $\EuScript{D}(\varepsilon_{\textsc{f}}) > 0$ for non-vanishing particle densities, the result in Eq.\,(\ref{ea52d}) implies that $\t{\b{\chi}}_{\sigma,\sigma}^{\X{(0)}}(\bm{0};0) < 0$, contradicting the fact that $\t{\b{\chi}}_{\sigma,\sigma}^{\X{(0)}}(\bm{0};0) \equiv 0$.\footnote{See footnote \raisebox{-1.0ex}{\normalsize{\protect\footref{noteo}}} on page \protect\pageref{OnAccountOf}.} Thus, as regards $\t{\b{\chi}}_{\sigma,\sigma}^{\X{(0)}}(\bm{k};z)$, in the thermodynamic limit the limits $\bm{k}\to \bm{0}$ and $z\to 0$ do \textsl{not} commute.

Having established the existence of the elements of the sequence $\{\b{\chi}_{\infty_j}(\bm{k})\| j\}$ for Hubbard-like models, we now proceed with establishing the existence of the sequence $\{P_{\sigma;\infty_j}(\bm{k})\| j\}$ of the coefficients corresponding to the asymptotic series of the function $\t{P}(\bm{k};z)$ according to
\begin{equation}\label{ea53}
\t{P}(\bm{k};z) \sim \sum_{j=1}^{\infty} \frac{P_{\infty_j}(\bm{k})}{z^j}\;\;\text{for}\;\; z\to\infty.
\end{equation}
To this end, we first note that from Eq.\,(\ref{ea3}) one has
\begin{equation}\label{ea54}
\h{P}(z) = \big(\h{I} + \h{\chi}(z) \h{v}\big)^{-1} \h{\chi}(z) \equiv \h{\chi}(z) \big(\h{I} + \h{v} \h{\chi}(z)\big)^{-1},
\end{equation}
which in the case at hand takes the form\,\footnote{The simplicity of the closed expression in Eq.\,(\protect\ref{ea55}) is owing to the spin-independence of the two-body interacting potential $\b{v}(\bm{k})$, whereby instead of having to deal with $\protect\t{\protect\b{\chi}}_{\sigma,\sigma'}(\bm{k};z)$ one deals with $\protect\t{\protect\b{\chi}}(\bm{k};z)$, Eq.\,(\protect\ref{ea34e1}). See footnote \raisebox{-1.0ex}{\normalsize{\protect\footref{notep}}} on p.\,\protect\pageref{TheAssumption}.}
\begin{equation}\label{ea55}
\t{P}(\bm{k};z) = \frac{\t{\b{\chi}}(\bm{k};z)}{1 + \b{v}(\bm{k}) \t{\b{\chi}}(\bm{k};z)}.
\end{equation}
Diagrammatically, the denominator of the expression on the RHS of this equality cancels the contributions of the improper polarisation diagrams to the function in the numerator.

With\,\footnote{Following Eqs\,(\protect\ref{ea40a}) and (\protect\ref{ea40h}), one has $\protect\b{\chi}_{\infty_j}(\bm{k}) = \protect\b{\chi}_{\infty_j}^+(\bm{k}) - \protect\b{\chi}_{\infty_j}^-(\bm{k})$, which on account of Eq.\,(\protect\ref{ea40j}) reduces to $\protect\b{\chi}_{\infty_j}(\bm{k}) = (1 + (-1)^j) \hspace{0.6pt}\protect\b{\chi}_{\infty_j}^+(\bm{k})$.} (\emph{cf.} Eq.\,(\ref{ea24}))
\begin{equation}\label{ea56}
\t{\b{\chi}}(\bm{k};z) \sim \sum_{j=1}^{\infty} \frac{\b{\chi}_{\infty_j}(\bm{k})}{z^j}\;\;\text{for}\;\; z\to\infty,
\end{equation}
where (\emph{cf.} Eqs\,(\ref{ea40}) and (\ref{ea34e3}))
\begin{equation}\label{ea56a}
\b{\chi}_{\infty_{j}}(\bm{k}) = \frac{1}{\Omega}\hspace{0.6pt}\langle\Psi_{N;0}\vert [(\h{L}^{j-1} \h{n}_{\bm{k}}^{\phantom{\dag}}), \hspace{0.6pt}\h{n}_{\bm{k}}^{\dag}]_{-} \vert\Psi_{N;0}\rangle,\;\; \forall j \in \mathds{N},
\end{equation}
one has
\begin{equation}\label{ea57}
\frac{1}{1 +  \b{v}(\bm{k})\t{\b{\chi}}(\bm{k};z)} \sim 1- \sum_{j=1}^{\infty} \frac{F_{\infty_j}(\bm{k})}{z^j}\;\; \text{for}\;\; z\to\infty,
\end{equation}
in which the coefficients $\{F_{\infty_j}(\bm{k}) \| j\}$ are recursively deduced from the following equalities:\,\footnote{Consult the algebraic details beginning from Eq.\,(2.102) of Ref.\,\protect\citen{BF19}.}
\begin{align}\label{ea58}
F_{\infty_1}(\bm{k}) &= \b{v}(\bm{k})\hspace{0.6pt} \b{\chi}_{\infty_1}(\bm{k}),\nonumber\\
F_{\infty_j}(\bm{k}) &= \b{v}(\bm{k}) \big(\b{\chi}_{\infty_j}(\bm{k}) - \sum_{j'=1}^{j-1} \b{\chi}_{\infty_{j'}}(\bm{k}) F_{\infty_{j-j'}}(\bm{k})\big),\;\; j\ge 2.
\end{align}
Multiplying the expressions on the RHSs of Eqs\,(\ref{ea56}) and (\ref{ea57}), on account of Eqs\,(\ref{ea53}) and (\ref{ea55}), one obtains
\begin{align}\label{ea59}
P_{\infty_1}(\bm{k}) &= \b{\chi}_{\infty_1}(\bm{k}),\nonumber\\
P_{\infty_j}(\bm{k}) &= \b{\chi}_{\infty_j}(\bm{k}) - \sum_{j'=1}^{j-1} \b{\chi}_{\infty_{j'}}(\bm{k})\hspace{0.6pt} F_{\infty_{j-j'}}(\bm{k}),\;\; j \ge 2.
\end{align}
Comparison of the equalities in Eqs\,(\ref{ea58}) and (\ref{ea59}), one observes that
\begin{equation}\label{ea60}
P_{\infty_j}(\bm{k}) \equiv \frac{F_{\infty_j}(\bm{k})}{\b{v}(\bm{k})},\;\; \forall j\in\mathds{N}.
\end{equation}
Taking account of the identity $\b{\chi}_{\infty_{2j-1}}(\bm{k}) \equiv 0$, $\forall j\in\mathds{N}$, Eq.\,(\ref{ea40k}), from the recursive expressions in Eq.\,(\ref{ea58}) and Eq.\,(\ref{ea60}) one obtains \cite{Note15}\footnote{As regards the identity $P_{\infty_2}(\bm{k}) \equiv \b{\chi}_{\infty_2}(\bm{k})$, see also Refs\,\protect\citen{BF99a} and \protect\citen{BF99}.}\refstepcounter{dummy}\label{FollowingEq}\footnote{Following Eq.\,(\protect\ref{ea3}), one has $\protect\t{\protect\b{\chi}}(\bm{k};z) = \protect\t{P}(\bm{k};z)/\big(1 - \protect\b{v}(\bm{k}) \protect\t{P}(\bm{k};z)\big)$. Comparing this expression with that in Eq.\,(\protect\ref{ea55}), one immediately infers that the expression for $\protect\b{\chi}_{\infty_j}(\bm{k})$ in terms of $\{P_{\infty_i}(\bm{k}) \| i=2, \dots, j\}$ is obtained from that for $P_{\infty_j}(\bm{k})$ in terms of $\{\protect\b{\chi}_{\infty_i}(\bm{k}) \| i=2, \dots, j\}$ by the exchange $\protect\b{\chi} \rightleftharpoons P$ of symbols and substitution of $\protect\b{v}(\bm{k})$ by $-\protect\b{v}(\bm{k})$. Thus, \emph{e.g.} \protect\cite{Note15}, $\protect\b{\chi}_{\infty_4}(\bm{k}) = P_{\infty_4}(\bm{k}) + \protect\b{v}(\bm{k}) (P_{\infty_2}(\bm{k}))^2$. \label{noteh1}}
\begin{align}\label{ea61}
P_{\infty_1}(\bm{k}) &\equiv 0,\;\; P_{\infty_2}(\bm{k}) \equiv \b{\chi}_{\infty_2}(\bm{k}),\;\;
P_{\infty_3}(\bm{k}) \equiv 0,\nonumber\\
P_{\infty_4}(\bm{k}) &\equiv \b{\chi}_{\infty_4}(\bm{k}) - \b{v}(\bm{k}) (\b{\chi}_{\infty_2}(\bm{k}))^2,\;\; P_{\infty_5}(\bm{k}) \equiv 0,\nonumber\\
P_{\infty_6}(\bm{k}) &\equiv \b{\chi}_{\infty_6}(\bm{k}) - 2 \b{v}(\bm{k}) \b{\chi}_{\infty_2}(\bm{k}) \b{\chi}_{\infty_4}(\bm{k}) + (\b{v}(\bm{k}))^2 (\b{\chi}_{\infty_2}(\bm{k}))^3,\; \text{\emph{etc.}}
\end{align}
With reference to the discussions following Eq.\,(\ref{e4r}) regarding the expression for the asymptotic coefficient $G_{\sigma;\infty_j}(\bm{k})$ in Eq.\,(\ref{e4p}), the expression in Eq.\,(\ref{ea56a}) reveals that for Hubbard-like models, described in terms of the Hamiltonian $\h{\mathcal{H}}$, the asymptotic coefficient $\b{\chi}_{\infty_{j}}(\bm{k})$ is bounded for arbitrary finite values of $j$. Since $P_{\infty_j}(\bm{k})$ is determined by $\{\b{\chi}_{\infty_2}(\bm{k}),\dots,\b{\chi}_{\infty_j}(\bm{k})\}$, it follows that for Hubbard-like models $P_{\infty_j}(\bm{k})$ is similarly bounded for arbitrary finite values of $j$. By the same reasoning as in the case of $\{G_{\sigma;\infty_j}(\bm{k})\| j\}$, \S\,\ref{sec.b.1}, for Hubbard-like models the positive sequence $\{P_{\infty_j}(\bm{k}) \| j\}$ can be shown to be similarly \textsl{determinate}.\footnote{See in particular the example to which the expression in Eq.\,(\protect\ref{e4oe}) corresponds.}

We point out that the expression in Eq.\,(\ref{ea56a}) makes explicit that $\b{\chi}_{\infty_j}(\bm{k})$ is a polynomial of order $j-1$ in the dimensionless coupling constant $\lambda$ of the two-body interaction potential, Eq.\,(\ref{e4r}). In the light of the expressions in Eq.\,(\ref{ea61}), the same applies to $P_{\infty_j}(\bm{k})$. Further, since $\b{\chi}_{\infty_j}(\bm{k})$ is defined in terms of the \textsl{interacting} $N$-particle GS $\vert\Psi_{N;0}\rangle$, there is a direct correspondence between the elements of the sequence $\{P_{\infty_j}(\bm{k}) \| j\}$ and those of the sequence $\{\t{P}^{\X{(\nu)}}(\bm{k};z)\| \nu\}$ corresponding to the perturbation series expansion of $\t{P}(\bm{k};z)$ in terms of the \textsl{interacting} Green functions $\{\t{G}_{\sigma}(\bm{k};z)\| \sigma\}$ and the bare two-body interaction potential, Eq.\,(\ref{ea6a}).\footnote{Consider the very similar remarks regarding $G_{\sigma;\infty_j}(\bm{k})$ following Eq.\,(\protect\ref{e4r}).}\refstepcounter{dummy}\label{SimilarToChi}\footnote{Like $\b{\chi}_{\infty_j}(\bm{k})$, $\b{\chi}^{\pm}_{\infty_j}(\bm{k})$ are polynomials of order $j-1$ in $\lambda$, Eq.\,(\protect\ref{ea40i}). On account of this fact, one may sum the perturbation series expansion for $\protect\t{\protect\b{\chi}}(\bm{k};z)$ on the basis of the function $\protect\t{\mathfrak{X}}(\bm{k};z,\varepsilon) \doteq \sum_{\nu=0}^{\infty} [\protect\t{\protect\b{\chi}}^{\protect\X{(\nu)}}(\bm{k};z)/\protect\b{\chi}_{\infty_{\nu+1}}^{+}(\bm{k})] \hspace{0.6pt}\varepsilon^{\nu}$ (\emph{cf.} Eq.\,(\protect\ref{e101})). All the considerations regarding the function $\protect\t{\mathfrak{G}}_{\sigma}(\bm{k};z,\varepsilon)$ in \S\,\protect\ref{sec.5} apply to the function $\protect\t{\mathfrak{X}}(\bm{k};z,\varepsilon)$, provided that in \S\,\protect\ref{sec.5} the function $A_{\sigma}(\bm{k};\varepsilon)$ be replaced by $\mathcal{A}_{\chi}(\bm{k};\varepsilon)$, Eqs\,(\protect\ref{ea40g1}) and (\protect\ref{ea40l}), the integrals with respect to $\varepsilon$ and $\varepsilon'$ over the interval $(-\infty,\infty)$ by integrals over the interval $[0,\infty)$, and the functions $\{G_{\sigma;\infty_j}(\bm{k})\| j\}$ by $\{\protect\b{\chi}_{\infty_j}^{+}(\bm{k})\| j\}$. We note that the perturbational contributions $\{\protect\t{\protect\b{\chi}}^{\protect\X{(\nu)}}(\bm{k};z)\| \nu\}$ can be recursively deduced from the perturbational contributions $\{\protect\t{P}^{\protect\X{(\nu)}}(\bm{k};z)\| \nu\}$ along the lines of Eqs\,(\protect\ref{ea57}) -- (\protect\ref{ea61}) (see footnote \raisebox{-1.0ex}{\normalsize{\protect\footref{noteh1}}} on p.\,\protect\pageref{FollowingEq}). \label{notei1}}

From Eq.\,(\ref{ea55}) one obtains
\begin{equation}\label{ea62}
\im[\t{P}(\bm{k};z)] = \frac{\im[\t{\b{\chi}}(\bm{k};z)]}{\vert 1 +  \b{v}(\bm{k})\t{\b{\chi}}(\bm{k};z)\vert^2},
\end{equation}
from which it follows that
\begin{equation}\label{ea63}
\sgn(\im[\t{P}(\bm{k};z)]) = \sgn(\im[\t{\b{\chi}}(\bm{k};z)]).
\end{equation}
Thus, in the light of Eq.\,(\ref{ea40e}),
\begin{equation}\label{ea64}
\mp\t{P}(\bm{k};z)\;\; \text{is Nevanlinna for}\; \re[z] \gtrless 0.
\end{equation}

\refstepcounter{dummyX}
\section{The classical moment problem in relation to
\texorpdfstring{$\protect\t{G}_{\sigma}(\bm{k};z)$}{} and
\texorpdfstring{$\protect\t{\Sigma}_{\sigma}(\bm{k};z)$}{}}
\phantomsection
\label{sab}
The details underlying the alternative proof in \S\,\ref{sec.3.2} of the uniform convergence of the series in Eq.\,(\ref{e4}) almost everywhere (\ae\footnote{Appendix \protect\ref{sae}.}) towards the exact self-energy $\t{\Sigma}_{\sigma}(\bm{k};z)$, which we present in this appendix, are of relevance to a range of problems -- some of which we encounter in this publication, justifying the relatively broad basis of the approach adopted in this appendix.\footnote{The recursion method and continued-fraction expansion as key elements of the classical moment problem, encountered in this appendix, have been utilised in the past in the context of electronic-structures calculations within mean-field frameworks, where the solution of an interacting many-body problem is reduced to the self-consistent solution of a set of time-independent one-particle Schr\"{o}dinger equations (or eigenvalue problems); for periodic solids (whether naturally periodic, or artificially so -- brought about by the use of super-cells whereby an aperiodic system is extended periodically into the entire space), each one-particle Schr\"{o}dinger equation in this set corresponds to an irreducible representation of the translation part (characterised by the Bravais lattice $\{\bm{R}_i\| i\}$) of the underlying space group \protect\cite{AM76} associated with a point $\bm{k}$ of the relevant $\protect\1BZ$. The aim had been to bypass the practical limitations of electronic-structures calculations based on \textsl{extended} plane waves -- as the basis functions of the said irreducible representations (\emph{i.e.} the basis functions of the one-particle Hilbert space spanned by the Bloch wave functions at each $\bm{k}\in\protect\1BZ$) by taking advantage of the localised nature of the atomic orbitals centred on each constituent atom of a solid. In fact, the Hubbard model \protect\cite{JH63}, which is a focus of the present publication, was born in part out of the realisation that in systems with narrow electronic bands near the chemical potential, corresponding largely to hybridised \textsl{localised} atomic valence orbitals, it is a necessity rather than a choice to approach the solid from the atomic limit, as opposed to the free-electron one \protect\cite{PWA59,PWA97}. For an overview of the recursion method and associated continued-fraction expansion as applied to one-particle Schr\"{o}dinger equations, we refer the reader to the contributions by Heine, \emph{et al.} in Ref.\,\protect\citen{HBHK80}. For some of the earlier experimental (describing line-widths and shapes) and theoretical uses of the moments of various physical spectral functions, the reader may further consult  Refs\,\protect\citen{LJFB43,vV48,PS50,HL67,WN72,NO80,NB89,EFNM91,EF93,EO94,EOMS94,BF02,BF07}.} In \S\,\ref{sec.b.1} we consider some aspects related to the single-particle Green function $\t{G}_{\sigma}(\bm{k};z)$. These aspects have direct bearing on the self-energy $\t{\Sigma}_{\sigma}(\bm{k};z)$, which function we consider in \S\,\ref{sec.b2}.

\refstepcounter{dummyX}
\subsection{The Green function}
\phantomsection
\label{sec.b.1}
We begin with the expression [Eqs\,(B.40) and (B.47) in Ref.\,\citen{BF07}]
\begin{equation}\label{e4e}
\t{G}_{\sigma}(\bm{k};z) = \int_{-\infty}^{\infty}\rd\varepsilon'\; \frac{A_{\sigma}(\bm{k};\varepsilon')}{z-\varepsilon'}
\end{equation}
for the Green function in terms of the associated one-particle spectral function
\begin{equation}\label{e4d}
A_{\sigma}(\bm{k};\varepsilon) \doteq \pm\frac{1}{\pi}\im[\t{G}_{\sigma}(\bm{k};\varepsilon\mp \ii 0^+),\;\; \varepsilon \in\mathds{R}.
\end{equation}
Following the equality in Eq.\,(\ref{e2}), one has [\emph{cf.} Eq.\,(B.41) in Ref.\,\citen{BF07}]
\begin{equation}\label{e4f}
A_{\sigma}(\bm{k};\varepsilon) \ge 0,\;\;\forall \bm{k}, \varepsilon.
\end{equation}
Owing to this property,
\begin{equation}\label{e4g}
\upgamma_{\sigma}(\bm{k};\varepsilon) \doteq \int_{-\infty}^{\varepsilon} \rd\varepsilon'\; A_{\sigma}(\bm{k};\varepsilon')
\end{equation}
is a non-decreasing function of $\varepsilon$ in terms of which the Green function $\t{G}_{\sigma}(\bm{k};z)$ in Eq.\,(\ref{e4e}) can be equivalently expressed as the following Stieltjes integral \cite{THH18,PRH50,APM11}:\footnote{See also \S\,376, p.\,538, in Ref.\,\protect\citen{EWH27}.}
\begin{equation}\label{e4j}
\t{G}_{\sigma}(\bm{k};z) = \int_{-\infty}^{\infty} \frac{\rd \upgamma_{\sigma}(\bm{k};\varepsilon')}{z-\varepsilon'}.
\end{equation}
Barring a countable set of energies $\{\varepsilon_{\sigma;l}(\bm{k})\| l\}$, which may be empty, at which the one-particle spectral function $A_{\sigma}(\bm{k};\varepsilon)$ is proportional to a $\delta$-function, for an arbitrary $\bm{k}$ the measure function \cite{PRH50,APM11} (concisely, \textsl{the measure}) $\upgamma_{\sigma}(\bm{k};\varepsilon)$ in Eq.\,(\ref{e4g}) is a \textsl{continuous} function of $\varepsilon$. For $\{\varepsilon_{\sigma;l}(\bm{k})\| l\}$ non-empty, the energies in the set are those of the one-particle excitations described in terms of well-defined Landau quasi-particles \cite{PN64,PN66}. In the cases where for a given $\bm{k}$ the function $A_{\sigma}(\bm{k};\varepsilon)$ is described \textsl{solely} in terms of $\delta$-functions, the associated measure $\upgamma_{\sigma}(\bm{k};\varepsilon)$ is a step-wise constant function of $\varepsilon$, discontinuously increasing from one constant value to a higher one on $\varepsilon$ increasing from below each $\varepsilon_{\sigma;l}(\bm{k})$ to above this excitation energy (\emph{cf.} Eqs\,(\ref{e6s}) and (\ref{e6t}) below).\refstepcounter{dummy}\label{ForInstance}\footnote{For instance, in the case of the `Hubbard atom' $\gamma(\varepsilon) = \hbar \{\Theta(\varepsilon + U/2) + \Theta(\varepsilon-U/2)\}/2$, Eq.\,(\protect\ref{exc1}) below. For the function $\upeta(u)$ in Eq.\,(\protect\ref{e4k}) below, one has $\upeta(u) = \gamma(\upvarepsilon_{\protect\X{0}} u)/(\upvarepsilon_{\protect\X{0}} \{1+ (U/(2\upvarepsilon_{\protect\X{0}}))^2\})$. \label{notej}} As regards the domain of variation of the measure $\upgamma_{\sigma}(\bm{k};\varepsilon)$, in the light of the expression in Eq.\,(\ref{e4g}) one clearly has $\upgamma_{\sigma}(\bm{k};\varepsilon)/\hbar \downarrow 0$ for $\varepsilon \downarrow -\infty$. Further, on account of the exact sum-rule [Eqs\,(B.39) and (B.41) in Ref.\,\citen{BF07}]
\begin{equation}\label{e4h}
\frac{1}{\hbar} \int_{-\infty}^{\infty} \rd\varepsilon'\; A_{\sigma}(\bm{k};\varepsilon') = 1,\;\; \forall \bm{k},
\end{equation}
from the expression in Eq.\,(\ref{e4g}) one infers that $\upgamma_{\sigma}(\bm{k};\varepsilon)/\hbar \uparrow 1$ for $\varepsilon\uparrow \infty$, $\forall\bm{k}$. Thus
\begin{equation}\label{e4i}
0\le \upgamma_{\sigma}(\bm{k};\varepsilon)/\hbar \le 1,\;\; \forall\bm{k}, \varepsilon.
\end{equation}
As indicated above, owing to $A_{\sigma}(\bm{k};\varepsilon) \ge 0$, Eq.\,(\ref{e4f}), $\upgamma_{\sigma}(\bm{k};\varepsilon)$ is a non-decreasing function of $\varepsilon$, beginning at $0$ for $\varepsilon = -\infty$ and steadily increasing and reaching $\hbar$ for $\varepsilon=\infty$. For $A_{\sigma}(\bm{k};\varepsilon)$ a function of bounded support, one for which $A_{\sigma}(\bm{k};\varepsilon) \equiv 0$, $\forall \varepsilon \le \varepsilon_{\mathrm{l}}$ and $\forall \varepsilon \ge \varepsilon_{\mathrm{u}}$ for finite values of $\varepsilon_{\mathrm{l}}$ and $\varepsilon_{\mathrm{u}}$,\refstepcounter{dummy}\label{ThisSpecification}\footnote{This specification does not rule out the possibility that $A_{\sigma}(\bm{k};\varepsilon) \equiv 0$ for, say, all $\varepsilon \in (\varepsilon_{\mathrm{l}}',\varepsilon_{\mathrm{u}}')$, where $\varepsilon_{\mathrm{l}} <\varepsilon_{\mathrm{l}}'$ and $\varepsilon_{\mathrm{u}}' <\varepsilon_{\mathrm{u}}$. Generally, in particular from the mathematical perspective, the support of $A_{\sigma}(\bm{k};\varepsilon)$ along the $\varepsilon$-axis, whether bounded or otherwise, need not be simply-connected. \label{notek1}} one has $\upgamma_{\sigma}(\bm{k};\varepsilon) \equiv 0$, $\forall \varepsilon \le \varepsilon_{\mathrm{l}}$, and $\upgamma_{\sigma}(\bm{k};\varepsilon) \equiv \hbar$, $\forall\varepsilon \ge \varepsilon_{\mathrm{u}}$.

Defining the function\,\refstepcounter{dummy}\label{Conversely}\footnote{Conversely, one has $\upgamma_{\sigma}(\bm{k};\varepsilon) = \upvarepsilon_{\X{0}} \int_{-\infty}^{\varepsilon/\upvarepsilon_{\protect\X{0}}} \mathrm{d}\upeta_{\sigma}(\bm{k};u)\hspace{0.6pt} (1+ u^2)$, which alternatively can be expressed as $\upgamma_{\sigma}'(\bm{k};\varepsilon) = (1+(\varepsilon/\upvarepsilon_{\protect\X{0}})^2) \hspace{0.6pt} \upeta_{\sigma}'(\bm{k};\varepsilon/\upvarepsilon_{\X{0}})$. \label{notew}}
\begin{equation}\label{e4k}
\upeta_{\sigma}(\bm{k};u) \doteq \frac{1}{\upvarepsilon_{\X{0}}} \int_{-\infty}^{\upvarepsilon_{\X{0}} u} \frac{\rd\upgamma_{\sigma}(\bm{k};\varepsilon)}{1+(\varepsilon/\upvarepsilon_{\X{0}})^2} \equiv \int_{-\infty}^{u} \rd u'\, \frac{A_{\sigma}(\bm{k};\upvarepsilon_{\X{0}} u')}{1 + {u'}^2},\;\;\forall \upvarepsilon_{\X{0}} > 0,
\end{equation}
from the expression in Eq.\,(\ref{e4j}) one obtains the following equivalent expression:
\begin{equation}\label{e4l}
-\t{G}_{\sigma}(\bm{k};z) = \int_{-\infty}^{\infty} \rd\upeta_{\sigma}(\bm{k};u)\,u + \int_{-\infty}^{\infty} \rd\upeta_{\sigma}(\bm{k};u)\,\frac{1+ u z/\upvarepsilon_{\X{0}}}{u -z/\upvarepsilon_{\X{0}}}.
\end{equation}
With $z$ having the dimensionality of energy (in the SI base units, $\llbracket z\rrbracket = \mathrm{J}$), the constant parameter $\upvarepsilon_{\X{0}}$ is to have similarly the dimensionality of energy in order for the ratio $z/\upvarepsilon_{\X{0}}$ to be dimensionless.\footnote{In the SI base units, one has $\llbracket\upgamma_{\sigma}(\bm{k};\varepsilon)\rrbracket = \mathrm{Js}$, and thus $\llbracket\upeta_{\sigma}(\bm{k};u)\rrbracket = \llbracket\protect\t{G}_{\sigma}(\bm{k};z)\rrbracket = \textrm{s}$. The variable $u$ is dimensionless.} In dealing with the single-band Hubbard Hamiltonian $\h{\mathcal{H}}$, Eq.\,(\ref{ex01bx}), one may naturally identify $\upvarepsilon_{\X{0}}$ with the on-site interaction energy $U$, or with the nearest-neighbour hopping integral $t$.\refstepcounter{dummy}\label{SeeFootnote}\footnote{See footnote \raisebox{-1.0ex}{\normalsize{\protect\footref{noteb1}}} on p.\,\protect\pageref{TheSingleParticle}. The hopping integral $t$ referred to here coincides with the $T_{i,j}$ associated with the nearest-neighbour lattice vectors $\bm{R}_i$ and $\bm{R}_j$. In practice, often $t$ is used as the unit of energy. \label{notec1}} The result in Eq.\,(\ref{e4l}) is interesting in that for the most general Nevanlinna function $\t{\varphi}(\zeta)$, $\zeta \in \mathds{C}$,\footnote{In contrast to $z$ that has the dimensionality of energy, here $\zeta$ is dimensionless, thus appropriately representing the dimensionless ratio $z/\upvarepsilon_{\protect\X{0}}$ in Eq.\,(\protect\ref{e4l}).} one has the following Riesz-Herglotz representation [App. C in Ref.\,\citen{HMN72}] [Ch.\,3 in Ref.\,\citen{NIA65}]:
\begin{equation}\label{e20h}
\t{\varphi}(\zeta) = \upmu\hspace{0.6pt}\zeta + \upnu + \int_{-\infty}^{\infty} \rd\upalpha(u)\,\frac{1+ u \zeta}{u-\zeta},
\end{equation}
where $\upmu$ and $\upnu$ are \textsl{real} constants and $\upalpha(u)$ is a bounded non-decreasing function of $u$ that in general need not be continuous; at the points where $\upalpha(u)$ is differentiable, one has [App. C in Ref.\,\citen{HMN72}]
\begin{equation}\label{e4kz}
\rd\upalpha(u) = \upalpha\hspace{0.6pt}'(u) \rd u,
\end{equation}
where $\upalpha\hspace{0.6pt}'(u) \doteq \rd \upalpha(u)/\rd u$. Further, to leading order $\t{\varphi}(\zeta)/\zeta \sim \upmu$ for $\zeta \to \infty$. With these specifications in mind, from the expression in Eq.\,(\ref{e20h}) one readily verifies that indeed (\emph{cf.} Eq.\,(\ref{e2}))
\begin{equation}\label{e4m}
\sgn(\im[\t{\varphi}(\zeta)]) = \sgn(\im[\zeta]),\;\; \im[\zeta] \not= 0,
\end{equation}
as expected of Nevanlinna functions of $\zeta$. Comparing the expression in Eq.\,(\ref{e20h}) with that in Eq.\,(\ref{e4l}), one observes that for the constants $\upmu$ and $\upnu$ corresponding to $-\t{G}_{\sigma}(\bm{k};z)$, which thus in principle depend on both $\bm{k}$ and $\sigma$, one has
\begin{equation}\label{e4m1}
\upmu_{\sigma}(\bm{k}) \equiv 0,\;\;\;
\upnu_{\sigma}(\bm{k}) = \int_{-\infty}^{\infty} \rd\upeta_{\sigma}(\bm{k};u)\,u \equiv
\frac{1}{\upvarepsilon_{\X{0}}} \int_{-\infty}^{\infty} \rd\upgamma_{\sigma}(\bm{k};\varepsilon)\,\frac{\varepsilon/\upvarepsilon_{\X{0}}}{1+(\varepsilon/\upvarepsilon_{\X{0}})^2}.
\end{equation}
One trivially verifies that for $z\to\infty$ to leading order the second term on the RHS of Eq.\,(\ref{e4l}) is equal to $-\upnu_{\sigma}(\bm{k})$, cancelling the first term on the RHS of Eq.\,(\ref{e4l}), in conformity with the fact that to leading order $\t{G}_{\sigma}(\bm{k};z) \sim \hbar/z$ for $z \to \infty$,\footnote{One has $\upvarepsilon_{\protect\X{0}}^{j} \int_{-\infty}^{\infty} \mathrm{d}\upeta_{\sigma}(\bm{k};u)\, (1 + u^2)\hspace{0.6pt} u^{j-1} = G_{\sigma;\infty_j}(\bm{k})$, $j\in \mathds{N}$ (\emph{cf.} Eqs\,(\protect\ref{e4o}), (\protect\ref{e4oa}), and (\protect\ref{e4ob})).} Eqs\,(\ref{e4n}), and (\ref{e4qa}) below.

In Ref.\,\citen{BF07} we have shown that since for Hubbard-like models $A_{\sigma}(\bm{k};\varepsilon)$ decays faster than any finite power of $1/\varepsilon$ as $\vert \varepsilon\vert \to \infty$, the coefficients $\{G_{\sigma;\infty_j}(\bm{k}) \| j\}$ of the asymptotic series expansion [Eq.\,(B.65) in Ref.\,\citen{BF07}]
\begin{equation}\label{e4n}
\t{G}_{\sigma}(\bm{k};z) \sim \frac{G_{\sigma;\infty_1}(\bm{k})}{z} + \frac{G_{\sigma;\infty_2}(\bm{k})}{z^2} + \frac{G_{\sigma;\infty_3}(\bm{k})}{z^3} + \dots,\;\;\text{for}\;\; z \to \infty,
\end{equation}
defined according to [Eq.\,(B.68) in Ref.\,\citen{BF07}]
\begin{equation}\label{e4o}
G_{\sigma;\infty_j}(\bm{k}) \doteq \int_{-\infty}^{\infty} \rd\varepsilon\; A_{\sigma}(\bm{k};\varepsilon) \, \varepsilon^{j-1},
\end{equation}
are bounded for arbitrary finite values of $j$.\footnote{For the general case, consult Ref.\,\protect\citen{BF02}.} Owing to the exact inequality in Eq.\,(\ref{e4f}) and the fact that $A_{\sigma}(\bm{k};\varepsilon) \not\equiv 0$, $\forall \bm{k}$, one has
\begin{equation}\label{e4od}
G_{\sigma;\infty_{2j+1}}(\bm{k}) > 0,\;\; \forall j\in \mathds{N}_{0},\, \forall \bm{k},
\end{equation}
where $\mathds{N}_{0} \equiv \mathds{N}\cup\{0\}$. From the equality in Eq.\,(\ref{e4h}) one observe that
\begin{equation}\label{e4qa}
G_{\sigma;\infty_1}(\bm{k}) = \hbar,\;\; \forall \bm{k},
\end{equation}
which is indeed positive. We note that the strict inequality in Eq.\,(\ref{e4od}) can be violated for $j >0$ only in the special case where (\emph{cf.} Eq.\,(\ref{e4h}))
\begin{equation}\label{e4oi}
A_{\sigma}(\bm{k};\varepsilon) = \hbar\hspace{0.6pt} \delta(\varepsilon),\;\; \forall \bm{k}.
\end{equation}
In contrast, the inequality in Eq.\,(\ref{e4od}) strictly applies for $j >0$ in the case of, for instance,
\begin{equation}\label{e4oj}
A_{\sigma}(\bm{k};\varepsilon) = \frac{\hbar}{2}\hspace{0.6pt} \big\{\delta(\varepsilon + \varepsilon_0) + \delta(\varepsilon -\varepsilon_0)\big\},\;\; \forall \bm{k},
\end{equation}
provided that $\varepsilon_0 \not=0$. Thus it applies for $\varepsilon_0 = 0^+$, to be distinguished from $\varepsilon_0 = 0$. In considering the `Hubbard atom' of spin-$\tfrac{1}{2}$ particles in this publication, \S\,\ref{sec3}, we encounter the one-particle spectral functions in Eqs\,(\ref{e4oi}) and (\ref{e4oj}); the spectral function in Eq.\,(\ref{e4oi}) corresponds to the Green function in Eq.\,(\ref{e15a}), and that in Eq.\,(\ref{e4oj}), with $\varepsilon_0$ identified with respectively $U/2$ and $\epsilon\hspace{0.4pt}U/2$, to the Green functions in respectively Eq.\,(\ref{e53})\footnote{See Eq.\,(\protect\ref{exc1}).} and Eq.\,(\ref{e44}).

The asymptotic series \cite{ETC65} in Eq.\,(\ref{e4n}) is deduced by substituting the geometric series $1/z + \varepsilon'/z^2 + \dots$ for the function  $1/(z-\varepsilon')$ on the RHS of Eq.\,(\ref{e4e}) and subsequently exchanging the order of summation and integration. With reference to the expression in Eq.\,(\ref{e4j}), one has
\begin{equation}\label{e4oa}
G_{\sigma;\infty_{j+1}}(\bm{k}) \equiv  \mu_{j},
\end{equation}
where
\begin{equation}\label{e4ob}
\mu_j \doteq \int_{-\infty}^{\infty} \rd\upgamma_{\sigma}(\bm{k};\varepsilon)\, \varepsilon^{j},\;\; j \in \mathds{N}_0.
\end{equation}
The equality $\mu_{j} = G_{\sigma;\infty_{j+1}}(\bm{k})$ underlines the direct link between the problems considered in the present publication and the classical \textsl{moment problem} \cite{NIA65,ST70,CPVWJ08,LW08}. Since here the problem of interest is defined over the infinite interval $(-\infty,\infty)$, the moment problem of interest coincides with that of Hamburger, to be contrasted with the Stieltjes moment problem corresponding to the semi-infinite interval $[0,\infty)$ \cite{NIA65,ST70,CPVWJ08,LW08}.\footnote{In appendix \protect\ref{sa}, in particular \S\,\protect\ref{sa.1}, where we deal with the density-density correlation function $\protect\t{\protect\b{\chi}}(\bm{k};z) \equiv \protect\t{P}^{\star}(\bm{k};z)$, the polarisation function $\protect\t{P}(\bm{k};z)$, and the dynamical screened-interaction potential $\protect\t{W}(\bm{k};z)$, we encounter moment problems of the Stieltjes type.} We remark however that the Hamburger moment problem of interest in this section can quite naturally be expressed as two Stieltjes moment problems, corresponding to semi-infinite intervals $(-\infty,\mu]$ and $[\mu,\infty)$, where $\mu$ stands for the chemical potential, Eq.\,(\ref{e1}). This follows from the fact that the one-particle Green function $\t{G}_{\sigma}(\bm{k};z)$ can be naturally expressed as [Eq.\,(B.1) in Ref.\,\citen{BF07}]\,\footnote{\emph{Cf.} Eq.\,(\protect\ref{e44}) and note that, with $T_{\protect\X{0}} = -U/2$, for the `Hubbard atom' of spin-$\tfrac{1}{2}$ particles at half-filling one has $\mu = 0$, \S\,\protect\ref{sec3.d}.}
\begin{equation}\label{e4oc}
\t{G}_{\sigma}(\bm{k};z) \equiv \t{G}_{\sigma}^-(\bm{k};z) + \t{G}_{\sigma}^+(\bm{k};z),
\end{equation}
where the functions $\t{G}_{\sigma}^{\mp}(\bm{k};z)$ are analytic in the regions $\re[z] \gtrless \mu$, $\forall \bm{k}$, in addition to being, like $\t{G}_{\sigma}(\bm{k};z)$, analytic away from the real axis of the $z$-plane.\footnote{The functions on the RHS of Eq.\,(\protect\ref{e4oc}) are not to be confused with \textsl{advanced} $\protect\t{G}_{\sigma}(\bm{k};\varepsilon -\protect\ii 0^+)$ and \textsl{retarded} $\protect\t{G}_{\sigma}(\bm{k};\varepsilon +\protect\ii 0^+)$ Green functions (\emph{cf.} Eqs\,(7.59), (7.62), and (7.63) in Ref.\,\protect\citen{FW03}, pp.\,75 and 77).} As a result, with $A_{\sigma}^{\mp}(\bm{k};\varepsilon)$ denoting the one-particle spectral function corresponding to $\t{G}_{\sigma}^{\mp}(\bm{k};z)$ (\emph{cf.} Eq.\,(\ref{e4d})), one has $A_{\sigma}^{\mp}(\bm{k};\varepsilon) \equiv 0$ for $\varepsilon \gtrless \mu$. Thus, given the spectral function $A_{\sigma}(\bm{k};\varepsilon)$, one deduces $A_{\sigma}^{\mp}(\bm{k};\varepsilon)$ according to the following prescription:\,\footnote{\emph{Cf.} Eq.\,(B.48) in Ref.\,\protect\citen{BF07}. With reference to Eq.\,(\protect\ref{e4g}) above, for the measures associated with $\protect\t{G}_{\sigma}^{\mp}(\bm{k};z)$ one has $\upgamma_{\sigma}^{-}(\bm{k};\varepsilon) = \int_{-\infty}^{\min(\varepsilon,\mu)} \mathrm{d}\varepsilon'\, A_{\sigma}(\bm{k};\varepsilon')$, and $\upgamma_{\sigma}^{+}(\bm{k};\varepsilon) = \int_{\min(\varepsilon,\mu)}^{\varepsilon} \mathrm{d}\varepsilon'\, A_{\sigma}(\bm{k};\varepsilon')$. Note that physically nothing is affected by adding $\varepsilon$-independent functions $a_{\sigma}^{\mp}(\bm{k})$ to $\upgamma_{\sigma}^{\mp}(\bm{k};\varepsilon)$, Eq.\,(\protect\ref{e4kz}). Since according to the above expressions $\upgamma_{\sigma}^{-}(\bm{k};\varepsilon) \equiv \upgamma_{\sigma}^{-}(\bm{k};\mu)$ for all $\varepsilon \ge \mu$, and $\upgamma_{\sigma}^{+}(\bm{k};\varepsilon) \equiv 0$ for all $\varepsilon \le \mu$, it appears natural to add $\upgamma_{\sigma}^{-}(\bm{k};\mu)$ to the above expression for $\upgamma_{\sigma}^{+}(\bm{k};\varepsilon)$, as in this way $\upgamma_{\sigma}^{-}(\bm{k};\varepsilon)$ and $\upgamma_{\sigma}^{+}(\bm{k};\varepsilon)$ match at $\varepsilon = \mu$.}
\begin{equation}\label{e4oc1}
A_{\sigma}^{\mp}(\bm{k};\varepsilon) \equiv A_{\sigma}^{}(\bm{k};\varepsilon)\hspace{0.6pt} \Theta(\pm\mu\mp\varepsilon).
\end{equation}

In the light of the considerations in this publication with regard to the perturbation series expansion of the self-energy $\t{\Sigma}_{\sigma}(\bm{k};z)$, it is crucially important that the function $G_{\sigma;\infty_j}(\bm{k})$ can be alternatively expressed as [Eq.\,(34) in Ref.\,\citen{BF02}] [Eq.\,(B.68) in Ref.\,\citen{BF07}]
\begin{equation}\label{e4p}
G_{\sigma;\infty_j}(\bm{k}) = \hbar\hspace{0.6pt} \langle\Psi_{N;\X{0}}\vert [\h{\mathcal{L}}^{j-1} \h{a}_{\bm{k};\sigma}^{\phantom{\dag}}, \h{a}_{\bm{k};\sigma}^{\dag}]_+ \vert\Psi_{N;\X{0}}\rangle,\;\; j\in \mathds{N},
\end{equation}
where $\vert\Psi_{N;\X{0}}\rangle$ stands for the $N$-particle uniform GS of the Hubbard-like Hamiltonian $\h{\mathcal{H}}$ describing the system under consideration, and $\h{\mathcal{L}}$ for the Liouville super-operator associated with $\h{\mathcal{H}}$, defined according to (\emph{cf.} Eq.\,(\ref{ea29}))
\begin{equation}\label{e4q}
\h{\mathcal{L}}\hspace{0.7pt} \h{a}_{\bm{k};\sigma} \doteq [\h{a}_{\bm{k};\sigma},\h{\mathcal{H}}]_-.
\end{equation}
Above, $\h{\mathcal{L}}^{j-1} \equiv \h{\mathcal{L}}\dots \h{\mathcal{L}}$ ($j-1$ times)\footnote{See Eq.\,(\protect\ref{ea29a}).} and further $[\phantom{.},\phantom{.}]_-$ denotes commutation and $[\phantom{.},\phantom{.}]_+$ anti-commutation. The expression in Eq.\,(\ref{e4p}) can be directly obtained from the Mori-Zwanzig representation of $\t{G}_{\sigma}(\bm{k};z)$ \cite{HM65,RWZ61,PF02}:\footnote{\emph{Cf.} Eq.\,(36), p.\,1440, in Ref.\,\protect\citen{BF02}.}
\begin{equation}\label{e4q1}
\t{G}_{\sigma}(\bm{k};z) = \hbar\hspace{0.6pt} \langle\Psi_{N;0}\vert [(z \h{1} - \h{\mathcal{L}})^{-1} \h{a}_{\bm{k};\sigma}^{\phantom{\dag}},\h{a}_{\bm{k};\sigma}^{\dag}]_{+}\vert\Psi_{N;0}\rangle.
\end{equation}
Compare with Eq.\,(\ref{ea34e6}). The equality in Eq.\,(\ref{e4p}) makes also explicit that for the uniform GSs of Hubbard-like models, the function $G_{\sigma;\infty_j}(\bm{k})$ is bounded for arbitrary finite values of $j$, in conformity with the statement preceding Eq.\,(\ref{e4n}) above with regard to the one-particle spectral function $A_{\sigma}(\bm{k};\varepsilon)$ decaying faster than any finite power of $1/\varepsilon$ for $\vert\varepsilon\vert\to\infty$. To clarify, for Hubbard-like models the nested anti-commutations in Eq.\,(\ref{e4p}) \textsl{cannot} result in an unbounded contribution for any finite value of $j$, as the underlying summands / integrands as well as the wave-vector region over which the relevant summations / integrations are carried out\,\footnote{Over the $\protect\1BZ$ in the cases where the underlying lattice is a Bravais one.} are bounded.

Expressing the full Hamiltonian $\h{\mathcal{H}}$ as
\begin{equation}\label{e4r}
\h{\mathcal{H}} \equiv \h{\mathcal{H}}_{\X{0}} + \lambda \h{\mathcal{H}}_{\X{1}},
\end{equation}
with $\h{\mathcal{H}}_{\X{0}}$ denoting the non-interacting part and $\lambda \h{\mathcal{H}}_{\X{1}}$ the interaction part of $\h{\mathcal{H}}$, with $\lambda$ signifying the dimensionless coupling constant of the two-body interaction potential,\footnote{Here $\lambda$ serves as a book-keeping device, which, when convenient, can be identified with unity.} from the expressions in Eqs\,(\ref{e4p}) and (\ref{e4q}) one immediately infers that $G_{\sigma;\infty_j}(\bm{k})$ is a polynomial of order $j-1$ in $\lambda$. This observation proves important to the considerations in this appendix. This is partly related to the expectation value in Eq.\,(\ref{e4p}) being with respect to the \textsl{interacting} $N$-particle uniform GS of $\h{\mathcal{H}}$, which establishes a one-to-one correspondence between the coefficients $\{ G_{\sigma;\infty_j}(\bm{k}) \| j\}$ and the coefficients of the asymptotic series expansion of the self-energy $\t{\Sigma}_{\sigma}(\bm{k};z)$ in the region $z\to\infty$, Eq.\,(\ref{e7b}) below, in terms of \textsl{skeleton} self-energy diagrams and the \textsl{interacting} one-particle Green functions $\{\t{G}_{\sigma}(\bm{k};z)\| \sigma\}$ \cite{BF02,BF07}. See Eqs\,(\ref{e7c}), (\ref{e7d}), and (\ref{e7e}) below.

In view of what follows, but also for illustration, let us consider the function $\phi_{\alpha}(\varepsilon) \doteq \vert\varepsilon\vert^{\alpha} \e^{-\vert\varepsilon\vert/\varepsilon_0}$, with $\alpha \ge 0$ and $\varepsilon_0 >0$ arbitrary \textsl{finite} constants (\emph{cf.} \S\,B.6.1 in Ref.\,\citen{BF07}). One has [\S\S\,6.1.1 and 6.1.39 in Ref.\,\citen{AS72}]
\begin{align}\label{e4oe}
s_{2 j} &\doteq \int_{-\infty}^{\infty} \rd\varepsilon\; \varepsilon^{2 j} \phi_{\alpha}(\varepsilon) = 2 \varepsilon_0^{2j+\alpha+1} \Gamma(2j+\alpha+1)\nonumber\\
&\sim 2 \sqrt{\pi/j} \e^{\alpha +1} \big(2\varepsilon_0 j/\mathrm{e}\big)^{2j + \alpha +1}\;\,\text{for}\;\, j\to \infty,
\end{align}
where $\Gamma(z)$ is the gamma function \cite{AS72} and $\e \equiv \ln^{-1}(1) = 2.718\dots$\,. Of interest is that\,\footnote{$\sqrt[x]{y} \equiv y^{1/x}$. Here we make use of $\big(2 \sqrt{\pi/j}\big)^{1/(2j)} \sim 1 - \ln(j)/(4 j) + \ln(4\pi)/(4 j) + \dots$ for $j\to\infty$.}
\begin{equation}\label{e4of}
\frac{1}{\sqrt[2j]{s_{2j}}} \sim \frac{\e}{2\varepsilon_0\hspace{0.7pt}j}\;\; \text{for}\;\; j\to\infty,
\end{equation}
whereby, for any finite $\nu_0\in \mathds{N}$,\footnote{With $\uppsi(z) \doteq \Gamma'(z)/\Gamma(z)$ denoting the digamma function [\S\,6.3.1, p.\,258, in Ref.\,\protect\citen{AS72}], for $\nu_0 \in \mathds{N}$ one has $\sum_{j=\nu_0}^{\nu} 1/j = \uppsi(\nu+1) - \uppsi(\nu_0) \sim \ln(\nu) - \uppsi(\nu_0) + 1/(2\nu) - 1/(12 \nu^2) + \dots$ as $\nu\to\infty$.}
\begin{equation}\label{e4of1}
\sum_{j=\nu_0}^{\nu} \frac{1}{\sqrt[2j]{s_{2j}}} \sim \frac{\e}{2\varepsilon_0}\hspace{0.6pt} \ln(\nu)\;\; \text{for}\;\; \nu\to\infty,
\end{equation}
leading to the fundamental observation that [p.\,85 in Ref.\,\citen{NIA65}]
\begin{equation}\label{e4og}
\sum_{j=1}^{\infty} \frac{1}{\sqrt[2j]{s_{2j}}} = \infty.
\end{equation}
The one-particle spectral function $A_{\sigma}(\bm{k};\varepsilon)$ of Hubbard-like models decaying like the above function $\phi_{\alpha}(\varepsilon)$ for $\vert\varepsilon\vert\to\infty$, with $0 \le \alpha < \infty$, from the result in Eq.\,(\ref{e4og}) and\,\refstepcounter{dummy}\label{WithReferenceTo}\footnote{With reference to Eq.\,(\protect\ref{e4ob}), one has $s_j = \mu_j$, $\forall j\in \mathds{N}_0$. \label{notei}} $s_j \equiv G_{\sigma;\infty_{j+1}}(\bm{k})$, Eq.\,(\ref{e4oa}), one infers that (see Eq.\,(\ref{e4od}))
\begin{equation}\label{e4oh}
\sum_{j=1}^{\infty} \frac{1}{\sqrt[2j]{G_{\sigma;\infty_{2j+1}}(\bm{k})}} = \infty,\;\;\forall\bm{k}.
\end{equation}
This result establishes the moment problem under consideration as being \textsl{determinate} \cite{NIA65,ST70,CPVWJ08},\footnote{In Ref.\,\protect\citen{ST70} one encounters also the term \textsl{determined} (as well as \textsl{indeterminate}) moment problem.} that is the sequence $\{G_{\sigma;\infty_j}(\bm{k}) \| j\}$ \textsl{essentially uniquely} (see later, p.\,\pageref{Followingtheabove}) determines the measure function \cite{PRH50,APM11} $\upgamma_{\sigma}(\bm{k};\varepsilon)$ in Eq.\,(\ref{e4ob}).

Following the defining expression in Eq.\,(\ref{e4o}), and equivalently Eqs\,(\ref{e4oa}) and (\ref{e4ob}), the strict inequality in Eq.\,(\ref{e4od}) reflects part of a more general property of the sequence $\{G_{\sigma;\infty_j}(\bm{k})\| j\}$, namely its \textsl{positiveness}, which \textsl{cannot} be immediately inferred from the expressions in Eqs\,(\ref{e4f}) and (\ref{e4o}), or equivalently Eq.\,(\ref{e4ob}). \textsl{Positive sequences} are important sequences in the context of the classical moment problem \cite{NIA65,ST70}. With $\{x_j\| j =0,1,\dots,m\}$ denoting the components of an arbitrary real $(m+1)$-vector $\bm{x}$, the sequence $\{ s_j\| j \in \mathds{N}_{0}\}$ is said to be \textsl{positive} if [Eq.\,[1.1] in Ref.\,\citen{NIA65}]
\begin{equation}\label{e4s}
\sum_{j,k=0}^{m} s_{j+k}\hspace{0.6pt} x_j x_k > 0,\;\; \forall m \in \mathds{N}_{0}.
\end{equation}
Defining the $(m+1)\times (m+1)$ real symmetric matrix $\mathbb{H}_m$, the Hankel moment matrix \cite{GM10},\refstepcounter{dummy}\label{InRef}\footnote{In Ref.\,\protect\citen{Note19} we present a set of programs (included in the Mathematica notebook accompanying the present publication) whereby for a given Hankel moment matrix the corresponding orthogonal polynomials $\{P_j(z)\| j \in \mathds{N}_0\}$ are generated and tested for orthonormality. A further program determines the associated polynomials $\{Q_j(z) \| j\in \mathds{N}_0\}$, with each $Q_j(z)$ \textsl{directly} calculated from $P_j(z)$, $\forall j\in \mathds{N}_0$, on the basis of the expression in Eq.\,(\protect\ref{e5xa}) below. These two sets of polynomials are the same as those determined on the basis of the three-term recurrence relation in Eq.\,(\protect\ref{e5y}) below, subject to the initial conditions in Eqs\,(\protect\ref{e5u}) and (\protect\ref{e5x}). We also present programs that \textsl{recursively} determine the latter polynomials by solving the above-mentioned three-term recurrence relation subject to the initial conditions in Eqs\,(\protect\ref{e5u}) and (\protect\ref{e5x}). For the relevance of these polynomials, see Eqs\,(\protect\ref{e5q}) and (\protect\ref{e5tb}) below, where $\protect\t{f}^{\protect\X{(n)}}(z)$ denotes the $n$th-order approximant of $\protect\t{f}(z)$, introduced in Eq.\,(\protect\ref{e5ka}) below. \label{noten1}}
\begin{equation}\label{e6c}
\mathbb{H}_m \doteq \begin{pmatrix}
s_0 & s_1 & \ldots & s_m \\
s_1 & s_2 & \ldots & s_{m+1} \\
\vdots & \vdots & \ddots & \vdots \\
s_m & s_{m+1} & \ldots & s_{2m} \end{pmatrix},
\end{equation}
it is observed that the expression on the LHS of Eq.\,(\ref{e4s}) is in fact the elliptic quadratic form $\mathscr{Q}(\bm{x})$ [\S\,4.11, p.\,326, in Ref.\,\citen{PP12}], defined according to
\begin{equation}\label{e6d}
\mathscr{Q}(\bm{x}) \doteq \bm{x}^{\textsc{t}} \cdot\mathbb{H}_m \bm{x} = \sum_{j,k=0}^{m} x_j \big(\mathbb{H}_m\big)_{j+1,k+1}\hspace{0.6pt} x_k \equiv \sum_{j,k=0}^{m} s_{j+k}\hspace{0.6pt} x_j x_k,
\end{equation}
where $\bm{x}^{\textsc{t}}$ denotes the transpose of the column vector $\bm{x}$, and the row and column indices of $\mathbb{H}_m$ are the elements of the set $\{1,2,\dots, m+1\}$. One has
\begin{equation}\label{e6da}
\big(\mathbb{H}_m\big)_{j,k} = s_{j+k-2}.
\end{equation}

For\refstepcounter{dummy}\label{Forthefollowing} the following considerations, it proves convenient to introduce $\mathbb{H}_m^{\X{X}}$ or $\mathbb{H}^{\X{X}} \equiv \mathbb{H}_{\infty}^{\X{X}}$, as the case may be, to denote the Hankel moment matrix corresponding to the problem associated with the function $X$; when suppressing $X$, we refer to generic Hankel matrices. Thus, in the case of considering the function $\t{G}_{\sigma}(\bm{k};z)$, one has to do with $\mathbb{H}_m^{\X{\t{\X{G}}_{\sigma}}}$, or $\mathbb{H}^{\X{\t{\X{G}}_{\sigma}}}$, and in the case of the function $\t{f}(z)$, with $\mathbb{H}_m^{\,\t{\!\X{f}}}$, or $\mathbb{H}^{\,\t{\!\X{f}}}$. In our considerations, the \textsl{negative} of the relevant functions, including $\t{f}(z)$, are Nevanlinna functions of $z$. For clarity, with reference to Eqs\,(\ref{e4oa}) and (\ref{e4ob}), the elements of the Hankel matrices $\mathbb{H}_m^{\X{\t{\X{G}}_{\sigma}}}$ and $\mathbb{H}^{\X{\t{\X{G}}_{\sigma}}}$ consist of the elements of the sequence $\{\mu_j\| j\in \mathds{N}_{0}\}$, where $\mu_j \equiv G_{\sigma;\infty_{j+1}}(\bm{k})$. As regards the elements of $\mathbb{H}_m^{\,\t{\!\X{f}}}$ and $\mathbb{H}^{\,\t{\!\X{f}}}$, see Eqs\,(\ref{e4da}) and (\ref{e4db}) below.\footnote{\emph{Cf.} Eq.\,(\protect\ref{e4n}).}

The condition in Eq.\,(\ref{e4s}) implies that the moment matrix $\mathbb{H}_{m}$ in Eq.\,(\ref{e6c}) is \textsl{positive definite}. In Refs\,\citen{NIA65} [p.\,1] and \citen{ST70} [p.\,viii] the positivity of the sequence $\{s_j \| j \in \mathds{N}_{0}\}$ is conditioned on [Theorem 1.2, p.\,5, in Ref.\,\citen{ST70}]
\begin{equation}\label{e6e}
\Delta_j \doteq \det\big(\mathbb{H}_j\big) > 0,\;\; \forall j \in \mathds{N}_{0}.
\end{equation}
For the  $(m+1)\times (m+1)$ matrix $\mathbb{H}_m$, the quantities $\{\Delta_0,\Delta_1,\dots,\Delta_m\}$ amount to the \textsl{leading} principal minors [Definition 1.2.5, p. 40, in Ref.\,\citen{HJ90}] of $\mathbb{H}_m$. The positivity of these principal minors are indeed necessary and sufficient for the positivity of the latter matrix \cite{HJ90}. Nonetheless, the positivity of \textsl{all} (and not merely the \textsl{leading}) principal minors of a general $n\times n$ real and symmetric (and similarly Hermitian) matrix $\mathbb{A}$ is often stated, under the rubric of the Sylvester criterion, as the necessary and sufficient condition for the positive definiteness of this matrix \cite{GTG01} (\textsl{all} principal sub-matrices of positive-definite matrices are positive-definite [Observation 7.1.2, p.\,397, in Ref.\,\citen{HJ90}]). In fact, however, the necessary and sufficient condition for the positive definiteness of the above $n\times n$ matrix $\mathbb{A}$ is the positivity of \textsl{any} $n$ \textsl{nested} principal minors of $\mathbb{A}$ (to be contrasted with the $n$ \textsl{leading} and \textsl{all} principal minors of $\mathbb{A}$) [Theorem 7.2.5, p.\,404, in Ref.\,\citen{HJ90}]. A corollary of this observation is that all diagonal elements of a positive definite matrix are positive. With reference to the inequality in Eq.\,(\ref{e4od}), one observes that for the moment matrices $\{\mathbb{H}_j^{\X{\t{\X{G}}_{\sigma}}}\| j\}$, associated with the sequence $\{G_{\sigma;\infty_j}(\bm{k})\| j\}$, this is indeed the case for \textsl{all} $\bm{k}$.\footnote{Barring the very specific case corresponding to the one-particle spectral function in Eq.\,(\protect\ref{e4oi}).}

We note that to an $n\times n$ matrix correspond $\binom{n}{k}$ principal sub-matrices of size $k\times k$, where $1\le k\le n$ [pp.\,40 and 479 in Ref.\,\citen{HJ90}]. It follows that to an $n\times n$ matrix correspond $\sum_{k=1}^{n} \binom{n}{k} = 2^n - 1$ principal minors, to be contrasted with $n$ \textsl{leading} principal minors.\refstepcounter{dummy}\label{WeRemarkThatVarious}\footnote{We remark that various minors of a matrix are related through the Sylvester identities [\S\,0.8.6, p.\,22, in Ref.\,\protect\citen{HJ90}], one of which, due to Aitken, reads thus [p.\,37 in Ref.\,\protect\citen{GM10}]: $(\Delta_m^j)^2 - \Delta_m^{j-1} \Delta_m^{j+1} + \Delta_{m+1}^{j-1} \Delta_{m-1}^{j+1} \equiv 0$, where $\Delta_m^j \doteq \det\big(\mathbb{H}_{m}^{[j]}\big)$, in which $\mathbb{H}_{m}^{[j]}$ is the $(m+1)\times (m+1)$ matrix obtained by adding the integer $j$ to the indices of all elements of the matrix $\mathbb{H}_m$ in Eq.\,(\protect\ref{e6c}). See Ref.\,\protect\citen{Note20}. \label{notet1}} One has $2^n - 1 > n$ for all $n > 1$.

For clarity, let $\mathbb{A}$ be an $n\times n$ symmetric (or Hermitian) matrix. With the row and column indices $i$ and $j$ of the elements $\{(\mathbb{A})_{i,j} \| i, j\}$ of this matrix ranging over the \textsl{ordered} set
\begin{equation}\label{e6e1}
\mathcal{I}_n \doteq \{1,2,\dots,n\},
\end{equation}
let $\upalpha$ denote an \textsl{ordered} set satisfying
\begin{equation}\label{e6e2}
\upalpha \subseteq \mathcal{I}_n.
\end{equation}
With $\upalpha$ consisting of $m$ elements (denoted by $\vert\upalpha\vert = m$), with $m \le n$, we denote the $m\times m$ \textsl{principal}\,\footnote{\textsl{Principal} here signifies $\mathbb{A}(\upalpha)$ as being a specific case of the more general sub-matrix $\mathbb{A}(\upalpha,\upbeta)$, characterised by $\upbeta \equiv \upalpha$ (\emph{i.e.}, $\mathbb{A}(\upalpha) \equiv \mathbb{A}(\upalpha,\upalpha)$), where $\upalpha$ and $\upbeta$ in $\mathbb{A}(\upalpha,\upbeta)$ denote the ordered sets of indices over which respectively the row and column indices of $\mathbb{A}(\upalpha,\upbeta)$ vary [\S\,0.7.1, p.\,17, in Ref.\,\citen{HJ90}].} sub-matrix of $\mathbb{A}$ the indices of whose elements range over $\upalpha$ by $\mathbb{A}(\upalpha)$. Following this definition, the $n$ \textsl{leading} principal sub-matrices of $\mathbb{A}$ are $\mathbb{A}_i \equiv \mathbb{A}(\upalpha_i)$, where the ordered set $\upalpha_i$, $i \in \{1,\dots,n\}$, is defined according to
\begin{equation}\label{e6e3}
\upalpha_1 \doteq \{1\},\;\; \upalpha_2 \doteq \{1,2\},\;\; \upalpha_i \doteq \{1,2,\dots,i\}\;\;\text{for}\;\; i > 2.
\end{equation}
The ordered set $\{\det(\mathbb{A}_1),\det(\mathbb{A}_2),\dots,\det(\mathbb{A}_n)\}$ is the sequence of the \textsl{leading} principal minors of $\mathbb{A}$.\footnote{\S\,7.2.5, p.\,404, in Ref.\,\protect\citen{HJ90}.}

As regards \textsl{all} principal minors of $\mathbb{A}$, let the ordered set $\upbeta$ satisfy the same condition as $\upalpha$ in Eq.\,(\ref{e6e2}). The principal sub-matrices of $\mathbb{A}$ comprise the set
\begin{equation}\label{e6e4}
\{\mathbb{A}(\upbeta_i)\| i=1,2,\dots,2^n -1\},
\end{equation}
where $\upbeta_i$ is a non-empty \textsl{ordered} subset of $\mathcal{I}_n$ satisfying
\begin{equation}\label{e6e5}
1 \le \vert\upbeta_i\vert \le n.
\end{equation}
For the specific case of $n = 3$, one has
\begin{align}\label{e6e6}
&\hspace{0.0cm} \upbeta_1 \doteq \{1\},\; \upbeta_2 \doteq \{2\},\; \upbeta_3 \doteq \{3\},\nonumber\\
&\hspace{0.0cm} \upbeta_{4} \doteq \{1,2\},\; \upbeta_{5} \doteq \{1,3\},\; \upbeta_{6} \doteq \{2,3\},\nonumber\\
&\hspace{0.0cm} \upbeta_{7} \doteq \{1,2,3\}.
\end{align}
With reference to Eq.\,(\ref{e6e3}), one has $\upbeta_1 = \upalpha_1$, $\upbeta_4 = \upalpha_2$, and $\upbeta_7 = \upalpha_3$. Identifying the present $3\times 3$ symmetric matrix $\mathbb{A}$ with the Hankel matrix $\mathbb{H}_2$, Eq.\,(\ref{e6c}), one has\,\footnote{$(s_i) \equiv s_i$.}
\begin{equation}\label{e6e7}
\mathbb{A}(\upbeta_1) = (s_0),\; \mathbb{A}(\upbeta_2) = (s_2),\; \mathbb{A}(\upbeta_3) = (s_4),
\end{equation}
\begin{equation}\label{e6e8}
\mathbb{A}(\upbeta_4) = \begin{pmatrix}
s_0 & s_1 \\
s_1 & s_2 \end{pmatrix}\!,\;
\mathbb{A}(\upbeta_5) = \begin{pmatrix}
s_0 & s_2 \\
s_2 & s_4 \end{pmatrix}\!,\;
\mathbb{A}(\upbeta_6) = \begin{pmatrix}
s_2 & s_3\\
s_3 & s_4 \end{pmatrix}\!,
\end{equation}
\begin{equation}\label{e6e9}
\mathbb{A}(\upbeta_7) = \begin{pmatrix}
s_0 & s_1 & s_2 \\
s_1 & s_2 & s_3 \\
s_2 & s_3 & s_4 \end{pmatrix}\!.
\end{equation}
With $\mathbb{H}_2$ positive definite, in the light of the above discussions, one has $\det(\mathbb{A}(\upbeta_i)) > 0$, $\forall i \in \{1,2,\dots,7\}$, to be contrasted with $\det(\mathbb{A}(\upalpha_i)) > 0$, $\forall i \in \{1,2,3\}$.

Following\refstepcounter{dummy}\label{Followingtheabove} the above discussions, a sequence such as $\{G_{\sigma;\infty_j}(\bm{k})\| j\}$ corresponding to the non-decreasing measure $\upgamma_{\sigma}(\bm{k};\varepsilon)$, Eqs\,(\ref{e4oa}) and (\ref{e4ob}), is necessarily \textsl{positive} \footnote{This sequence need not be infinite. See the discussions centred on Eq.\,(\protect\ref{e4oi}) above as well as the relationships in Eq.\,(\protect\ref{e5kb}) below.} [Theorem 2.1.1, p.\,30, in Ref.\,\citen{NIA65}] [Theorem 2.1, p.\,27, in Ref.\,\citen{ST70}] (see later).\footnote{See in particular Eq.\,(\protect\ref{e5kb}) below.} By the same theorem, any \textsl{positive} sequence can be identified as coinciding with the sequence of the moments of a non-decreasing real measure function \cite{PRH50,APM11} of the kind encountered above, however such function is in general \textsl{not} uniquely determined by the sequence. In this context, the uniqueness of a measure function is specified up to an arbitrary constant at all its points of continuity [p.\,vi in Ref.\,\citen{NIA65} and p.\,xii in Ref.\,\citen{ST70}]. Two measures that are different however are in this sense identical, are referred to as \textsl{essentially identical} or \textsl{substantially identical} \cite{NIA65,ST70}. Defining \textsl{uniqueness} for measure functions in this way reflects the fundamental property in Eq.\,(\ref{e4kz}).

The above-mentioned fuzziness in the uniqueness of measure functions, which is inherent to the classical moment problem, does \textsl{not} concern us however, since the function of interest in our considerations is \textsl{not} the measure function $\upgamma_{\sigma}(\bm{k};\varepsilon)$ itself, but its derivative with respect to $\varepsilon$, Eq.\,(\ref{e4g}), rendering the above-mentioned arbitrariness (\emph{i.e.} the constant difference at all points of continuity) irrelevant. Aside from this fact, insofar as the considerations regarding the convergence behaviour of the series in Eq.\,(\ref{e4}) are concerned, the \textsl{exact} Green function $\t{G}_{\sigma}(\bm{k};z)$, and therefore its associated spectral function $A_{\sigma}(\bm{k};\varepsilon)$, is assumed as given, not to be reconstructed from the elements of the sequence $\{G_{\sigma;\infty_j}(\bm{k})\| j\}$. Later in this appendix, \S\,\ref{sec.b2}, we shall encounter the measure function $\upsigma_{\sigma}(\bm{k};\varepsilon)$ associated with the self-energy $\t{\Sigma}_{\sigma}(\bm{k};z)$, Eqs\,(\ref{e20c}) and (\ref{e20d}). Similar arguments as in the case of $\upgamma_{\sigma}(\bm{k};\varepsilon)$ apply to $\upsigma_{\sigma}(\bm{k};\varepsilon)$.

As\refstepcounter{dummy}\label{AsRegards} regards the positivity of $\mathbb{H}_m^{\X{\t{G}_{\sigma}}}$ referred to above in relation to the sequence $\{G_{\sigma;\infty_j}(\bm{k})\| j\}$, we point out that this is the case for arbitrary finite values of $m$ provided that the associated measure $\upgamma_{\sigma}(\bm{k};\varepsilon)$, Eq.\,(\ref{e4g}), has infinite number of `points of increase'\,\footnote{For instance, these points comprise the set $\{\lambda_j\| j\}$ in the case of the spectral function in Eq.\,(\protect\ref{e6r}) below: with the measure $\upgamma_{n+1}(\varepsilon,\protect\ptau)$ as presented in Eq.\,(\protect\ref{e6s}), one observes that, for $\gamma_j > 0$, Eq.\,(\protect\ref{e6m}), on $\varepsilon$ increasing from $\lambda_j - 0^+$ to $\lambda_j + 0^+$ indeed $\upgamma_{n+1}(\varepsilon,\protect\ptau)$ increases by $\gamma_j$, Eq.\,(\protect\ref{e6t}).} along the $\varepsilon$-axis.\footnote{See the discussions regarding the equality sign implied by the $\geqq$ on the RHS of Eq.\,[2.2], p.\,30, in Ref.\,\protect\citen{NIA65}.} This is the case particularly when $\upgamma_{\sigma}(\bm{k};\varepsilon)$ is \textsl{continuous} \ae along the $\varepsilon$-axis. Notably, $\mathbb{H}_m^{\X{\t{G}_{\sigma}}}$ fails to be positive definite (and thus is \textsl{ positive semi-definite}) for arbitrary $m$ in the case of the `Hubbard atom', \S\,\ref{sec3.b}, whose corresponding relevant measure $\upgamma(\varepsilon)$ consists solely of \textsl{two} `points of increase', at $\varepsilon = \mp U/2$, Eqs\,(\ref{e53}) and (\ref{exc1}).\footnote{See footnote \raisebox{-1.0ex}{\normalsize{\protect\footref{notej}}} on page \protect\pageref{ForInstance}. That the $\mathbb{H}_m^{\protect\X{\t{G}_{\sigma}}}$ corresponding to the `Hubbard atom' fails to be positive definite for arbitrary $m$, can be directly established on the basis of the expressions in Eqs\,(\protect\ref{e6a}) -- (\protect\ref{e6b}) below, making use of the exact results in Eq.\,(\protect\ref{e7xa}).} Therefore, in what follows the inequalities `$\dots > 0$' are to be viewed as possibly representing `$\dots \ge 0$' for systems in the \textsl{local} limit. More explicitly, the `$\dots > 0$' in the following are to become `$\dots \ge 0$' when the local limit is effected prior to calculating the Green function (and the self-energy). We have opted for the use of `$>$', in preference to `$\ge$', throughout this appendix by the consideration that for realistic macroscopic interacting systems the equality as implied by `$\ge$' at best applies only over a subset of measure zero of the $\bm{k}$ space.

Following the above considerations, the positivity of $\{G_{\sigma;\infty_j}(\bm{k})\| j\}$ implies an infinite hierarchy\,\footnote{Without loss of generality, in the following we assume that this is an infinite sequence.} of inequalities to be satisfied by the elements of this sequence, which are additional to those described by the inequality in Eq.\,(\ref{e4od}). A minimal hierarchy is obtained from the inequality in Eq.\,(\ref{e6e}). Limiting oneself to $\mathbb{H}_n^{\t{\X{G}}_{\sigma}}$, with $n$ a finite integer, an equivalent hierarchy of inequalities is obtained by considering \textsl{any} of the \textsl{nested} principal minors of $\mathbb{H}_n^{\t{\X{G}}_{\sigma}}$. With $s_j \equiv G_{\sigma;\infty_{j+1}}(\bm{k})$ for all $j \in \mathds{N}_0$, Eqs\,(\ref{e4o}) and (\ref{e4oa}),\footnote{See also footnote \raisebox{-1.0ex}{\normalsize{\protect\footref{notei}}} on page \protect\pageref{WithReferenceTo}.}  in the specific case of $n=2$, for which the symmetric Hankel matrix $\mathbb{H}_n^{\t{\X{G}}_{\sigma}}$ is a $3\times 3$ one, following Eq.\,(\ref{e6e7}) and $\det(\mathbb{A}(\upbeta_i) > 0$, $i=1,2,3$, one has $G_{\sigma;\infty_{j}}(\bm{k}) > 0$, $j = 1, 3, 5$, in conformity with Eq.\,(\ref{e4od}). With $G_{\sigma;\infty_1}(\bm{k}) = \hbar$, Eq.\,(\ref{e4qa}), following Eq.\,(\ref{e6e8}) and $\det(\mathbb{A}(\upbeta_i)) > 0$, $i=4,5,6$, one further has \cite{Note10}
\begin{align}\label{e6a}
&\hspace{0.0cm}-G_{\sigma;\infty_2}^2(\bm{k}) +\hbar\hspace{0.6pt} G_{\sigma;\infty_3}(\bm{k}) > 0,\\
\label{e6a1}
&\hspace{0.0cm}-G_{\sigma;\infty_3}^2(\bm{k}) +\hbar\hspace{0.6pt} G_{\sigma;\infty_5}(\bm{k}) > 0,\\
\label{e6a2}
&\hspace{0.0cm}-G_{\sigma;\infty_4}^2(\bm{k}) + G_{\sigma;\infty_3}(\bm{k})\hspace{0.6pt} G_{\sigma;\infty_5}(\bm{k})> 0,
\end{align}
and, following Eq.\,(\ref{e6e9}) and $\det(\mathbb{A}(\upbeta_7)) > 0$ \cite{Note10},
\begin{align}\label{e6b}
-G_{\sigma;\infty_3}^3(\bm{k}) &+ 2 G_{\sigma;\infty_2}(\bm{k}) G_{\sigma;\infty_3}(\bm{k}) G_{\sigma;\infty_4}(\bm{k}) -\hbar\hspace{0.6pt} G_{\sigma;\infty_4}^2(\bm{k}) \nonumber\\
&- G_{\sigma;\infty_2}^2(\bm{k}) G_{\sigma;\infty_5}(\bm{k}) + \hbar\hspace{0.6pt} G_{\sigma;\infty_3}(\bm{k}) G_{\sigma;\infty_5}(\bm{k}) > 0.
\end{align}
We note in passing that the function $-G_{\sigma;\infty_2}^2(\bm{k}) +\hbar\hspace{0.6pt} G_{\sigma;\infty_3}(\bm{k})$ in Eq.\,(\ref{e6a}) coincides with $\hbar^3$ times the coefficient of the leading-order term, decaying like $1/z$, in the asymptotic series expansion of $\t{\Sigma}_{\sigma}(\bm{k};z) -\Sigma_{\sigma}^{\textsc{hf}}(\bm{k})$ for $z\to\infty$, Eqs\,(\ref{e7b}), (\ref{e7ba}), and (\ref{e7c}) below. In the light of Eqs\,(\ref{e5kc}) and (\ref{e5kd}) below, this is not a coincidence.

We\refstepcounter{dummy}\label{WeRemarkThat} remark that in the cases of p-h symmetric $N$-particle GSs\,\footnote{Such as the half-filled GS of the single-band Hubbard Hamiltonian for spin-$\tfrac{1}{2}$ particles on a bipartite lattice. See the paragraph beginning with Eq.\,(\protect\ref{eac6a}) in \S\,\protect\ref{sd32}.} (ESs) and $\mu = 0$, for $\bm{k}$ over some subset of the underlying $\bm{k}$-space the coefficients $\{G_{\sigma;\infty_{2j}}(\bm{k})\| j \in \mathds{N}\}$, Eq.\,(\ref{e4o}), are identically vanishing.\footnote{See the discussions following Eq.\,(\protect\ref{eac6m}) below. We should emphasise that the equality in Eq.\,(\protect\ref{eac6m}) is specific to the cases of spin-unpolarised GSs (ESs).} In such cases, for $\bm{k}$ an element of the said subset the positivity conditions (corresponding to the positivity of \textsl{all} principal minors of $\mathbb{H}^{\X{\t{\X{G}}_{\sigma}}} \equiv \mathbb{H}_{\infty}^{\X{\t{\X{G}}_{\sigma}}}$) satisfied by the elements of the positive sequence $\{G_{\sigma;\infty_{2j+1}}(\bm{k})\| j \in \mathds{N}_{0}\}$ are greatly simplified.\footnote{This simplification applies to \textsl{all} $\bm{k}$ in the local / atomic limit.} This is seen to be indeed the case on identifying with zero the functions $G_{\sigma;\infty_2}(\bm{k})$ and $G_{\sigma;\infty_4}(\bm{k})$ in Eqs\,(\ref{e6a}), (\ref{e6a2}), and (\ref{e6b}). For instance, in this case (and in part on account of $G_{\sigma;\infty_3}(\bm{k}) > 0$, Eq.\,(\ref{e4od})) the relationship in Eq.\,(\ref{e6b}) reduces to $-G_{\sigma;\infty_3}^2(\bm{k}) +\hbar\hspace{0.6pt} G_{\sigma;\infty_5}(\bm{k}) > 0$, which coincides with that in Eq.\,(\ref{e6a1}). We note that the GS of the `Hubbard atom' of spin-$\tfrac{1}{2}$ particles at half-filling is p-h symmetric, with the corresponding $\mu =0$, \S\,\ref{sec3.d}, however, in the light of the expressions in Eq.\,(\ref{e7xa}) below, in this case one has $-G_{\infty_3}^2 +\hbar\hspace{0.6pt} G_{\infty_5} = 0$.\footnote{See the remarks regarding the positivity of $\mathbb{H}_m^{\protect\X{\protect\t{G}_{\sigma}}}$ on page \protect\pageref{AsRegards}.}

On account of $-\t{G}_{\sigma}(\bm{k};z)$ being a Nevanlinna function of $z$, and with $\t{G}_{\sigma}(\bm{k};z)$ possessing the asymptotic series expansion in Eq.\,(\ref{e4n}), it can be rigorously established that $G_{\sigma;\infty_1}(\bm{k}) = 0$ for any $\bm{k}$ implies $\t{G}_{\sigma}(\bm{k};z) = 0$ for all $z$ [Lemma 2.3, p.\,26, in Ref.\,\citen{ST70}]. The fact that $G_{\sigma;\infty_1}(\bm{k}) > 0$, $\forall \bm{k}$, Eq.\,(\ref{e4qa}), and not merely $G_{\sigma;\infty_1}(\bm{k}) \not\equiv 0$, is directly related to $-\t{G}_{\sigma}(\bm{k};z)$ being a non-constant Nevanlinna function of $z$\cite{ST70}. Identifying for conciseness, as well as the convenience of later reference, the function $\t{G}_{\sigma}(\bm{k};z)$ with $\t{f}(z)$, the asymptotic series expansion in Eq.\,(\ref{e4n}) can be written as (\emph{cf.} Eq.\,(\ref{e4oa}))
\begin{equation}\label{e4da}
\t{f}(z) \sim \alpha_0 + \frac{\mu_0}{z} + \frac{\mu_1}{z^2} + \dots\;\; \text{for}\;\; z\to\infty,
\end{equation}
where
\begin{equation}\label{e4db}
\alpha_0 = 0.
\end{equation}
Despite $\alpha_0 = 0$, we have introduced $\alpha_0$ in Eq.\,(\ref{e4da}) both for uniformity of notation (\emph{cf.} Eqs\,\,(\ref{e5c}) and (\ref{e5h}) below) and in anticipation of our later, \S\,\ref{sec.b2}, identification of $\t{f}(z)$ with $\t{\Sigma}_{\sigma}(\bm{k};z)$, in which case the corresponding $\alpha_0$ is generally non-vanishing (\emph{cf.} Eqs\,(\ref{e7b}), and (\ref{e7ba}) below). In the light of Eq.\,(\ref{e20h}), one has $\alpha_0 \in \mathds{R}$.

The function $\t{f}(z)$, with $-\t{f}(z)$ a Nevanlinna function, can be expressed as [p.\,30 in Ref.\,\citen{ST70}]
\begin{equation}\label{e5b}
\t{f}(z) = \alpha_0 + \frac{\beta_0}{z - \t{f}_1(z)} \;\Longleftrightarrow\; \t{f}_{1}(z) = z + \frac{\beta_0}{\alpha_0 - \t{f}(z)},
\end{equation}
where $\beta_0$ is a constant\,\footnote{That is, independent of $z$. In general, $\beta_0$ may be a function of both $\bm{k}$ and $\sigma$.} for which, in the light of the asymptotic expression in Eq.\,(\ref{e4da}), one has (see Eq.\,(\ref{e4qa}))
\begin{equation}\label{e5cb}
\beta_0 = \mu_0 \equiv G_{\sigma;\infty_1}(\bm{k}) \equiv \hbar.
\end{equation}
The function $-\t{f}_1(z)$ in Eq.\,(\ref{e5b}) is a Nevanlinna function, with $\t{f}_1(z)$ behaving as (\emph{cf.} Eq.\,(\ref{e4da}))
\begin{equation}\label{e5c}
\t{f}_1(z) \sim \alpha_1 + \frac{\mu_0^{\X{(1)}}}{z} + \frac{\mu_1^{\X{(1)}}}{z^2} + \dots\;\;\text{for}\;\; z\to \infty,
\end{equation}
where, as in the case of $\t{G}_{\sigma}(\bm{k};z) \equiv \t{f}(z)$, one has $\t{f}_1(z) = \alpha_1$, $\forall z$, in the event of $\mu_0^{\X{(1)}} = 0$. Since $-\t{f}_1(z)$ is a Nevanlinna function, one has $\alpha_1 \in \mathds{R}$ (\emph{cf.} Eq.\,(\ref{e20h})), and further $\mu_0^{\X{(1)}} \ge 0$, so that in the cases where  $\mu_0^{\X{(1)}} \not= 0$ one necessarily has  $\mu_0^{\X{(1)}} > 0$ (\emph{cf.} Eqs\,(\ref{e4da}), and (\ref{e5cb})).

From the Dyson equation, Eq.\,(\ref{e4a}), corresponding to a single-band many-body Hamiltonian $\h{\mathcal{H}}$, such as the single-band \textsl{Hubbard} Hamiltonian, Eq.\,(\ref{ex01bx}), described in terms of the non-interacting one-particle energy dispersion $\varepsilon_{\bm{k}}$, one obtains [Eq.\,(2.4) in Ref.\,\citen{BF07}]
\begin{equation}\label{e5d}
\t{G}_{\sigma}(\bm{k};z) = \frac{\hbar}{z -\varepsilon_{\bm{k}} -\hbar\t{\Sigma}_{\sigma}(\bm{k};z)}.
\end{equation}
With the above Nevanlinna function $-\t{f}(z)$ identified with $-\t{G}_{\sigma}(\bm{k};z)$, comparison of the expression in Eq.\,(\ref{e5d}) with those in Eqs\,(\ref{e4da}) -- (\ref{e5c})  reveals that\,\footnote{With reference to Eq.\,(\protect\ref{e2}), wherein the function $\protect\t{X}_{\sigma}(\bm{k};z)$ is specified following Eq.\,(\protect\ref{e1}), one observes that indeed the negative of the function $\protect\t{f}_1(z)$ in Eq.\,(\protect\ref{e5f}) is a Nevanlinna function of $z$.}
\begin{equation}\label{e5f}
\t{f}_{1}(z) \equiv \t{E}_{\sigma}(\bm{k};z) \doteq \varepsilon_{\bm{k}} + \hbar\t{\Sigma}_{\sigma}(\bm{k};z),
\end{equation}
and
\begin{equation}\label{e5e}
\alpha_1 \equiv \varepsilon_{\bm{k}} + \hbar\Sigma_{\sigma}^{\textsc{hf}}(\bm{k}),
\end{equation}
where $\Sigma_{\sigma}^{\textsc{hf}}(\bm{k}) \equiv \Sigma_{\sigma}^{\textsc{hf}}(\bm{k};[\{\t{G}_{\sigma'}\}])$ stands for the \textsl{exact} Hartree-Fock self-energy \cite{BF02,BF07,BF13}\footnote{See also \S\,\protect\ref{sec.3a.1}.} evaluated in terms of the exact interacting one-particle Green functions $\{\t{G}_{\sigma}(\bm{k};z)\| \sigma\}$ [App. A in Ref.\,\citen{BF13}]. In arriving at the equality in Eq.\,(\ref{e5e}), we have used the fact that to leading order $\t{\Sigma}_{\sigma}(\bm{k};z) \sim \Sigma_{\sigma}^{\textsc{hf}}(\bm{k})$ for $z\to \infty$, Eqs\,(\ref{e7b}) and (\ref{e7ba}) below.\footnote{One analogously obtains $\mu_0^{\protect\X{(1)}} = \hbar\Sigma_{\sigma;\infty_1}(\bm{k})$, where $\Sigma_{\sigma;\infty_1}(\bm{k})$ as expressed in terms of $G_{\sigma;\infty_2}(\bm{k})$ and $G_{\sigma;\infty_3}(\bm{k})$ is presented in Eq.\,(\protect\ref{e7c}) below. Thus, following Eq.\,(\protect\ref{e5i}) below, $\beta_1 = \hbar\Sigma_{\sigma;\infty_1}(\bm{k})$ (\emph{cf.} Eq.\,(\protect\ref{e5ib}) below).} For later reference, on identifying $\t{f}(z)$ with $\t{G}_{\sigma}(\bm{k};z)$, as above, from the expressions in Eqs\,(\ref{e4n}), and (\ref{e5b}) one deduces that [\emph{cf.} Eq.\,(2.48) in Ref.\,\citen{BF13}]
\begin{equation}\label{e5ea}
G_{\sigma;\infty_2}(\bm{k}) = G_{\sigma;\infty_1}(\bm{k}) \alpha_1 \equiv \hbar \big(\varepsilon_{\bm{k}} + \hbar\Sigma_{\sigma}^{\textsc{hf}}(\bm{k})\big).
\end{equation}

The function $\t{f}_1(z)$ considered above is the first of the sequence of auxiliary functions $\{\t{f}_j(z)\| j \in \mathds{N}\}$, with $-\t{f}_j(z)$ a Nevanlinna function of $z$, satisfying the recursive relationship [p.\,31 in Ref.\,\citen{ST70}]
\begin{equation}\label{e5g}
\t{f}_j(z) = \alpha_j + \frac{\beta_j}{z - \t{f}_{j+1}(z)} \; \Longleftrightarrow\; \t{f}_{j+1}(z) = z + \frac{\beta_j}{\alpha_j - \t{f}_j(z)},
\end{equation}
where $\t{f}_{j}(z)$, $\forall j\in\mathds{N}$, behaves as (\emph{cf.} Eq.\,(\ref{e5c}))
\begin{equation}\label{e5h}
\t{f}_{j}(z) \sim \alpha_{j} + \frac{\mu_0^{\X{(j)}}}{z} +  \frac{\mu_1^{\X{(j)}}}{z^2} + \dots\;\;\text{for}\;\; z\to\infty.
\end{equation}
As in the case of $j=1$, in the event of $\mu_0^{\X{(j)}} = 0$ for an arbitrary value of the integer $j$, one has $\t{f}_{j}(z) = \alpha_{j}$, $\forall z$. One immediately infers that (\emph{cf.} Eq.\,(\ref{e5cb}))
\begin{equation}\label{e5i}
\beta_j = \mu_0^{\X{(j)}}.
\end{equation}
Note that when $\t{f}_{j}(z) \equiv \alpha_j$ for some $j \in\mathds{N}$, one has $\t{f}_{j'}(z) \equiv 0$ for \textsl{all} $j' > j$ [Lemma 2.3, p.\,26, in Ref.\,\citen{ST70}]. As a result, in particular $\alpha_{j'} = 0$ and $\beta_{j'} = 0$ for \textsl{all} $j' > j$ (\emph{cf.} Eq.\,(\ref{e5kb}) below). Note further that since $-\t{f}_j(z)$ is a Nevanlinna function of $z$, one has $\alpha_j \in \mathds{R}$ (\emph{cf.} Eq.\,(\ref{e20h})), and in addition $\mu_0^{\X{(j)}} \ge 0$, $\forall j\in\mathds{N}$. It follows that in the cases where $\mu_0^{\X{(j)}} \not= 0$, one necessarily has $\mu_0^{\X{(j)}} > 0$, $\forall j\in\mathds{N}$. In the light the equality in Eq.\,(\ref{e5i}), the same applies to $\beta_j$, $\forall j\in\mathds{N}$. We remark that [p.\,5 in Ref.\,\citen{NIA65}] \cite{Note19}\,\refstepcounter{dummy}\label{TheDeviation}\footnote{The deviation of the result in Eq.\,(\protect\ref{e5ia}) from that for $a_{k,k+1} = b_k$ in Ref.\,\protect\citen{NIA65}, pp.\,4 and 5, is due to the deviation of the three-term recurrence relation in Eq.\,(\protect\ref{e5y}) below form that on p.\,4 of Ref.\,\protect\citen{NIA65}. Note that $\lambda$ in the latter reference is the equivalent of the complex variable $z$ in our present considerations. Further, for the  $\beta_{0}$ in Eq.\,(\protect\ref{e5cb}) one has $\llbracket\beta_0\rrbracket =\mathrm{Js}$, while for the $\beta_j$, $\forall j\in\mathds{N}$, $\llbracket\beta_j\rrbracket = \mathrm{J}^2$. Therefore, despite the notational similarity, $\beta_0$ is \textsl{not} to be viewed as $\beta_j$ with $j$ identified with $0$. Note that $\llbracket G_{\sigma;\infty_j}(\bm{k})\rrbracket = \mathrm{J}^{\protect\X{j}}\mathrm{s}$ (following $\llbracket A_{\sigma}(\bm{k};\varepsilon)\rrbracket = \mathrm{s}$), whereby $\llbracket\Delta_j \rrbracket = \mathrm{J}^{\protect\X{(j+1)^2}}\mathrm{s}^{\protect\X{j+1}}$. \label{notem1}} (\emph{cf.} Eq.\,(\ref{e6e}))
\begin{equation}\label{e5ia}
\beta_j = \frac{\Delta_{j-2}\hspace{0.6pt} \Delta_{j}}{\Delta_{j-1}^2},\;\; j \in \mathds{N},
\end{equation}
where by convention $\Delta_{-1} = 1$.\footnote{See the footnote on p.\,3 of Ref.\,\protect\citen{NIA65}.} Thus insofar as $\t{G}_{\sigma}(\bm{k};z)$ is concerned, one, for instance, has\,\footnote{Using $\Delta_{-1} = 1$, $\Delta_0 = \det(\mathbb{A}(\upbeta_1)) = G_{\sigma;\infty_1}(\bm{k}) \equiv \hbar$, Eq.\,(\protect\ref{e4qa}), and $\Delta_1 = \det(\mathbb{A}(\upbeta_4))$, Eq.\,(\ref{e6a}).}
\begin{equation}\label{e5ib}
\beta_1 = \frac{1}{\hbar^2} \big(\hspace{-1.2pt}-G_{\sigma;\infty_2}^2(\bm{k}) + \hbar\hspace{0.6pt} G_{\sigma;\infty_3}(\bm{k})\big),
\end{equation}
which, in the light of Eq.\,(\ref{e6a}), is indeed positive.

Carrying out the above recursive procedure, one arrives at the following continued-fraction expansion for $\t{f}(z)$ [p.\,31 in Ref.\,\citen{ST70}]:
\begin{equation}\label{e5k}
\t{f}(z) = \alpha_0 + \cfrac{\beta_0}{z -\alpha_1 -\cfrac{\beta_1}{z - \alpha_2 -\dots}} \equiv \alpha_0 - \EuScript{K}_{j=1}^{\infty} \Big(\frac{-\beta_{j-1}}{z -\alpha_{j}}\Big),
\end{equation}
where on the RHS we have used the convenient notation adopted in Refs\,\citen{CPVWJ08} and \citen{LW08}. In the light of the identity in Eq.\,(\ref{e5f}), this procedure also leads to a continued-fraction expansion of the self-energy $\t{\Sigma}_{\sigma}(\bm{k};z)$. Explicitly, in view of Eqs\,(\ref{e5cb}) and (\ref{e5e}), and taking into account that $\alpha_0 = 0$ in the case of the Green function, Eq.\,(\ref{e4db}), from Eq.\,(\ref{e5k}) one has
\begin{equation}\label{e5kc}
\t{G}_{\sigma}(\bm{k};z) = -\EuScript{K}_{j=1}^{\infty} \Big(\frac{-\beta_{j-1}}{z -\alpha_{j}}\Big) \equiv \cfrac{\hbar}{z - \varepsilon_{\bm{k}} -\hbar\Sigma_{\sigma}^{\textsc{hf}}(\bm{k}) + \EuScript{K}_{j=1}^{\infty} \Big(\frac{-\beta_{j}}{z -\alpha_{j+1}}\Big)},
\end{equation}
whereby
\begin{equation}\label{e5kd}
\t{\Sigma}_{\sigma}(\bm{k};z) = \Sigma_{\sigma}^{\textsc{hf}}(\bm{k}) -\frac{1}{\hbar}\hspace{0.6pt} \EuScript{K}_{j=1}^{\infty} \Big(\frac{-\beta_{j}}{z -\alpha_{j+1}}\Big),
\end{equation}
where we have used Eq.\,(\ref{e5d}). Insofar as the convergence of the continued-fraction expansion in Eq.\,(\ref{e5kc}) (Eq.\,(\ref{e5kd})) is concerned, this is established by the result in Eq.\,(\ref{e4oh}) above (Eq.\,(\ref{e7m}) below).\footnote{Whether the moment problem associated with the moments $\{\mu_0,\mu_1,\dots\}$ and the function $\protect\t{f}(z)$ in Eq.\,(\protect\ref{e4da}) is \textsl{determinate} or \textsl{indeterminate} stands in a direct relationship with whether the continued-fraction expansion in Eq.\,(\protect\ref{e5k}) is convergent or divergent [p.\,viii and Theorem 2.10, p.\,51, in Ref.\,\protect\citen{ST70}]. Note that the circle $C_{n+1}(z)$ in Ref.\,\protect\citen{ST70} (defined on p.\,48 herein) denotes the reflection into the real axis of the circle $\mathsf{K}_{n+1}(z)$ in Ref.\,\protect\citen{NIA65} and, further, that $P_n(z)$ and $Q_n(z)$ in the former reference denote respectively the $Q_n(z)$ and $P_n(z)$ of the latter reference.} We note in passing that the continued fractions of the type in Eqs\,(\ref{e5kc}) and (\ref{e5kd}) are classified as `$J$-fractions' in Ref.\,\citen{CPVWJ08} (see for instance p.\,37 herein). For later reference, in view of Eq.\,(\ref{e5kd}), we introduce
\begin{equation}\label{e5ke}
\t{\mathcal{E}}_{\sigma}(\bm{k};z) \doteq \hbar \big(\t{\Sigma}_{\sigma}(\bm{k};z) -\Sigma_{\sigma}^{\textsc{hf}}(\bm{k})\big) \equiv -\EuScript{K}_{j=1}^{\infty} \Big(\frac{-\beta_{j}}{z -\alpha_{j+1}}\Big).
\end{equation}

We shall have occasion in the following, as well as in the main text, to consider the $n$th-order approximant of the function $\t{f}(z)$, defined as \cite{Note16} (\emph{cf.} Eq.\,(\ref{e5k}))
\begin{equation}\label{e5ka}
\t{f}^{\X{(n)}}(z) \doteq \alpha_0 - \EuScript{K}_{j=1}^{n} \Big(\frac{-\beta_{j-1}}{z-\alpha_{j}}\Big).
\end{equation}
This approximant naturally identically coincides with $\t{f}(z)$ for $\beta_n =0$, which, as we have indicated following Eq.\,(\ref{e5i}) above, implies $\beta_j = 0$ for \textsl{all} $j\ge n$. It is important to note that\,\footnote{With reference to Eq.\,(\protect\ref{e5ia}), when $\Delta_{k} > 0$, for $-1 \le  k < j$, one clearly has $\Delta_j > 0 \Leftrightarrow \beta_j > 0$.} [Theorems 2.3 and 2.4, pp.\,31 and 32, and the footnotes on pp.\,32 and 47 of Ref.\,\citen{ST70}]
\begin{align}\label{e5kb}
&\hspace{0.4cm}\beta_0 > 0,\; \dots, \;\beta_m > 0,\; \beta_{m+1} = \beta_{m+2} = \dots = 0 \nonumber\\
&\hspace{-0.6cm}\Longleftrightarrow\, \Delta_0 > 0,\; \dots, \;\Delta_m > 0,\; \Delta_{m+1} = \Delta_{m+2} = \dots = 0,
\end{align}
where $\Delta_j$ is defined in Eq.\,(\ref{e6e}). This correspondence applies also for the case of $m=\infty$, which is to say that $\beta_j > 0$ for \textsl{all} $j \in \mathds{N}_{0}$ implies $\Delta_j > 0$ for \textsl{all} $j \in\mathds{N}_{0}$, and \emph{vice versa}. In the cases where $\beta_{m+1} = 0$ for some finite $m$, the continued-fraction expansion of $\t{f}(z)$ in Eq.\,(\ref{e5k}) is terminating, otherwise non-terminating. In the latter case, the question of the convergence of the expansion in Eq.\,(\ref{e5k}) arises, which in the light of the results in Eq.\,(\ref{e4oh}) above and Eq.\,(\ref{e7m}) below does not concern us in this appendix. Irrespective of whether or not the expansion in Eq.\,(\ref{e5k}) is terminating, one can consider the \textsl{truncated} or \textsl{reduced} moment problem associated with the finite set of moments $\{\mu_0,\mu_1,\dots,\mu_{2n-1}\}$,\footnote{See Eqs\,(\protect\ref{e5td}), (\protect\ref{e6jc}), and (\protect\ref{e6p}) below. The $\varsigma$ in the latter equation is specified in Eq.\,(\protect\ref{e6q}).} Eqs\,(\ref{e4ob}) and (\ref{e4da}), where $n$ is some finite integer \cite{NIA65,ST70}. As will become evident below, considerations regarding truncated moment problems are relevant to our discussions in view of their relationship to the perturbation series expansion in Eq.\,(\ref{e4}) (see in particular \S\,\ref{sec.3.2.1}).

For the discussions of this appendix, it proves convenient to employ some further functions considered in Ref.\,\citen{CPVWJ08}. Let [\S\,1.1, p.\,9, and \S\,2.2, p.\,30, in Ref.\,\citen{CPVWJ08}]
\begin{equation}\label{e5l}
\mathsf{s}_{\hspace{0.4pt}0}(z,w) \doteq \alpha_0 + w,\;\; \mathsf{s}_j(z,w) \doteq \frac{\beta_{j-1}}{z-\alpha_{j} -w},\;\; j\in\mathds{N},
\end{equation}
and
\begin{equation}\label{e5m}
\mathsf{S}_0(z,w) \doteq \mathsf{s}_{\hspace{0.4pt}0}(z,w),\;\; \mathsf{S}_j(z,w) \doteq \mathsf{S}_{j-1}(z,\mathsf{s}_j(z,w)),\;\; j \in \mathds{N}.
\end{equation}
Clearly, the left-most expression in Eq.\,(\ref{e5b}) can be equivalently written as
\begin{equation}\label{e5n}
\t{f}(z) = \mathsf{S}_1(z,\t{f}_1(z)-\alpha_1),
\end{equation}
and, more generally,
\begin{equation}\label{e5o}
\t{f}(z) = \mathsf{S}_j(z,\t{f}_j(z)-\alpha_j),\;\; \forall j \in\mathds{N}.
\end{equation}
Further,\footnote{\emph{Cf.} Eq.\,(\protect\ref{e6jc}) below.}
\begin{equation}\label{e5p}
\t{f}^{\X{(n)}}(z) \equiv \mathsf{S}_n(z,0).
\end{equation}
Note that $\t{f}_j(z) - \alpha_j = o(1)$ in the asymptotic region $z\to \infty$ [Lemma 2.3, p.\,26, in Ref.\,\citen{ST70}], Eqs\,(\ref{e5c}) and (\ref{e5h}). More explicitly, on the basis of the remarks following Eq.\,(\ref{e5i}), unless $\t{f}_j(z) - \alpha_j \equiv 0$, one has $\t{f}_j(z) - \alpha_j = O(1/z)$ for $z\to\infty$, Eq.\,(\ref{e5h}).\footnote{For the definitions of the order symbols $o$ and $O$ \protect\cite{ETC65,WW62,EWH26}, see appendix \protect\ref{sae}.} In the light of the identity in Eq.\,(\ref{e5p}), these observations reveal the accuracy with which the function $\t{f}^{\X{(n)}}(z)$, for finite values of $n$, describes the function $\t{f}(z)$ in the region $z\to\infty$, where $\t{f}^{\X{(n)}}(z)$ is obtained from $\mathsf{S}_n(z,\t{f}_n(z)-\alpha_n)$ through replacing the small-amplitude function $\t{f}_n(z)-\alpha_n$ by zero.

For completeness, we note that in Ref.\,\citen{EFNM91} we have with considerable advantage used the truncated continued-fraction expansions of the screened interaction potential $\t{W}(z)$, appendix \ref{sa}, and the self-energy in the so-called $GW$ approximation \cite{HL69} (see also Ref.\,\citen{FEDvH94}, and \S\,8.9, p.\,205, in Ref.\,\citen{BF99a}).\footnote{For a fundamental shortcoming of the $GW$ approximation of the self-energy, we refer the reader to \S\,4, p.\,1499, of Ref.\,\protect\citen{BF02}.} The \textsl{terminating function} $\mathsf{t}(z)$ as encountered in Eq.\,(16) of Ref.\,\citen{EFNM91} takes approximate account of the equivalent of the function $\t{f}_n(z)-\alpha_n$ discussed in the previous paragraph. As should be evident from the above considerations, like $-\t{f}_n(z)$ the function $-\mathsf{t}(z)$ is to be a Nevanlinna function of $z$. A relatively detailed discussion of terminating functions can be found in Ref.\,\citen{CPVWJ08} under the heading of \textsl{tail sequence}.

The function $\t{f}^{\X{(n)}}(z)$ introduced in Eq.\,(\ref{e5ka}) can be expressed as \cite{Note16} [p.\,23 in Ref.\,\citen{NIA65}]\,\footnote{For the reason that we describe in footnote \raisebox{-1.0ex}{\normalsize{\protect\footref{noteq}}} on p.\,\protect\pageref{HereWe} below, our notation here differs from that in Ref.\,\protect\citen{NIA65} (p.\,23). To be consistent with the continued-fraction expansions in Eq.\,(\protect\ref{e5k}) \emph{et seq.}, one has to employ the recurrence relation $Y_{k+1} = (z - a_{k+1}) Y_k - \beta_k Y_{k-1}$ subject to the initial conditions given in Eqs\,(\protect\ref{e5t}) and (\protect\ref{e5ta}) below, to be contrasted with the initial conditions in Eq.\,[1.37] of Ref.\,\protect\citen{NIA65}.}
\begin{equation}\label{e5q}
\t{f}^{\X{(n)}}(z) = \alpha_0 + \beta_0 \hspace{0.6pt} \frac{\mathcal{N}_n(z)}{\mathcal{M}_n(z)},
\end{equation}
where $\mathcal{N}_n(z)$ and $\mathcal{M}_n(z)$ are \textsl{monic} polynomials\,\footnote{The coefficient of the leading term of a \textsl{monic} polynomial is equal to unity [Definition 4.5.4, p.\,182, in Ref.\,\protect\citen{AA14}].} of the form
\begin{equation}\label{e5s}
\mathcal{N}_n(z) = z^{n-1} + b_1 z^{n-1} + \dots + b_{n-1},
\end{equation}
\begin{equation}\label{e5r}
\mathcal{M}_n(z) = z^{n} + a_1 z^{n-1} + \dots + a_n,
\end{equation}
subject to the conditions [Eq.\,[1.37], p.\,23, in Ref.\,\citen{NIA65}]
\begin{equation}\label{e5t}
\mathcal{N}_0(z) = 0,\; \mathcal{N}_1(z) = 1,
\end{equation}
\begin{equation}\label{e5ta}
\mathcal{M}_0(z) = 1,\; \mathcal{M}_1(z) = z -\alpha_1.
\end{equation}
With reference to the equalities in Eqs\,(\ref{e5cb}), and (\ref{e5e}), one observes that for $n=1$ the expression in Eq.\,(\ref{e5q}) as applied to the Green function, for which $\alpha_0=0$, Eq.\,(\ref{e4db}), results in the approximation of this function within the framework of the \textsl{exact} Hartree-Fock theory.\footnote{For some relevant details, see \S\,\ref{sec.3a.1}.} We have discussed some significant properties of this approximate Green function in Ref.\,\citen{BF13}.

One demonstrates that [p.\,24 in Ref.\,\citen{NIA65}]
\begin{equation}\label{e5tb}
\frac{\mathcal{N}_n(z)}{\mathcal{M}_n(z)} \equiv \frac{Q_n(z)}{P_n(z)},
\end{equation}
where $\{P_j(z)\| j\in \mathds{N}_{0}\}$ and $\{Q_j(z)\| j\in \mathds{N}_{0}\}$ are orthogonal polynomials of respectively first and second kind, to be introduced below, Eqs\,(\ref{e5y}), (\ref{e5u}), and (\ref{e5x}). We thus write the expression in Eq.\,(\ref{e5q}) in the following equivalent form \cite{Note16}:
\begin{equation}\label{e5tc}
\t{f}^{\X{(n)}}(z) = \alpha_0 + \beta_0 \hspace{0.8pt} \frac{Q_n(z)}{P_n(z)}.
\end{equation}
Since $Q_n(z)$ and $P_n(z)$ are respectively $(n-1)$th-order and $n$th-order polynomials of $z$, one has the exact asymptotic expression [Eq.\,[1.34b], p.\,22, in Ref.\,\citen{NIA65}]
\begin{equation}\label{e5td}
\t{f}^{\X{(n)}}(z) \sim \alpha_0 + \frac{\mu_0}{z} + \frac{\mu_1}{z^2} + \dots + \frac{\mu_{2n-2}}{z^{2n-1}} + O(1/z^{2n})\;\; \text{for}\;\; z\to\infty,
\end{equation}
where, in the light of the expression in Eq.\,(\ref{e4da}), the moments $\{\mu_0,\mu_1,\dots,\mu_{2n-2}\}$ are those of the \textsl{exact} moment problem, to be distinguished from the truncated moment problem under discussion.

For the following considerations, it is relevant to note that from  the expressions in Eqs\,(\ref{e5b}) and (\ref{e5g}) one obtains the exact result [Eq.\,(2.38), p.\,47, in Ref.\,\citen{ST70}]
\begin{equation}\label{e5te}
\t{f}(z) = \alpha_0 + \beta_0\hspace{0.8pt} \frac{Q_{n+1}(z) - (\t{f}_{n+1}(z) -\alpha_{n+1})\hspace{0.6pt} Q_{n}(z)}{P_{n+1}(z) - (\t{f}_{n+1}(z)-\alpha_{n+1})\hspace{0.6pt} P_{n}(z)},
\end{equation}
where the function $\t{f}_{n+1}(z)$ is recursively determined on the basis of the right-most equality in Eq.\,(\ref{e5g}), with the function $\t{f}_1(z)$ specific to the one-particle Green function given in Eq.\,(\ref{e5f}). Clearly, for $\t{f}_{n+1}(z) \equiv \alpha_{n+1}$ the expression on the RHS of Eq.\,(\ref{e5te}) correctly reduces to the rational function $\t{f}^{\X{(n+1)}}(z)$, Eq.\,(\ref{e5tc}).\footnote{See the remarks following Eqs\,(\ref{e5i}) and (\ref{e5ka}) above.}

The coefficients $\{\alpha_j\| j\in\mathds{N}\}$ and $\{\beta_j\| j\in\mathds{N}\}$ encountered above, with $\alpha_j \in \mathds{R}$ and $\beta_j > 0$,\footnote{See the remarks following Eq.\,(\ref{e5i}) above.} constitute the following symmetric Jacobi matrix \cite{NIA65,GM10},\footnote{For Jacobi matrices the interested is referred to Ref.\,\protect\citen{GK02}.} which is central to the moment problem under discussion:\,\refstepcounter{dummy}\label{HereWe}\footnote{Here we deviate from the notation in Ref.\,\protect\citen{NIA65}. The reason for this is that the continued fractions in Eq.\,(\protect\ref{e5k}) \emph{et seq.} rely on the notational conventions of Ref.\,\protect\citen{ST70}, where, incidentally, Jacobi matrices are not explicitly referred to. Our notation here coincides with that of Ref.\,\protect\citen{GM10} (see Ch.\,2 herein). The three-term recurrence relation in Eq.\,(\protect\ref{e5y}) below also coincides with that in Eq.\,(2.8) of Ref.\,\protect\citen{GM10}, however in contrast to here, in the latter reference only one solution of this recurrence relation is considered and subject to initial conditions that differ from either of the initial conditions in Eqs\,(\protect\ref{e5u}) and (\protect\ref{e5x}) below. The choice of the latter initial conditions is strictly tied to the continued-fractions expansions in Eq.\,(\protect\ref{e5k}) \emph{et seq.} \label{noteq}}
\begin{equation}\label{e5j}
\mathscr{J} \doteq \begin{pmatrix}
\alpha_1 & \sqrt{\beta_1} & 0 & 0 & 0 & \ldots \\
\sqrt{\beta_1} & \alpha_2 & \sqrt{\beta_2} & 0 & 0 & \ldots \\
0 & \sqrt{\beta_2} & \alpha_3 & \sqrt{\beta_3} & 0 & \ldots \\
\vdots & \vdots & \vdots & \vdots & \vdots & \ddots \end{pmatrix},\;\;\;\alpha_j \in\mathds{R},\; \beta_j > 0.
\end{equation}
In considering the Green function $\t{G}_{\sigma}(\bm{k};z)$, we denote the relevant Jacobi matrix by $\mathscr{J}^{\X{\t{G}_{\sigma}}}$, described in terms of the coefficients $\{\alpha_j \| j\in\mathds{N}\}$ and $\{\beta_j\| j\in\mathds{N}\}$ encountered in Eq.\,(\ref{e5kc}). With reference to the expressions in Eqs\,(\ref{e5kc}) and (\ref{e5ke}), the Jacobi matrix $\mathscr{J}^{\X{\t{\mathcal{E}}_{\sigma}}}$ is evidently a sub-matrix of $\mathscr{J}^{\X{\t{G}_{\sigma}}}$.

Following\refstepcounter{dummy}\label{FollowingAkhiezer} Akhiezer [p.\,28 in Ref.\,\citen{NIA65}], we denote the symmetric Jacobi matrix obtained by removing the first $p$ rows and $p$ columns of the Jacobi matrix $\mathscr{J}$ in  Eq.\,(\ref{e5j}) by $\mathscr{J}_p$ and refer to it as the ``$p$-th abbreviated matrix'' ``in relation to the matrix $\mathscr{J}$''. Thus $\mathscr{J}^{\X{\t{\mathcal{E}}_{\sigma}}}$ is the \textsl{first abbreviated matrix} in relation to $\mathscr{J}^{\X{\t{G}_{\sigma}}}$. Of particular significance is that $\mathscr{J}$ and $\mathscr{J}_p$ are of the same \textsl{type} [p.\,28 in Ref.\,\citen{NIA65}], of either the $\mathfrak{C}$ type or the $\mathfrak{D}$ type [Definition 1.3.2, p.\,19, in Ref.\,\citen{NIA65}].\footnote{See the considerations in \S\,\protect\ref{sec.3.2.2}.} The moment problems associated with type-$\mathfrak{C}$ Jacobi matrices are \textsl{indeterminate} and those associated with type-$\mathfrak{D}$ Jacobi matrices \textsl{determinate} [Corollary 2.2.4, p.\,41, in Ref.\,\citen{NIA65}]. According to Carleman [p.\,85 in Ref.\,\citen{NIA65}], the result in Eq.\,(\ref{e4oh}) implies that $\mathscr{J}^{\X{\t{G}_{\sigma}}}$ is a type-$\mathfrak{D}$ Jacobi matrix. It thus follows that $\mathscr{J}^{\X{\t{\mathcal{E}}_{\sigma}}}$ is also a type-$\mathfrak{D}$ Jacobi matrix. Below we show that $\mathscr{J}^{\X{\t{\mathcal{E}}_{\sigma}}} = \mathscr{J}^{\X{\t{\Sigma}_{\sigma}}}$, Eq.\,(\ref{e5z}) below.\footnote{For now, with reference to Eqs\,(\protect\ref{e7b}) and (\protect\ref{e7ba}) below, note that since $\Sigma_{\sigma}^{\textsc{hf}}(\bm{k})$ is independent of $z$, one has $\mathcal{E}_{\sigma;\infty_j}(\bm{k}) \equiv \hbar\Sigma_{\sigma;\infty_j}(\bm{k})$, $\forall j\in\mathds{N}$.} Thus in the case of the $N$-particle uniform GSs of Hubbard-like models the moment problem associated with $\mathscr{J}^{\X{\t{\Sigma}_{\sigma}}}$ is like that associated with $\mathscr{J}^{\X{\t{G}_{\sigma}}}$ \textsl{determinate}. The result in Eq.\,(\ref{e7m}) below explicitly confirms the validity of this observation.

We shall discuss the consequence of the moment problem associated with $\mathscr{J}^{\X{\t{\Sigma}_{\sigma}}}$ being \textsl{determinate} later in this appendix, \S\,\ref{sec.b2}, and for now suffice to mention that the moment problem associated with $\mathscr{J}^{\X{\t{G}_{\sigma}}}$ being \textsl{determinate} implies amongst others that the measure $\upgamma_{\sigma}(\bm{k};\varepsilon)$ in Eq.\,(\ref{e4ob}) is essentially uniquely determined by the \textsl{positive sequence} $\{G_{\sigma;\infty_j}(\bm{k}) \| j\}$. It further implies that polynomials of $z$ are dense in the inner-product space $L_{\upgamma_{\sigma}}^2$ associated with the measure $\upgamma_{\sigma}(\bm{k};\varepsilon)$ [Theorem and Corollary 2.3.3, p.\,45, in Ref.\,\citen{NIA65}], implying that functions inside the space $L_{\upgamma_{\sigma}}^2$ can be arbitrary accurately described in terms of polynomials of $z$. The latter property is however not unique to \textsl{determinate} moment problems; following Riesz [Theorem 2.3.2, p.\,43, in Ref.\,\citen{NIA65}], it also applies to so-called \textsl{$N$-extremal} [p.\,43 in Ref.\,\citen{NIA65}] measure functions of \textsl{indeterminate} moment problems.

Any symmetric Jacobi matrix $\mathscr{J}$ of the kind presented in Eq.\,(\ref{e5j}), that is with $\alpha_j \in\mathds{R}$ and $\beta_j > 0$, corresponds to a set of orthogonal polynomials $\{P_j(z)\| j\}$, referred to as polynomials of the first kind, and an associated set of polynomials $\{Q_j(z)\| j\}$, referred to as polynomials of the second kind,\footnote{Whether or not the polynomials $\{Q_j(z)\| j\}$ are orthogonal plays no direct role in the considerations of the moment problem. With reference to Eq.\,(\protect\ref{e5xa}) below, it is however evident that the measure function with respect to which $\{Q_j(z)\| j\}$ are orthogonal is not the same as that with respect to which $\{P_j(z)\| j\}$ are orthogonal. We note that according to a general theorem [Theorem 2.8, p.\,12, in Ref.\,\protect\citen{GM10}] for the polynomials satisfying a recurrence relation of the kind in Eq.\,(\protect\ref{e5y}) below, there exists a measure function with respect to which they are orthogonal. These observations are exemplified by the details in footnote \raisebox{-1.0ex}{\normalsize{\protect\footref{noter1}}} on p.\,\protect\pageref{NoteThat}.} which satisfy the three-term recurrence relation \cite{Note19}\,\footnote{See the remarks in footnote \raisebox{-1.0ex}{\normalsize{\protect\footref{noteq}}} on p.\,\protect\pageref{HereWe} regarding the symmetric Jacobi matrix in Eq.\,(\protect\ref{e5j}).}
\begin{equation}\label{e5y}
\sqrt{\beta_{j}} X_{j-1}(z) + \alpha_{j+1} X_{j}(z) + \sqrt{\beta_{j+1}} X_{j+1}(z) = z X_{j}(z),\;\; j \in \mathds{N},
\end{equation}
subject to the following initial conditions [p.\,8 in Ref.\,\citen{NIA65}]:
\begin{equation}\label{e5u}
P_0(z) =\frac{1}{\sqrt{s_0}},\;\; P_1(z) = \frac{z-\alpha_1}{\sqrt{s_0\hspace{0.6pt} \beta_1}},
\end{equation}
\begin{equation}\label{e5x}
Q_0(z) = 0,\;\; Q_1(z) = \frac{1}{\sqrt{s_0\hspace{0.6pt} \beta_1}},
\end{equation}
where $s_0$ is the $0$th-order moment of the relevant measure function, Eq.\,(\ref{e6c}).\footnote{In \textsl{this} section where the relevant measure function is $\upgamma_{\sigma}(\bm{k};\varepsilon)$, $s_0 = G_{\sigma;\infty_1}(\bm{k}) \equiv \hbar$, Eqs\,(\protect\ref{e4qa}), (\protect\ref{e4oa}), and (\protect\ref{e4ob}). In \S\,\protect\ref{sec.b2} where the relevant measure function is $\upsigma_{\sigma}(\bm{k};\varepsilon)$, $s_0 = \Sigma_{\sigma;\infty_1}(\bm{k})$, Eqs\,(\protect\ref{e20c}), (\protect\ref{e7j}), and (\protect\ref{e7k}). Note that, in the former case $\llbracket s_0\rrbracket = \mathrm{Js}$, while in the latter $\llbracket s_0\rrbracket = \mathrm{Js}^{-1}$.}\footnote{Following Eqs\,(\protect\ref{e5e}) and (\ref{e5ib}), one has $\llbracket P_0(z)\rrbracket = \llbracket P_1(z)\rrbracket = \mathrm{J}^{-1/2}\mathrm{s}^{-1/2}$ and $\llbracket Q_1(z)\rrbracket = \mathrm{J}^{-3/2}\mathrm{s}^{-1/2}$, which appropriately result in $\llbracket \beta_0\hspace{0.6pt} Q_1(z)/P_1(z)\rrbracket = \mathrm{s}$, Eq.\,(\protect\ref{e5tc}). We recall that $\llbracket\beta_0\rrbracket = \mathrm{Js}$ and $\llbracket\beta_j\rrbracket = \mathrm{J}^2$, $\forall j\in \mathds{N}$ (see footnote \raisebox{-1.0ex}{\normalsize{\protect\footref{notem1}}} on p.\,\protect\pageref{TheDeviation}). In the light of Eqs\,(\protect\ref{e5s}), (\protect\ref{e5r}), and (\protect\ref{e5tb}), note that $\llbracket Q_n(z)/P_n(z) \rrbracket = \llbracket 1/z\rrbracket$, so that for $\llbracket z\rrbracket = \mathrm{J}$ one must indeed have $\llbracket Q_n(z)/P_n(z) \rrbracket = \mathrm{J}^{-1}$, $\forall n$.} Denoting this measure function, associated with the orthogonal polynomials $\{P_j(z) \| j\}$, by $\upvartheta(\varepsilon)$,\,\footnote{One thus has $\int_{-\infty}^{\infty} \mathrm{d}\upvartheta(\varepsilon)\, P_i(\varepsilon) P_j(\varepsilon) = \delta_{i,j}$.} the polynomial $Q_j(z)$ can be directly deduced from $P_j(z)$ according to the following expression [\emph{cf.} Eq.\,[1.14], p.\,8, in Ref.\,\protect\citen{NIA65}] \cite{Note19}:
\begin{equation}\label{e5xa}
Q_{j}(z) = \frac{1}{s_0} \int_{-\infty}^{\infty} \mathrm{d}\upvartheta(\varepsilon)\, \frac{P_j(z) - P_j(\varepsilon)}{z-\varepsilon},\;\; \forall j \in \mathds{N}_0.
\end{equation}
One verifies that this expression indeed reproduces the polynomials in Eq.\,(\ref{e5x}) on the basis of those in Eq.\,(\ref{e5u}). We have already met the polynomials $\{P_j(z)\| j\}$ and $\{Q_j(z)|| j\}$ in Eqs\,(\ref{e5tb}) and (\ref{e5tc}). Denoting the polynomials associated with $\mathscr{J}_p$ by $\{P_j^{\X{(p)}}(z)\}$ and $\{Q_j^{\X{(p)}}(z)\}$, one naturally has $P_j(z) \equiv P_j^{\X{(0)}}(z)$ and $Q_j(z) \equiv Q_j^{\X{(0)}}(z)$. In the light of the discussions in the previous paragraph, one observes that with the expression in Eq.\,(\ref{e5tc}), wherein $\alpha_0$ is identified with $0$ (following Eq.\,(\ref{e4db})), describing the $n$th-order approximant of the Green function $\t{G}_{\sigma}(\bm{k};z)$, one obtains the $n$th-order approximant of $\t{\mathcal{E}}_{\sigma}(\bm{k};z)$, Eq.\,(\ref{e5ke}), on replacing  $\beta_0$ with $\beta_1$ and the polynomials $P_n(z)$ and $Q_n(z)$ with respectively $P_n^{\X{(1)}}(z)$ and $Q_n^{\X{(1)}}(z)$.\footnote{The initial conditions specific to $P_n^{\protect\X{(1)}}(z)$ and $Q_n^{\protect\X{(1)}}(z)$ are the same as those in Eqs\,(\protect\ref{e5u}) and (\protect\ref{e5x}) except that $\alpha_1$ and $\sqrt{\beta_1}$ herein are to be replaced by respectively $\alpha_2$ and $\sqrt{\beta_2}$ (compare the first row of the Jacobi matrix $\mathscr{J}$ in Eq.\,(\protect\ref{e5j}) with that of the corresponding $\mathscr{J}_1$). We note that whereas $P_1(z)$ and $Q_1(z)$ depend on $s_0$, the ratio $Q_1(z)/P_1(z)$ does not. In fact, and importantly, the ratio $Q_n(z)/P_n(z)$ as encountered in Eq.\,(\protect\ref{e5tc}) is independent of $s_0$ for arbitrary values of $n\in \mathds{N}$.} Note that since $\t{\mathcal{E}}_{\sigma}(\bm{k};z) = o(1)$ for $z\to\infty$, Eqs\,(\ref{e7b}) and (\ref{e7ba}) below, the $\alpha_0$ associated with $\t{\mathcal{E}}_{\sigma}(\bm{k};z)$ is, like that associated with $\t{G}_{\sigma}(\bm{k};z)$, vanishing. Denoting the $n$th-order approximant of $\t{\mathcal{E}}_{\sigma}(\bm{k};z)$ by $\t{\mathcal{E}}_{\sigma}^{\X{(n)}}(\bm{k};z)$, following Eq.\,(\ref{e5ke}) the $n$th-order approximant of the self-energy $\t{\Sigma}_{\sigma}(\bm{k};z)$ is equal to $\Sigma_{\sigma}^{\textsc{hf}}(\bm{k}) +\hbar^{-1} \t{\mathcal{E}}_{\sigma}^{\X{(n)}}(\bm{k};z)$. In view of the expression for $\alpha_1$ in Eq.\,(\ref{e5e}), for the $n$th-order approximant of the function $\t{E}_{\sigma}(\bm{k};z)$ as defined in Eq.\,(\ref{e5f}) one has $\t{E}_{\sigma}^{\X{(n)}}(\bm{k};z) = \alpha_1 + \t{\mathcal{E}}_{\sigma}^{\X{(n)}}(\bm{k};z)$. Summarising,
\begin{equation}\label{e5z}
\mathscr{J}^{\X{\t{E}_{\sigma}}} = \mathscr{J}^{\X{\t{\Sigma}_{\sigma}}} =\mathscr{J}^{\X{\t{\mathcal{E}}_{\sigma}}}.
\end{equation}

For the following considerations in this appendix it is important to point out that the condition $P_0(z) = 1/\sqrt{s_0}$ in Eq.\,(\ref{e5u}) has important consequences for any expression that relies on the Christoffel-Darboux formula [\S\,22.12.1, p.\,785, in Ref.\,\citen{AS72}].\footnote{See also Ref.\,\protect\citen{NIA65}. For a detailed discussion of this formula, consult \S\,3.2, p.\,42, in Ref.\,\protect\citen{GS75}.} In Ref.\,\citen{NIA65}, on which our considerations in this appendix rely as main reference source regarding the moment problem, the Christoffel-Darboux formula is used under the implicit assumption that the orthogonal polynomials $\{P_j(z)\| j\in \mathds{N}_0\}$ are normalised equally.\footnote{Under this assumption, the Christoffel-Darboux formula takes the form as in Theorem (3.2.3), p.\,43, in Ref.\,\protect\citen{GS75}, to be contrasted with the formula in \S\,22.12.1, p.\,785, in Ref.\,\protect\citen{AS72}.} While this is the case for $P_j(z)$, $\forall j\in \mathds{N}$, this is not the case insofar as the function $P_0(z) \equiv 1$ is concerned. Denoting the measure function under consideration by $\upvartheta(\varepsilon)$ (\emph{cf.} Eq.\,(\ref{e5xa})), which may be identified with the main of measure functions encountered in this appendix,\footnote{Explicitly, the measures $\upgamma_{\sigma}(\bm{k};\varepsilon)$, Eq.\,(\protect\ref{e4g}), and $\upsigma_{\sigma}(\bm{k};\varepsilon)$, Eq.\,(\protect\ref{e20c}) below. Interestingly, for the measure $\upgamma(\varepsilon)$ in Eq.\,(\protect\ref{e6qg}) below one has $\int_{-1}^{1} \mathrm{d}\upgamma(\varepsilon)\, T_0^2(\varepsilon) = 1$, and  $\int_{-1}^{1} \mathrm{d}\upgamma(\varepsilon)\, T_j^2(\varepsilon) = 1/2$, $\forall j \in \mathds{N}$. Note that in \S\,\protect\ref{sabx}, $\llbracket \varepsilon\rrbracket = \llbracket z\rrbracket = 1$, that is $\varepsilon$ and $z$ are dimensionless.}\footnote{See footnote \raisebox{-1.0ex}{\normalsize{\protect\footref{noten1}}} on p.\,\protect\pageref{InRef}.} one has\,\footnote{Unless $s_0 = 1$, Eq.\,[1.3] on p.\,3 of Ref.\,\protect\citen{NIA65} is incorrect.}
\begin{equation}\label{e6ja}
\int_{-\infty}^{\infty} \mathrm{d}\upvartheta(\varepsilon)\, P_j^2(\varepsilon) = \left\{\begin{array}{ll} s_0, & j = 0,\\ 1, & j \in \mathds{N},\\
\end{array}\right. \;\;\;\;\; \text{(Assuming $P_0(z) \equiv 1$)}
\end{equation}
where $s_0$ is the positive constant encountered in the Hankel moment matrix in Eq.\,(\ref{e6c}). In the case of the moment problem associated with $\t{G}_{\sigma}(\bm{k};z)$, one has $s_0 = \hbar$, Eqs\,(\ref{e4qa}), (\ref{e4oa}), and (\ref{e4ob}), and in the case of that associated with $\t{\Sigma}_{\sigma}(\bm{k};z)$,  $s_0 = \Sigma_{\sigma;\infty_1}(\bm{k})$, Eqs\,(\ref{e7j}), (\ref{e7k}), (\ref{e7la}), and (\ref{e7c}) below. The expressions in Ref.\,\citen{NIA65} that rely on the Christoffel-Darboux formula remain applicable irrespective of whether the $s_0$ on the RHS of Eq.\,(\ref{e6ja}) is equal to $1$ or the functions $\{P_j(z)\| j\in\mathds{N}\}$ are normalised such that the $1$ on the RHS of Eq.\,(\ref{e6ja}) is replaced by $s_0$.\refstepcounter{dummy}\label{OneCanReadily}\footnote{One can readily appreciate this fact by considering the first expression in Eq.\,(\protect\ref{e6ub}) below, where inspection shows that $\rho_{n+1}^{\mathrm{c}}(z)$ does not depend on $s_0$. \emph{This observation is not to be misconstrued as implying that we could have suppressed the $s_0$ in Eqs\,(\ref{e5u}) and (\ref{e5x}), and thus in Eq.\,(\protect\ref{e6ub}), as by doing so the relevant expression for $\rho_{n+1}^{\mathrm{c}}(z)$ (as well as the right-most expression for $\gamma_j$ in Eq.\,(\protect\ref{e6n}) below) would have looked functionally differently.} In contrast, in the case $\int_{-\infty}^{\infty} \mathrm{d}\upvartheta(\varepsilon)\, P_j^2(\varepsilon) = s_0$, $\forall j\in \mathds{N}_0$, the positive constant $s_0$ (which may or may not be equal to $1$) is merely a multiplicative constant on both sides of the Christoffel-Darboux formula [\S\,22.12.1, p.\,785, in Ref.\,\protect\citen{AS72}], which can be discarded. In \S\,\protect\ref{sabx} we highlight these facts. \label{noteo1}}

Given the \textsl{orthogonal} polynomials $\{P_j(z)\| j\}$ of first kind and the associated polynomials $\{Q_j(z)\| j \}$ of the second kind (and, more generally, $\{P_j^{\X{(p)}}(z)\| j\}$ and $\{Q_j^{\X{(p)}}(z)\| j\}$), one introduces the following \textsl{quasi-orthogonal} polynomials $\{P_j(z,\ptau)\| j\}$ and the associated polynomials $\{Q_j(z,\ptau)\| j\}$ ($\{P_j^{\X{(p)}}(z,\ptau)\| j\}$ and $\{Q_j^{\X{(p)}}(z,\ptau)\| j\}$) [pp.\,10, 11, 21, and 22 in Ref.\,\citen{NIA65}]:
\begin{equation}\label{e6j}
X_j(z,\ptau) \doteq X_j(z) - \ptau X_{j-1}(z).
\end{equation}
For an arbitrary real $\ptau \in (-\infty,\infty)$, the rational function $Q_{n+1}(z,\ptau)/P_{n+1}(z,\ptau)$ allows for the following exact representation [Eq.\,[2.4], p.\,30, in Ref.\,\citen{NIA65}]:
\begin{equation}\label{e6k}
\t{w}_{n+1}(z,\ptau) \doteq -\frac{Q_{n+1}(z,\ptau)}{P_{n+1}(z,\ptau)} = -\sum_{j=1}^{n+1} \frac{\gamma_j}{z -\lambda_j},
\end{equation}
where $\lambda_j \equiv \lambda_j^{\X{(n+1)}}(\ptau)$ denotes the $j$th of the $n+1$ ordered zeros of $P_{n+1}(z,\ptau)$, $\ptau \in\mathds{R}$, satisfying the strict inequalities [\S\,4.1, p.\,20 in Ref.\,\citen{NIA65}]
\begin{equation}\label{e6l}
\lambda_1 < \lambda_2 < \dots < \lambda_{n+1}.
\end{equation}
Following $P_{n+1}(\lambda_j,\ptau) = 0 \Leftrightarrow \ptau = P_{n+1}(\lambda_j)/P_{n}(\lambda_j)$,\footnote{See Theorem 1.2.2, p.\,10, in Ref.\,\protect\citen{NIA65}.} one obtains that\,\footnote{Up to the prefactor $1/s_0$, the last equality in Eq.\,(\protect\ref{e6n}) coincides with that in Eq.\,[1.33b] of Ref.\,\protect\citen{NIA65} (p.\,22). The latter equality is obtained on the basis of the Liouville-Ostrogradskii and Christoffel-Darboux formulae [Eqs\,[1.15] and [1.17], p.\,9, in Ref.\,\protect\citen{NIA65}]. With reference to the Christoffel-Darboux formula in \S\,22.12.1, p.\,785, of Ref.\,\protect\citen{AS72}, here we have used $k_{n+1}/k_n = 1/\sqrt{\beta_{n+1}}$, Eqs\,(\protect\ref{e7wd}), (\protect\ref{e7wex}), and (\protect\ref{e7wgx}), together with $x-y = \delta \Leftrightarrow x = y + \delta$ with $\delta \to 0$, and insofar as the Liouville-Ostrogradskii formula is concerned, here we have used Eq.\,(\protect\ref{e7wc}). With $P_{n+1}'(\lambda_j,\protect\ptau) \equiv \partial P_{n+1}(z,\protect\ptau)/\partial z\vert_{z=\lambda_j}$, we note that $\gamma_j$ is appropriately dimensionless, in accord with the fact that $\llbracket \protect\t{w}_{n+1}(z,\protect\ptau)\rrbracket = \llbracket 1/z\rrbracket$, Eqs\,(\protect\ref{e5s}), (\protect\ref{e5r}), (\protect\ref{e5tb}), and (\protect\ref{e6k}). See the remarks in footnote \raisebox{-1.0ex}{\normalsize{\protect\footref{noteo1}}} on p.\,\protect\pageref{OneCanReadily}.} [pp.\,21 and 22 in Ref.\,\citen{NIA65}]
\begin{equation}\label{e6n}
\gamma_j = \frac{Q_{n+1}(\lambda_j,\ptau)}{P_{n+1}'(\lambda_j,\ptau)} \equiv \frac{P_n(\lambda_j) Q_{n+1}(\lambda_j) - P_{n+1}(\lambda_j) Q_n(\lambda_j)}{P_n(\lambda_j) P_{n+1}'(\lambda_j) - P_{n+1}(\lambda_j) P_n'(\lambda_j)} \equiv \frac{1}{s_0} \sum_{k=0}^{n} \frac{1}{\vert P_k(\lambda_j)\vert^2},
\end{equation}
establishing the inequality
\begin{equation}\label{e6m}
\gamma_j \equiv \gamma_j^{\X{(n+1)}}(\ptau) > 0,\;\; \ptau \in\mathds{R}.
\end{equation}
The significance of the function $\t{w}_{n+1}(z,\ptau)$ becomes apparent by comparing its defining expression in Eq.\,(\ref{e6k}) with the exact expression for $\t{f}(z)$ in Eq.\,(\ref{e5te}); one observes that
\begin{equation}\label{e6jd}
-\beta_0\hspace{0.6pt} \t{w}_{n+1}(z,\ptau) + \alpha_0 = \left. \t{f}(z)\right|_{\t{f}_{n+1}(z) -\alpha_{n+1} \to\ptau},
\end{equation}
where the $\t{f}(z)$ on the RHS stands for the exact expression on the RHS of Eq.\,(\ref{e5te}). With reference to the expressions in Eqs\,(\ref{e5ka}), (\ref{e5n}), (\ref{e5o}), and (\ref{e5tc}), one further observes that [p.\,47 in Ref.\,\citen{ST70}]\,\footnote{\emph{Cf.} Eqs\,(\protect\ref{e5p}) and (\protect\ref{e119a}).}
\begin{equation}\label{e6jc}
-\beta_0\hspace{0.6pt} \t{w}_{n+1}(z,\ptau) + \alpha_0\equiv \left.\t{f}^{\X{(n+1)}}(z)\right|_{\alpha_{n+1} \to \alpha_{n+1} + \ptau} \equiv \mathsf{S}_{n+1}(z,\ptau).
\end{equation}
Following Eq.\,(\ref{e6j}),
\begin{equation}\label{e6qa}
\frac{Q_{n+1}(z,\ptau)}{P_{n+1}(z,\ptau)} = \left\{ \begin{array}{ll} Q_n(z)/P_n(z), & \ptau = \pm\infty,\\ \\
Q_{n+1}(z)/P_{n+1}(z), & \ptau = 0, \end{array}\right.
\end{equation}
which, with reference to Eq.\,(\ref{e6k}), leads to [p.\,14, in Ref.\,\citen{NIA65}]
\begin{equation}\label{e6jb}
\t{w}_{n+1}(z,\pm\infty) \equiv \t{w}_{n}(z,0).
\end{equation}

In analogy with the expression in Eq.\,(\ref{e4d}), for the spectral function $\mathcal{A}_{n+1}(\varepsilon,\ptau)$ corresponding to the function $-\t{w}_{n+1}(z,\ptau)$, Eq.\,(\ref{e6k}), one has
\begin{equation}\label{e6r}
\mathcal{A}_{n+1}(\varepsilon,\ptau) = \sum_{j=1}^{n+1} \gamma_j\hspace{0.8pt} \delta(\varepsilon-\lambda_j),
\end{equation}
which, on account of the inequality in Eq.\,(\ref{e6m}), in non-negative. With reference to Eq.\,(\ref{e4g}), to the spectral function $\mathcal{A}_{n+1}(\varepsilon,\ptau)$ corresponds a measure function  \cite{PRH50,APM11} $\upgamma_{n+1}(\varepsilon,\ptau)$, defined according to\,\footnote{For finite values of $n$, the measure $\upgamma_{n+1}(\varepsilon,\ptau)$ clearly depends on $\ptau$ through the dependence of $ \gamma_j$ and $\lambda_j$ on $\ptau$.}
\begin{equation}\label{e6s}
\upgamma_{n+1}(\varepsilon,\ptau) \doteq\int_{-\infty}^{\varepsilon} \rd\varepsilon'\; \mathcal{A}_{n+1}(\varepsilon',\ptau) = \sum_{j=1}^{n+1} \gamma_j\hspace{0.8pt} \Theta(\varepsilon-\lambda_j),
\end{equation}
which is a non-decreasing stepwise-constant function of $\varepsilon$. One can thus express the function $\t{w}_{n+1}(z,\ptau)$ in Eq.\,(\ref{e6k}) in the following equivalent form:\,\footnote{The sign difference between the expression in Eq.\,(\protect\ref{e6ua}) and that in, for instance, Eq.\,(\protect\ref{e4j}) signifies the fact that while $\t{w}_{n+1}(z,\ptau)$ is a Nevanlinna function of $z$, it is not $\t{G}_{\sigma}(\bm{k};z)$ but $-\t{G}_{\sigma}(\bm{k};z)$ that is similarly a Nevanlinna function of $z$.}
\begin{equation}\label{e6ua}
\t{w}_{n+1}(z,\ptau) = -\int_{-\infty}^{\infty} \frac{\rd\upgamma_{n+1}(\varepsilon',\ptau)}{z-\varepsilon'}.
\end{equation}
One clearly has [p.\,31 in Ref.\,\citen{NIA65}]
\begin{equation}\label{e6t}
\upgamma_{n+1}(\lambda_j+0^+,\ptau) - \upgamma_{n+1}(\lambda_j-0^+,\ptau) = \gamma_j,\;\; j\in \{1, 2, \dots, n+1\}.
\end{equation}

Defining  $s_j \equiv s_j^{\X{(n+1)}}(\ptau)$ as (\emph{cf.} Eq.\,(\ref{e4ob}))
\begin{equation}\label{e6u}
s_j \doteq \int_{-\infty}^{\infty} \rd\upgamma_{n+1}(\varepsilon,\ptau)\,\varepsilon^j,
\end{equation}
from the expression in Eq.\,(\ref{e6ua}) one obtains
\begin{equation}\label{e4pa}
-\t{w}_{n+1}(z,\ptau) \sim \frac{s_0}{z} + \frac{s_1}{z^2} + \dots\;\; \text{for}\;\; z\to\infty.
\end{equation}
An identical asymptotic expression is obtained from the right-most expression in Eq.\,(\ref{e6k}) through employing the geometric expansion $1/z + \lambda_j/z^2 + \dots$ for $1/(z -\lambda_j)$, from which one infers that [Eq.\,[2.5], p.\,31, in Ref.\,\citen{NIA65}]
\begin{equation}\label{e6o}
s_j = \sum_{k=1}^{n+1} \gamma_k^{\phantom{j}} \lambda_k^j.
\end{equation}
With reference to Eqs\,(\ref{e4n}), (\ref{e4da}), and (\ref{e4db}), and in the light of the discussions in the paragraph following Eq.\,(\ref{e6d}) above (p.\,\pageref{Forthefollowing}), the sequence $\{s_j\| j\in\mathds{N}_{0}\}$ determines the Hankel moment matrix $\mathbb{H}_m^{\X{-}\t{\X{w}}_{\X{n+1}}}$ associated with $-\t{w}_{n+1}(z,\ptau)$. Note that $\t{w}_{n+1}(z,\ptau)$ is Nevanlinna, not $-\t{w}_{n+1}(z,\ptau)$.

Since the polynomials $Q_j(z)$ and $P_j(z)$, $j=n,n+1$, in terms of which the polynomials $Q_{n+1}(z,\ptau)$ and $P_{n+1}(z,\ptau)$ are determined, Eq.\,(\ref{e6j}), correspond to the Jacobi matrix $\mathscr{J}^{\X{\t{G}_{\sigma}}}$, it can be shown that for \textsl{all} finite real values of $\ptau$ one has [p.\,23 in Ref.\,\citen{NIA65}]
\begin{equation}\label{e6oa}
s_j = \mu_j,\;\; j \in \{0,1,\dots,2n+\varsigma\},
\end{equation}
where $\mu_j$ is the moment in Eq.\,(\ref{e4ob}) (see Eq.\,(\ref{e4oa})), and
\begin{equation}\label{e6q}
\varsigma = \left\{\begin{array}{ll} 0, & \ptau \not=0, \pm\infty,\\
 1, & \ptau = 0. \end{array}\right.
\end{equation}
Similarly, in considering the above-mentioned polynomials corresponding to the Jacobi matrix $\mathscr{J}^{\X{\t{\Sigma}_{\sigma}}}$, the moment $s_j$ in Eq.\,(\ref{e6oa}) is to be equated with the moment $\b{\mu}_j$ as defined in Eq.\,(\ref{e7k}) below (see Eq.\,(\ref{e7j})). In the light of the equality in Eq.\,(\ref{e6oa}), one thus has the exact result (\emph{cf.} Eq.\,(\ref{e6k})) [Eqs\,[1.34a] and [1.34b], p.\,22, in Ref.\,\citen{NIA65}]\,\footnote{See also the remark following Eq.\,[1.33a], p.\,21, in Ref.\,\protect\citen{NIA65}.}
\begin{equation}\label{e6p}
-\t{w}_{n+1}(z,\ptau) \equiv \frac{Q_{n+1}(z,\ptau)}{P_{n+1}(z,\ptau)} \sim \frac{\mu_0}{z} +\frac{\mu_1}{z^2} + \dots +\frac{\mu_{2n+\varsigma}}{z^{2n+1+\varsigma}} + O\big(1/z^{2n+2+\varsigma}\big)\;\; \text{for}\;\; z\to \infty.
\end{equation}
We note that the expressions in Eqs\,(\ref{e6u}), (\ref{e6o}), and (\ref{e6oa}) constitute the Gauss \textsl{quadrature formula} [Eq.\,[1.32], p.\,21, in Ref.\,\citen{NIA65}], [Theorem 5.5.4, p.\,99, in Ref.\,\citen{CPVWJ08}] [Ch. 6 in Ref.\,\citen{GM10}].

Following Eqs\,(\ref{e6o}) and (\ref{e6oa}), one has
\begin{equation}\label{e7a}
\sum_{k=1}^{n+1} \gamma_k = \mu_0,\;\; \forall n \in \mathds{N}_0,
\end{equation}
which, on account of the inequality in Eq.\,(\ref{e6m}), is positive, and independent of the real parameter $\ptau$. Recall that in the case of the Green function $\t{G}_{\sigma}(\bm{k};z)$, one has $\mu_0 = \hbar$, Eq.\,(\ref{e5cb}). With reference to Eq.\,(\ref{e6s}) and in the light of Eqs\,(\ref{e6l}) and (\ref{e7a}), it follows that $\upgamma_{n+1}(\varepsilon,\ptau) \equiv 0$ for $\varepsilon < \lambda_1$ and $\upgamma_{n+1}(\varepsilon,\ptau) = \mu_0$ for $\varepsilon > \lambda_{n+1}$. Thus $\upgamma_{n+1}(\varepsilon,\ptau)$ is \textsl{bounded} for arbitrary values of $n$, in addition to being a non-decreasing function of $\varepsilon$ (\emph{cf.} Eq.\,(\ref{e4i})). For an illustrative example, see \S\,\ref{sabx} below.

From Eqs\,(\ref{e5tc}), (\ref{e6k}), and (\ref{e6qa}), one has (\emph{cf.} Eq.\,(\ref{e6jc}))
\begin{equation}\label{e6qb}
-\beta_0\hspace{0.6pt} \t{w}_{n+1}(z,\ptau) +\alpha_0 = \left\{\begin{array}{ll} \t{f}^{\X{(n)}}(z),  & \ptau=\pm\infty,\\ \\
\t{f}^{(n+1)}(z),  & \ptau=0. \end{array}\right.
\end{equation}
The above observations, based on the expressions in Eqs\,(\ref{e5td}), (\ref{e4pa}), (\ref{e6oa}), and (\ref{e6p}), reveal the interesting fact that even for $\ptau\not= 0, \pm\infty$ one has
\begin{equation}\label{e6qe}
-\beta_0\hspace{0.6pt} \t{w}_{n+1}(z,\ptau) +\alpha_0 = \t{f}^{(n+1)}(z) + O(1/z^{2n+2})\;\; \text{for}\;\; z\to\infty.
\end{equation}
This equality is relevant in that according to a theorem due to Hellinger [Theorem 1.2.3, p.\,11, in Ref.\,\citen{NIA65}] [Theorem 2.7, p.\,48, in Ref.\,\citen{ST70}], for a fixed but arbitrary $z$, with $\im[z] \gtrless 0$, the function $\t{w}_{n+1}(z,\ptau)$ maps out a circle (or, circular contour) $\mathsf{K}_{n+1}(z)$ in the complex half-plane $\im[w] \gtrless 0$ on the real parameter $\ptau$ in Eq.\,(\ref{e6k}) traversing the real axis.\footnote{While this circle [\emph{Kreis} in German] is generally denoted by $\mathsf{K}_{n+1}(z)$, it is common practice also to denote the region of the complex $w$-plane enclosed by this circle by $\mathsf{K}_{n+1}(z)$. Generally, but not invariably, the context clarifies the sense in which $\mathsf{K}_{n+1}(z)$ is used. This ambiguity becomes immaterial for $n\to\infty$ in the cases where the radius of $\mathsf{K}_{n+1}(z)$ converges to zero. See the discussions centred on Eqs\,(\protect\ref{e7wb}) and (\protect\ref{e7wf}) where we consider the function $\protect\t{w}_{n+1}(z,\zeta)$ as a M\"{o}bius transformation \protect\cite{TN00} in the complex $\zeta$-plane.} The possible convergence of the radius of this circle to zero for $n\to\infty$ implies the uniform convergence of the function $\t{f}^{\X{(n)}}(z)$ to $\t{f}(z) \equiv -\beta_0\hspace{0.6pt} \t{w}(z) + \alpha_0$ in the region $\im[z]\not=0$, where $\t{w}(z)$ is the $\ptau$-independent limit of the function $\t{w}_{n+1}(z,\ptau)$ for $n\to\infty$ (see the following paragraph as well as \S\,\ref{sec.b1a}).\footnote{As regards the $\protect\ptau$-independence of $\protect\t{w}_{n+1}(z,\protect\ptau)$ in the limit $n\to\infty$, see footnote \raisebox{-1.0ex}{\normalsize{\protect\footref{notel1}}} on p.\,\protect\pageref{TheIndependence}.} In this limit, the measure $\upgamma_{n+1}(\varepsilon,\ptau)$ in Eq.\,(\ref{e6s}) also uniformly converges to the \textsl{essentially unique} measure function $\upgamma(\varepsilon)$ [p.\,31 in Ref.\,\citen{NIA65}] (see p.\,\pageref{Followingtheabove} above).

Denoting the radius of the above-mentioned \textsl{circle} $\mathsf{K}_{n+1}(z)$ by $\rho_{n+1}^{\textrm{c}}(z)$, and the location of its centre in the $w$-plane by $w_{n+1}^{\textrm{c}}(z)$, one has\,\footnote{For a discussion of the expressions in Eq.\,(\protect\ref{e6ub}), see \S\,\protect\ref{sec.b1a} below. Note that while for any finite value of $n$ the function $-\protect\t{w}_{n+1}(z,\protect\ptau)$ depends on $\protect\ptau$, the functions $\rho_{n+1}^{\textrm{c}}(z)$ and $w_{n+1}^{\textrm{c}}(z)$ do not. For this, see the remark following Eq.\,(\protect\ref{e6uh}) below. We note that the $s_0$ in the denominator of the expression for $\rho_{n+1}^{\textrm{c}}(z)$ is directly related to the normalisation of $P_0(\varepsilon)$ to unity, Eq.\,(\protect\ref{e5u}). See the remarks in footnote \raisebox{-1.0ex}{\normalsize{\protect\footref{noteo1}}} on p.\,\protect\pageref{OneCanReadily}.} [Eqs\,[1.20] and [1.21], pp. 11 and 12, in Ref.\,\citen{NIA65}]
\begin{equation}\label{e6ub}
\rho_{n+1}^{\textrm{c}}(z) = \frac{1}{\vert z -z^*\vert}\hspace{1.0pt} \frac{1}{s_0 \sum_{j=0}^{n} \vert P_j(z)\vert^2},\;\;\; w_{n+1}^{\textrm{c}}(z) = -\frac{Q_{n+1}(z) P_{n}^*(z) - Q_{n}(z) P_{n+1}^*(z)}{P_{n+1}(z) P_{n}^*(z) - P_n(z) P_{n+1}^*(z)}.
\end{equation}
One observes that for $\im[z] \not= 0$ and $\sum_{j=0}^{n} \vert P_j(z)\vert^2 \to \infty$ as $n\to \infty$, the radius $\rho_{n+1}^{\textrm{c}}(z)$ of $\mathsf{K}_{n+1}(z)$ approaches zero for $n\to \infty$ [p.\,49 in Ref.\,\citen{NIA65}]. Since the function $\t{w}_{n+1}(z,\zeta)$ maps the half-plane $\im[\zeta] \gtrless 0$, for $\im[z] \lessgtr 0$, into the \textsl{interior} of $\mathsf{K}_{n+1}(z)$ [p.\,48 in Ref.\,\citen{ST70}], the property $\rho_{n+1}^{\textrm{c}}(z) \to 0$ for $n\to\infty$ entails that the moment problem under consideration is \textsl{determinate}. In this connection, the significance of the half-plane $\im[\zeta] \gtrless 0$ for $\im[z] \lessgtr 0$ lies in the fact that, with the function $-\t{f}_j(z)$, Eq.\,(\ref{e5g}), being Nevanlinna for all $j \in \mathds{N}$, the imaginary part of the function
\begin{equation}\label{e6uc}
\zeta(z) \doteq \t{f}_{n+1}(z) -\alpha_{n+1}
\end{equation}
in Eq.\,(\ref{e5te})\,\footnote{\emph{Cf.} Eq.\,(\protect\ref{e6jd}).} is positive / negative for $\im[z] \lessgtr 0$. Whether or not $\lim_{n\to\infty} \rho_{n+1}^{\textrm{c}}(z) = 0$, the region enclosed by the circle $\mathsf{K}_{n+2}(z)$ is contained in that enclosed by $\mathsf{K}_{n+1}(z)$, with the two circles meeting at one point in the $w$-plane [p.\,13 in Ref.\,\citen{NIA65}]. Further, since convergence of $\sum_{j=0}^{\infty} \vert P_j(z)\vert^2$ at \textsl{any} $z$ in the finite part of the $z$-plane, with $\im[z]\not=0$, implies \textsl{uniform} convergence of $\sum_{j=0}^{\infty} \vert P_j(z)\vert^2$ for \textsl{all} $z$ in the finite part of the $z$-plane away from the real axis [Theorem 1.3.2, \S\,3.2, p.\,16, in Ref.\,\citen{NIA65}], it follows that in the cases where $\sum_{j=0}^{n} \vert P_j(z)\vert^2 \to \infty$ as $n\to \infty$ for \textsl{any} $z$, with $\im[z]\not=0$, this series necessarily diverges over the entire finite part of the $z$-plane away from the real axis.

From the above discussions, we are led to the conclusion that in the cases where $\rho_{n+1}^{\textrm{c}}(z) \to 0$, $\im[z] \not=0$, for $n\to\infty$, the convergence of $\t{w}_{n+1}(z,\ptau)$ to the limit function $\t{w}(z) \doteq \t{w}_{\infty}(z,\ptau)$ is \textsl{uniform} for \textsl{all} $z \in \mathds{C}$ away from the real axis, implying the same for the function $\t{f}^{\X{(n)}}(z)$ in relation to the function $\t{f}(z)$. \emph{This uniformity of convergence is a crucial property that ultimately establishes the uniformity of convergence of the series in Eq.\,(\ref{e4}) for almost all $\bm{k}$ and $z$.} For completeness, we remark that  $\rho_{n+1}^{\textrm{c}}(z) \to 0$ as $n\to \infty$ for \textsl{all} $z$, $\im[z] \not= 0$, if at least one of the two sums $\sum_{j=0}^n \vert P_j(0)\vert^2$ and $\sum_{j=0}^n \vert Q_j(0)\vert^2$ is divergent [Theorem 2.17, p.\,68, in Ref.\,\citen{ST70}] [p.\,84 in Ref.\,\citen{NIA65}]. The criterion introduced in Eq.\,(\ref{e4og}), which is independent of $z$, is intimately related to the latter theorem.

As we have discussed above, the measure function $\upgamma_{n+1}(\varepsilon,\ptau)$ in Eq.\,(\ref{e6s}) is \textsl{bounded} for all $n$. As a consequence, the sequence $\{\t{w}_n(z,\ptau) \| n \}$ is a bounded sequence over the finite part of the $z$-plane away from the real axis. This sequence is therefore \textsl{locally bounded} [p.\,148 in Ref.\,\citen{RR98}] (also referred to as \textsl{uniformly bounded} [p.\,415 in Ref.\,\citen{AIM65}]) in the mentioned region of the $z$-plane. In the case where it converges, corresponding to $\lim_{n\to\infty} \rho_{n+1}^{\textrm{c}}(z) = 0$, the limiting function $\t{w}(z)$ is an accumulation point in the complex $w$-plane. Thus, following the Vitali (or Stieltjes-Vitali) theorem [p.\,150 in Ref.\,\citen{RR98}] [\S\,5.21, p.\,168, in Ref.\,\citen{ECT52}] [\S\,87, Theorem 17.18, p.\,417, in Ref.\,\citen{AIM65}] the sequence $\{\t{w}_n(z,\ptau) \| n \}$ converges \textsl{compactly} [p.\,322 in Ref.\,\citen{RR91}] over the finite part of the $z$-plane away from the real axis. Here `\textsl{compact} convergence' refers to the same property as `\textsl{uniform} convergence' [Ch. 11 in Ref.\,\citen{RR91}] (see Theorem 17.18, p.\,417, in Ref.\,\citen{AIM65}). For clarity, in the latter reference `\textsl{normal} convergence'  [p.\,322 in Ref.\,\citen{RR91}] stands for the `\textsl{absolute} convergence' elsewhere [\S\,2.32, p.\,18, in Ref.\,\citen{WW62}]. In this connection, note that $\t{w}_n(z,\ptau)$ is \textsl{analytic} in the region $\im[z] \not= 0$ of the complex $z$-plane for all $n$, implying that $\t{w}_n(z,\ptau)$ is indeed \textsl{meromorphic}\,\footnote{``A function is said to be \textsl{meromorphic} in a region if it is analytic in the region except at a finite number of poles.'' [\S\,3.2 in Ref.\,\protect\citen{ECT52}] The \textsl{pole set} $P(\t{w}_n(z,\ptau))$ (referred to in the definitions of \textsl{compact} and \textsl{normal} convergence) in the region $\im[z] \not=0$ is therefore \emph{a priori} empty for all $n$.} in this region, as required by the Vitali theorem referred to above.\footnote{For the \textsl{compact family of analytic functions}, see \S\,86, p.\,411, in Ref.\,\protect\citen{AIM65}. A family $\mathcal{F}$ of analytic functions in $\mathcal{D}$ is \textsl{conditionally compact} if every sequence $\{\phi_n(z)\| n\}$ contains a subsequence $\{\phi_{n_k}(z)\| n_k\}$ which is uniformly convergent inside $\mathcal{D}$. The limit of a (conditionally) compact sequence of analytic functions in $\mathcal{D}$ is analytic [p.\,412 in Ref.\,\protect\citen{AIM65}]. We note that the property `locally bounded' in Ref.\,\protect\citen{RR98} [p.\,148] is the same as the property `uniformly bounded' in Ref.\,\protect\citen{AIM65} [p.\,415].}

\refstepcounter{dummyX}
\subsubsection{Discussion}
\phantomsection
\label{sec.b1a}
The fact that the sum $\sum_{j=0}^n \vert P_j(z)\vert^2$ in the denominator of the expression for $\rho_{n+1}^{\textrm{c}}(z)$ in Eq.\,(\ref{e6ub}) is multiplied by $\vert z - z^*\vert$ implies that, when necessary, the limit $\protect\im[z] \to 0$ is \textsl{in general} to be taken \textsl{after} taking the limit $n\to \infty$.\footnote{As we shall demonstrate below, insofar as the function $-\protect\t{w}_{n+1}(z,\protect\ptau)$ is concerned, $z$ can be safely identified with $\varepsilon \in\mathds{R}$, for any value of $n \in \mathds{N}$, provided that $\varepsilon$ is located in an open interval of $\mathds{R}$ over which $A_{\sigma}(\bm{k};\varepsilon)$ is identically vanishing; if such regions exist, the series $-\protect\t{w}_{\infty}(\varepsilon,\protect\ptau)$ is uniformly convergent for $\varepsilon$ over these regions.} To elaborate on this remark, it is convenient first to introduce a simplified notation.

Let $\{z_i\| i=1,2,3,4\}$ be complex numbers, each representing an appropriate polynomial such as encountered in Eq.\,(\ref{e6ub}) for a specific $z \in \mathds{C}$.\refstepcounter{dummy}\label{Explicitly}\footnote{Explicitly, $z_1 \equiv Q_{n+1}(z)$, $z_2 \equiv Q_{n}(z)$, $z_3 \equiv P_{n+1}(z)$, and $z_4 \equiv P_n(z)$. Since the coefficients of the polynomials $\{P_j(z)\| j\}$ and $\{Q_j(z)\| j\}$ are real constants, one has $P_j(z) \in \mathds{R}$ for $\protect\im[z] = 0$, $\forall j$, and more generally $P_j(z^*) = P_j^*(z)$, $\forall z\in\mathds{C}$. \label{notej1}} One has the following identity:
\begin{equation}\label{e6ud}
-\frac{z_1 -\ptau z_2}{z_3 - \ptau z_4} \equiv - \frac{z_1 z_4^* - z_2 z_3^*}{z_3 z_4^* - z_4 z_3^*} + \frac{z_1 z_4 - z_2 z_3}{z_3 z_4^* - z_4 z_3^*}\hspace{1.0pt} z_5,
\end{equation}
where
\begin{equation}\label{e6ud1}
z_5 \doteq\frac{z_3^* -\ptau z_4^*}{z_3 - \ptau z_4}.
\end{equation}
For $\ptau \in \mathds{R}$ one has $\vert z_5\vert = 1$ so that it can be expressed as
\begin{equation}\label{e6ue}
z_5 = \e^{\ii\theta_5},\;\; \theta_5 \in \mathds{R}\;\; \text{for}\;\, \ptau\in \mathds{R}.
\end{equation}
Thus, using
\begin{equation}\label{e6uf}
\frac{z_1 z_4 - z_2 z_3}{z_3 z_4^* - z_4 z_3^*} \equiv \left|\frac{z_1 z_4 - z_2 z_3}{z_3 z_4^* - z_4 z_3^*}\right|\hspace{1.0pt} \e^{\ii \phi},\;\; \phi\in\mathds{R},
\end{equation}
the identity in Eq.\,(\ref{e6ud}) can be written as
\begin{equation}\label{e6ug}
-\frac{z_1 -\ptau z_2}{z_3 - \ptau z_4} \equiv - \frac{z_1 z_4^* - z_2 z_3^*}{z_3 z_4^* - z_4 z_3^*} +  \left|\frac{z_1 z_4 - z_2 z_3}{z_3 z_4^* - z_4 z_3^*}\right|\hspace{1.0pt} \e^{\ii\varphi},
\end{equation}
where
\begin{equation}\label{e6uh}
\varphi \doteq \phi + \theta_5,
\end{equation}
which is real for $\ptau \in \mathds{R}$. Note that the dependence of the function in Eq.\,(\ref{e6ug}) on $\ptau$ is fully taken up by the phase $\varphi$, through the dependence of $\theta_5$ on $\ptau$.\refstepcounter{dummy}\label{TheIndependence}\footnote{The independence of $\protect\t{w}_{\infty}(z,\protect\ptau)$ from $\ptau$, referred to following Eq.\,(\protect\ref{e6qe}), follows from the fact that the amplitude multiplying the function $\protect\e^{\protect\ii\varphi}$ on the RHS of Eq.\,(\protect\ref{e6ug}) vanishes for $n\to\infty$. \label{notel1}} The expressions in Eq.\,(\ref{e6ub}) are based on the identity in Eq.\,(\ref{e6ug}) in which the amplitude on the RHS multiplying $\e^{\ii\varphi}$ corresponds to the radius $\rho_{n+1}^{\textrm{c}}(z)$ in Eq.\,(\ref{e6ub}); the simplification in the expression for $\rho_{n+1}^{\textrm{c}}(z)$ in the latter equation has been brought about through the application of in particular the Liouville-Ostrogradskii formula, Eq.\,(\ref{e7wc}).\footnote{With reference to footnote \raisebox{-1.0ex}{\normalsize{\protect\footref{notej1}}} on p.\,\pageref{Explicitly}, the function on the LHS of Eq.\,(\protect\ref{e6ug}) represents the function $-\protect\t{w}_{n+1}(z,\protect\ptau)$, Eq.\,(\protect\ref{e6k}). Thus, the first term on the RHS of Eq.\,(\protect\ref{e6ug}) represents the function $w_{n+1}^{\textrm{c}}(z)$ in Eq.\,(\protect\ref{e6ub}).}

Clearly, the identity in Eq.\,(\ref{e6ud}) breaks down for $z_3 z_4^* = z_4 z_3^* \equiv (z_3 z_4^*)^*$. This is the case in particular for $z_3, z_4 \in \mathds{R}$. With reference to the remarks in footnote \raisebox{-0.7ex}{\normalsize{\protect\footref{notej1}}} on p.\,\pageref{Explicitly}, one has $z_3, z_4 \in \mathds{R}$ for $\im[z] = 0$. Thus, \emph{the functions in Eq.\,(\ref{e6ub}) are to be used only for $\im[z] \not=0$}; for investigating the behaviours of these functions along the real energy axis, they are therefore to be considered in the limit $\im[z] \to 0$,\footnote{In practice, for instance along the lines of Ref.\,\protect\citen{DvHF92}.} and \textsl{not} at $\im[z] = 0$.

It is important to realise that for systems for which $A_{\sigma}(\bm{k};\varepsilon)$ is of bounded support, that is systems for which $A_{\sigma}(\bm{k};\varepsilon) \equiv 0$ for $\varepsilon \le \varepsilon_{\mathrm{l}}$ and $\varepsilon \ge \varepsilon_{\mathrm{u}}$, where $\varepsilon_{\mathrm{l}}$ and $\varepsilon_{\mathrm{u}}$ are some \textsl{finite} energies,\footnote{With reference to footnote \raisebox{-1.0ex}{\normalsize{\protect\footref{notek1}}} on p.\,\protect\pageref{notek1}, should $A_{\sigma}(\bm{k};\varepsilon)\equiv 0$ for, say, all $\varepsilon \in (\varepsilon_{\mathrm{l}}',\varepsilon_{\mathrm{u}}')$, where $\varepsilon_{\mathrm{l}} < \varepsilon_{\mathrm{l}}'$ and $\varepsilon_{\mathrm{u}}' < \varepsilon_{\mathrm{u}}$, along the same lines as described in this section the series for $-\protect\t{w}_{\infty}(z,\protect\ptau)$ can be shown to be uniformly convergent for all $z \in (\varepsilon_{\mathrm{l}}',\varepsilon_{\mathrm{u}}')$. The reason that we avoid explicitly dealing with finite supports that are not simply-connected is merely convenience of notation; whereas in the case at hand the maximum (minimum) value of $\varepsilon_{\mathrm{l}}$ ($\varepsilon_{\mathrm{u}}$) can be identified with the infimum (supremum) of the set $\{\lambda_j^{\protect\X{(n+1)}}(\protect\ptau)\| 1 \le j \le n+1\}$ for $n=\infty$, in a case where the support is not simply-connected the latter infinite set is to be split into as many infinite subsets as the number of connected pieces of which the support consists, each with its own limiting points (band edges).} the series for\,\footnote{The function $-\protect\t{w}_{\infty}(\varepsilon,\protect\ptau)$, $\varepsilon\in\mathds{R}$, is to be distinguished from $-w_{\infty}(\varepsilon,\protect\ptau)$ (without tilde), Eq.\,(\protect\ref{e1}).} $-\t{w}_{\infty}(\varepsilon,\ptau)$ is \textsl{uniformly convergent} for all
\begin{equation}\label{e6ui}
\varepsilon \in \mathsf{D},
\end{equation}
where\,\footnote{Taking account of the ranking in Eq.\,(\protect\ref{e6l}), wherein $\lambda_j \equiv \lambda_j^{\protect\X{(n+1)}}(\protect\ptau)$ (the $j$th zero of the quasi-orthogonal polynomial $P_{n+1}(z,\protect\ptau)$ for arbitrary $\protect\ptau \in \mathds{R}$), we note that $\lambda_1$ ($\lambda_{n+1}$) approaches $\varepsilon_{\mathrm{l}}$ ($\varepsilon_{\mathrm{u}}$) from above (below) for $n\to\infty$. For this, consult Theorem 1.2.2, p.\,10, and the remark following Eq.\,[1.33b], p.\,22, in Ref.\,\protect\citen{NIA65}. It follows that for the cases where $A_{\sigma}(\bm{k};\varepsilon)$ is of bounded support and $\gamma_j \equiv \gamma_j^{\protect\X{(n+1)}}(\protect\ptau) > 0$ for all $j \le n+1$, $\forall n$ (for $n=\infty$, $\gamma_j^{\protect\X{(n+1)}}(\protect\ptau)$ may possibly be vanishing for $j$ over a finite subset of $\mathds{N}$), the maximum (minimum) value of $\varepsilon_{\mathrm{l}}$ ($\varepsilon_{\mathrm{u}}$) is an accumulation point of the set $\{\lambda_j \| 1\le j\le n+1\}$ for $n=\infty$. This is consistent with the theorem of Bolzano regarding bounded infinite sequences of real numbers [\S\,2.21, pp.\,12 and 13, in Ref.\,\protect\citen{WW62}].}
\begin{equation}\label{e6uj}
\mathsf{D} \doteq \{ \varepsilon \| \varepsilon < \varepsilon_{\mathrm{l}},\, \varepsilon > \varepsilon_{\mathrm{u}}\}.
\end{equation}
The validity of this statement is established as follows. Let
\begin{equation}\label{e6uk}
\mathsf{D}_{\updelta} \doteq \{ \varepsilon \| \varepsilon < \varepsilon_{\mathrm{l}} -\updelta,\, \varepsilon > \varepsilon_{\mathrm{u}}+ \updelta\},
\end{equation}
where $\updelta \ge 0$. Clearly, $\mathsf{D} = \mathsf{D}_0$. For any \textsl{positive} $\updelta$,
\begin{equation}\label{e6um}
\frac{\gamma_j}{\vert \varepsilon - \lambda_j\vert} < \EuScript{M}_j,\;\; \forall \varepsilon\in \mathsf{D}_{\updelta},\;\forall j\in \{1,2,\dots, n+1\},
\end{equation}
where\,\footnote{Since $\lambda_j \in \mathds{R}$, $\forall j$, for $\vert\hspace{-1.0pt}\protect\im[z]\vert > \updelta_0 >0$ one can safely replace the $\updelta$ in Eq.\,(\protect\ref{e6un}) by $\updelta_0$. In this way, the following details amount to a proof for the unform convergence of the series for $-\protect\t{w}_{\infty}(z,\protect\ptau)$ in the region $\vert\hspace{-1.0pt}\protect\im[z]\vert > \updelta_0$ of the $z$-plane (for this, the $\gamma_j/\vert\varepsilon - \lambda_j\vert$ on the LHS of Eq.\,(\protect\ref{e6um}) is to be replaced by $\gamma_j/\vert z - \lambda_j\vert$, and the condition $\varepsilon\in \mathsf{D}_{\updelta}$ by $\vert\hspace{-1.0pt}\protect\im[z]\vert > \updelta_0 >0$).}
\begin{equation}\label{e6un}
\EuScript{M}_j \doteq \frac{\gamma_j}{\updelta},
\end{equation}
which is \textsl{independent} of $z$. Following Eq.\,(\ref{e7a}), one has
\begin{equation}\label{e6uo}
\sum_{j=1}^{n+1} \EuScript{M}_j = \frac{\mu_0}{\updelta} < \infty,\;\; \forall n \in\mathds{N}_0,\; \updelta >0.
\end{equation}
On the strength of Weierstrass' comparison test [\S\,3.34, p.\,49, in Ref.\,\citen{WW62}], the series for $-\t{w}_{\infty}(\varepsilon,\ptau)$, Eq.\,(\ref{e6k}), is hereby shown to converge uniformly for \textsl{all} $\varepsilon \in \mathsf{D}_{\updelta}$. Because of the strict inequalities in the defining expression in Eq.\,(\ref{e6uk}), this result equally applies for $\varepsilon \in \mathsf{D}$. In \S\,\ref{sabx} we consider an illustrative example for which $\varepsilon_{\mathrm{l}} = -1$ and $\varepsilon_{\mathrm{u}} = 1$. In the Mathematical notebook accompanying this publication, we provide some programs for inspecting and visualising the dependence of the convergence behaviour of the approximant $\t{w}_{n}(z,\ptau)$ of the function $\t{w}(z)$ considered in \S\,\ref{sabx} on the location of $z$ in the $z$-plane, in particular for $z = \varepsilon + \ii \eta$ with $\varepsilon$ in the range $[-1,1]$ and outside this range.

\refstepcounter{dummyX}
\subsection{The self-energy}
\phantomsection
\label{sec.b2}
On the basis of the analyticity of $\t{\Sigma}_{\sigma}(\bm{k};z)$ in the entire region $\im[z]\not=0$ of the $z$-plane, and the fact that for $z \to\infty$ to leading order the exact self-energy satisfies the asymptotic relationship $\t{\Sigma}_{\sigma}(\bm{k};z) \sim \Sigma_{\sigma}^{\textsc{hf}}(\bm{k})$, with $\Sigma_{\sigma}^{\textsc{hf}}(\bm{k}) \in\mathds{R}$ denoting the exact Hartree-Fock self-energy, Eqs\,(\ref{e7b}) and (\ref{e5e}), by the Cauchy theorem [\S\,5.2, p.\,85, in Ref.\,\citen{WW62}] one obtains (\S\,B.5 in Ref.\,\citen{BF07})
\begin{equation}\label{e20a}
\t{\Sigma}_{\sigma}(\bm{k};z) = \Sigma_{\sigma}^{\textsc{hf}}(\bm{k}) + \int_{-\infty}^{\infty} \rd\varepsilon'\; \frac{B_{\sigma}(\bm{k};\varepsilon')}{z - \varepsilon'},\;\; \im[z] \not=0,
\end{equation}
where (\emph{cf.} Eq.\,(\ref{e4d}))
\begin{equation}\label{e20b}
B_{\sigma}(\bm{k};\varepsilon) \doteq \pm\frac{1}{\pi}\im\big[\t{\Sigma}_{\sigma}(\bm{k};\varepsilon \mp \ii 0^+)\big].
\end{equation}
On account of the equality in Eq.\,(\ref{e2}), one has (\emph{cf.} Eq.\,(\ref{e4f}))
\begin{equation}\label{e20i}
B_{\sigma}(\bm{k};\varepsilon) \ge 0,\;\; \forall \bm{k}, \varepsilon.
\end{equation}
Although this inequality is \textsl{not} relevant to the analyticity of the function on the RHS of Eq.\,(\ref{e20a}) in the region $\im[z] \not= 0$ [\S\,5.32 in Ref.\,\citen{WW62}] [Theorem 17.19, p.\,418, in Ref.\,\citen{AIM65}], it can be explicitly shown that it is a \textsl{necessary} condition for the satisfaction of the equality in Eq.\,(\ref{e2}) by the self-energy; violation of the inequality in Eq.\,(\ref{e20i}) over a non-zero-measure subset of the real $\varepsilon$ axis leads to the failure of the function $-\t{\Sigma}_{\sigma}(\bm{k};z)$, described according to the expression in Eq.\,(\ref{e20a}), as being a Nevanlinna function of $z$ [App.\,C in Ref.\,\citen{HMN72}] [Ch.\,3 in Ref.\,\citen{NIA65}].

On general grounds [\S\,3.9, p.\,1492, in Ref.\,\citen{BF02}] [\S\,B.5 in Ref.\,\citen{BF07}], one can show that\,\footnote{For the small-$o$ symbol see for instance \S\,5 in Ref.\,\protect\citen{EWH26}. See also appendix \protect\ref{sae}.} $B_{\sigma}(\bm{k};\varepsilon) = o(1/\varepsilon)$ for $\vert\varepsilon\vert\to \infty$. In the case of the uniform GSs of Hubbard-like models, one can more explicitly demonstrate that $B_{\sigma}(\bm{k};\varepsilon)$ decays faster than any finite power of $1/\varepsilon$ for $\vert\varepsilon\vert\to \infty$ [\S\,B.6 in Ref.\,\citen{BF07}]. This follows from the general expression for the one-particle spectral function $A_{\sigma}(\bm{k};\varepsilon)$ in terms of the real and imaginary parts of the self-energy [Eq.\,(B.43) in Ref.\,\citen{BF07}], on the basis of which one has\,\footnote{This result is consistent with $\llbracket A_{\sigma}(\bm{k};\varepsilon)\rrbracket = \llbracket \protect\t{G}_{\sigma}(\bm{k};z)\rrbracket =\mathrm{s}$ and $\llbracket B_{\sigma}(\bm{k};\varepsilon)\rrbracket = \llbracket\protect\t{\Sigma}_{\sigma}(\bm{k};z)\rrbracket =\mathrm{s}^{-1}$.}
\begin{equation}\label{e20j}
B_{\sigma}(\bm{k};\varepsilon) \sim \frac{1}{\hbar^2}\hspace{1.0pt} \varepsilon^2\hspace{0.2pt} A_{\sigma}(\bm{k};\varepsilon)\;\;\text{for}\;\; \vert\varepsilon\vert \to \infty.
\end{equation}
We shall return to the behaviour of the function $B_{\sigma}(\bm{k};\varepsilon)$ in the asymptotic region $\vert\varepsilon\vert \to \infty$ later in this section.

The inequality in Eq.\,(\ref{e20i}) implies that the function (\emph{cf.} Eq.\,(\ref{e4g}))
\begin{equation}\label{e20c}
\upsigma_{\sigma}(\bm{k};\varepsilon) \doteq \int_{-\infty}^{\varepsilon} \rd\varepsilon'\; B_{\sigma}(\bm{k};\varepsilon')
\end{equation}
is a non-decreasing function of $\varepsilon$ over $\mathds{R}$. In the light of the remarks in the previous paragraph, for Hubbard-like models $\upsigma_{\sigma}(\bm{k};\varepsilon)$ is further \textsl{bounded} for all $\varepsilon \in \mathds{R}$.\footnote{With reference to Eqs\,(\protect\ref{e7i}) and (\protect\ref{e7c}) below, for $\varepsilon$ increasing from $-\infty$ towards $+\infty$, the measure $\upsigma_{\sigma}(\bm{k};\varepsilon)$ increases (not necessarily monotonically) from $0$ towards $\Sigma_{\sigma;\infty_1}(\bm{k}) \equiv \big(-G_{\sigma;\infty_2}^2(\bm{k}) + \hbar G_{\sigma;\infty_3}(\bm{k})\big)/\hbar^3 > 0$, $\forall\bm{k}$. Compare with Eq.\,(\protect\ref{e4i}), where the $\hbar$ stands for $G_{\sigma;\infty_1}(\bm{k})$, Eq.\,(\protect\ref{e4qa}).} Following the defining expression in Eq.\,(\ref{e20c}), one can thus re-write the equality in Eq.\,(\ref{e20a}) in the following alternative form, in terms of an Stieltjes integral \cite{THH18} (\emph{cf.} Eq.\,(\ref{e4j})):
\begin{equation}\label{e20d}
\t{\Sigma}_{\sigma}(\bm{k};z) = \Sigma_{\sigma}^{\textsc{hf}}(\bm{k}) + \int_{-\infty}^{\infty} \frac{\rd\upsigma_{\sigma}(\bm{k};\varepsilon')}{z -\varepsilon'},\;\; \im[z]\not=0.
\end{equation}

Defining the function\,\footnote{In the SI base units, one has $\llbracket\upsigma_{\sigma}(\bm{k};\varepsilon)\rrbracket = \mathrm{Js^{-1}}$, and $\llbracket \uptau_{\sigma}(\bm{k};u)\rrbracket = \llbracket\protect\t{\Sigma}_{\sigma}(\bm{k};z)\rrbracket = \mathrm{s^{-1}}$. The variable $u$ is dimensionless.} (\emph{cf.} Eq.\,(\ref{e4k}))
\begin{equation}\label{e20e}
\uptau_{\sigma}(\bm{k};u) \doteq \frac{1}{\upvarepsilon_{\X{0}}} \int_{-\infty}^{\upvarepsilon_{\X{0}} u} \frac{\rd\upsigma_{\sigma}(\bm{k};\varepsilon)}{1+(\varepsilon/\upvarepsilon_{\X{0}})^2} \equiv \int_{-\infty}^{u} \rd u'\; \frac{B_{\sigma}(\bm{k};\upvarepsilon_{\X{0}} u')}{1 + {u'}^2},\;\; \forall \upvarepsilon_{\X{0}} > 0,
\end{equation}
from the expression in Eq.\,(\ref{e20d}) one obtains (\emph{cf.} Eq.\,(\ref{e4l}))
\begin{equation}\label{e20g}
-\t{\Sigma}_{\sigma}(\bm{k};z) = -\Sigma_{\sigma}^{\textsc{hf}}(\bm{k}) +\int_{-\infty}^{\infty} \rd\uptau_{\sigma}(\bm{k};u)\, u + \int_{-\infty}^{\infty} \rd\uptau_{\sigma}(\bm{k};u)\; \frac{1 + u z/\upvarepsilon_{\X{0}}}{u - z/\upvarepsilon_{\X{0}}}.
\end{equation}
Comparing the above expression with the Riesz-Herglotz representation for the Nevanlinna function $\t{\varphi}(z)$ in Eq.\,(\ref{e20h}), for the constants $\upmu_{\sigma}(\bm{k})$ and $\upnu_{\sigma}(\bm{k})$ corresponding to $-\t{\Sigma}_{\sigma}(\bm{k};z)$, one obtains (\emph{cf.} Eq.\,(\ref{e4m1}))
\begin{align}\label{e20f}
\upmu_{\sigma}(\bm{k}) \equiv 0,\;\; \upnu_{\sigma}(\bm{k}) &= -\Sigma_{\sigma}^{\textsc{hf}}(\bm{k})
+\int_{-\infty}^{\infty} \rd\uptau_{\sigma}(\bm{k};u)\, u\nonumber\\
&\equiv -\Sigma_{\sigma}^{\textsc{hf}}(\bm{k}) + \frac{1}{\upvarepsilon_{\X{0}}}\int_{-\infty}^{\infty} \rd\upsigma_{\sigma}(\bm{k};\varepsilon)\; \frac{\varepsilon/\upvarepsilon_{\X{0}}}{1+(\varepsilon/\upvarepsilon_{\X{0}})^2}.
\end{align}
One verifies that for $z\to\infty$ to leading order the third term on the RHS of Eq.\,(\ref{e20g}) cancels the second term, so that indeed to leading order the expression in Eq.\,(\ref{e20g}) is in conformity with the expressions in Eqs\,(\ref{e7b}) and (\ref{e7ba}) below.

Proceeding as in the case of the Green function, \S\,\ref{sec.b.1}, on employing the geometric series expansion $1/(z-\varepsilon') = 1/z + \varepsilon'/z^2 + \dots$ in the integral on the RHS of Eq.\,(\ref{e20a}) and changing the order of summation and integration, one obtains the following asymptotic series expansion (\emph{cf.} Eq.\,(\ref{e4n})) [Eqs\,(B.71) and (B.72) in Ref.\,\citen{BF07}]:
\begin{equation}\label{e7b}
\t{\Sigma}_{\sigma}(\bm{k};z) \sim \Sigma_{\sigma;\infty_0}(\bm{k}) + \frac{\Sigma_{\sigma;\infty_1}(\bm{k})}{z} + \frac{\Sigma_{\sigma;\infty_2}(\bm{k})}{z^2} + \dots\;\; \text{for}\;\;\ z \to\infty,
\end{equation}
where (see Eq.\,(\ref{e5e}))
\begin{equation}\label{e7ba}
\Sigma_{\sigma;\infty_0}(\bm{k}) \equiv  \Sigma_{\sigma}^{\textsc{hf}}(\bm{k}),
\end{equation}
and (\emph{cf.} Eq.\,(\ref{e4o}))
\begin{equation}\label{e7i}
\Sigma_{\sigma;\infty_j}(\bm{k}) \doteq \int_{-\infty}^{\infty} \rd\varepsilon\; B_{\sigma}(\bm{k};\varepsilon) \,\varepsilon^{j-1},\;\; \forall j\ge 1.
\end{equation}
Compare with the expressions in Eqs\,(B.87) and (B.88) in Ref.\,\citen{BF07}. On the basis of the expression in Eq.\,(\ref{e20d}), one can write (\emph{cf.} Eqs\,(\ref{e4oa}) and (\ref{e4ob}))
\begin{equation}\label{e7j}
\Sigma_{\sigma;\infty_{j+1}}(\bm{k}) \equiv \b{\mu}_j,
\end{equation}
where
\begin{equation}\label{e7k}
\b{\mu}_j \doteq \int_{-\infty}^{\infty} \rd\upsigma_{\sigma}(\bm{k};\varepsilon)\, \varepsilon^j.
\end{equation}
One clearly has (\emph{cf.} Eq.\,(\ref{e4od}))
\begin{equation}\label{e7l}
\Sigma_{\sigma;\infty_{2j+1}}(\bm{k}) > 0,\;\; \forall j \in \mathds{N}_{0}.
\end{equation}
With reference to Eqs\,(\ref{e5c}) and (\ref{e5f}), one further has
\begin{equation}\label{e7la}
\Sigma_{\sigma;\infty_j}(\bm{k}) = \frac{1}{\hbar}\hspace{0.6pt} \mu_{j-1}^{\X{(1)}},\;\; \forall j \in \mathds{N}.
\end{equation}
As will become evident below, the boundedness of $G_{\sigma;\infty_j}(\bm{k})$ for arbitrary finite values of $j$ in the case of Hubbard-like models implies the same for the $\Sigma_{\sigma;\infty_j}(\bm{k})$ corresponding to these models. Thus, with reference to the expression in Eq.\,(\ref{e7i}), it follows that for these models the function $B_{\sigma}(\bm{k};\varepsilon)$ must indeed decay faster than any finite power of $1/\varepsilon$ for $\vert\varepsilon\vert\to\infty$. This observation, based on general grounds, is in conformity with the expression in Eq.\,(\ref{e20j}). With reference to the remarks centred on  Eq.\,(\ref{e4oi}), the \textsl{strict} inequality in Eq.\,(\ref{e7l}) is violated for all $j \in\mathds{N}$ in the specific case where $B_{\sigma}(\bm{k};\varepsilon) \propto \delta(\varepsilon)$.

We note that since the sequence $\{\Sigma_{\sigma;\infty_j}(\bm{k}) \| j\in \mathds{N}\}$ corresponds to the non-decreasing measure function $\upsigma_{\sigma}(\bm{k};\varepsilon)$, it is similar to $\{G_{\sigma;\infty_j}(\bm{k}) \| j\in \mathds{N}\}$ a \textsl{positive sequence} (see the discussions centred on Eq.\,(\ref{e4s}) above).\footnote{As regards the positivity of $\mathbb{H}_m^{\protect\t{\Sigma}_{\sigma}}$, the same remarks as presented on p.\,\protect\pageref{AsRegards} regarding $\mathbb{H}_m^{\protect\t{G}_{\sigma}}$ apply here. A case in point, concerning the `Hubbard atom', is discussed in \S\,\protect\ref{sec.b2.1} below.} Further, for the function $A_{\sigma}(\bm{k};\varepsilon)$ decaying faster than any finite power of $1/\varepsilon$ in the region $\vert\varepsilon\vert \to \infty$, on the basis of the asymptotic expression in Eq.\,(\ref{e20j}) and along the same lines as leading to the result in Eq.\,(\ref{e4oh}), one arrives at the following fundamental result:
\begin{equation}\label{e7m}
\sum_{j=1}^{\infty} \frac{1}{\sqrt[2j]{\Sigma_{\sigma;\infty_{2j+1}}(\bm{k})}} = \infty,\;\;\forall\bm{k},
\end{equation}
implying that the positive sequence $\{\Sigma_{\sigma;\infty_j}(\bm{k}) \| j\in \mathds{N}\}$ is associated with a \textsl{determinate} moment problem. For completeness, following the asymptotic expression in Eq.\,(\ref{e20j}), the relevant function to be considered in dealing with the spectral function $B_{\sigma}(\bm{k};\varepsilon)$ is $\phi_{\alpha+2}(\varepsilon)$, for which the same $\alpha$-independent \textsl{leading-order} asymptotic expression on the RHS of Eq.\,(\ref{e4of}) applies. Hence, the result in Eq.\,(\ref{e4og}) is indeed satisfied also for the sequence $\{s_{2j}\| j\}$ associated with the function $\phi_{\alpha+2}(\varepsilon)$.

For completeness, with reference to the remarks following Eq.\,(\ref{e4ob}), the moment problem as encountered in this section is also of the Hamburger type. Similarly as in the case of the Green function $\t{G}_{\sigma}(\bm{k};z)$, the moment problem associated with $\t{\Sigma}_{\sigma}(\bm{k};z)$ can be naturally expressed as two Stieltjes moment problems. This is related to the fact that $\t{\Sigma}_{\sigma}(\bm{k};z)$ can be naturally expressed as (\emph{cf.} Eq.\,(\ref{e4oc}))
\begin{equation}\label{e7ma}
\t{\Sigma}_{\sigma}(\bm{k};z) \equiv \t{\Sigma}_{\sigma}^-(\bm{k};z) + \t{\Sigma}_{\sigma}^+(\bm{k};z),
\end{equation}
where the functions $\t{\Sigma}_{\sigma}^{\X{\mp}}(\bm{k};z)$ are analytic in the regions $\re[z] \gtrless \mu$, $\forall \bm{k}$, in addition to being like $\t{\Sigma}_{\sigma}(\bm{k};z)$ analytic away from the real axis of the $z$-plane. As a consequence, with $B_{\sigma}^{\mp}(\bm{k};\varepsilon)$ denoting the spectral function corresponding to $\t{\Sigma}_{\sigma}^{\X{\mp}}(\bm{k};z)$ (\emph{cf.} Eq.\,(\ref{e20b})), one has $B_{\sigma}^{\mp}(\bm{k};\varepsilon) \equiv 0$ for $\varepsilon \gtrless \mu$. For the spectral function $B_{\sigma}(\bm{k};\varepsilon)$ given, one deduces $B_{\sigma}^{\mp}(\bm{k};\varepsilon)$ according to the same relationship as in Eq.\,(\ref{e4oc1}).\footnote{Thus, with reference to Eq.\,(\protect\ref{e20c}), for the measures associated with $\protect\t{\Sigma}_{\sigma}^{\mp}(\bm{k};z)$ one has $\upsigma_{\sigma}^{-}(\bm{k};\varepsilon) = \int_{-\infty}^{\min(\varepsilon,\mu)} \mathrm{d}\varepsilon'\, B_{\sigma}(\bm{k};\varepsilon')$, and $\upsigma_{\sigma}^{+}(\bm{k};\varepsilon) = \int_{\min(\varepsilon,\mu)}^{\varepsilon} \mathrm{d}\varepsilon'\, B_{\sigma}(\bm{k};\varepsilon')$.}

We proceed now with establishing a direct bridge between the moment problem that we have introduced thus far in the present section with that introduced in \S\,\ref{sec.b.1}. By doing so, we establish a direct relationship between the moment problem associated with the positive sequence $\{\Sigma_{\sigma;\infty_j}(\bm{k})\| j\}$ and the perturbation series expansion of the self-energy $\t{\Sigma}_{\sigma}(\bm{k};z)$ in terms of \textsl{skeleton} self-energy diagrams and the \textsl{interacting} one-particle Green functions $\{\t{G}_{\sigma}(\bm{k};z)\| \sigma\}$.

From the Dyson equation in Eq.\,(\ref{e4a}), one has
\begin{equation}\label{e7f}
\t{\Sigma}_{\sigma}(\bm{k};z) = 1/\t{G}_{\X{0};\sigma}(\bm{k};z) - 1/\t{G}_{\sigma}(\bm{k};z),
\end{equation}
where, following Eq.\,(\ref{e5d}) and with $G_{\sigma;\infty_1}(\bm{k})$ and $G_{\sigma;\infty_2}(\bm{k})$ as given in respectively Eqs\,(\ref{e4qa}) and (\ref{e5ea}), one has
\begin{equation}\label{e7da}
\t{G}_{\X{0};\sigma}(\bm{k};z) = \frac{\hbar}{z - \varepsilon_{\bm{k}}} \equiv \frac{\hbar}{z -G_{\sigma;\infty_2}(\bm{k})/\hbar + \hbar\Sigma_{\sigma}^{\textsc{hf}}(\bm{k})},
\end{equation}
where $\Sigma_{\sigma}^{\textsc{hf}}(\bm{k})$ is the exact Hartree-Fock self-energy.\footnote{See \S\,\protect\ref{sec.3a.1}.} Employing the right-most expression in Eq.\,(\ref{e7da}) for $\t{G}_{\X{0};\sigma}(\bm{k};z)$ and the asymptotic series expansion in Eq.\,(\ref{e4n}) for $\t{G}_{\sigma}(\bm{k};z)$, from the equality in Eq.\,(\ref{e7f}) one arrives at the asymptotic series expansion for $\t{\Sigma}_{\sigma}(\bm{k};z)$ corresponding to $z\to\infty$ [\S\,3.3 in Ref.\,\citen{BF02}] [\S\S\,B.3-B.7 in Ref.\,\citen{BF07}]. Identifying the coefficient of $1/z^j$, $j\in\mathds{N}$, in this series with that in Eq.\,(\ref{e7b}), for the  $\Sigma_{\sigma;\infty_j}(\bm{k})$ corresponding to $j=1,2,3$ one obtains \cite{Note11} [Eqs\,(77) and (78), p.\,1449, in Ref.\,\citen{BF02}] [Eqs\,(B.71)-(B.73), and Eqs\,(B.89), (B.91), and (B.92) in Ref.\,\citen{BF07}]:
\begin{equation}\label{e7c}
\Sigma_{\sigma;\infty_1}(\bm{k}) = \frac{1}{\hbar^3} \big(-G_{\sigma;\infty_2}^2(\bm{k}) + \hbar G_{\sigma;\infty_3}(\bm{k})\big),
\end{equation}
\begin{equation}\label{e7d}
\Sigma_{\sigma;\infty_2}(\bm{k}) = \frac{1}{\hbar^4} \big(G_{\sigma;\infty_2}^3(\bm{k}) - 2\hbar\hspace{0.6pt} G_{\sigma;\infty_2}(\bm{k}) G_{\sigma;\infty_3}(\bm{k}) + \hbar^2\hspace{0.6pt} G_{\sigma;\infty_4}(\bm{k})\big),
\end{equation}
\begin{align}\label{e7e}
\Sigma_{\sigma;\infty_3}(\bm{k}) = \frac{1}{\hbar^5} \big(-G_{\sigma;\infty_2}^4(\bm{k}) &+ 3 \hbar\hspace{0.6pt} G_{\sigma;\infty_2}^2(\bm{k}) G_{\sigma;\infty_3}(\bm{k}) -\hbar^2\hspace{0.6pt} G_{\sigma;\infty_3}^2(\bm{k})\nonumber\\
&- 2 \hbar^2\hspace{0.6pt} G_{\sigma;\infty_2}(\bm{k}) G_{\sigma;\infty_4}(\bm{k}) + \hbar^3\hspace{0.6pt} G_{\sigma;\infty_5}(\bm{k})\big).
\end{align}
As we have indicated above, $\{\Sigma_{\sigma;\infty_j}(\bm{k})\| j \in \mathds{N}\}$ is similar $\{G_{\sigma;\infty_j}(\bm{k})\| j \in \mathds{N}\}$ a \textsl{positive sequence}. Thus, provided that the associated measure $\upsigma_{\sigma}(\bm{k};\varepsilon)$, Eq.\,(\ref{e20c}), has infinite number of `points of increase' along the $\varepsilon$ axis,\footnote{See similar remarks regarding $\mathbb{H}_{m}^{\protect\t{\protect\X{G}}_{\sigma}}$, with $s_j \equiv G_{\sigma;\infty_{j+1}}(\bm{k})$, on p.\,\pageref{AsRegards}.} the associated Hankel moment matrix $\mathbb{H}_{m}^{\t{\X{\Sigma}}_{\sigma}}$, Eq.\,(\ref{e6c}), with $s_j \equiv \Sigma_{\sigma;\infty_{j+1}}(\bm{k})$, is positive definite for arbitrary finite values of $m$.\footnote{In the light of the earlier considerations with regard to the Nevanlinna functions $\t{f}(z)$ and $\t{f}_1(z)$ in Eqs\,(\ref{e5b}) and (\ref{e5f}), \textsl{and} the fact that the sequence $\{G_{\sigma;\infty_j}(\bm{k})\| j\in \mathds{N}\}$ is positive, the positivity of the sequence $\{\Sigma_{\sigma;\infty_j}(\bm{k})\| j\in \mathds{N}\}$, discussed above, should not come as a surprise.}

The above expressions reveal that in general $\Sigma_{\sigma;\infty_j}(\bm{k})$ consists of a combination of the asymptotic coefficients  $\{G_{\sigma;\infty_2}(\bm{k}), \dots, G_{\sigma;\infty_{j+2}}(\bm{k})\}$, for all $j \in \mathds{N}$ (recall that $G_{\sigma;\infty_1}(\bm{k}) \equiv \hbar$, Eq.\,(\ref{e4qa})) \cite{BF02,BF07}. The elements of the latter sequence corresponding to Hubbard-like models being bounded for arbitrary finite values of $j$, it follows that for these models $\Sigma_{\sigma;\infty_j}(\bm{k})$ is similarly bounded for arbitrary finite values of $j$. In view of Eq.\,(\ref{e7i}), this implies that indeed for these models the spectral function $B_{\sigma}(\bm{k};\varepsilon)$, Eq.\,(\ref{e20b}), decays faster than any finite power of $1/\varepsilon$ for $\vert\varepsilon\vert\to\infty$, in conformity with the asymptotic expression in Eq.\,(\ref{e20j}) and the earlier observation with regard to the behaviour of the function $A_{\sigma}(\bm{k};\varepsilon)$ corresponding to Hubbard-like models for $\vert\varepsilon\vert\to\infty$.\footnote{Consider the function $\phi_{\alpha}(\varepsilon)$ employed in Eq.\,(\protect\ref{e4oe}).} This observation leads one to an explicit proof similar to that culminating in the result in Eq.\,(\ref{e4oh}) that the moment problem associated with the self-energy is a \textsl{determinate} one, Eq.\,(\ref{e7m}).

Further,\refstepcounter{dummy}\label{FurtherTheFact} the fact that $G_{\sigma;\infty_j}(\bm{k})$ is a polynomial of order $j-1$ in $\lambda$ for all $j \in \mathds{N}$, where $\lambda$ denotes the dimensionless coupling constant of the two-body interaction potential (see the discussions centred around Eq.\,(\ref{e4r})), leads to the important observation that \cite{BF02,BF07} $\Sigma_{\sigma;\infty_j}(\bm{k})$ is a polynomial of order $j+1$ in $\lambda$.\footnote{In the light of the expression in Eq.\,(\protect\ref{e4p}) in which $\vert\Psi_{N;\protect\X{0}}\rangle$ is the $N$-particle GS of $\protect\h{\mathcal{H}}\equiv \protect\h{\mathcal{H}}_{\protect\X{0}} + \protect\h{\mathcal{H}}_{\protect\X{1}}$, the coefficients of the two polynomials referred to here are functions/functionals of the \textsl{full} interaction potential, corresponding to $\lambda=1$. This observation is relevant to our explicit reference below to the \textsl{skeleton} self-energy diagrams evaluated in terms of the \textsl{interacting} Green functions $\{\protect\t{G}_{\sigma}(\bm{k};z)\| \sigma\}$.}\footnote{To ascertain that $\Sigma_{\sigma;\infty_j}(\bm{k})$ is indeed a polynomial of order $j+1$ in $\lambda$, consider the expressions in Eqs\,(\protect\ref{e7c}), (\protect\ref{e7d}), and (\protect\ref{e7e}). For instance, the function $\Sigma_{\sigma;\infty_1}(\bm{k})$ in Eq.\,(\ref{e7c}) consists of a linear superposition of the \textsl{square} of a first-order polynomial of $\lambda$ and a second-order polynomial of $\lambda$, resulting in $\Sigma_{\sigma;\infty_1}(\bm{k})$ being indeed a second-order polynomial of $\lambda$. Note that since $\Sigma_{\sigma;\infty_1}(\bm{k})$ is (fully) determined by $\protect\t{\Sigma}_{\sigma}^{\protect\X{(2)}}(\bm{k};z)$, Eq.\,(\protect\ref{e7h}), the zeroth- and first-order terms of the just-mentioned two second-order polynomials of $\lambda$, insofar as they are non-vanishing, must cancel.} Hence, $\Sigma_{\sigma;\infty_j}(\bm{k})$, $j\in \mathds{N}$, is \textsl{fully} determined in terms of \textsl{skeleton} self-energy diagrams (evaluated in terms of the \textsl{interacting} Green functions $\{\t{G}_{\sigma}(\bm{k};z) \| \sigma\}$) of order $2$ up to and including order $j+1$.\footnote{The lowest order is $2$ owing to the fact that $\t{\Sigma}_{\sigma}^{\protect\X{(\nu)}}(\bm{k};z)$ depends on $z$ only for $\nu\ge 2$, Eq.\,(\protect\ref{e7r}).} Consequently, for any finite value of $j \in \mathds{N}$, the asymptotic coefficient $\Sigma_{\sigma;\infty_j}(\bm{k})$ is fully determined by the contributions of the elements of the finite sequence $\{\Sigma_{\sigma}^{\X{(2)}}(\bm{k};z),\dots, \Sigma_{\sigma}^{\X{(j+1)}}(\bm{k};z)\}$ according to [Eq.\,(B.105) in Ref.\,\citen{BF07}]\,\footnote{See also Table VI in Ref.\,\protect\citen{BF07}. \label{noteb}} (\emph{cf.} Eq.\,(\ref{e4}))
\begin{equation}\label{e7h}
\Sigma_{\sigma;\infty_j}(\bm{k}) = \sum_{\nu=2}^{j+1} \Sigma_{\sigma;\infty_j}^{\X{(\nu)}}(\bm{k}),\;\; \forall j\in \mathds{N}.
\end{equation}
This implies that for \textsl{all} $\nu \ge 2$ [Eqs\,(B.103) and (B.104) in Ref.\,\citen{BF07}]
\begin{equation}\label{e7g}
\t{\Sigma}_{\sigma}^{\X{(\nu)}}(\bm{k};z) \sim \frac{\t{\Sigma}_{\sigma;\infty_{\nu-1}}^{\X{(\nu)}}(\bm{k})}{z^{\nu-1}} + \frac{\t{\Sigma}_{\sigma;\infty_{\nu}}^{\X{(\nu)}}(\bm{k})}{z^{\nu}} + \dots\;\; \text{as}\;\; z\to \infty,
\end{equation}
which is to say that $\t{\Sigma}_{\sigma}^{\X{(\nu)}}(\bm{k};z)$ decays at the slowest like $1/z^{\nu-1}$ for $z\to\infty$, as the decay of $\t{\Sigma}_{\sigma}^{\X{(\nu)}}(\bm{k};z)$ like $1/z^{\nu-2}$ for $z\to\infty$ would imply contribution of $\t{\Sigma}_{\sigma}^{\X{(\nu)}}(\bm{k};z)$ to $\Sigma_{\sigma;\infty_{\nu-2}}(\bm{k})$, in contradiction to the equality in Eq.\,(\ref{e7h}).\footref{noteb}

\refstepcounter{dummyX}
\subsubsection{Discussion}
\phantomsection
\label{sec.b2.1}
Following the first equality in Eq.\,(\ref{e38}), for the `Hubbard atom' of spin-$\tfrac{1}{2}$ particles one has (\emph{cf.} Eq.\,(\ref{e112}))
\begin{equation}\label{e7xa}
G_{\infty_{2j}} \equiv 0,\;\; G_{\infty_{2j+1}} = \hbar \Big(\frac{U}{2}\Big)^{2j},\;\; j \in \mathds{N}_{0}.
\end{equation}
On the basis of the expressions in Eqs\,(\ref{e7c}), (\ref{e7d}), and (\ref{e7e}), one thus obtains
\begin{equation}\label{e7xb}
\Sigma_{\infty_1} = \frac{U^2}{4\hbar},\;\; \Sigma_{\infty_2} = \Sigma_{\infty_3} = 0,
\end{equation}
in conformity with the exact expression for $\t{\Sigma}(z)$ in Eq.\,(\ref{e25}), which in fact implies that $\Sigma_{\infty_j} = 0$ for \textsl{all} $j > 1$. As regards in particular $\Sigma_{\infty_{2j+1}} = 0$, $\forall j \in \mathds{N}$, see the remark following Eq.\,(\ref{e7la}).

On account of the equality in Eq.\,(\ref{e7h}), one has
\begin{align}\label{e7xc}
\Sigma_{\infty_2} = 0 &\,\iff\, \Sigma_{\infty_2}^{\X{(2)}} + \Sigma_{\infty_2}^{\X{(3)}} = 0,\\
\label{e7xd}
\Sigma_{\infty_3} = 0 &\,\iff\, \Sigma_{\infty_3}^{\X{(2)}} + \Sigma_{\infty_3}^{\X{(3)}} + \Sigma_{\infty_3}^{\X{(4)}} = 0.
\end{align}
By p-h symmetry, one has $\Sigma_{\infty_j}^{\X{(3)}} = 0$, $\forall j \in \mathds{N}$,\footnote{See \S\,\protect\ref{sec.d52}.} so that from the right-most equality in Eq.\,(\ref{e7xc}) one obtains
\begin{equation}\label{e7xe}
\Sigma_{\infty_2}^{\X{(2)}} = 0,
\end{equation}
and from that in Eq.\,(\ref{e7xd})
\begin{equation}\label{e7xf}
\Sigma_{\infty_3}^{\X{(4)}} = -\Sigma_{\infty_3}^{\X{(2)}}.
\end{equation}
The equality in Eq.\,(\ref{e7xe}) is in conformity with the asymptotic expression in Eq.\,(\ref{e50a}). Following the latter expression, the equality in Eq.\,(\ref{e7xf}) implies that
\begin{equation}\label{e7xg}
\Sigma_{\infty_3}^{\X{(4)}} = -\frac{9 U^4}{16\hbar}.
\end{equation}
Explicit calculations in \S\,\ref{sec.d53} of the contributions of the $4$th-order skeleton self-energy diagrams in terms of the interacting Green function reveal violation of the equality in Eq.\,(\ref{e7xg}): we obtain $\Sigma_{\infty_3}^{\X{(4)}} = -45 U^4/(128\hbar)$, Eq.\,(\ref{ex0f}), instead of $\Sigma_{\infty_3}^{\X{(4)}} = -9 U^4/(16\hbar)$. \emph{This implies the breakdown of the equality in Eq.\,(\ref{e7h}) in the case of the `Hubbard atom' under consideration.} To clarify, this equality is based on the consideration that $\Sigma_{\sigma}^{\X{(\nu)}}(\bm{k};z)/U^{\nu}$ does not \textsl{explicitly} depend on $U$, whereby the upper bound of the sum on the RHS of Eq.\,(\ref{e7h}) is equal to $j+1$. The $4$th-order self-energy contribution in Eq.\,(\ref{e52b}), with $\epsilon =1$, is clearly in violation of this assumption. On this account, some contributions from higher-order skeleton self-energy diagrams (beginning from the $6$th order) evaluated in terms of the interacting Green function are to compensate for the deviation of $-45 U^4/(128\hbar)$ from $-9 U^4/(16\hbar)$. On the basis of the detailed diagrammatic calculations presented in this publication, in particular in appendix \ref{sacx}, one can convince oneself of the fact that, for instance, the direct proportionality of the last two terms contributing to the $4$th-order self-energy contribution in Eq.\,(\ref{e52b}), with $\epsilon=1$, to respectively $U^3$ and $U^2$ is due to the strict \textsl{locality} of the underlying interacting one-particle Green function. For a general $\nu$th-order skeleton self-energy diagram evaluated in terms of the interacting Green function, the \textsl{non-dispersive} character of the one-particle excitation energies $\pm U/2$ results in some denominators reducing the power of the interaction energy $U$ in the numerator, thus changing the $U^{\nu}$ multiplying the corresponding terms into $U^{\nu-1}$, $U^{\nu-2}$, \emph{etc.} For a macroscopic system, outside the atomic limit (in the case of the Hubbard Hamiltonian, for non-vanishing hopping terms), such change of $U^{\nu}$ into $U^{\nu-1}$, $U^{\nu-2}$, \emph{etc.}, can obtain only over a zero-measure subset of the underlying $\1BZ$.

\refstepcounter{dummyX}
\subsection{On the uniform convergence}
\phantomsection
\label{sec.b2.2}
Having established that the positive sequence $\{\Sigma_{\sigma;\infty_j}(\bm{k}) \| j\}$ exists for Hubbard-like models and corresponds to a \textsl{determinate} moment problem, Eq.\,(\ref{e7m}), by considering for these models the \textsl{truncated} problem associated with the moments $\{\b{\mu}_0,\b{\mu}_1,\dots, \b{\mu}_{2n-1}\}$, Eqs\,(\ref{e7j}) and (\ref{e7k}), and introducing the measure \cite{PRH50,APM11} $\upsigma_{n+1}(\varepsilon,\ptau)$ describing the function\,\footnote{Here bars over symbols mark them as corresponding to the self-energy $\protect\t{\Sigma}_{\sigma}(\bm{k};z)$, distinguishing them from those in \S\,\protect\ref{sec.b.1} that correspond to the Green function $\protect\t{G}_{\sigma}(\bm{k};z)$.} (\emph{cf.} Eq.\,(\ref{e6k}))
\begin{equation}\label{e7n}
\t{\b{w}}_{n+1}(z,\ptau) \doteq -\frac{\b{Q}_{n+1}(z,\ptau)}{\b{P}_{n+1}(z,\ptau)}
\end{equation}
for an arbitrary value of the real parameter $\ptau$ as (\emph{cf.} Eq.\,(\ref{e6ua}))
\begin{equation}\label{e7q}
\t{\b{w}}_{n+1}(z,\ptau) = -\int_{-\infty}^{\infty} \frac{\rd\upsigma_{n+1}(\varepsilon',\ptau)}{z-\varepsilon'},
\end{equation}
following the Vitali theorem [p.\,150 in Ref.\,\citen{RR98}] [\S\,5.21, p.\,168, in Ref.\,\citen{ECT52}] one demonstrates that the sequence $\{\t{\b{w}}_{n}(z,\ptau)\| n\}$ converges \textsl{compactly} [p.\,322 in Ref.\,\citen{RR91}] for $z$ over the finite part of the $z$-plane away from the real axis [Ch. 11 in Ref.\,\citen{RR98}].\footnote{Since $\protect\b{Q}_{n+1}(z,\protect\ptau)$ ($\protect\b{P}_{n+1}(z,\protect\ptau)$) is a polynomial of order $n$ $(n+1)$ in $z$, the question of convergence of $\t{\b{w}}_{n+1}(z,\ptau)$ at the point of infinity of the $z$-plane is devoid of meaning.} Here, as in the case of the sequence $\{\t{w}_{n}(z,\ptau) \| n\}$, Eq.\,(\ref{e6k}), \textsl{compact} convergence of $\{\t{\b{w}}_{n}(z,\ptau)\| n\}$ over the region $\im[z] \not= 0$ amounts to the \textsl{uniform} convergence of $\{\t{\b{w}}_n(z,\ptau)\| n\}$ for $z$ over this region.

Since, following Eqs\,(\ref{e7j}) and (\ref{e7h}), in the case at hand $\b{\mu}_j$, $j \in \mathds{N}_{0}$, is directly associated with the perturbation series expansion of $\t{\Sigma}_{\sigma}(\bm{k};z)$ in terms of \textsl{skeleton} self-energy diagrams (and the exact \textsl{interacting} one-particle Green functions $\{\t{G}_{\sigma}(\bm{k};z)\| \sigma\}$) up to order $j+2$ in $\lambda$, the uniform convergence of $\{\t{\b{w}}_n(z,\ptau)\| n\}$ towards the \textsl{unique} function $\t{\b{w}}(z) \equiv \t{\b{w}}_{\infty}(z,\ptau)$ for $z$ over the region $\im[z]\not= 0$ (established above) implies uniform convergence of the series in Eq.\,(\ref{e4}) over the region $\im[z] \not= 0$, for all $\bm{k}$ where $\t{\Sigma}_{\sigma}(\bm{k};z)$ is \textsl{bounded} and \textsl{continuous}. This conclusion relies on a crucial observation presented in \S\,\ref{sec.3.2.1}, namely that even while the counterpart of the function $\t{f}^{\X{(n)}}(z)$ considered in this appendix, that is $\t{\mathfrak{S}}_{\sigma}^{\X{(n)}}(\bm{k};z)$, may not be a Nevanlinna function of $z$ for an arbitrary finite value of $n \in \mathds{N}$, its analyticity in the region $\im[z] \not= 0$ of the $z$-plane ensures that the deviation of the convergence property of $\t{\mathfrak{S}}_{\sigma}^{\X{(n)}}(\bm{k};z)$ from that of $\t{f}^{\X{(n)}}(z)$ for $n\to\infty$ manifests itself over a countable set of points on the real axis of the $z$-plane. Interestingly, the uniform convergence of $-\t{\mathfrak{S}}_{\sigma}^{\X{(n)}}(\bm{k};z)$ a.e. towards the Nevanlinna function $-\t{\Sigma}_{\sigma}(\bm{k};z)$ for $n\to\infty$ implies that, if not Nevanlinna, the degree of deviation of $-\t{\mathfrak{S}}_{\sigma}^{\X{(n)}}(\bm{k};z)$ from a Nevanlinna function of $z$ uniformly diminishes for $n\to\infty$.

\refstepcounter{dummyX}
\subsection{An illustrative example}
\phantomsection
\label{sabx}
In this section we consider a simple model that will illustrate some of the observations made in the previous sections of this appendix. In this model, we identify the polynomials $P_n(z)$ and $Q_n(z)$, such as encountered in Eq.\,(\ref{e5tb}), with respectively the $n$th-order Chebyshev polynomial of the first kind $T_n(z)$ and the $(n-1)$th-order Chebyshev polynomial of the second kind $U_{n-1}(z)$ [Ch. 22 in Ref.\,\citen{AS72}] [\S\,2.4, p.\,29, in Ref.\,\citen{GS75}]. In this connection, we note that the polynomials $T_n(z)$ and $U_n(z)$ satisfy the same three-term recurrence relation as the polynomials $P_n(z)$ and $Q_n(z)$ considered in this appendix, Eq.\,(\ref{e5y}) \cite{Note14}; the initial conditions are however different.\refstepcounter{dummy}\label{TheChebyshev}\footnote{The Chebyshev polynomials $T_n(z)$ and $U_n(z)$ satisfy the three-term recurrent relation $\tfrac{1}{2} y_{j-1}(z) + \tfrac{1}{2} y_{j+1}(z) = z y_j(z)$ [Table 22.7, p.\,782, in Ref.\,\protect\citen{AS72}], subject to the initial conditions $T_0(z) = 1$, $T_1(z) = z$, and $U_0(z) = 1$, $U_1(z) = 2z$ [Table 22.4, p.\,777, in Ref.\,\protect\citen{AS72}] \protect\cite{Note17}. Using the explicit expression $U_n(\cos(\theta)) = \sin((n+1)\theta)/\sin(\theta)$ [\S\,22.3.16, p.\,776, in Ref.\,\protect\citen{AS72}], one obtains $U_{-n}(z) = -U_{n-2}(z)$, whereby $Q_0(z) \equiv U_{-1}(z) = 0$. Further, $Q_1(z) \equiv U_0(z) = 1$ [Table 22.4, p.\,777, in Ref.\,\protect\citen{AS72}]. Compare these initial conditions with those in Eqs\,(\protect\ref{e5u}) and (\ref{e5x}). Note that the latter initial conditions are partly tied to the initial conditions in Eqs\,(\protect\ref{e5t}) and (\protect\ref{e5ta}) and partly to the choice $P_0(z) \equiv 1/\sqrt{s_0}$, ensuring normalisation of $P_0(z)$ to unity with regard to the relevant measure function (\emph{cf.} Eq.\,(\protect\ref{e6ja})). \label{notep1}} For convenience, \emph{in this section we consider $\varepsilon$ and $z$ as being dimensionless quantities,} in contrast to elsewhere in the publication where $\llbracket \varepsilon \rrbracket = \llbracket z \rrbracket = \mathrm{J}$. For later reference, in \textsl{this} section \cite{Note12} (\emph{cf.} Eqs\,(\ref{e6j}) and (\ref{e6k}))
\begin{equation}\label{e6qi}
\t{w}_{n+1}(z,\ptau) \doteq -\frac{U_{n}(z) - \ptau U_{n-1}(z)}{T_{n+1}(z) - \ptau T_{n}(z)}.
\end{equation}

Assuming $\ptau\not=0, \pm\infty$, for the case at hand from the defining expression in Eq.\,(\ref{e6qi}) one obtains \cite{Note12} (\emph{cf.} Eq.\,(\ref{e6gi}) below)
\begin{align}\label{e6qc}
-\t{w}_{4+1}(z,\ptau) &\sim \frac{1}{z} + \frac{1/2}{z^3} + \frac{3/2^3}{z^5} + \frac{5/2^4}{z^7} + \frac{35/2^7}{z^9}  \!\stackrel{\downarrow}{+}\! \frac{\ptau/2^8}{z^{10}}+ \frac{(125+\ptau^2)/2^9}{z^{11}} \nonumber\\
&\hspace{6.5cm} + \dots\;\, \text{for}\;\; z\to\infty,\\
\label{e6qd}
-\t{w}_{5+1}(z,\ptau) &\sim \frac{1}{z} + \frac{1/2}{z^3} + \frac{3/2^3}{z^5} + \frac{5/2^4}{z^7} + \frac{35/2^7}{z^9} + \frac{63/2^8}{z^{11}} \!\stackrel{\downarrow}{+}\! \frac{\ptau/2^{10}}{z^{12}}
+ \frac{(461+\ptau^2)/2^{11}}{z^{13}}\nonumber\\
&\hspace{6.5cm}  + \dots\;\, \text{for}\;\; z\to\infty,
\end{align}
where for clarity we have marked the locations beyond which terms of order higher than $1/z^{2n+1+\varsigma}$ appear in the asymptotic expansions, where $\varsigma$ is defined in Eq.\,(\ref{e6q}). One observes that both expressions in Eqs\,(\ref{e6qc}) and (\ref{e6qd}) are in conformity with the expression in Eq.\,(\ref{e6p}), with the coefficients of the terms up to and including order $1/z^{2n+1+\varsigma}$ independent of $\ptau$. Further, on identifying $\ptau$ with zero, the erroneous term decaying like $1/z^{10}$ ($1/z^{12}$) in Eq.\,(\ref{e6qc}) (Eq.\,(\ref{e6qd})) appropriately disappears (\emph{cf.} Eq.\,(\ref{e6gi}) below), and the coefficient of that decaying like $1/z^{11}$ ($1/z^{13}$) reduces to $125/2^9 = 0.2441\ldots$ ($461/2^{11} = 0.2250\ldots$), to be contrasted with the exact coefficient $63/2^8 = 0.2460\ldots$ ($231/2^{10} = 0.2255\ldots$) in Eq.\,(\ref{e6gi}) below, in conformity with Eqs\,(\ref{e6q}) and (\ref{e6p}).

\begin{figure}[t!]
\psfrag{x}[c]{\LARGE $\varepsilon$}
\psfrag{y}[c]{\LARGE $\protect\im[-\protect\t{w}(\varepsilon -\protect\ii \eta)]$, $\protect\im[-\protect\t{w}_n(\varepsilon -\protect\ii \eta,\zeta)]$}
\centerline{\includegraphics[angle=0, width=0.62\textwidth]{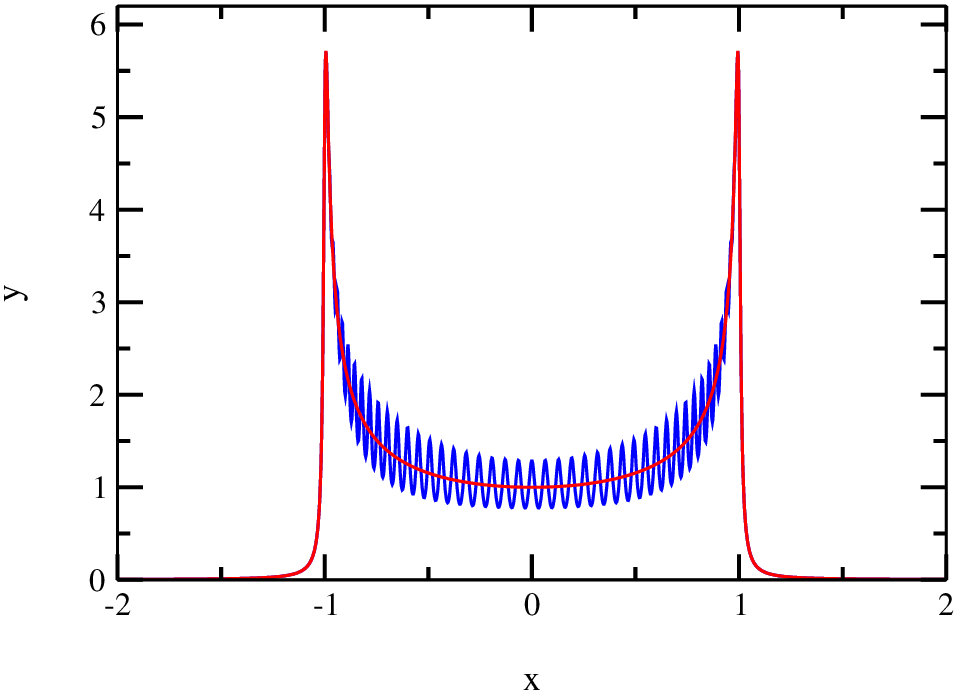}}
\caption{(Colour) The imaginary parts of the functions $-\protect\t{w}(z)$ (red), Eq.\,(\protect\ref{e6gh}), and $-\protect\t{w}_n(z,\zeta)$ (blue), Eq.\,(\protect\ref{e6qi}), for $z =\varepsilon -\protect\ii \eta$, with $\eta = 0.01$, $n=50$, and $\zeta = 2\protect\ii$. On the present scale, the two functions are indistinguishable in the regions $\varepsilon \lesssim -1$ and $\varepsilon \gtrsim +1$ (see \S\,\protect\ref{sec.b1a}). The considerable deviation of $\protect\im[-\protect\t{w}_n(z,\zeta)]$ from $\protect\im[-\protect\t{w}(z)]$ in the interval $-1 < \varepsilon < +1$ is due to the fixed value of $\zeta$, independent of $z$; identifying $\zeta$ with an appropriate function of $z$ and $n$ would result in $-\protect\t{w}_n(z,\zeta)$ identically coinciding with the function $-\protect\t{w}(z)$, Eqs\,(\protect\ref{e5te}) and (\protect\ref{e6jd}). The oscillatory behaviour of the function $\protect\im[-\protect\t{w}_n(\varepsilon -\protect\ii \eta,\zeta)]$ for $-1 \lesssim \varepsilon \lesssim 1$ is appreciated by considering the right-most expression in Eq.\,(\protect\ref{e6k}). Note that here appropriately $\protect\im[z] < 0$ and $\protect\im[\zeta] > 0$. }
\label{f13}
\end{figure}

\begin{figure}[t!]
\psfrag{x}[c]{\LARGE $\varepsilon$}
\psfrag{y}[c]{\LARGE $\protect\re[-\protect\t{w}(\varepsilon -\protect\ii \eta)]$, $\protect\re[-\protect\t{w}_n(\varepsilon -\protect\ii \eta,\zeta)]$}
\centerline{\includegraphics[angle=0, width=0.6\textwidth]{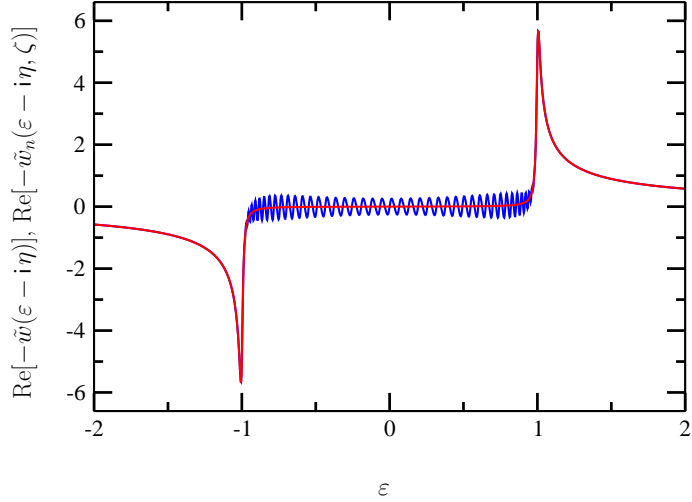}}
\caption{(Colour) The real parts of the functions $-\protect\t{w}(z)$ (red), Eq.\,(\protect\ref{e6gh}), and $-\protect\t{w}_n(z,\zeta)$ (blue), Eq.\,(\protect\ref{e6qi}), for $z =\varepsilon -\protect\ii \eta$, with $\eta = 0.01$, $n=50$, and $\zeta = 2\protect\ii$. On the present scale, the two functions are indistinguishable in the regions $\varepsilon \lesssim -1$ and $\varepsilon \gtrsim +1$ (see \S\,\protect\ref{sec.b1a}). The oscillatory behaviour of the function $\protect\re[-\protect\t{w}_n(\varepsilon -\protect\ii \eta,\zeta)]$ for $-1 \lesssim \varepsilon \lesssim 1$ is appreciated by considering the right-most expression in Eq.\,(\protect\ref{e6k}). Note that for $z\to\infty$ to leading order both $-\protect\t{w}_n(z,\zeta)$ and $-\protect\t{w}(z)$ are to decay like $1/z$, Eqs\,(\protect\ref{e6qc}), (\protect\ref{e6qd}) and (\protect\ref{e6gi}). Consult the caption of Fig.\,\protect\ref{f13} for further details.}
\label{f14}
\end{figure}

The function $-\t{w}_{n+1}(z,\ptau)$, Figs\,\ref{f13} and \ref{f14}, considered in this section corresponds to the truncated moment-expansion problem associated with the function (\emph{cf.} Eq.\,(\ref{e6ua}))
\begin{equation}\label{e6qf}
-\t{w}(z) = \int_{-1}^{1}\frac{\rd\upgamma(\varepsilon)}{z-\varepsilon},
\end{equation}
where\,\refstepcounter{dummy}\label{NoteThat}\footnote{Note that the polynomials $\{T_n(x)\| n\}$ and $\{U_n(x)\| n\}$ are orthogonal over the interval $[-1,1]$ with respect to the weight functions $\mathsf{w}_{\textsc{t}}(x) \equiv 1/\sqrt{1-x^2}$ and $\mathsf{w}_{\textsc{u}}(x) \equiv \sqrt{1-x^2}$ [Table 22.2, p.\,774, in Ref.\,\protect\citen{AS72}]. They are however \textsl{not} normalised to unity: with $\langle f,g\rangle_{\textsc{x}} \doteq \int_{-1}^{1} \mathrm{d}x\, \mathsf{w}_{\textsc{x}}(x) f(x) g(x)$, one has $\langle T_0,T_0\rangle_{\textsc{t}} = \pi$, $\langle T_n,T_n\rangle_{\textsc{t}} = \pi/2$, $\forall n \in\mathds{N}$, and $\langle U_n,U_n\rangle_{\textsc{u}} = \pi/2$, $\forall n \in \mathds{N}_0$ [Table 22.2, p.\,774, in Ref.\,\protect\citen{AS72}]. One further has $U_n(x) = \tfrac{1}{\pi} \int_{-1}^{1} \mathrm{d}y\, \mathsf{w}_{\textsc{t}}(y) \big(T_{n+1}(x)-T_{n+1}(y)\big)/(x-y)$ for $-1\le x \le 1$, $\forall n \in \mathds{N}_0$, where $\pi \equiv \langle T_{0},T_{0}\rangle_{\textsc{t}} \equiv s_0$, Eq.\,(\ref{e6ja}). This expression for $U_n(x)$ is the properly indexed and normalised version of the expression in Eq.\,(\protect\ref{e5xa}); re-indexing has been required because in this section $U_n(z)$ stands for $Q_{n+1}(z)$ (\emph{cf.} Eqs\,(\protect\ref{e6k}) and (\protect\ref{e6qi})). \label{noter1}}
\begin{equation}\label{e6qg}
\upgamma(\varepsilon) \equiv \frac{1}{\pi} \arcsin(\varepsilon) + C,\;\; \varepsilon \in [-1,1],
\end{equation}
in which $C$ is an arbitrary finite real constant. Following Eq.\,(\ref{e4kz}), one thus has
\begin{equation}\label{e6gh}
-\t{w}(z) \equiv \int_{-1}^{1} \frac{\rd\varepsilon}{\pi\sqrt{1-\varepsilon^2}}\frac{1}{z-\varepsilon} = \frac{1}{z} \frac{1}{\sqrt{1-1/z^2}},\;\; \im[z] \not= 0.
\end{equation}
The function on the RHS has two branch points at $z = \pm 1$ and a branch cut over the interval $[-1,1]$, as it should. Further, $\t{w}(z)$ is a Nevanlinna function, satisfying the equality $\sgn(\im[\t{w}(z)]) = \sgn(\im[z])$ for $\im[z]\not=0$. It is tempting to express the function on the RHS of Eq.\,(\ref{e6gh}) as $1/\sqrt{z^2 -1}$, however this function has an additional branch point at the point of infinity of the $z$-plane, and thus an additional branch-cut discontinuity covering the imaginary axis of this plane.\refstepcounter{dummy}\label{ThisIsEasily}\footnote{This is easily verified by drawing two vectors from $\pm 1$ to $z$, making evident that on $z$ crossing the imaginary axis of the $z$-plane, $\zeta \doteq z^2 -1$ crosses the negative axis of the $\zeta$-plane (at which point $\vert\hspace{-1.0pt}\arg(\zeta)\vert$ exceeds $\pi$), \emph{i.e.} the branch line of $\sqrt{\zeta}$ covering the interval $[-\infty,0]$. This is \textsl{not} the case for $\zeta \doteq 1-1/z^2$. Regarding the branch point of $1/\sqrt{z^2 -1}$ at the point of infinity of the $z$-plane, for $z\to \infty$ one has $1/\sqrt{z^2 -1} \sim \sqrt{\zeta}$, where $\zeta \doteq 1/z^2 \equiv (1/z)^2$. The mentioned branch point at the point of infinity of the $z$-plane is hereby seen to coincide with the branch point of the square-root function $\sqrt{\zeta}$ at $\zeta = 0$. The \textsl{entire} imaginary axis of the $z$-plane being a branch cut of the function $1/\sqrt{z^2 -1}$ is also hereby seen to be tied to the fact that $\zeta$ is equal to the \textsl{square} of $1/z$. \label{noter}}

On the basis of the right-most expression in Eq.\,(\ref{e6gh}), one obtains (\emph{cf.} Eqs\,(\ref{e6qc}) and (\ref{e6qd}))
\begin{equation}\label{e6gi}
-\t{w}(z) \sim \frac{1}{z} + \frac{1/2}{z^3} + \frac{3/2^3}{z^5} + \frac{5/2^4}{z^7} + \frac{35/2^7}{z^9} + \frac{63/2^8}{z^{11}} +\frac{231/2^{10}}{z^{13}} + \dots\;\; \text{for}\;\; z\to \infty.\;\;\;
\end{equation}
On the basis of the middle expression in Eq.\,(\ref{e6gh}), for the coefficient $c_j$ of $1/z^j$ in the above asymptotic series one has (\emph{cf.} Eq.\,(\ref{e7i}))
\begin{equation}\label{e6gj}
c_j  =  \int_{-1}^{1} \rd\varepsilon\; \frac{\varepsilon^{j-1}}{\pi\sqrt{1-\varepsilon^2}} \equiv \frac{(1 - (-1)^j) \hspace{0.6pt}\Gamma(j/2)}{2 \sqrt{\pi}\hspace{0.6pt} \Gamma((j+1)/2)},\;\; j\in \mathds{N},
\end{equation}
where $\Gamma(z)$ is the gamma function [\S\,6.1.1, p.\,255, in Ref.\,\citen{AS72}]. One verifies that indeed $c_1 = 1$, $c_2 = 0$, $c_3 = 1/2$, \emph{etc.} With reference to Eqs\,(\ref{e4ob}), (\ref{e6u}), and (\ref{e7k}), one has $c_j = s_{j-1}$. On account of [\S\,6.1.47, p.\,257, in Ref.\,\citen{AS72}]
\begin{equation}\label{e6gk}
s_{2j} = \frac{\Gamma(j+1/2)}{\sqrt{\pi}\hspace{0.6pt} \Gamma(j+1)} \sim \frac{1}{\sqrt{\pi j}}
\Big(1 -\frac{1}{8 j} + \frac{1}{128 j^2} + \dots\big) \;\; \text{for}\;\; j\to\infty,
\end{equation}
one obtains
\begin{equation}\label{e6gl}
\frac{1}{\sqrt[2j]{s_{2j}}} \sim 1 + \frac{\ln(j)}{4 j} + \frac{\ln(\pi)}{4 j} +\dots \;\;\text{for}\;\; j\to\infty,
\end{equation}
whereby, for any finite $\nu_0 \in \mathds{N}$, $\sum_{j = \nu_0}^{\nu} 1/\sqrt[2j]{s_{2j}} \sim \nu$ as $\nu\to\infty$. Thus (\emph{cf.} Eq.\,(\ref{e4og}))
\begin{equation}\label{e6gm}
\sum_{j=1}^{\infty} \frac{1}{\sqrt[2j]{s_{2j}}} = \infty.
\end{equation}
Hence the moment problem at hand is \textsl{determinate} [p.\,85 in Ref.\,\citen{NIA65}].

With reference to the observations in \S\,\protect\ref{sec.b1a}, to quantify the convergence behaviour of the function $\t{w}_{n+1}(z,\ptau)$ in Eq.\,(\ref{e6qi}) over the $z$-plane towards $\t{w}(z)$, Eq.\,(\ref{e6gh}), for $n\to\infty$, one may use the radius $\rho_{n+1}^{\textrm{c}}(z)$ of the circle $\mathsf{K}_{n+1}(z)$ as introduced in Eq.\,(\ref{e6ub}). Since [Table 22.2, p.\,774, in Ref.\,\citen{AS72}] (\emph{cf.} Eq.\,(\ref{e6ja}))\,\footnote{Note that $T_0(\varepsilon) \equiv 1$ [Table 22.4, p.\,777, in Ref.\,\protect\citen{AS72}]. Insofar as our present considerations are concerned, the significant aspect revealed by the results in Eq.\,(\protect\ref{e6gn}) is that the norms of $T_0(\varepsilon)$ and $T_j(\varepsilon)$, $\forall j\in\mathds{N}$, with respect to the measure $\upgamma(\varepsilon)$, Eq.\,(\protect\ref{e6qg}), are \textsl{unequal}.}
\begin{equation}\label{e6gn}
\int_{-1}^{1} \mathrm{d}\upgamma(\varepsilon)\, T_j^2(\varepsilon) = \left\{\begin{array}{ll}1, & j=0,\\
1/2, & j \in \mathds{N},\end{array}\right.
\end{equation}
the expression for $\rho_{n+1}^{\mathrm{c}}(z)$ in Eq.\,(\ref{e6ub}), based as it is on the Christoffel-Darboux formula [\S\,22.12.1, p.\,785, in Ref.\,\citen{AS72}] under the condition of equally-normalised orthogonal polynomials $\{P_j(z)\| j \in \mathds{N}_0\}$, is \textsl{not} applicable. Using the general expression for $\rho_{n+1}^{\mathrm{c}}(z)$  [p.\,12 in Ref.\,\citen{NIA65}], for the radius of the $\mathsf{K}_{n+1}(z)$ corresponding to the function $\t{w}_{n+1}(z)$ in Eq.\,(\ref{e6qi}) one has\,\footnote{Since $T_j(\varepsilon)$ is a real polynomial for $\varepsilon \in\mathds{R}$, here we are using the identity $T_j^*(z) \equiv T_j(z^*)$.}
\begin{equation}\label{e6go}
\rho_{n+1}^{\mathrm{c}}(z) \equiv \left|\frac{U_n(z) T_n(z) - U_{n-1}(z) T_{n+1}(z)}{T_{n+1}(z) T_n(z^*) - T_n(z) T_{n+1}(z^*)} \right|.
\end{equation}
On account of the Liouville-Ostrogradskii formula \cite{NIA65},\,\refstepcounter{dummy}\label{TheLS}\footnote{The Liouville-Ostrogradskii formula in Eq.\,(\protect\ref{e7wc}) stands for $P_{n}(z) Q_{n+1}(z) - P_{n+1}(z) Q_{n}(z) = 1/(s_0 \sqrt{\beta_{n+1}})$. This result is based on $P_j(z)$ being normalised to unity for all $j\in \mathds{N}_0$. In order to apply the latter formula to the case at hand, with reference to Eqs\,(\protect\ref{e6ja}) and (\protect\ref{e6gn}), the latter $s_0$ should be identified with $2$. With reference to the details in footnote \raisebox{-1.0ex}{\normalsize{\protect\footref{notep1}}} on p.\,\protect\pageref{TheChebyshev}, one observes that in the case at hand $\sqrt{\beta_j} = 1/2$, $\forall j \in \mathds{N}$. In this way, one arrives at the result in Eq.\,(\protect\ref{e6gp}). \label{noteq1}}\footnote{One alternatively arrives at the identity in Eq.\,(\protect\ref{e6gp}) by making use of $T_n(x) \equiv U_n(x) - x U_{n-1}(x)$ [\S\,22.5.6, p.\,777, in Ref.\,\protect\citen{AS72}] and the identity $U_n^2(x) - U_{n-1}(x) U_{n+1}(x) \equiv 1$. The latter is a trigonometric identity, based on the definition $U_n(x) \doteq \sin((n+1)\theta)/\sin(\theta)$, where $x =\cos(\theta)$ [\S\,22.3.16, p.\,776, in Ref.\,\protect\citen{AS72}].}
\begin{equation}\label{e6gp}
U_n(z) T_n(z) - U_{n-1}(z) T_{n+1}(z) \equiv 1.
\end{equation}
Making use of the Christoffel-Darboux formula,\footnote{In the Christoffel-Darboux formula [\S\,22.12.1, p.\,785, in Ref.\,\protect\citen{AS72}] as applied to the Chebyshev polynomials $\{T_j(z)\| j\}$ considered in this section, $h_0 = 1$ and $h_j = 1/2$, $\forall j \in \mathds{N}$ (\emph{cf.} Eq.\,(\protect\ref{e6gn})). Further, for the coefficient $k_n$ of $z^n$ in $T_n(z)$ one has $k_n = 2^{n-1}$ [\S\,22.3.6, p.\,775, in Ref.\,\protect\citen{AS72}].} one obtains
\begin{align}\label{e6gq}
T_{n+1}(z) T_n(z^*) - T_n(z) T_{n+1}(z^*) &= (z - z^*) \big\{T_0(z) T_0(z^*) + 2 \sum_{j=1}^{n} T_j(z) T_j(z^*)\big\}\nonumber\\
&\equiv (z-z^*) \big\{ 1 + 2 \sum_{j=1}^{n} \vert T_j(z)\vert^2 \big\},\;\; \forall n \ge 1.
\end{align}
One trivially verifies that for $n = 0$ the LHS of Eq.\,(\ref{e6gq}) is identical to $z-z^*$. One can therefore use the expression on the RHS of Eq.\,(\ref{e6gq}) also for $n=0$ by discarding the sum on the RHS by convention. Hence
\begin{equation}\label{e6gr}
\rho_{n+1}^{\mathrm{c}}(z) = \frac{1}{\vert z-z^*\vert}\frac{1}{1 + 2 \sum_{j=1}^{n} \vert T_j(z)\vert^2}.
\end{equation}
The expression for $\rho_{n+1}^{\mathrm{c}}(z)$ in Eq.\,(\ref{e6gr}) is clearly functionally different from that in Eq.\,(\ref{e6ub}).\footnote{This is the case even on identifying the $s_0$ in Eq.\,(\ref{e6ub}) with $1$, as in Ref.\,\protect\citen{NIA65}.}

In the Mathematica notebook accompanying this publication, we provide a set of programs for visually investigating the convergence behaviour of the function $\t{w}_{n+1}(z,\ptau)$ in Eq.\,(\ref{e6qi}) for increasing values of $n$ at arbitrary points of the $z$-plane. In these programs, the real variable $\ptau$ can be made complex, as in the calculations underlying the results depicted in blue in Figs\,\ref{f13} and \ref{f14} and in the considerations of in particular \S\,\ref{sec.3.2.1} where $\ptau$ is explicitly replaced by the complex variable $\zeta$.\footnote{See in particular the discussions centred on Eq.\,(\protect\ref{e7wh}). See also Eq.\,(\protect\ref{e6jd}).}

\refstepcounter{dummyX}
\section{On the self-consistent dynamical mean-field theory on a Bethe lattice in \texorpdfstring{$d=\infty$}{}}
\phantomsection
\label{sd}
Here we consider a well-known local theory and demonstrate that while the negative of the one-particle Green function in this theory is appropriately a Nevanlinna function of $z$, the negative of the associated self-energy is not.\footnote{See footnote \raisebox{-1.0ex}{\normalsize{\protect\footref{notee1}}} on p.\,\protect\pageref{ThisObservation}.}

Let
\begin{equation}\label{ed1a}
\Phi_a(\varepsilon) \doteq \left\{
\begin{array}{lc}\displaystyle \frac{2}{\pi a^2} \sqrt{a^2 -\varepsilon^2}, & \vert\varepsilon\vert \le a,\\ \\
0, & \vert\varepsilon\vert \ge a. \\ \end{array} \right.
\end{equation}
For the (local) self-energy $\t{\Sigma}(z)$ in the framework of the dynamical mean-field theory (DMFT) on a Bethe lattice with infinite coordination $\EuScript{Z} = 2 d$ and corresponding to the (semi-circular) density of states
\begin{equation}\label{ed1ax}
\mathcal{D}(\varepsilon) = \Phi_{2t}(\varepsilon)
\end{equation}
and $\mu=0$ (half-filling), one has [Eqs\,(7), (13), (22), (23) in Ref.\,\citen{GKKR96}] \cite{SK98,GK00}
\begin{equation}\label{ed10}
\t{\Sigma}(z) = z - t^2 \t{G}(z) -1/\t{G}(z),
\end{equation}
where $\t{G}(z)$ is the (local) one-particle Green function, and $t > 0$ the hopping integral. \emph{Here (as in the remaining part of thus appendix) we have identified $\hbar$ with unity.} With (\emph{cf.} Eq.\,(\ref{e4d}))
\begin{equation}\label{ed4}
A(\varepsilon) \doteq \mp \frac{1}{\pi} \im[\t{G}(\varepsilon \pm \ii 0^+)],
\end{equation}
one has the spectral representation (\emph{cf.} Eq.\,(\ref{e4e}))
\begin{equation}\label{ed5}
\t{G}(z) = \int_{-\infty}^{\infty} \rd\varepsilon'\, \frac{A(\varepsilon')}{z -\varepsilon'},
\end{equation}
making explicit that $\t{G}(z)$ is analytic over the region $\im[z] \not= 0$. Since $A(\varepsilon) \ge 0$, the expression in Eq.\,(\ref{ed5}) makes further explicit that $-\t{G}(z)$ is a Nevanlinna function of $z$, Eq.\,(\ref{e2}). Other relevant expressions as in \S\,\ref{sec.b2} can be similarly introduced here.

Since the function $\t{G}(z)$ has no zero in the finite part of the $z$-plane away from region $\im[z] =0$, a fact that can be directly established from the expression in Eq.\,(\ref{ed5}),\footnote{With $z \equiv x + \protect\ii y$, $x, y \in \mathds{R}$, from Eq.\,(\protect\ref{ed5}) one obtains $\protect\im[\protect\t{G}(z)] = -y \int_{-\infty}^{\infty} \mathrm{d}\varepsilon'\, A(\varepsilon')/\big((x -\varepsilon')^2 + y^2\big)$, which, since $A(\varepsilon) \ge 0$, $\forall\varepsilon$, is non-vanishing for \textsl{all} $y \not= 0, \infty$. One observes that $\mathrm{sgn}(\protect\im[-\protect\t{G}(z)]) = \mathrm{sgn}(\protect\im[z])$. See footnote \raisebox{-1.0ex}{\normalsize{\protect\footref{notee1}}} on p.\,\protect\pageref{ThisObservation}.} it follows that the self-energy $\t{\Sigma}(z)$ in Eq.\,(\ref{ed10}) is like $\t{G}(z)$ analytic everywhere in the region $\im[z] \not=0$. In contrast to $\t{G}(z)$ however, the function $\t{\Sigma}(z)$ in Eq.\,(\ref{ed10}) does \textsl{not} in general satisfy the relationship in Eq.\,(\ref{e2}). In other words, in contrast to $-\t{G}(z)$, $-\t{\Sigma}(z)$ is not in general a Nevanlinna function of $z$. This is directly surmised by considering the following expression, obtained from that in Eq.\,(\ref{ed10}):
\begin{equation}\label{ed11}
\im[\t{\Sigma}(z)] = \im[z] + \Big(\frac{1}{\vert\t{G}(z)\vert^2} - t^2\Big) \im[\t{G}(z)].
\end{equation}
Clearly, the $\im[z]$ on the RHS is a contributing factor to the violation of the relationship in Eq.\,(\ref{e2}) by the self-energy in Eq.\,(\ref{ed10}), although it does \textsl{not} by itself constitute a sufficient condition for this violation. In contrast, for $\im[z] \not=0$ the inequality
\begin{equation}\label{ed11a}
\vert \t{G}(z)\vert^2 \ge 1/t^2
\end{equation}
amounts to a sufficient condition for the violation of the equality in Eq.\,(\ref{e2}) by the self-energy in Eq.\,(\ref{ed10}). At first glance, the inequality in Eq.\,(\ref{ed11a}) should be satisfied in \textsl{some} regions of the complex $z$-plane for sufficiently large $\vert t\vert$, notably for $z$ close to those regions of the real energy axis where $\vert\t{G}(z)\vert$ is large, if not unbounded. That this is indeed the case, is confirmed by numerical calculations on two models, to be discussed below. Before doing so, it is instructive to consider the region of large $\vert z\vert$ in the complex $z$-plane. Assuming the function $A(\varepsilon)$ to decay faster than any finite power of $1/\varepsilon$ for $\vert\varepsilon\vert\to\infty$, from the expression in Eq.\,(\ref{ed5}) one obtains the asymptotic series (\emph{cf.} Eq.\,(\ref{e4n}))
\begin{equation}\label{e10z}
\t{G}(z)\sim \sum_{j=1}^{\infty} \frac{G_{\infty_j}}{z^j}\;\; \text{for}\;\; z \to \infty,
\end{equation}
where (\emph{cf.} Eq.\,(\ref{e4o}))
\begin{equation}\label{e10y}
G_{\infty_j} \doteq \int_{-\infty}^{\infty} \rd\varepsilon\; \varepsilon^{j-1} A(\varepsilon).
\end{equation}
One has $G_{\infty_1} = 1$ by the normalization to unity of $A(\varepsilon)$, Eq.\,(\ref{e4h}). For p-h symmetric cases, as relevant to the considerations in the present appendix, $G_{\infty_{2j}} = 0$ for \textsl{all} $j \in \mathds{N}$.\footnote{\emph{Cf.} Eq.\,(\ref{e7xa}), and see the relevant remarks on p.\,\protect\pageref{WeRemarkThat}.} One however has $G_{\infty_{2j-1}} > 0$ for \textsl{all} $j \in \mathds{N}$, irrespective of whether or not $A(\varepsilon)$ is symmetric with respect to $\varepsilon = 0$.\footnote{Exception to this rule concerns the case where the one-particle spectral function is of the type presented in Eq.\,(\protect\ref{e4oi}).} Thus, following the expressions in Eqs\,(\ref{ed10}) and (\ref{e10z}), for the p-h symmetric problem at hand to leading order one has\,\footnote{With reference to Eq.\,(\protect\ref{e20j}), we note that \textsl{finite-order} asymptotic series in the asymptotic region $z\to\infty$ with respect to the asymptotic sequence $\{1,1/z, 1/z^2,\dots\}$ are not capable of describing exponentially-decaying behaviour of functions in the region $z\to\infty$. This is reflected in the property that for $z = \varepsilon + \protect\ii \eta$, $\vert\varepsilon\vert\to \infty$, the leading asymptotic term in the left-most expression in Eq.\,(\protect\ref{e10x}) becomes identically vanishing on identifying $\eta$ with $0$. This is naturally an exact result in the case of $A(\varepsilon)$ being of bounded support, otherwise only an exact result up to an exponentially small correction. Consult in particular \S\,I.5, p.\,16, of Ref.\,\protect\citen{RBD73}.}
\begin{equation}\label{e10x}
\im[\t{\Sigma}(z)] \sim \frac{t^2 -G_{\infty_3}}{\vert z\vert^2}\hspace{0.6pt}\im[z] \;\Longrightarrow\; \sgn(\im[\t{\Sigma}(z)]) = \sgn(t^2 - G_{\infty_3}) \hspace{-0.4pt}\im[z]\;\; \text{as}\;\; z \to \infty.
\end{equation}
From the right-most expression, one observes that for $t^2 > G_{\infty_3}$ the function $\t{\Sigma}(z)$ violates that condition in Eq.\,(\ref{e2}) in the asymptotic region $z \to\infty$. We conclude that \emph{the function $-\t{\Sigma}(z)$ considered here, Eq.\,(\ref{ed10}), cannot in general be a Nevanlinna function of $z$.} This is invariably the case for sufficiently large values of $\vert t\vert$, in violation of the fact that the exact $-\t{\Sigma}_{\sigma}(\bm{k};z)$ is a Nevanlinna function of $z$ \cite{BF07}.\footnote{See Eq.\,(\protect\ref{e20g}) and the preceding considerations.}

Now we proceed with examining the above observation on a simple model. To this end, in view of $\int_{-\infty}^{\infty} \rd\varepsilon\, \Phi_a(\varepsilon) = 1$, with $\Phi_a(\varepsilon)$ as defined  in Eq.\,(\ref{ed1a}), we introduce
\begin{equation}\label{ed2}
A(\varepsilon) \doteq \frac{1}{t} \Phi_{2}(\varepsilon/w t) +\frac{1}{2} (1-w) \big(\Phi_{U/2-\Delta}(\varepsilon-U/2)+ \Phi_{U/2-\Delta}(\varepsilon+U/2)\big),
\end{equation}
where $U$ is the on-site interaction energy, $4wt$ the total width of the spectral function in the mid-gap region, centred at $\varepsilon=0$, and $U-2\Delta$ the total width of the peak in the spectral function centred at $\varepsilon= \mp U/2$, corresponding to the lower/upper Hubbard band.\footnote{Compare with Fig.\,1 in Ref.\,\protect\citen{SK98} with the $a$ herein identified with $2$. Compare also with Fig.\,22, p.\,54, in Ref.\,\protect\citen{GKKR96}. In Ref.\,\protect\citen{SK98}, the spectral peaks corresponding to the lower and upper Hubbard bands comprise the $\rho_{h}(\epsilon)$ in Eq.\,(6).} For obvious reasons, we assume $U/2 > \Delta > 2 w t$. One has (\emph{cf.} Eq.\,(\ref{e4h}))
\begin{equation}\label{ed3}
\int_{-\infty}^{\infty} \rd\varepsilon\, A(\varepsilon) = 1,
\end{equation}
and
\begin{equation}\label{ed3c}
\int_{-\Delta}^{\Delta} \rd\varepsilon\, A(\varepsilon) \equiv \int_{-2 w t}^{2 w t} \rd\varepsilon\; A(\varepsilon)= w.
\end{equation}
We note that, as required for the metallic state of the half-filled case, at Fermi energy $\varepsilon_{\textsc{f}} = 0$ we have \cite{SK98} [Eq.\,(19) in Ref.\,\citen{EMH89}]
\begin{equation}\label{ed3d}
A(0) = \frac{1}{\pi t}.
\end{equation}

With\,\footnote{We note that the expression for $\t{D}(\zeta)$ given in Eq.\,(22) of Ref.\,\protect\citen{GKKR96} is \textsl{incorrect}, however the corresponding expression for $R(G)$, given in the same equation, is correct. The error is due to taking the incorrect branch of the two-valued square-root function. In the notation of Ref.\,\protect\citen{GKKR96}, for the correct function $\protect\t{D}(\zeta)$ one has $\protect\t{D}(\zeta) = (\zeta - \protect\ii\mathpzc{s} \sqrt{4 t^2 - z^2})/(2 t^2)$, where $\mathpzc{s} \doteq\sgn(\im[\zeta])$. This expression coincides with that on the RHS of Eq.\,(\protect\ref{ed3a}) for $a = 2 t$. Although $\sqrt{-1} = \protect\ii\hspace{0.2pt}$, here identification of $\protect\ii \sqrt{4 t^2 - z^2}$ with $\sqrt{z^2 -4 t^2}$ is incorrect; this identification results in the selection of the incorrect branch of the square-root function when $\protect\sgn(\protect\re[z]) =  -\protect\sgn(\protect\im[z])$. See footnote \raisebox{-1.0ex}{\normalsize{\protect\footref{noter}}} on p.\,\pageref{ThisIsEasily}.}
\begin{equation}\label{ed3a}
\t{\Psi}_a(z) \doteq \int_{-\infty}^{\infty} \rd\varepsilon'\; \frac{\Phi_a(\varepsilon')}{z -\varepsilon'} \equiv  \frac{2 z}{a^2} \Big(1 -\sqrt{1-\frac{a^2}{z^2}}\Big),
\end{equation}
from the expressions in Eqs\,(\ref{ed5}) and (\ref{ed2}) one obtains\,\footnote{Note that for $\protect\t{\Psi}_a(z)$, as defined in Eq.\,(\ref{ed3a}), one has $\int_{-\infty}^{\infty} \mathrm{d}\varepsilon'\, \Phi_a(\varepsilon' + \varepsilon_0)/(z-\varepsilon') = \protect\t{\Psi}_{a}(z+\varepsilon_0)$, $\forall\varepsilon_0 \in \mathds{R}$, and $\int_{-\infty}^{\infty} \mathrm{d}\varepsilon'\, \Phi_a(\alpha \varepsilon')/(z-\varepsilon') = \protect\sgn(\alpha)\hspace{0.6pt} \protect\t{\Psi}_{a}(\alpha z)$, $\forall \alpha \in \mathds{R}\backslash\{0\}$.}
\begin{equation}\label{ed3b}
\t{G}(z) = \frac{1}{t}\hspace{1.0pt} \t{\Psi}_2(z/w t) + \frac{1}{2} (1-w) \Big(\t{\Psi}_{U/2-\Delta}(z-U/2) + \t{\Psi}_{U/2-\Delta}(z+U/2) \Big).
\end{equation}
The square-root function in Eq.\,(\ref{ed3a}) is the principal branch of the function $\sqrt{z}$, defined according to $\sqrt{z} \doteq \sqrt{\vert z\vert} + \tfrac{1}{2} \ii \arg(z)$, where $\vert\arg(z)\vert < \pi$. The function $\t{\Psi}_{a}(z)$ has two branch points at $z=\pm a$ and a branch-cut covering the open interval $(-a,a)$. From the first equality in Eq.\,(\ref{ed3a}), making use of
\begin{equation}\label{ed8}
\frac{1}{\varepsilon - \varepsilon' \pm \ii 0^+} = \mathcal{P}\Big(\frac{1}{\varepsilon -\varepsilon'}\Big) \mp \pi \!\ii\hspace{0.6pt} \delta(\varepsilon-\varepsilon'),
\end{equation}
one obtains
\begin{equation}\label{ed7}
\t{\Psi}_a(\varepsilon + \ii 0^+) -\t{\Psi}_a(\varepsilon - \ii 0^+) = -2\pi\!\ii\hspace{0.4pt} \Phi_a(\varepsilon).
\end{equation}
The RHS of this equality is indeed non-vanishing only for $\varepsilon \in (-a,a)$. From the explicit expression for $\t{\Psi}_a(z)$ on the RHS of Eq.\,(\ref{ed3a}), one verifies that the function $\t{G}(z)$ in Eq.\,(\ref{ed3b}) is indeed analytic for \textsl{all} $\im[z] \not=0$. Further, this function satisfies the condition in Eq.\,(\ref{e2}), so that indeed $-\t{G}(z)$ is a Nevanlinna function of $z$.

For the $G_{\infty_3}$, Eq.\,(\ref{e10y}), associated with the function $\t{G}(z)$ in Eq.\,(\ref{ed3b}) one has (\emph{cf.} Eq.\,(\ref{e10y}))
\begin{equation}\label{ed8a}
G_{\infty_3} = w^3 t^2 + \frac{1-w}{4} \big(\frac{5}{4} U^2 - \Delta U + \Delta^2\big).
\end{equation}
With reference to the expression in Eq.\,(\ref{e10x}), assuming $0\le w < 1$, one thus has
\begin{equation}\label{ed8b}
t^2 > G_{\infty_3}\, \iff\, t^2 \ge \frac{1-w}{4 (1-w^3)} \big(\frac{5}{4} U^2 -\Delta U +\Delta^2\big).
\end{equation}
The set of values of the parameters $(t,U,\Delta,w)$ for which the right-most inequality in Eq.\,(\ref{ed8b}) is satisfied is clearly non-empty.

\begin{figure}[t!]
\psfrag{x}[c]{\huge $\varepsilon$}
\psfrag{y}[c]{\huge $\t{G}(\varepsilon + \protect\ii\eta)$}
\centerline{\includegraphics[angle=0, width=0.62\textwidth]{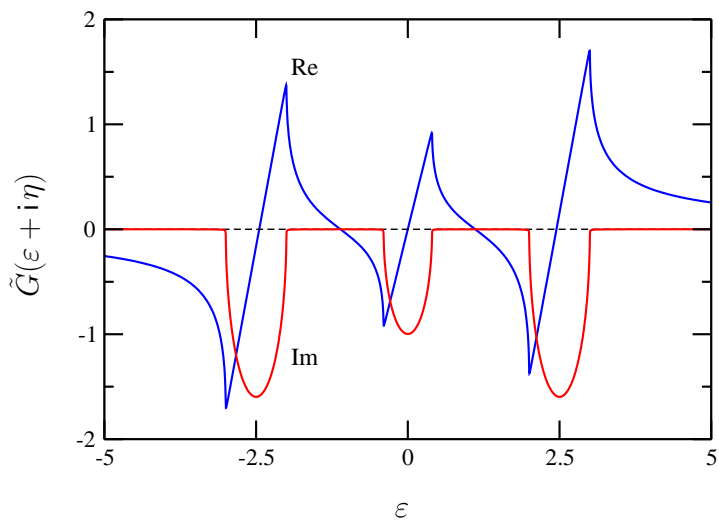}}
\caption{(Colour) The real (solid blue) and the imaginary (solid red) part of the Green function $\t{G}(\varepsilon+\protect\ii\eta)$, Eq.\,(\protect\ref{ed3b}), as functions of $\varepsilon$ for $t=1$, $U=5$, $\Delta =2$, $w = 1/5$ and $\eta = 0.001$ (in some atomic units). Note that, following the expression in Eq.\,(\protect\ref{ed4}), $\protect\im[\t{G}(\varepsilon+\protect\ii\eta)] = -\pi A(\varepsilon)$ for $\eta = 0^+$, where the $A(\varepsilon)$ as adopted in the present calculation is that presented in Eq.\,(\protect\ref{ed2}). The horizontal broken line marks the zero level. Note that indeed $\t{G}(\varepsilon+\protect\ii\eta)$ satisfies the relationship in Eq.\,(\protect\ref{e2}). The function $\vert\t{G}(\varepsilon+\protect\ii\eta)\vert^2$ is displayed in Fig.\,\protect\ref{f2} below (green broken line).}
\label{f1}
\end{figure}

\begin{figure}[t!]
\psfrag{x}[c]{\huge $\varepsilon$}
\psfrag{y}[c]{\huge $\t{\Sigma}(\varepsilon + \protect\ii\eta)$}
\centerline{\includegraphics[angle=0, width=0.62\textwidth]{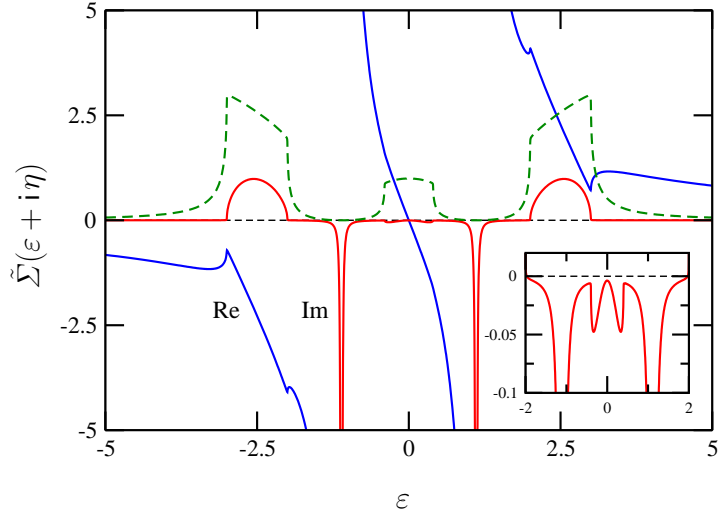}}
\caption{(Colour) The real (solid blue) and the imaginary (solid red) part of the self-energy $\t{\Sigma}(\varepsilon+\protect\ii\eta)$, Eq.\,(\protect\ref{ed10}), corresponding to the Green function in Eq.\,(\protect\ref{ed3b}) as functions of $\varepsilon$ (in some atomic units) for the same parameters as described in the caption of Fig.\,\protect\ref{f1}. The inset is a magnification of the details of $\im[\t{\Sigma}(\varepsilon+\protect\ii\eta)]$ over a limited range of $\varepsilon$. One clearly observes that $\t{\Sigma}(z)$ fails to satisfy the equality in Eq.\,(\protect\ref{e2}) (since here $\eta >0$, the latter equality demands $\im[\t{\Sigma}(\varepsilon+\protect\ii\eta)] \le 0$, $\forall\varepsilon \in \mathds{R}$). With reference to the expression in Eq.\,(\protect\ref{ed11}), by considering the function $\vert \t{G}(\varepsilon+\protect\ii\eta)\vert^2$ (broken green), one observes that this failure indeed takes place where $\vert \t{G}(\varepsilon + \protect\ii\eta)\vert^2$ is relatively large. Further, the singularities in $\t{\Sigma}(\varepsilon+\protect\ii\eta)$ at $\varepsilon \approx \pm 1.11$ correspond to the zeros of $\protect\t{G}(\varepsilon+\protect\ii 0^+)$ at these energies, apparent in Fig.\,\protect\ref{f1}.}
\label{f2}
\end{figure}

\begin{figure}[t!]
\centerline{\includegraphics[angle=0, width=0.62\textwidth]{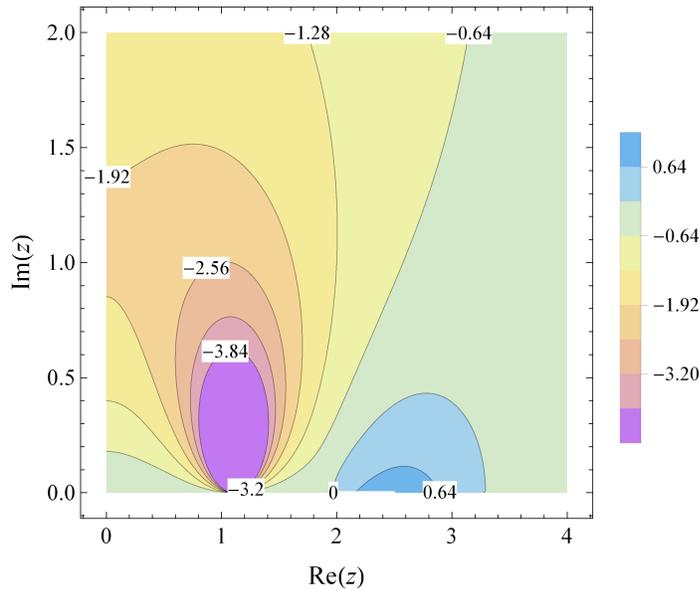}}
\caption{(Colour) A \protect\refstepcounter{dummy}\label{ContourPlot}contour plot of the function $\protect\im[\protect\t{\Sigma}(z)]$, with $\protect\t{\Sigma}(z)$, Eq.\,(\protect\ref{ed10}), corresponding to the Green function in Eq.\,(\protect\ref{ed3b}), in a section of the first quadrant of the complex $z$-plane. The underlying parameters are the same as those specified in the caption of Fig.\,\protect\ref{f1}. The region where $-\t{\Sigma}(z)$ violates the Nevanlinna property (\emph{i.e.} where $\protect\im[\protect\t{\Sigma}(z)] > 0$) is clearly visible. Compare with Fig.\,\protect\ref{f2}.}
\label{f3}
\end{figure}

In Fig.\,\ref{f1} we present the real and imaginary parts of the Green function $\t{G}(z)$ in Eq.\,(\ref{ed3b}) for $z = \varepsilon +\ii\eta$ with $\eta$ a small positive constant. One observes that this function clearly satisfies the property in Eq.\,(\ref{e2}). This is however not the case for the corresponding self-energy $\t{\Sigma}(z)$ obtained on the basis of the expression in Eq.\,(\ref{ed10}), Fig.\,\ref{f2}, which, in the light of the above discussions, does not come as a surprise. In Fig.\,\ref{f2} we also display the function $\vert\t{G}(\varepsilon+\ii\eta)\vert^2$, which is relevant from the perspective of the inequality in Eq.\,(\ref{ed11a}); the violation of the property in Eq.\,(\ref{e2}) is seen to take place over the range of values of $\varepsilon$ where $\vert\t{G}(\varepsilon+\ii\eta)\vert^2$ is greater than $1/t^2$ (here equal to $1$). In order to show that violation of the property in Eq.\,(\ref{e2}) is not restricted to the close neighbourhood of the real axis of the $z$-plane, in Fig.\,\ref{f3}, p.\,\pageref{ContourPlot}, we present a colour-coded contour plot of $\im[\t{\Sigma}(z)]$ over a region of the first quadrant of the $z$-plane, clearly showing the finite part of this region of the $z$-plane over which the Nevanlinna property, Eq.\,(\ref{e2}), is violated by $-\t{\Sigma}(z)$.

A noteworthy aspect made visible by the functions displayed in Fig.\,\ref{f1} is that at two points $\varepsilon = \mp\varepsilon_0$ in the intervals $(-\Delta,-2wt)$ and $(+2wt,+\Delta)$, where $\im[\t{G}(\varepsilon+\ii 0^+)] \equiv 0$, or $A(\varepsilon) \equiv 0$, the function $\re[\t{G}(\varepsilon+\ii 0^+)]$ is zero. From Fig.\,\ref{f2} one observes that indeed at each of these two points the function $\im[\t{\Sigma}(\varepsilon+\ii \eta)]$ appears to be described by a $\delta$-function singularity.\footnote{The apparent finite widths of the peaks centred at $\varepsilon = \mp\varepsilon_0$ is due to the finite value of the $\eta$ used in the present calculations.} That these singularities are indeed $\delta$ functions follows from the zeros of $\t{G}(\varepsilon + \ii 0^+)$ at $\varepsilon = \mp\varepsilon_0$ being simple ones. Investigation of the behaviour $\t{G}(\varepsilon + \ii 0^+)$ in the neighbourhoods of $\varepsilon = \mp\varepsilon_0$ reveals that $-\partial\t{G}(\varepsilon + \ii 0^+)/\partial t\vert_{\varepsilon=\mp\varepsilon_0}$ is positive and an \textsl{increasing} function of $t$, whereby, on account of $\delta(u x) = \delta(x)/\vert u\vert$, the amplitudes of the $\delta$-functions at $\varepsilon = \mp\varepsilon_0$ are to be \textsl{decreasing} functions of $t$. Thus, while for $0 < t < \Delta/2w$ one expects to have
\begin{equation}\label{ed9a}
-\int_{-\Delta}^{\Delta} \mathrm{d}\varepsilon\, \im[\t{\Sigma}(\varepsilon+\ii 0^+)] > C,
\end{equation}
where $C$ is a finite $t$-dependent positive constant, one similarly expects the LHS of Eq.\,(\ref{ed9a}) to be a \textsl{decreasing} function of $t$. Numerically, we have obtained
\begin{equation}\label{ed9b}
-\int_{-\Delta}^{\Delta} \mathrm{d}\varepsilon\, \im[\t{\Sigma}(\varepsilon+\ii 0^+)] \sim a - b\hspace{0.6pt} t^2,\;\; a, b > 0,
\end{equation}
which is to be contrasted with the expression in Eq.\,(13) of Ref.\,\citen{SK98}. Remarkably, as $t$ increases, $-\im[\t{\Sigma}(\varepsilon+\ii 0^+)]$ becomes negative for $2 t w\lesssim \vert\varepsilon\vert \le \Delta$, as a result of which the LHS of Eq.\,(\ref{ed9b}) becomes negative for $t > t_0$, with $t_0$ a finite constant.\footnote{In the light of Eq.\,(\protect\ref{e10x}), one expects that $t_0 = O(G_{\infty_3}^{1/2})$. The expression for the $G_{\infty_3}$ corresponding to the Green function in Eq.\,(\protect\ref{ed3b}) is given in Eq.\,(\ref{ed8a}), in which for $t_0$ finite and $w\to 0$ the first term on the RHS may be neglected, and $(1-w)$ replaced by $1$. For the parameters indicated in the caption of Fig.\,\protect\ref{f1}, one obtains $G_{\infty_3}^{1/2} \approx 2.249$ (solving the equation $G_{\infty_3} = t^2$, one arrives at $t_0 = \tfrac{1}{2}\big((1-w) (5 U^2/4 + \Delta^2  -\Delta U)/(1-w^3)\big)^{1/2}$, yielding for the said parameters $t_0 \approx 2.256$), to be contrasted with the exact result $t_0 \approx 4.953$, obtained by numerically solving for the zero of the LHS of Eq.\,(\protect\ref{ed9b}) as a function of $t$.}\footnote{Note that in order for the first contribution on the RHS of Eq.\,(\protect\ref{ed2}) not touch or overlap with the remaining contributions, one must have $0 < 2 w t < \Delta$, implying $t \in (0, \Delta/2w)$.} Interestingly, before $t$ reaches $t_0$, the slope of $\re[\t{G}(\varepsilon+\ii 0^+)]$ for $\varepsilon$ in the neighbourhood of $\varepsilon = 0$ becomes negative (\emph{cf.} Fig.\,\ref{f1}), whereby the above-mentioned two $\delta$-functions disappear and $\im[\t{\Sigma}(\varepsilon+\ii 0^+)]$ turns into a positive function (violating the equality in Eq.\,(\ref{e2})) that is nowhere vanishing inside the interval $(-2w t, +2wt)$.

In order to show that the above observations are not due to the specific form of the function $\Phi_a(\varepsilon)$ in Eq.\,(\ref{ed1a}), we remark that very similar observations are made by adopting the following definition for the latter function:\,\footnote{Like the function in Eq.\,(\protect\ref{ed1a}), that in Eq.\,(\protect\ref{ed1}) satisfies $\int_{-\infty}^{\infty} \mathrm{d}\varepsilon\, \Phi_a(\varepsilon) = 1$, $\forall a > 0$.}
\begin{equation}\label{ed1}
\Phi_a(\varepsilon) \doteq \left\{\begin{array}{lc} (a+\varepsilon)/a^2, & -a \le \varepsilon \le 0, \\
(a-\varepsilon)/a^2, & \phantom{-}0\le \varepsilon \le a,\\
0, & \phantom{-0\le}\vert \varepsilon\vert \ge a,\\ \end{array}\right.
\end{equation}
for which one has (\emph{cf.} Eq.\,(\ref{ed3a}))
\begin{equation}\label{ed6}
\t{\Psi}_a(z) = \frac{(z-a) \ln(z-a) + (z+a) \ln(z+a) -2 z \ln(z)}{a^2},
\end{equation}
where $\ln(z) = \ln\vert z\vert + \ii \arg(z)$, $\vert \arg(z)\vert < \pi$, denotes the principal branch of the logarithm function. The function $\t{\Psi}_a(z)$ in Eq.\,(\ref{ed6}) has branch points at $z = \pm a$ and a branch-cut covering the open interval $(-a,a)$. Assuming $U/2 > \Delta > \pi w t$, for the present choice of the function $\Phi_a(\varepsilon)$, we model the one-particle spectral function as\,\footnote{With $\Phi_a(\varepsilon)$ as defined in Eq.\,(\protect\ref{ed1}), one has $\Phi_{a}(0) = 1/a$, so that indeed the equality in Eq.\,(\protect\ref{ed3d}) applies also to the function $A(\varepsilon)$ in Eq.\,(\protect\ref{ed6a}). For this function, one further has $A(\varepsilon) \ge 0$, $\forall w \in (0,1]$, and $\int_{-\infty}^{\infty} \mathrm{d}\varepsilon\, A(\varepsilon) = 1$.} (\emph{cf.} Eq.\,(\ref{ed2}))
\begin{equation}\label{ed6a}
A(\varepsilon) \doteq \frac{1}{t}\hspace{1.0pt} \Phi_{\pi}(\varepsilon/w t) +\frac{1}{2} (1-w) \big(\Phi_{U/2-\Delta}(\varepsilon-U/2)+ \Phi_{U/2-\Delta}(\varepsilon+U/2)\big),
\end{equation}
resulting in (\emph{cf.} Eq.\,(\ref{ed3b}))
\begin{equation}\label{ed9}
\t{G}(z) = \frac{1}{t}\hspace{1.0pt} \t{\Psi}_{\pi}(z/w t) + \frac{1}{2} (1-w) \Big(\t{\Psi}_{U/2-\Delta}(z-U/2) + \t{\Psi}_{U/2-\Delta}(z + U/2)\Big).
\end{equation}
From this expression one can directly verify that $\t{G}(z)$ is analytic for all $\im[z] \not=0$ and that it satisfies the equality in Eq.\,(\ref{e2}). Although the corresponding self-energy $\t{\Sigma}(z)$, Eq.\,(\ref{ed10}), is analytic for $\im[z] \not=0$, similarly to the self-energy corresponding to the Green function in Eq.\,(\ref{ed3b}) it fails to satisfy the relationship in Eq.\,(\ref{e2}) for arbitrary $z \in \mathds{C}\backslash\mathds{R}$.

\refstepcounter{dummyX}
\section{Diagrammatica}
\phantomsection
\label{sacx}
In \refstepcounter{dummy}\label{InThisAppendix}this appendix we consider all skeleton\,\footnote{Also known as  $G$-skeleton and two-particle irreducible, 2PI.} self-energy diagrams of up to and including $4$th order as appropriate to normal GSs (ESs) of general models of spin-$\mathsf{s}$ particles\,\footnote{Unless we indicate otherwise, the particles considered in this appendix are fermions. This assumption is reflected in our use of such term as `fermion loop' and the minus sign associated with the contribution of each such loop to a self-energy diagram. We further assume that the GSs (ESs) under consideration are normal, as opposed to superconductive \protect\cite{BF19,BF21b}.} interacting through a two-body potential that may or may not be spin dependent. Therefore, unless we indicate otherwise, the directed solid lines in these diagrams represent the elements of the $(2\mathsf{s}+1)\times (2\mathsf{s}+1)$ matrix $\mathbb{G}$ of the \textsl{exact} one-particle Green functions $\{G_{\sigma,\sigma'} \| \sigma,\sigma'\}$, where $G_{\sigma,\sigma'} = (\mathbb{G})_{\sigma,\sigma'}$. At places, as in the main body of this publication, we explicitly assume $\mathbb{G}$ to be diagonal, where we denote the diagonal elements of this matrix by $\{G_{\sigma}\| \sigma\}$ \cite{FW03,BF19}. Explicitly, for $\mathbb{G}$ diagonal, $G_{\sigma,\sigma'} = G_{\sigma}\hspace{0.6pt}\delta_{\sigma,\sigma'}$. For $\mathbb{G}$ a full matrix, each diagram represents a contribution to a specified element of the full $(2\mathsf{s}+1)\times (2\mathsf{s}+1)$ self-energy matrix $\mathbb{\Sigma}$, for which one has $(\mathbb{\Sigma})_{\sigma,\sigma'} = \Sigma_{\sigma,\sigma'}$. Denoting the bare two-body interaction potential mediating between particles $1$ and $2$ by $\mathsf{v}_{\sigma_1,\sigma_1';\sigma_2,\sigma_2'}(\bm{r}_1,\bm{r}_2)$, satisfying\,\footnote{See Fig.\,\protect\ref{f4} below, p.\,\protect\pageref{TwoInternal}.}
\begin{equation}\label{ea3c}
\mathsf{v}_{\sigma_1,\sigma_1';\sigma_2,\sigma_2'}(\bm{r}_1,\bm{r}_2) \equiv \mathsf{v}_{\sigma_2,\sigma_2';\sigma_1,\sigma_1'}(\bm{r}_2,\bm{r}_1),
\end{equation}
and focussing on the spin-dependence of this potential, a number of different categories of two-body potentials are conceivable of which we consider three characterised by the following specifications:\,\footnote{Below, $v_{\sigma_1,\sigma_2}$ stands for $v_{\sigma_1,\sigma_2}(\bm{r}_1,\bm{r}_2)$, and similarly, $v$ for $v(\bm{r}_1,\bm{r}_2)$. The Coulomb potential, $v_{\textsc{c}}(\|\bm{r}_1-\bm{r}_2\|)$, employed in first-principles calculations on physical systems, falls into category (3). This potential is both isotropic and translation invariant, as reflected in its dependence on $\|\bm{r}_1-\bm{r}_2\|)$. In the considerations of this appendix these properties are tacitly assumed, this as reflected in the Fourier transforms of the two-body potentials under consideration depending on $\|\bm{q}\|$ and $\|\bm{k}\|$ (see Eqs\,(\protect\ref{ex01a}) and (\protect\ref{ex01}), as well as Fig.\,\protect\ref{f4}, p.\,\protect\pageref{TwoInternal}).}
\vspace{0.2cm}
\begin{itemize}
\item[(1)] \protect\refstepcounter{dummy}\label{ThreeCases}$\mathsf{v}_{\sigma_1,\sigma_1';\sigma_2,\sigma_2'}$ depends non-trivially on $\sigma_1,\sigma_1', \sigma_2$, and $\sigma_2'$ (\emph{cf.} Eq.\,(\ref{ex01a}) below),
\item[(2)] $\mathsf{v}_{\sigma_1,\sigma_1';\sigma_2,\sigma_2'} = v_{\sigma_1,\sigma_2}\hspace{1.0pt}\delta_{\sigma_1,\sigma_1'} \delta_{\sigma_2,\sigma_2'}$, where $v_{\sigma_1,\sigma_2}$ depends non-trivially on $\sigma_1$ and $\sigma_2$ (\emph{cf.} Eq.\,(\ref{ea34k}) and Eqs\,(\ref{ex01h}) -- (\ref{ex01k}) below), and
\item[(3)] $\mathsf{v}_{\sigma_1,\sigma_1';\sigma_2,\sigma_2'} = v\hspace{1.0pt} \delta_{\sigma_1,\sigma_1'} \delta_{\sigma_2,\sigma_2'}$ (\emph{cf.} Eqs\,(\ref{ex01b}), (\ref{ex03}), and (\ref{ex01bx}) below).
\end{itemize}
\vspace{0.2cm}
In the most general case the self-energy $\mathbb{\Sigma}$ is a full $(2\mathsf{s}+1)\times (2\mathsf{s}+1)$ matrix, even where $\mathbb{G}$ is diagonal.\refstepcounter{dummy}\label{ConsideringSpin}\footnote{Considering spin-$\tfrac{1}{2}$ particles, this is not the case when, \emph{e.g.}, $\mathsf{v}_{\sigma_1,\sigma_1';\sigma_2,\sigma_2'} = v_0\hspace{1.0pt}\mathbb{1}(1)_{\sigma_1,\sigma_1'} \mathbb{1}(2)_{\sigma_2,\sigma_2'} + v_1\hspace{1.0pt} \vec{\mathbb{\sigma}}(1)_{\sigma_1,\sigma_1'} \bm{\cdot} \vec{\mathbb{\sigma}}(2)_{\sigma_2,\sigma_2'}$, where $\mathbb{1}$ is the $2\times 2$ unit matrix, and $\vec{\mathbb{\sigma}} \equiv \{\mathbb{\sigma}^{\mathrm{x}}, \mathbb{\sigma}^{\mathrm{y}}, \mathbb{\sigma}^{\mathrm{z}}\}$ the $3$-vector of the $2\times 2$ Pauli matrices [p.\,104 in Ref.\,\protect\citen{FW03}] (we should point out that in Ref.\,\protect\citen{FW03} it has only been shown that making use of a diagonal $\mathbb{G}_{\protect\X{0}}$, to \textsl{first order} in the present two-body potential the Green matrix $\mathbb{G}$ is diagonal -- see below). Thus, $\mathbb{1}(1)_{\sigma_1,\sigma_1'} \mathbb{1}(2)_{\sigma_2,\sigma_2'} = \delta_{\sigma_1,\sigma_1'} \delta_{\sigma_2,\sigma_2'}$. Denoting the second part of the present potential by $\mathsf{v}'_{\sigma_1,\sigma_1';\sigma_2,\sigma_2'}$, one has $\mathsf{v}'_{\uparrow,\uparrow;\uparrow,\uparrow} = -\mathsf{v}'_{\uparrow,\uparrow;\downarrow,\downarrow} = \tfrac{1}{2} \mathsf{v}'_{\uparrow,\downarrow;\downarrow,\uparrow} = \tfrac{1}{2} \mathsf{v}'_{\downarrow,\uparrow;\uparrow,\downarrow} = -\mathsf{v}'_{\downarrow,\downarrow;\uparrow,\uparrow} = \mathsf{v}'_{\downarrow,\downarrow;\downarrow,\downarrow} = v_1$. For other combinations of the spin indices $\{\sigma_1,\sigma_1',\sigma_2,\sigma_2'\}$ the potential $\mathsf{v}'_{\sigma_1,\sigma_1';\sigma_2,\sigma_2'}$ is identically vanishing. The two contributions $\mathsf{v}'_{\uparrow,\downarrow;\downarrow,\uparrow}$ and $\mathsf{v}'_{\downarrow,\uparrow;\uparrow,\downarrow}$, which mediate the \textsl{exchange} of the spin indices of two particles with opposite spins on interacting, fundamentally distinguish the present $\mathsf{v}'_{\sigma_1,\sigma_1';\sigma_2,\sigma_2'}$ from a two-body potential in category (2). Owing to the symmetry $\mathsf{v}_{\sigma_1,\sigma_1';\sigma_2,\sigma_2'}(\bm{r}_1,\bm{r}_2) \equiv \mathsf{v}_{\sigma_2,\sigma_2';\sigma_1,\sigma_1'}(\bm{r}_2,\bm{r}_1)$, the latter two contributions make identical contributions to the many-body Hamiltonian. Denoting the contribution of these two terms to the many-body Hamiltonian by $\delta\protect\h{H}$, formally the many-body Hamiltonian of the system under consideration can be written as $\protect\h{H} + \delta\protect\h{H}$, where $\protect\h{H}$ denotes the many-body Hamiltonian describing systems in terms of the two-body potentials in category (2) (\emph{cf.} Eq.\,(\protect\ref{ea34k})). Interestingly (and in accord with the observation based on the first-order calculation in Ref.\,\protect\citen{FW03}, referred to above), one can show that, similarly to $\protect\h{H}$, $\delta\protect\h{H}$ commutes with $\protect\h{N}_{\sigma}$, $\sigma = \uparrow, \downarrow$, so that the many-body Hamiltonian corresponding to the two-body potential under investigation indeed commutes with the $z$-component of the total-spin operator $\protect\h{S}^{\mathrm{z}} \equiv \hbar (\protect\h{N}_{\uparrow} - \protect\h{N}_{\downarrow})/2$. See footnote \raisebox{-1.0ex}{\normalsize{\protect\footref{notem}}} on p.\,\protect\pageref{notem}. \label{notes1}} Only in the cases where the two-body potential falls in one of the categories (2) and (3) is $\mathbb{\Sigma}$ diagonal when is $\mathbb{G}$ diagonal. Although for $\mathbb{G}$ diagonal the contribution of a diagram corresponding to $\Sigma_{\sigma,\sigma'}$ is identically vanishing for $\sigma\not=\sigma'$ when the underlying two-body potential is in one of the latter two categories, we find it convenient in such case to view each self-energy diagram as contributing to a specified diagonal element $\Sigma_{\sigma} \equiv (\mathbb{\Sigma})_{\sigma,\sigma}$ of $\mathbb{\Sigma}$. The above observations equally apply on replacing $\mathbb{G}$ by its non-interacting counterpart $\mathbb{G}_{\X{0}}$. In this connection, similarly to $\{G_{\sigma}\| \sigma\}$, $\{G_{\X{0};\sigma}\| \sigma\}$ denote the diagonal elements of $\mathbb{G}_{\X{0}}$.

The number of skeleton self-energy diagrams can be read off from the following counting polynomial due to Molinari and Manini [Eq.\,(17) in Ref.\,\citen{MM06}]:\footnote{We note that $\EuScript{C}(0) = 1$ signifies that the expression in Eq.\,(\protect\ref{eac3a}) corresponds to $\Sigma' \equiv \Sigma - \Sigma^{\textsc{h}}$ (for the $\EuScript{C}(y)$ corresponding to $\Sigma$ one has $\EuScript{C}(0) = 2$). See \S\S\,2.6 and 2.8 in Ref.\,\protect\citen{BF19}.}\footnote{The self-energy diagrams and their numbers to be dealt with in this appendix can be calculated with the aid of the programs, written in the Mathematica programming language \protect\cite{SW16}, incorporated in appendices B and C of Ref.\,\protect\citen{BF19}.}
\begin{equation}\label{eac3a}
\EuScript{C}(y) \equiv 1 + (1 + \ell) y + (4 + 5\ell + \ell^2) y^2 + (27 + 40 \ell + 14 \ell^2 + \ell^3) y^3 + \dots,
\end{equation}
where the power of $y$ indicates $\nu-1$, with $\nu$ the order of the perturbation expansion, and the power of $\ell$ the number of the fermion loops in the relevant set of diagrams. Thus, there are $\mathcal{N}_2 = 2$ second-order skeleton self-energy diagrams, of which one without and one with a fermion loop, depicted in Fig.\,\ref{f7} below, p.\,\pageref{SecondOrderS} (diagrams (2.1) and (2.2)). Further, there are $\mathcal{N}_3 = 4+5+1 = 10$ third-order ($4$ without loop, $5$ with a single loop, and $1$ with two loops), and $\mathcal{N}_4 = 27+40+14+1= 82$ fourth-order skeleton self-energy diagrams. The expression in Eq.\,(\protect\ref{eac3a}) [Eq.\,(14) in Ref.\,\protect\citen{MM06}], is to be contrasted with that in Eq.\,(11) of Ref.\,\citen{LGM05},\footnote{See Ref.\,\protect\citen{PH07}.} which concerns the number of \textsl{proper}, or one-particle irreducible (1PI), self-energy diagrams, including skeleton and non-skeleton ones, at each order of the perturbation theory. According to this expression, at for instance the $4$th order there are $\mathscr{N}_4' = 189$ proper self-energy diagrams.\footnote{We have included the program (in the programming language of Mathematica \protect\cite{SW16}) that determines the 1PI self-energy diagrams (and their numbers, for any given order $\nu$) without tadpole insertion in version-2 (v2) of Ref.\,\protect\citen{BF19}, \S\,C.1. The new (sub-) programs included in the latter section are named \texttt{Snu1PI}, \texttt{S1PI}, \texttt{Snux1PI}, \texttt{Snu1PInoTP}, \texttt{CountL}, and \texttt{Tadpole}.}  We note that the considerations in Ref.\,\citen{LGM05} are based on the assumption that the `non-interacting' Green function takes account of the \textsl{exact} Hartree self-energy $\Sigma^{\textsc{h}}$,\footnote{For a two-body potential in category (1), specified above (p.\,\protect\pageref{ThreeCases}), the Hartree self-energy $\mathbb{\Sigma}^{\textsc{h}}$ is a full $(2\mathsf{s}+1)\times (2\mathsf{s}+1)$ matrix, even when $\mathbb{G}$ is diagonal (use of $\Sigma^{\textsc{h}}$ in the main text is for simplicity and uniformity of notation). Whether or not $\mathbb{G}$ is diagonal, $\mathbb{\Sigma}^{\textsc{h}}$ is diagonal for two-body potentials in categories (2) and (3); for two-body potentials in category (2) the diagonal elements $\Sigma_{\sigma}^{\textsc{h}}$ and $\Sigma_{\sigma'}^{\textsc{h}}$ differ in general for $\sigma \not=\sigma'$, however for those in category (3) $\Sigma_{\sigma}^{\textsc{h}}=\Sigma_{\sigma'}^{\textsc{h}}$, $\forall\sigma,\sigma'$. Further, for two-body potentials in categories (2) and (3) $\mathbb{\Sigma}^{\textsc{h}}$ depends only on the diagonal elements of $\mathbb{G}$, irrespective of whether or not $\mathbb{G}$ is diagonal. More explicitly, for potentials in category (3) $\mathbb{\Sigma}^{\textsc{h}}$ depends only on the diagonal trace of $\mathbb{G}$. At zero temperature the diagonal element $G_{\sigma} \equiv G_{\sigma,\sigma}$ of $\mathbb{G}$, as encountered in the expressions for the elements of the matrix $\mathbb{\Sigma}^{\textsc{h}}$, is equal to $\protect\ii n_{\sigma}$, where $n_{\sigma}(\bm{r})$ is the number density of the particles with spin index $\sigma$ in the $N$-particle GS of the system under consideration. It follows that (whether or not $\mathbb{G}$ is diagonal) for two-body potentials in category (2) $\Sigma_{\sigma}^{\textsc{h}}$ can in principle be calculated exactly within the framework of the spin-density functional theory (sDFT) of the Hohenberg-Kohn-sham density-functional formalism \protect\cite{KS65,HK64,ED11} and for two-body potentials in category (3) the same can in principle be achieved within the framework of the density functional theory (DFT) of the same formalism. For some relevant details, see Refs\,\protect\citen{BF97,BF97b}, and \protect\citen{BF99}.} whereby the diagrams containing the tadpole insertions are discarded. In the absence of the exact  $\Sigma^{\textsc{h}}$, the 1PI diagrams containing tadpole insertions cannot be unconditionally discarded.\footnote{In the absence of the exact $\Sigma^{\textsc{h}}$, the difference between the exact and the approximate Hartree self-energy is to be taken into account as a \textsl{local} perturbation along the lines of Refs\,\protect\citen{BF97} and \protect\citen{BF97b}. In these references the relevant local perturbation is related to the local exchange-correlation potential. In Ref.\,\protect\citen{BF97b} also account is taken of a \textsl{non-local} perturbation associated with the exchange-correlation vector potential.} The number of 1PI self-energy diagrams, including those with tadpole insertions, are considerably larger than those without these. Denoting the total number of $\nu$th-order 1PI self-energy diagrams by $\mathscr{N}_{\nu}$, one has $\mathscr{N}_1 = 2$, $\mathscr{N}_2 = 6$, $\mathscr{N}_3 = 42$, $\mathscr{N}_4 = 414$, \emph{etc.},\footnote{We have included the programs (in the programming language of Mathematica \protect\cite{SW16}) that calculate these numbers in version-2 (v2) of Ref.\,\protect\citen{BF19}, \S\,C.1.} to be contrasted with $\mathscr{N}_1' = 1$, $\mathscr{N}_2' = 3$,\footnote{See Fig.\,\protect\ref{f7}, p.\,\protect\pageref{SecondOrderS}, below.} $\mathscr{N}_3' = 20$, $\mathscr{N}_4' = 189$, \emph{etc.}

Instead of drawing, we specify the relevant self-energy diagrams in a notation that is both concise and useful for symbolic computation, \S\,\ref{sx1}.\footnote{With reference to the considerations in Ref.\,\protect\citen{BF19}, provision of the elements of the symmetric group $S_{2\nu}$ (or their [lexicographic] ranks) associated with the $\nu$th-order self-energy diagrams, along with the integers $r$ and $s$ associated with the external vertices, would suffice. As we have indicated at the outset (footnote \raisebox{-1.0ex}{\normalsize{\protect\footref{notes}}} on p.\,\protect\pageref{WorkOnThe}), we have not attempted to rewrite the original text of the present publication, which dates from early May 2015, in the light of the developments reported in Ref.\,\protect\citen{BF19}. To make contact with the latter publication, where appropriate we shall however in the following present, in footnotes, the relevant elements of $S_{4}$, $S_{6}$, and $S_{8}$, along with the pertinent values for the integers $r$ and $s$.}  For the specific case of the Hubbard Hamiltonian for spin-$\tfrac{1}{2}$ particles, both the entire set and a proper subset of skeleton self-energy diagrams suffice to calculate the self-energy, each of the two sets corresponding to a different representation of this Hamiltonian.\footnote{The two sets of diagrams correspond to the perturbation series expansions in powers of respectively the $\protect\h{\mathcal{H}}_1$ in Eq.\,(\protect\ref{ex01bx}) and the $\hspace{0.28cm}\protect\h{\hspace{-0.28cm}\mathpzc{H}}_1$ in Eq.\,(\protect\ref{ex01f}); note the $\sum_{\sigma,\sigma'}$ ($\sum_{\sigma}$) in the former (latter) equation. The diagrams corresponding to the perturbation series expansion in Eq.\,(2.106) for the one-particle Green function in Eq.\,(2.87) in terms of the $N_{\nu}$ and $D_{\nu}$ in Eqs\,(2.88) and (2.89) of Ref.\,\protect\citen{BF19} are in powers of $\protect\h{\mathsf{H}}_1$, Eq.\,(\protect\ref{ex01k}). The simplified expressions for the $N_{\nu}$ and $D_{\nu}$ in Eqs\,(2.96) and (2.97) of Ref.\,\protect\citen{BF19} (see also appendix D herein) generate the terms associated with the diagrams corresponding to the perturbation series expansion of the one-particle Green function in powers of  $\hspace{0.28cm}\protect\h{\hspace{-0.28cm}\mathpzc{H}}_1$.} In \S\,\ref{sd2} we present and discuss these equivalent representations of the Hubbard Hamiltonian for spin-$\tfrac{1}{2}$ particles that clearly are \textsl{not} equivalent insofar as the perturbation series expansion of in particular the self-energy is concerned. The extant literature is mostly silent on the issues that we discuss in \S\,\ref{sd2}, and in the cases where they are not, the explanations are incidental and sometimes open to different interpretations.

In this appendix we further present and describe a short but powerful program,\footnote{Program \texttt{Equiv}, p.\,\protect\pageref{Equiv}.} written in the Mathematica programming language \cite{SW16},\footnote{$\copyright$ 2021 \textsf{All methods, algorithms and programs presented in this appendix, as well as elsewhere in this publication, are intellectual property of the author. Any commercial use of these without his written permission is strictly prohibited. All academic and non-commercial uses of the codes in this publication, or modifications thereof, must be appropriately cited. The same restrictions apply to the contents of the Mathematica$^{\protect\X{\circledR}}$ notebook that we publish alongside this publication.}} that on the basis of a \textsl{symbolic} computation determines whether or not two proper self-energy diagrams, which may or may not be skeleton, are equivalent, \S\,\ref{sd3}. Here by \textsl{equivalence} of two proper self-energy diagrams of a given order $\nu$ we refer to the equality of the algebraic contributions of these diagrams up to a determinate constant multiplying factor of either sign. For two equivalent diagrams, the value of this constant is fully determined by the number of the particle loops in the relevant diagrams and a signature $\varsigma \in \{-1,+1\}$, which the above-mentioned program returns as output. The partitioning of a set of proper self-energy diagrams of a given order into non-overlapping sets of equivalent diagrams (in the sense just described) will in general substantially reduce the computational overhead in the perturbational many-body calculations.

The title of this appendix is borrowed from the classic work by Veltman \protect\cite{MV94}.

\refstepcounter{dummyX}
\subsection{Specifying the self-energy diagrams}
\begin{figure}[t!]
\centerline{\includegraphics*[angle=0, width=0.67\textwidth]{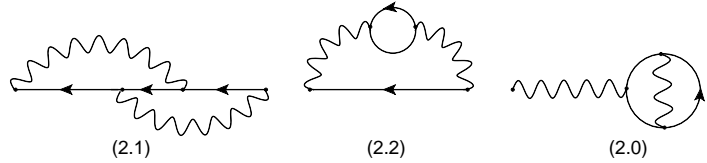}}
\caption{(2.1) and (2.2): \protect\refstepcounter{dummy}\label{SecondOrderS}Second-order skeleton self-energy diagrams. These diagrams are the only ones contributing to the self-energy in the second order when the directed solid lines represent either the interacting Green function or the non-interacting Green function that takes explicit account of the first-order Hartree-Fock self-energy $\Sigma^{\textsc{hf}} \equiv \Sigma^{\textsc{h}} + \Sigma^{\textsc{f}}$. In the event of the directed solid line representing the non-interacting Green function that takes explicit account of the first-order Hartree self-energy $\Sigma^{\textsc{h}}$ (represented by the tadpole diagram), in second order the non-skeleton but \textsl{proper} self-energy diagram (2.0) is also to be taken into account. The solid wavy lines represent the two-body interaction potential. With reference to \S\,\protect\ref{sd2}, in the case of the Hubbard Hamiltonian for spin-$\tfrac{1}{2}$ particles there are two alternative ways of interpreting the Feynman diagrams, according to one of which the contribution of diagram (2.1) is equal to minus one-half of the contribution of diagram (2.2), and according to the other the diagram (2.1) is discarded (on account of it containing at least one interaction line whose end-points are located on a continuous set of Green-function lines) and diagram (2.2) takes half its conventional value (see footnote \raisebox{-1.0ex}{\normalsize{\protect\footref{notet}}} on p.\,\pageref{WithReference}).}
\label{f7}
\end{figure}

\phantomsection
\label{sx1}
We consider a Hamiltonian for particles with spin, interacting through an isotropic two-body potential \cite{FW03,NO98,LW60}. A proper (or 1PI) self-energy diagram of order $\nu$ corresponding to this Hamiltonian consists of $2\nu$ vertices (of which $2\nu-2$ internal and $2$ external), $\nu$ wavy lines representing the interaction potential, and $2\nu-1$ directed solid lines, each representing a one-particle Green function.\refstepcounter{dummy}\label{ForSpin}\footnote{As we have indicated at the outset of this appendix, for spin-$\mathsf{s}$ particles each directed solid line generally represents an element $G_{\sigma,\sigma'}$ of the $(2\mathsf{s} + 1) \times (2\mathsf{s}+1)$ Green matrix $\mathbb{G}$ \protect\cite{FW03,BF19}; at times such line may also represent an element $G_{\protect\X{0};\sigma,\sigma'}$ of the non-interacting Green matrix $\mathbb{G}_{\protect\X{0}}$. The rules regarding the transcription of the self-energy diagrams into mathematical expressions specify the way in which the \textsl{internal} spin indices are to be summed over, effecting the multiplication of the underlying Green matrices in a specific order \protect\cite{FW03}. Later in this appendix we explicitly assume that $\mathbb{G}$ (and $\mathbb{G}_{\protect\X{0}}$) is diagonal (see \S\,2.2.1 in Ref.\,\protect\citen{BF19}). Under this assumption, for the two-body potentials in categories (2) and (3), specified above -- p.\,\protect\pageref{ThreeCases}, the self-energy matrix $\mathbb{\Sigma}$ is also diagonal. For $\mathbb{G}$ ($\mathbb{G}_{\protect\X{0}}$) and $\mathbb{\Sigma}$ diagonal, determination of the contributions of the self-energy diagrams is simplified. The prescriptions in \S\,\protect\ref{sd21} below, which concern calculation of the perturbational contributions to the diagonal element $\Sigma_{\sigma}$, are the outcomes of having carried out the above-mentioned multiplication of the relevant Green matrices in appropriate order. \label{notef1}} Each \textsl{internal} vertex of a self-energy diagram is attached to (i) one end of an interaction line, (ii) one ingoing and (iii) one outgoing Green-function line.\footnote{An \textsl{ingoing} (\textsl{outgoing}) Green-function line points towards (away from) the relevant vertex (Fig.\,\protect\ref{f4}, p.\,\protect\pageref{TwoInternal}, below).} Barring the Hartree diagram and the diagrams associated with it,\footnote{To be explicit, in the cases where the Green-function lines in the proper self-energy diagrams under consideration represent the elements of the exact Green matrix $\mathbb{G}$, there is one such diagram, describing the \textsl{exact} Hartree self-energy $\mathbb{\Sigma}^{\textsc{h}}$ (see \S\,\protect\ref{sec.3a.1}). In the cases where the mentioned lines represent the elements of a non-interacting Green matrix $\mathbb{G}_{\protect\X{0}}$, there are infinite number of such diagrams, each arising from the perturbation series expansion of the $\mathbb{G}$ in the diagram describing the exact $\mathbb{\Sigma}^{\textsc{h}}$ in terms of $\mathbb{G}_{\protect\X{0}}$. The $2$nd-order non-skeleton self-energy diagram $(2.0)$ in Fig.\,\protect\ref{f7}, p.\,\pageref{SecondOrderS}, is one such diagram.} which have one external vertex and which we do \textsl{not} explicitly consider in this appendix, the two \textsl{external} vertices of a proper self-energy diagram are each attached to (i$'$) one end of an interaction line, and to (ii$''$) either an \textsl{internal} ingoing or an \textsl{internal} outgoing Green-function line. In some proper self-energy diagrams the latter ingoing and outgoing internal Green-function lines are the same line, as in diagram $(2.2)$ in Fig.\,\ref{f7}, p.\,\pageref{SecondOrderS}, below.

We\refstepcounter{dummy}\label{WeMarkThe} mark the $2\nu$ vertices of a $\nu$th-order self-energy diagram by the numbers $1'$, $2'$, \dots, $(2\nu)'$. Barring the Hartree and the associated self-energy diagrams (see above), we invariably reserve the number $1'$ ($2'$) for the external vertex attached to an \textsl{internal} outgoing (ingoing) Green-function line.\footnote{See Figs\, \protect\ref{f15} and \protect\ref{f16}, pp.\,\protect\pageref{DiagramA1} and \protect\pageref{DiagramA2}, below.}\footnote{We use primed integers for marking the vertices of diagrams in order to retain unprimed integers for use as subscripts of the elements of the set $\{x_i \| i=1,2,\dots, 3\nu-1\}$.} Thus, \emph{unless we indicate otherwise, the self-energies in the main body of this appendix are perturbational contributions to $\Sigma(2',1')$.}\refstepcounter{dummy}\label{UsingTheConvention}\footnote{Using the convention as employed in appendices B and C of Ref.\,\protect\citen{BF19} (see for instance Eq.\,(B.20) and Figs\,2 and 3 herein), here $1'$ ($2'$) would take the role of $r$ ($s$). However, unless $1'$ and $2'$ are connected by a line representing the two-body interaction potential, $(r,s) = (1',2')$ conflicts with the conventions of Ref.\,\protect\citen{BF19}, where the integers associated with the vertices of this potential are to be contiguous (in the form $(2j-1,2j)$, $j \in\{1,2,\dots,\nu\}$; \emph{cf.} Eq.\,(2.88) in Ref.\,\protect\citen{BF19}, where one encounters $\mathsf{v}(1,2) \dots \mathsf{v}(2\nu-1,2\nu)$). To bypass this conflict, in presenting the relevant $2\nu$-permutations below, we will have to rely on a different numbering scheme for the vertices than that adopted in the \textsl{main body} of this appendix. Consequently, we shall have to specify also the relevant pairs $(r,s)$. As in appendix C of Ref.\,\protect\citen{BF19}, $(r,s)$ can be one of the three pairs $(1',1')$, $(1',2')$, and $(1',3')$. Since in this appendix $\Sigma^{\textsc{h}}$ and $\Sigma^{\textsc{f}}$ will not be explicitly dealt with, in the \textsl{footnotes} only the pair $(r, s) = (1', 3')$ will be encountered. This is to be contrasted with the pair $(r, s) = (1', 2')$ \textsl{throughout} the \textsl{main body} of this appendix. \label{notek}} In self-energy diagrams, as in other Feynman diagrams, the Green-function lines are invariably directed, in contrast to the interaction lines. In evaluating the contributions of these diagrams in the energy-momentum space, the interaction lines are also to be directed, representing the direction of the energy-momentum transferred by the interaction potential. Use of directed interaction lines\,\footnote{As in Figs\, \protect\ref{f15} and \protect\ref{f16}, pp.\,\protect\pageref{DiagramA1} and \protect\pageref{DiagramA2}, below.} is necessary even in the cases where the two-body interaction potentials are local both in time and space.\footnote{As regards non-locality in time, this is the case when the self-energy is expanded in terms of the dynamically screened interaction potential, described in appendix \protect\ref{sa}.} For this reason, in representing the self-energy diagrams with the aim of their quantitative treatment -- symbolically, algebraically or numerically, we label the totality of $(2\nu-1) + \nu = 3\nu -1$ lines in a $\nu$th-order proper self-energy diagram by the \textsl{signed} labels $x_1$, $x_2$, \dots, $x_{3\nu-1}$, with the set $\{x_1,x_2,\dots,x_{2\nu-1}\}$ reserved for the Green-function lines, and the set $\{x_{2\nu},x_{2\nu+1},\dots,x_{3\nu-1}\}$ for the interaction lines; the order in which the elements of the latter two sets of signed labels are assigned to the relevant lines is immaterial. By \textsl{singed} label we mean that $x_j$, $\forall j$, corresponds to the direction assigned to the line with which it is associated; thus, $-x_j$ corresponds to the direction opposite to that assigned to the latter line.

In\refstepcounter{dummy}\label{InTheFollowing} the following, we denote the directed Green-function line connecting the vertices $i'$ and $j'$ in this direction, by $\langle i', j'\rangle$ (to be distinguished from $\langle j',i'\rangle$),\footnote{The solid line $\langle i', j'\rangle$ thus represents $G(j',i')$. Clearly, $G(j',i') \not\equiv G(i',j')$.} and the interaction line connecting the vertices $k'$ and $l'$, by either $(k',l')$ or $(l',k')$; we denote the interaction line \textsl{directed} in the direction of $k'$ to $l'$ by $\prec\! k',l'\!\succ$ (to be distinguished from $\prec\! l',k'\!\succ$). In view of the above-indicated convention with regard to the vertices $1'$ and $2'$, they will \textsl{only} appear as $\langle 1',j'\rangle$ and $\langle i',2'\rangle$. In other words, $1'$ will only be the \textsl{left} index in $\langle i',j'\rangle$, and $2'$ the \textsl{right} index. It will prove convenient to impose the same restrictions on $1'$ and $2'$ in representing \textsl{directed} interaction lines. That is, $1'$ and $2'$ will \textsl{by convention} only appear as $\prec\! 1',l'\!\succ$ and $\prec\! k',2'\!\succ$. \emph{Unless we indicate otherwise, this convention will apply in the remaining part of this appendix.}\footnote{See \emph{e.g.} Eqs\,(\protect\ref{eac3}), (\protect\ref{eac4}) and Eqs\,(\protect\ref{ed15g}), (\protect\ref{ed15h}), and the associated Figs\,\protect\ref{f15} and \protect\ref{f16}, pp.\,\protect\pageref{DiagramA1} and \protect\pageref{DiagramA2}, below.} This convention does \textsl{not} strictly apply to undirected interaction lines, even though in the following we adhere to this convention also in denoting undirected interaction lines; \textsl{strictly}, the interaction line connecting the vertices $k'$ and $l'$ can be arbitrarily denoted by either $(k',l')$ or $(l',k')$.

\refstepcounter{dummyX}
\subsubsection{The 2nd-order skeleton self-energy diagrams}
\phantomsection
\label{sd11}
As we have indicated earlier, at $2$nd-order there are two skeleton self-energy diagrams, which we depict in Fig.\,\ref{f7}, marked by (2.1) and (2.2). Below we specify these in the notation introduced above, \S\,\ref{sx1}.

\emph{The diagram with no loop} --- Defining the diagrammatic component\,\footnote{Throughout, the subscripts of the symbols denoting the main diagrammatic components (the root diagrams) signify the number of the particle loops, and the superscripts the order of the associated self-energy diagrams.}
\begin{equation}\label{eac7a}
A_0^2 \doteq \langle 1',3'\rangle,\; \langle 3',4'\rangle,\; \langle 4',2'\rangle,
\end{equation}
one has\,\footnote{With reference to the remark in footnote \raisebox{-1.0ex}{\normalsize{\protect\footref{notek}}} on page \protect\pageref{UsingTheConvention}, with $(r,s) = (1', 3')$, in the array notation the $4$-permutation associated with $\Sigma_{\sigma}^{\protect\X{(2.1)}}(3',1')$ reads \protect\cite{BF19} $(3', 4', 2', 1')$, summarily denoted by $\Sigma_{\sigma}^{\protect\X{(2.1)}}(3',1'): (3', 4', 2', 1')$, and more concisely by $\Sigma_{\sigma}^{\protect\X{(2.1)}}: (3', 4', 2', 1')$. Note that the self-energy in Eq.\,(\protect\ref{eac7b}) is $\Sigma^{\protect\X{(2.1)}}(2',1')$.}
\begin{equation}\label{eac7b}
\;\Sigma_{\sigma}^{\X{(2.1)}}:\;\; A_0^2,\; (1',4'),\; (3',2').
\end{equation}

\emph{The diagram with one loop} --- Defining the diagrammatic component
\begin{equation}\label{eac7c}
A_1^2 \doteq \langle 1',2'\rangle,\; \langle 3',4'\rangle,\; \langle 4',3'\rangle,
\end{equation}
one has\,\footnote{$\Sigma_{\sigma}^{\protect\X{(2.2)}}: (3', 4', 1', 2')$. Recall that this contribution, as well as those in the following \textsl{footnotes}, correspond to $\Sigma(3',1')$; those in the \textsl{main body} of this appendix correspond to $\Sigma(2',1')$.}
\begin{equation}\label{eac7d}
\;\Sigma_{\sigma}^{\X{(2.2)}}:\;\; A_1^2,\; (1',4'),\; (3',2').
\end{equation}
Note that, in the above equations $1'$ ($2'$) invariably appears as the first (second) index.

For illustration and later reference, furnishing the two diagrams considered above with the earlier-mentioned \textsl{signed} indices $\{x_1, x_2, \dots, x_5\}$, one has (see Fig.\,\ref{f15}, p.\,\pageref{DiagramA1}, below):
\begin{equation}\label{eac3}
\Sigma_{\sigma}^{\X{(2.1)}}:\;\; x_1 = \langle 1',3'\rangle,\; x_2 = \langle 3',4'\rangle,\; x_3 = \langle 4',2'\rangle,\; x_4 = \prec\! 1',4'\!\succ,\; x_5 = \prec\! 3',2'\!\succ,
\end{equation}
\begin{equation}\label{eac4}
\Sigma_{\sigma}^{\X{(2.2)}}:\;\; x_1 = \langle 1',2'\rangle,\; x_2 = \langle 3',4'\rangle,\; x_3 = \langle 4',3'\rangle,\; x_4 = \prec\! 1',4'\!\succ,\; x_5 = \prec\! 3',2'\!\succ.
\end{equation}
The notation as in Eqs\,(\ref{eac3}) and (\ref{eac4}) is especially suited to symbolic computations on diagrams, to be considered in \S\,\ref{sd3} below.

\refstepcounter{dummyX}
\subsubsection{The 3rd-order skeleton self-energy diagrams}
\phantomsection
\label{sd12}
At third order there are $10$ skeleton self-energy diagrams, of which $4$ without loop, $5$ with a single loop, and $1$ with two loops, Eq.\,(\ref{eac3a}), which we specify below.

\emph{The diagrams with no loop} --- The $4$ $3$rd-order skeleton self-energy diagrams containing no loop, to be presented below, consist of the following diagrammatic component:
\begin{equation}\label{eac8}
A_0^3 \doteq  \langle 1',3'\rangle,\; \langle 3',4'\rangle,\; \langle 4',5'\rangle,\; \langle 5',6'\rangle,\; \langle 6',2'\rangle.
\end{equation}
One has\,\footnote{$\Sigma_{\sigma}^{\protect\X{(3.1)}}: (3', 6', 5', 2', 4', 1')$, $\Sigma_{\sigma}^{\protect\X{(3.2)}}: (3', 4', 5', 6', 2', 1')$, $\Sigma_{\sigma}^{\protect\X{(3.3)}}: (3', 6', 5', 1', 2', 4')$, $\Sigma_{\sigma}^{\protect\X{(3.4)}}: (3', 5', 2', 6', 4', 1')$.}
\begin{equation}\label{eac8a}
\;\Sigma_{\sigma}^{\X{(3.1)}}:\;\; A_0^3,\; (1',4'),\; (3',6'),\; (5',2'),
\end{equation}
\begin{equation}\label{eac8b}
\;\Sigma_{\sigma}^{\X{(3.2)}}:\;\; A_0^3,\; (1',5'),\; (3',6'),\; (4',2'),
\end{equation}
\begin{equation}\label{eac8c}
\;\Sigma_{\sigma}^{\X{(3.3)}}:\;\; A_0^3,\; (1',5'),\; (4',6'),\; (3',2'),
\end{equation}
\begin{equation}\label{eac8d}
\;\Sigma_{\sigma}^{\X{(3.4)}}:\;\; A_0^3,\; (1',6'),\; (3',5'),\; (4',2').
\end{equation}

\emph{The diagrams with one loop} --- The $5$ $3$rd-order skeleton self-energy diagrams containing a single loop, to be presented below, consist of the following diagrammatic components:
\begin{align}\label{eac9}
A_1^3 &\doteq  \langle 1',2'\rangle,\; \langle 3',4'\rangle,\; \langle 4',5'\rangle,\; \langle 5',6'\rangle,\; \langle 6',3'\rangle,\nonumber \\
B_1^3 &\doteq  \langle 1',3'\rangle,\; \langle 3',2'\rangle,\; \langle 4',5'\rangle,\; \langle 5',6'\rangle,\; \langle 6',4'\rangle,\nonumber\\
C_1^3 &\doteq \langle 1',3'\rangle,\; \langle 3',4'\rangle,\; \langle 4',2'\rangle,\; \langle 5',6'\rangle,\; \langle 6',5'\rangle.
\end{align}
One has\,\footnote{$\Sigma^{\protect\X{(3,5)}} : (3', 5', 1', 6', 4', 2')$, $\Sigma^{\protect\X{(3,6)}}: (3', 6', 5', 2', 1', 4')$, $\Sigma^{\protect\X{(3,7)}}: (3', 4', 5', 6', 1', 2')$, $\Sigma^{\protect\X{(3,8)}}: (3', 5', 2', 6', 1', 4')$, $\Sigma^{\protect\X{(3,9)}}: (3', 6', 5', 1', 4', 2')$.}
\begin{equation}\label{eac9a}
\;\Sigma_{\sigma}^{\X{(3.5)}}:\;\; A_1^3,\; (1',5'),\; (4',6'),\; (3',2'),
\end{equation}
\begin{equation}\label{eac9b}
\;\Sigma_{\sigma}^{\X{(3.6)}}:\;\; B_1^3,\; (1',6'),\; (3',5'),\; (4',2'),
\end{equation}
\begin{equation}\label{eac9c}
\;\Sigma_{\sigma}^{\X{(3.7)}}:\;\; B_1^3,\; (1',4'),\; (3',5'),\; (6',2'),
\end{equation}
\begin{equation}\label{eac9d}
\;\Sigma_{\sigma}^{\X{(3.8)}}:\;\; C_1^3,\; (1',4'),\; (3',6'),\; (5',2'),
\end{equation}
\begin{equation}\label{eac9e}
\;\Sigma_{\sigma}^{\X{(3.9)}}:\;\; C_1^3,\; (1',6'),\; (4',5'),\; (3',2').
\end{equation}

\emph{The diagram with two loops} --- Defining
\begin{equation}\label{eac10}
A_2^3 \doteq \langle 1',2'\rangle,\; \langle 3',4'\rangle,\; \langle 4',3'\rangle,\; \langle 5',6'\rangle,\; \langle 6',5'\rangle,
\end{equation}
one has\,\footnote{$\Sigma_{\sigma}^{\protect\X{(3.10)}}: (3', 6', 1', 5', 4', 2')$.}
\begin{equation}\label{eac10a}
\;\Sigma_{\sigma}^{\X{(3.10)}}:\;\; A_2^3,\; (1',6'),\; (4',5'),\; (3',2').
\end{equation}

\refstepcounter{dummyX}
\subsubsection{The 4th-order skeleton self-energy diagrams}
\phantomsection
\label{sd13}
At fourth order, there are $82$ skeleton self-energy diagrams, consisting of $27$ diagrams without loop, $40$ with a single loop, $14$ with two loops, and $1$ with three loops, Eq.\,(\ref{eac3a}), which we specify below.

\emph{The diagrams with no loop} --- The $27$ $4$th-order skeleton self-energy diagrams that contain no loop, to be presented below, have the following diagrammatic component in common:
\begin{equation}\label{e60}
A_0^4 \doteq \langle 1',3'\rangle,\;  \langle 3',4'\rangle,\; \langle 4',5'\rangle,\;  \langle 5',6' \rangle,\;  \langle 6',7'\rangle,\; \langle 7',8'\rangle,\; \langle 8',2'\rangle.
\end{equation}
One has\,\footnote{$\Sigma_{\sigma}^{\protect\X{(4.1)}}: (3', 7', 5', 8', 2', 4', 6', 1')$, $\Sigma_{\sigma}^{\protect\X{(4.2)}}: (3', 7', 5', 2', 4', 8', 6', 1')$, $\Sigma_{\sigma}^{\protect\X{(4.3)}}: (3', 5', 2', 8', 7', 4', 6', 1')$, $\Sigma_{\sigma}^{\protect\X{(4.4)}}: (3', 6', 5', 8', 2', 7', 4', 1')$, $\Sigma_{\sigma}^{\protect\X{(4.5)}}: (3', 8', 5', 2', 7', 4', 6', 1')$, $\Sigma_{\sigma}^{\protect\X{(4.6)}}: (3', 8', 5', 6', 7', 1', 2', 4')$, $\Sigma_{\sigma}^{\protect\X{(4.7)}}: (3', 4', 5', 7', 2', 8', 6', 1')$, $\Sigma_{\sigma}^{\protect\X{(4.8)}}: (3', 7', 5', 6', 2', 8', 4', 1')$, $\Sigma_{\sigma}^{\protect\X{(4.9)}}: (3', 6', 5', 8', 7', 4', 2', 1')$, $\Sigma_{\sigma}^{\protect\X{(4.10)}}: (3', 8', 5', 1', 7', 2', 6', 4')$, $\Sigma_{\sigma}^{\protect\X{(4.11)}}: (3', 5', 2', 7', 4', 8', 6', 1')$, $\Sigma_{\sigma}^{\protect\X{(4.12)}}: (3', 5', 2', 8', 7', 1', 4', 6')$, $\Sigma_{\sigma}^{\protect\X{(4.13)}}: (3', 8', 5', 1', 7', 4', 2', 6')$, $\Sigma_{\sigma}^{\protect\X{(4.14)}}: (3', 8', 5', 7', 4', 2', 6', 1')$, $\Sigma_{\sigma}^{\protect\X{(4.15)}}: (3', 8', 5', 2', 7', 1', 4', 6')$, $\Sigma_{\sigma}^{\protect\X{(4.16)}}: (3', 6', 5', 2', 7', 8', 4', 1')$, $\Sigma_{\sigma}^{\protect\X{(4.17)}}: (3', 4', 5', 8', 7', 1', 2', 6')$, $\Sigma_{\sigma}^{\protect\X{(4.18)}}: (3', 4', 5', 6', 7', 8', 2', 1')$, $\Sigma_{\sigma}^{\protect\X{(4.19)}}: (3', 7', 5', 8', 2', 1', 4', 6')$, $\Sigma_{\sigma}^{\protect\X{(4.20)}}: (3', 8', 5', 6', 7', 2', 4', 1')$, $\Sigma_{\sigma}^{\protect\X{(4.21)}}: (3', 6', 5', 8', 7', 1', 4', 2')$, $\Sigma_{\sigma}^{\protect\X{(4.22)}}: (3', 6', 5', 7', 4', 8', 2', 1')$, $\Sigma_{\sigma}^{\protect\X{(4.23)}}: (3', 8', 5', 7', 4', 1', 2', 6')$, $\Sigma_{\sigma}^{\protect\X{(4.24)}}: (3', 7', 5', 1', 2', 8', 6', 4')$, $\Sigma_{\sigma}^{\protect\X{(4.25)}}: (3', 5', 2', 6', 7', 8', 4', 1')$, $\Sigma_{\sigma}^{\protect\X{(4.26)}}: (3', 4', 5', 8', 7', 2', 6', 1')$, $\Sigma_{\sigma}^{\protect\X{(4.27)}}: (3', 6', 5', 1', 7', 8', 2', 4')$.}
\begin{equation}\label{e60a}
\;\Sigma_{\sigma}^{\X{(4.1)}}:\;\; A_0^4,\; (1',7'),\; (3',6'),\; (5',8'),\; (4',2'),
\end{equation}
\begin{equation}\label{e60b}
\;\Sigma_{\sigma}^{\X{(4.2)}}:\;\; A_0,\; (1',6'),\; (3',5'),\; (4',8'),\; (7',2'),
\end{equation}
\begin{equation}\label{e60e}
\;\Sigma_{\sigma}^{\X{(4.3)}}:\;\; A_0^4,\; (1',8'),\; (3',6'),\; (5',7'),\; (4',2'),
\end{equation}
\begin{equation}\label{e60d}
\;\Sigma_{\sigma}^{\X{(4.4)}}:\;\; A_0^4,\; (1',7'),\; (3',5'),\; (6',8'),\; (4',2'),
\end{equation}
\begin{equation}\label{e60c}
\;\Sigma_{\sigma}^{\X{(4.5)}}:\;\; A_0^4,\; (1',4'),\; (3',7'),\; (6',8'),\; (5',2'),
\end{equation}
\begin{equation}\label{e60f}
\;\Sigma_{\sigma}^{\X{(4.6)}}:\;\; A_0^4,\; (1',6'),\; (3',8'),\; (5',7'),\; (4',2'),
\end{equation}
\begin{equation}\label{e60g}
\;\Sigma_{\sigma}^{\X{(4.7)}}:\;\; A_0^4,\; (1',7'),\; (3',5'),\; (4',8'),\; (6',2'),
\end{equation}
\begin{equation}\label{e60h}
\;\Sigma_{\sigma}^{\X{(4.8)}}:\;\; A_0^4,\; (1',7'),\; (3',6'),\; (4',8'),\; (5',2'),
\end{equation}
\begin{equation}\label{e60i}
\;\Sigma_{\sigma}^{\X{(4.9)}}:\;\; A_0^4,\; (1',6'),\; (3',7'),\; (5',8'),\; (4',2'),
\end{equation}
\begin{equation}\label{e60j}
\Sigma_{\sigma}^{\X{(4.10)}}:\;\; A_0^4,\; (1',5'),\; (4',7'),\; (6',8'),\; (3',2'),
\end{equation}
\begin{equation}\label{e60k}
\Sigma_{\sigma}^{\X{(4.11)}}:\;\; A_0^4,\; (1',8'),\; (3',5'),\; (4',7'),\; (6',2'),
\end{equation}
\begin{equation}\label{e60l}
\Sigma_{\sigma}^{\X{(4.12)}}:\;\; A_0^4,\; (1',8'),\; (3',7'),\; (4',6'),\; (5',2'),
\end{equation}
\begin{equation}\label{e60m}
\Sigma_{\sigma}^{\X{(4.13)}}:\;\; A_0^4,\; (1',6'),\; (4',8'),\; (5',7'),\; (3',2'),
\end{equation}
\begin{equation}\label{e60n}
\Sigma_{\sigma}^{\X{(4.14)}}:\;\; A_0^4,\; (1',4'),\; (3',6'),\; (5',8'),\; (7',2'),
\end{equation}
\begin{equation}\label{e60o}
\Sigma_{\sigma}^{\X{(4.15)}}:\;\; A_0^4,\; (1',5'),\; (3',8'),\; (4',7'),\; (6',2'),
\end{equation}
\begin{equation}\label{e60p}
\Sigma_{\sigma}^{\X{(4.16)}}:\;\; A_0^4,\; (1',5'),\; (3',7'),\; (4',8'),\; (6',2'),
\end{equation}
\begin{equation}\label{e60q}
\Sigma_{\sigma}^{\X{(4.17)}}:\;\; A_0^4,\; (1',6'),\; (3',8'),\; (4',7'),\; (5',2'),
\end{equation}
\begin{equation}\label{e60r}
\Sigma_{\sigma}^{\X{(4.18)}}:\;\; A_0^4,\; (1',6'),\; (3',7'),\; (4',8'),\; (5',2'),
\end{equation}
\begin{equation}\label{e60s}
\Sigma_{\sigma}^{\X{(4.19)}}:\;\; A_0^4,\; (1',7'),\; (3',8'),\; (4',6'),\; (5',2'),
\end{equation}
\begin{equation}\label{e60t}
\Sigma_{\sigma}^{\X{(4.20)}}:\;\; A_0^4,\; (1',4'),\; (3',7'),\; (5',8'),\; (6',2'),
\end{equation}
\begin{equation}\label{e60u}
\Sigma_{\sigma}^{\X{(4.21)}}:\;\; A_0^4,\; (1',4'),\; (3',8'),\; (5',7'),\; (6',2'),
\end{equation}
\begin{equation}\label{e60v}
\Sigma_{\sigma}^{\X{(4.22)}}:\;\; A_0^4,\; (1',5'),\; (3',6'),\; (4',8'),\; (7',2'),
\end{equation}
\begin{equation}\label{e60w}
\Sigma_{\sigma}^{\X{(4.23)}}:\;\; A_0^4,\; (1',5'),\; (3',8'),\; (4',6'),\; (7',2'),
\end{equation}
\begin{equation}\label{e60x}
\Sigma_{\sigma}^{\X{(4.24)}}:\;\; A_0^4,\; (1',7'),\; (4',6'),\; (5',8'),\; (3',2'),
\end{equation}
\begin{equation}\label{e60y}
\Sigma_{\sigma}^{\X{(4.25)}}:\;\; A_0^4,\; (1',8'),\; (3',6'),\; (4',7'),\; (5',2'),
\end{equation}
\begin{equation}\label{e60z}
\Sigma_{\sigma}^{\X{(4.26)}}:\;\; A_0^4,\; (1',5'),\; (3',7'),\; (6',8'),\; (4',2'),
\end{equation}
\begin{equation}\label{e60za}
\Sigma_{\sigma}^{\X{(4.27)}}:\;\; A_0^4,\; (1',6'),\; (4',7'),\; (5',8'),\; (3',2').
\end{equation}

\emph{The diagrams with one loop} --- The $40$ $4$th-order skeleton self-energy diagrams containing one loop, to be presented below, have the following diagrammatic components in common:
\begin{align}\label{e90}
&\hspace{0.0cm}A_1^4 \doteq \langle 1',2'\rangle,\; \langle 3',4'\rangle,\; \langle 4',5'\rangle,\; \langle 5',6'\rangle,\;\langle 6',7'\rangle,\; \langle 7',8'\rangle,\; \langle 8',3'\rangle,\;\nonumber \\
&\hspace{0.0cm}B_1^4 \doteq \langle 1',3'\rangle,\; \langle 3',2'\rangle,\; \langle 4',5'\rangle,\; \langle 5',6'\rangle,\;\langle 6',7'\rangle,\; \langle 7',8'\rangle,\; \langle 8',4'\rangle, \nonumber \\
&\hspace{0.0cm}C_1^4 \doteq \langle 1',3'\rangle,\; \langle 3',4'\rangle,\; \langle 4',2'\rangle,\; \langle 5',6'\rangle,\; \langle 6',7'\rangle,\; \langle 7',8'\rangle,\; \langle 8',5'\rangle,\nonumber \\
&\hspace{0.0cm}D_1^4 \doteq \langle 1',3'\rangle,\; \langle 3',4'\rangle,\; \langle 4',5'\rangle,\; \langle 5',2'\rangle,\; \langle 6',7'\rangle,\; \langle 7',8'\rangle,\; \langle 8',6'\rangle, \nonumber\\
&\hspace{0.0cm}E_1^4 \doteq \langle 1',3'\rangle,\; \langle 3',4'\rangle,\; \langle 4',5'\rangle,\; \langle 5',6'\rangle,\; \langle 6',2'\rangle,\; \langle 7',8'\rangle,\; \langle 8',7'\rangle.
\end{align}
One has\,\footnote{$\Sigma_{\sigma}^{\protect\X{(4.28)}}: (3', 7', 1', 6', 4', 8', 5', 2')$, $\Sigma_{\sigma}^{\protect\X{(4.29)}}: (3', 8', 1', 7', 4', 2', 6', 5')$, $\Sigma_{\sigma}^{\protect\X{(4.30)}}: (3', 7', 1', 5', 8', 2', 4', 6')$, $\Sigma_{\sigma}^{\protect\X{(4.31)}}: (3', 6', 1', 8', 4', 7', 5', 2')$, $\Sigma_{\sigma}^{\protect\X{(4.32)}}: (3', 6', 7', 2', 4', 8', 1', 5')$, $\Sigma_{\sigma}^{\protect\X{(4.33)}}: (3', 4', 7', 5', 8', 2', 1', 6')$, $\Sigma_{\sigma}^{\protect\X{(4.34)}}: (3', 8', 7', 6', 4', 2', 1', 5')$, $\Sigma_{\sigma}^{\protect\X{(4.35)}}: (3', 7', 5', 8', 1', 4', 6', 2')$, $\Sigma_{\sigma}^{\protect\X{(4.36)}}: (3', 6', 7', 5', 8', 4', 1', 2')$, $\Sigma_{\sigma}^{\protect\X{(4.37)}}: (3', 8', 5', 6', 1', 7', 2', 4')$, $\Sigma_{\sigma}^{\protect\X{(4.38)}}: (3', 8', 5', 2', 7', 4', 1', 6')$, $\Sigma_{\sigma}^{\protect\X{(4.39)}}: (3', 4', 5', 6', 7', 8', 1', 2')$, $\Sigma_{\sigma}^{\protect\X{(4.40)}}: (3', 8', 5', 6', 7', 2', 1', 4')$, $\Sigma_{\sigma}^{\protect\X{(4.41)}}: (3', 6', 5', 8', 7', 4', 1', 2')$, $\Sigma_{\sigma}^{\protect\X{(4.42)}}: (3', 6', 5', 2', 7', 8', 1', 4')$, $\Sigma_{\sigma}^{\protect\X{(4.43)}}: (3', 4', 5', 8', 7', 2', 1', 6')$, $\Sigma_{\sigma}^{\protect\X{(4.44)}}: (3', 7', 2', 6', 4', 8', 1', 5')$, $\Sigma_{\sigma}^{\protect\X{(4.45)}}: (3', 7', 5', 1', 4', 8', 6', 2')$, $\Sigma_{\sigma}^{\protect\X{(4.46)}}: (3', 5', 2', 8', 7', 4', 1', 6')$, $\Sigma_{\sigma}^{\protect\X{(4.47)}}: (3', 5', 2', 6', 7', 8', 1', 4')$, $\Sigma_{\sigma}^{\protect\X{(4.48)}}: (3', 7', 5', 8', 2', 4', 1', 6')$, $\Sigma_{\sigma}^{\protect\X{(4.49)}}: (3', 8', 5', 1', 7', 2', 4', 6')$, $\Sigma_{\sigma}^{\protect\X{(4.50)}}: (3', 7', 5', 6', 2', 8', 1', 4')$, $\Sigma_{\sigma}^{\protect\X{(4.51)}}: (3', 8', 5', 2', 7', 1', 6', 4')$, $\Sigma_{\sigma}^{\protect\X{(4.52)}}: (3', 4', 5', 8', 7', 1', 6', 2')$, $\Sigma_{\sigma}^{\protect\X{(4.53)}}: (3', 6', 5', 1', 7', 8', 4', 2')$, $\Sigma_{\sigma}^{\protect\X{(4.54)}}: (3', 8', 5', 7', 4', 2', 1', 6')$, $\Sigma_{\sigma}^{\protect\X{(4.55)}}: (3', 6', 5', 7', 4', 8', 1', 2')$, $\Sigma_{\sigma}^{\protect\X{(4.56)}}: (3', 7', 5', 2', 4', 8', 1', 6')$, $\Sigma_{\sigma}^{\protect\X{(4.57)}}: (3', 4', 5', 7', 2', 8', 1', 6')$, $\Sigma_{\sigma}^{\protect\X{(4.58)}}: (3', 5', 2', 7', 4', 8', 1', 6')$, $\Sigma_{\sigma}^{\protect\X{(4.59)}}: (3', 7', 5', 1', 2', 8', 4', 6')$, $\Sigma_{\sigma}^{\protect\X{(4.60)}}: (3', 5', 2', 8', 7', 1', 6', 4')$, $\Sigma_{\sigma}^{\protect\X{(4.61)}}: (3', 8', 5', 1', 7', 4', 6', 2')$, $\Sigma_{\sigma}^{\protect\X{(4.62)}}: (3', 7', 5', 8', 2', 1', 6', 4')$, $\Sigma_{\sigma}^{\protect\X{(4.63)}}: (3', 8', 5', 6', 7', 1', 4', 2')$, $\Sigma_{\sigma}^{\protect\X{(4.64)}}: (3', 6', 5', 8', 7', 1', 2', 4')$, $\Sigma_{\sigma}^{\protect\X{(4.65)}}: (3', 8', 5', 7', 4', 1', 6', 2')$, $\Sigma_{\sigma}^{\protect\X{(4.66)}}: (3', 6', 5', 8', 2', 7', 1', 4')$, $\Sigma_{\sigma}^{\protect\X{(4.67)}}: (3', 6', 5', 8', 7', 2', 4', 1')$.}
\begin{equation}\label{e90a}
\Sigma_{\sigma}^{\X{(4.28)}}:\; A_1^4,\; (1',6'),\; (4',8'),\; (5',7'),\; (3',2'),
\end{equation}
\begin{equation}\label{e90b}
\Sigma_{\sigma}^{\X{(4.29)}}:\; A_1^4,\; (1',6'),\; (4',7'),\; (5',8'),\; (3',2'),
\end{equation}
\begin{equation}\label{e90c}
\Sigma_{\sigma}^{\X{(4.30)}}:\; A_1^4,\; (1',5'),\; (4',7'),\; (6',8'),\; (3',2'),
\end{equation}
\begin{equation}\label{e90d}
\Sigma_{\sigma}^{\X{(4.31)}}:\; A_1^4,\; (1',7'),\; (4',6'),\; (5',8'),\; (3',2'),
\end{equation}
\begin{equation}\label{e90e}
\Sigma_{\sigma}^{\X{(4.32)}}:\; B_1^4,\; (1',8'),\; (3',6'),\; (5',7'),\; (4',2'),
\end{equation}
\begin{equation}\label{e90f}
\Sigma_{\sigma}^{\X{(4.33)}}:\; B_1^4,\; (1',4'),\; (3',6'),\; (5',7'),\; (8',2'),
\end{equation}
\begin{equation}\label{e90g}
\Sigma_{\sigma}^{\X{(4.34)}}:\; B_1^4,\; (1',7'),\; (3',6'),\; (5',8'),\; (4',2'),
\end{equation}
\begin{equation}\label{e90h}
\Sigma_{\sigma}^{\X{(4.35)}}:\; B_1^4,\; (1',7'),\; (3',5'),\; (6',8'),\; (4',2'),
\end{equation}
\begin{equation}\label{e90i}
\Sigma_{\sigma}^{\X{(4.36)}}:\; B_1^4,\; (1',4'),\; (3',5'),\; (6',8'),\; (7',2'),
\end{equation}
\begin{equation}\label{e90j}
\Sigma_{\sigma}^{\X{(4.37)}}:\; B_1^4,\; (1',4'),\; (3',6'),\; (5',8'),\; (7',2'),
\end{equation}
\begin{equation}\label{e91i}
\Sigma_{\sigma}^{\X{(4.38)}}:\; C_1^4,\; (1',8'),\; (3',7'),\; (4',6'),\; (5',2'),
\end{equation}
\begin{equation}\label{e91j}
\Sigma_{\sigma}^{\X{(4.39)}}:\; C_1^4,\; (1',5'),\; (3',6'),\; (4',7'),\; (8',2'),
\end{equation}
\begin{equation}\label{e91k}
\Sigma_{\sigma}^{\X{(4.40)}}:\; C_1^4,\; (1',7'),\; (3',6'),\; (4',8'),\; (5',2'),
\end{equation}
\begin{equation}\label{e91l}
\Sigma_{\sigma}^{\X{(4.41)}}:\; C_1^4,\; (1',7'),\; (3',8'),\; (4',6'),\; (5',2'),
\end{equation}
\begin{equation}\label{e91m}
\Sigma_{\sigma}^{\X{(4.42)}}:\; C_1^4,\; (1',8'),\; (3',6'),\; (4',7'),\; (5',2'),
\end{equation}
\begin{equation}\label{e91n}
\Sigma_{\sigma}^{\X{(4.43)}}:\; C_1^4,\; (1',5'),\; (3',7'),\; (4',6'),\; (8',2'),
\end{equation}
\begin{equation}\label{e90k}
\Sigma_{\sigma}^{\X{(4.44)}}:\; C_1^4,\; (1',4'),\; (3',7'),\; (6',8'),\; (5',2'),
\end{equation}
\begin{equation}\label{e90l}
\Sigma_{\sigma}^{\X{(4.45)}}:\; C_1^4,\; (1',7'),\; (4',5'),\; (6',8'),\; (3',2'),\;
\end{equation}
\begin{equation}\label{e90m}
\Sigma_{\sigma}^{\X{(4.46)}}:\; D_1^4,\; (1',5'),\; (3',8'),\; (4',7'),\; (6',2'),
\end{equation}
\begin{equation}\label{e90n}
\Sigma_{\sigma}^{\X{(4.47)}}:\; D_1^4,\; (1',5'),\; (3',6'),\; (4',7'),\; (8',2'),
\end{equation}
\begin{equation}\label{e90o}
\Sigma_{\sigma}^{\X{(4.48)}}:\; D_1^4,\; (1',4'),\; (3',8'),\; (5',7'),\; (6',2'),
\end{equation}
\begin{equation}\label{e90p}
\Sigma_{\sigma}^{\X{(4.49)}}:\; D_1^4,\; (1',8'),\; (4',7'),\; (5',6'),\; (3',2'),
\end{equation}
\begin{equation}\label{e90q}
\Sigma_{\sigma}^{\X{(4.50)}}:\; D_1^4,\; (1',4'),\; (3',6'),\; (5',7'),\; (8',2'),
\end{equation}
\begin{equation}\label{e90r}
\Sigma_{\sigma}^{\X{(4.51)}}:\; D_1^4,\; (1',8'),\; (3',5'),\; (4',7'),\; (6',2'),
\end{equation}
\begin{equation}\label{e90s}
\Sigma_{\sigma}^{\X{(4.52)}}:\; D_1^4,\; (1',6'),\; (3',5'),\; (4',7'),\; (8',2'),
\end{equation}
\begin{equation}\label{e90t}
\Sigma_{\sigma}^{\X{(4.53)}}:\; D_1^4,\; (1',6'),\; (4',7'),\; (5',8'),\; (3',2'),
\end{equation}
\begin{equation}\label{e90u}
\Sigma_{\sigma}^{\X{(4.54)}}:\; D_1^4,\; (1',8'),\; (3',7'),\; (5',6'),\; (4',2'),
\end{equation}
\begin{equation}\label{e90v}
\Sigma_{\sigma}^{\X{(4.55)}}:\; D_1^4,\; (1',6'),\; (3',7'),\; (5',8'),\; (4',2'),
\end{equation}
\begin{equation}\label{e90w}
\Sigma_{\sigma}^{\X{(4.56)}}:\; E_1^4,\; (1',4'),\; (3',8'),\; (6',7'),\; (5',2'),
\end{equation}
\begin{equation}\label{e90x}
\Sigma_{\sigma}^{\X{(4.57)}}:\; E_1^4,\; (1',5'),\; (3',8'),\; (6',7'),\; (4',2'),
\end{equation}
\begin{equation}\label{e90y}
\Sigma_{\sigma}^{\X{(4.58)}}:\; E_1^4,\; (1',6'),\; (3',8'),\; (5',7'),\; (4',2'),
\end{equation}
\begin{equation}\label{e90z}
\Sigma_{\sigma}^{\X{(4.59)}}:\; E_1^4,\; (1',5'),\; (4',8'),\; (6',7'),\; (3',2'),
\end{equation}
\begin{equation}\label{e91a}
\Sigma_{\sigma}^{\X{(4.60)}}:\; E_1^4,\; (1',6'),\; (3',5'),\; (4',8'),\; (7',2'),
\end{equation}
\begin{equation}\label{e91b}
\Sigma_{\sigma}^{\X{(4.61)}}:\; E_1^4,\; (1',8'),\; (4',6'),\; (5',7'),\; (3',2'),
\end{equation}
\begin{equation}\label{e91c}
\Sigma_{\sigma}^{\X{(4.62)}}:\; E_1^4,\; (1',5'),\; (3',6'),\; (4',8'),\; (7',2'),
\end{equation}
\begin{equation}\label{e91d}
\Sigma_{\sigma}^{\X{(4.63)}}:\; E_1^4,\; (1',8'),\; (3',6'),\; (5',7'),\; (4',2'),
\end{equation}
\begin{equation}\label{e91e}
\Sigma_{\sigma}^{\X{(4.64)}}:\; E_1^4,\; (1',4'),\; (3',6'),\; (5',8'),\; (7',2'),
\end{equation}
\begin{equation}\label{e91f}
\Sigma_{\sigma}^{\X{(4.65)}}:\; E_1^4,\; (1',8'),\; (3',6'),\; (4',7'),\; (5',2'),
\end{equation}
\begin{equation}\label{e91g}
\Sigma_{\sigma}^{\X{(4.66)}}:\; E_1^4,\; (1',5'),\; (3',8'),\; (4',6'),\; (7',2'),
\end{equation}
\begin{equation}\label{e91h}
\Sigma_{\sigma}^{\X{(4.67)}}:\; E_1^4,\; (1',8'),\; (3',5'),\; (6',7'),\; (4',2').
\end{equation}

\emph{The diagrams with two loops} --- The $14$ $4$th-order skeleton self-energy diagrams containing two loops, have the following diagrammatic components in common:
\begin{align}\label{e92}
&\hspace{0.0cm}A_2^4 \doteq \langle 1',2'\rangle,\;  \langle 3',4'\rangle,\; \langle 4',3'\rangle,\; \langle 5',6'\rangle,\;  \langle 6',7'\rangle,\;  \langle 7',8'\rangle,\; \langle 8',5'\rangle, \nonumber\\
&\hspace{0.0cm}B_2^4 \doteq \langle 1',2'\rangle,\;  \langle 3',4'\rangle,\; \langle 4',5'\rangle,\; \langle 5',3'\rangle,\;  \langle 6',7'\rangle,\;  \langle 7',8'\rangle,\; \langle 8',6'\rangle, \nonumber\\
&\hspace{0.0cm}C_2^4 \doteq \langle 1',3'\rangle,\;  \langle 3',2'\rangle,\; \langle 4',5'\rangle,\; \langle 5',6'\rangle,\;  \langle 6',4'\rangle,\;  \langle 7',8'\rangle,\; \langle 8',7'\rangle,\nonumber \\
&\hspace{0.0cm}D_2^4 \doteq \langle 1',3'\rangle,\;  \langle 3',4'\rangle,\; \langle 4',2'\rangle,\; \langle 5',6'\rangle,\;  \langle 6',5'\rangle,\;  \langle 7',8'\rangle,\; \langle 8',7'\rangle.
\end{align}
One has\,\footnote{$\Sigma_{\sigma}^{\protect\X{(4.68)}}: (3', 7', 1', 5', 4', 8', 6', 2')$, $\Sigma_{\sigma}^{\protect\X{(4.69)}}: (3', 8', 1', 6', 4', 7', 5', 2')$, $\Sigma_{\sigma}^{\protect\X{(4.70)}}: (3', 5', 1', 8', 4', 7', 6', 2')$, $\Sigma_{\sigma}^{\protect\X{(4.71)}}: (3', 6', 1', 7', 4', 8', 5', 2')$, $\Sigma_{\sigma}^{\protect\X{(4.72)}}: (3', 8', 1', 7', 4', 2', 5', 6')$, $\Sigma_{\sigma}^{\protect\X{(4.73)}}: (3', 6', 7', 5', 8', 2', 1', 4')$, $\Sigma_{\sigma}^{\protect\X{(4.74)}}: (3', 6', 7', 8', 4', 2', 1', 5')$, $\Sigma_{\sigma}^{\protect\X{(4.75)}}: (3', 7', 5', 8', 1', 2', 6', 4')$, $\Sigma_{\sigma}^{\protect\X{(4.76)}}: (3', 6', 5', 8', 1', 7', 2', 4')$, $\Sigma_{\sigma}^{\protect\X{(4.77)}}: (3', 8', 5', 2', 1', 7', 6', 4')$, $\Sigma_{\sigma}^{\protect\X{(4.78)}}: (3', 4', 5', 8', 1', 7', 6', 2')$, $\Sigma_{\sigma}^{\protect\X{(4.79)}}: (3', 6', 5', 8', 7', 2', 1', 4')$, $\Sigma_{\sigma}^{\protect\X{(4.80)}}: (3', 7', 2', 5', 4', 8', 1', 6')$, $\Sigma_{\sigma}^{\protect\X{(4.81)}}: (3', 7', 5', 1', 4', 8', 2', 6')$.}
\begin{equation}\label{e92a}
\Sigma_{\sigma}^{\X{(4.68)}}:\; A_2^4,\; (1',7'),\; (4',5'),\; (6',8'),\; (3',2'),
\end{equation}
\begin{equation}\label{e92b}
\Sigma_{\sigma}^{\X{(4.69)}}:\; A_2^4,\; (1',4'),\; (3',7'),\; (6',8'),\; (5',2'),
\end{equation}
\begin{equation}\label{e92c}
\Sigma_{\sigma}^{\X{(4.70)}}:\; A_2^4,\; (1',7'),\; (3',6'),\; (4',8'),\; (5',2'),
\end{equation}
\begin{equation}\label{e92d}
\Sigma_{\sigma}^{\X{(4.71)}}:\; B_2^4,\; (1',8'),\; (4',7'),\; (5',6'),\; (3',2'),
\end{equation}
\begin{equation}\label{e92e}
\Sigma_{\sigma}^{\X{(4.72)}}:\; B_2^4,\; (1',8'),\; (4',6'),\; (5',7'),\; (3',2'),
\end{equation}
\begin{equation}\label{e92f}
\Sigma_{\sigma}^{\X{(4.73)}}:\; C_2^4,\; (1',8'),\; (3',5'),\; (6',7'),\; (4',2'),
\end{equation}
\begin{equation}\label{e92g}
\Sigma_{\sigma}^{\X{(4.74)}}:\; C_2^4,\; (1',8'),\; (3',4'),\; (6',7'),\; (5',2'),
\end{equation}
\begin{equation}\label{e92h}
\Sigma_{\sigma}^{\X{(4.75)}}:\; C_2^4,\; (1',4'),\; (3',5'),\; (6',8'),\; (7',2'),
\end{equation}
\begin{equation}\label{e92i}
\Sigma_{\sigma}^{\X{(4.76)}}:\; C_2^4,\; (1',5'),\; (3',4'),\; (6',8'),\; (7',2'),
\end{equation}
\begin{equation}\label{e92j}
\Sigma_{\sigma}^{\X{(4.77)}}:\; C_2^4,\; (1',4'),\; (3',8'),\; (6',7'),\; (5',2'),
\end{equation}
\begin{equation}\label{e92k}
\Sigma_{\sigma}^{\X{(4.78)}}:\; C_2^4,\; (1',5'),\; (3',8'),\; (6',7'),\; (7',2'),
\end{equation}
\begin{equation}\label{e92l}
\Sigma_{\sigma}^{\X{(4.79)}}:\; D_2^4,\; (1',8'),\; (3',6'),\; (4',7'),\; (5',2'),
\end{equation}
\begin{equation}\label{e92m}
\Sigma_{\sigma}^{\X{(4.80)}}:\; D_2^4,\; (1',4'),\; (3',8'),\; (6',7'),\; (5',2'),
\end{equation}
\begin{equation}\label{e92n}
\Sigma_{\sigma}^{\X{(4.81)}}:\; D_2^4,\; (1',8'),\; (4',5'),\; (6',7'),\; (3',2').
\end{equation}

\emph{The diagram with three loops} --- Defining
\begin{equation}\label{e93}
A_3^4  \doteq \langle 1',2'\rangle,\;  \langle 3',4'\rangle,\; \langle 4',3'\rangle,\; \langle 5',6'\rangle,\;  \langle 6',5'\rangle,\;  \langle 7',8'\rangle,\; \langle 8',7'\rangle,
\end{equation}
one has\,\footnote{$\Sigma_{\sigma}^{\protect\X{(4.82)}}: (3', 8', 1', 5', 4', 7', 6', 2')$.}
\begin{equation}\label{e93a}
\Sigma_{\sigma}^{\X{(4.82)}}:\; A_3^4,\; (1',8'),\; (4',5'),\; (6',7'),\; (3',2').
\end{equation}

\refstepcounter{dummyX}
\subsection{On the self-energy diagrams in the energy-momentum representation}
\phantomsection
\label{sd2}
Here we consider the mathematical expressions associated with proper self-energy diagrams corresponding to uniform GSs or ESs of single-band Hamiltonians in general and of the single-band Hubbard Hamiltonian in particular. As we have indicated earlier in this appendix, the extant relevant literature is not explicit about some of the aspects to be discussed in this section and insofar as there are some remarks regarding these, they are incidental and sometimes liable to different interpretations. The discrepancy arises from the neglected fact that \emph{in general different sets of self-energy diagrams correspond to equivalent representations of a given Hamiltonian,}\footnote{The same applies to diagrams corresponding to any other correlation function.} \emph{and in particular when the sets of the self-energy diagrams are identical, the functions associated with individual diagrams crucially depend on the specifics of the underlying representation of the Hamiltonian.} In general, these distinctions are not relevant in exact and therefore often formal treatments, where account is formally taken of \textsl{all} diagrams, however they are very relevant in approximate treatments in practical applications, where only restricted sets of diagrams, albeit even to arbitrary high order, are taken into account.

\emph{Unless we indicate otherwise, in this section and the remaining part of this appendix (as in the body of this publication) we assume the Green matrices $\mathbb{G}_{\X{0}}$ and $\mathbb{G}$ and the self-energy matrix $\mathbb{\Sigma}$ to be diagonal. Further, the self-energy diagrams that we consider correspond to the diagonal elements $\{\Sigma_{\sigma}(\bm{k};\varepsilon)\| \sigma\}$ of $\mathbb{\Sigma}$ expressed in terms of the diagonal elements $\{G_{\X{0};\sigma}(\bm{k};\varepsilon) \| \sigma\}$ or $\{G_{\sigma}(\bm{k};\varepsilon) \| \sigma\}$.}

\begin{figure}[t!]
\centerline{
\includegraphics*[angle=0, width=0.4\textwidth]{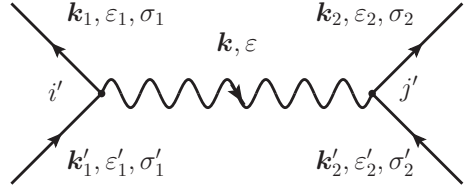}}
\caption{Two \protect\refstepcounter{dummy}\label{TwoInternal}internal vertices $i'$ and $j'$ in a Feynman diagram, associated with the Hamiltonian $\h{\mathscr{H}}$ in Eq.\,(\protect\ref{ex01a}), connected by the directed interaction line $\prec\hspace{-1.4pt} i',j'\hspace{-1.6pt}\succ$ representing the two-body potential $\b{\mathsf{v}}_{\sigma_1,\sigma_1';\sigma_2,\sigma_2'}(\|\bm{k}\|)$. By the conservation of energy-momentum, at vertex $i'$ ($j'$) one has $\varepsilon_1 = \varepsilon_1' - \varepsilon$ and $\bm{k}_1 =\bm{k}_1' - \bm{k}$ ($\varepsilon_2 = \varepsilon_2' + \varepsilon$ and $\bm{k}_2 = \bm{k}_2' + \bm{k}$). In this appendix we consider the following cases: (i) $\b{\mathsf{v}}_{\sigma_1,\sigma_1';\sigma_2,\sigma_2'}(\|\bm{k}\|) \equiv \b{v}(\|\bm{k}\|)\hspace{0.6pt} \delta_{\sigma_1,\sigma_1'}\hspace{0.4pt} \delta_{\sigma_2,\sigma_2'}$, Eqs\,(\protect\ref{ex01b}) and (\protect\ref{ex03}); (ii) $\b{\mathsf{v}}_{\sigma_1,\sigma_1';\sigma_2,\sigma_2'}(\|\bm{k}\|) \equiv \b{v}_{\sigma_1,\sigma_2}(\|\bm{k}\|)\hspace{0.6pt} \delta_{\sigma_1,\sigma_1'}\hspace{0.4pt} \delta_{\sigma_2,\sigma_2'}$, Eq.\,(\protect\ref{ex01h}) (\emph{cf.} Eqs\,(\protect\ref{ea34k}) and (\protect\ref{ea34l})), where in particular $\b{v}_{\sigma_1,\sigma_2}(\|\bm{k}\|) \propto (1-\delta_{\sigma_1,\sigma_2})$, Eq.\,(\protect\ref{ex01j}). On suppressing the energies $\varepsilon$, $\varepsilon_j, \varepsilon_j'$, $j=1,2$, and identifying $\bm{k}$ with $-\bm{q}$, $(\bm{k}_1,\sigma_1)$ with $(\bm{k}+\bm{q},\sigma)$, $(\bm{k}_2,\sigma_2)$ with $(\bm{p}-\bm{q}, \sigma'')$, $(\bm{k}_2',\sigma_2')$ with $(\bm{p},\sigma''')$, and $(\bm{k}_1',\sigma_1')$ with $(\bm{k},\sigma')$, this diagram can also be viewed as representing the interaction part of the Hamiltonian $\h{\mathscr{H}}$ in Eq.\,(\protect\ref{ex01a}) by identifying the directed straight lines marked by $(\bm{k}_1,\sigma_1)$, $(\bm{k}_2,\sigma_2)$, $(\bm{k}_2',\sigma_2')$, and $(\bm{k}_1',\sigma_1')$ with respectively $\h{a}_{\bm{k}+\bm{q};\sigma}^{\protect\X{\dag}}$, $\h{a}_{\bm{p}-\bm{q};\sigma''}^{\protect\X{\dag}}$, $\h{a}_{\bm{p};\sigma'''}^{\phantom{\protect\X{\dag}}}$, and $\h{a}_{\bm{k};\sigma'}^{\protect\phantom{\protect\X{\dag}}}$. Note that the left-to-right ordering of the operators in the interaction part of $\h{\mathscr{H}}$ corresponds to the clockwise ordering of the directed lines, beginning from the line marked by $(\bm{k}_1,\sigma_1) = (\bm{k}+\bm{q},\sigma)$. These observations are instructive, providing an immediate insight into the fact that in the diagrammatic expansions, for instance for the proper self-energy, corresponding to the Hubbard Hamiltonian $\hspace{0.28cm}\protect\h{\hspace{-0.28cm}\mathpzc{H}}$ in Eq.\,(\protect\ref{ex01f}) below (which is specific to spin-$\tfrac{1}{2}$ particles) the spin indices associated with the Green-function lines passing through the two ends of an interaction line are \textsl{invariably} each other's complements, $\sigma$ and $\protect\b{\sigma}$.}
\label{f4}
\end{figure}

We begin with the single-band Hamiltonian (see Fig.\,\ref{f4})
\begin{equation}\label{ex01a}
\h{\mathscr{H}} = \sum_{\bm{k},\sigma} \varepsilon_{\bm{k}}\,
\h{a}_{\bm{k};\sigma}^{\dag} \h{a}_{\bm{k};\sigma}^{\phantom{\dag}} + \frac{1}{2 \Omega}
\sum_{\substack{\sigma,\sigma'\\ \sigma'',\sigma'''}} \sum_{\bm{k}, \bm{p}, \bm{q}} \b{\mathsf{v}}_{\sigma,\sigma';\sigma'',\sigma'''}(\|\bm{q}\|)\,
\h{a}_{\bm{k}+\bm{q};\sigma}^{\dag} \h{a}_{\bm{p}-\bm{q};\sigma''}^{\dag}
\h{a}_{\bm{p};\sigma'''}^{\phantom{\dag}} \h{a}_{\bm{k};\sigma'}^{\phantom{\dag}},
\end{equation}
described in terms of the non-interacting energy dispersion $\varepsilon_{\bm{k}}$ and the canonical creation and annihilation operators, respectively $\h{a}_{\bm{k};\sigma}^{\X{\dag}}$ and $\h{a}_{\bm{k};\sigma}^{\phantom{\X{\dag}}}$, in the Schr\"{o}dinger picture (\emph{cf.} Eq.\,(\ref{eg3c})). In Eq.\,(\ref{ex01a}), $\Omega$ stands for the volume of the system and the function $\b{\mathsf{v}}_{\sigma,\sigma';\sigma'',\sigma'''}(\|\bm{q}\|)$ for the Fourier transform of the isotropic two-body interaction potential $\mathsf{v}_{\sigma,\sigma';\sigma'',\sigma'''}(\|\bm{r}-\bm{r}'\|)$, or $\mathsf{v}_{\sigma,\sigma';\sigma'',\sigma'''}(\|\bm{R}_i-\bm{R}_j\|)$, depending on whether the system described by $\h{\mathscr{H}}$ is defined over $\mathds{R}^d$ or a lattice $\{\bm{R}_i\| i\}$ embedded in $\mathds{R}^d$. As we have indicated at the outset of this publication (p.\,\pageref{AssumingUniformGSs}), for simplicity we assume $\{\bm{R}_i\| i\}$ to be a Bravais lattice \cite{AM76}. In the case of the system under consideration being defined over $\mathds{R}^d$ ($\{\bm{R}_i\| i =1,2,\dots, N_{\textsc{s}}\}$), the summations with respect to $\bm{k}$, $\bm{p}$ and $\bm{q}$ in Eq.\,(\ref{ex01a}) are over the entire discrete set of wave vectors in $\mathds{R}^d$ (the first Brillouin zone \cite{AM76}, $\1BZ$, associated with $\{\bm{R}_i\| i\}$) that conform with the box boundary condition. For lattice models, $\h{a}_{\bm{k};\sigma}^{\phantom{\dag}}$ (and its Hermitian conjugate $\h{a}_{\bm{k};\sigma}^{\dag}$) is periodic, with the $\1BZ$ the fundamental region of periodicity. This periodicity is explicitly accounted for by the Fourier representation
\begin{equation}\label{ex02}
\h{a}_{\bm{k};\sigma} = \frac{1}{\sqrt{N_{\textsc{s}}\phantom{|}\!}} \sum_{i = 1}^{N_{\textsc{s}}} \e^{+\ii\bm{k}\cdot \bm{R}_i} \h{c}_{i;\sigma},
\end{equation}
which is the inverse of the representation in Eq.\,(\ref{eg3a}). The relevant periodicity can be enforced by identifying, for instance, $\h{a}_{\bm{k}+\bm{q};\sigma}^{\X{\dag}}$, $\bm{k}, \bm{q} \in \1BZ$, with $\h{a}_{\bm{k}+\bm{q}+\bm{K}_{0};\sigma}^{\X{\dag}}$, where the reciprocal-lattice vector $\bm{K}_{\X{0}}$ is to ensure that $\bm{k}+\bm{q}+\bm{K}_{0} \in \1BZ$.

We now focus on systems described in terms of spin-independent two-body interaction potentials, for which one has [Eq.\,(9.17), p.\,104, in Ref.\,\citen{FW03}]
\begin{equation}\label{ex01b}
\b{\mathsf{v}}_{\sigma,\sigma';\sigma'',\sigma'''}(\|\bm{q}\|) \equiv \b{v}(\|\bm{q}\|)\hspace{0.6pt} \delta_{\sigma,\sigma'}\hspace{0.6pt} \delta_{\sigma'',\sigma'''},
\end{equation}
transforming the Hamiltonian $\h{\mathscr{H}}$ in Eq.\,(\ref{ex01a}) into the following simpler Hamiltonian (\emph{cf.} Eq.\,(A1) in Ref.\,\citen{BF13}):
\begin{equation}\label{ex01}
\h{H} = \sum_{\bm{k},\sigma} \varepsilon_{\bm{k}}\,
\h{a}_{\bm{k};\sigma}^{\dag} \h{a}_{\bm{k};\sigma}^{\phantom{\dag}} + \frac{1}{2 \Omega}
\sum_{\sigma,\sigma'} \sum_{\bm{k}, \bm{p}, \bm{q}} \b{v}(\|\bm{q}\|)\,
\h{a}_{\bm{k}+\bm{q};\sigma}^{\dag} \h{a}_{\bm{p}-\bm{q};\sigma'}^{\dag}
\h{a}_{\bm{p};\sigma'}^{\phantom{\dag}} \h{a}_{\bm{k};\sigma}^{\phantom{\dag}},
\end{equation}
With $\b{v}(\|\bm{q}\|)$ having the dimensionality of energy times volume (in the SI base units, $\llbracket \b{v}\rrbracket = \mathrm{Jm}^d$),\footnote{This on account of $\protect\b{v}(\|\bm{q}\|)  \doteq \int_{\protect\b{\Omega}} \mathrm{d}^dr\, \protect\e^{-\protect\ii \bm{q}\cdot \bm{r}} v(\|\bm{r}\|)$ (\emph{cf.} Eq.\,(\protect\ref{ea39c})), where $\protect\b{\Omega}\subseteq \mathds{R}^d$ denotes the region of volume $\Omega$ occupied by the system under consideration, and the fact that $\llbracket v(\|\bm{r}\|)\rrbracket = \mathrm{J}$.} the Hubbard Hamiltonian $\h{\mathcal{H}}$ described in terms of the on-site interaction energy $U$ is obtained from the Hamiltonian $\h{H}$ in Eq.\,(\ref{ex01}) by means of the substitution \cite{BF04,BF13}
\begin{equation}\label{ex03}
\b{v}(\|\bm{q}\|) \rightharpoonup \frac{\Omega U}{N_{\textsc{s}}}.
\end{equation}
Explicitly, one has (\emph{cf.} Eq.\,(1.1) in Ref.\,\citen{BF13})
\begin{equation}\label{ex01bx}
\h{\mathcal{H}} = \sum_{\bm{k},\sigma} \varepsilon_{\bm{k}}\,
\h{a}_{\bm{k};\sigma}^{\dag} \h{a}_{\bm{k};\sigma}^{\phantom{\dag}} + \frac{U}{2 N_{\textsc{s}}}
\sum_{\sigma,\sigma'} \sum_{\bm{k}, \bm{p}, \bm{q}}
\h{a}_{\bm{k}+\bm{q};\sigma}^{\dag} \h{a}_{\bm{p}-\bm{q};\sigma'}^{\dag}
\h{a}_{\bm{p};\sigma'}^{\phantom{\dag}} \h{a}_{\bm{k};\sigma}^{\phantom{\dag}}.
\end{equation}
It follows that \emph{the diagrammatic expansion, and the rules associated with the underlying diagrams, of the self-energy corresponding to the $N$-particle GS (or ESs) of the Hubbard Hamiltonian $\h{\mathcal{H}}$ identically coincide with those of the more general Hamiltonian $\h{H}$ in Eq.\,(\ref{ex01})}.\footnote{Contrast this observation with the following: ``Lastly, there are no Fock contributions because the Hubbard interaction acts between different spin species only.'' [Ref.\,\protect\citen{GJMNN03}, following Eq.\,(3.4)].} Nonetheless, perturbational calculations regarding the Hubbard Hamiltonian for spin-$\tfrac{1}{2}$ particles are greatly simplified by removing a redundancy, to be specified below, from the Hamiltonian in Eq.\,(\ref{ex01bx}) in the case of these particles.\footnote{This redundancy is also removed in \S\,2.2.7 of Ref.\,\protect\citen{BF19} (see also appendix D herein).}

Denoting\refstepcounter{dummy}\label{DenotingTheHubbard} the Hubbard Hamiltonian for spin-$\tfrac{1}{2}$ particles from which the above-mentioned redundancy has been removed by $\hspace{0.28cm}\h{\hspace{-0.28cm}\mathpzc{H}}$, Eq.\,(\ref{ex01f}) below, the set of the self-energy diagrams associated with this Hamiltonian consists of a \textsl{proper} subset of the self-energy diagrams associated with $\h{\mathcal{H}}$, \S\,\ref{sd21} below. To make this statement precise, we need to elaborate on the notions of `set' and `subset', which for the purpose of this section are ambiguous in regard to diagrams with fermion loops.\footnote{A fermion \textsl{loop} is comprised of a contiguous set of fermion lines of matching directions.} This ambiguity is removed by viewing any self-energy diagram associated with $\h{\mathcal{H}}$ and comprised of $\ell$ fermion loops, with $\ell > 0$, as consisting of an explicit sum of $2^{\ell}$ self-energy diagrams (\emph{cf.} Eq.\,(\ref{ex04a}) below),\footnote{Recall that here we are focussing on spin-$\tfrac{1}{2}$ particles.} with each loop in each diagram corresponding to fermions of definite spin index. We recall that in dealing with the self-energy diagrams associated with either $\h{H}$ or $\h{\mathcal{H}}$, the latter sum is conventionally accounted for by declaring evaluation of the sums over the spin indices associated with the fermion loops as being part of the rules whereby contributions of the self-energy diagrams are determined,\footnote{One may think of a summation convention, similar to the Einstein summation convention, that, insofar as diagrammatic calculations are concerned, is to apply to the spin indices associated with fermion loops (this convention is indeed adopted in \emph{e.g.} Ref.\,\protect\citen{FW03}). Suspending this summation convention requires that all relevant summands be explicitly written out. The notion of `subset' as relevant here relies on the understanding that all diagrams corresponding to different summands are explicitly displayed. Aside from the difference of the spin indices associated with different loops, these diagrams outwardly look identical.} \S\,\ref{sd21}. The notions of `set' and `subset' as meant here are based on the understanding that in dealing with the self-energy diagrams associated with $\h{\mathcal{H}}$, all Green-function lines are to have definite spin indices and \textsl{no} summation over the spin indices of the fermion loops is implicit in the specification of the contributions of the self-energy diagrams. Of the $2^{\ell}$ diagrams consisting of $\ell$ loops that thus conventionally are represented by a single self-energy diagram associated with $\h{\mathcal{H}}$, \textsl{at most} only one specific diagram corresponds to $\hspace{0.28cm}\h{\hspace{-0.28cm}\mathpzc{H}}$, \S\,\ref{sd21}. Later we shall specify the unique characteristic of this specific diagram. There are other self-energy diagrams corresponding to $\h{\mathcal{H}}$ that do not correspond to $\hspace{0.28cm}\h{\hspace{-0.28cm}\mathpzc{H}}$, which we shall also specify later in this section.

To remove the above-mentioned redundancy from the Hubbard Hamiltonian $\h{\mathcal{H}}$ corresponding to spin-$\tfrac{1}{2}$ particles, we express the equality in Eq.\,(\ref{ex01bx}) as
\begin{equation}\label{ex01c}
\h{\mathcal{H}} = \h{\mathcal{H}}_{\X{0}} + \h{\mathcal{H}}_{\X{1}}
\end{equation}
and focus on the interaction part $\h{\mathcal{H}}_1$. Making use of the Fourier representation in Eq.\,(\ref{ex02}) and the closure relation
\begin{equation}\label{ex01ca}
\frac{1}{N_{\textsc{s}}} \sum_{\bm{k}\in \fbz} \e^{\ii \bm{k} \cdot (\bm{R}_i - \bm{R}_j)} = \delta_{i,j},
\end{equation}
one readily obtains that
\begin{equation}\label{ex01d}
\h{\mathcal{H}}_{\X{1}} = \frac{U}{2} \sum_{\sigma,\sigma'} \sum_{i=1}^{N_{\textsc{s}}} \h{c}_{i;\sigma}^{\dag} \h{c}_{i;\sigma'}^{\dag} \h{c}_{i;\sigma'}^{\phantom{\dag}} \h{c}_{i;\sigma}^{\phantom{\dag}}.
\end{equation}
For spin-$\tfrac{1}{2}$ particles, the expression in Eq.\,(\ref{ex01d}) can be equivalently written as
\begin{equation}\label{ex01e}
\h{\mathcal{H}}_{\X{1}} =  \frac{U}{2} \sum_{\sigma} \sum_{i=1}^{N_{\textsc{s}}} \h{c}_{i;\sigma}^{\dag} \h{c}_{i;\b{\sigma}}^{\dag} \h{c}_{i;\b{\sigma}}^{\phantom{\dag}} \h{c}_{i;\sigma}^{\phantom{\dag}} + \frac{U}{2} \sum_{\sigma} \sum_{j=1}^{N_{\textsc{s}}} \h{c}_{i;\sigma}^{\dag} \h{c}_{i;\sigma}^{\dag} \h{c}_{i;\sigma}^{\phantom{\dag}} \h{c}_{i;\sigma}^{\phantom{\dag}} \equiv \h{\mathcal{H}}_{\X{1}}' + \h{\mathcal{H}}_{\X{1}}''.
\end{equation}
On account of the canonical anti-commutation relation $[\h{c}_{i;\sigma},\h{c}_{i;\sigma'}]_+ = \h{0}$, one has $\h{c}_{i;\sigma} \h{c}_{i;\sigma} = - \h{c}_{i;\sigma} \h{c}_{i;\sigma}$, or $\h{c}_{i;\sigma} \h{c}_{i;\sigma} = \h{0}$, implying that $\h{\mathcal{H}}_{\X{1}}'' = \h{0}$, and therefore,\footnote{For later reference, following the equality in Eq.\,(\protect\ref{e12b}), one has $\protect\h{\mathcal{H}}_{\protect\X{1}} = U \sum_{j=1}^{N_{\textsc{s}}} \protect\h{n}_{i;\uparrow} \protect\h{n}_{i;\downarrow}$.} following the canonical anti-commutation relation $[\h{c}_{i;\sigma}^{\phantom{\dag}},\h{c}_{i;\sigma'}^{\dag}]_+ = \delta_{\sigma,\sigma'}$ (\emph{cf.} Eqs\,(\ref{e12}) and (\ref{e12a})),
\begin{equation}\label{ex01fy}
\h{\mathcal{H}}_{\X{1}} =  \frac{U}{2}  \sum_{\sigma} \sum_{j=1}^{N_{\textsc{s}}}\h{c}_{i;\sigma}^{\dag} \h{c}_{i;\b{\sigma}}^{\dag} \h{c}_{i;\b{\sigma}}^{\phantom{\dag}} \h{c}_{i;\sigma}^{\phantom{\dag}} \equiv \frac{U}{2} \sum_{\sigma} \sum_{j=1}^{N_{\textsc{s}}} \h{n}_{i;\sigma} \h{n}_{i;\b{\sigma}}.
\end{equation}
Using the Fourier representation in Eq.\,(\ref{eg3a}), for the Hubbard Hamiltonian one thus obtains
\begin{equation}\label{ex01fx}
\h{\mathcal{H}} = \hspace{0.28cm}\h{\hspace{-0.28cm}\mathpzc{H}},\;\;\;\;\; \text{(For spin-$\tfrac{1}{2}$ particles)}
\end{equation}
where
\begin{equation}\label{ex01f}
\hspace{0.28cm}\h{\hspace{-0.28cm}\mathpzc{H}} \doteq  \sum_{\bm{k},\sigma} \varepsilon_{\bm{k}}\,
\h{a}_{\bm{k};\sigma}^{\dag} \h{a}_{\bm{k};\sigma}^{\phantom{\dag}} + \frac{U}{2 N_{\textsc{s}}}
\sum_{\sigma} \sum_{\bm{k}, \bm{p}, \bm{q}}
\h{a}_{\bm{k}+\bm{q};\sigma}^{\dag} \h{a}_{\bm{p}-\bm{q};\b{\sigma}}^{\dag}
\h{a}_{\bm{p};\b{\sigma}}^{\phantom{\dag}} \h{a}_{\bm{k};\sigma}^{\phantom{\dag}}.
\end{equation}
The expression on the RHS of Eq.\,(\ref{ex01f}) is to be contrasted with that on the RHS of Eq.\,(\ref{ex01bx}). The simplification thus achieved is not \emph{a priori} evident from the latter expression, since in contrast to the site-representation where one has the product $\h{c}_{i;\sigma} \h{c}_{i;\sigma}$, with the two operators annihilating fermions with the same spin index on the same lattice site, whereby $\h{c}_{i;\sigma} \h{c}_{i;\sigma} = \h{0}$ (reflecting the Pauli exclusion principle), in the wave-vector representation one has $\h{a}_{\bm{p};\sigma} \h{a}_{\bm{k};\sigma}$, with the two operators annihilating fermions with the same spin index however in general with different momenta, so that in general $\h{a}_{\bm{p};\sigma} \h{a}_{\bm{k};\sigma} \not= \h{0}$. In this connection, we emphasize that while $\h{\mathcal{H}}_1'' = \h{0}$, irrespective of the type of the representation adopted, in general
\begin{equation}\label{ex01g}
\h{a}_{\bm{k}+\bm{q};\sigma}^{\dag} \h{a}_{\bm{p}-\bm{q};\sigma}^{\dag}
\h{a}_{\bm{p};\sigma}^{\phantom{\dag}} \h{a}_{\bm{k};\sigma}^{\phantom{\dag}} \not= \h{0}.
\end{equation}

The inequality in Eq.\,(\ref{ex01g}) clarifies the reason for the observation to be made later in this appendix, \S\,\ref{sd4}, that at any given order of the perturbation theory corresponding to the Hubbard Hamiltonian $\h{\mathcal{H}}$ for spin-$\tfrac{1}{2}$ particles,\footnote{Here $\protect\h{\mathcal{H}}$ is to be distinguished from $\hspace{0.28cm}\protect\h{\hspace{-0.28cm}\mathpzc{H}}$.} in the energy-momentum representation the contributions of the \textsl{individual} self-energy diagrams associated with $\h{\mathcal{H}}_{\X{1}}''$ are \textsl{not} vanishing, in contrast to their \textsl{total} contribution, which is indeed identically vanishing.\footnote{It should be evident that since at the $\nu$th-order of the perturbation series expansion one encounters the product $\protect\h{\mathcal{H}}_{\protect\X{1}}(t_1) \dots \protect\h{\mathcal{H}}_{\protect\X{1}}(t_{\nu})$, where $\protect\h{\mathcal{H}}_{\protect\X{1}}(t_i)$ denotes $\protect\h{\mathcal{H}}_{\protect\X{1}}$ in the interaction representation at time $t = t_i$ [Eq.\,(6.5), p.\,54, and Eq.\,(9.5), p.\,96, in Ref.\,\protect\citen{FW03}], `diagrams associated with $\protect\h{\mathcal{H}}_{\protect\X{1}}''$' are also in part due to $\protect\h{\mathcal{H}}_{\protect\X{1}}'$.} These diagrams whose contributions at any given order of the perturbation theory add up to zero, are \textsl{not} present in the perturbation series expansion of the self-energy corresponding to $\hspace{0.28cm}\h{\hspace{-0.28cm}\mathpzc{H}}$. It is tempting to assume that these diagrams were present, however each made a vanishing contribution. This assumption is both unfounded and demonstrably false, \S\,\ref{sd4} (see also the earlier relevant discussion, on p.\,\pageref{DenotingTheHubbard}). In order to maintain the same set of self-energy diagrams that occur in the perturbation series expansion of the self-energy corresponding to $\h{\mathcal{H}}$ for spin-$\tfrac{1}{2}$ particles, with the property that each self-energy diagram that does not occur in the perturbation series expansion corresponding to the Hubbard Hamiltonian $\hspace{0.28cm}\h{\hspace{-0.28cm}\mathpzc{H}}$ makes an identically vanishing contribution by itself, it is necessary to employ an alternative representation of the Hubbard Hamiltonian, namely $\h{\mathsf{H}}$, which we shall introduce later in this section, Eq.\,(\ref{ex01k}) below.

We\refstepcounter{dummy}\label{WeAreNowInAPosition} are now in a position to specify the characteristics of the self-energy diagrams corresponding to particles with spin index $\sigma$ and associated with the Hubbard Hamiltonian $\hspace{0.28cm}\h{\hspace{-0.28cm}\mathpzc{H}}$ for spin-$\tfrac{1}{2}$ particles. In these diagrams, the two ends of each interaction line correspond to complementary spin indices, $\sigma$ and $\b{\sigma}$, reflecting the property that in the interaction part of the Hamiltonian in Eq.\,(\ref{ex01f}) one has the pair\,\footnote{The $\sigma$ here is a summation variable and is therefore not to be confused with the $\sigma$ in for instance $\Sigma_{\sigma}^{\protect\X{(\nu)}}(\bm{k};\varepsilon)$.} $(\h{a}_{\bm{k}+\bm{q};\sigma}^{\X{\dag}},  \h{a}_{\bm{k};\sigma}^{\phantom{\X{\dag}}})$ in combination with the pair $(\h{a}_{\bm{p}-\bm{q};\b{\sigma}}^{\X{\dag}},\h{a}_{\bm{p};\b{\sigma}}^{\phantom{\X{\dag}}})$ (see also Fig.\,\ref{f4}, p.\,\pageref{TwoInternal}); in contrast, in the Hamiltonian in Eq.\,(\ref{ex01}) one \textsl{additionally} has the pair $(\h{a}_{\bm{p}-\bm{q};\sigma}^{\X{\dag}},\h{a}_{\bm{p};\sigma}^{\phantom{\X{\dag}}})$, in consequence of the summation with respect to $\sigma'$. As a result,\refstepcounter{dummy}\label{WeNoteThat}\footnote{We note that Eq.\,(\protect\ref{ea2}) applies to the cases where the two-body interaction potential is spin-independent (category (3) above -- p.\,\protect\pageref{ThreeCases}), as in the case of $\protect\h{\mathcal{H}}$ in Eq.\,(\protect\ref{ex01bx}). In dealing with such Hamiltonian as $\protect\h{\mathsf{H}}$, Eqs\,(\protect\ref{ex01j}) and (\protect\ref{ex01k}) below (where the two-body potential is in category (2) -- p.\,\protect\pageref{ThreeCases}), for uniform GSs (ESs) one has $\protect\b{W}_{\sigma,\sigma'}(\bm{k};\varepsilon) = U_{\sigma,\sigma'} + U^2 \protect\b{\chi}_{\protect\b{\sigma},\protect\b{\sigma}'}(\bm{k};\varepsilon)$, whereby  $\protect\b{W}_{\sigma,\protect\b{\sigma}}(\bm{k};\varepsilon) = U + U^2 \protect\b{\chi}_{\protect\b{\sigma},\sigma}(\bm{k};\varepsilon)$ and $\protect\b{W}_{\sigma,\sigma}(\bm{k};\varepsilon) = U^2 \protect\b{\chi}_{\protect\b{\sigma},\protect\b{\sigma}}(\bm{k};\varepsilon)$ (see \S\,3.2 in Ref.\,\protect\citen{BF19}). With reference to the final remarks in the caption of Fig.\,\protect\ref{f4}, p.\,\protect\pageref{TwoInternal}, in dealing with the Hubbard Hamiltonian $\hspace{0.28cm}\h{\hspace{-0.28cm}\mathpzc{H}}$, Eq.\,(\protect\ref{ex01f}), the relevant proper self-energy diagrams cannot contain those two-body interaction contributions associated with $\protect\b{W}_{\sigma,\protect\b{\sigma}}$ that mediate between the vertices of a contiguous set of Green-function lines (we emphasise that proper self-energy diagrams containing such interaction lines are \textsl{absent} and the term `\textsl{ruled out}' under (i) and (ii) below refers to this \textsl{absence}; in contrast, in dealing with the Hamiltonian $\protect\h{\mathsf{H}}$ such diagrams are present, however each makes an identically vanishing contribution). Similarly, these proper self-energy diagrams cannot contain those two-body interaction contributions associated with $\protect\b{W}_{\sigma,\sigma}$ that mediate between two disjoint sets of contiguous Green-function lines associated with complementary spin indices (\emph{i.e.} with spin indices $\sigma$ and $\protect\b{\sigma}$). It is not difficult to verify that, in dealing with $\hspace{0.28cm}\h{\hspace{-0.28cm}\mathpzc{H}}$, a cascade of loops connecting the external vertices of a polarisation diagram associated with $\protect\b{\chi}_{\protect\b{\sigma},\sigma}$ ($\protect\b{\chi}_{\protect\b{\sigma},\protect\b{\sigma}}$) is necessarily comprised of an \textsl{even} (\textsl{odd}) number of loops. Note that diagrams contributing the $\protect\b{\chi}_{\sigma,\sigma'} \equiv \protect\b{P}_{\sigma,\sigma'}^{\star}$, as opposed to those contributing to $\protect\b{P}_{\sigma,\sigma'}$, may be both proper and improper. Of the two proper polarisation diagrams presented in Fig.\,\protect\ref{f12}, p.\,\protect\pageref{Two2ndO}, diagram (a) consists of a single loop connecting the external vertices, and diagram (b) of a cascade of two loops of this type. \label{noted1}}
\vspace{0.1cm}
\begin{itemize}
\item[(i)] self-energy\protect\refstepcounter{dummy}\label{SelfEnergyDiagrams} diagrams containing interaction lines whose ends are identified with two vertices of a \textsl{contiguous} set of Green-function lines are \textsl{ruled out}, and, similarly,
\item[(ii)] self-energy diagrams in which two interaction lines link two vertices of a \textsl{contiguous} set of Green-function lines through an intermediate cascade of an \textsl{even} number of loops [p.\,110 in Ref.\,\citen{FW03}] are \textsl{ruled out}.\footnote{The chain of the two interaction lines and the mentioned polarisation insertions thus account for a contribution to the dynamically-screened interaction potential, considered in appendix \protect\ref{sa}.}
\end{itemize}
\vspace{0.1cm}
As a consequence of (i), there are \textsl{no} sums to be evaluated over the spin indices of the fermion loops (the minus sign associated with each loop [p.\,98 in Ref.\,\citen{FW03}] remains),\footnote{A \textsl{loop} is any closed train of directed Green-function lines, which may pass through any number of vertices.} each of which\,\footnote{\emph{Cf.} Eq.\,(\protect\ref{ex04a}) below.} in the case of the perturbation expansion associated with the Hamiltonian $\h{\mathcal{H}}$ (to be distinguished from $\hspace{0.28cm}\h{\hspace{-0.28cm}\mathpzc{H}}\hspace{0.4pt}$) \textsl{and} unpolarised states of spin-$\tfrac{1}{2}$ particles gives rise to a factor of $2\times \frac{1}{2} + 1 = 2$, appendix \ref{sa}. We emphasise once more that \emph{diagrams that are \textsl{ruled out} simply do \textsl{not} occur in the perturbation series expansion under discussion, as a direct consequence of the structure of the Hamiltonian $\hspace{0.28cm}\h{\hspace{-0.28cm}\mathpzc{H}}$ in Eq.\,(\ref{ex01f}).} This is strictly different from the case of these diagrams occurring in the diagrammatic expansion under discussion, however having identically-vanishing contribution, either individually or collectively. As we have indicated earlier in this section, at each order of the perturbation theory the diagrams that occur in the perturbation series expansion corresponding to $\h{\mathcal{H}}$ but not in that corresponding to $\hspace{0.28cm}\h{\hspace{-0.28cm}\mathpzc{H}}$, collectively make \textsl{no} contribution. Explicit calculations, to be discussed later in this appendix, \S\,\ref{sd4}, clearly illustrate these facts.

In order to maintain the diagrammatic structure of the perturbation series associated with $\h{\mathcal{H}}$, however with the diagrams that do not occur in the perturbation series associated with $\hspace{0.28cm}\h{\hspace{-0.28cm}\mathpzc{H}}$ making vanishing contribution diagram-by-diagram (as opposed to collectively -- see above), one should return to the general single-band Hamiltonian $\h{\mathscr{H}}$ in Eq.\,(\ref{ex01a}) and employ the following relationship, instead of that in Eq.\,(\ref{ex01b}) [\emph{cf.} Eq.\,(9.19), p.\,104, in Ref.\,\citen{FW03}]:\,\footnote{Compare also with Eqs\,(1.20) and (1.21), p.\,16, in Ref.\,\protect\citen{PN66}.}
\begin{equation}\label{ex01h}
\b{\mathsf{v}}_{\sigma,\sigma';\sigma'',\sigma'''}(\|\bm{q}\|) \equiv \b{v}_{\sigma,\sigma''}(\|\bm{q}\|)\hspace{0.6pt} \delta_{\sigma,\sigma'}\hspace{0.6pt} \delta_{\sigma'',\sigma'''}\hspace{0.6pt}.
\end{equation}
With reference to Eq.\,(\ref{ex03}), in the case of the single-band Hubbard Hamiltonian this implies the substitution
\begin{equation}\label{ex01i}
\b{v}_{\sigma,\sigma'}(\|\bm{q}\|) \rightharpoonup \frac{\Omega U_{\sigma,\sigma'}}{N_{\textsc{s}}},
\end{equation}
where for spin-$\tfrac{1}{2}$ particles\,\footnote{Compare with Eq.\,(2.71) in Ref.\,\protect\citen{BF19}.}
\begin{equation}\label{ex01j}
U_{\sigma,\sigma'} \doteq (1-\delta_{\sigma,\sigma'})\hspace{0.6pt} U,
\end{equation}
so that $U_{\sigma,\sigma} = 0$ and $U_{\sigma,\b{\sigma}} = U$. Defining the Hamiltonian
\begin{equation}\label{ex01k}
\h{\mathsf{H}} \doteq \sum_{\bm{k},\sigma} \varepsilon_{\bm{k}}\,
\h{a}_{\bm{k};\sigma}^{\dag} \h{a}_{\bm{k};\sigma}^{\phantom{\dag}} + \frac{1}{2 N_{\textsc{s}}}
\sum_{\sigma,\sigma'} \sum_{\bm{k}, \bm{p}, \bm{q}} U_{\sigma,\sigma'}\,
\h{a}_{\bm{k}+\bm{q};\sigma}^{\dag} \h{a}_{\bm{p}-\bm{q};\sigma'}^{\dag}
\h{a}_{\bm{p};\sigma'}^{\phantom{\dag}} \h{a}_{\bm{k};\sigma}^{\phantom{\dag}},
\end{equation}
one observes that (\emph{cf.} Eqs\,(\ref{ex01c}) and (\ref{ex01e}))
\begin{equation}\label{ex01l}
\h{\mathsf{H}} = \h{\mathcal{H}} - \h{\mathcal{H}}_{\X{1}}'' \equiv \h{\mathcal{H}}_{\X{0}} + \h{\mathcal{H}}_{\X{1}}' = \hspace{0.28cm}\h{\hspace{-0.28cm}\mathpzc{H}}.\;\;\;\; \text{(For spin-$\tfrac{1}{2}$ particles)}
\end{equation}
The interaction part of the Hamiltonian $\h{\mathsf{H}}$ involving a double sum over spin indices, the perturbation series expansion of the self-energy associated with this Hamiltonian gives rise to the same self-energy diagrams as the Hubbard Hamiltonian $\h{\mathcal{H}}$ in Eq.\,(\ref{ex01bx}), and the more general Hamiltonian $\h{H}$ in Eq.\,(\ref{ex01}). With reference to the equality in Eq.\,(\ref{ex01j}), one immediately appreciates that for spin-$\tfrac{1}{2}$ particles any diagram associated with $\h{\mathsf{H}}$ that does not occur in the perturbation series expansion of the self-energy associated with $\hspace{0.28cm}\h{\hspace{-0.28cm}\mathpzc{H}}$ (see above, p.\,\pageref{WeAreNowInAPosition}), has an identically-vanishing contribution by itself.

Unless\refstepcounter{dummy}\label{UnlessWeIndicateOtherwise} we indicate otherwise, \emph{in the following, when considering the self-energy contributions corresponding to the single-band Hubbard Hamiltonian in the energy-momentum space, these contributions correspond to the Hubbard Hamiltonian $\h{\mathcal{H}}$ in Eq.\,(\ref{ex01bx}), evaluated on the basis of the general rules applicable to the Hamiltonian $\h{H}$ in Eq.\,(\ref{ex01}), \S\,\ref{sd21}. Thus spin sums over the spin indices of the fermion loops are evaluated and no diagrams are discarded as not occurring in the diagrammatic expansion, in contrast to the case where $\hspace{0.28cm}\h{\hspace{-0.28cm}\mathpzc{H}}$, Eq.\,(\ref{ex01f}), is the underlying Hamiltonian. When the self-energy contributions in the energy-momentum space are evaluated on the basis of the perturbation series expansion corresponding to the Hamiltonian $\h{\mathsf{H}}$ in Eq.\,(\ref{ex01k}), we either explicitly indicate this, or supplement the relevant contributions with the subscript $\h{\mathsf{H}}$ (at places we shall also use $\hspace{0.28cm}\h{\hspace{-0.28cm}\mathpzc{H}}$ as subscript with similar intention).} In dealing with the half-filled `Hubbard atom' for spin-$\tfrac{1}{2}$ particles, \S\,\ref{sd4}, we write, for instance, $\t{\Sigma}_{\sigma}^{\X{(4.31)}}(z)\vert_{\X{\h{\mathsf{H}}}}$, which is identically vanishing,\footnote{Thus, the diagram associated with $\protect\t{\Sigma}_{\sigma}^{\protect\X{(4.31)}}(z)$ \textsl{does not appear} in the diagrammatic expansion of the self-energy associated with $\hspace{0.28cm}\protect\h{\hspace{-0.28cm}\mathpzc{H}}$, Eq.\,(\protect\ref{ex01f}).} in contrast to $\t{\Sigma}_{\sigma}^{\X{(4.31)}}(z)$ which is non-vanishing, Eq.\,(\ref{ex31}).

Before concluding this subsection, we mention that of the skeleton self-energy diagrams specified in \S\,\ref{sx1}, for spin-$\tfrac{1}{2}$ particles \textsl{only} the ones associated with the following self-energy contributions occur in the perturbation series expansion of the self-energy corresponding to the Hamiltonian $\hspace{0.28cm}\h{\hspace{-0.28cm}\mathpzc{H}}$ in Eq.\,(\ref{ex01f}):
\begin{align}\label{ex01m}
\text{$2$nd order:}\; &\Sigma_{\sigma}^{\X{(2.2)}},\\
\label{ex01n}
\text{$3$rd order:}\; &\Sigma_{\sigma}^{\X{(3.6)}},\; \Sigma_{\sigma}^{\X{(3.7)}},\\
\label{ex01o}
\text{$4$th order:}\; &\Sigma_{\sigma}^{\X{(4.38)}},\; \Sigma_{\sigma}^{\X{(4.39)}},\; \Sigma_{\sigma}^{\X{(4.40)}},\; \Sigma_{\sigma}^{\X{(4.41)}},\; \Sigma_{\sigma}^{\X{(4.42)}},\; \Sigma_{\sigma}^{\X{(4.43)}},\; \Sigma_{\sigma}^{\X{(4.70)}},\; \Sigma_{\sigma}^{\X{(4.79)}},\;  \Sigma_{\sigma}^{\X{(4.82)}}.\nonumber\\
\end{align}
Compare the number of the above diagrammatic contributions at the orders indicated, that is $1$, $2$, and $9$, with respectively $2$, $10$, and $82$, specified at the outset of the present appendix, p.\,\pageref{InThisAppendix}.

\refstepcounter{dummyX}
\subsection{Contributions of the self-energy diagrams in the energy-momentum representation}
\phantomsection
\label{sd21}
Here we first consider the contribution $\Sigma_{\sigma}^{\X{(\nu.j)}}(\bm{k};\varepsilon)$ of the $j$th $\nu$th-order proper self-energy diagram corresponding to the GS of the Hamiltonian $\h{H}$ in Eq.\,(\ref{ex01}), described in terms of the spin-independent isotropic two-body potential $\b{v}(\|\bm{q}\|)$.\footnote{With some slight modification, the considerations of this section equally apply to the contributions of the self-energy diagrams corresponding to uniform ESs.} The integer $j$ takes a value between and including $1$ and one of the relevant numbers indicated at the outset of the present appendix; for instance, in considering the skeleton self-energy diagrams, $j \in \{1,2,\dots, \mathcal{N}_{\nu}\}$, where $\mathcal{N}_1 =2$, $\mathcal{N}_3 = 10$, \emph{etc}. We assume that the underlying GS is such that the corresponding one-particle Green function, and thus the self-energy, is diagonal in the spin space \cite{BF19}, with $\t{G}_{\sigma}(\bm{k};z)$ denoting a diagonal element, where $\sigma$ takes $2\mathsf{s}+1$ distinct values for spin-$\mathsf{s}$ particles. This assumption implies that in the $\nu$th-order self-energy diagram under consideration, contributing to $\Sigma_{\sigma}^{\X{(\nu)}}(\bm{k};\varepsilon)$, all Green-function lines correspond to particles with spin index $\sigma$ (\emph{i.e.} the \textsl{external} spin index), except those Green-function lines that form loops -- if the diagram under consideration has any, with all Green-function lines in a given loop corresponding to particles with the same spin index.\refstepcounter{dummy}\label{WithRefTo}\footnote{With reference to the remarks in footnote \raisebox{-1.0ex}{\normalsize{\protect\footref{notef1}}} on p.\,\protect\pageref{ForSpin}, we note that the prescriptions specified in this section assume that the $(2\mathsf{s}+1) \times (2\mathsf{s}+1)$ matrix $\mathbb{G}$, or $\mathbb{G}_{\protect\X{0}}$, is diagonal, and that the multiplication of the Green matrices in the expressions corresponding to self-energy diagrams have been carried out. As a result, in these prescriptions one only encounters the diagonal elements $\{G_{\sigma} \| \sigma\}$, or $\{G_{\protect\X{0};\sigma} \| \sigma\}$, depending on whether the diagrams under consideration are in terms of the interacting or the non-interacting Green functions. \label{noteg1}} With $\ell$ denoting the number of loops in the self-energy diagram under consideration, for $\ell > 0$ we denote the set of the spin indices associated with the Green-function lines in the loops by $\{\sigma_1,\dots,\sigma_{\ell}\}$. These indices are to be summed over, Eq.\,(\ref{ex04a}) below.

As a special case, we consider the $\Sigma_{\sigma}^{\X{(\nu.j)}}(\bm{k};\varepsilon)$ corresponding to the GS of the single-band Hubbard Hamiltonian $\h{\mathcal{H}}$ in Eq.\,(\ref{ex01bx}) described in terms of the two-body interaction potential $\b{v}(\|\bm{q}\|) \equiv \Omega U/N_{\textsc{s}}$, Eq.\,(\ref{ex03}), which is \textsl{independent} of $\bm{q}$. This leads us to the details concerning the contributions of the proper self-energy diagrams associated with the alternative representation of this Hamiltonian, namely the Hamiltonian $\h{\mathsf{H}}$ in Eq.\,(\ref{ex01k}), despite the fact that $\h{\mathsf{H}}$ is described in terms of the spin-dependent two-body interaction potential energy $U_{\sigma,\sigma'}$, Eq.\,(\ref{ex01j}). The observations regarding the proper self-energy diagrams associated with $\h{\mathsf{H}}$ directly lead to insights regarding the proper self-energy diagrams associated with yet another alternative representation of $\h{\mathcal{H}}$, namely the Hubbard Hamiltonian $\hspace{0.28cm}\h{\hspace{-0.28cm}\mathpzc{H}}$ in Eq.\,(\ref{ex01f}).

In the light of the above descriptions, with reference to the considerations described in, \emph{e.g.}, Ref.\,\citen{FW03} (culminating in the rules specified on pp.\,102-105 herein),\footnote{See the remark following Eq.\,(\protect\ref{ex05}) below.} one has\,\footnote{The analytic continuation of $\Sigma_{\sigma}^{\protect\X{(\nu.j)}}(\bm{k};\varepsilon)$ into the complex $z$-plane, to obtain the function $\protect\t{\Sigma}_{\sigma}^{\protect\X{(\nu.j)}}(\bm{k};z)$, is to be effected only after the evaluation of the integrals with respect to $\{\varepsilon_i\| i = 1,\dots,\nu\}$. Otherwise, $\varepsilon \rightharpoonup z$, $\protect\im[z] \not=0$, is likely to cause the arguments of some Green functions to cross some underlying branch cuts that are not to be crossed. Compare the functions $P_{\epsilon}^{\X{(0)}}(\varepsilon)$ and $\lim_{\beta\to\infty}\t{\mathscr{P}}^{\X{(0)}}_{\epsilon}(z)$ in respectively Eq.\,(\protect\ref{e79}) and Eq.\,(\protect\ref{e83}) and the fundamental difference between the two as reflected in Eq.\,(\protect\ref{e85}).}
\begin{equation}\label{ex04}
\Sigma_{\sigma}^{\X{(\nu.j)}}(\bm{k};\varepsilon) = \frac{(-1)^{\nu+\ell}}{\hbar^{\nu}} \frac{1}{\Omega} \sum_{\bm{k}_1} \dots \frac{1}{\Omega}\sum_{\bm{k}_{\nu}} \int_{-\infty}^{\infty}\frac{\rd\varepsilon_1}{2\pi\!\ii\hbar} \dots \ \frac{\rd\varepsilon_{\nu}}{2\pi\!\ii\hbar}\; F_{\sigma}^{(\nu.j)}(\bm{k},\varepsilon;\bm{k}_1,\varepsilon_1;\dots;\bm{k}_{\nu},\varepsilon_{\nu}),
\end{equation}
where the function $F_{\sigma}^{\X{(\nu.j)}}$ consists of a product of $2\nu-1$ one-particle Green functions and $\nu$ interaction potentials $\b{v}$ whose relevant arguments are such that at each of the $2\nu$ vertices of the $\nu$th-order diagram under consideration energy and momentum are conserved, \S\,\ref{sd3} (see Fig.\,\ref{f4}, p.\,\pageref{TwoInternal}).\footnote{In Ref.\,\protect\citen{BF19} we present both an algorithm and a program, written in the Mathematica programming language \protect\cite{SW16}, for the determination of the function $F_{\sigma}^{\protect\X{(\nu.j)}}$ in explicit form.} For $\ell=0$, all the mentioned $2\nu-1$ one-particle Green functions correspond to particles with spin index $\sigma$. For $\ell > 0$, one however has
\begin{equation}\label{ex04a}
F_{\sigma}^{(\nu.j)}(\bm{k},\varepsilon;\bm{k}_1,\varepsilon_1;\dots;\bm{k}_{\nu},\varepsilon_{\nu}) = \sum_{\sigma_1, \dots, \sigma_{\ell}} \mathcal{F}_{\sigma;\sigma_1,\dots,\sigma_{\ell}}^{(\nu.j)}(\bm{k},\varepsilon;\bm{k}_1,\varepsilon_1;\dots;
\bm{k}_{\nu},\varepsilon_{\nu}).
\end{equation}
In the case of spin-unpolarised GSs of spin-$\mathsf{s}$ fermions, the sums on the RHS of this expression result in the numerical factor $(2\mathsf{s}+1)^{\ell}$ times the function $\mathcal{F}_{\sigma;\X{\sigma_1,\dots,\sigma_{\ell}}}^{\X{(\nu.j)}}$ for an arbitrary choice for the indices $\{\sigma_1,\dots,\sigma_{\ell}\}$.

The similarity of the single-band Hubbard Hamiltonian $\h{\mathcal{H}}$ in Eq.\,(\ref{ex01bx}) with the more general single-band Hamiltonian $\h{H}$ in Eq.\,(\ref{ex01}) implies that the above considerations are directly applicable to the contributions of the $\nu$th-order proper self-energy diagrams corresponding to $\h{\mathcal{H}}$. One has however the following simplification:
\begin{equation}\label{ex06}
F_{\sigma}^{(\nu.j)}(\bm{k},\varepsilon;\bm{k}_1,\varepsilon_1;\dots;\bm{k}_{\nu},\varepsilon_{\nu}) = \Big(\frac{\Omega U}{N_{\textsc{s}}}\Big)^{\nu}\hspace{0.8pt} \b{F}_{\sigma}^{(\nu.j)}(\bm{k},\varepsilon;\bm{k}_1,\varepsilon_1;\dots;\bm{k}_{\nu},\varepsilon_{\nu}),
\end{equation}
where the function $\b{F}_{\sigma}^{\X{(\nu.j)}}$ is independent of $U$ in the case of the diagram under consideration being evaluated in terms $\{G_{\X{0};\sigma}\|\sigma\}$, and an \textsl{implicit} function of $U$ in the case of this diagram being evaluated in terms of $\{G_{\sigma}\|\sigma\}$. For the Hubbard Hamiltonian $\h{\mathcal{H}}$ one thus arrives at
\begin{align}\label{ex07}
\Sigma_{\sigma}^{\X{(\nu.j)}}(\bm{k};\varepsilon) &= \frac{(-1)^{\nu+\ell} U^{\nu}}{\hbar^{\nu}} \frac{1}{N_{\textsc{s}}} \sum_{\bm{k}_1\in\fbz} \dots \frac{1}{N_{\textsc{s}}}\sum_{\bm{k}_{\nu}\in\fbz} \int_{-\infty}^{\infty}\frac{\rd\varepsilon_1}{2\pi\!\ii\hbar} \dots  \frac{\rd\varepsilon_{\nu}}{2\pi\!\ii\hbar}\;\nonumber\\
&\hspace{5.0cm} \times \b{F}_{\sigma}^{(\nu.j)}(\bm{k},\varepsilon;\bm{k}_1,\varepsilon_1;\dots;\bm{k}_{\nu},\varepsilon_{\nu}).
\end{align}
Since $G_{\sigma}(\bm{k};\varepsilon)$ (or $G_{\X{0};\sigma}(\bm{k};\varepsilon)$, as the case may be) has the dimensionality of time (\emph{cf.} Eq.\,(\ref{e25})), \emph{i.e.} $\llbracket G_{\sigma}(\bm{k};\varepsilon)\rrbracket = \mathrm{s}$ in the SI base units,\footnote{See appendix \protect\ref{sae}.} one has $\llbracket \b{F}_{\sigma}^{\X{(\nu.j)}}\rrbracket = \mathrm{s}^{\X{2\nu-1}}$. Thus for the function $\Sigma_{\sigma}^{\X{(\nu.j)}}(\bm{k};\varepsilon)$ in Eq.\,(\ref{ex07}), as well as that in Eq.\,(\ref{ex04}), it follows that $\llbracket \Sigma_{\sigma}^{\X{(\nu.j)}}(\bm{k};\varepsilon)\rrbracket = \mathrm{s}^{\X{-1}}$, so that $\llbracket\hbar\Sigma_{\sigma}^{\X{(\nu.j)}}(\bm{k};\varepsilon)\rrbracket = \mathrm{J}$, as befits a perturbational contribution to $\hbar\Sigma_{\sigma}(\bm{k};\varepsilon)$.

The Hamiltonian $\h{\mathsf{H}}$ in Eq.\,(\ref{ex01k}) is similar to the Hamiltonian $\h{\mathcal{H}}$ in Eq.\,(\ref{ex01bx}), except that the interaction potential energy in $\h{\mathsf{H}}$, that is $U_{\sigma,\sigma'}$, is spin dependent. Focusing on spin-$\tfrac{1}{2}$ particles ($\mathsf{s} = \tfrac{1}{2}$), due to the nature of this interaction potential, Eq.\,(\ref{ex01j}), the Green-function lines at the two vertices of this potential, Fig.\,\ref{f4}, p.\,\pageref{TwoInternal}, must correspond to particles of opposite spin index, or the contribution of the self-energy diagram under discussion is identically vanishing. This implies that the contribution of any proper self-energy diagram in which two vertices on a contiguous set of Green-function lines are connected by an interaction line, is identically vanishing. More generally, the contribution of any proper self-energy diagram in which two vertices on a contiguous set of Green-function lines are connected by two interaction lines mediated by a polarisation insertion\,\footnote{Appendix \protect\ref{sa}.} containing an \textsl{even} number of loops in cascade, is identically vanishing.\footnote{The order $\nu$ of a diagram of this type is therefore equal to or greater than $3$.} Since polarisation insertions are comprised of particle loops,\footnote{See Figs\,\protect\ref{f10}, \protect\ref{f12}, and \protect\ref{f7}, pp.\,\protect\pageref{A4thOr}, \protect\pageref{Two2ndO}, and \protect\pageref{SecondOrderS}, above, and Figs\,\protect\ref{f15} and \protect\ref{f16}, pp.\,\protect\pageref{DiagramA1} and \protect\pageref{DiagramA2}, below.} with each loop consisting of a \textsl{closed} contiguous set of Green-function lines, by the same reasoning as above in order for a polarisation insertion not be identically vanishing, it is required that no interaction line and/or two interaction lines mediated by a polarisation insertion containing an \textsl{even} number of loops in cascade be attached to more than one vertex of a loop contained in the mentioned polarisation insertion. These requirements result in a drastic reduction\,\footnote{This reduction is algebraically realised within the diagram-free framework of Ref.\,\protect\citen{BF19}. See \S\,2.2.7 and appendix D herein.} of the number of self-energy diagrams that are not identically vanishing: diagrams whose contributions are not identically vanishing must contain loops, that is polarisation insertions, which in turn are subject to the described restrictions on their connectivity.

With reference to Eqs\,(\ref{ex04a}) and (\ref{ex06}), in the light of the above observations,\footnote{See also \S\,\protect\ref{sd2}.} the contribution of the diagram associated with $\Sigma_{\sigma}^{\X{(\nu.j)}}(\bm{k};\varepsilon)$ and specific to the Hamiltonian $\h{\mathsf{H}}$ in Eq.\,(\ref{ex01k}) is either identically vanishing, or corresponds to
\begin{equation}\label{ex04b}
\b{F}_{\sigma}^{(\nu.j)}(\bm{k},\varepsilon;\bm{k}_1,\varepsilon_1;\dots;\bm{k}_{\nu},\varepsilon_{\nu}) = \b{\mathcal{F}}_{\sigma;\sigma_1,\dots,\sigma_{\ell}}^{(\nu.j)}(\bm{k},\varepsilon;\bm{k}_1,\varepsilon_1;
\dots;\bm{k}_{\nu},\varepsilon_{\nu}),\;\;\ell > 0,
\end{equation}
where $\{\sigma_1,\dots,\sigma_{\ell}\}$ is a \textsl{fixed} configuration of the spin indices associated with the $\ell$ loops of the diagram under consideration, determined by the external spin index $\sigma$. Note that the case of $\Sigma_{\sigma}^{\X{(\nu.j)}}(\bm{k};\varepsilon)$ being identically vanishing corresponds to the situation in which, for the given external spin index $\sigma$, all possible $2^{\ell}$ configurations of $\{\sigma_1,\dots,\sigma_{\ell}\}$ amount to conflicting spin indices for at least one vertex of the diagram associated with $\Sigma_{\sigma}^{\X{(\nu.j)}}(\bm{k};\varepsilon)$ (for an illustrative example, see below). Further, the contribution of any $\nu$th-order proper self-energy diagram, $\forall\nu \in \mathds{N}$, corresponding to $\h{\mathsf{H}}$ is identically vanishing for $\ell=0$, following the fact that in such diagram the ends of an interaction line is necessarily attached to two vertices of a contiguous set of Green-function lines.

For illustration, consider diagram (A) in Fig.\,\ref{f15}, p.\,\pageref{DiagramA1}, below. With $\sigma = \uparrow$, the spin indices associated with the external vertices $1'$ and $2'$ are equal to $\uparrow$. The vertices $3'$ and $4'$ corresponding to a contiguous set of Green-function lines linking $1'$ to $2'$, it follows that the spin indices associated with $3'$ and $4'$ must also be equal to $\uparrow$. On the other hand, the vertex $1'$ ($2'$) being connected to vertex $4'$ ($3'$) by means of an interaction line, the spin index associated with vertex $4'$ ($3'$) must be equal to $\downarrow$. This conflicting situation implies that the contribution of diagram (A) is identically vanishing. This is however demonstrably not the case for diagram (A) corresponding to the perturbation series expansion associated with the Hubbard Hamiltonian $\h{\mathcal{H}}$. Considering diagram (B) in Fig.\,\ref{f15} below, p.\,\pageref{DiagramA1}, \textsl{no} conflict as in the case of diagram (A) arises in dealing with $\h{\mathsf{H}}$; clearly, for the case where the spin indices associated with the external vertices of diagram (B) are equal to $\uparrow$ ($\downarrow$), the spin indices associated with the vertices of the polarisation loop in this diagram must be $\downarrow$ ($\uparrow$) in order for the contribution of this diagram not be identically vanishing. There is therefore no summation over the two spin states of the polarisation loop in diagram (B) (\emph{cf.} Eqs\,(\ref{ex04a}) and (\ref{ex04b})).

Regarding the self-energy diagrams associated with the Hubbard Hamiltonian $\hspace{0.28cm}\h{\hspace{-0.28cm}\mathpzc{H}}$ in Eq.\,(\ref{ex01f}), for the external spin index $\sigma$ given, the totality of the $\nu$th-order diagrams contributing to $\Sigma_{\sigma}(\bm{k};\varepsilon)$ is comprised of those corresponding to the Hamiltonian $\h{\mathsf{H}}$ that are not identically vanishing; those self-energy diagrams associated with $\h{\mathsf{H}}$ that are identically vanishing (on account of the mechanism specified above) do \textsl{not} appear in the diagrammatic expansion of the self-energy corresponding to $\hspace{0.28cm}\h{\hspace{-0.28cm}\mathpzc{H}}$. Leaving aside the fundamental distinction between a diagram not existing and a diagram existing however making an identically vanishing contribution, quantitatively the perturbation series expansions of the self-energy on the basis of the representations $\hspace{0.28cm}\h{\hspace{-0.28cm}\mathpzc{H}}$ and $\h{\mathsf{H}}$ of the single-band Hubbard Hamiltonian may be considered as being equivalent \textsl{diagram-by-diagram}. In this connection, \emph{we emphasise that the total $\nu$th-order self-energy $\Sigma_{\sigma}^{\X{(\nu)}}(\bm{k};\varepsilon)$ contribution, as opposed to the $j$th contribution $\Sigma_{\sigma}^{\X{(\nu.j)}}(\bm{k};\varepsilon)$ to the latter, is independent of whether one employs $\h{\mathcal{H}}$, $\h{\mathsf{H}}$, or $\hspace{0.28cm}\h{\hspace{-0.28cm}\mathpzc{H}}$.}

In closing this section, the following two remarks are in order. Firstly, in the thermodynamic limit one has\,\footnote{Under some general assumptions regarding the behaviour of the function $f(\bm{k})$, for $\Omega\to\infty$ the leading correction to the RHS of Eq.\,(\protect\ref{ex05}) is of the order of $1/\Omega^{1/d}$, to be followed by a correction of the order of $1/\Omega^{2/d}$, \emph{etc.} [\S\,5.3.19 in Ref.\,\protect\citen{BF07}]. Here we have assumed a degree of isotropy in the shape of the region $\protect\b{\Omega}$, whereby $\Omega^{1/d}$ represents the order of magnitude of the length of the system in all spatial directions. These results, which are deduced from the Euler-Maclaurin summation formula [\S\,23.1.30, p.\,806, in Ref.\,\protect\citen{AS72}], are consequential to the order of the limits $\Omega\to\infty$ and $d\to\infty$.}
\begin{equation}\label{ex05}
\frac{1}{\Omega} \sum_{\bm{k}} f(\bm{k}) = \int \frac{\mathrm{d}^d k}{(2\pi)^d}\; f(\bm{k}).
\end{equation}
Applying the above equality to the expression on the RHS of Eq.\,(\ref{ex04}), for $d=3$ and $\omega_j = \varepsilon_j/\hbar$ one in particular recovers the prescription under point 7 on p.\,103 of Ref.\,\citen{FW03} referred to above. In this connection, note that since a proper self-energy diagram in terms of $\{G_{\X{0};\sigma}\| \sigma\}$ contributing to $\Sigma_{\sigma}^{\X{(\nu)}}(\bm{k};\varepsilon)$ is deduced from a Green-function diagram in terms of  $\{G_{\X{0};\sigma}\| \sigma\}$ contributing to $G_{\sigma}^{\X{(\nu)}}(\bm{k};\varepsilon)$ by amputating its two external Green-function lines, each representing the function $G_{\X{0};\sigma}(\bm{k};\varepsilon)$, it follows that the multiplicative numerical factor associated with a $\nu$th-order proper self-energy diagram is the same as that associated with the corresponding Green-function diagram.

Secondly, one has
\begin{equation}\label{ex08x}
 \frac{1}{N_{\textsc{s}}} \sum_{\bm{k}\in\fbz} 1 = 1,
\end{equation}
so that in the atomic / local limit, where the function $\b{F}_{\sigma}^{(\nu.j)}$ does \textsl{not} depend on $\bm{k}_1$, \dots, $\bm{k}_{\nu}$ (as well as $\bm{k}$), the appropriate expression corresponding to the self-energy diagram under consideration coincides with that in Eq.\,(\ref{ex07}), with the multiple normalised sums
\begin{equation}\label{ex09}
\frac{1}{N_{\textsc{s}}} \sum_{\bm{k}_1\in\fbz} \dots \frac{1}{N_{\textsc{s}}}\sum_{\bm{k}_{\nu}\in\fbz} \nonumber
\end{equation}
herein suppressed. In other cases, in the thermodynamic limit one employs
\begin{equation}\label{ex08}
\frac{1}{N_{\textsc{s}}} \sum_{\bm{k}\in \fbz} f(\bm{k}) = \Omega_{\textrm{u}} \int_{\fbz} \frac{\mathrm{d}^d k}{(2\pi)^d}\; f(\bm{k}),
\end{equation}
where
\begin{equation}\label{ex08a}
\Omega_{\textrm{u}} \doteq \frac{\Omega}{N_{\textsc{s}}},
\end{equation}
the volume per lattice site. For a $d$-dimensional hypercube with lattice parameter $a$, one has $\Omega_{\textrm{u}} = a^d$, so that by expressing, for this lattice, the wave vectors $\bm{k}_1$, \dots, $\bm{k}_{\nu}$ in units of $1/a$, $\Omega_{\textrm{u}}$ can be identified with unity. In this case, one has
\begin{equation}\label{ex09x}
\1BZ = (-\pi,\pi] \times \dots \times (-\pi,\pi].\;\;\;\; \text{($d$ times)}
\end{equation}

\refstepcounter{dummyX}
\subsection{Symbolic computation}
\phantomsection
\label{sd3}
The method of symbolic computation that we introduce and illustrate in this section\,\footnote{$\copyright$ 2021 \textsf{All methods, algorithms and programs presented in this section, as well as elsewhere in this publication, are intellectual property of the author. Any commercial use of these without his written permission is strictly prohibited. All academic and non-commercial uses of the codes in this publication, or modifications thereof, must be appropriately cited. The same restrictions apply to the contents of the Mathematica$^{\protect\X{\circledR}}$ notebook that we publish alongside this publication.}} aims at establishing the possible algebraic equivalence (up to determinate multiplicative constants of either sign) of the contributions of the proper self-energy diagrams of the same order that are topologically inequivalent \emph{without calculating their contributions, whether algebraically or numerically}. This method is based on the observation that contributions of two topologically-inequivalent self-energy diagrams of the same order may prove to be identical (up to the above-mentioned multiplicative constant) through a judicious choice of the energy-momentum vectors associated with the lines of one of the diagrams, while ensuring that this choice remains in conformity with the principle of conservation of energy-momentum at the vertices of this diagram.\footnote{Reference here solely to `energy-momentum vectors' reflects the fact that in this section we explicitly deal with spin-independent interaction potentials and unpolarised GSs (ESs). As we briefly indicate below, in a more general setting account has to be taken of more details than the energy-momentum conservation at the vertices of the self-energy diagrams.} The approach suggested itself to us in the course of the calculations underlying the algebraic expressions presented in \S\,\ref{sd4} below; the wealth of topologically-inequivalent however algebraically `equivalent' skeleton self-energy diagrams of the same order as considered in \S\,\ref{sd4} hinted at a systematic reason for this `equivalence'. Naturally, some `equivalence' was to be expected on account of the particle-hole symmetry of the half-filled GS of the `Hubbard atom' for spin-$\tfrac{1}{2}$ particles dealt with in \S\,\ref{sd4}.

The method of symbolic computation to be presented below is applicable to spin-unpolarised GSs (ESs)\,\footnote{Fully-spin-polarised GSs (ESs) can be similarly dealt with. This is relevant specifically for electrons in the quantum-Hall region \protect\cite{ZFE16,BF97b}, where they are subject to strong external magnetic field.} corresponding to the Hubbard Hamiltonian $\h{\mathcal{H}}$ in Eq.\,(\ref{ex01bx}).\footnote{These limitations apply also to the \textsl{non-interacting} GSs (ESs), for the case the diagrams are evaluated in terms of $\t{G}_{\protect\X{0};\sigma}(\bm{k};z)$ ($\t{\mathscr{G}}_{\protect\X{0};\sigma}(\bm{k};z)$).}\footnote{We recall that (see footnote \raisebox{-1.0ex}{\normalsize{\protect\footref{notef1}}} on p.\,\protect\pageref{ForSpin}), unless we indicate otherwise, here we assume that the one-particle Green matrices $\mathbb{G}$ and $\mathbb{G}_{\protect\X{0}}$, as well as the self-energy matrix $\mathbb{\Sigma}$, are diagonal in the spin space, so that our considerations are confined to the diagonal elements $G_{\sigma} \equiv (\mathbb{G})_{\sigma,\sigma}$, $G_{\protect\X{0};\sigma} \equiv (\mathbb{G}_{\protect\X{0}})_{\sigma,\sigma}$, and $\Sigma_{\sigma} \equiv (\mathbb{\Sigma})_{\sigma,\sigma}$.}\refstepcounter{dummy}\label{FromThePerspective}\footnote{From the perspective of the symbolic computations to be introduced in this section, the representation of the Hubbard Hamiltonian in Eq.\,(\protect\ref{ex01bx}) is to be strictly distinguished from the representations in Eqs\,(\protect\ref{ex01f}) and (\protect\ref{ex01k}). The factor $\mathsf{g} = 2\mathsf{s} + 1$ as considered in this section is tied to the double sum with respect to $\sigma$ and $\sigma'$ in Eq.\,(\protect\ref{ex01bx}) and the spin independence of the two-body interaction potential energy, Eq.\,(\protect\ref{ex03}); clearly the two-body interaction potential in Eq.\,(\protect\ref{ex01k}) is spin dependent. \label{noteu}} By spin-unpolarised GSs (ESs) we refer to those for which $\t{G}_{\sigma}(\bm{k};z)$ ($\t{\mathscr{G}}_{\sigma}(\bm{k};z)$) is \textsl{independent} of $\sigma$, where $\sigma$ takes $2\hspace{0.6pt}\mathsf{s} + 1$ different values for spin-$\mathsf{s}$ particles.\footnote{In this publication, generally $\mathsf{s} = \tfrac{1}{2}$.} This method is equally applicable when the alternative representations $\hspace{0.28cm}\h{\hspace{-0.28cm}\mathpzc{H}}$, Eq.\,(\ref{ex01f}), and $\h{\mathsf{H}}$, Eq.\,(\ref{ex01k}), of $\h{\mathcal{H}}$ are employed, provided that the diagrams considered are those in the set generated by the perturbation series expansion corresponding to $\hspace{0.28cm}\h{\hspace{-0.28cm}\mathpzc{H}}$; any diagram outside this set makes an identically-vanishing contribution when dealing with $\h{\mathsf{H}}$. On account of the equality in Eq.\,(\ref{ex04b}) in relation to that in Eq.\,(\ref{ex04a}), \refstepcounter{dummy}\label{InDealingWithH}\emph{in dealing with $\hspace{0.28cm}\h{\hspace{-0.28cm}\mathpzc{H}}$ and $\h{\mathsf{H}}$ the quantity $\mathsf{g} \doteq 2\mathsf{s} +1$, to be encountered below, \S\,\ref{sd31}, is to be identified with $1$.}

The above-mentioned limitations of the symbolic computation method to be introduced in this section are not fundamental, rather are adopted solely for keeping the presentation and the programming transparent. They can be relaxed so as to make the method suitable to the more general Hamiltonians $\h{\mathscr{H}}$, Eq.\,(\ref{ex01a}), and $\h{H}$, Eq.\,(\ref{ex01}), and to the more general GSs (ESs) than the spin unpolarised ones considered here. The method can be trivially extended for dealing with the spin-unpolarised\,\footnote{As well as fully-spin-polarised GSs (ESs). See above.} GSs (ESs) of the Hamiltonian $\h{H}$ in Eq.\,(\ref{ex01}) by symbolically considering the norms of the wave vectors associated with the signed labels $\{x_{2\nu},\dots,x_{3\nu-1}\}$, each index representing an energy-momentum, returned as output of program \texttt{Equiv}, p.\,\pageref{Equiv}, and establishing whether these are a permutation of the norms of the wave vectors associated with the signed labels $\{x_{2\nu},\dots,x_{3\nu-1}\}$ corresponding to the diagram A to be specified below, \S\,\ref{sd31}.\footnote{With reference to the first illustrative example in \S\,\protect\ref{sd33} below, following Eq.\,(\protect\ref{ed15z}), which corresponds to $\nu=2$, one has $\{x_4,x_5\} = \{p-p_1,p_1-p_2\}$. For the wave vector associated with $x_4$ one thus has $\bm{k}- \bm{k}_1$, and for that associated with $x_5$, $\bm{k}_1 - \bm{k}_2$, where $\bm{k}_i \equiv \bm{p}_i/\hbar$. In the two outputs of \texttt{Equiv} presented in Eq.\,(\protect\ref{ed15e}), one has $\{x_4,x_5\} = \{p-p_1, p-p_1\}$ and $\{x_4,x_5\} = \{p_1-p_2, p_1-p_2\}$. Since neither $\{\|\bm{k}-\bm{k}_1\|,\|\bm{k}-\bm{k}_1\|\}$ nor $\{\|\bm{k}_1-\bm{k}_2\|,\|\bm{k}_1-\bm{k}_2\|\}$ is a permutation of $\{\|\bm{k}-\bm{k}_1\|,\|\bm{k}_1-\bm{k}_2\|\}$, for the $\bm{q}$-dependent potential energy $\protect\b{v}(\|\bm{q}\|)$ diagrams A and B in the illustrative example under consideration are \textsl{not} equivalent in the sense considered in this section. This observation remains intact when the underlying spin-unpolarised GS (ES) is p-h symmetric; the set $\{x_4,x_5\}$ pertaining to the $6$ distinct outputs of program \texttt{Equiv} (to which we refer in \S\,\protect\ref{sd33}, p.\,\protect\pageref{WithoutGoing}), corresponding to the input parameters \texttt{all = 1} and \texttt{phs = 1}, read as follows: $\{p-p_1,p-p_1\}$, $\{p_1-p_2,p_1-p_2\}$, $\{p-p_1,p-p_1\}$, $\{p+p_2,p+p_2\}$, $\{p+p_2,p+p_2\}$, and $\{p_1-p_2,p_1-p_2\}$.}

Extension of the method of symbolic computation under discussion for dealing with the general Hamiltonian $\h{\mathscr{H}}$ in Eq.\,(\ref{ex01a}), and the less general Hamiltonian $\h{\EuScript{H}}$ in terms of the spin-dependent two-body potential in Eq.\,(\ref{ex01h}),\footnote{The indices $\sigma$ and $\sigma''$ in Eq.\,(\protect\ref{ex01h}) may refer to pseudo-spins. For instance, in coupled layers of two-dimensional electron systems in the quantum-Hall regime (where electron spins are fully polarised), these are layer indices. See part III, p.\,389, in Ref.\,\protect\citen{ZFE16}.} is straightforwardly realised, it requiring to define each signed label $x_i$, \ref{sx1}, as representing not only energy-momentum, but also spin indices. For instance, in dealing with the less general Hamiltonian $\h{\EuScript{H}}$, $x_i$ is to represent a $3$-vector, with its first component, $x_i^{\X{1}}$, serving as the $x_i$ in the considerations of the present section, \S\,\ref{sd3}, namely representing the energy-momentum, and the second and third components, respectively $x_i^{\X{2}}$ and $x_i^{\X{3}}$, representing the spin indices associated with the ends of the line to which $x_i$ is assigned, whether a Green-function line or an interaction line.\footnote{For the specific case of spin-$\tfrac{1}{2}$ particles, $x_i^{\protect\X{2}}, x_i^{\protect\X{3}} \in \{-1/2,+1/2\}$, or simply $x_i^{\protect\X{2}}, x_i^{\protect\X{3}} \in \{-1,+1\}$.} In dealing with spin-unpolarised GSs (ESs), the second and third components of $x_i$, with $i\in \{1,2,\dots,2\nu-1\}$ (corresponding to Green-function lines), are equal. For illustration, let the $3$-vector $x_l$, $l \in \{1,2,\dots,2\nu-1\}$, be assigned to the directed line $\langle i,j\rangle$ representing $G(j,i)$, \S\,\ref{sx1}. In the energy-momentum representation $\langle i,j\rangle$ symbolising $G_{\sigma_j,\sigma_i}(\bm{k}_l,\varepsilon_l)$, one has $x_l^{\X{1}} = (\varepsilon_l,\hbar\bm{k}_l)$, $x_l^{\X{2}} = \sigma_j$, and $x_l^{\X{3}} = \sigma_i$, where $\bm{k}_l$ stands for a linear combination of the external wave vector $\bm{k}$ and the internal wave vectors $\{\bm{k}_1,\bm{k}_2,\dots,\bm{k}_{\nu}\}$, and $\varepsilon_l$ for a linear combination of the external energy parameter $\varepsilon$ and the internal energy parameters $\{\varepsilon_1,\varepsilon_2,\dots,\varepsilon_{\nu}\}$ (\emph{cf.} Eq.\,(\ref{ex04})); these linear combinations guarantee the conservation of energy-momentum at the vertices $i$ and $j$ of the proper self-energy diagram under consideration. Similarly as regards the directed interaction lines, such as $\prec\hspace{-0.6pt}i,j\hspace{-0.6pt}\succ$ and the associated $x_l$, $l \in \{2\nu,\dots,3\nu-1\}$, except that in this case in the Fourier space the association of $x_l$ with $\b{v}_{\sigma_j,\sigma_i}(\|\bm{k}_l\|)$ entails $x_l^{\X{1}} = \hbar\bm{k}_l$, $x_l^{\X{2}} = \sigma_j$, and $x_l^{\X{3}} = \sigma_i$. In an analogous way, for dealing with the Hamiltonian $\h{\mathscr{H}}$, Eq.\,(\ref{ex01a}), $x_i$ is to represent a $5$-vector, as in this case two spin indices are associated with each end of an interaction line (see Fig.\,\ref{f4}, p.\,\pageref{TwoInternal}).\footnote{With reference to Eq.\,(\protect\ref{ex01a}), in this case, for $x_l$ assigned to the directed interaction line representing $\protect\b{\mathsf{v}}_{\sigma,\sigma';\sigma'',\sigma'''}(\|\bm{q}\|)$, one has $x_l^{\protect\X{1}} = \hbar\bm{q}$, $x_l^{\protect\X{2}} = \sigma$, $x_l^{\protect\X{3}} = \sigma'$, $x_l^{\protect\X{4}} = \sigma''$, and $x_l^{\protect\X{5}} = \sigma'''$.}

Despite the above-mentioned limitations of the formalism to be explicitly considered in this section, program \texttt{Equiv} on p.\,\pageref{Equiv} below can be fruitfully used in dealing with Hamiltonians $\h{\mathscr{H}}$, Eq.\,(\ref{ex01a}), and $\h{\EuScript{H}}$.\footnote{Recall that $\protect\h{\EuScript{H}}$ is defined as $\protect\h{\mathscr{H}}$ subject to the condition in Eq.\,(\protect\ref{ex01h}).} This on account of the fact that search for the equivalence of two diagrams corresponding to the latter Hamiltonians (possibly under the relaxation of the assumption of the underlying GSs (ESs) being spin-unpolarised) can be limited to those determined by program \texttt{Equiv} as being equivalent; those that are found not to be equivalent by \texttt{Equiv} for the spin-unpolarised GSs (ESs) of the Hamiltonian $\h{\mathcal{H}}$, \textsl{cannot} be equivalent for the GSs (ESs), whether spin-unpolarised or otherwise, of the Hamiltonians $\h{\mathscr{H}}$ and $\h{\EuScript{H}}$.

The symbolic-computation method that we introduce in this section can be straightforwardly modified and made applicable for dealing with the diagrams associated with other many-body correlation functions, such as the polarisation function considered in appendix \ref{sa}.

\refstepcounter{dummyX}
\subsubsection{Generalities}
\phantomsection
\label{sd31}
Before going into the details of the above-mentioned method of symbolic computation on \textsl{proper} (\emph{i.e.} 1PI) self-energy diagrams, which may or may not be \textsl{skeleton} (\emph{i.e.} 2PI),\footnote{They must however be in terms of the bare two-body interaction potential $v$, unless the energy-momentum dependence of the dynamic interaction potential $W$, appendix \protect\ref{sa}, is neglected. Such neglect in general amounts to an uncontrolled approximation.} below we first present an outline of the method.

Considering a set of $\nu$th-order proper self-energy diagrams, which may consist of all or a proper subset of such diagrams, we select from this set an arbitrary diagram, which we call diagram A, and proceed with determining the diagrams in the set that up to determinate constants of proportionality are equivalent with diagram A. For spin-$\mathsf{s}$ particles, the latter proportionality constants are of the form $\varsigma (-\mathsf{g})^{\delta\ell}$, where $\varsigma \in \{-1,+1\}$, $\mathsf{g} = 2\mathsf{s} +1$, and $\delta\ell \equiv \ell_{\textsc{b}} - \ell_{\textsc{a}}$, with $\ell_{\textsc{b}/\textsc{a}}$ denoting the number of closed fermion loops in diagram B$/$A ($\delta\ell$ may thus be positive, zero, and negative), Eq.\,(\ref{ex07});\refstepcounter{dummy}\label{InTheMore}\footnote{In the more general setting where the one-particle Green functions corresponding to different spin indices are not identified from the outset, \S\,\protect\ref{sd3}, whereby the relevant summations corresponding to the spin indices associated with the internal vertices of the self-energy diagrams are not evaluated implicity (resulting in the powers of $\mathsf{g}$), the proportionality factors are of the form $\varsigma (-1)^{\delta\ell}$. \label{notev}} generally $\varsigma = 1$, however for particle-hole (p-h) symmetric GSs (ESs) one may have $\varsigma = -1$.\footnote{The program \texttt{Equiv}, to be introduced below (p.\,\protect\pageref{Equiv}), determines the value of $\varsigma$. Program \texttt{Equiv} calls a second program named \texttt{Verify}, p.\,\protect\pageref{Verify}, that in the p-h symmetric cases (specified by assigning a non-zero integer to the input variable \texttt{phs}) verifies the validity of the calculations by means of performing a set of affine transformations of the independent energy-momentum vectors associated with the internal vertices of the diagrams.} Thus, for a set consisting of $\mathpzc{n}$ proper self-energy diagrams, one will be considering $\mathpzc{n}-1$ pairs of A-B diagrams, where A is fixed and B variable. Following the completion of this process, one will have determined a class of diagrams that are `equivalent' with diagram A, which we call class $\mathcal{A}_1$. This class may consist of only the diagram A itself, but also of all diagrams of the initial set of diagrams. In the case of not all diagrams falling into $\mathcal{A}_1$, we choose a diagram \textsl{outside} $\mathcal{A}_1$ and call it diagram A and proceed as above, and thus arrive at a class of diagrams that are `equivalent' with the new diagram A. This class we call class $\mathcal{A}_2$. Note that the B diagrams in the A-B pairs considered in determining the latter class are similarly to the new diagram A taken from the set of diagrams complementary to the class $\mathcal{A}_1$, this on account of the symmetric nature of the equivalence relations in general\,\footnote{For \textsl{equivalence relations}, consult \S\,1.2.1, p.\,8, in Ref.\,\protect\citen{AA14}.} and of the one considered here in particular, namely that when A is `equivalent' to B, so is B `equivalent' to A.

Proceeding as above, the original set of $\mathpzc{n}$ proper self-energy diagrams will be subdivided into $\mathpzc{n}_{\hspace{0.4pt}\textrm{c}}$ non-overlapping \textsl{classes} of diagrams $\mathcal{A}_1$, \dots, $\mathcal{A}_{\mathpzc{n}_{\hspace{0.4pt}\textrm{c}}}$ whose cardinal numbers we denote by $\vert\mathcal{A}_1\vert$, \dots, $\vert\mathcal{A}_{\mathpzc{n}_{\hspace{0.4pt}\textrm{c}}}\vert$, for which one has $\sum_{i=1}^{\mathpzc{n}_{\hspace{0.4pt}\textrm{c}}} \vert\mathcal{A}_i\vert = \mathpzc{n}$. One may have $\mathpzc{n}_{\hspace{0.4pt}\textrm{c}} = 1$ (corresponding to $\vert\mathcal{A}_1\vert = \mathpzc{n}$), but also $\mathpzc{n}_{\hspace{0.4pt}\textrm{c}} = \mathpzc{n}$ (corresponding to $\vert\mathcal{A}_i\vert = 1$, $\forall i$). The actual (analytical and/or numerical) calculation of the contributions of the set of $\mathpzc{n}$ proper self-energy diagrams of interest will thus be reduced to that of $\mathpzc{n}_{\hspace{0.4pt}\textrm{c}}$ representative diagrams of the classes of `equivalent' diagrams: one diagram from each class, with the contributions of the remaining diagrams in each class being trivially obtained through the multiplication of the contribution of the representative diagram with the relevant constants of the form $\varsigma (-\mathsf{g})^{\delta\ell}$, referred to above.

\refstepcounter{dummyX}
\subsubsection{Characterisation of the self-energy diagrams}
\phantomsection
\label{sd32}
Here\refstepcounter{dummy}\label{HereWeDescribe} we describe the way in which we characterise the $\nu$th-order proper self-energy diagrams A and B, referred to above, for the purpose of testing for their `equivalence'\,\footnote{That is, \textsl{equivalence} of their corresponding analytic expressions up to a determinate constant multiplying factor, which may be equal to unity. With this in mind, henceforth we shall generally use the word \textsl{equivalence} without marking it by the quotation marks.} by means of a \textsl{symbolic} computation -- for which no explicit \textsl{floating-point} arithmetic operations are carried out. As will become evident below, in our approach diagrams A and B are characterised differently: whereas we characterise diagram A by means of a $(2\nu-1)$-vector $\vec{\bm{\xi}}$, the characterisation of diagram B relies on two non-square sparse \cite{SP84} matrices $\mathbb{M}_{\textsc{l}}$ (of size $2\nu \times (3\nu-1)$) and $\mathbb{M}_{\textsc{s}}$ (of size $\nu \times (2\nu-1)$) whose entries consist of the elements of the set $\{-1,0,+1\}$.\footnote{In program \texttt{Equiv}, p.\,\protect\pageref{Equiv}, \texttt{ML} and \texttt{MS} represent respectively $\mathbb{M}_{\textsc{l}}$ and $\mathbb{M}_{\textsc{s}}$.} Here the subscript \textsc{l} (\textsc{s}) stands for `Large' (`Small'). The matrices $\mathbb{M}_{\textsc{l}}$ and $\mathbb{M}_{\textsc{s}}$ are similar, however \textsl{not} identical, to \textsl{incidence} matrices \cite{FH69,GY04} of directed graphs, or \textsl{digraphs}, as encountered in graph theory\cite{FH69}.\footnote{Some of the algorithms and the associated Mathematica programs that we present in Ref.\,\protect\citen{BF19} make use of graph theory.} The difference between the two types of matrices can be overcome by modifying the rules to be introduced in this appendix through which the relevant matrix equations (resulting in the above-mentioned matrices $\mathbb{M}_{\textsc{l}}$ and $\mathbb{M}_{\textsc{s}}$) are set up. Such modification is however \textsl{not} advantageous in the context of investigating the equivalence of the self-energy diagrams.\footnote{See the discussions following Eq.\,(\protect\ref{eac5d}) below.}

The above-mentioned characterisations rely on the principle of conservation of the energy-momentum at the vertices of the self-energy diagrams, which for proper self-energy diagrams of order $\nu$ gives rise to $2\nu$ linear equations.\footnote{Recall that a $\nu$th-order proper self-energy diagram is comprised of $2\nu$ vertices (of which $2$ external), $\nu$ interaction lines, and $2\nu-1$ Green-function lines. There are therefore $3\nu-1$ lines in this diagram. Considering an \textsl{imaginary} directed line that leaves the external vertex $2'$ and enters the external vertex $1'$, the total number of lines amounts to $3\nu$, with the latter \textsl{imaginary} line carrying the energy-momentum $p \equiv (\varepsilon,\hbar\bm{k})$. This \textsl{imaginary} line is depicted by a broken line in Fig.\,3, appendix C, of Ref.\,\protect\citen{BF19} (in this figure, the vertex number $3$ ($1$) takes the place of $2'$ ($1'$)). See also Figs\,\protect\ref{f15} and \protect\ref{f16} below, pp.\,\protect\pageref{DiagramA1} and \protect\pageref{DiagramA2}, where the above-mentioned \textsl{imaginary} line is not depicted. \label{notec}} This number is reduced to $2\nu-1$ on account of the requirement that the external energy-momentum $p \equiv (\varepsilon, \hbar\bm{k})$ entering the self-energy diagram\,\footnote{Representing the self-energy contribution $\Sigma^{\protect\X{(\nu.j)}}(2',1')$, \S\,\protect\ref{sx1}. For the superscript $(\nu.j)$, see Eqs\,(\protect\ref{ex1a}), (\protect\ref{ex1b}), and (\protect\ref{ex1c}) below.} at vertex $1'$ coincide with that leaving this diagram at vertex $2'$.\footnote{Pictorially, the energy-momentum $p$ is transferred through the \textsl{imaginary} line referred to in footnote \raisebox{-1.0ex}{\normalsize{\protect\footref{notec}}} above.} Assigning the energy-momentum vectors $\{p_1,p_2,\dots,p_{3\nu-1}\}$ to the $3\nu-1$ lines in a $\nu$th-order proper self-energy diagram, that is by equating the singed index $x_i$, \S\,\ref{sx1}, with $p_i$, $\forall i\in \{1,2,\dots,3\nu-1\}$, where $p_i \equiv (\varepsilon_i,\hbar\bm{k}_i)$, only $\nu$ of these can be chosen freely,\footnote{These $\nu$ energy-momentum vectors are integrated in evaluating the contribution of the corresponding proper self-energy diagram at the energy-momentum $p$ (\emph{cf.} Eq.\,(\protect\ref{ex07})).} the remaining $2\nu-1$ energy-momentum vectors being fixed by the above-mentioned $2\nu-1$ linear equations.\footnote{Solution of this set of linear equations is \textsl{not} unique, as evidenced by the different solutions obtained by program \texttt{Equiv}, p.\,\protect\pageref{Equiv}, on setting the input parameter \texttt{all} equal to a \textsl{non-zero} integer.} Thus, a general $\nu$th-order proper self-energy diagram can be characterised by means of a $(3\nu-1)$-vector whose specific $\nu$ components consist of the elements of $\{p_1,p_2,\dots, p_{\nu}\}$, and the remaining $2\nu-1$ components of the solutions of the above-mentioned set of $2\nu-1$ linear equations, with each solution $p_j$, $j = \nu+1, \nu+2, \dots, 3\nu-1$, expressed as a linear combination (with \textsl{integer} coefficients) of the elements of the set $\{p, p_1, p_2, \dots, p_{\nu}\}$. Since by assumption, \S\,\ref{sd3}, the bare two-body interaction potential is independent of energy-momentum, it follows that diagram A in the restricted framework considered in this section can be adequately characterised by means of a $(2\nu-1)$-vector deduced from the above-mentioned $(3\nu-1)$-vector through leaving out the components associated with the energy-momentum vectors of the interaction lines. This reduced $(2\nu-1)$-vector coincides, up to an ordering of its components, with the vector $\vec{\bm{\xi}}$ (referred to above) that we shall explicitly specify below.

For completeness, we point out that the interrelationship between Feynman diagrams and graphs has long since been recognised and investigated \cite{NN71,DK00}. The formalism in this section has arisen from the necessities of the investigations underlying the present publication.\footnote{Due to circumstances resulting in relatively wide gaps between our publications, the transitions between the present publication and Refs.\,\citen{BF19} and \protect\citen{BF21b} are not seamless. We hope to amend this shortcoming in a future publication, providing in this a set of programs that seamlessly and without any manual intervention algebraically compute various many-body correlation functions (including the simplifications that are partly the subject of the present section, \emph{i.e.} \S\,\protect\ref{sd3}) for normal, superfluid, and superconductive states.} It is interesting to note that theoretical and computational aspects related to sparse matrices, in particular their decomposition into computationally convenient forms, and whether or not these are in turn sparse, have long-standing and well-established connections to graph theory \cite{DJR72,SP84}.

We\refstepcounter{dummy}\label{WeAreNow} are now in a position to present a more explicit description of the way in which we specify the pair of the proper self-energy diagrams A and B, \S\,\ref{sd31}, with an eye on the subsequent test of these diagrams for equivalence. To this end, we identify the energy-momentum $(\varepsilon,\hbar\bm{k})$ entering and leaving both diagrams, A and B, at respectively $1'$ and $2'$, with $p$, to be denoted by the \textsl{symbol} \texttt{p} in the program \texttt{Equiv} to be introduced below, p.\,\pageref{Equiv}.\footnote{Any other \textsl{symbol} will be equally valid, such as \texttt{e} (which is a natural choice in dealing with the Hubbard model in the \textsl{atomic} limit), however this \textsl{symbol} must be used consistently and \textsl{never} assigned any numerical value (symbolic computations in which \texttt{p}, or \texttt{e}, is equated with a number are devoid of meaning in the context of the considerations of this section, \emph{i.e.} \S\,\protect\ref{sd3}). We therefore strongly recommend that before its first use, \texttt{p} be cleared, using \texttt{Clear[p]}, or \texttt{ClearAll[p]}. The same applies to all symbols that are to be treated as such, as opposed to \textsl{variables} to which values are assigned in advance of computation (such as \texttt{phs} and \texttt{all}, p.\,\protect\pageref{Equiv}).} Following the above considerations, we specify diagram A by means of the $(2\nu-1)$-vector $\vec{\bm{\xi}}$ defined as
\begin{equation}\label{eac4a}
(\vec{\bm{\xi}})_i = x_i,\;\; i\in \{1,2,\dots,2\nu-1\},
\end{equation}
where \textsl{by convention} (\emph{cf.} Eqs\,(\ref{ed15}) and (\ref{ed15i}) below)
\begin{equation}\label{eac4b}
x_i \equiv p_i,\;\; i\in \{1,2,\dots,\nu\}.
\end{equation}
As we have indicated earlier, each $x_i$, $\nu < i \le 3\nu-1$, is a solution of a set of $2\nu-1$ linear equations and is described as a linear combination (with integer coefficients) of the elements of the set $\{p, p_1,p_2,\dots,p_{\nu}\}$ (\emph{cf.} Eqs\,(\ref{ed15}) and (\ref{ed15i}) below). In the light of the specifications in \S\,\ref{sx1}, one observes that the components of $\vec{\bm{\xi}}$ correspond to the signed labels associated with the Green-function lines in diagram A.

Having specified diagram A in terms of the vector $\vec{\bm{\xi}}$, Eq.\,(\ref{eac4a}), we proceed with the specification of diagram B. As in the case of diagram A, for this we rely on the conventions introduced in \S\,\ref{sx1} regarding the labeling of the vertices and lines of the self-energy diagrams. In particular, as in the case of diagram A, all interaction lines in diagram B must be directed, even though by assumption the underlying two-body interaction potential is independent of energy-momentum. By the nature of the proper self-energy diagrams, always one \textsl{internal} Green-function line leaves (enters) the external vertex $1'$ ($2'$). Since further by the convention laid down in \S\,\ref{sx1} one has the index $1'$ ($2'$) in the first (second) entries of the directed lines specified by $\langle i',j'\rangle$ and $\prec\hspace{-3.0pt} i',j'\hspace{-3.0pt}\succ$ (\emph{cf.} Eqs\,(\ref{eac3}) and (\ref{eac4})), for the energy-momentum conservation equations corresponding to vertices $1'$ and $2'$ one invariably has
\begin{subequations}\label{eac5}
\begin{align}
\label{eac5a}
& 1':\hspace{0.2cm} x_i + x_j = p, \\
& 2':\hspace{0.2cm} x_k + x_l = p,
\label{eac5b}
\end{align}
where $x_i$ and $x_j$ ($x_k$ and $x_l$) stand for the signed labels associated with the \textsl{internal} lines that leave (enter) vertex $1'$ ($2'$).\footnote{These details are relevant in that by the adopted conventions the pre-factors of $x_i$, $x_j$, $x_k$ and $x_l$ in Eqs\,(\protect\ref{eac5a}) and (\protect\ref{eac5b}) are equal to $+1$. This aspect is reflected in the fact that each one of the first two rows of matrix $\mathbb{M}_{\textsc{l}}$, referred to in \S\,\protect\ref{sd32}, consists solely of \textsl{two} non-zero entries equal to $1$ (\emph{cf.} Eqs\,(\protect\ref{ed15b}) and (\protect\ref{ed15k}) below).} \emph{For uniformity of notation, by convention we assume $i < j$ and $k < l$}.\footnote{Compare with the convention spelled out following Eq.\,(\protect\ref{eac5f}) below.} We note that depending on the nature of diagram B, $k$ or $l$ need not be different from $i$ or $j$. For instance, in the case of the Fock self-energy diagram one has $i=k =1$ and $j=l = 2$.

Displaying the vertex numbers $1'$ and $2'$ to the left of the equations in Eqs\,(\ref{eac5a}) and (\ref{eac5b}) serves to indicate that the primed numbers associated with vertices of diagram B correspond to the row index of the $2\nu \times (3\nu-1)$ matrix $\mathbb{M}_{\textsc{l}}$, referred to above. With this in mind, the subscripts of the relevant signed labels in the set $\{x_1, x_2, \dots, x_{3\nu-1}\}$ correspond to the column index of the latter matrix.\footnote{It thus becomes apparent why $\protect\mathbb{M}_{\protect\textsc{l}}$ is a $2\nu \times (3\nu-1)$ matrix.} In this light, the first (second) row of the matrix $\mathbb{M}_{\textsc{l}}$ has all its entries equal to $0$, except those at the columns indexed $i$ and $j$ ($k$ and $l$), which are equal to $1$ (\emph{cf.} Eqs\,(\ref{ed15b}) and (\ref{ed15k}) below). For symbolic computation, it is of vital importance that the latter $0$ and $1$ be strictly distinguished from respectively $0.0$ and $1.0$. With reference to our earlier remark concerning the similarity between the matrix $\mathbb{M}_{\textsc{l}}$ and the \textsl{incidence} matrix of a \textsl{digraph} (p.\,\pageref{HereWeDescribe}), and that these two are \textsl{not} identical, we note that in an incidence matrix $-1$ ($+1$) is used to signify an edge directed \textsl{towards} (\textsl{from}) a vertex \cite{GY04}. To comply with this convention, we would have to write the energy-momentum conservation equations in Eq\,(\ref{eac5a}) and (\ref{eac5b}) as
\begin{align}
\label{eac5c}
& 1':\hspace{0.32cm} x_i + x_j - p = 0, \\
& 2':\hspace{0.0cm} -x_k - x_l + p = 0.
\label{eac5d}
\end{align}
\end{subequations}
These equations are consistent with our convention with regard to \textsl{directed} Green-function and interaction lines (as stated in the paragraph immediately preceding \S\,\ref{sd11}, p.\,\pageref{InTheFollowing}), although they deviate from the convention underlying Eqs\,(\ref{eac5a}) and (\ref{eac5b}), according to which the program \texttt{Equiv},\footnote{Written in the Mathematica programming language \protect\cite{SW16}.} to be discussed below (p.\,\pageref{Equiv}), has been set up. Disregarding the latter consideration, the equations in Eqs\,(\ref{eac5c}) and (\ref{eac5d}) importantly assume the two external Green-function lines that have been amputated from the Green-function diagram underlying the self-energy diagram under consideration (that is, diagram B) as being part of this diagram (viewed as a graph), these lines carrying the energy-momentum $p$ into and out of the self-energy diagram.\footnote{This aspect is in conformity with the conventions adopted in appendices B and C of Ref.\,\protect\citen{BF19}.} Already the fact that matrix $\mathbb{M}_{\textsc{l}}$ does \textsl{not} take account of the energy-momentum $p$, suffices to show that it differs from a graph-theoretical \textsl{incidence} matrix. Even neglecting this fact, as will become evident from the following, any attempt to bring matrix $\mathbb{M}_{\textsc{l}}$ into conformity with an \textsl{incidence} matrix is unnecessarily restrictive and cumbersome; attempts of this kind only amount to a wasteful effort of maintaining some equations as they are and multiplying others by $-1$.

Following the above specification of the first two rows of the integer matrix $\mathbb{M}_{\textsc{l}}$, the remaining $2\nu-2$ rows of this matrix are fully specified by considering the $2\nu-2$ equations describing the conservation of the energy-momentum at the $2\nu-2$ \textsl{internal} vertices of the self-energy diagram B. Our adopted conventions, as implemented in the Mathematica program \texttt{Equiv}, p.\,\pageref{Equiv}, require that these equations be expressed in homogenised form, that is that the three\,\footnote{This number, \emph{i.e.} three, coincides with the total number of lines (two Green-function lines and one interaction line) meeting at any \textsl{internal} vertex of a (proper) self-energy diagram. For this, consider the fundamental vertices $i'$ and $j'$ in Fig.\,\protect\ref{f4} above, p.\,\protect\pageref{TwoInternal}.} relevant signed labels $x_i$, $x_j$ and $x_k$, $i,j,k \in \{1,2,\dots, 2\nu\}$, each representing an energy-momentum vector, stand to the left and $0$ to the right of the equality sign.\footnote{With reference to our earlier remark concerning the \textsl{incidence} matrices of digraphs, note that this equation remains valid by multiplying it by $-1$.} The coefficients of $x_i$, $x_j$ and $x_k$, which may be either $-1$ or $+1$, in the equation corresponding to the \textsl{internal} vertex $i'$ of diagram B, where $i' \in \{3', 4', \dots, (2\nu)'\}$, fully determine the three non-vanishing entries of the $i'$th row of the matrix $\mathbb{M}_{\textsc{l}}$; all entries are equal to $0$, except the entries at the columns indexed $i$, $j$, and $k$. Since either two lines enter and one line leaves, or one line enters and two lines leave an internal vertex, it follows that in principle each of the last $2\nu-2$ rows of the matrix $\mathbb{M}_{\textsc{l}}$ consists of either two $+1$s and one $-1$, \textsl{or} of two $-1$s and one $+1$. In this connection, considering the \textsl{internal} vertex $i' \in \{3',4',\dots, (2\nu)'\}$, one in general has
\begin{equation}\label{eac5e}
i':\hspace{0.32cm} \varsigma_i x_i + \varsigma_j x_j + \varsigma_k x_k = 0,
\end{equation}
where $\{x_i, x_j, x_k\}$ are the signed labels (representing energy-momentum vectors) associated with the three \textsl{directed} lines\,\footnote{With reference to Fig.\,\protect\ref{f4} above, p.\,\protect\pageref{TwoInternal}, two of these lines represent the one-particle Green function, and one the two-body interaction potential.} meeting at the internal vertex $i'$, and where two of the three variables $\{\varsigma_i, \varsigma_j, \varsigma_k\}$ are equal to $+1$ ($-1$) and the third one equal to $-1$ ($+1$). Since the equation in Eq.\,(\ref{eac5e}) is homogenous and $\varsigma_l \in \{-1,+1\}$, $l=i, j, k$, by multiplying both sides of Eq.\,(\ref{eac5e}) with, say, $\varsigma_i$, one obtains the equivalent equation
\begin{equation}\label{eac5f}
i':\hspace{0.32cm} x_i + \varsigma_j' x_j + \varsigma_k' x_k = 0,
\end{equation}
where $\varsigma_l' \doteq \varsigma_i \varsigma_l \in \{-1,+1\}$, $l= j, k$. \emph{In the following we adopt the convention in Eq.\,(\ref{eac5f}), and for uniformity of notation assume that $i < j < k$} (\emph{cf.} Eqs\,(\ref{ed15a}) and (\ref{ed15j}) below). Note that $\varsigma_j' = - \varsigma_k'$, so that, in view of Eq.\,(\ref{eac5f}), the $i'$th row of the matrix $\mathbb{M}_{\textsc{l}}$ consists of two $+1$s (the first of which in the $i$th column) and one $-1$, with the remaining elements $0$ (\emph{cf.} Eqs\,(\ref{ed15b}) and (\ref{ed15k}) below). Further, by the definition of $\varsigma_l'$, $l=j, k$, $\varsigma_j' = 1$ ($\varsigma_k' = 1$) if the direction, in regard to the vertex $i'$, of the line associated with $x_j$ ($x_k$) is the same\,\footnote{That is, both directed towards or away from $i'$.} as that of the line associated with $x_i$.

To specify the $\nu \times (2\nu-1)$ matrix $\mathbb{M}_{\textsc{s}}$, referred to above, as we have hinted earlier, the $\nu$ conservation equations underlying this matrix are to be selected out of the $2\nu$ equations associated with the rows of the $2\nu \times (3\nu-1)$ matrix $\mathbb{M}_{\textsc{l}}$ in the range $\{3,4,\dots, 2\nu\}$. The choice must be from the equations involving the elements of $\{x_{2\nu}, x_{2\nu+2},\dots, x_{3\nu-1}\}$, which by convention, \S\,\ref{sx1}, correspond to directed \textsl{interaction} lines.\footnote{Note for instance that the $i$th component of the $(2\nu-1)$-vector $\vec{\xi}$ is equal to $x_i$, $i=1,2,\dots, 2\nu-1$, Eq.\,(\protect\ref{eac4b}), and corresponds to a Green-function line in diagram A.} After having made this selection, we express each relevant equation as
\begin{equation}\label{eac6}
i:\;\;\; x_{2\nu-1+i} = \varsigma_j\hspace{0.4pt} x_j + \varsigma_k\hspace{0.4pt} x_k,\;\; i \in\{1,2,\dots,\nu\},
\end{equation}
where $j, k \in \{1,2,\dots,2\nu-1\}$ and $\varsigma_j, \varsigma_k \in \{-1,+1\}$. Here $i$ refers to the $i$th row, and $j$ and $k$ refer to the $j$th and $k$th columns of the $\nu \times (2\nu-1)$ matrix $\mathbb{M}_{\textsc{s}}$. Thus, the entries of each row of this matrix are equal to $0$, except the entries at the columns indexed $j$ and $k$ where they are equal to respectively $\varsigma_j$ and $\varsigma_k$ (\emph{cf.} Eqs\,(\protect\ref{ed15d}) and (\ref{ed15m}) below). As in the case of $\mathbb{M}_{\textsc{l}}$, here also the symbols $0$ and $1$ are strictly to be distinguished from their numerically-equivalent representations $0.0$ and $1.0$.

Before introducing the program \texttt{Equiv}, p.\,\pageref{Equiv}, referred to above, we briefly discuss the case of particle-hole (p-h) symmetric GSs (ESs) from the limited perspective that the systems under consideration are spin-unpolarised, \S\,\ref{sd31}, meaning that the underlying one-particle Green functions $\{\t{G}_{\sigma}(\bm{k};z) \| \sigma\}$ do not explicitly depend on $\sigma$.\footnote{For a more general treatment, consult \S\,4.5.2., p.\,171, in Ref.\,\citen{PF03} (\S\,4.5 of this reference deals with various symmetries; time-reversal symmatry/invariance, which is also relevant to a more general consideration, is dealt with in detail in \S\,3.7, p.\,104, of this reference). Consult also Refs\,\protect\citen{EF13} and \protect\citen{JI16}.} For the specific case of spin-$\tfrac{1}{2}$ particles, one thus has $\t{G}_{\uparrow}(\bm{k};z) \equiv \t{G}_{\downarrow}(\bm{k};z)$. We take our cue from the one-particle Green function corresponding to a non-interacting GS (ESs),\footnote{The properties in Eqs\,(\protect\ref{eac6e}) and (\protect\ref{eac6f}) below are inherited by the \textsl{interacting} Green function at arbitrary finite order of the perturbation theory.} for which one has (\emph{cf.} Eq.\,(\ref{e7da}))\,\footnote{We continue to use the spin index $\sigma$ in order to avoid introduction of a new symbol.}
\begin{equation}\label{eac6a}
\t{G}_{\X{0};\sigma}(\bm{k};z) = \frac{\hbar}{z -\varepsilon_{\bm{k}}}.
\end{equation}
By time-reversal symmetry,
\begin{equation}\label{eac6b}
\varepsilon_{-\bm{k}} \equiv \varepsilon_{\bm{k}},
\end{equation}
whereby
\begin{equation}\label{eac6c}
\t{G}_{\X{0};\sigma}(-\bm{k};z) \equiv \t{G}_{\X{0};\sigma}(\bm{k};z).
\end{equation}
The p-h symmetry of the $N$-particle GS (ES) under consideration implies existence of a point $\bm{k}_0$ in the underlying $\bm{k}$-space for which, and for the corresponding chemical potential $\mu$, one has\,\footnote{For instance, in the case of the single-band Hubbard model for spin-$\tfrac{1}{2}$ particles on a square lattice of lattice constant $a$ and with the particles restricted to nearest-neighbour hopping -- quantified by the hopping integral $t$, for the non-interacting energy dispersion $\varepsilon_{\bm{k}}$ one has $\varepsilon_{\bm{k}} = -2t\hspace{0.6pt} (\cos(a k_x) + \cos(a k_y))$, where $(k_x, k_y)$ are the Cartesian coordinates of $\bm{k}$. In this case, the GS is p-h symmetric at half-filling, for which $\mu=0$. With $\bm{k} = \bm{0}$ identified with the centre of the relevant $\protect\1BZ$, for $\bm{k}_0 \in \protect\1BZ$ one has $\bm{k}_0 = (\varsigma_x \pi/a, \varsigma_y\pi/a)$, where $\varsigma_{\alpha} \in \mathds{R}$ and $\vert\varsigma_{\alpha}\vert = 1$ for $\alpha = x, y$. One generally takes $\bm{k}_0 = (\pi/a,\pi/a)$, on account of $\protect\1BZ \equiv (-\pi/a, \pi/a] \times (-\pi/a, \pi/a]$, Eq.\,(\protect\ref{ex09x}).}
\begin{equation}\label{eac6d}
\mu - \varepsilon_{\bm{k}} = \varepsilon_{\bm{k}+\bm{k}_0} - \mu,\;\; \forall \bm{k},
\end{equation}
whereby
\begin{equation}\label{eac6e}
\t{G}_{\X{0};\sigma}(-\bm{k};-z) = - \t{G}_{\X{0};\sigma}(\bm{k}_0 -\bm{k};z + 2\mu),\;\;\forall\bm{k},
\end{equation}
which, following Eq.\,(\ref{eac6c}), can be equivalently written as
\begin{equation}\label{eac6f}
\t{G}_{\X{0};\sigma}(-\bm{k};-z) = - \t{G}_{\X{0};\sigma}(\bm{k} -\bm{k}_{0};z + 2\mu),\;\;\forall\bm{k}.
\end{equation}
With
\begin{equation}\label{eac6g}
p \doteq (z,\hbar\bm{k}),
\end{equation}
on defining
\begin{equation}\label{eac6h}
\t{\mathsf{G}}_{\X{0};\sigma}(p) \doteq \t{G}_{\X{0};\sigma}(\bm{k};z),
\end{equation}
the equality in Eq.\,(\ref{eac6f}) can be written as
\begin{equation}\label{eac6i}
\t{\mathsf{G}}_{\X{0};\sigma}(-p) = - \t{\mathsf{G}}_{\X{0};\sigma}(p+p_0),
\end{equation}
where
\begin{equation}\label{eac6j}
p_0 \doteq (2\mu, -\hbar\bm{k}_0).
\end{equation}
The equalities in Eqs\,(\ref{eac6f}) and (\ref{eac6i}) apply also to the one-particle Green functions corresponding to spin-unpolarised p-h symmetric \textsl{interacting} uniform GSs (ESs), a fact that can be established order-by-order within the framework of the many-body perturbation theory for $\t{G}_{\sigma}(\bm{k};z)$ ($\t{\mathscr{G}}_{\sigma}(\bm{k};z)$). For later reference, we therefore write
\begin{equation}\label{eac6l}
\t{G}_{\sigma}(-\bm{k};-z) = - \t{G}_{\sigma}(\bm{k} -\bm{k}_{0};z + 2\mu) \, \Longleftrightarrow\, \t{\mathsf{G}}_{\sigma}(-p) = - \t{\mathsf{G}}_{\sigma}(p+p_0).
\end{equation}
It is interesting to note that, following Eq.\,(\ref{e4d}), the equality on the left implies that\,\footnote{Writing the equality in Eq.\,(\protect\ref{eac6m}) equivalently as $A_{\sigma}(\bm{k};\varepsilon) = A_{\sigma}(-\bm{k}-\bm{k}_0;-\varepsilon+2\mu)$, on the basis of time-reversal symmetry one equivalently has $A_{\sigma}(\bm{k};\varepsilon) = A_{\sigma}(\bm{k}+\bm{k}_0;-\varepsilon+2\mu)$. For $\mu = 0$ this equality coincides with the symmetry relation in Eq.\,(18) of Ref.\,\protect\citen{GEH00}, in which $\bm{Q}$ coincides with our $\bm{k}_0$. With $\bm{k}_0 = (\pi,\pi)$ (in the units where the lattice constant $a=1$), the symmetry relation in Eq.\,(\protect\ref{eac6n}) below can be gleaned from the data in the right-hand panels of Fig.\,1 in Ref.\,\protect\citen{GEH00}. See also the comparable numerical data in Ref.\,\protect\citen{OS04}.}
\begin{equation}\label{eac6m}
A_{\sigma}(-\bm{k};-\varepsilon) = A_{\sigma}(\bm{k}-\bm{k}_0;\varepsilon+2\mu),\;\;\forall\bm{k},\; \forall\varepsilon\in \mathds{R},
\end{equation}
from which, and by time-reversal symmetry, it follows that
\begin{equation}\label{eac6n}
A_{\sigma}(\varsigma\bm{k}_0/2;\mu-\varepsilon) = A_{\sigma}(\varsigma'\bm{k}_0/2;\mu+\varepsilon),\;\;\varsigma,\varsigma' \in \{-1,+1\},\; \forall\varepsilon\in \mathds{R}.
\end{equation}
On observes that for p-h symmetric GSs (ESs) the single-particle spectral function $A_{\sigma}(\bm{k};\varepsilon)$ is a symmetric function of $\varepsilon$ with respect to $\varepsilon = \mu$ at $\bm{k}= \pm\bm{k}_0/2$. This result applies for all $\bm{k}= \mathbb{\alpha}_i\bm{k}_0/2$, where $\mathbb{\alpha}_i$ is an element of the set $\{\mathbb{\alpha}_i \| i\}$ of the point-group symmetry operations of the underlying lattice. In the specific case of $\mu = 0$, the function $A_{\sigma}(\bm{k};\varepsilon)$ is therefore an \textsl{even} function of $\varepsilon$ for $\bm{k}\in \{\mathbb{\alpha}_i\bm{k}_0/2\| i\}$. Thus, with reference to Eq.\,(\ref{e4o}), in this case $G_{\sigma;\infty_{2j}}(\bm{k}) \equiv 0$, $\forall j \in \mathds{N}$, for $\bm{k} \in\{\mathbb{\alpha}_i \bm{k}_0/2\| i\}$.

For\refstepcounter{dummy}\label{ForParticleHole} p-h symmetric GSs (ESs),\footnote{Specified by equating the variable \texttt{phs} with a non-vanishing integer.} the program \texttt{Equiv}, p.\,\pageref{Equiv}, reverses with respect to the origin the $2\nu-1$ energy-momentum vectors associated with the internal Green-function lines of diagram A in all admissible ways and checks for the equivalence of the corresponding diagrams with diagram B. In doing so, the program \texttt{Verify}, p.\,\pageref{Verify}, verifies whether on the basis of the correspondence in Eq.\,(\ref{eac6l}) the resulting $p_0$ in the relevant arguments can be eliminated by means of a consistent set of affine transformations of the energy-momentum $(1+d)$-vectors\,\footnote{As we have indicated earlier, in the \textsl{local} limit one deals with $1$-vectors, $\{\varepsilon_1,\varepsilon_2,\dots, \varepsilon_{\nu}\}.$} $\{p_1,p_2,\dots,p_{\nu}\}$ that are integrated over in evaluating the contribution of diagram A. The constant $\varsigma$ introduced in \S\,\ref{sd31} amounts to $(-1)^{\uprho}$, where $\uprho$ is the number of the reversals $(\vec{\bm{\xi}})_i \to -(\vec{\bm{\xi}})_i$ effected, with each $-1$ accounting for the minus sign on the RHS of the right-most equality in Eq.\,(\ref{eac6l}).\footnote{More precisely, let $(\vec{\bm{\xi}})_i \to -(\vec{\bm{\xi}})_i $ for $i\in \mathcal{I}$, where $\mathcal{I} \subseteq \{1,2,\dots,2\nu-1\}$. One has $\uprho = \vert\mathcal{I}\vert$, the cardinal number of $\mathcal{I}$.}

We are now in a position to introduce the program \texttt{Equiv}, referred to above, which has been written in the Mathematica programming language \cite{SW16}. This program, which is fully parallelisable, determines whether the $\nu$th-order proper self-energy diagram B is equivalent with the $\nu$th-order proper self-energy diagram A. As we have indicated earlier, these diagrams may or may not be skeleton.\footnote{By restricting oneself to skeleton self-energy diagrams, the parts in program \texttt{Equiv}, p.\,\protect\pageref{Equiv}, involving \texttt{DeleteDuplocates}, \texttt{SameQ}, and the vector \texttt{h} become redundant. Operations involving the latter are necessary in dealing with proper \textsl{non-skeleton} self-energy diagrams, this on account of the non-trivial powers of one-particle Green functions in the algebraic expressions associated with these diagrams.} With the input parameter \texttt{all} $\not=$ \texttt{0} (see later) and in the absence of p-h symmetry,\footnote{That is, for \texttt{phs = 0}.} for \textsl{skeleton} self-energy diagrams of order $\nu$ the arithmetic complexity of the computations is of the order of $(2\nu-1)!$. This arithmetic complexity is \textsl{increased} to $2^{2\nu -1} (2\nu-1)!$ when allowing for p-h symmetry. For \textsl{non-skeleton} proper self-energy diagrams, the latter complexity is reduced to $2^{\nu'} (2\nu-1)!$, where $\nu' < 2\nu -1$.

The above arithmetic complexities can be substantially reduced by setting the input parameter \texttt{all} equal to \texttt{0}.\footnote{Ideally, the factor $(2\nu-1)!$ in the previous paragraph drops out. In this connection, it may prove useful to randomize the selection of the $(2\nu-1)$-permutation operations in \texttt{Equiv}, instead of sequentially moving through the sequence of these permutations arranged according to a fixed order (as is the case in the present implementation of \texttt{Equiv}).} By doing so, a loop inside \texttt{Equiv} that searches for all allowable permutations (and energy-momentum reversals in the case of \texttt{phs} $\not=$ \texttt{0}) of the components of $\vec{\bm{\xi}}$ for which the algebraic expressions corresponding to diagrams A and B coincide (up to a determinate multiplicative constant), is terminated on finding the first relevant permutation (and the possible appropriate energy-momentum reversals).

Regarding the input parameters of \texttt{Equiv}, the parameters \texttt{ML} and \texttt{MS} represent the \textsl{integer} matrices $\mathbb{M}_{\textsc{l}}$ and $\mathbb{M}_{\textsc{s}}$ defined above; \texttt{nu} stands for the \textsl{integer} $\nu$, denoting the order of the self-energy diagrams A and B under investigation; \texttt{p} is a \textsl{symbolic character}, with no numerical value assigned to it, representing the external energy-momentum $p = (\varepsilon,\hbar\bm{k})$ in general and the external energy $p = \varepsilon$ in the local / atomic limit; \texttt{xi} represents the $(2\nu-1)$-vector $\vec{\bm{\xi}}$, characterising diagram A, Eq.\,(\ref{eac4a});\footnote{While in principle unnecessary, we have defined \texttt{p} as an input parameter of \texttt{Equiv} by the consideration that in the local limit one might be inclined to define $\vec{\bm{\xi}}$ in terms of \texttt{e}, \texttt{e1}, \dots, \texttt{enu}, in which case one would need to pass this information to program \texttt{Equiv}.} by convention, the first $\nu$ components of $\vec{\bm{\xi}}$ are the \textsl{symbolic characters} \texttt{p1}, \dots, \texttt{pnu}, Eqs\,(\ref{eac4a}) and (\ref{eac4b}), with \texttt{pj}, \texttt{j} $=$ \texttt{1, 2, \dots, nu}, representing the \textsl{internal} energy-momentum vector $p_j = (\varepsilon_j,\hbar\bm{k}_j)$, and each of the remaining $\nu-1$ components consists of an appropriate linear combination of \texttt{p}, \texttt{p1},\dots, \texttt{pnu};\,\footnote{Each component of $\vec{\bm{\xi}}$ is an element of the set of signed labels $\{x_1,x_2,\dots,x_{2\nu-1}\}$; it is thus associated with a Green-function line in diagram A.} \texttt{phs} is an \textsl{integer}, with $0$ specifying that \textsl{no} p-h symmetry is to be assumed, and any other integer specifying the contrary assumption; and finally, \texttt{all} specifies whether (\texttt{all} $\not=$ \texttt{0}) or not (\texttt{all} $=$ \texttt{0}) all possible allowable indexing of the internal lines of diagram B are to be determined on establishing the equivalence of diagram B with diagram A.

{\footnotesize\refstepcounter{dummy}\label{Equiv}
\begin{verbatim}

(* Program `Equiv' whose input parameters are specified in the main text. *)
ClearAll[Equiv];
Equiv[ML_, MS_, nu_, p_, xi_, phs_, all_] :=
  Module[{i, j, k, l, m, n, n1, nf, q, v, xo, y, s, sig, g, h, T, T0,
    T1, T2, S0, SIG, ONE, Vr},
   Print[Style["Program Equiv", Gray, Bold, 16]];
   Print[Style[DateString[], Gray]];
   Print["\[Nu]: ", nu, ". Particle-hole symmetry? ",
    If[phs == 0, Style[No, Red], Style[Yes, Red]],
    " | Search for all possibilities? ",
    If[all == 0, Style[No, Red], Style[Yes, Red]]];
   n = 2 nu - 1; n1 = n; nf = n!;
   v = Join[{p, p}, Table[0, {i, 1, n - 1}]];
   ONE = Table[1, {i, 1, n}];
   Do[h[i] = i, {i, 1, n}];
   If[phs == 0, 99, (S0 = DeleteDuplicates[xi];
     n1 = Count[S0, Except[-1000]];
     If[n1 == n,
      98, (Do[Do[
         If[SameQ[xi[[i]], S0[[k]]], (h[i] = k), 97], {k, 1, n1}], {i,
          1, n}])])];
   T0 = Range[n]; If[phs == 0, (m = 1), (m = 2^n1)];
   Print["Number of Green functions \[Nu]\[CloseCurlyQuote] = 2\[Nu] \
- 1 (\!\(\*SuperscriptBox[\(2\), \(\(\[Nu]\)\(\[CloseCurlyQuote]\)\)]\
\)): ", Style[n1, Red], " (", 2^n1, ")"]; q = 0;
   Do[If[nu > 3, (PrintTemporary["\[RightArrow] l = ", l, "/", m]),
     96];
    s = IntegerDigits[l - 1, 2, n1];
    Do[T = Table[(-1)^s[[h[i]]] xi[[i]], {i, 1, n}];
     T1 = Table[T[[Permutations[T0][[j, i]]]], {i, 1, n}];
     T2 = Simplify[MS.T1]; xo = Join[T1, T2]; y = ML.xo; g = True;
     Do[g = g && SameQ[Simplify[v[[i]] - y[[i]]], 0], {i, 1, n}];
     If[g, (q = q + 1; sig = 1;
       Do[sig = sig*(-1)^(s[[h[i]]]), {i, 1, n}];
       SIG = Table[(-1)^s[[h[i]]], {i, 1, n}];
       Print["\!\(\*
StyleBox[\"\[DoubleLongRightArrow]\",\nFontColor->RGBColor[0, 0.67, \
0]]\) l, j = ", l, ", ", j, ". \!\(\*OverscriptBox[
StyleBox[\"\[Stigma]\",\nFontWeight->\"Plain\"], \
\(\[RightVector]\)]\) =(\!\(\*SubscriptBox[\(\[Stigma]\), \(1\)]\), \
\!\(\*SubscriptBox[\(\[Stigma]\), \(2\)]\),..., \!\(\*SubscriptBox[\(\
\[Stigma]\), \(2  \[Nu] - 1\)]\)): ", SIG, ". \[Stigma]: ",
        Style[sig, Red], "."];
       If[SIG != ONE, (Vr = Verify[xi, SIG];
         Print[Style["Verification \[DoubleRightArrow]",
           Darker[Green], Plain, 12],
          " {\!\(\*SubscriptBox[\(\[Alpha]\), \(1\)]\), \
\!\(\*SubscriptBox[\(\[Alpha]\), \(2\)]\), ..., \!\(\*SubscriptBox[\(\
\[Alpha]\), \(\[Nu]\)]\)}: ", Vr]), 100];
       Do[
        Print[Subscript[x, i], ": ", Style[xo[[i]], Blue]], {i, 1, n}];
       Do[Print[Subscript[x, i], ": ", xo[[i]]], {i, n + 1, n + nu}];
       If[all == 0, Goto[end], 95]), 94], {j, 1, nf}], {l, 1, m}];
   Label[end];
   Print["Number of distinct outputs: ", Style[q, Red], ". (If \!\(\*
StyleBox[\"0\",\nFontColor->RGBColor[1, 0, 0]]\), diagrams A and B \
are not equivalent.)"];
   Print[Style[DateString[], Gray]]; xo;];

\end{verbatim}}

{\footnotesize\refstepcounter{dummy}\label{Verify}
\begin{verbatim}

(* Program `Verify' whose input parameters are specified in the main text. *)
ClearAll[Verify];
Verify[xia_, sig_] :=
  Module[{len, nu, xo, y, y1, xo1, xo2, x, x1, x2, ze, i, p0, fx,
    fp}, len = Length[xia]; nu = (len + 1)/2;
   xo = Table[0, {i, 1, len}];
   fx := Function[{i},
     ToExpression["x" <> ToString[#]] & /@ Range[i]];
   fp := Function[{i},
     ToExpression["p" <> ToString[#]] & /@ Range[i]];
   x = fx[nu]; y = fp[nu];
   y1 = Table[y[[i]] ->  y[[i]] + x[[i]] "p0", {i, 1, nu}];
   Do[If[sig[[i]] == -1, xo[[i]] = xia[[i]] + "p0",
     xo[[i]] = xia[[i]]], {i, 1, len}]; xo1 = xo /. y1;
   xo2 = Simplify[(xo1 - xia)/"p0"]; ze = Table[0, {i, 1, len}];
   x1 = Solve[xo2 == ze, x, Integers];
   x2 = Table[x1[[1, i, 2]], {i, 1, nu}]; x2];

\end{verbatim}}

The most relevant output of \texttt{Equiv} is a sequential list of the signed labels
\begin{equation}
\{x_1,x_2, \dots, x_{3\nu-1}\},\nonumber
\end{equation}
with each variable followed by the relevant energy-momentum vector with which it is to be identified. These output signed labels are those associated with the lines of diagram B, and the energy-momentum vectors assigned to $\{x_1,x_2,\dots,x_{2\nu-1}\}$ are up to a possible minus sign\,\footnote{This possible minus sign can only occur for \texttt{phs} $\not=$ \texttt{0}.} (resulting from energy-momentum reversals) the components of the vector $\vec{\bm{\xi}}$ in terms of which diagram A has been specified; the energy-momentum vectors assigned to $\{x_{2\nu},x_{2\nu+1},\dots,x_{3\nu-1}\}$ correspond to the interaction lines of diagram B, with each energy-momentum consisting of a linear combination (with integer coefficients) of the elements of the set $\{p, p_1,p_2,\dots,p_{\nu}\}$.\footnote{\texttt{Equiv} prints the energy-momentum vectors associated with $\{x_1,x_2,\dots,x_{2\nu-1}\}$, corresponding to Green-function lines, in blue, and those associated with $\{x_{2\nu},x_{2\nu+1},\dots,x_{3\nu-1}\}$, corresponding to interaction lines, in black.} A further relevant output is the $(2\nu-1)$-vector $\vec{\bm{\varsigma}}$ the sign of whose $i$th component\,\footnote{One has $(\vec{\bm{\varsigma}})_i \in \{-1,+1\}$.} $(\vec{\bm{\varsigma}})_i$ signifies whether the $i$th component of the $(2\nu-1)$-vector $\vec{\bm{\xi}}$ has been reversed or not. The parameter $\varsigma \equiv (-1)^{\uprho}$, referred to in \S\,\ref{sd31} as well as in the present section, p.\,\pageref{ForParticleHole}, is also a significant output of \texttt{Equiv}.

The input of the program \texttt{Verify}, p.\,\pageref{Verify}, called within \texttt{Equiv}, consists of the vectors $\vec{\bm{\xi}}$ and $\vec{\bm{\varsigma}}$, and its output is a $\nu$-vector $\vec{\bm{\upalpha}}$ whose components $\{\upalpha_1,\upalpha_2,\dots,\upalpha_{\nu}\}$ define the affine transformations\,\footnote{The vector $\vec{\bm{\upalpha}}$ is \textsl{not} unique, unless $\upalpha_i$, $\forall i \in \{1,2,\dots,\nu\}$, is restricted to the set $\{-1,0,+1\}$.}
\begin{equation}\label{eac6k}
p_i \rightharpoonup p_i + \upalpha_i\hspace{0.6pt} p_0,\;\; i \in \{1,2,\dots,\nu\},
\end{equation}
that render the algebraic expression corresponding to diagram A (or B),\footnote{Note that \textsl{all} outputs, including $\{\upalpha_1,\upalpha_2,\dots,\upalpha_{\nu}\}$, correspond to the cases where A and B have been established to be algebraically equivalent, \S\S\,\protect\ref{sd3} and \protect\ref{sd31}.} subsequent to the application of the right-most equality in Eq.\,(\ref{eac6l}), independent of $p_0$. Existence of such vector $\vec{\bm{\upalpha}}$ implies that up to the constant multiplicative factor $\varsigma (-\mathsf{g})^{\delta\ell}$, \S\,\ref{sd31}, diagrams A and B are indeed equivalent in the p-h symmetric case.

\begin{figure}[t!]
\psfrag{x}[c]{\huge $\varepsilon$}
\psfrag{y}[c]{\huge $\t{G}(\varepsilon + \protect\ii\eta)$}
\centerline{\includegraphics[angle=0, width=0.62\textwidth]{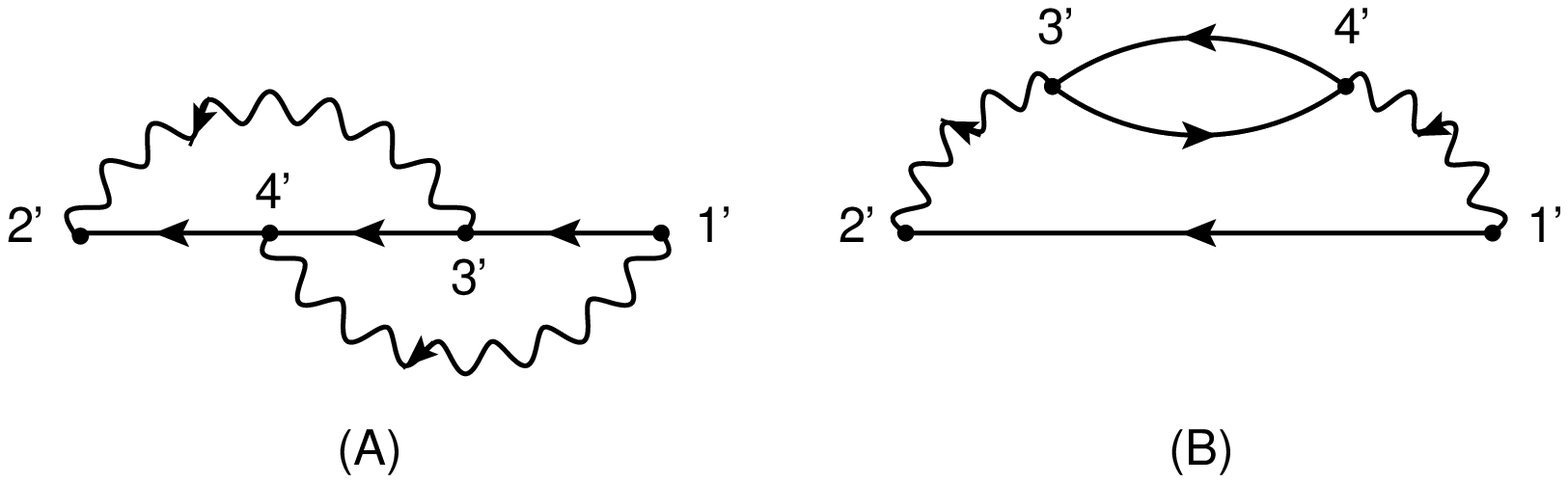}}
\caption{Diagram \protect\refstepcounter{dummy}\label{DiagramA1}A corresponds to the $2$nd-order skeleton self-energy $\Sigma_{\sigma}^{\protect\X{(2.1)}}(2',1')$ as specified in Eq.\,(\protect\ref{eac3}), and diagram B to the $2$nd-order skeleton self-energy $\Sigma_{\sigma}^{\protect\X{(2.2)}}(2',1')$ as specified in Eq.\,(\protect\ref{eac4}). Note that the wavy lines, representing the two-body interaction potential, are directed, similar to the solid lines representing the one-particle Green function. All directions are in accordance with the specifications in Eqs\,(\protect\ref{eac3}) and (\protect\ref{eac4}).}
\label{f15}
\end{figure}

\refstepcounter{dummyX}
\subsubsection{Illustrative examples}
\phantomsection
\label{sd33}
Here we illustrate the use of the program \texttt{Equiv}, p.\,\pageref{Equiv}, described above, through its application to two examples.

As the first illustrative example, we consider the second-order self-energy diagrams displayed in Fig.\,\ref{f15},\footnote{Barring some details, these are reproductions of the diagrams (2.1) and (2.2) in Fig.\,\protect\ref{f7}, p.\,\protect\pageref{SecondOrderS}.} identifying the diagram associated with $\Sigma^{\X{(2.1)}}(2',1')$, Eq.\,(\ref{eac3}), as diagram A, and that associated with $\Sigma^{\X{(2.2)}}(2',1')$, Eq.\,(\ref{eac4}), as diagram B. Following the conventions in \S\,\ref{sx1}, on account of the conservation of the energy-momentum at the $4$ vertices of diagram A in Fig.\,\ref{f15}, taking into account that the energy-momentum $p$ entering the self-energy diagram at vertex $1'$ coincides with that leaving at vertex $2'$, we obtain
\begin{equation}\label{ed15z}
x_1 = p_1,\; x_2 = p_2,\; x_3 = p - p_1 + p_2,\; x_4 = p - p_1,\; x_5 = p_1 - p_2,
\end{equation}
where $p_1$ and $p_2$ denote the internal energy-momentum vectors assigned to the signed labels $x_1$ and $x_2$, respectively, Eq.\,(\ref{eac3}). With reference to Eqs\,(\ref{eac4a}) and (\ref{eac4b}), for the $3$-vector $\vec{\bm{\xi}}$, specifying diagram A, one thus has
\begin{equation}\label{ed15}
\vec{\bm{\xi}} = \begin{pmatrix} p_1 \\ p_2 \\ p - p_1 + p_2\end{pmatrix}.
\end{equation}

On account of the specifications in Eq.\,(\ref{eac4}), for diagram B in Fig.\,\ref{f15} one has the following $2\nu=4$ energy-momentum conservation equations, expressed in accordance with the conventions in Eqs\,(\ref{eac5a}), (\ref{eac5b}), and (\ref{eac5f}):
\begin{align}\label{ed15a}
&1':\;\; x_1 + x_4 = p,\nonumber\\
&2':\;\; x_1 + x_5 = p,\nonumber\\
&3':\;\; x_2 - x_3 + x_5 = 0,\nonumber\\
&4':\;\; x_2 - x_3 + x_4 = 0.
\end{align}
For the $2\nu \times (3\nu-1) = 4\times 5$ matrix $\mathbb{M}_{\textsc{l}}$ one thus has (below $\b{1} \equiv -1$)
\begin{equation}\label{ed15b}
\mathbb{M}_{\textsc{l}} = \begin{pmatrix}
1 &  0 &   0     & 1 & 0 \\
1 &  0 &   0     & 0 & 1 \\
0 &  1 & \b{1} & 0 & 1 \\
0 &  1 & \b{1} & 1 & 0 \end{pmatrix}.
\end{equation}
Expressing the equalities associated with $4'$ and $3'$ in Eq.\,(\ref{ed15a}) in accordance with the convention in Eq.\,(\ref{eac6}), one has
\begin{align}\label{ed15c}
&\hspace{0.0cm} 1:\;\; x_4 = -x_2 + x_3,\nonumber\\
&\hspace{0.0cm} 2:\;\; x_5 = -x_2 + x_3,
\end{align}
from which for the $\nu \times (2\nu-1) = 2 \times 3$ matrix $\mathbb{M}_{\textsc{s}}$ one obtains
\begin{equation}\label{ed15d}
\mathbb{M}_{\textsc{s}} = \begin{pmatrix}
0 & \b{1} & 1 \\
0 & \b{1} & 1 \end{pmatrix}.
\end{equation}

To test the equivalence of the diagrams A and B in Fig.\,\ref{f15} in the absence of p-h symmetry (specified by \texttt{phs = 0}), the following instructions are to be sequentially executed:

{\footnotesize
\begin{verbatim}

(* Illustrative example. I *)
Clear[p, p1, p2, xi, ML, MS];
nu = 2 (*Order of the diagrams*);
phs = 0 (*0=No p-h symmetry; 1=p-h symmetry*);
all = 1 (*0, Do not search for all; 1, Search for all*);
xi = {p1, p2, p - p1 + p2};
Print["\!\(\*OverscriptBox[
StyleBox[\"\[Xi]\",\nFontWeight->\"Bold\"], \(\[RightVector]\)]\) = ", MatrixForm[xi]];
ML = SparseArray[{{1, 1} -> 1, {1, 4} -> 1, {2, 1} -> 1, {2, 5} -> 1,
       {3, 2} -> 1, {3, 3} -> -1, {3, 5} -> 1, {4, 2} -> 1, {4, 3} -> -1,
       {4, 4} -> 1}];
Print["\!\(\*SubscriptBox[\(\[DoubleStruckCapitalM]\), \(L\)]\) = ", MatrixForm[ML]];
MS = SparseArray[{{1, 2} -> -1, {1, 3} -> 1, {2, 2} -> -1, {2, 3} -> 1}];
Print["\!\(\*SubscriptBox[\(\[DoubleStruckCapitalM]\), \(S\)]\) = ", MatrixForm[MS]];
Equiv[ML, MS, nu, p, xi, phs, all]

\end{verbatim}}

\noindent
The following two distinct sets of \textsl{signed} indices $\{x_1, x_2, x_3, x_4, x_5\}$ are the outputs of \texttt{Equiv}:\,\footnote{Absence of output would imply that under the condition specified by \texttt{phs}, the relevant diagrams A and B were not equivalent. In the case at hand, \textsl{two} outputs (there would have been \textsl{one} output if \texttt{all} had been set equal to \texttt{0}) implies that diagrams A and B in Fig.\,\protect\ref{f15}, p.\,\protect\pageref{DiagramA1}, are `equivalent'. Since $\varsigma = 1$ and $\delta\ell \equiv \ell_{\textsc{b}} - \ell_{\textsc{a}} = 1$, for spin-$\mathsf{s}$ particles the contribution of diagram B is equal to that of diagram A times $-(2\mathsf{s}+1)$, \S\,\protect\ref{sd3}.}
\begin{align}\label{ed15e}
&\hspace{0.0cm}(1):\; x_1 = p_1,\; x_2 = p_2,\; x_3 = p - p_1 + p_2,\nonumber\\
&\hspace{0.9cm} x_4 = p - p_1,\; x_5 = p - p_1,\nonumber \\
&\hspace{0.0cm}(2):\; x_1 = p - p_1 + p_2,\; x_2 = p_2,\; x_3 = p_1,\nonumber\\
&\hspace{0.9cm} x_4 = p_1 - p_2,\; x_5 = p_1 - p_2.
\end{align}
Since \texttt{phs = 0} (absence of p-h symmetry), naturally the signature $\varsigma$ associated with both outputs is equal to $1$. With reference to Eq.\,(\ref{eac4}) and Fig.\,\ref{f15}, p.\,\protect\pageref{DiagramA1}, one immediately verifies that for $p \equiv (\varepsilon,\hbar\bm{k})$ both outputs indeed satisfy the required energy-momentum conservations at the $4$ vertices of diagram B, associated with $\Sigma_{\sigma}^{\X{(2.2)}}(\bm{k};\varepsilon)$. What the above two outputs reveal is that in the absence of p-h symmetry there are two different ways in which the internal lines of diagram B can be indexed whereby the algebraic expression associated with this diagram up to a determinate multiplicative constant (for which see the following paragraph) identically coincides with the algebraic expression associated with diagram A, corresponding to $\Sigma_{\sigma}^{\X{(2.1)}}(\bm{k};\varepsilon)$, as indexed according to the equalities in Eq.\,(\ref{ed15z}) and partially represented by the $3$-vector $\vec{\bm{\xi}}$ in Eq.\,(\ref{ed15}).\footnote{For \textsl{equivalence} of diagrams, see the description at the outset of this section, \emph{i.e.} \protect\S\,\ref{sd3}.}

Without\refstepcounter{dummy}\label{WithoutGoing} going into details, we remark that for the p-h symmetric case (that is, for the case where \texttt{phs} $\not=$ \texttt{0}, in conjunction with \texttt{all} $\not=$ \texttt{0}), program \texttt{Equiv} yields $6$ distinct outputs, instead of the $2$ corresponding to the case of \texttt{phs} $=0$ considered above.\footnote{Note that for the `equivalence' of diagrams A and B a \textsl{single} output suffices. Thus, for determining whether or not diagrams A and B are `equivalent', it suffices to execute program \texttt{Equiv} with \texttt{all} $=$ \texttt{0}.} The output parameter $\varsigma$ in all $6$ cases is equal to $1$. For the case of the half-filled `Hubbard atom' of spin-$\tfrac{1}{2}$ particles, which is p-h symmetric, see Eq.\,(\ref{ex2b}) below, and note that in this case, where $\delta\ell \equiv \ell_{\textsc{b}} -\ell_{\textsc{a}} = 1$, one indeed has $\varsigma (-\mathsf{g})^{\delta\ell} = -2$, \S\,\ref{sd31}.

\begin{figure}[t!]
\psfrag{x}[c]{\huge $\varepsilon$}
\psfrag{y}[c]{\huge $\t{G}(\varepsilon + \protect\ii\eta)$}
\centerline{\includegraphics[angle=0, width=0.74\textwidth]{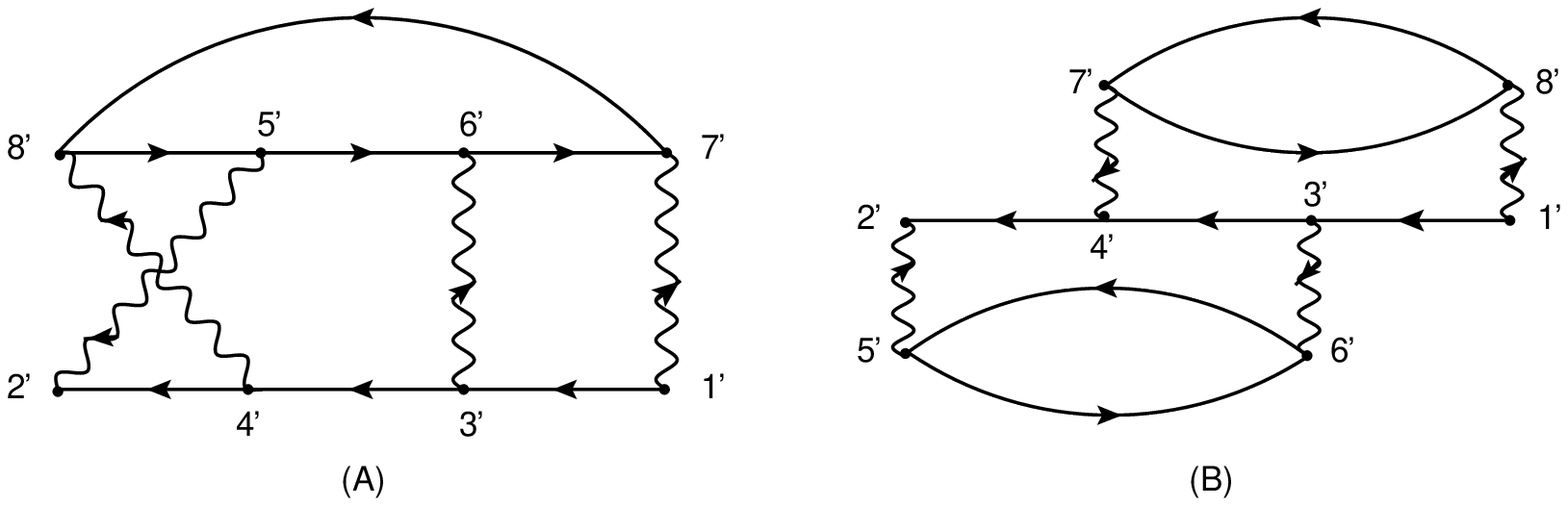}}
\caption{Diagram \protect\refstepcounter{dummy}\label{DiagramA2}A corresponds to the $4$th-order skeleton self-energy $\Sigma_{\sigma}^{\protect\X{(4.40)}}(2',1')$ as specified in Eq.\,(\protect\ref{ed15g}), and diagram B to the $4$th-order skeleton self-energy $\Sigma_{\sigma}^{\protect\X{(4.79)}}(2',1')$ as specified in Eq.\,(\protect\ref{ed15h}). Note that the wavy lines, representing the two-body interaction potential, are directed, similar to the solid lines representing the one-particle Green function. All directions are in accordance with the specifications in Eqs\,(\protect\ref{ed15g}) and (\protect\ref{ed15h}).}
\label{f16}
\end{figure}

As the second illustrative example, we consider the two fourth-order skeleton self-energy diagrams displayed in Fig.\,\ref{f16}, identifying the diagram associated with $\Sigma^{\X{(4.40)}}(2',1')$, Eq.\,(\ref{e91k}), as diagram A, and that associated with $\Sigma^{\X{(4.79)}}(2',1')$, Eq.\,(\ref{e92l}), as diagrams B. Following the prescriptions in \S\,\ref{sx1} and in analogy with the specifications in Eqs\,(\ref{eac3}) and (\ref{eac4}), we specify the present diagrams A and B as follows (see Fig.\,\ref{f16}):
\begin{align}\label{ed15g}
\hspace{-0.6cm}\Sigma_{\sigma}^{\X{(4.40)}}:\;\; x_1 &= \langle 1',3'\rangle,\; x_2 = \langle 3',4'\rangle,\; x_3 = \langle 4',2'\rangle,\; x_4 = \langle 5',6'\rangle,\; x_5 = \langle 6',7'\rangle, \nonumber\\
x_6 &= \langle 7',8'\rangle,\; x_7 = \langle 8',5'\rangle,\nonumber\\
x_8 &= \prec\! 1',7'\!\succ,\; x_9 = \prec\! 3',6'\!\succ,\; x_{10} = \prec\! 4',8'\!\succ,\;
x_{11} = \prec\! 5',2'\!\succ,
\end{align}
\begin{align}\label{ed15h}
\hspace{-0.6cm}\Sigma_{\sigma}^{\X{(4.79)}}:\;\; x_1 &= \langle 1',3'\rangle,\; x_2 = \langle 3',4'\rangle,\; x_3 = \langle 4',2'\rangle,\; x_4 = \langle 5',6'\rangle,\; x_5 = \langle 6',5'\rangle, \nonumber\\
x_6 &= \langle 7',8'\rangle,\; x_7 = \langle 8',7'\rangle,\nonumber\\
x_8 &= \prec\! 1',8'\!\succ,\; x_9 = \prec\! 3',6'\!\succ,\; x_{10} = \prec\! 7',4'\!\succ,\;
x_{11} = \prec\! 5',2'\!\succ.
\end{align}
Following the specification in Eq.\,(\ref{ed15g}), on account of the conservation of the energy-momentum at the $8$ vertices of diagram A in Fig.\,\ref{f16}, taking into account that the energy-momentum $p$ entering the self-energy diagram at vertex $1'$ coincides with that leaving at vertex $2'$, we obtain
\begin{align}\label{ed15x}
&x_1 = p_1,\; x_2 = p_2,\; x_3 = p_3,\; x_4 = p_4,\; x_5 = p_1 - p_2 + p_4,\nonumber\\
&x_6 = p - p_2 + p_4,\; x_7 = p - p_3 + p_4,\nonumber\\
&x_8 = p - p_1,\; x_9 = p_1 - p_2,\; x_{10} = p_2 - p_3,\; x_{11} = p - p_3.
\end{align}
With reference to Eqs\,(\ref{eac4a}) and (\ref{eac4b}), diagram A is thus specified by the following $7$-vector:
\begin{equation}\label{ed15i}
\vec{\bm{\xi}} = \begin{pmatrix}
p_1 \\ p_2 \\ p_3 \\ p_4 \\ p_1 - p_2 + p_4 \\ p - p_2 + p_4 \\ p - p_3 + p_4
\end{pmatrix}.
\end{equation}

On account of the specifications in Eq.\,(\ref{ed15h}), for diagram B in Fig.\,\ref{f16} one has the following $2\nu = 8$ energy-momentum-conservation equations, expressed in accordance with the conventions in Eqs\,(\ref{eac5a}), (\ref{eac5b}), and (\ref{eac5f}):
\begin{align}\label{ed15j}
&1':\;\; x_1 + x_8 = p,\nonumber\\
&2':\;\; x_3 + x_{11} = p,\nonumber\\
&3':\;\; x_1 - x_2 - x_9 = 0,\nonumber\\
&4':\;\; x_2 - x_3 + x_{10} = 0,\nonumber\\
&5':\;\; x_4 - x_5 + x_{11} = 0,\nonumber\\
&6':\;\; x_4 - x_5 + x_9 = 0,\nonumber\\
&7':\;\; x_6 - x_7 + x_{10} = 0,\nonumber\\
&8':\;\; x_6 - x_7 + x_8 = 0.
\end{align}
For the $2\nu \times (3\nu-1) = 8 \times 11$ matrix $\mathbb{M}_{\textsc{l}}$ one thus has (below $\b{1} \equiv -1$)
\setcounter{MaxMatrixCols}{11}
\begin{equation}\label{ed15k}
\mathbb{M}_{\textsc{l}} = \begin{pmatrix}
1 & 0 & 0 & 0 & 0 & 0 & 0 & 1 & 0 & 0 & 0 \\
0 & 0 & 1 & 0 & 0 & 0 & 0 & 0 & 0 & 0 & 1 \\
1 & \b{1} & 0 & 0 & 0 & 0 & 0 & 0 & \b{1} & 0 & 0 \\
0 & 1 & \b{1} & 0 & 0 & 0 & 0 & 0 & 0 & 1 & 0 \\
0 & 0 & 0 & 1 & \b{1} & 0 & 0 & 0 & 0 & 0 & 1 \\
0 & 0 & 0 & 1 & \b{1} & 0 & 0 & 0 & 1 & 0 & 0 \\
0 & 0 & 0 & 0 & 0 & 1 & \b{1} & 0 & 0 & 1 & 0 \\
0 & 0 & 0 & 0 & 0 & 1 & \b{1} & 1 & 0 & 0 & 0 \end{pmatrix}.
\end{equation}
Expressing the equalities associated with $8'$, $6'$, $7'$ and $5'$ in Eq.\,(\ref{ed15j}) in accordance with the convention in Eq.\,(\ref{eac6}), one has
\begin{align}\label{ed15l}
&1:\;\; x_8 = -x_6 + x_7,\nonumber\\
&2:\;\; x_9 = -x_4 + x_5,\nonumber\\
&3:\;\; x_{10} = -x_6 + x_7,\nonumber\\
&4:\;\; x_{11} = -x_4 + x_5,
\end{align}
from which for the  $\nu \times (2\nu-1) = 4 \times 7$ matrix $\mathbb{M}_{\textsc{s}}$ one obtains
\begin{equation}\label{ed15m}
\mathbb{M}_{\textsc{s}} = \begin{pmatrix}
0 & 0 & 0 & 0 & 0 & \b{1} & 1 \\
0 & 0 & 0 & \b{1} & 1 & 0 & 0 \\
0 & 0 & 0 & 0 & 0 & \b{1} & 1 \\
0 & 0 & 0 & \b{1} & 1 & 0 & 0 \end{pmatrix}.
\end{equation}

To test the equivalence of the diagrams A and B in Fig.\,\ref{f16} in the absence of p-h symmetry (specified by \texttt{phs = 0}), the following instructions are to be sequentially executed:

{\footnotesize
\begin{verbatim}

(* Illustrative example. II *)
Clear[p, p1, p2, p3, p4, xi, ML, MS];
nu = 4 (*Order of the diagrams*);
phs = 0 (*0= No p-h symmetry; 1 = p-h symmetry*);
all = 1 (*0, Do not search for all; 1, Search for all*);
xi = {p1, p2, p3, p4, p1 - p2 + p4, p - p2 + p4, p - p3 + p4};
Print["\!\(\*OverscriptBox[
StyleBox[\"\[Xi]\",\nFontWeight->\"Bold\"], \(\[RightVector]\)]\) = ", MatrixForm[xi]];
ML = SparseArray[{{1, 1} -> 1, {1, 8} -> 1, {2, 3} -> 1, {2, 11} -> 1,
        {3, 1} -> 1, {3, 2} -> -1, {3, 9} -> -1, {4, 2} -> 1, {4, 3} -> -1,
        {4, 10} -> 1, {5, 4} -> 1, {5, 5} -> -1, {5, 11} -> 1, {6, 4} -> 1,
        {6, 5} -> -1, {6, 9} -> 1, {7, 6} -> 1, {7, 7} -> -1, {7, 10} -> 1,
        {8, 6} -> 1, {8, 7} -> -1, {8, 8} -> 1}];
Print["\!\(\*SubscriptBox[\(\[DoubleStruckCapitalM]\), \(L\)]\) = ", MatrixForm[ML]];
MS = SparseArray[{{1, 6} -> -1, {1, 7} -> 1, {2, 4} -> -1, {2, 5} -> 1,
        {3, 6} -> -1, {3, 7} -> 1, {4, 4} -> -1, {4, 5} -> 1}];
Print["\!\(\*SubscriptBox[\(\[DoubleStruckCapitalM]\), \(S\)]\) = ", MatrixForm[MS]];
Equiv[ML, MS, nu, p, xi, phs, all]

\end{verbatim}}

\noindent
The sequential execution of the above instructions generates no output for the set $\{x_1,x_2,\dots,x_{11}\}$ of the signed labels, Eq.\,(\ref{ed15h}), implying that the diagrams A and B in Fig.\,\ref{f16}, p.\,\protect\pageref{DiagramA2}, are \textsl{not} equivalent in the absence of p-h symmetry. In contrast, by changing \texttt{phs = 0} into a non-zero integer (using for instance \texttt{phs = 1}), it generates $8$ distinct sets for $\{x_1,x_2,\dots,x_{11}\}$, implying that in the presence of p-h symmetry the diagrams A and B under consideration are equivalent.\footnote{We emphasise once more that a single output would suffice to establish the equivalence of diagrams A and B. For the purpose of establishing this equivalence, it is therefore advisable to set \texttt{all} equal to \texttt{0}.} The signature $\varsigma \equiv (-1)^{\uprho}$ in all $8$ cases is equal to $-1$. In the light of $\delta\ell \equiv \ell_{\textsc{b}} - \ell_{\textsc{a}} = 2 -1 = 1$, for spin-$\tfrac{1}{2}$ particles (that is for $\mathsf{s} = \tfrac{1}{2}$ and thus $\mathsf{g} = 2$) the corresponding $\varsigma (-\mathsf{g})^{\delta\ell}$, \S\,\ref{sd31}, is equal to $2$. One thus arrives at
\begin{equation}\label{ed15n}
\Sigma_{\sigma}^{\X{(4.79)}}(\bm{k};\varepsilon) \equiv +2\Sigma_{\sigma}^{\X{(4.40)}}(\bm{k};\varepsilon).\;\;\; \text{(For p-h symmetric GSs of spin-$\tfrac{1}{2}$ particles)}
\end{equation}
This result is to be compared with the equalities in Eqs\,(\ref{ex40}) and (\ref{ex79}) below for the specific case of the half-filled Hubbard atom for spin-$\tfrac{1}{2}$ particles whose GS is indeed p-h symmetric. With reference to the discussions in \S\,\ref{sd2}, p.\,\pageref{UnlessWeIndicateOtherwise}, the result in Eq.\,(\ref{ed15n}) can also be written as follows:\,\footnote{See footnote \raisebox{-1.0ex}{\normalsize{\protect\footref{noteu}}} on p.\,\protect\pageref{FromThePerspective} and footnote \raisebox{-1.0ex}{\normalsize{\protect\footref{notev}}} on p.\,\protect\pageref{InTheMore}. See also the remark concerning $\mathsf{g} = 1$ on p.\,\protect\pageref{InDealingWithH}.}
\begin{equation}\label{ed15o}
\left.\Sigma_{\sigma}^{\X{(4.79)}}(\bm{k};\varepsilon)\right|_{\h{\mathsf{H}}} \equiv \left. \Sigma_{\sigma}^{\X{(4.40)}}(\bm{k};\varepsilon)\right|_{\h{\mathsf{H}}}.\;\;\; \text{(For p-h symmetric GSs of spin-$\tfrac{1}{2}$ particles)}
\end{equation}
This result in implicit in the diagrammatic expression displayed in Fig.\,3, p.\,496, of Ref.\,\citen{GJMNN03}, since the first (second) diagram in this figure coincides with our present diagram B (A). With reference to Eq.\,(\ref{ex01o}), we note that the diagrams associated with $\Sigma_{\sigma}^{\X{(4.40)}}(\bm{k};\varepsilon)$ and $\Sigma_{\sigma}^{\X{(4.79)}}(\bm{k};\varepsilon)$ do occur in the perturbation series expansion of the self-energy corresponding to the Hubbard Hamiltonian $\hspace{0.28cm}\h{\hspace{-0.28cm}\mathpzc{H}}$, Eq.\,(\ref{ex01f}). One has
\begin{equation}\label{ed15p}
\left.\Sigma_{\sigma}^{\X{(4.79)}}(\bm{k};\varepsilon)\right|_{\hspace{0.28cm}\h{\hspace{-0.28cm}\mathpzc{H}}} \equiv \left. \Sigma_{\sigma}^{\X{(4.40)}}(\bm{k};\varepsilon)\right|_{\hspace{0.28cm}\h{\hspace{-0.28cm}\mathpzc{H}}}.\;\;\; \text{(For p-h symmetric GSs of spin-$\tfrac{1}{2}$ particles)}
\end{equation}
The functions in Eqs\,(\ref{ed15o}) and (\ref{ed15p}) identically coincide.

For completeness, below we present the first three of the above-mentioned $8$ sets for $\{x_1,x_2,\dots,x_{11}\}$, Eq.\,(\ref{ed15h}), generated by \texttt{Equiv} for \texttt{phs} $\not=$ \texttt{0}:
\begin{align}\label{ed15px}
&\hspace{0.2cm}(1):\; x_1 = -p_4,\; x_2 = -p_2,\; x_3 = p - p_2 + p_4,\; x_4 = p_1 - p_2 + p_4,\; x_5 = p_1,\nonumber\\
&\hspace{1.1cm}  x_6 = -p + p_3 - p_4,\; x_7 = p_3,\nonumber\\
&\hspace{1.1cm}  x_8 = p + p_4,\; x_9 = p_2 - p_4,\; x_{10} = p + p_4,\; x_{11} = p_2 - p_4,\\
&\hspace{0.2cm}(2):\; x_1 = p - p_2 + p_4,\; x_2 = -p_2,\; x_3 = - p_4,\; x_4 = -p + p_3 - p_4,\; x_5 = p_3, \nonumber\\
&\hspace{1.1cm}x_6 = p_1 - p_2 + p_4,\; x_7 = p_1,\nonumber\\
&\hspace{1.1cm} x_8 = p_2 - p_4,\; x_9 = p + p_4,\; x_{10} = p_2 - p_4,\; x_{11} = p + p_4,\nonumber\\
&\hspace{0.2cm}(3):\; x_1 = -p_4 ,\; x_2 = -p_2,\; x_3 = p - p_2 + p_4,\; x_4 = p_1 - p_2 + p_4,\; x_5 = p_1,\nonumber\\
&\hspace{1.1cm} x_6 = -p_3,\; x_7 = p- p_3 + p_4,\nonumber\\
&\hspace{1.1cm} x_8 = p + p_4,\; x_9 = p_2 - p_4,\; x_{10} = p + p_4,\; x_{11} = p_2 - p_4.
\end{align}
For the corresponding $4$-vectors $\vec{\bm{\upalpha}} = \{\upalpha_1,\upalpha_2,\upalpha_3,\upalpha_4\}$, Eq.\,(\ref{eac6k}), one has (below $\b{1} \equiv -1$)
\begin{align}\label{ed15q}
&\hspace{0.2cm}(1):\; \upalpha_1 = 0,\; \upalpha_2 = \b{1},\; \upalpha_3 = 0,\; \upalpha_4 = \b{1},\nonumber\\
&\hspace{0.2cm}(2):\; \upalpha_1 = 0,\; \upalpha_2 = \b{1},\; \upalpha_3 = 0,\; \upalpha_4 = \b{1},\nonumber\\
&\hspace{0.2cm}(3):\; \upalpha_1 = 0,\; \upalpha_2 = \b{1},\; \upalpha_3 = \b{1},\; \upalpha_4 = \b{1}.
\end{align}

\refstepcounter{dummyX}
\subsection{The half-filled `Hubbard atom' of spin-\texorpdfstring{$\tfrac{1}{2}$}{} particles}
\phantomsection
\label{sd4}
Here we consider the skeleton self-energy diagrams corresponding to the atomic limit of the half-filled GS of the Hubbard Hamiltonian $\h{\mathcal{H}}$, Eq.\,(\ref{ex01bx}), for spin-$\tfrac{1}{2}$ particles. We explicitly deal with the $2$nd-, $3$rd- and $4$th-order skeleton self-energy diagrams, evaluating the contributions of these in terms of the exact interacting Green function $\t{G}(z)$, Eq.\,(\ref{e25}). In this connection, and in the light of the discussions centred on the expression in Eq.\,(\ref{e14a}), by identifying the atomic energy $T_{\X{0}}$ with $-\hbar \Sigma^{\textsc{hf}} \equiv -U/2$, the first-order self-energy as calculated in terms of $\t{G}(z)$ is identically vanishing. More generally, by the p-h symmetry of $\t{G}(z)$, in the case at hand $\t{\Sigma}^{\X{(2\nu+1)}}(z) \equiv 0$ for \textsl{all} $\nu \in \mathds{N}_{0}$ (see Eq.\,(\ref{e4})). We emphasise that this result does \textsl{not} apply to the contributions of the individual $(2\nu+1)$th-order self-energy diagrams, but to their combined contribution. This is attested by the explicit results concerning the $3$rd-order skeleton self-energy diagrams in \S\,\ref{sec.d52}.

The results corresponding to the $4$th-order self-energy diagrams to be presented in this section are closely related to those by Gebhard \emph{et al.} \cite{GJMNN03} regarding the half-filled Hubbard model for spin-$\tfrac{1}{2}$ particles in `infinite dimensions'. The calculations by the latter authors being based on non-interacting Green functions $\{\t{G}_{\X{0};\sigma}(z) \| \sigma\}$, described in terms of a semi-circular density of states, they take account of both skeleton and non-skeleton self-energy diagrams. In this section we shall also deal with $4$th-order \textsl{non-skeleton} self-energy diagrams, the reason for which will be clarified in due place below. Since for the one-particle spectral function $A(\varepsilon)$ associated with the Green function $\t{G}(z)$ in Eq.\,(\ref{e25}) one has\,\footnote{See Eqs\,(\protect\ref{e4d}) and (\protect\ref{e4oj}).}
\begin{equation}\label{exc1}
A(\varepsilon) = \frac{\hbar}{2}\hspace{0.6pt} \big\{ \delta(\varepsilon+U/2) + \delta(\varepsilon-U/2)\big\},
\end{equation}
one observes that for $U\to 0$, the function $A(\varepsilon)$ may be viewed as a limiting case of the semi-circular density of states $A(\varepsilon)$ underlying the calculations in Ref.\,\citen{GJMNN03}, with the width $\mathsf{W}$ of $A(\varepsilon)$ approaching zero (see the discussions centred on Eq.\,(\ref{e44})). We note that the Hubbard Hamiltonian underlying the calculations in Ref.\,\citen{GJMNN03} is the grand-canonical Hamiltonian $\hspace{0.28cm}\h{\hspace{-0.28cm}\mathpzc{K}}\doteq \hspace{0.28cm}\h{\hspace{-0.28cm}\mathpzc{H}} - \mu \h{N}$, Eq.\,(\protect\ref{e27k}), associated with the Hubbard Hamiltonian $\hspace{0.28cm}\h{\hspace{-0.28cm}\mathpzc{H}}$ for spin-$\frac{1}{2}$ particles, defined in Eq.\,(\ref{ex01f}). See therefore the relevant discussions on p.\,\pageref{WeAreNowInAPosition}, \S\,\ref{sd2}, regarding the self-energy diagrams associated with $\hspace{0.28cm}\h{\hspace{-0.28cm}\mathpzc{H}}$, as well as \S\,\ref{sd21}.

We introduce the function
\begin{equation}\label{ex0}
\t{\Phi}_{\upalpha,\upbeta}(z,U) \doteq \frac{U}{8\hspace{0.6pt} \hbar}\Big\{ \Big(\frac{U}{z - \upalpha\hspace{0.4pt} U/2}\Big)^{\upbeta} -\Big(\frac{-U}{z + \upalpha\hspace{0.4pt} U/2}\Big)^{\upbeta}\Big\},
\end{equation}
in terms of which we express the self-energy contributions to be encountered below. In this connection, we note that\footnote{We had been hoping to obtain a general expression for $\protect\t{\Sigma}_{\sigma}^{\protect\X{(\nu)}}(z)$, for arbitrary $\nu$, in terms of the function $\protect\t{\Phi}_{\upalpha,\upbeta}(z,U)$. We have however not investigated this possibility in any exhaustive fashion. One notes that for $\nu = 2$, $\upalpha = 3$ and $\upbeta = 1$, Eq.\,(\protect\ref{ex2c}), and that for $\nu = 4$, $\upalpha \in \{3,5\}$ and $\upbeta \in \{1,2\}$, Eq.\,(\protect\ref{ex0e}). For $\nu > 2$ one may in general have superpositions, with \textsl{rational} coefficients, of the functions $\{\protect\t{\Phi}_{\upalpha,\upbeta}(z) \| \upalpha,\upbeta\}$, with $\upalpha \in\{\nu-1,\nu+1\}$ and $\upbeta\in \{1,2,\dots,\nu-2\}$. Note that the \textsl{opposite values} of the pre-factors of $\protect\t{\Phi}_{5,1}(z,U)$ and $\protect\t{\Phi}_{3,1}(z,U)$ in Eq.\,(\protect\ref{ex0e}) reflect the asymptotic expression in Eq.\,(\protect\ref{e7g}) and the asymptotic expression in Eq.\,(\protect\ref{ex0b}) corresponding to $\upbeta=1$.} (\emph{cf.} Eq.\,(\ref{e25}))
\begin{equation}\label{ex0c}
\t{\Sigma}(z) \equiv \frac{U^2}{4\hbar z} = \t{\Phi}_{0,1}(z,U).
\end{equation}
For later reference, we note that
\begin{equation}
\label{ex0b}
\t{\Phi}_{\upalpha,\upbeta}(z,U) \sim \left\{\begin{array}{ll}
\displaystyle \frac{U}{2 \hbar}\Big\{ \frac{U}{2 z} + \upalpha^2 \Big(\frac{U}{2 z}\Big)^3 + \dots\Big\}, & \upbeta = 1, \\ \\
\displaystyle \frac{2\upalpha U}{\hbar}\Big\{ \Big(\frac{U}{2 z}\Big)^3 + 2 \upalpha^2 \Big(\frac{U}{2 z}\Big)^5 + \dots\Big\}, & \upbeta = 2, \\ \\
\displaystyle \frac{2 U}{\hbar} \Big\{ \Big(\frac{U}{2 z}\Big)^3 + 6\upalpha^2 \Big(\frac{U}{2 z}\Big)^5 +\dots\Big\}, & \upbeta =3,\end{array}\right. \; \text{for}\;\; z\to\infty.
\end{equation}

Below we consider the perturbational contributions $\{\t{\Sigma}_{\sigma}^{\X{(\nu)}}(z)\| \nu=2,3,4\}$, obtained according to the following expressions (see Eq.\,(\ref{eac3a})):\,\footnote{In the following we restore the spin index $\sigma$ to the symbols denoting the self-energy and the perturbational contributions to it, thus deviating from the simplified notation adopted in \S\S\,\protect\ref{sec3.a}, \protect\ref{sec3.b}, \protect\ref{sec3.f}, \protect\ref{s4xa}, and \protect\ref{s4xc} in regard to these functions associated with the `Hubbard atom'.}
\begin{equation}\label{ex1a}
\t{\Sigma}_{\sigma}^{\X{(2)}}(z) = \sum_{j=1}^{2} \t{\Sigma}_{\sigma}^{\X{(2.j)}}(z),
\end{equation}
\begin{equation}\label{ex1b}
\t{\Sigma}_{\sigma}^{\X{(3)}}(z) = \sum_{j=1}^{10} \t{\Sigma}_{\sigma}^{\X{(3.j)}}(z),
\end{equation}
\begin{equation}\label{ex1c}
\t{\Sigma}_{\sigma}^{\X{(4)}}(z) = \sum_{j=1}^{82} \t{\Sigma}_{\sigma}^{\X{(4.j)}}(z),
\end{equation}
where $\t{\Sigma}_{\sigma}^{\X{(3)}}(z) \equiv 0$, as indicated above. The relevance of considering the $3$rd-order self-energy contribution $\t{\Sigma}_{\sigma}^{\X{(3)}}(z)$ lies in the insight that it provides regarding the way in which the latter identity is realised. Further, following Eqs\,(\ref{ex01m}), (\ref{ex01n}), and (\ref{ex01o}), one has\,\footnote{For illustration, $\protect\t{\Sigma}_{\sigma}^{\protect\X{(2.2)}}(z) \vert_{\protect\h{\eta}} = \protect\t{\Phi}_{3,1}(z,U)$, which, in the light of the equality in Eq.\,(\protect\ref{ex1d}), is to be contrasted with the equalities in Eqs\,(\protect\ref{ex2b}) and (\protect\ref{ex2c}).}
\begin{equation}\label{ex1d}
\t{\Sigma}_{\sigma}^{\X{(2)}}(z) \equiv \left.\t{\Sigma}_{\sigma}^{\X{(2.2)}}(z) \right|_{\h{\eta}},
\end{equation}
\begin{equation}\label{ex1e}
\t{\Sigma}_{\sigma}^{\X{(3)}}(z) \equiv \left.\t{\Sigma}_{\sigma}^{\X{(3.6)}}(z) \right|_{\h{\eta}} + \left.\t{\Sigma}_{\sigma}^{\X{(3.7)}}(z) \right|_{\h{\eta}},
\end{equation}
\begin{align}\label{ex1f}
\t{\Sigma}_{\sigma}^{\X{(4)}}(z) &\equiv \left.\t{\Sigma}_{\sigma}^{\X{(4.38)}}(z) \right|_{\h{\eta}} + \left.\t{\Sigma}_{\sigma}^{\X{(4.39)}}(z) \right|_{\h{\eta}} + \left.\t{\Sigma}_{\sigma}^{\X{(4.40)}}(z) \right|_{\h{\eta}} + \left.\t{\Sigma}_{\sigma}^{\X{(4.41)}}(z) \right|_{\h{\eta}} + \left.\t{\Sigma}_{\sigma}^{\X{(4.42)}}(z) \right|_{\h{\eta}}\nonumber\\
 &+ \left.\t{\Sigma}_{\sigma}^{\X{(4.43)}}(z) \right|_{\h{\eta}} + \left.\t{\Sigma}_{\sigma}^{\X{(4.70)}}(z) \right|_{\h{\eta}} +
\left.\t{\Sigma}_{\sigma}^{\X{(4.79)}}(z) \right|_{\h{\eta}} + \left.\t{\Sigma}_{\sigma}^{\X{(4.82)}}(z) \right|_{\h{\eta}},
\end{align}
where, Eqs\,(\ref{ex01f}) and (\ref{ex01k}),
\begin{equation}\label{ex1g}
\h{\eta} = \hspace{0.28cm}\h{\hspace{-0.28cm}\mathpzc{H}},\; \h{\mathsf{H}}.
\end{equation}

\refstepcounter{dummyX}
\subsubsection{The 2nd-order self-energy contributions}
\phantomsection
\label{sec.d51}
On the basis of the details presented in \S\,\ref{sd21}, for the contributions of the $2$nd-order skeleton self-energy diagrams as specified in \S\,\ref{sd11}, in terms of the interacting Green function $\t{G}(z)$ in Eq.\,(\ref{e25}), one obtains
\begin{align}
\label{ex2a}
\t{\Sigma}_{\sigma}^{\X{(2.1)}}(z) &= -\t{\Phi}_{3,1}(z,U),\\
\label{ex2b}
\t{\Sigma}_{\sigma}^{\X{(2.2)}}(z) &= -2 \t{\Sigma}_{\sigma}^{\X{(2.1)}}(z) \equiv 2\hspace{0.8pt}\t{\Phi}_{3,1}(z,U).
\end{align}
Thus, following Eq.\,(\ref{ex1a}),
\begin{equation}\label{ex2c}
\t{\Sigma}_{\sigma}^{\X{(2)}}(z) = \t{\Phi}_{3,1}(z,U).
\end{equation}
This result identically coincides with what one would have obtained by discarding the diagram (2.1) in Fig.\,\ref{f7}, p.\,\protect\pageref{SecondOrderS} (\emph{cf.} Eq.\,(\ref{ex1d})), \textsl{and} relaxing the spin sum\,\footnote{See Eq.\,(\protect\ref{ex04a}), and compare with Eq.\,(\protect\ref{ex04b}).} involved in the evaluation of the contribution of the diagram (2.2) in Fig.\,\ref{f7} from which the factor $2$ on the RHS of Eq.\,(\ref{ex2b}) arises. \emph{We note that $\t{\Sigma}_{\sigma}^{\X{(2)}}(z)$ is a Nevanlinna function of $z$.}

From the expression in Eq.\,(\ref{ex0b}) one observes that (\emph{cf.} Eqs\,(\ref{e25}) and (\ref{e50a}))
\begin{equation}\label{ex2d}
\t{\Sigma}_{\sigma}^{\X{(2)}}(z) \sim \frac{U^2}{4\hbar\hspace{0.6pt}z} + \frac{9 U^4}{16\hbar\hspace{0.6pt} z^3} + \dots \equiv \t{\Sigma}_{\sigma}(z) +  \frac{9 U^4}{16\hbar\hspace{0.6pt} z^3} + \dots.\;\; \text{for}\;\; z\to \infty.
\end{equation}
The fact that $\t{\Sigma}_{\sigma}^{\X{(2)}}(z)$ to leading order decays like $1/z$ for $z\to\infty$ is in conformity with the exact result in Eq.\,(\ref{e7g}). Since $\t{\Sigma}^{\X{(3)}}(z) \equiv 0$, \S\,\ref{sec.d52}, on account of the latter exact result it is to be expected that (i) the perturbatively-calculated self-energy (in terms of skeleton self-energy diagrams and the exact Green function $\t{G}(z)$) does not contain any contribution decaying like $1/z^2$ for $z\to\infty$, and (ii) the leading-order asymptotic contribution of $\t{\Sigma}_{\sigma}^{\X{(4)}}(z)$ in the region $z\to\infty$ cancels the next-to-leading-order term in the asymptotic series expansion of $\t{\Sigma}_{\sigma}^{\X{(2)}}(z)$ in this region. In \S\,\ref{sec.d53} we explicitly establish that while indeed to leading order $\t{\Sigma}_{\sigma}^{\X{(4)}}(z)$ decays like $1/z^3$ for $z\to\infty$ (in conformity with the exact result in Eq.\,(\ref{e7g})), the coefficient of the relevant term is \textsl{not} equal to $-9 U^4/(16\hbar)$, but equal to $5/8$th of this. The reason for this lies in the \textsl{locality} of the problem at hand, leading to the violation of Eq.\,(\ref{e7h}) for $j \ge 3$; in the local limit, $\Sigma_{\infty_j}$ for $j\ge 3$ may be determined by the contributions of $\Sigma_{\infty_j}^{\X{(\nu)}}$ with $\nu$ indefinitely large, and not as in Eq.\,(\ref{e7h}) with $\nu \in\{2,3,\dots, j+1\}$ (see the discussions following Eqs\,(\ref{e50a}) and (\ref{e52b})). We shall return to this problem in \S\,\ref{sec.d53}.

\refstepcounter{dummyX}
\subsubsection{The 3rd-order self-energy contributions}
\phantomsection
\label{sec.d52}
Similarly as in the previous section, and on the basis of the details presented in \S\,\ref{sd21}, for the $3$rd-order self-energy contributions corresponding to the diagrams specified in \S\,\ref{sd12} and the GS of the `Hubbard atom', we obtain
\begin{align}
\label{ex3a}
\t{\Sigma}_{\sigma}^{\X{(3.1)}}(z) &= +\frac{1}{4} \t{\Phi}_{3,2}(z,U) -\frac{1}{4} \t{\Phi}_{3,1}(z,U), \\
\label{ex3b}
\t{\Sigma}_{\sigma}^{\X{(3.2)}}(z) &= -\t{\Sigma}_{\sigma}^{\X{(3.1)}}(z), \phantom{\frac{U^2}{123}} \\
\label{ex3c}
\t{\Sigma}_{\sigma}^{\X{(3.3)}}(z) &= +\t{\Sigma}_{\sigma}^{\X{(3.1)}}(z), \phantom{\frac{U^2}{123}} \\
\label{ex3d}
\t{\Sigma}_{\sigma}^{\X{(3.4)}}(z) &= +\t{\Sigma}_{\sigma}^{\X{(3.1)}}(z), \phantom{\frac{U^2}{123}} \\
\label{ex3e}
\t{\Sigma}_{\sigma}^{\X{(3.5)}}(z) &= -2\t{\Sigma}_{\sigma}^{\X{(3.1)}}(z), \phantom{\frac{U^2}{123}} \\
\label{ex3f}
\t{\Sigma}_{\sigma}^{\X{(3.6)}}(z) &= -2\t{\Sigma}_{\sigma}^{\X{(3.1)}}(z), \phantom{\frac{U^2}{123}} \\
\label{ex3g}
\t{\Sigma}_{\sigma}^{\X{(3.7)}}(z) &= +2\t{\Sigma}_{\sigma}^{\X{(3.1)}}(z), \phantom{\frac{U^2}{123}} \\
\label{ex3h}
\t{\Sigma}_{\sigma}^{\X{(3.8)}}(z) &= -2\t{\Sigma}_{\sigma}^{\X{(3.1)}}(z), \phantom{\frac{U^2}{123}} \\
\label{ex3i}
\t{\Sigma}_{\sigma}^{\X{(3.9)}}(z) &= -2\t{\Sigma}_{\sigma}^{\X{(3.1)}}(z), \phantom{\frac{U^2}{123}} \\
\label{ex3j}
\t{\Sigma}_{\sigma}^{\X{(3.10)}}(z) &= +4\t{\Sigma}_{\sigma}^{\X{(3.1)}}(z). \phantom{\frac{U^2}{123}}
\end{align}
Thus, following the expression in Eq.\,(\ref{ex1b}), one arrives at the expected result
\begin{equation}\label{ex3k}
\t{\Sigma}_{\sigma}^{\X{(3)}}(z) \equiv 0.
\end{equation}
Following Eq.\,(\ref{ex1e}), one similarly has (see Eqs\,(\ref{ex3f}) and (\ref{ex3g}))
\begin{equation}\label{ex3l}
\t{\Sigma}_{\sigma}^{\X{(3)}}(z) = -\t{\Sigma}_{\sigma}^{\X{(3.1)}}(z) + \t{\Sigma}_{\sigma}^{\X{(3.1)}}(z) \equiv 0.
\end{equation}
Note that while $\t{\Sigma}_{\sigma}^{\X{(3.6)}}(z) = -2 \t{\Sigma}_{\sigma}^{\X{(3.1)}}(z)$, Eq.\,(\ref{ex3f}), one has $\t{\Sigma}_{\sigma}^{\X{(3.6)}}(z)\vert_{\h{\eta}} = -\t{\Sigma}_{\sigma}^{\X{(3.1)}}(z)$, and while $\t{\Sigma}_{\sigma}^{\X{(3.7)}}(z) = +2 \t{\Sigma}_{\sigma}^{\X{(3.1)}}(z)$, Eq.\,(\ref{ex3g}), one has $\t{\Sigma}_{\sigma}^{\X{(3.7)}}(z)\vert_{\h{\eta}} = \t{\Sigma}_{\sigma}^{\X{(3.1)}}(z)$ (\emph{cf.} Eqs\,(\ref{ex04a}) and (\ref{ex04b})).

\refstepcounter{dummyX}
\subsubsection{The 4th-order self-energy contributions}
\phantomsection
\label{sec.d53}
Along the same lines as in the previous two sections, for the contributions of the $4$th-order skeleton self-energy diagrams specified in \S\,\ref{sd13}, corresponding to the atomic limit of the half-filled GS of the Hubbard Hamiltonian $\h{\mathcal{H}}$ for spin-$\tfrac{1}{2}$ particles, Eq.\,(\ref{ex01bx}),\footnote{See \S\,\protect\ref{sec3.b}.} one obtains
\begin{align}
\label{ex1}
\t{\Sigma}_{\sigma}^{\X{(4.1)}}(z) &= +\frac{9}{64}\hspace{0.8pt} \t{\Phi}_{5,1}(z,U) -\frac{1}{16}\hspace{0.8pt} \t{\Phi}_{3,3}(z,U) -\frac{9}{64}\hspace{0.8pt} \t{\Phi}_{3,1}(z,U), \\
\label{ex2}
\t{\Sigma}_{\sigma}^{\X{(4.2)}}(z) &= -\frac{1}{16}\hspace{0.8pt} \t{\Phi}_{3,3}(z,U) +\frac{3}{32}\hspace{0.8pt} \t{\Phi}_{3,2}(z,U) -\frac{3}{32}\hspace{0.8pt} \t{\Phi}_{3,1}(z,U), \\
\label{ex3}
\t{\Sigma}_{\sigma}^{\X{(4.3)}}(z) &= -\frac{9}{32}\hspace{0.8pt} \t{\Phi}_{5,1}(z,U) + \hspace{2.2pt}\frac{1}{8}\hspace{2.2pt} \t{\Phi}_{3,2}(z,U) +\frac{3}{16}\hspace{0.8pt} \t{\Phi}_{3,1}(z,U), \\
\label{ex4}
\t{\Sigma}_{\sigma}^{\X{(4.4)}}(z) &= +\t{\Sigma}_{\sigma}^{\X{(4.1)}}(z), \phantom{\frac{U^2}{123}} \\
\label{ex5}
\t{\Sigma}_{\sigma}^{\X{(4.5)}}(z) &= +\t{\Sigma}_{\sigma}^{\X{(4.2)}}(z), \phantom{\frac{U^2}{123}} \\
\label{ex6}
\t{\Sigma}_{\sigma}^{\X{(4.6)}}(z) &= -\t{\Sigma}_{\sigma}^{\X{(4.1)}}(z), \phantom{\frac{U^2}{123}} \\
\label{ex7}
\t{\Sigma}_{\sigma}^{\X{(4.7)}}(z) &= -\t{\Sigma}_{\sigma}^{\X{(4.3)}}(z), \phantom{\frac{U^2}{123}} \\
\label{ex8}
\t{\Sigma}_{\sigma}^{\X{(4.8)}}(z) &= -\t{\Sigma}_{\sigma}^{\X{(4.1)}}(z), \phantom{\frac{U^2}{123}} \\
\label{ex9}
\t{\Sigma}_{\sigma}^{\X{(4.9)}}(z) &= -\t{\Sigma}_{\sigma}^{\X{(4.1)}}(z), \phantom{\frac{U^2}{123}} \\
\label{ex10}
\t{\Sigma}_{\sigma}^{\X{(4.10)}}(z) &= +\t{\Sigma}_{\sigma}^{\X{(4.3)}}(z), \phantom{\frac{U^2}{123}} \\
\label{ex11}
\t{\Sigma}_{\sigma}^{\X{(4.11)}}(z) &= +\t{\Sigma}_{\sigma}^{\X{(4.3)}}(z), \phantom{\frac{U^2}{123}} \\
\label{ex12}
\t{\Sigma}_{\sigma}^{\X{(4.12)}}(z) &= +\t{\Sigma}_{\sigma}^{\X{(4.2)}}(z), \phantom{\frac{U^2}{123}} \\
\label{ex13}
\t{\Sigma}_{\sigma}^{\X{(4.13)}}(z) &= +\t{\Sigma}_{\sigma}^{\X{(4.2)}}(z), \phantom{\frac{U^2}{123}} \\
\label{ex14}
\t{\Sigma}_{\sigma}^{\X{(4.14)}}(z) &= +\t{\Sigma}_{\sigma}^{\X{(4.1)}}(z), \phantom{\frac{U^2}{123}} \\
\label{ex15}
\t{\Sigma}_{\sigma}^{\X{(4.15)}}(z) &= +\t{\Sigma}_{\sigma}^{\X{(4.3)}}(z), \phantom{\frac{U^2}{123}} \\
\label{ex16}
\t{\Sigma}_{\sigma}^{\X{(4.16)}}(z) &= -\t{\Sigma}_{\sigma}^{\X{(4.3)}}(z), \phantom{\frac{U^2}{123}} \\
\label{ex17}
\t{\Sigma}_{\sigma}^{\X{(4.17)}}(z) &= -\t{\Sigma}_{\sigma}^{\X{(4.3)}}(z), \phantom{\frac{U^2}{123}} \\
\label{ex18}
\t{\Sigma}_{\sigma}^{\X{(4.18)}}(z) &= +\t{\Sigma}_{\sigma}^{\X{(4.2)}}(z), \phantom{\frac{U^2}{123}} \\
\label{ex19}
\t{\Sigma}_{\sigma}^{\X{(4.19)}}(z) &= -\t{\Sigma}_{\sigma}^{\X{(4.1)}}(z), \phantom{\frac{U^2}{123}} \\
\label{ex20}
\t{\Sigma}_{\sigma}^{\X{(4.20)}}(z) &= -\t{\Sigma}_{\sigma}^{\X{(4.1)}}(z), \phantom{\frac{U^2}{123}} \\
\label{ex21}
\t{\Sigma}_{\sigma}^{\X{(4.21)}}(z) &= +\t{\Sigma}_{\sigma}^{\X{(4.1)}}(z), \phantom{\frac{U^2}{123}} \\
\label{ex22}
\t{\Sigma}_{\sigma}^{\X{(4.22)}}(z) &= -\t{\Sigma}_{\sigma}^{\X{(4.1)}}(z), \phantom{\frac{U^2}{123}} \\
\label{ex23}
\t{\Sigma}_{\sigma}^{\X{(4.23)}}(z) &= +\t{\Sigma}_{\sigma}^{\X{(4.1)}}(z), \phantom{\frac{U^2}{123}} \\
\label{ex24}
\t{\Sigma}_{\sigma}^{\X{(4.24)}}(z) &= +\t{\Sigma}_{\sigma}^{\X{(4.3)}}(z), \phantom{\frac{U^2}{123}} \\
\label{ex25}
\t{\Sigma}_{\sigma}^{\X{(4.25)}}(z) &= -\t{\Sigma}_{\sigma}^{\X{(4.3)}}(z), \phantom{\frac{U^2}{123}} \\
\label{ex26}
\t{\Sigma}_{\sigma}^{\X{(4.26)}}(z) &= -\t{\Sigma}_{\sigma}^{\X{(4.3)}}(z), \phantom{\frac{U^2}{123}} \\
\label{ex27}
\t{\Sigma}_{\sigma}^{\X{(4.27)}}(z) &= -\t{\Sigma}_{\sigma}^{\X{(4.3)}}(z), \phantom{\frac{U^2}{123}} \\
\label{ex28}
\t{\Sigma}_{\sigma}^{\X{(4.28)}}(z) &= -2 \t{\Sigma}_{\sigma}^{\X{(4.2)}}(z), \phantom{\frac{U^2}{123}} \\
\label{ex29}
\t{\Sigma}_{\sigma}^{\X{(4.29)}}(z) &= +2 \t{\Sigma}_{\sigma}^{\X{(4.3)}}(z), \phantom{\frac{U^2}{123}} \\
\label{ex30}
\t{\Sigma}_{\sigma}^{\X{(4.30)}}(z) &= -2 \t{\Sigma}_{\sigma}^{\X{(4.3)}}(z), \phantom{\frac{U^2}{123}} \\
\label{ex31}
\t{\Sigma}_{\sigma}^{\X{(4.31)}}(z) &= -2 \t{\Sigma}_{\sigma}^{\X{(4.3)}}(z), \phantom{\frac{U^2}{123}} \\
\label{ex32}
\t{\Sigma}_{\sigma}^{\X{(4.32)}}(z) &= -2 \t{\Sigma}_{\sigma}^{\X{(4.3)}}(z), \phantom{\frac{U^2}{123}} \\
\label{ex33}
\t{\Sigma}_{\sigma}^{\X{(4.33)}}(z) &= +2 \t{\Sigma}_{\sigma}^{\X{(4.3)}}(z), \phantom{\frac{U^2}{123}} \\
\label{ex34}
\t{\Sigma}_{\sigma}^{\X{(4.34)}}(z) &= -2 \t{\Sigma}_{\sigma}^{\X{(4.1)}}(z), \phantom{\frac{U^2}{123}} \\
\label{ex35}
\t{\Sigma}_{\sigma}^{\X{(4.35)}}(z) &= -2 \t{\Sigma}_{\sigma}^{\X{(4.1)}}(z), \phantom{\frac{U^2}{123}} \\
\label{ex36}
\t{\Sigma}_{\sigma}^{\X{(4.36)}}(z) &= +2 \t{\Sigma}_{\sigma}^{\X{(4.1)}}(z), \phantom{\frac{U^2}{123}} \\
\label{ex37}
\t{\Sigma}_{\sigma}^{\X{(4.37)}}(z) &= +2 \t{\Sigma}_{\sigma}^{\X{(4.1)}}(z), \phantom{\frac{U^2}{123}} \\
\label{ex38}
\t{\Sigma}_{\sigma}^{\X{(4.38)}}(z) &= -2 \t{\Sigma}_{\sigma}^{\X{(4.2)}}(z), \phantom{\frac{U^2}{123}} \\
\label{ex39}
\t{\Sigma}_{\sigma}^{\X{(4.39)}}(z) &= -2 \t{\Sigma}_{\sigma}^{\X{(4.2)}}(z), \phantom{\frac{U^2}{123}} \\
\label{ex40}
\t{\Sigma}_{\sigma}^{\X{(4.40)}}(z) &= +2 \t{\Sigma}_{\sigma}^{\X{(4.1)}}(z), \phantom{\frac{U^2}{123}} \\
\label{ex41}
\t{\Sigma}_{\sigma}^{\X{(4.41)}}(z) &= +2 \t{\Sigma}_{\sigma}^{\X{(4.1)}}(z), \phantom{\frac{U^2}{123}} \\
\label{ex42}
\t{\Sigma}_{\sigma}^{\X{(4.42)}}(z) &= +2 \t{\Sigma}_{\sigma}^{\X{(4.3)}}(z), \phantom{\frac{U^2}{123}} \\
\label{ex43}
\t{\Sigma}_{\sigma}^{\X{(4.43)}}(z) &= +2 \t{\Sigma}_{\sigma}^{\X{(4.3)}}(z), \phantom{\frac{U^2}{123}} \\
\label{ex44}
\t{\Sigma}_{\sigma}^{\X{(4.44)}}(z) &= -2 \t{\Sigma}_{\sigma}^{\X{(4.2)}}(z), \phantom{\frac{U^2}{123}} \\
\label{ex45}
\t{\Sigma}_{\sigma}^{\X{(4.45)}}(z) &= -2 \t{\Sigma}_{\sigma}^{\X{(4.2)}}(z), \phantom{\frac{U^2}{123}} \\
\label{ex46}
\t{\Sigma}_{\sigma}^{\X{(4.46)}}(z) &= -2 \t{\Sigma}_{\sigma}^{\X{(4.3)}}(z), \phantom{\frac{U^2}{123}} \\
\label{ex47}
\t{\Sigma}_{\sigma}^{\X{(4.47)}}(z) &= +2 \t{\Sigma}_{\sigma}^{\X{(4.3)}}(z), \phantom{\frac{U^2}{123}} \\
\label{ex48}
\t{\Sigma}_{\sigma}^{\X{(4.48)}}(z) &= -2 \t{\Sigma}_{\sigma}^{\X{(4.1)}}(z), \phantom{\frac{U^2}{123}} \\
\label{ex49}
\t{\Sigma}_{\sigma}^{\X{(4.49)}}(z) &= -2 \t{\Sigma}_{\sigma}^{\X{(4.3)}}(z), \phantom{\frac{U^2}{123}} \\
\label{ex50}
\t{\Sigma}_{\sigma}^{\X{(4.50)}}(z) &= +2 \t{\Sigma}_{\sigma}^{\X{(4.1)}}(z), \phantom{\frac{U^2}{123}} \\
\label{ex51}
\t{\Sigma}_{\sigma}^{\X{(4.51)}}(z) &= -2 \t{\Sigma}_{\sigma}^{\X{(4.3)}}(z), \phantom{\frac{U^2}{123}} \\
\label{ex52}
\t{\Sigma}_{\sigma}^{\X{(4.52)}}(z) &= +2 \t{\Sigma}_{\sigma}^{\X{(4.3)}}(z), \phantom{\frac{U^2}{123}} \\
\label{ex53}
\t{\Sigma}_{\sigma}^{\X{(4.53)}}(z) &= +2 \t{\Sigma}_{\sigma}^{\X{(4.3)}}(z), \phantom{\frac{U^2}{123}} \\
\label{ex54}
\t{\Sigma}_{\sigma}^{\X{(4.54)}}(z) &= -2 \t{\Sigma}_{\sigma}^{\X{(4.1)}}(z), \phantom{\frac{U^2}{123}} \\
\label{ex55}
\t{\Sigma}_{\sigma}^{\X{(4.55)}}(z) &= +2 \t{\Sigma}_{\sigma}^{\X{(4.1)}}(z), \phantom{\frac{U^2}{123}} \\
\label{ex56}
\t{\Sigma}_{\sigma}^{\X{(4.56)}}(z) &= -2 \t{\Sigma}_{\sigma}^{\X{(4.2)}}(z), \phantom{\frac{U^2}{123}} \\
\label{ex57}
\t{\Sigma}_{\sigma}^{\X{(4.57)}}(z) &= +2 \t{\Sigma}_{\sigma}^{\X{(4.3)}}(z), \phantom{\frac{U^2}{123}} \\
\label{ex58}
\t{\Sigma}_{\sigma}^{\X{(4.58)}}(z) &= -2 \t{\Sigma}_{\sigma}^{\X{(4.3)}}(z), \phantom{\frac{U^2}{123}} \\
\label{ex59}
\t{\Sigma}_{\sigma}^{\X{(4.59)}}(z) &= -2 \t{\Sigma}_{\sigma}^{\X{(4.3)}}(z), \phantom{\frac{U^2}{123}} \\
\label{ex60}
\t{\Sigma}_{\sigma}^{\X{(4.60)}}(z) &= -2 \t{\Sigma}_{\sigma}^{\X{(4.2)}}(z), \phantom{\frac{U^2}{123}} \\
\label{ex61}
\t{\Sigma}_{\sigma}^{\X{(4.61)}}(z) &= -2 \t{\Sigma}_{\sigma}^{\X{(4.2)}}(z), \phantom{\frac{U^2}{123}} \\
\label{ex62}
\t{\Sigma}_{\sigma}^{\X{(4.62)}}(z) &= +2 \t{\Sigma}_{\sigma}^{\X{(4.1)}}(z), \phantom{\frac{U^2}{123}} \\
\label{ex63}
\t{\Sigma}_{\sigma}^{\X{(4.63)}}(z) &= +2 \t{\Sigma}_{\sigma}^{\X{(4.1)}}(z), \phantom{\frac{U^2}{123}} \\
\label{ex64}
\t{\Sigma}_{\sigma}^{\X{(4.64)}}(z) &= -2 \t{\Sigma}_{\sigma}^{\X{(4.1)}}(z), \phantom{\frac{U^2}{123}} \\
\label{ex65}
\t{\Sigma}_{\sigma}^{\X{(4.65)}}(z) &= -2 \t{\Sigma}_{\sigma}^{\X{(4.1)}}(z), \phantom{\frac{U^2}{123}} \\
\label{ex66}
\t{\Sigma}_{\sigma}^{\X{(4.66)}}(z) &= -2 \t{\Sigma}_{\sigma}^{\X{(4.1)}}(z), \phantom{\frac{U^2}{123}} \\
\label{ex67}
\t{\Sigma}_{\sigma}^{\X{(4.67)}}(z) &= -2 \t{\Sigma}_{\sigma}^{\X{(4.1)}}(z), \phantom{\frac{U^2}{123}} \\
\label{ex68}
\t{\Sigma}_{\sigma}^{\X{(4.68)}}(z) &= +4 \t{\Sigma}_{\sigma}^{\X{(4.2)}}(z), \phantom{\frac{U^2}{123}} \\
\label{ex69}
\t{\Sigma}_{\sigma}^{\X{(4.69)}}(z) &= +4 \t{\Sigma}_{\sigma}^{\X{(4.2)}}(z), \phantom{\frac{U^2}{123}} \\
\label{ex70}
\t{\Sigma}_{\sigma}^{\X{(4.70)}}(z) &= +4 \t{\Sigma}_{\sigma}^{\X{(4.3)}}(z), \phantom{\frac{U^2}{123}} \\
\label{ex71}
\t{\Sigma}_{\sigma}^{\X{(4.71)}}(z) &= +4 \t{\Sigma}_{\sigma}^{\X{(4.3)}}(z), \phantom{\frac{U^2}{123}} \\
\label{ex72}
\t{\Sigma}_{\sigma}^{\X{(4.72)}}(z) &= -4 \t{\Sigma}_{\sigma}^{\X{(4.3)}}(z), \phantom{\frac{U^2}{123}} \\
\label{ex73}
\t{\Sigma}_{\sigma}^{\X{(4.73)}}(z) &= +4 \t{\Sigma}_{\sigma}^{\X{(4.1)}}(z), \phantom{\frac{U^2}{123}} \\
\label{ex74}
\t{\Sigma}_{\sigma}^{\X{(4.74)}}(z) &= -4 \t{\Sigma}_{\sigma}^{\X{(4.1)}}(z), \phantom{\frac{U^2}{123}} \\
\label{ex75}
\t{\Sigma}_{\sigma}^{\X{(4.75)}}(z) &= -4 \t{\Sigma}_{\sigma}^{\X{(4.1)}}(z), \phantom{\frac{U^2}{123}} \\
\label{ex76}
\t{\Sigma}_{\sigma}^{\X{(4.76)}}(z) &= +4 \t{\Sigma}_{\sigma}^{\X{(4.1)}}(z), \phantom{\frac{U^2}{123}} \\
\label{ex77}
\t{\Sigma}_{\sigma}^{\X{(4.77)}}(z) &= +4 \t{\Sigma}_{\sigma}^{\X{(4.3)}}(z), \phantom{\frac{U^2}{123}} \\
\label{ex78}
\t{\Sigma}_{\sigma}^{\X{(4.78)}}(z) &= -4 \t{\Sigma}_{\sigma}^{\X{(4.3)}}(z), \phantom{\frac{U^2}{123}} \\
\label{ex79}
\t{\Sigma}_{\sigma}^{\X{(4.79)}}(z) &= +4 \t{\Sigma}_{\sigma}^{\X{(4.1)}}(z), \phantom{\frac{U^2}{123}} \\
\label{ex80}
\t{\Sigma}_{\sigma}^{\X{(4.80)}}(z) &= +4 \t{\Sigma}_{\sigma}^{\X{(4.2)}}(z), \phantom{\frac{U^2}{123}} \\
\label{ex81}
\t{\Sigma}_{\sigma}^{\X{(4.81)}}(z) &= +4 \t{\Sigma}_{\sigma}^{\X{(4.2)}}(z), \phantom{\frac{U^2}{123}} \\
\label{ex82}
\t{\Sigma}_{\sigma}^{\X{(4.82)}}(z) &= -8 \t{\Sigma}_{\sigma}^{\X{(4.2)}}(z). \phantom{\frac{U^2}{123}}
\end{align}
Thus, following Eq.\,(\ref{ex1c}),
\begin{equation}\label{ex0e}
\t{\Sigma}_{\sigma}^{\X{(4)}}(z) = -\frac{27}{64}\hspace{0.8pt} \Phi_{5,1}(z,U) + \frac{3}{32}\hspace{0.8pt} \Phi_{3,2}(z,U) + \frac{27}{64}\hspace{0.8pt} \Phi_{3,1}(z,U).
\end{equation}
Following Eq.\,(\ref{ex1f}), for $\h{\eta} = \hspace{0.28cm}\h{\hspace{-0.28cm}\mathpzc{H}}$, $\h{\mathsf{H}}$ one has
\begin{align}\label{ex0g}
\t{\Sigma}_{\sigma}^{\X{(4)}}(z) &= -\t{\Sigma}_{\sigma}^{\X{(4.2)}}(z) - \t{\Sigma}_{\sigma}^{\X{(4.2)}}(z) + \t{\Sigma}_{\sigma}^{\X{(4.1)}}(z) + \t{\Sigma}_{\sigma}^{\X{(4.1)}}(z) + \t{\Sigma}_{\sigma}^{\X{(4.3)}}(z)\nonumber\\
&+ \t{\Sigma}_{\sigma}^{\X{(4.3)}}(z) + \t{\Sigma}_{\sigma}^{\X{(4.3)}}(z) + \t{\Sigma}_{\sigma}^{\X{(4.1)}}(z) - \t{\Sigma}_{\sigma}^{\X{(4.2)}}(z)\nonumber\\
&\equiv +3 \t{\Sigma}_{\sigma}^{\X{(4.1)}}(z) - 3\t{\Sigma}_{\sigma}^{\X{(4.2)}}(z) + 3\t{\Sigma}_{\sigma}^{\X{(4.3)}}(z),
\end{align}
which on account of the expressions in Eqs\,(\ref{ex1}), (\ref{ex2}), and (\ref{ex3}) results in the same expression as that on the RHS of Eq.\,(\ref{ex0e}). \emph{We note that $\t{\Sigma}_{\sigma}^{\X{(4)}}(z)$ is \textsl{not} a Nevanlinna function of $z$.} It is however clearly analytic in the region $\im[z] \not=0$ of the $z$-plane and satisfies the reflection property $\t{\Sigma}_{\sigma}^{\X{(4)}}(z^*) = \t{\Sigma}_{\sigma}^{\X{(4)}*}(z)$ in Eq.\,(\ref{e3b}).

From the expression in Eq.\,(\ref{ex0e}), one deduces that
\begin{equation}\label{ex0f}
\t{\Sigma}_{\sigma}^{\X{(4)}}(z) \sim -\frac{45 U^4}{128\hbar\hspace{0.6pt} z^3} - \frac{837 U^6}{256\hbar\hspace{0.6pt} z^5} - \dots \;\; \text{for}\;\;\ z\to \infty,
\end{equation}
in conformity with the general asymptotic expression in Eq.\,(\ref{e7g}). Nonetheless, one observes that the leading-order term on the RHS of Eq.\,(\ref{ex0f}) does not fully compensate the next-to-leading order term on the RHS of Eq.\,(\ref{ex2d}). This would be necessary, since the asymptotic series expansion of the exact self-energy, Eq.\,(\ref{e25}), for $z\to\infty$ consists of a single term, decaying like $1/z$. One notes that $45/128 = (5/8) (9/16)$, that is the leading-order term on the RHS of Eq.\,(\ref{ex0f}) is $5/8$th of its expected value from the perspective of Eq.\,(\ref{ex2d}). For clarity, let us assume that we had not calculated the function $\t{\Sigma}_{\sigma}^{\X{(4)}}(z)$. On account of the exact result in Eq.\,(\ref{e7h}), one has
\begin{equation}\label{ex0h}
\Sigma_{\sigma;\infty_3} = \Sigma_{\sigma;\infty_3}^{\X{(2)}} + \Sigma_{\sigma;\infty_3}^{\X{(3)}} + \Sigma_{\sigma;\infty_3}^{\X{(4)}}.
\end{equation}
Following Eq.\,(\ref{e25}), $\Sigma_{\sigma;\infty_3} \equiv 0$. Since $\t{\Sigma}_{\sigma}^{\X{(3)}}(z) \equiv 0$, Eq.\,(\ref{ex3k}), one further has $ \Sigma_{\sigma;\infty_3}^{\X{(3)}} = 0$. Thus the equality in Eq.\,(\ref{ex0h}) results in
\begin{equation}\label{ex0i}
\Sigma_{\sigma;\infty_3}^{\X{(4)}} = -\Sigma_{\sigma;\infty_3}^{\X{(2)}} \equiv -\frac{9 U^4}{16\hbar},
\end{equation}
where the last equality follows from the asymptotic expression in Eq.\,(\ref{ex2d}). The result in Eq.\,(\ref{ex0i}) contradicts that in Eq.\,(\ref{ex0f}) obtained through the explicit calculation of $\t{\Sigma}_{\sigma}^{\X{(4)}}(z)$. We have therefore established that \emph{in the case of the half-filled GS of the Hubbard Hamiltonian for spin-$\tfrac{1}{2}$ particles in the atomic limit, the equality in Eq.\,(\ref{e7h}) fails for at least $j = 3$, Eq.\,(\ref{ex0h}); we expect this failure to be a permanent feature for all $j \ge 3$.} To be more explicit, in the local limit, at the very least the RHS of Eq.\,(\ref{ex0h}) is to be complemented by $\Sigma_{\sigma;\infty_3}^{\X{(6)}}$, the asymptotic coefficient $\Sigma_{\sigma;\infty_3}^{\X{(5)}}$ being in the case at hand vanishing by p-h symmetry.

Additional light on the above observation is shed by considering the \textsl{proper} $4$th-order self-energy diagrams as evaluated in terms of the non-interacting Green function $\t{G}_{\X{0}}(z)$.\footnote{In the light of the observation in \S\,\protect\ref{s4xa}, in terms of $\protect\t{G}_{\epsilon}(z)$, with $\epsilon >0$, effecting the limit $\epsilon\downarrow 0$ subsequent to the evaluation of the perturbational contributions.} To this end, by employing the Hubbard Hamiltonian $\hspace{0.28cm}\h{\hspace{-0.28cm}\mathpzc{H}}$, Eq.\,(\ref{ex01f}), or $\h{\mathsf{H}}$, Eq.\,(\ref{ex01k}), only three additional diagrams (\textsl{non-skeleton} ones) are to be considered. Denoting the self-energy contributions associated with these three diagrams by $\Sigma_{\sigma}^{\X{(4.83)}}$, $\Sigma_{\sigma}^{\X{(4.84)}}$, and $\Sigma_{\sigma}^{\X{(4.85)}}$, following the procedure described in \S\,\ref{sx1}, with
\begin{align}\label{ex0j}
\mathsf{A}_2^4 &\doteq \langle 1',2'\rangle,\; \langle 3',4'\rangle,\; \langle 4',5'\rangle,\; \langle 5',6'\rangle,\; \langle 6',3'\rangle,\; \langle 7',8'\rangle,\; \langle 8',7'\rangle, \nonumber\\
\mathsf{B}_2^4 &\doteq \langle 1',3'\rangle,\; \langle 3',4'\rangle,\; \langle 4',2'\rangle,\; \langle 5',6'\rangle,\; \langle 6',5'\rangle,\; \langle 7',8'\rangle,\; \langle 8',7'\rangle,
\end{align}
one has\,\footnote{$\Sigma_{\sigma}^{\protect\X{(4.83)}}: (3', 4', 1', 5', 7', 8', 2', 6')$, $\Sigma_{\sigma}^{\protect\X{(4.84)}}: (3', 7', 1', 2', 4', 8', 5', 6')$, $\Sigma_{\sigma}^{\protect\X{(4.85)}}: (3', 4', 5', 2', 7', 8', 1', 6')$.}
\begin{equation}\label{ex0k}
\Sigma_{\sigma}^{\X{(4.83)}}:\; \mathsf{A}_2^4,\; (1',4'),\; (5',8'),\; (6',7'),\; (3',2'),
\end{equation}
\begin{equation}\label{ex0l}
\Sigma_{\sigma}^{\X{(4.84)}}:\; \mathsf{A}_2^4,\; (1',3'),\; (6',8'),\; (5',7'),\; (4',2'),
\end{equation}
\begin{equation}\label{ex0m}
\Sigma_{\sigma}^{\X{(4.85)}}:\; \mathsf{B}_2^4,\; (1',8'),\; (3',6'),\; (4',5'),\; (7',2').
\end{equation}
These three non-skeleton proper self-energy diagrams coincide with respectively the middle, the lower and the upper diagram in Fig.\,5, p.\,498, of Ref.\,\citen{GJMNN03}. For uniformity of presentation, but without loss of generality (as will become apparent below), we evaluate the contributions of these diagrams in terms the \textsl{interacting} Green function $\t{G}(z)$, Eq.\,(\ref{e25}), obtaining the contributions associated with $\t{G}_{\X{0}}$ (better  $\t{G}_{\X{0^+}}$) on the basis of the prescription in Eq.\,(\ref{ex0s}) below. \emph{Similarly as in other cases, we calculate the contributions of these diagrams under the assumption that they correspond to the Hubbard Hamiltonian $\h{\mathcal{H}}$, Eq.\,(\ref{ex01bx}), for spin-$\tfrac{1}{2}$ particles, \S\,\ref{sd2}, accounting for the underlying Hamiltonian being $\hspace{0.28cm}\h{\hspace{-0.28cm}\mathpzc{H}}$ by means of an appropriate multiplicative factor in Eq.\,(\ref{ex0q}) below.} We obtain
\begin{align}
\label{ex0n}
\t{\Sigma}_{\sigma}^{\X{(4.83)}}(z) &= +\frac{9}{16}\hspace{0.8pt} \t{\Phi}_{5,1}(z,U) - \frac{1}{8}\hspace{0.8pt} \t{\Phi}_{3,2}(z,U) - \frac{9}{16}\hspace{0.8pt} \t{\Phi}_{3,1}(z,U), \\
\label{ex0o}
\t{\Sigma}_{\sigma}^{\X{(4.84)}}(z) &= +\t{\Sigma}_{\sigma}^{\X{(4.83)}},\phantom{\frac{U^2}{123}} \\
\label{ex0p}
\t{\Sigma}_{\sigma}^{\X{(4.85)}}(z) &= +\t{\Sigma}_{\sigma}^{\X{(4.83)}}.\phantom{\frac{U^2}{123}}
\end{align}

With $\delta\t{\Sigma}_{\sigma}^{\X{(4)}}(z)$ denoting the contributions of all $4$th-order \textsl{non-skeleton} diagrams as evaluated (\textsl{inappropriately}) in terms of the interacting Green function $\t{G}(z)$, following the considerations in \S\,\ref{sd2}, one has (\emph{cf.} Eqs\,(\ref{ex1d}), (\ref{ex1e}) and (\ref{ex1f}))
\begin{align}\label{ex0q}
\delta\t{\Sigma}_{\sigma}^{\X{(4)}}(z) &= \left.\t{\Sigma}_{\sigma}^{\X{(4.83)}}(z)\right|_{\h{\eta}} + \left.\t{\Sigma}_{\sigma}^{\X{(4.84)}}(z)\right|_{\h{\eta}} + \left.\t{\Sigma}_{\sigma}^{\X{(4.85)}}(z)\right|_{\h{\eta}} = \frac{3}{4} \t{\Sigma}_{\sigma}^{\X{(4.83)}}(z)\nonumber\\
&\equiv +\frac{27}{64}\hspace{0.8pt}\t{\Phi}_{5,1}(z,U) - \frac{3}{32}\hspace{0.8pt} \t{\Phi}_{3,2}(z,U) - \frac{27}{64}\hspace{0.8pt}\t{\Phi}_{3,1}(z,U),
\end{align}
where the pre-factor $\tfrac{1}{4}$ following the second equality accounts for the fact that the diagram corresponding to $\t{\Sigma}_{\sigma}^{\X{(4.83)}}(z)$ consists of two fermion loops, the sum over the spin indices of which (accounting for a factor of $2^2$ in the case of spin unpolarised states of spin-$\tfrac{1}{2}$ particles) must be undone in going from $\t{\Sigma}_{\sigma}^{\X{(4.83)}}(z)\vert_{\h{\eta}}$ to $\t{\Sigma}_{\sigma}^{\X{(4.83)}}(z)$ (\emph{cf.} Eqs\,(\ref{ex04a}) and (\ref{ex04b})).

With reference to Eq.\,(\ref{ex0e}), the result in Eq.\,(\ref{ex0q}) reveals that
\begin{equation}\label{ex0r}
\delta\t{\Sigma}_{\sigma}^{\X{(4)}}(z) \equiv -\t{\Sigma}_{\sigma}^{\X{(4)}}(z).
\end{equation}
To clarify the significance of this result, we should first recall that while the function $\t{\Sigma}_{\sigma}^{\X{(4)}}(z) + \delta\t{\Sigma}_{\sigma}^{\X{(4)}}(z)$ takes account of \textsl{all} $4$th-order proper self-energy diagrams (including both skeleton and non-skeleton ones), the contributions of these diagrams are inappropriately evaluated in terms of the \textsl{interacting} Green function $\t{G}(z)$, instead of the non-interacting one $\t{G}_{\X{0}}(z)$ (more precisely, the Green function $\t{G}_{\epsilon}(z)$, Eq.\,(\ref{e44}), in the limit $\epsilon = 0^+$, \S\,\ref{s4xa}). The contributions of the diagrams corresponding to $\t{\Sigma}_{\sigma}^{\X{(4)}}(z)$ and $\delta\t{\Sigma}_{\sigma}^{\X{(4)}}(z)$ as evaluated in terms of $\t{G}_{\X{0}}(z)$ are deduced on the basis of the following general expression:\,\footnote{See the remark in the paragraph following Eq.\,(\protect\ref{ex0u}) below.}
\begin{equation}\label{ex0s}
\t{\mathsf{S}}^{\X{(\nu)}}(z,U;[G_{\X{0}}]) = U^{\nu} \lim_{U\downarrow 0} \frac{\t{\mathsf{S}}^{\X{(\nu)}}(z,U;[G])}{U^{\nu}},
\end{equation}
where $\t{\mathsf{S}}^{\X{(\nu)}}(z,U;[G])$ represents the functions $\t{\Sigma}_{\sigma}^{\X{(\nu)}}(z)$ and $\delta\t{\Sigma}_{\sigma}^{\X{(\nu)}}(z)$ as evaluated in terms $\t{G}(z)$, and $\t{\mathsf{S}}^{\X{(\nu)}}(z,U;[G_{\X{0}}])$ the same function as evaluated in terms $\t{G}_{\X{0}}(z)$. In this way, following the asymptotic series expansion in Eq.\,(\ref{ex0f}), which identically coincides with the asymptotic series expansion of the same function for $U\to 0$, and Eq.\,(\ref{ex0r}), one has
\begin{equation}\label{ex0t}
\t{\Sigma}_{\sigma}^{\X{(4)}}(z;[G_{\X{0}}]) = -\frac{45 U^4}{128\hbar z^3},\;\;\; \delta\t{\Sigma}_{\sigma}^{\X{(4)}}(z;[G_{\X{0}}]) = +\frac{45 U^4}{128\hbar z^3},
\end{equation}
whereby
\begin{equation}\label{ex0u}
\t{\Sigma}_{\sigma}^{\X{(4)}}(z;[G_{\X{0}}]) + \delta\t{\Sigma}_{\sigma}^{\X{(4)}}(z;[G_{\X{0}}]) \equiv 0,
\end{equation}
in conformity with the expressions in Eqs\,(\ref{e55}) and (\ref{e49a}).

We remark that in the light of the deviation of the leading-order asymptotic term on the RHS Eq.\,(\ref{ex0f}) from the expected result, namely $-9 U^4/(16\hbar z^3)$, Eq.\,(\ref{ex0i}), the limit on the RHS of Eq.\,(\ref{ex0s}) will in general be unbounded for $\nu \ge 6$ with $\t{\mathsf{S}}^{\X{(\nu)}}(z,U;[G])$ representing the total contribution of the $\nu$th-order skeleton self-energy diagrams. That this is not the case for $\nu=4$ is directly related to the fact that the leading-order asymptotic term on the RHS of Eq.\,(\ref{ex2d}) identically coincides with the expected exact result (\emph{cf.} Eq.\,(\ref{e49a})), which rules out the possibility of contributions corresponding to skeleton self-energy diagrams of order $\nu \ge 4$ contributing to $\Sigma_{\sigma;\infty_1}^{\X{(2)}}$ (and also $\Sigma_{\sigma;\infty_2}^{\X{(2)}}$) in the process of the perturbational series expansion of $\t{G}(z)$ in terms of $\t{G}_{\X{0}}(z)$.\footnote{That is, the expansion of the $\protect\t{G}(z)$ in terms of which the contributions of the relevant skeleton self-energy diagrams have been calculated.} \hfill$\square$

\refstepcounter{dummyX}
\section{List of acronyms and mathematical symbols (not exhaustive)}
\phantomsection
\label{sae}

\scriptsize{
\label{sax}
\begin{longtable}{ll}
\ae & almost everywhere (\emph{i.e.} excluding subsets of measure zero of such sets as $\mathds{R}$, $\mathds{C}$,\\
{} & and the $\protect\1BZ$ corresponding to a \textsl{macroscopic} Bravais lattice)\\
\textrm{cc} & Complex conjugate (\emph{e.g.} $f + \textrm{cc} \equiv f + f^*)$\\
DMFT & Dynamical mean-field theory \\
ESs & Ensemble of states; in this publication, the grand-canonical \textsl{thermal} equilibrium ones\\
GS / GSs & Ground state, Ground-state / Ground states \\
\textsc{h} / \textsc{f} / \textsc{hf} & As superscripts, Hartree / Fock / Hartree-Fock \\
`Hubbard atom' & Unless explicitly indicated otherwise, the system of spin-$\tfrac{1}{2}$ particles described in\\
{} & terms of the single-band Hubbard Hamiltonian at half-filling in the atomic limit \\
KMS & Kubo-Martin-Schwinger\\
LHS / RHS & Left-hand side / Right-hand side\\
p-h & Particle-hole (as in \textsl{particle-hole symmetric})\\
PFD & Partial-fraction decomposition \\
RPA & Random-phase approximation\\
1PI & One-particle irreducible (diagrams representing $\Sigma$ are 1PI, those representing $\Sigma^{\star}$\\
{} & are not in general)\\
2PI & Two-particle irreducible ($G$-skeleton)\\ \\
App. & Appendix\\
Ch. & Chapter\\
p. / pp. & Page / Pages\\
Ref. / Refs & Reference / References \\
\S\, / \S\S & Section / Sections\\ \\
$\mathds{C}$ & Set of complex numbers, the complex plane\\
$\mathds{N}$ & Set of positive integers, $\{1,2,3, \dots\}$\\
$\mathds{N}_{0}$ & Set of non-negative integers, $\mathds{N} \cup \{0\} = \{0,1,2,\dots\}$\\
$\mathds{R}$ & Set of real numbers\\
$\mathds{Z}$ & Set of negative, zero, and positive integers, $\{\dots,-2,-1,0,1,2,\dots\}$\\
$\1BZ$ & The first Brillouin zone in the reciprocal space\\ \\
$\mathrm{e}$ & $\ln^{-1}(1) = 2.718\dots$\\
$\protect\ii$ & $\sqrt{-1}$, the imaginary unit, to be distinguished from the integer-valued index $i$ \\ \\
$\llbracket a\rrbracket $ & The dimensionality of the quantity $a$ in the SI base units. \emph{E.g.}, $\llbracket \hbar \rrbracket = \mathrm{Js}$,\\
{} & Joule-second, $\llbracket\t{G}_{\sigma}(\bm{k};z) \rrbracket = \mathrm{s}$, second, and $\llbracket\t{\Sigma}_{\sigma}(\bm{k};z) \rrbracket = \mathrm{s}^{-1}$, inverse-second \\
$\| \bm{a} \|$ & $\|\bm{a}\|_{\X{2}}$, the norm of the vector $\bm{a}$ in $\mathds{R}^d$\\
$\wedge$ & Logical \textsl{and}; $p_1 \wedge p_2$ is true only if both propositions $p_1$ and $p_2$ are true\\
$\forall$ & For all\\
$a \uparrow b$ ($a \downarrow b$) & The real quantity $a$ approaching the real quantity $b$ from below (above)\\
$A\backslash B$ & The subset of set $A$ from which the elements of set $B$ have been removed. \emph{E.g.},\\
${}$ & $\mathds{N} = \mathds{N}_0\backslash \{0\}$, or, more concisely, $\mathds{N} = \mathds{N}_0\backslash 0$\\
$\protect\b{i}$ & $-i$ (\emph{e.g.} $\protect\b{1} \equiv -1$)\\
$o$ & Order symbol: $\upphi(z) = o(\uppsi(z))$ signifies that to leading order $\upphi(z)/\uppsi(z) \sim 0$\\
{} & in the relevant asymptotic region, in this publication the region $z\to\infty$ \\
$O$ & Order symbol: $\upphi(z) = O(\uppsi(z))$ signifies that $\upphi(z)/\uppsi(z)$ is bounded for all\\
{} & relevant values of $z$\\
$[p/q]$ & Denotes the $[p/q]$ Pad\'{e} approximant of a function, where $p, q \in \mathds{N}_0$ [\S\,\ref{sec.5}]\\
$\mathrm{sgn}$ & The sign function: $\mathrm{sgn}(x) = \pm 1$ for $x \gtrless 0$\\
$\doteq$ & Equality \textsl{by definition}\\
$\sim$ & Symbolises asymptotic relation: $\upphi(z) \sim \uppsi(z)$ signifies that $\upphi(z)/\uppsi(z)$ is to leading \\
{} & order equal to $1$ is the asymptotic region, in this publication the region $z\to\infty$\\
$\lfloor{\,}\rfloor$ & The floor function: $\lfloor x\rfloor =$ the greatest integer less than or equal to $x$. \emph{E.g.},\\ {} & $\lfloor 1.3\rfloor = 1$ \\
$\lceil{\,}\rceil$ & The ceiling function: $\lceil x\rceil =$ the smallest integer greater than or equal to $x$.\\
{} & \emph{E.g.}, $\lceil 1.3\rceil = 2$\\
$[\phantom{.},\phantom{.}]_{\mp}$ & Commutation / Anti-commutation: $[a,b]_{\mp} = a b \mp b a$\\ \\
$\binom{n}{k}$ & The binomial coefficient $\frac{n!}{(n-k)!\hspace{0.4pt} k!}$\\
$\sqrt[x]{y}$ & $y^{1/x}$ \\ \\
$\langle i', j'\rangle$ & A \textsl{directed solid line} (in the direction of $i'$ to $j'$) connecting the vertices marked\\
{} & by $i'$ and $j'$, representing the one-particle Green function $G(j',i')$ in a Feynman\\
{} & diagram. The primes are \textsl{marks} appended to the integers $i$ and $j$ (as in \emph{e.g.} $2'$),\\
{} & to be transferred to the variables with which they are associated; with $i = \bm{r}_i,t_i.\sigma_i$,\\
{} & one thus has $i' = \bm{r}_i',t_i',\sigma_i'$\\
$(i',j')$ & An \textsl{undirected} \textsl{wavy line} connecting the vertices marked as $i'$ and $j'$, representing\\
{} & the two-body interaction potential $v(i',j')$, in a Feynman diagram\\
$\prec\hspace{-1.4pt} i',j'\hspace{-1.6pt}\succ$ & A \textsl{directed wavy line} (in the direction of $i'$ to $j'$) connecting the vertices marked \\
{} & by $i'$ and $j'$ in a Feynman diagram \\ \\
$\h{0}$ & The Fock-space zero, $0 \times \h{1}$ \\
$\h{1}$ & The identity operator in the Fock space \\
$\bbalpha_j$ & Matrix representation of the $j$th point-group operation of the space group of the\\
{} & Bravais lattice $\{\bm{R}_i\| i\}$\\
$\h{a}_{\bm{k};\sigma}^{\dag}$ $\,/\,$ $\h{a}_{\bm{k};\sigma}^{\phantom{\dag}}$ & Canonical creation / annihilation operator in the $\bm{k}$-space (in the Schr\"{o}dinger picture)\\
$A_{\sigma}(\bm{k};\varepsilon)$ & The single-particle spectral function corresponding to $\t{G}_{\sigma}(\bm{k};z)$. In the local / atomic\\
{} & limit, $\bm{k}$ is suppressed. For the `Hubbard atom', also $\sigma$ is generally suppressed\\
$\beta$ & `Inverse temperature' $1/(k_{\textsc{b}} T)$, where $k_{\textsc{b}} =$ constant of Boltzmann, $T=$ absolute\\
{} & temperature. Not to be confused with $\upbeta$\\
$\h{c}_{i;\sigma}^{\dag}$ $\,/\,$ $\h{c}_{i;\sigma}^{\phantom{\dag}}$ & Canonical site creation / annihilation operator (in the Schr\"{o}dinger picture), with\\
{} & $i=$ site index, \emph{i.e.} the index of the lattice site $\bm{R}_i$\\
$\mathfrak{C}$, $\mathfrak{D}$ & Two types of symmetric Jacobi matrices. The classical moment problems associated\\
{} & with $\mathfrak{C}$- ($\mathfrak{D}$-) type Jacobi matrices are \textsl{indeterminate} (\textsl{determinate})\\
$d$ & Dimension of the spatial space (a subspace of the Euclidean space $\mathds{R}^d$) into which\\
{} & the system is confined; mostly encountered in $\mathrm{d}^dr$ to denote the integration measure\\
{} & in $\mathds{R}^d$\\
$\h{D}_i$ & Site-double-occupancy operator for spin-$\tfrac{1}{2}$ particles: $\h{n}_{i;\uparrow} \h{n}_{i;\downarrow}$\\
$\h{D}$ & Total double-occupancy operator $\sum_{i=1}^{N_{\textsc{s}}} \h{D}_i$\\
$D_i$ & Site-double-occupancy: $\langle \h{D}_i\rangle = \mathrm{Tr}\{\h{\varrho}_{\textsc{g}}\hspace{0.4pt} \h{D}_i\}$, where $\mathrm{Tr}$ = trace over the states of the\\
{} & relevant Fock space\\
$\mathcal{D}$ & Site-double-occupancy $D_i$ corresponding to uniform thermal ensemble of states\\
$\EuScript{D}(\varepsilon)$ & Spin-resolved density of the single-particle states, \emph{i.e.} short-hand for $\EuScript{D}_{\sigma}(\varepsilon)$ when\\
{} & $\EuScript{D}_{\uparrow}(\varepsilon) \equiv \EuScript{D}_{\downarrow}(\varepsilon)$\\
$\epsilon$ & A dimensionless real parameter varying between $0$ and $1$\\
$\varepsilon$ & Real energy parameter \\
$\varepsilon_{\textsc{f}}$ & Fermi energy \\
$\varepsilon_{\bm{k}}$ & The non-interacting one-particle energy dispersion\\
$\varepsilon_{\bm{k};\sigma}^{\textsc{x}}$ & The one-particle energy dispersion $\varepsilon_{\bm{k}} + \hbar \Sigma_{\sigma}^{\textsc{x}}(\bm{k})$, where $\textsc{x} = \textsc{h}$, $\textsc{f}$, $\textsc{hf}$\\
$\t{E}_{\sigma}(\bm{k};z)$, $\t{E}_{\sigma}$ & $\varepsilon_{\bm{k}} + \hbar\t{\Sigma}_{\sigma}(\bm{k};z)$\\
$\t{\mathcal{E}}_{\sigma}(\bm{k};z)$, $\t{\mathcal{E}}_{\sigma}$ & $\hbar (\t{\Sigma}_{\sigma}(\bm{k};z)- \Sigma_{\sigma}^{\textsc{hf}}(\bm{k}))$\\
$\h{\upepsilon}(z)$, $\h{\upepsilon}$ & The single-particle operator associated with the dielectric function\\
$e_s$ & The (neutral-) excitation energy $E_{N;s}- E_{N;0} \ge 0$ \\
$E_{N;0}$ & The $N$-particle ground-state energy, that is the eigenvalue of $\h{H}$ corresponding\\
{} & to $\vert\Psi_{N;0}\rangle$\\
$E_{M;s}$ & The eigenvalue of $\h{H}$ corresponding to the $M$-particle eigenstate $\vert\Psi_{M;s}\rangle$\\
$\upeta_{\sigma}(\bm{k};u)$ & An alternative measure function associated with $\t{G}_{\sigma}(\bm{k};z)$; $u$ is dimensionless\\
$\t{f}(z)$ & $-\t{f}(z)$ is a Nevanlinna function of $z$, with $\t{f}(z)$ variously identified with $\t{G}_{\sigma}(\bm{k};z)$\\
{} & and $\t{\Sigma}_{\sigma}(\bm{k};z)$\\
$\t{f}^{\X{(n)}}(z)$ & The $n$th-order approximant of $\t{f}(z)$ [Eqs\,(\protect\ref{e5ka}), (\protect\ref{e5q}), and (\protect\ref{e5tc})]\\
$\vert\Phi_{N;0}\rangle$ & The non-interacting counterpart of $\vert\Psi_{N;0}\rangle$\\
$G_{\sigma}(\bm{k};\varepsilon)$ & The `physical' zero-temperature one-particle Green function ($\varepsilon \in\mathds{R}$) corresponding\\
{} & to particles with spin index $\sigma$. In the local / atomic limit, $\bm{k}$ is suppressed. For the\\
{} & `Hubbard atom', also $\sigma$ is generally suppressed\\
$G_{\protect\X{0};\sigma}(\bm{k};\varepsilon)$ & The non-interacting counterpart of $G_{\sigma}(\bm{k};\varepsilon)$\\
$\mathbb{G}$ & For spin-$\mathsf{s}$ particles, the $(2\mathsf{s}+1)\times (2\mathsf{s}+1)$ matrix of which $G_{\sigma} \equiv G_{\sigma,\sigma}$, where\\
{} & $G_{\sigma,\sigma'} \equiv (\mathbb{G})_{\sigma,\sigma'}$, is a diagonal element [\S\,\protect\ref{sx1}]\\
$\mathbb{G}_{\X{0}}$ & Similar to $\mathbb{G}$, except that $G_{\X{0};\sigma} \equiv G_{\X{0};\sigma,\sigma}$, where $G_{\X{0};\sigma,\sigma'} \equiv (\mathbb{G}_{\X{0}})_{\sigma,\sigma'}$\\
$\t{G}_{\sigma}(\bm{k};z)$, $\t{G}_{\sigma}$ & The analytic continuation of $G_{\sigma}(\bm{k};\varepsilon)$ into the complex $z$-plane\\
$\h{G}_{\sigma}(z)$, $\h{G}_{\sigma}$ & The single-particle operator associated with the one-particle Green function\\
$\t{G}_{\protect\X{0};z}(\bm{k};\varepsilon)$ & The non-interacting counterpart of $\t{G}_{\sigma}(\bm{k};z)$\\
$G_{\epsilon}(\varepsilon)$ & The physical one-particle Green function corresponding to the `Hubbard atom',\\
{} & with $\epsilon U$ denoting the `on-site' interaction energy; thus $\epsilon = 0$ ($1$) specifies the\\
{} & non-interacting (interacting) `Hubbard atom'\\
$\t{G}_{\epsilon}(z)$ & The analytic continuation of $G_{\epsilon}(\varepsilon)$ into the $z$-plane\\
$\t{\mathscr{G}}_{\sigma}(\bm{k};z)$ & The thermal one-particle Green function within the Matsubara formalism; the\\
{} & counterpart of $\t{G}_{\sigma}(\bm{k};z)$\\
$\t{\mathscr{G}}_{\protect\X{0};\sigma}(\bm{k};z)$ & The non-interacting counterpart of $\t{\mathscr{G}}_{\sigma}(\bm{k};z)$\\
$\t{\mathscr{G}}_{\epsilon}(z)$ & The thermal counterpart of $\t{G}_{\epsilon}(z)$\\
$\mathpzc{G}_{\epsilon}(\tau)$ & The counterpart of $\t{\mathscr{G}}_{\epsilon}(z)$ in the imaginary-time domain, with $-\hbar\beta \le \tau\le \hbar\beta$\\
$\mathpzc{G}_{\epsilon}'(\tau)$ & $\mathpzc{G}_{\epsilon}\big(\Lambda_{\hbar\beta}(\tau)\big)$, the periodic extension of $\mathpzc{G}_{\epsilon}(\tau)$ over the entire $\tau$-axis\\
$G_{\sigma;\infty_j}(\bm{k})$ & The coefficient of $1/z^j$ in the asymptotic series expansion of $\t{G}_{\sigma}(\bm{k};z)$ for $z\to\infty$\\
$\t{\mathfrak{G}}_{\sigma}(\bm{k};z,\varepsilon)$ & An auxiliary function underlying a \textsl{moment-constant method} of summation of\\
{} & the perturbation series for $\t{G}_{\sigma}(\bm{k};z)$ [Eq.\,(\protect\ref{e101}) -- \S\,\protect\ref{sec.5}]\\
$\upgamma_{\sigma}(\bm{k};\varepsilon)$ & The measure function associated with $\t{G}_{\sigma}(\bm{k};z)$, denoted by $\upgamma_{\sigma}(\varepsilon)$ or $\upgamma(\varepsilon)$ in the\\
{}& atomic / local limit\\
$\h{H}$ & A general many-body Hamiltonian in the second-quantisation representation\\
{} & [Eqs\,(\protect\ref{ea34k}) and (\protect\ref{ex01})]\\
$\h{\mathscr{H}}$ & Similar to $\h{H}$, however corresponding to a more general two-body interaction\\
{} & potential [Eq.\,(\protect\ref{ex01a})]\\
$\EuScript{H}$ & A many-body Hamiltonian more general than $\h{H}$ and less general than $\h{\mathscr{H}}$,\\
{} & defined in terms of the two-body potential in Eq.\,(\protect\ref{ex01h})\\
$\h{H}_{\protect\X{0}}$ & The `non-interacting' counterpart of $\h{H}$; it may take account of some mean-field\\
{} & effects\\
$\h{\mathcal{H}}$, $\hspace{0.28cm}\h{\hspace{-0.28cm}\mathpzc{H}}$, $\h{\mathsf{H}}$ & Three different but equivalent representations of the single-band Hubbard \\
{} & Hamiltonian for spin-$\tfrac{1}{2}$ particles. The three are however \textsl{not} equivalent within\\
{} & the framework of the many-body perturbation theory, with the understanding\\
{} & that this inequivalence is a conceptual, and not a quantitative, one as regards\\
{} & $\hspace{0.28cm}\h{\hspace{-0.28cm}\mathpzc{H}}$ and $\h{\mathsf{H}}$ [Eqs\,(\protect\ref{ex01bx}), (\protect\ref{ex01f}), and (\protect\ref{ex01k})]\\
$\h{\mathcal{H}}_{\X{0}}$ & The non-interacting part of $\h{\mathcal{H}}$, which is shared by $\hspace{0.28cm}\h{\hspace{-0.28cm}\mathpzc{H}}$ and $\h{\mathsf{H}}$\\
$\h{\EuScript{H}}^{\textsc{x}}$ & The many-body Hamiltonian $\sum_{\bm{k},\sigma} \hbar\Sigma_{\sigma}^{\textsc{x}}(\bm{k})\hspace{0.6pt} \h{a}_{\bm{k};\sigma}^{\dag} \h{a}_{\bm{k};\sigma}^{\phantom{\dag}}$, where $\textsc{x} = \textsc{h}, \textsc{hf}$ [\S\,\protect\ref{sec.3a.1}]\\
$\mathbb{H}_m$ & The $(m+1) \times (m+1)$ Hankel moment matrix (real and symmetric) \\
$\mathbb{H}$ & The Hankel moment matrix $\mathbb{H}_m$ with $m=\infty$\\
$\mathbb{H}_m^{\textsc{x}}$, $\mathbb{H}^{\textsc{x}}$ & The Hankel moment matrices corresponding to function $\textsc{x}$, $\textsc{x} = \t{G}_{\sigma}$, $\t{\Sigma}_{\sigma}$, \emph{etc.}\\
$\h{I}$ & The identity operator in the one-particle Hilbert space\\
$\mathscr{J}$ & The Jacobi matrix (real and symmetric)\\
$\mathscr{J}_p$ & The $p$th abbreviated Jacobi matrix in relation to $\mathscr{J}$, obtained by removing\\
${}$ & the first $p$ rows and $p$ columns of $\mathscr{J}$ \\
$\mathscr{J}^{\textsc{x}}$ & The Jacobi matrix corresponding to function $\textsc{x}$, $\textsc{x} = \t{G}_{\sigma}$, $\t{\Sigma}_{\sigma}$, \emph{etc.} \\
$\bm{k}$, $\bm{p}$, $\bm{q}$ & Wave vectors in $\mathds{R}^d$ or subsets herein conform the box boundary condition\\
$\h{\mathcal{K}}$ & The thermodynamic Hamiltonian, $\h{\mathcal{H}} -\mu\h{N}$\\
$\mathsf{K}_n(z)$ & The circle (and commonly the region it encloses) on the complex $w$-plane\\
{} & traversed by the mapping $\t{w}_n(z,\ptau)$ on the parameter $\ptau$ traversing the real\\
{} & axis from $-\infty$ to $+\infty$\\
$\EuScript{K}_{j=1}^n(a_j/b_j)$ & An $n$th-order continued fraction. \emph{E.g.}, $\EuScript{K}_{j=1}^3(a_j/b_j) = a_1/(b_1 + a_2/(b_2 + a_3/b_3))$ \\
$\vec{\bm{\xi}}$ & A $(2\nu-1)$-vector characterising a $\nu$th-order proper self-energy diagram,\\
{} & referred to as `diagram A' in symbolic computation [\S\,\ref{sd3}]\\
$\h{\chi}$ & The single-particle operator associated with the density-density correlation function\\
$\chi$ & The density-density correlation function; the same as the \textsl{improper} polarisation\\
{} & function $P^{\star}$. One has $\chi = \sum_{\sigma,\sigma'} \chi_{\sigma,\sigma'}$\\
$\chi^{\X{(0)}}$ & The zeroth-order density-density correlation function. One has  $\chi^{\X{(0)}} = \sum_{\sigma} \chi_{\sigma,\sigma}^{\X{(0)}}$\\
$\t{\b{\chi}}(\bm{k};z)$ & Density-density correlation function in the Fourier space; bar (tilde) has bearing\\
{} & on $\bm{k}$ ($z$)\\
$\b{\chi}_{\infty_j}(\bm{k})$ & The coefficient of $1/z^j$ in the asymptotic series expansion of $\t{\b{\chi}}(\bm{k};z)$ for $z\to\infty$;\\
{} & one has $\b{\chi}_{\infty_{2j+1}}(\bm{k}) \equiv 0$ for $j\in\mathds{N}_0$\\
$\chi_{\sigma,\sigma'}$ & The spin-dependent density-density correlation function\\
$\chi_{\sigma,\sigma'}^{\X{(0)}}$ & The zeroth-order density-density correlation function, identical to $P_{\sigma,\sigma'}^{\X{(0)}}$. One\\
{} & has $\chi_{\sigma,\sigma'}^{\X{(0)}} = \chi_{\sigma,\sigma}^{\X{(0)}}\hspace{0.6pt} \delta_{\sigma,\sigma'} \equiv \chi_{\sigma}^{\X{(0)}}\hspace{0.6pt} \delta_{\sigma,\sigma'}$\\
$\t{\mathfrak{X}}(\bm{k};z,\varepsilon)$ & An auxiliary function underlying a \textsl{moment-constant method} of summation of\\
{} & the perturbation series for $\t{\b{\chi}}(\bm{k};z)$ [footnote \raisebox{-1.0ex}{\normalsize{\protect\footref{notei1}}} on p.\,\protect\pageref{SimilarToChi} -- App. \protect\ref{sa}]\\
$\ell$ & A non-negative integer signifying the number of (fermion) loops in a Feynman\\
{} & diagram\\
$\ell_{\textsc{x}}$ & The number of (fermion) loops in `diagram \textsc{x}', where $\textsc{x} = \textsc{a}, \textsc{b}$ [\S\,\protect\ref{sd3}]\\
$\lambda$ & Dimensionless coupling-constant of the two-body interaction potential, with\\
{} & $\lambda=1$ corresponding to full interaction\\
$\Lambda_a(x)$ & $\Lambda_{a}(x) \doteq x - 2 a (\lfloor (x-a)/(2a)\rfloor +1)$, to be used for the purpose of periodic\\ {} & extension of functions (such as $\mathpzc{G}_{\epsilon}(\tau)$, leading to $\mathpzc{G}_{\epsilon}(\Lambda_{\hbar\beta}(\tau)) \equiv \mathpzc{G}_{\epsilon}'(\tau)$) to the\\
{} & entire $\tau$-axis\\
$\h{L}$ $\,/\,$ $\h{\mathcal{L}}$ & The Liouville super-operator associated with $\h{H}\hspace{1.2pt}/\hspace{1.2pt}\h{\mathcal{H}}$\\
$\mathsf{m}$ & The bare particle mass\\
$\mathbb{M}_{\textsc{l}}$ & A $2\nu \times (3\nu-1)$ integer matrix characterising a $\nu$th-order proper self-energy\\
{} & diagram, referred to as `diagram B' in symbolic computation  [\S\,\ref{sd3}]\\
$\mathbb{M}_{\textsc{s}}$ & A $\nu \times (2\nu-1)$ integer matrix characterising a $\nu$th-order proper self-energy\\
{} & diagram, referred to as `diagram B' in symbolic computation  [\S\,\ref{sd3}]\\
$\mu$ & The chemical potential corresponding to $\b{N} = N$, where $\b{N}$ stands for the\\
{} & ensemble average of the number of particles \\
$\mu'$ & Shifted chemical potential: $\h{\mathcal{H}} -\mu \h{N} \equiv \h{\mathcal{H}}' -\mu' \h{N}$, where $\h{\mathcal{H}}' \doteq \h{\mathcal{H}} -\kappa \h{N}$ and\\
{} & $\mu' \doteq \mu - \kappa$, $\kappa \in \mathds{R}$ [\S\,\ref{sec.3a.1}]\\
$\h{n}_{\sigma}(\bm{r})$ & Number-density operator $\h{\psi}_{\sigma}^{\dag}(\bm{r}) \h{\psi}_{\sigma}^{\phantom{\dag}}(\bm{r})$ corresponding to particles with spin\\
{} & index $\sigma$\\
$\h{n}(\bm{r})$ & Total number-density operator, $\sum_{\sigma} \h{n}_{\sigma}(\bm{r})$\\
$\h{n}_{i;\sigma}$ & Site occupation-number operator $\h{c}_{i;\sigma}^{\dag} \h{c}_{i;\sigma}^{\phantom{\dag}}$ corresponding to particles with spin\\
{} & index $\sigma$, with $i$ referring to the lattice vector $\bm{R}_i$\\
$\h{n}_{\bm{k};\sigma}$ & Fourier transform of $\h{n}_{\sigma}(\bm{r})$\\
$\mathsf{n}_{\sigma}(\bm{k})$ & Interacting GS momentum-distribution function corresponding to particles\\
{} & with spin index $\sigma$\\
$\mathsf{n}_{\sigma}^{\X{(0)}}(\bm{k})$ & Non-interacting GS momentum-distribution function corresponding to particles\\
{} & with spin index $\sigma$\\
$\h{N}$ & Total-number operator\\
$N$ & Number of particles in the GS $\vert\Psi_{N;0}\rangle$ of the system\\
$N_{\textsc{s}}$ & Number of lattice sites\\
$\nu$ & Generally, the order of the perturbation theory, as in $\t{\Sigma}_{\sigma}^{\X{(\nu)}}(\bm{k};z)$\\
$\omega_m$ $\,/\,$ $\nu_m$ & Fermionic / Bosonic Matsubara frequency; $\omega_m = (2m +1) \pi/(\hbar\beta)$, $\nu_m = 2m \pi/(\hbar\beta)$,\\
{} & where $m\in \mathds{Z}$\\
$\Omega$ & `Volume' of the system. For a $d$-cube of side length $L$, $\Omega = L^d$\\
$\b{\Omega}$ & The space in $\mathds{R}^d$ into which the system with `volume' $\Omega$ is confined\\
$p$ & Energy-momentum $(\varepsilon,\bm{p})$; may represent $(\varepsilon,\hbar\bm{k})$ and $(z,\hbar\bm{k})$, where $\varepsilon\in\mathds{R}$, $z\in\mathds{C}$\\
$\h{\mathsf{P}}$ & Total-momentum operator\\
$P$ & The \textsl{proper} polarisation function \\
$P^{\X{(\nu)}}$ & The total contribution of all proper polarisation diagrams of order $\nu$ to $P$\\
$P_{\sigma,\sigma'}$ & The spin-dependent proper polarisation function. One has $P = \sum_{\sigma,\sigma'} P_{\sigma,\sigma'}$ \\
$P^{\star}$ & The \textsl{improper} polarisation function, or the density-density correlation function;\\
{} & the same as $\chi$ \\
$P_{\sigma,\sigma'}^{\star}$ & The spin-dependent improper polarisation function. One has $P^{\star} = \sum_{\sigma,\sigma'} P_{\sigma,\sigma'}^{\star}$ \\
$\t{\mathscr{P}}^{\X{(0)}}$ & The $0$th-order thermal polarisation function; the counterpart of $P^{\X{(0)}}$ \\
$P_j(z)$ & A $j$th-order polynomial in the set $\{P_j(z)\| j\in\mathds{N}_0\}$ of orthogonal polynomials\\
{} & directly corresponding to the positive sequence $\{s_0,s_1,s_2,\dots\}$ determining the\\
{} & Hankel moment matrix $\mathbb{H}_m$, $\forall m \in \mathds{N}_0$ [Eq.\,(\protect\ref{e6c})] \protect\cite{Note19}\\
$\h{\psi}_{\sigma}^{\dag}(\bm{r})$ $\,/\,$ $\h{\psi}_{\sigma}^{\phantom{\dag}}(\bm{r})$  & Creation / annihilation field operator in the Schr\"{o}dinger picture\\
$\vert\Psi_{N;0}\rangle$ & The normalised $N$-particle ground state of the many-body Hamiltonian $\h{H}$\\
{} & corresponding to eigenenergy $E_{N;0}$\\
$\vert\Psi_{M;s}\rangle $ & An $M$-particle normalised eigenstate of $\h{H}$ corresponding to eigenenergy $E_{M;s}$,\\ {} & with $s$ a compound index and $s=0$ by convention signifying the ground state\\
$Q_j(z)$ & A $(j-1)$th-order polynomial from the set $\{Q_j(z)\| j\in \mathds{N}_0\}$ of polynomials \protect\cite{Note19}\\
{} & associated with the set $\{P_j(z)\|j\in\mathds{N}_0\}$ of orthogonal polynomials corresponding\\
{} & to the positive sequence $\{s_0,s_1,s_2,\dots\}$\\
$\bm{R}_i$ & The $i$th vector of the Bravais lattice $\{\bm{R}_i \| i = 1,2,\dots, N_{\textsc{s}}\}$\\
$\h{\varrho}_{\textsc{g}}$ & The grand-canonical statistical operator, $\e^{-\beta \h{\mathcal{K}}}/\mathrm{Tr}[\e^{-\beta \h{\mathcal{K}}}]$\\
$\sigma$ & Spin index, corresponding to the $z$-component of the single-particle spin operator. \\
{} & For spin-$\tfrac{1}{2}$ particles, $\sigma \in \{\uparrow,\downarrow\}$ \\
$\b{\sigma}$ & For spin-$\tfrac{1}{2}$ particles, the complement of $\sigma$; thus for $\sigma =\uparrow (\downarrow)$, $\b{\sigma} = \downarrow (\uparrow)$\\
$\vec{\mathbb{\sigma}}$ & The Cartesian $3$-vector $\{\mathbb{\sigma}^{\mathrm{x}},\mathbb{\sigma}^{\mathrm{y}},\mathbb{\sigma}^{\mathrm{z}}\}$ of the $2\times 2$ Pauli matrices\\
$\upsigma_{\sigma}(\bm{k};\varepsilon)$ & The measure function associated with $\t{\Sigma}_{\sigma}(\bm{k};z)$, denoted by $\upsigma_{\sigma}(\varepsilon)$ or $\upsigma(\varepsilon)$ in the\\
{}& atomic / local limit\\
$\mathsf{s}_j$, $\mathsf{S}_j$ & Two recursively-determined functions of $z, w\in \mathds{C}$. In particular, one has\\
{} & $\t{f}(z)=\mathsf{S}_j(z, \t{f}_j(z)-\alpha_j)$, $\forall j\in \mathds{N}$, and $\t{f}^{\protect\X{(n)}}(z) = \mathsf{S}_n(z,0)$, where $\t{f}_j(z)-\alpha_j = o(1)$\\
{} & for $z\to\infty$. The constant $\alpha_j \in \mathds{R}$ and the function $\t{f}_j(z)$ are also recursively\\
{} & determined. Like $-\t{f}(z)$, $-\t{f}_j(z)$ is a Nevanlinna functions of $z$ [App. \ref{sab}]\\
$\Sigma^{\textsc{h}}(\bm{k})$ & The Hartree self-energy; may also be denoted as $\Sigma_{\sigma}^{\textsc{h}}(\bm{k})$ [\S\,\protect\ref{sec.3a.1}]\\
$\Sigma_{\sigma}^{\textsc{x}}(\bm{k})$ & The Hartree / Fock / Hartree-Fock self-energy for $\textsc{x} = \textsc{h}$ / $\textsc{f}$ / $\textsc{hf}$ [\S\,\protect\ref{sec.3a.1}]\\
$\Sigma_{\sigma}(\bm{k};\varepsilon)$ & The `physical' zero-temperature self-energy ($\varepsilon \in\mathds{R}$) corresponding to particles with\\
{} & spin index $\sigma$. In the local / atomic limit, $\bm{k}$ is suppressed. For the `Hubbard atom',\\
{} & also $\sigma$ is generally suppressed\\
$\mathbb{\Sigma}$ & Similar to $\mathbb{G}$, except that $\Sigma_{\sigma} \equiv \Sigma_{\sigma,\sigma}$, where $\Sigma_{\sigma,\sigma'} \equiv (\mathbb{\Sigma})_{\sigma,\sigma'}$\\
$\mathbb{\Sigma}^{\textsc{x}}$ & Similar to $\mathbb{\Sigma}$, except that $\Sigma_{\sigma}^{\textsc{x}} \equiv \Sigma_{\sigma,\sigma}^{\textsc{x}}$, where $\Sigma_{\sigma,\sigma'}^{\textsc{x}} \equiv (\mathbb{\Sigma}^{\textsc{x}})_{\sigma,\sigma'}$\\
$\t{\Sigma}_{\sigma}(\bm{k};z)$, $\t{\Sigma}_{\sigma}$ & The analytic continuation of $\Sigma_{\sigma}(\bm{k};\varepsilon)$ into the complex $z$-plane\\
$\h{\Sigma}_{\sigma}(z)$, $\h{\Sigma}_{\sigma}$ & The single-particle operator associated with the self-energy\\
$\t{\Sigma}_{\sigma}^{\X{(\nu)}}(\bm{k};z)$ & The total contribution of all skeleton self-energy diagrams of order $\nu$ to $\t{\Sigma}_{\sigma}(\bm{k};z)$\\
{} & in terms of the bare two-body potential $v$\\
$\t{\Sigma}_{\sigma}^{\prime\hspace{0.4pt}\X{(\nu)}}(\bm{k};z)$ & The same as $\t{\Sigma}_{\sigma}^{\X{(\nu)}}(\bm{k};z)$, however in terms of the dynamical screened interaction\\
{} & potential $W$ [App. \ref{sa}]\\
$\t{\Sigma}_{\sigma}^{\X{(\nu.j)}}(\bm{k};z)$ & The contribution of the $j$th $\nu$th-order skeleton self-energy diagram; for \emph{e.g.} $\nu=4$,\\
{} & $j \in \{1,2,\dots, 82\}$\\
$\Sigma_{\sigma;\infty_j}(\bm{k})$ & The coefficient of $1/z^j$ in the asymptotic series expansion of $\t{\Sigma}_{\sigma}(\bm{k};z)$ for $z\to\infty$\\
$\t{\mathfrak{S}}_{\sigma}^{\protect\X{(n)}}(\bm{k};z)$ & Partial sum of the leading $\mathcal{I}(n) +1$ terms of the ordered sequence $\{\t{\Sigma}_{\sigma}^{\X{(\nu)}}(\bm{k};z)\| \nu\}$,\\
{} & where $\mathcal{I}(n) \ge 2n+\varsigma$, with $\varsigma \in \{0,1\}$. The relationship between this function and\\
{} & the exact self-energy $\t{\Sigma}_{\sigma}(\bm{k};z)$ is similar (\textsl{not} identical) to that between $\t{f}(z)$ and\\
{} &  its $n$th-order approximant $\t{f}^{\protect\X{(n)}}(z)$, where $-\t{f}(z)$ is like $-\t{\Sigma}_{\sigma}(\bm{k};z)$ a Nevanlinna\\
{} & function of $z$\\
$\t{\mathscr{S}}_{\sigma}(\bm{k};z)$ & The thermal self-energy; the counterpart of $\t{\Sigma}_{\sigma}(\bm{k};z)$\\
$\t{\mathscr{S}}_{\sigma}^{\X{(\nu)}}(\bm{k};z)$ & The thermal counterpart of $\t{\Sigma}_{\sigma}^{\X{(\nu)}}(\bm{k};z)$\\
$\t{\mathfrak{S}}_{\sigma}'(\bm{k};z,\varepsilon)$ & An auxiliary function underlying a \textsl{moment-constant method} of summation of the\\
{} & perturbation series for $\t{\Sigma}_{\sigma}'(\bm{k};z) \doteq \t{\Sigma}_{\sigma}(\bm{k};z)- \Sigma_{\sigma}^{\textsc{hf}}(\bm{k})$ [footnote \raisebox{-1.0ex}{\normalsize{\protect\footref{notey}}} on p.\,\protect\pageref{InDealingWith} -- \S\,\protect\ref{sec.5}]\\
$\mathcal{T}$ & Fermion time-ordering operator\\
$t$ $\,/\,$ $\tau$ & Real / Imaginary time ($t, \tau \in \mathds{R}$)\\
$\ptau$ & A real parameter varying over $(-\infty,\infty)$; not to be confused with the imaginary\\
{} & time $\tau$ [appendix \protect\ref{sab}; see Eq.\,(\protect\ref{e6j}) \emph{et seq.}]\\
$\uptau(\bm{r})$ & Single-particle kinetic-energy operator. \emph{E.g.}, $\uptau(\bm{r}) = -\tfrac{\hbar^2}{2\mathsf{m}} \nabla_{\bm{r}}^2$, where $\mathsf{m} =$ bare\\
{} & particle mass \\
$\uptau_{\sigma}(\bm{k};u)$ & An alternative measure function associated with $\t{\Sigma}_{\sigma}(\bm{k};z)$; $u$ is dimensionless\\
$\upvartheta(\varepsilon)$ & A measure function that may denote $\upgamma_{\sigma}(\bm{k};\varepsilon)$ or $\upsigma_{\sigma}(\bm{k};\varepsilon)$ in considering the\\
{} & \textsl{orthogonal} polynomials $\{P_j(z)\| j\in \mathds{N}_0\}$; one has $\int_{-\infty}^{\infty} \mathrm{d}\upvartheta(\varepsilon)\, P_i(\varepsilon) P_j(\varepsilon) = 0$ for\\
{} &  $i\not=j$; generally, where $P_0(z) \equiv 1$, $\int_{-\infty}^{\infty} \mathrm{d}\upvartheta(\varepsilon)\, P_0^2(\varepsilon) = s_0$ and $\int_{-\infty}^{\infty} \mathrm{d}\upvartheta(\varepsilon)\, P_j^2(\varepsilon) = 1$,\\ {} & $\forall j \in \mathds{N}$ [\emph{cf.} Eq.\,(\protect\ref{e6ja})], however in this publication, where $P_0(z) \equiv 1/\sqrt{s_0}$,\\
{} & $\int_{-\infty}^{\infty} \mathrm{d}\upvartheta(\varepsilon)\, P_i(\varepsilon) P_j(\varepsilon)= \delta_{i,j}$, $\forall i, j \in \mathds{N}_0$, so that in this publication the set\\
{} & $\{P_j(z)\| j\}$ is \textsl{orthonormal}\\
$\upsilon$ & $\epsilon\hspace{0.6pt}U/2$. For $\epsilon = 0\hspace{0.4pt}(1)$ the function $\t{G}_{\epsilon}(z)$ coincides with the noninteracting\\
{} & (interacting) one-particle Green function of the `Hubbard atom'\\
$U$ & The one-site interaction energy in the Hubbard Hamiltonian \\
$u(\bm{r})$ & Local external potential\\
$v_{\sigma,\sigma'}(\bm{r},\bm{r}')$ & Spin-dependent two-body interaction potential, satisfying $v_{\sigma,\sigma'}(\bm{r},\bm{r}')\equiv v_{\sigma',\sigma}(\bm{r}',\bm{r})$\\
$\b{v}_{\sigma,\sigma'}(\|\bm{k}\|)$ & Fourier transform of the two-body potential $v_{\sigma,\sigma'}$, assumed to be isotropic\\
$\h{v}$ & The single-particle operator associated with the two-body interaction potential $v$\\
$\b{v}(\|\bm{k}\|)$ & Fourier transform of the two-body interaction potential $v$, assumed to be isoropic\\
$\h{W}$ & The single-particle operator associated with the dynamical screened interaction\\
{} & potential\\
$\t{W}(\bm{k};z)$ & The dynamical screened interaction potential in the Fourier space\\
$z$ & Complex energy\\
$\zeta$ & Complex number, which may be dimensionless or have, like $z$, the dimensionality\\
{} & of energy; it may also be a complex function of $z$, \emph{i.e.} $\zeta \equiv\zeta(z)$\\
$\zeta_m$ & $\ii\hbar\omega_m + \mu$, $m\in\mathds{Z}$\\
$\EuScript{Z}$ & Coordination number, \emph{i.e.} the number of the nearest neighbours of a lattice\\
{} & point in a Bravais lattice; for a $d$-cubic lattice, $\EuScript{Z} = 2 d$\\
{} & {\normalsize $\hfill\Box$}
\end{longtable}
}

\end{appendix}

\refstepcounter{dummyX}
\addcontentsline{toc}{section}{References}

\end{document}